%% file: s03_ew.tex
\documentclass[11pt]{report}

\usepackage{Lep2Rep}
\usepackage{gg}
\usepackage{ff}
\usepackage{4f_s03}
\usepackage{smat}
\usepackage{gc}
\usepackage{mw}

\usepackage[english]{babel}
\usepackage{graphicx,rotating}
\usepackage{a4p,here}
\usepackage{cite,mcite,epsfig}

\renewcommand{\Huge}{\huge}
\newcommand{\updates}[1]%
 {\fbox{\parbox{\linewidth}{\textbf{Updates with respect to summer 2002:}\\#1}}}

\input{rotate}

\begin{document}
\flushbottom
\begin{titlepage}
\begin{center}
\Large {EUROPEAN ORGANIZATION FOR NUCLEAR RESEARCH}
\end{center}

\begin{flushright}
       CERN-EP/2003-091 \\
       LEPEWWG/2003-02  \\
       ALEPH   2003-017 PHYSIC 2003-005 \\
       DELPHI  2003-072 PHYS 937 \\
       L3 Note 2825   \\
       OPAL PR 392    \\
       SLD-Physics-Note-78\\
       hep-ex/0312023 \\
       {\bf 4 December 2003}
\end{flushright}

\begin{center}
\boldmath
\Huge {\bf A Combination of Preliminary \\
               Electroweak Measurements and \\
            Constraints on the Standard Model\\[.5cm]

}
\unboldmath

\vspace*{0.5cm}
\Large {\bf
The LEP Collaborations\footnote{The LEP Collaborations each take
responsibility for the preliminary results of their own experiment.}
 ALEPH, DELPHI, L3, OPAL,\\
    the LEP Electroweak Working Group,\footnote{%
WWW access at {\tt http://www.cern.ch/LEPEWWG}

The members of the 
LEP Electroweak Working Group 
who contributed significantly to this
note are: \\
D.~Abbaneo,         %
J.~Alcaraz,         %
G.~Anagnostou,      %
P.~Antilogus,       %
A.~Bajo-Vaquero,    %
P.~Bambade,         %
E.~Barberio,        %
A.~Blondel,         %
D.~Bourilkov,       %
R.~Bruneliere,      %
P.~Checchia,        %
R.~Chierici,        %
R.~Clare,           %
J.~D'Hondt,         %
B.~de~la~Cruz,      %
P.~de~Jong,         %
G.~Della~Ricca,     %
M.~Dierckxsens,     %
D.~Duchesneau,      %
G.~Duckeck,         %
M.~Elsing,          %
C.~Goy,             %
M.W.~Gr\"unewald,   %
A.~Gurtu,           %
J.B.~Hansen,        %
R.~Hawkings,        %
J.~Holt,            %
St.~Jezequel,       %
R.W.L.~Jones,       %
T.~Kawamoto,        %
N.~Kjaer,           %
T.~Kr\"amer,        %
E.~Lan{\c c}on,     %
W.~Liebig,          %
L.~Malgeri,         %
M.~Martinez,        %
F.~Matorras,        %
S.~Mele,            %
E.~Migliore,        %
M.N.~Minard,        %
K.~M\"onig,         %
S.~Natale,          %
R.~Ofierzynski,     %
C.~Parkes,          %
U.~Parzefall,       %
M.~Pepe-Altarelli,  %
B.~Pietrzyk,        %
G.~Quast,           %
P.~Renton,          %
S.~Riemann,         %
K.~Sachs,           %
A.~Straessner,      %
D.~Strom,           %
R.~Tenchini,        %
F.~Teubert,         %
M.A.~Thomson,       %
S.~Todorova-Nova,   %
E.~Tournefier,      %
A.~Valassi,         %
A.~Venturi,         %
H.~Voss,            %
C.P.~Ward,          %
N.K.~Watson,        %
P.S.~Wells,         %
St.~Wynhoff.        %
}\\
the SLD Electroweak and Heavy Flavour Groups\footnote{%
N.~de Groot,        %
P.C.~Rowson,        %
B.~Schumm,          %
D.~Su.              %
}\\
}
\vskip 0.5cm
\large\textbf{Prepared from Contributions of the LEP and SLD
  Experiments \\
to the 2003 Summer Conferences.}\\
\end{center}
\vfill
\begin{abstract}
  This note presents a combination of published and preliminary
  electroweak results from the four LEP collaborations and the SLD
  collaboration which were prepared for the 2003 summer conferences.
  Averages from $\Zzero$ resonance results are derived for hadronic
  and leptonic cross sections, the leptonic forward-backward
  asymmetries, the $\tau$ polarisation asymmetries, the $\bb$ and
  $\cc$ partial widths and forward-backward asymmetries and the $\qq$
  charge asymmetry.  Above the $\Zzero$ resonance, averages are
  derived for di-fermion cross sections and forward-backward
  asymmetries, photon-pair, W-pair, Z-pair, single-W and single-Z
  cross sections, electroweak gauge boson couplings, W mass and width
  and W decay branching ratios.  Also, an investigation of the
  interference of photon and Z-boson exchange is presented, and colour
  reconnection and Bose-Einstein correlation analyses in W-pair
  production are combined.  The main changes with respect to the
  experimental results presented in summer 2002 are updates to the
  mass of the W boson, four-fermion cross sections and gauge
  couplings, all measured at LEP-2, and the LEP heavy-flavour results
  measured at the Z pole.
  
  The results are compared with precise electroweak measurements from
  other experiments, notably the recent final result on the
  electroweak mixing angle determined in neutrino-nucleon scattering
  by the NuTeV collaboration and the new result in atomic parity
  violation in Caesium.  The parameters of the Standard Model are
  evaluated, first using the combined LEP electroweak measurements,
  and then using the full set of electroweak results.
\end{abstract}
\end{titlepage}
\setcounter{page}{3}
\renewcommand{\thefootnote}{\arabic{footnote}}
\setcounter{footnote}{0}

\chapter{Introduction}
\label{sec-Intro}

This paper presents an update of combined results on electroweak
parameters by the four LEP experiments and SLD using published and
preliminary measurements, superseding previous
analyses\cite{bib-EWEP-02}.  Results derived from the $\Zzero$
resonance are based on data recorded until the end of 1995 for the LEP
experiments and 1998 for SLD.
Since 1996 LEP has run at energies above the W-pair production
threshold.  In 2000, the final year of data taking at LEP, the total
delivered luminosity was as high as in 1999; the maximum
centre-of-mass energy attained was close to 209~\GeV\ although most of
the data taken in 2000 was collected at 205 and 207~\GeV.  By the end
of $\LEPII$ operation, a total integrated luminosity of approximately
700\pb\ per experiment has been recorded above the Z resonance.

The $\LEPI$ (1990-1995) $\Zzero$-pole measurements consist of the
hadronic and leptonic cross sections, the leptonic forward-backward
asymmetries, the $\tau$ polarisation asymmetries, the $\bb$ and $\cc$
partial widths and forward-backward asymmetries and the $\qq$ charge
asymmetry.  The measurements of the left-right cross section
asymmetry, the $\bb$ and $\cc$ partial widths and
left-right-forward-backward asymmetries for b and c quarks from SLD
are treated consistently with the LEP data.  Many technical aspects of
their combination are described in References~\citen{LEPLS},
\citen{ref:lephf} and references therein.

The $\LEPII$ (1996-2000) measurements are di-fermion cross sections
and forward-backward asymmetries; di-photon production, W-pair,
Z-pair, single-W and single-Z production cross sections, and
electroweak gauge boson self couplings.  W boson properties, like
mass, width and decay branching ratios are also measured. New studies
on photon/Z interference in fermion-pair production as well as on
colour reconnection and Bose-Einstein correlations in W-pair
production are presented.

Several measurements included in the combinations are still
preliminary. 

This note is organised as follows:
\begin{description}
\item [Chapter~\ref{sec-LS}] $\Zzero$ line shape and leptonic
  forward-backward asymmetries;
\item [Chapter~\ref{sec-TP}] $\tau$ polarisation;
\item [Chapter~\ref{sec-ALR}] Measurement of polarised asymmetries at SLD;
\item [Chapter~\ref{sec-HF}] Heavy flavour analyses;
\item [Chapter~\ref{sec-QFB}] Inclusive hadronic charge asymmetry;
\item [Chapter~\ref{sec-GG}] Photon-pair production at energies above the Z;
\item [Chapter~\ref{sec-FF}] Fermion-pair production at energies above the Z;
\item [Chapter~\ref{sec-smat}] Photon/Z-boson interference;
\item [Chapter~\ref{sec-4F}] W and four-fermion production;
\item [Chapter~\ref{sec-GC}] Electroweak gauge boson self couplings;
\item [Chapter~\ref{sec-CR}] Colour reconnection in W-pair events;
\item [Chapter~\ref{sec-BE}] Bose-Einstein correlations in W-pair events;
\item [Chapter~\ref{sec-MW}] W-boson mass and width;
\item [Chapter~\ref{sec-eff}] Interpretation of the Z-pole results
  in terms of effective couplings of the neutral weak current;
\item [Chapter~\ref{sec-MSM}] Interpretation of all results, also
  including results from neutrino interaction and atomic parity
  violation experiments as well as from CDF and D\O\  in terms of
  constraints on the Standard Model
\item [Chapter~\ref{sec-Conc}] Conclusions including prospects for the future.
\end{description}
To allow a quick assessment, a box highlighting the updates is given
at the beginning of each chapter.

\boldmath
\chapter{$\Zzero$ Lineshape and Lepton Forward-Backward Asymmetries}\label{sec-LS-SM}
\label{sec-LS}
\unboldmath

\updates{ Unchanged w.r.t. summer 2000: All experiments have published
  final results which enter in the combination.  The final combination
  procedure is used, the obtained averages are final. }

\noindent
The results presented here are based on the full \LEPI{} data set.
This includes the data taken during the
energy scans in 1990 and 1991 in the range\footnote{In this note
  $\hbar=c=1$.}  $|\roots-\MZ|<3$~\GeV{}, the data collected at the
$\Zzero$ peak in 1992 and 1994 and the precise energy scans in
1993 and 1995 ($|\roots-\MZ|<1.8$~\GeV{}).  
The total event statistics are given in Table~\ref{tab-LSstat}.
Details of the individual analyses can be found in 
References~\citen{ALEPHLS,DELPHILS,L3LS,OPALLS}. 

\begin{table}[hbtp]
\begin{center}\begin{tabular}{lr} %
\begin{minipage}[b]{0.49\textwidth}
\begin{center}\begin{tabular}{r||rrrr||r}
  \multicolumn{6}{c}{$\qq$}  \\
\hline
   year & A &   D  &   L  &  O  & all \\
\hline
'90/91  & 433 &  357 &  416 &  454 &  1660\\
'92     & 633 &  697 &  678 &  733 &  2741\\
'93     & 630 &  682 &  646 &  649 &  2607\\
'94     &1640 & 1310 & 1359 & 1601 &  5910\\
'95     & 735 &  659 &  526 &  659 &  2579\\
\hline
 total  & 4071 & 3705 & 3625 & 4096 & 15497\\
\end{tabular}\end{center}
\end{minipage}
   &
\begin{minipage}[b]{0.49\textwidth}
\begin{center}\begin{tabular}{r||rrrr||r}
  \multicolumn{6}{c}{$\leptlept$} \\
\hline
   year & A &   D  &   L  &  O  & all \\
\hline
'90/91  &  53 &  36 &  39  &  58  &  186 \\
'92     &  77 &  70 &  59  &  88  &  294 \\
'93     &  78 &  75 &  64  &  79  &  296 \\
'94     & 202 & 137 & 127  & 191  &  657 \\
'95     &  90 &  66 &  54  &  81  &  291 \\
\hline
total   & 500 & 384 & 343  & 497  & 1724 \\
\end{tabular}\end{center}
\end{minipage} \\
\end{tabular} \end{center}
\caption[Recorded event statistics]{
The $\qq$ and $\leptlept$ event statistics, in units of $10^3$, used
for the analysis of the $\Zzero$ line shape and lepton forward-backward
asymmetries by the experiments ALEPH (A), DELPHI (D), L3
(L) and OPAL (O).
}
\label{tab-LSstat}
\end{table}

For the averaging of results the LEP experiments provide a standard
set of 9 parameters describing the information contained in hadronic
and leptonic cross sections and leptonic forward-backward
asymmetries.  These parameters are
convenient for fitting and averaging since they have small
correlations. They are:
\begin{itemize}
\item The mass $\MZ$ and total width $\GZ$ of the Z boson, where
  the definition is based on the Breit-Wigner denominator
  $(s-\MZ^2+is\GZ/\MZ)$ with $s$-dependent width~\cite{ref:QEDCONV}.
\item The hadronic pole cross section of Z exchange:
\begin{equation}
\shad\equiv{12\pi\over\MZ^2}{\Gee\Ghad\over\GZ^2}\,.
\end{equation}
Here $\Gee$ and $\Ghad$ are the partial widths of the $\Zzero$ for
decays into electrons and hadrons.

\item The ratios:
\begin{equation}\label{eqn-sighad}
 \Ree\equiv\Ghad/\Gee, \;\; \Rmu\equiv\Ghad/\Gmumu \;\mbox{and}\;
\Rtau\equiv\Ghad/\Gtautau.
\end{equation}
Here $\Gmumu$ and $\Gtautau$ are the partial widths of the $\Zzero$
for the decays $\Ztomumu$ and $\Ztotautau$.  Due to the mass of
the $\tau$ lepton, a difference of 0.2\% is expected between the
values for $\Ree$ and $\Rmu$, and the value for $\Rtau$, even under
the assumption of lepton universality~\cite{ref:consoli}.
\item The pole asymmetries, $\Afbze$, $\Afbzm$ and $\Afbzt$, for the
  processes $\eeee$, $\eemumu$ and $\eetautau$. In terms of the real
  parts of the effective vector and axial-vector neutral current
  couplings of fermions, $\gvf$ and $\gaf$, the pole asymmetries
  are expressed as
\begin{equation}
\label{eqn-apol}
\Afbzf \equiv {3\over 4} \cAe\cAf
\end{equation}
with
\begin{equation}
\label{eqn-cAf}
\cAf\equiv\frac{2\gvf \gaf} {\gvf^{2}+\gaf^{2}}\ = 2 \frac{\gvf/\gaf} {1+(\gvf/\gaf)^{2}}\,.
\end{equation}
\end{itemize}
The imaginary parts of the vector and axial-vector coupling constants
as well as real and imaginary parts of the photon vacuum polarisation
are taken into account explicitly in the fitting formulae and are fixed to
their Standard Model values.
The fitting procedure takes into account the effects of initial-state
radiation~\cite{ref:QEDCONV} to 
${\cal O}(\alpha^3)$~\cite{ref:Jadach91,ref:Skrzypek92,ref:Montagna96}, as
well as the $t$-channel and  the $s$-$t$ interference contributions in the case 
of $\ee$ final states.

The set of 9 parameters does not describe hadron and
lepton-pair production completely, because it does not include the
interference of the $s$-channel $\Zzero$ exchange with the $s$-channel
$\gamma$ exchange.  For the results presented in this section and used
in the rest of the note, the $\gamma$-exchange contributions and the
hadronic $\gamma\Zzero$ interference terms are fixed to their Standard
Model values.  The leptonic $\gamma\Zzero$ interference terms are
expressed in terms of the effective couplings.

\begin{table}[tp] \begin{center}{\small
\begin {tabular} {lr|@{\,}r@{\,}r@{\,}r@{\,}r@{\,}r@{\,}r@{\,}r@{\,}r@{\,}r}
\hline %
\multicolumn{2}{c|}{~}& \multicolumn{9}{c}{correlations} \\
\multicolumn{2}{c|}{~} & $\MZ$ & $\GZ$ & $\shad$ &
     $\Ree$ &$\Rmu$ & $\Rtau$ & $\Afbze$ & $\Afbzm$ & $\Afbzt$ \\
\hline %
\multicolumn{2}{l}{ $\pzz \chi^2/N_{\rm df}\,=\, 169/176$}& 
                                      \multicolumn{9}{c}{ALEPH} \\
\hline %
 $\MZ$\,[\GeV{}]\hspace*{-.5pc} & 91.1891 $\pm$ 0.0031     &   
  1.00 \\
 $\GZ$\,[\GeV]\hspace*{-2pc}  &  2.4959 $\pm$ 0.0043     & 
  .038 & ~1.00 \\ 
 $\shad$\,[nb]\hspace*{-2pc}  &  41.558 $\pm$ 0.057$\pz$ &   
 $-$.091 & $-$.383 & ~1.00 \\
 $\Ree$        &  20.690 $\pm$ 0.075$\pz$ &   
  .102  &~.004 & ~.134 & ~1.00 \\
 $\Rmu$        &  20.801 $\pm$ 0.056$\pz$ &   
 $-$.003 & ~.012 & ~.167 & ~.083 & ~1.00 \\ 
 $\Rtau$       &  20.708 $\pm$ 0.062$\pz$ &   
 $-$.003 & ~.004 & ~.152 & ~.067 & ~.093 & ~1.00 \\ 
 $\Afbze$      &  0.0184 $\pm$ 0.0034     &   
 $-$.047 & ~.000 & $-$.003 & $-$.388 & ~.000 & ~.000 & ~1.00 \\ 
 $\Afbzm$      &  0.0172 $\pm$ 0.0024     &   
 .072 & ~.002 & ~.002 & ~.019 & ~.013 & ~.000 & $-$.008 & ~1.00 \\ 
 $\Afbzt$      &  0.0170 $\pm$ 0.0028     &   
 .061 & ~.002 & ~.002 & ~.017 & ~.000 & ~.011 & $-$.007 & ~.016 & ~1.00 \\
    ~          & \multicolumn{2}{c}{~}                \\[-0.5pc]
\hline %
\multicolumn{2}{l}{$\pzz \chi^2/N_{\rm df}\,=\, 177/168$} & 
                                        \multicolumn{9}{c}{DELPHI} \\
\hline %
 $\MZ$\,[\GeV{}]\hspace*{-.5pc}   &  91.1864 $\pm$ 0.0028    &
 ~1.00 \\ 
 $\GZ$\,[\GeV]\hspace*{-2pc}   &  2.4876 $\pm$ 0.0041     &
 ~.047 & ~1.00 \\ 
 $\shad$\,[nb]\hspace*{-2pc}   &  41.578 $\pm$ 0.069$\pz$ &
 $-$.070 & $-$.270 & ~1.00 \\ 
 $\Ree$        &  20.88  $\pm$ 0.12$\pzz$ &
 ~.063 & ~.000 & ~.120 & ~1.00 \\ 
 $\Rmu$        &  20.650 $\pm$ 0.076$\pz$ &
 $-$.003 & $-$.007 & ~.191 & ~.054 & ~1.00 \\ 
 $\Rtau$       &  20.84  $\pm$ 0.13$\pzz$ &
 ~.001 & $-$.001 & ~.113 & ~.033 & ~.051 & ~1.00 \\ 
 $\Afbze$      &  0.0171 $\pm$ 0.0049     &
 ~.057 & ~.001 & $-$.006 & $-$.106 & ~.000 & $-$.001 & ~1.00 \\ 
 $\Afbzm$      &  0.0165 $\pm$ 0.0025    &
 ~.064 & ~.006 & $-$.002 & ~.025 & ~.008 & ~.000 & $-$.016 & ~1.00 \\ 
 $\Afbzt$      &  0.0241 $\pm$ 0.0037     & 
 ~.043 & ~.003 & $-$.002 & ~.015 & ~.000 & ~.012 & $-$.015 & ~.014 & ~1.00 \\
    ~          & \multicolumn{2}{c}{~}                \\[-0.5pc]
\hline %
\multicolumn{2}{l}{$\pzz \chi^2/N_{\rm df}\,=\, 158/166 $}  & 
                                    \multicolumn{9}{c}{L3} \\
\hline %
 $\MZ$\,[\GeV{}]\hspace*{-.5pc}   &  91.1897 $\pm$ 0.0030      & 
 ~1.00 \\ 
 $\GZ$\,[\GeV]\hspace*{-2pc}   &   2.5025 $\pm$ 0.0041      & 
 ~.065 & ~1.00 \\ 
 $\shad$\,[nb]\hspace*{-2pc}   &   41.535 $\pm$ 0.054$\pz$  & 
 ~.009 & $-$.343 & ~1.00 \\ 
 $\Ree$        &   20.815  $\pm$ 0.089$\pz$ & 
 ~.108 & $-$.007 & ~.075 & ~1.00 \\ 
 $\Rmu$        &   20.861  $\pm$ 0.097$\pz$ & 
 $-$.001 & ~.002 & ~.077 & ~.030 & ~1.00 \\ 
 $\Rtau$       &   20.79 $\pz\pm$ 0.13$\pzz$& 
 ~.002 & ~.005 & ~.053 & ~.024 & ~.020 & ~1.00 \\ 
 $\Afbze$      &   0.0107 $\pm$ 0.0058      & 
 $-$.045 & ~.055 & $-$.006 & $-$.146 & $-$.001 & $-$.003 & ~1.00 \\ 
 $\Afbzm$      &   0.0188 $\pm$ 0.0033      & 
 ~.052 & ~.004 & ~.005 & ~.017 & ~.005 & ~.000 & ~.011 & ~1.00 \\ 
 $\Afbzt$      &   0.0260 $\pm$ 0.0047      & 
 ~.034 & ~.004 & ~.003 & ~.012 & ~.000 & ~.007 & $-$.008 & ~.006 & ~1.00 \\
    ~          & \multicolumn{2}{c}{~}                \\[-0.5pc]
\hline %
\multicolumn{2}{l}{$\pzz \chi^2/N_{\rm df}\,=\, 155/194 $} & 
                                      \multicolumn{9}{c}{OPAL} \\
\hline %
 $\MZ$\,[\GeV{}]\hspace*{-.5pc}   & 91.1858 $\pm$ 0.0030 &
 ~1.00 \\ 
 $\GZ$\,[\GeV]\hspace*{-2pc}   & 2.4948  $\pm$ 0.0041 & 
 ~.049 & ~1.00 \\ 
 $\shad$\,[nb]\hspace*{-2pc}   & 41.501 $\pm$ 0.055$\pz$ & 
 ~.031 & $-$.352 & ~1.00 \\ 
 $\Ree$        & 20.901 $\pm$ 0.084$\pz$ &
 ~.108 & ~.011 & ~.155 & ~1.00 \\ 
 $\Rmu$        & 20.811 $\pm$ 0.058$\pz$ &
 ~.001 & ~.020 & ~.222 & ~.093 & ~1.00 \\ 
 $\Rtau$       & 20.832 $\pm$ 0.091$\pz$ &
 ~.001 & ~.013 & ~.137 & ~.039 & ~.051 & ~1.00 \\ 
 $\Afbze$      & 0.0089 $\pm$ 0.0045 &
 $-$.053 & $-$.005 & ~.011 & $-$.222 & $-$.001 & ~.005 & ~1.00 \\ 
 $\Afbzm$     & 0.0159 $\pm$ 0.0023 &
 ~.077 & $-$.002 & ~.011 & ~.031 & ~.018 & ~.004 & $-$.012 & ~1.00 \\ 
 $\Afbzt$      & 0.0145 $\pm$ 0.0030 & 
 ~.059 & $-$.003 & ~.003 & ~.015 & $-$.010 & ~.007 & $-$.010 & ~.013 & ~1.00 \\
\hline %
\end{tabular}}%
\caption[Nine parameter results]{\label{tab-ninepar}
Line Shape and asymmetry parameters from fits to the data of the four
LEP experiments and their correlation coefficients. }
\end{center}
\end{table} 

The four sets of nine parameters provided by the LEP experiments are
presented in Table~\ref{tab-ninepar}.  For performing the average over
these four sets of nine parameters, the overall covariance matrix is
constructed from the covariance matrices of the individual LEP
experiments and taking into account common systematic
errors~\cite{LEPLS}.  The common systematic errors include theoretical
errors as well as errors arising from the uncertainty in the LEP beam
energy.  The beam energy uncertainty contributes an uncertainty of
$\pm1.7~\MeV$ to $\MZ$ and $\pm1.2~\MeV$ to $\GZ$. In addition, the
uncertainty in the centre-of-mass energy spread of about $\pm1~\MeV$
contributes $\pm0.2~\MeV$ to $\GZ$.  The theoretical error on
calculations of the small-angle Bhabha cross section is
$\pm$0.054\,\%\cite{bib-lumthopal} for OPAL and
$\pm$0.061\,\%\cite{bib-lumth99} for all other experiments, and
results in the largest common systematic uncertainty on $\shad$.  QED
radiation, dominated by photon radiation from the initial state
electrons, contributes a common uncertainty of $\pm$0.02\,\% on
$\shad$, of $\pm0.3$~\MeV{} on $\MZ$ and of $\pm0.2$~\MeV{} on $\GZ$.
The contribution of $t$-channel diagrams and the $s$-$t$ interference
in $\Zzero\ra\ee$ leads to an additional theoretical uncertainty
estimated to be $\pm0.024$ on $\Ree$ and $\pm0.0014$ on $\Afbze$,
which are fully anti--correlated.  Uncertainties from the
model-independent parameterisation of the energy dependence of the
cross section are almost negligible, if the definitions of
Reference\,\cite{bib-PCP99} are applied. Through unavoidable remaining
Standard Model assumptions, dominated by the need to fix the
$\gamma$-$\Zzero$ interference contribution in the $\qq$ channel,
there is some small dependence of $\pm 0.2$ \MeV{} of $\MZ$ on the
Higgs mass, $\MH$ (in the range 100 \GeV{} to 1000 \GeV{}) and the
value of the electromagnetic coupling constant. Such ``parametric''
errors are negligible for the other results. The combined
parameter set and its correlation matrix are given in
Table~\ref{tab-zparavg}.

\begin{table}[htb]\begin{center}
\begin {tabular} {lr|r@{\,}r@{\,}r@{\,}r@{\,}r@{\,}r@{\,}r@{\,}r@{\,}r}
\hline %
\multicolumn{2}{c|} {without lepton universality} & 
                                    \multicolumn{9}{l}{~~~correlations} \\
\hline %
\multicolumn{2}{c|}{$\pzz \chi^2/N_{\rm df}\,=\,32.6/27 $} &
   $\MZ$ & $\GZ$ & $\shad$ &
     $\Ree$ &$\Rmu$ & $\Rtau$ & $\Afbze$ & $\Afbzm$ & $\Afbzt$ \\
\hline %
 $\MZ$ [\GeV{}]  & 91.1876$\pm$ 0.0021 &
 ~1.00 \\
 $\GZ$ [\GeV]  & 2.4952 $\pm$ 0.0023 &
 $-$.024 & ~1.00 \\ 
 $\shad$ [nb]  & 41.541 $\pm$ 0.037$\pz$ &
 $-$.044 & $-$.297 & ~1.00 \\ 
 $\Ree$        & 20.804 $\pm$ 0.050$\pz$ &
 ~.078 & $-$.011 & ~.105 & ~1.00 \\ 
 $\Rmu$        & 20.785 $\pm$ 0.033$\pz$ & 
 ~.000 & ~.008 & ~.131 & ~.069 & ~1.00 \\ 
 $\Rtau$       & 20.764 $\pm$ 0.045$\pz$ &  
 ~.002 & ~.006 & ~.092 & ~.046 & ~.069 & ~1.00 \\ 
 $\Afbze$      & 0.0145 $\pm$ 0.0025 &
 $-$.014 & ~.007 & ~.001 & $-$.371 & ~.001 & ~.003 & ~1.00 \\ 
 $\Afbzm$      & 0.0169 $\pm$ 0.0013 &
 ~.046 & ~.002 & ~.003 & ~.020 & ~.012 & ~.001 & $-$.024 & ~1.00 \\ 
 $\Afbzt$      & 0.0188 $\pm$ 0.0017 &
 ~.035 & ~.001 & ~.002 & ~.013 & $-$.003 & ~.009 & $-$.020 & ~.046 & ~1.00 \\ 
\multicolumn{3}{c}{~}\\[-0.5pc]
\multicolumn{2}{c} {with lepton universality} \\
\hline %
\multicolumn{2}{c|}{$\pzz \chi^2/N_{\rm df}\,=\,36.5/31 $}  & 
   $\MZ$ & $\GZ$ & $\shad$ & $\Rl$ &$\Afbzl$ \\
\hline %
 $\MZ$ [\GeV{}]  & 91.1875$\pm$ 0.0021$\pz$    &
 ~1.00 \\ 
 $\GZ$ [\GeV]  & 2.4952 $\pm$ 0.0023$\pz$    &
 $-$.023  & ~1.00 \\ 
 $\shad$ [nb]  & 41.540 $\pm$ 0.037$\pzz$ &
 $-$.045 & $-$.297 &  ~1.00 \\ 
 $\Rl$         & 20.767 $\pm$ 0.025$\pzz$ &
 ~.033 & ~.004 & ~.183 & ~1.00 \\ 
 $\Afbzl$      & 0.0171 $\pm$ 0.0010   & 
 ~.055 & ~.003 & ~.006 & $-$.056 &  ~1.00 \\ 
\hline %
\end{tabular} 
\caption[]{
  Average line shape and asymmetry parameters from the data of the
  four LEP experiments,  without and with the
  assumption of lepton universality. }
\label{tab-zparavg}
\end{center}
\end{table}

\begin{figure}[p]
\vspace*{-0.6cm}
\begin{center}
  \mbox{\includegraphics[width=0.9\linewidth]{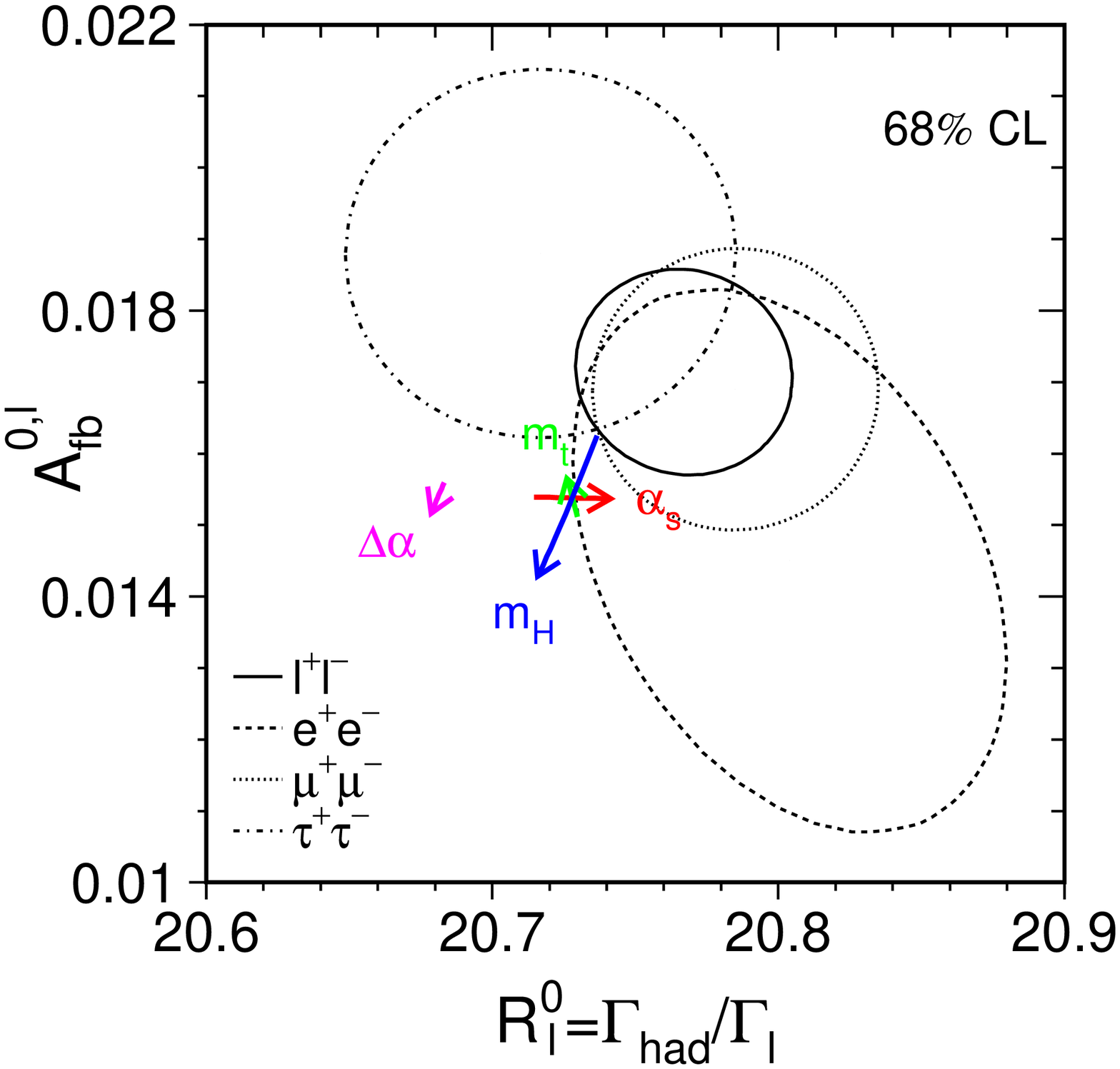}}
\end{center}
\caption[]{
  Contours of 68\% probability in the $\RZ$-$\Afbzl$ plane. For better
  comparison the results for the $\tau$ lepton are corrected to
  correspond to the massless case.  The $\SM$ prediction for
  $\MZ=91.1875$~\GeV{}, $\Mt=174.3$~\GeV{}, $\MH=300$~\GeV{}, and
  $\alfmz=0.118$ is also shown. The lines with arrows correspond to
  the variation of the $\SM$ prediction when $\Mt$, $\MH$, $\alfmz$
  and $\Delta\alpha_{\mathrm{had}}^{(5)}(\MZ^2)$ are varied in the
  intervals $\Mt=174.3\pm5.1$~\GeV{}, $\MH=300^{+700}_{-186}~\GeV$,
  $\alfmz=0.118\pm0.002$ and
  $\Delta\alpha_{\mathrm{had}}^{(5)}(\MZ^2)=0.02761\pm0.00036$,
  respectively. The arrows point in the direction of increasing values
  of $\Mt$, $\MH$, $\alfas$ and
  $\Delta\alpha_{\mathrm{had}}^{(5)}(\MZ^2)$.  }
\label{fig-LU} 
\end{figure}

If lepton universality is assumed, the set of 9 parameters  
is reduced to a set of 5 parameters. $\RZ$ is defined as 
$\RZ\equiv\Ghad/\Gll$, where $\Gll$ refers to the partial 
$\Zzero$ width for the decay into a pair of massless charged 
leptons.  The data of each of the four LEP experiments are 
consistent with lepton universality (the difference in $\chi^2$ 
over the difference in d.o.f.{} with and without the assumption 
of lepton universality is 3/4, 6/4, 5/4 and 3/4 for ALEPH, DELPHI, 
L3 and OPAL, respectively). The lower part of Table~\ref{tab-zparavg} 
gives the combined result and the corresponding correlation matrix.  
Figure~\ref{fig-LU} shows, for each lepton species and for the 
combination assuming lepton universality, the resulting 68\% 
probability contours in the $\RZ$-$\Afbzl$ plane. Good agreement
is observed. 

For completeness the partial decay widths of the $\Zzero$ boson 
are listed in Table~\ref{tab-widths}, although 
they are more correlated than the ratios given in 
Table~\ref{tab-zparavg}. The leptonic pole cross-section, 
$\sll$, defined as
\begin{eqnarray}
\sll & \equiv & {12\pi\over\MZ^2}{\Gll^2\over\GZ^2} \, ,
\end{eqnarray}
in analogy to $\shad$, is shown in the last line of the Table.
Because QCD final state corrections appear twice in the 
denominator via $\GZ$, $\sll$ has a higher sensitivity to $\alpha_s$ 
than $\shad$ or $\Rl$, where the dependence on QCD corrections is 
only linear.

\begin{table}[hbtp] \begin{center} 
\begin{tabular} {lr|r@{\,}r@{\,}r@{\,}r}
\hline %
 \multicolumn{2}{c|}{without lepton universality} &
                            \multicolumn{4}{l}{~~~correlations} \\
 & & $\Ghad$ & $\Gee$ & $\Gmumu$ & $\Gtautau$ \\
\hline %
$\Ghad$ [\MeV]     & 1745.8$\pz\pm$2.7$\pz\pzz$ & ~1.00 \\
$\Gee$ [\MeV]      & 83.92$\pm$0.12$\pzz$& $-$0.29 & ~1.00 \\ 
$\Gmumu$ [\MeV]    & 83.99$\pm$0.18$\pzz$& ~0.66 & $-$0.20 & ~1.00 \\
$\Gtautau$ [\MeV]  & 84.08$\pm$0.22$\pzz$&  0.54 & $-$0.17 & ~0.39 & ~1.00 \\  
\hline %
\multicolumn{6}{c}{~} \\[-0.5pc]
\hline %
 \multicolumn{2}{c|}{with    lepton universality} &
                            \multicolumn{4}{l}{~~~correlations} \\
 & & $\Ginv$ & $\Ghad$ & $\Gll$ &  \\
\hline %
$\Ginv$ [\MeV]     & $\pz$499.0$\pzz\pm$1.5$\pz\pzz$    & ~1.00 \\
$\Ghad$ [\MeV]     & 1744.4$\pzz\pm$2.0$\pz\pzz$   & $-$0.29 & ~1.00 \\
$\Gll$ [\MeV]       & 83.984$\pm$0.086$\pz$  & ~0.49 & ~0.39 & ~1.00 \\
\hline %
$\Ginv/\Gll$        & {$\pz$5.942\pz$\pm$0.016$\pz$} &    \\
\hline %
$\sll$ [nb]      & {2.0003$\pm$0.0027} &  \\
\hline %
\end{tabular} 
\caption[]{
  Partial decay widths of the $\Zzero$ boson, derived from the results of the
  9-parameter averages in Table~\ref{tab-zparavg}. In the
  case of lepton universality, $\Gll$ refers to the partial $\Zzero$ width for
  the decay into a pair of massless charged leptons.  }
\label{tab-widths}
\end{center}
\end{table}
\boldmath
\section{Number of Neutrino Species}
\label{sec-Nnu}
\unboldmath

An important aspect of our measurement concerns the information
related to $\Zzero$ decays into invisible channels. Using the results
of Table~\ref{tab-zparavg}, the ratio of
the $\Zzero$ decay width into invisible particles and the leptonic
decay width is determined:
\begin{eqnarray}
\Ginv / \Gll & = & 5.942\pm 0.016\,.
\end{eqnarray}
The Standard Model value for the ratio of the partial widths to
neutrinos and charged leptons is:
\begin{eqnarray}
(\Gnn / \Gll)_{\mathrm{SM}} & = & 1.9912\pm 0.0012\,.
\end{eqnarray}
The central value is evaluated for $\MZ=91.1875$~\GeV{} 
and the error quoted accounts for
a variation of $\Mt$ in the range $\Mt=174.3\pm5.1~\GeV$ and a
variation of $\MH$ in the range $100~\GeV \le \MH \le 1000~\GeV$.  
The number of light neutrino
species is given by the ratio of the two expressions listed above:
\begin{eqnarray}
\Nnu & = & 2.9841\pm 0.0083,
\end{eqnarray}
which is two standard deviations below the value of 3 expected from 3
observed fermion families.

Alternatively, one can assume 3 neutrino species and determine the
width from additional invisible decays of the Z.  This yields
\begin{eqnarray}
  \Delta\Ginv & = & -2.7 \pm 1.6\ \MeV.
\end{eqnarray}
The measured total width
is below the Standard Model expectation.  
If a conservative approach is taken to limit the result to only
positive values of $\Delta\Ginv$ and   normalising the probability for
$\Delta\Ginv\ge0$ to be unity, then the resulting 95\% CL upper limit on
additional invisible decays of the Z is
\begin{eqnarray}
  \Delta\Ginv & < & 2.0\ \MeV.
\end{eqnarray}

The theoretical error on the luminosity\cite{bib-lumth99} constitutes
a large part of the uncertainties on $\Nnu$ and $\Delta\Ginv$.

\boldmath
\chapter{The $\tau$ Polarisation}
\label{sec-TP}
\unboldmath

\updates{ Unchanged w.r.t. summer 2002: All experiments have published
  final results which enter the combination. The final combination
  procedure is used, the obtained averages are final. }

\noindent
The longitudinal $\tau$ polarisation $\cal {P}_{\tau}$ of $\tau$
pairs produced 
in $\Zzero$ decays is defined as
\begin{eqnarray}
{\cal P}_{\tau} & \equiv & 
\frac{\sigma_{\mathrm{R}} - \sigma_{\mathrm{L}}}
{\sigma_{\mathrm{R}} + \sigma_{\mathrm{L}}} \, ,
\end{eqnarray}
where $\sigma_{\mathrm{R}}$ and $\sigma_{\mathrm{L}}$ are the $\tau$-pair cross sections for
the production of a right-handed and left-handed $\tau^-$,
respectively. The distribution of $\ptau$ as a function of the polar
scattering angle $\theta$ between the $\mathrm{e}^-$ and the $\tau^-$,
at $\roots = \MZ$, is given by
\begin{eqnarray}
\label{eqn-taupol}
{\cal P}_{\tau}(\cos\theta) & = &
 - \frac{\cAt(1+\cos^2\theta) + 2    \cAe\cos\theta}
             {1+\cos^2\theta  + 2\cAt\cAe\cos\theta} \, ,
\end{eqnarray}
with $\cAe$ and $\cAt$ as defined in Equation~(\ref{eqn-cAf}).
Equation~(\ref{eqn-taupol}) is valid for pure Z exchange.  The effects of
$\gamma$ exchange, $\gamma$-$\Zzero$ interference and electromagnetic
radiative corrections in the initial and final states
are taken into account in the experimental analyses.  In
particular, these corrections account for the $\sqrt{s}$ dependence of
the $\tau$ polarisation, which is important because
the off-peak data are included in the event samples for all
experiments.
When averaged over all production angles $\cal {P}_{\tau}$ is a
measurement of $\cAt$. As a function of $\cos\theta$, $\cal
{P}_{\tau}(\cos\theta)$ provides nearly independent determinations of
both $\cAt$ and $\cAe$, thus allowing a test of the universality of
the couplings of the $\Zzero$ to $\mathrm{e}$ and $\tau$.

Each experiment makes separate $\ptau$ measurements using the five
$\tau$ decay modes e$\nu \overline{\nu}$, $\mu\nu \overline{\nu}$,
$\pi\nu$, $\rho\nu$ and
$a_{1}\nu$\cite{bib-ALEPHTAU,bib-DELPHITAUnew,bib-L3TAUfin,bib-OPALTAU}.
The $\rho\nu$ and $\pi\nu$ are the most sensitive channels,
contributing weights of about $40\%$ each in the average.  DELPHI and
L3 also use an inclusive hadronic analysis. The combination is made
using the results from each experiment already averaged over the
$\tau$ decay modes.

\section{Results}

Tables~\ref{tab-tau1} and~\ref{tab-tau2} show the most recent results
for $\cAt$ and $\cAe$ obtained by the four LEP
collaborations\cite{bib-ALEPHTAU,bib-DELPHITAUnew,bib-L3TAUfin,bib-OPALTAU}
and their combination.  Although the sizes of the event samples used by
the four experiments are roughly equal, smaller errors are quoted by
ALEPH. This is largely associated with the higher angular granularity
of the ALEPH electromagnetic calorimeter.  Common systematic errors
arise from uncertainties in radiative corrections (decay radiation) in
the $\pi\nu$ and $\rho\nu$ channels, and in the modelling of the
$a_{1}$ decays\cite{bib-EWPPE187}.  These errors and their
correlations need further investigation, but are already taken into
account in the combination (see also Reference~\citen{bib-L3TAUfin}).
The statistical
correlation between the extracted values of $\cAt$ and $\cAe$ is small
($\le$ 5\%). %

The average values for $\cAt$ and $\cAe$:
\begin{eqnarray}
  \cAt & = & 0.1439 \pm 0.0043 \\
  \cAe & = & 0.1498 \pm 0.0049 \,,
\end{eqnarray}
with a correlation of 0.012, are compatible, in good agreement with
neutral-current lepton universality.  This combination is performed
including the small common systematic errors between $\cAt$ and $\cAe$
within each experiment and between experiments.  Assuming
$\mathrm{e}$-$\tau$ universality, the values for $\cAt$ and $\cAe$ can
be combined.  The combined result of $\cAt$ and $\cAe$ is:
\begin{eqnarray}
  \cAl & = & 0.1465 \pm 0.0033 \,,
\end{eqnarray}
where the error includes a systematic component of 0.0016.

\begin{table}[htbp]
\renewcommand{\arraystretch}{1.15}
\begin{center}
\begin{tabular}{|ll||c|}
\hline
Experiment & & $\cAt$ \\
\hline
\hline
ALEPH  &(90 - 95), final       & $0.1451\pm0.0052\pm0.0029$  \\
DELPHI &(90 - 95), final       & $0.1359\pm0.0079\pm0.0055$  \\
L3     &(90 - 95), final       & $0.1476\pm0.0088\pm0.0062$  \\
OPAL   &(90 - 95), final       & $0.1456\pm0.0076\pm0.0057$  \\
\hline
\hline
LEP Average &  final           & $0.1439\pm0.0035\pm0.0026$  \\
\hline
\end{tabular}
\end{center}
\caption[]{
  LEP results for $\cAt$.  The first error is statistical and the
  second systematic. }
\label{tab-tau1}
\end{table}
\begin{table}[htbp]
\renewcommand{\arraystretch}{1.15}
\begin{center}
\begin{tabular}{|ll||c|}
\hline
Experiment & & $\cAe$ \\
\hline
\hline
ALEPH   &(90 - 95), final       & $0.1504\pm0.0068\pm0.0008$  \\
DELPHI  &(90 - 95), final       & $0.1382\pm0.0116\pm0.0005$  \\
L3      &(90 - 95), final       & $0.1678\pm0.0127\pm0.0030$  \\
OPAL    &(90 - 95), final       & $0.1454\pm0.0108\pm0.0036$  \\
\hline
\hline
LEP Average &  final            & $0.1498\pm0.0048\pm0.0009$  \\
\hline
\end{tabular}
\end{center}
\caption[]{
  LEP results for $\cAe$.  The first error is statistical and the
  second systematic.  }
\label{tab-tau2}
\end{table}

\boldmath
\chapter{Measurement of polarised lepton asymmetries at SLC}
\unboldmath
\label{sec-ALR}

\updates{ Unchanged w.r.t. summer 2000: SLD has published final
  results for \ALR{} and the leptonic left-right forward-backward
  asymmetries.  }

\noindent
The measurement of the left-right cross section asymmetry ($\ALR$) by
SLD\cite{ref:sld-s00} at the SLC provides a systematically precise,
statistics-dominated determination of the coupling $\cAe$, and is
presently the most precise single measurement, with the smallest
systematic error, 
of this quantity.  In principle the analysis is
straightforward: one counts the numbers of Z bosons produced by left
and right longitudinally polarised electrons, forms an asymmetry, and
then divides by the luminosity-weighted e$^-$ beam polarisation
magnitude (the e$^+$ beam is not polarised):
\begin{equation}
  \label{eq:ALR}
  \ALR = \frac{N_{\mathrm{L}} - N_{\mathrm{R}}}%
              {N_{\mathrm{L}} + N_{\mathrm{R}}}%
         \frac{1}{P_{\mathrm{e}}}.
\end{equation}
Since the advent of high polarisation ``strained lattice'' GaAs
photo-cathodes (1994), the average electron
polarisation at the interaction point has been in the range 73\% to 77\%.
The method requires no
detailed final state event identification ($\ee$ final state events
are removed, as are non-Z backgrounds) and is insensitive to all
acceptance and efficiency effects.  The small total systematic error
of  0.64\% relative 
is
dominated by the 0.50\%
relative 
systematic error in the determination of the e$^-$ polarisation.
The 
relative 
statistical error on $\ALR$ is about 1.3\%.

The precision Compton polarimeter detects beam electrons
that are scattered by photons from a circularly polarised laser.
Two additional polarimeters that are sensitive to the Compton-scattered
photons and which are operated in the absence of positron beam, 
have verified the precision polarimeter result and are used to set a
calibration uncertainty of 0.4\% relative.
In 1998, a dedicated experiment was performed in order to test directly
the expectation that accidental polarisation of the positron beam was
negligible; the e$^+$ polarisation was found to be consistent with 
zero ($-0.02\pm 0.07$)\%.

The $\ALR$ analysis includes several very small corrections. The
polarimeter result is corrected for higher order QED and
accelerator related effects, a total of
($-0.22\pm0.15$)\% relative for 1997/98 data. 
The event asymmetry is
corrected for backgrounds and accelerator asymmetries, a total of
($+0.15\pm0.07$)\% relative, for 1997/98 data.

The translation of the $\ALR$ result to a ``pole'' value is a ($-2.5\pm0.4$)\%
relative shift, where the uncertainty arises from the precision of the
centre-of-mass
energy
determination.  
This small error due to the beam energy measurement 
reflects the results
of a scan of the Z peak used to calibrate the energy spectrometers 
to $\MZ$ from LEP data. 
The pole value, $\ALRz$, is equivalent to a measurement
of $\cAe$.

The 2000  result is included in a running average of all
of the SLD $\ALR$ measurements (1992, 1993, 1994/1995, 1996, 1997 and 1998).
This updated result for $\ALRz$ ($\cAe$)
is $0.1514 \pm 0.0022$.
In addition, the left-right forward-backward asymmetries for leptonic
final states are measured\cite{ref:sld-asym}.  From these, the parameters $\cAe$,
$\cAm$ and $\cAt$ can be determined.  The results are
$\cAe = 0.1544 \pm 0.0060$, $\cAm = 0.142 \pm 0.015$ and $\cAt =
0.136 \pm 0.015$. 
The lepton-based result for $\cAe$ can be combined 
with the $\ALRz$ result to yield
$\cAe = 0.1516 \pm 0.0021$, 
including small correlations in the systematic errors.
The correlation of this measurement with  
$\cAm$ and $\cAt$ is 
indicated in Table~\ref{tab:corr-As}.

Assuming lepton universality, the $\ALR$ result and the results on the
leptonic left-right forward-backward asymmetries 
can be combined, while accounting
for small
correlated systematic errors, yielding
\begin{equation}
 \cAl = 0.1513 \pm 0.0021.
\end{equation}

\begin{table}[h]
\centering
\begin{tabular}{c|ccc} 
       & $\cAe$ & $\cAm$ & $\cAt$ \\
\hline
$\cAe$ &  1.000  \\
$\cAm$ &  0.038 & 1.000 \\
$\cAt$ &  0.033 & 0.007  & 1.000 \\
\end{tabular}
\caption{Correlation coefficients  between $\cAe$,
$\cAm$ and $\cAt$}
\label{tab:corr-As}
\end{table}

\boldmath
\chapter{Results from b and c Quarks}
\label{sec-HF}
\unboldmath

\input{s03_hf} %

\boldmath
\chapter{The Hadronic Charge Asymmetry $\avQfb$}
\label{sec-QFB}
\unboldmath

\updates{ Unchanged w.r.t. summer 2002: All experiments have published
  final results which enter the combination. The final combination
  procedure is used, the obtained averages are final.}

\noindent
The LEP experiments ALEPH\cite{ALEPHcharge1996},
DELPHI\cite{DELPHIcharge}, L3\cite{ref:ljet} and OPAL\cite{OPALcharge}
provide measurements of the hadronic charge asymmetry based on
the mean difference in jet charges measured in the forward and
backward event hemispheres, $\avQfb$. DELPHI also provides a
related measurement of the total charge asymmetry by making a charge
assignment on an event-by-event basis and performing a likelihood
fit\cite{DELPHIcharge}.  The experimental values quoted for the
average forward-backward charge difference, $\avQfb$, cannot be
directly compared as some of them include detector dependent effects
such as acceptances and efficiencies.  Therefore the effective
electroweak mixing angle, $\swsqeffl$, as defined in
Section~\ref{sec-SW}, is used as a means of combining the experimental
results summarised in Table~\ref{partab}.

\begin{table}[htb]
\begin{center}
\renewcommand{\arraystretch}{1.1}
\begin{tabular}{|ll||c|}
\hline
Experiment & & $\swsqeffl$ \\
\hline
\hline
ALEPH & (90-94), final & $0.2322\pm0.0008\pm0.0011$ \\
DELPHI& (91-91), final & $0.2345\pm0.0030\pm0.0027$ \\
L3    & (91-95), final & $0.2327\pm0.0012\pm0.0013$ \\
OPAL  & (90-91), final & $0.2326\pm0.0012\pm0.0029$ \\
\hline
\hline
LEP Average  &           & $0.2324\pm0.0012$ \\
\hline
\end{tabular}
\caption[]{
  Summary of the determination of $\swsqeffl$ from inclusive hadronic
  charge asymmetries at LEP. For each experiment, the first error is
  statistical and the second systematic. The latter, amounting to
  0.0010 in the average, is dominated by fragmentation and decay
  modelling uncertainties.  }
\label{partab}
\end{center}
\end{table}

The dominant source of systematic error arises from the modelling of
the charge flow in the fragmentation process for each flavour. All
experiments measure the required charge properties for $\Zzero\ra\bb$
events from the data. ALEPH also determines the charm charge
properties from the data. The fragmentation model implemented in the
JETSET Monte Carlo program\cite{JETSET} is used by all experiments as
reference; the one of the HERWIG Monte Carlo program\cite{HERWIG} is
used for comparison. The JETSET fragmentation parameters are varied to
estimate the systematic errors. The central values chosen by the
experiments for these parameters are, however, not the same. 
The smaller of
the two fragmentation errors in any pair of results is treated as
common to both.  The present average of $\swsqeffl$ from $\avQfb$ and
its associated error are not very sensitive to the treatment of common
uncertainties.
The ambiguities due to QCD corrections may cause changes in the
derived value of $\swsqeffl$. These are, however, well below the
fragmentation uncertainties and experimental errors. The effect of
fully correlating the estimated systematic uncertainties from this
source between the experiments has a negligible effect upon the
average and its error.

There is also some correlation between these results and those for
$\Abb$ using jet charges. The dominant source of correlation is again
through uncertainties in the fragmentation and decay models used. The
typical correlation between the derived values of $\swsqeffl$ from
the $\avQfb$ and the $\Abb$ jet charge measurements is estimated
to be about 20\% to 25\%. This leads to only a small change in the
relative weights for the $\Abb$ and $\avQfb$ results when averaging
their $\swsqeffl$ values (Section~\ref{sec-SW}). 
Thus, the correlation between $\avQfb$ and $\Abb$ from
jet charge has little impact on the overall Standard Model fit,
and is neglected at present.

\boldmath
\chapter{Photon-Pair Production at \LEPII}
\label{sec-GG}
\unboldmath

\updates{ Unchanged w.r.t. summer 2002: ALEPH, L3 and OPAL have
  provided final results for the complete LEP-2 dataset, DELPHI up to
  1999 data and preliminary results for the 2000 data. }

\input{gg}

\boldmath
\chapter{Fermion-Pair Production at \LEPII}
\label{sec-FF}
\unboldmath

\updates{ The measurements of cross sections and asymmetries are
  unchanged. The results derived for 2nd and 3rd generation
  leptoquarks have been corrected. }

\input{ff}

\boldmath
\chapter{Investigation of the Photon/Z-Boson Interference}
\label{sec-smat}
\unboldmath

\updates{ Unchanged w.r.t. summer 2002: Results are preliminary. }

\input{smat}

\boldmath
\chapter{W and Four-Fermion Production at \LEPII}
\label{sec-4F}
\unboldmath

\updates{ The WW cross-section, $\rww$ and the W branching ratios
  combinations are updated including the final \Delphi\ results. The
  determination of $|\mathrm{V}_{\mathrm{cs}}|$
  is updated with new inputs from the PDG 2002 \\
  The ZZ cross-section and $\rzz$ combinations are updated accounting
  for the final \Delphi\,
  \Ltre\ and \Opal\ results. \\
  The Zee cross-section and the corresponding $\rzee$ combinations are
  updated with the
  first ever input by \Aleph\ in the hadronic channel. \\
  The WW$\gamma$ cross-section combination is updated with the \Opal\ 
  inputs and by rescaling the \Delphi\ and \Ltre\ results to better
  match the agreed LEP signal definition.  All combinations are
  preliminary. }

\input{4f_s03}

\boldmath
\chapter{Electroweak Gauge Boson Self Couplings}
\label{sec-GC}
\unboldmath

\updates{ Updated combinations of the charged TGCs \gz, \kg\ and \lg\ 
  in single and multi-parameter fits are presented.  Updated results
  for the QGCs from the \ZZgg\ vertex, \acl\ and \azl, are given as
  well.  The combinations of neutral TGCs $h_i^V$ and $f_i^V$ include
  an updated $f_i^V$ combination.}

\input{gc}

\boldmath
\chapter{Colour Reconnection in W-Pair Events}
\label{sec-CR}
\unboldmath

\updates{ Unchanged w.r.t. summer 2002: Results are preliminary. }

\input{fsi_cr}

\boldmath
\chapter{Bose-Einstein Correlations in W-Pair Events}
\label{sec-BE}
\unboldmath

\updates{ New preliminary results from ALEPH and OPAL using "mixing
  method" are used. Preliminary results from DELPHI have been updated.
  The L3 results are final and published.  }

\input{be}

\boldmath
\chapter{W-Boson Mass and Width at \LEPII}
\label{sec-MW}
\unboldmath

\updates{ New preliminary results on the mass of the W boson from
  ALEPH are used in the combination.  }

\input{mw}

\boldmath
\chapter{Effective Couplings of the Neutral Weak Current}
\label{sec-eff}
\unboldmath

\updates{Updated preliminary and published measurements as discussed
  in the previous chapters are taken into account. Results are
  preliminary.}

\boldmath
\section{The Coupling Parameters $\cAf$}
\label{sec-AF}
\unboldmath

The coupling parameters $\cAf$ are defined in terms of the effective
vector and axial-vector neutral current couplings of fermions
(Equation~(\ref{eqn-cAf})).  The LEP measurements of the
forward-backward asymmetries of charged leptons (Chapter~\ref{sec-LS})
and b and c quarks (Chapter~\ref{sec-HF}) determine the products
$\Afbzf=\frac{3}{4}\cAe\cAf$ (Equation~(\ref{eqn-apol})). The LEP
measurements of the $\tau$ polarisation (Chapter~\ref{sec-TP}),
$\ptau(\cos\theta)$, determine $\cAt$ and $\cAe$ separately
(Equation~(\ref{eqn-taupol})).  Owing to polarised beams at SLC, SLD
measures the coupling parameters directly with the left-right and
forward-backward left-right asymmetries (Chapters~\ref{sec-ALR}
and~\ref{sec-HF}).

Table~\ref{tab-AF-L} shows the results for the leptonic coupling
parameter $\cAl$ from the LEP and SLD measurements, assuming lepton
universality. 

\begin{table}[tbp]
\begin{center}
\renewcommand{\arraystretch}{1.15}
\begin{tabular}{|c||c|c|r|}
\hline
                   & $\cAl$            & Cumulative Average & $\chi^2$/d.o.f.\\
\hline
\hline
$\Afbzl$           & $0.1512\pm0.0042$ &                    &                \\
$\ptau $           & $0.1465\pm0.0033$ & $0.1482\pm 0.0026$ &  0.8/1         \\
\hline
$\cAl$ (SLD)       & $0.1513\pm0.0021$ & $0.1501\pm 0.0016$ &  1.6/2          \\
\hline
\end{tabular}
\end{center}
\caption[]{
  Determination of the leptonic coupling parameter $\cAl$ assuming
  lepton universality. The second column lists the $\cAl$ values
  derived from the quantities listed in the first column. The third
  column contains the cumulative averages of the $\cAl$ results up to
  and including this line.
  The $\chi^2$ per degree of freedom for the cumulative
  averages is given in the last column.  }
\label{tab-AF-L}
\end{table}
\begin{table}[tbp]
\begin{center}
\renewcommand{\arraystretch}{1.15}
\begin{tabular}{|c||c|c|c|c|}
\hline
         &    LEP                  & SLD  & LEP+SLD  & \mcc{Standard} \\
         & ($\cAl=0.1482\pm0.0026$)&      & ($\cAl=0.1501\pm0.0016$) & \mcc{Model fit}\\
\hline
\hline
$\cAb$   & $0.898\pm0.021$    & $0.925 \pm 0.020$  & $0.903\pm0.013$ & 0.935 \\
$\cAc$   & $0.631\pm0.033$    & $0.670 \pm 0.026$  & $0.653\pm0.020$ & 0.668 \\
\hline
\end{tabular}
\end{center}
\caption[]{
  Determination of the quark coupling parameters $\cAb$ and $\cAc$
  from LEP data alone (using the LEP average for $\cAl$), from SLD
  data alone, and from LEP+SLD data (using the LEP+SLD average for
  $\cAl$) assuming lepton universality. }
\label{tab-AF-Q}
\end{table}

Using the measurements of $\cAl$ one can extract $\cAb$ and $\cAc$
from the LEP measurements of the b and c quark asymmetries.  The SLD
measurements of the left-right forward-backward asymmetries for b and
c quarks are direct determinations of $\cAb$ and $\cAc$.
Table~\ref{tab-AF-Q} shows the results on the quark coupling
parameters $\cAb$ and $\cAc$ derived from LEP measurements
(Equations~\ref{eqn-hf4}) and SLD measurements separately, and from
the combination of LEP+SLD measurements (Equation~\ref{eqn-hf6}).

\begin{figure}[htbp]
  \begin{center}
    \leavevmode
   \mbox{
    \includegraphics[width=0.49\textwidth]{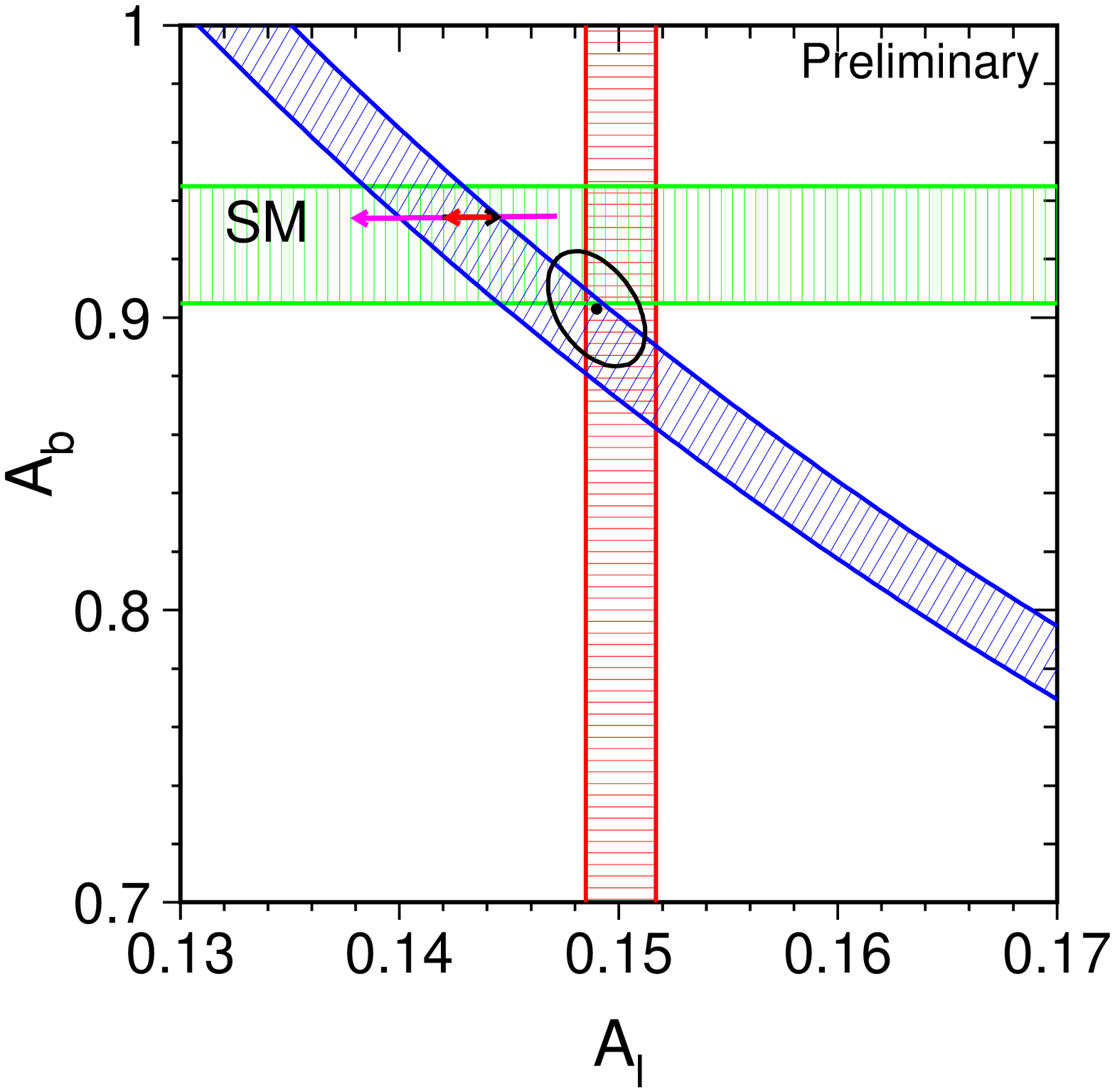}
    \includegraphics[width=0.49\textwidth]{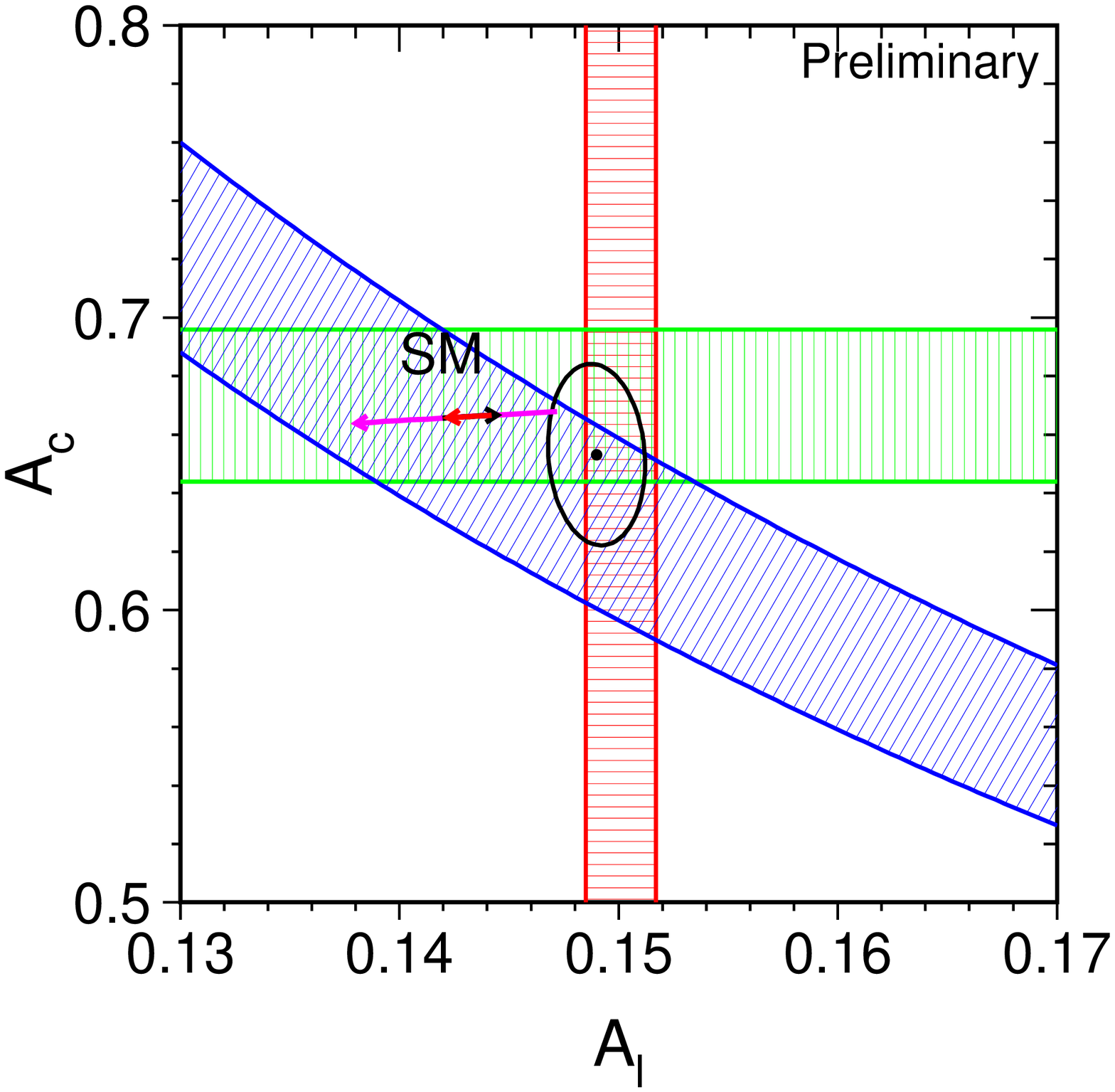}
    }
    \caption{The measurements of the combined LEP+SLD $\cAl$ (vertical
      band), SLD $\cAb$,$\cAc$ (horizontal bands) and LEP
      $\Afbzb$,$\Afbzc$ (diagonal bands), compared to the Standard
      Model expectations (arrows).  The arrow pointing to the left
      shows the variation in the $\SM$ prediction for $\MH$ in the
      range $300^{+700}_{-186}$ \GeV, and the arrow pointing to the
      right for $\Mt$ in the range $174.3 \pm 5.1$ \GeV.  Varying the
      hadronic vacuum polarisation by $\dalhad=0.02761\pm0.00036$
      yields an additional uncertainty on the Standard Model
      prediction, oriented in direction of the Higgs-boson arrow and
      size corresponding to the top-quark arrow.  Also shown is the
      68\% confidence level contour for the two asymmetry parameters
      resulting from the joint analyses.  Although the $\Afbzb$
      measurements prefer a high Higgs mass, the Standard Model fit to
      the full set of measurements prefers a low Higgs mass, for
      example because of the influence of $\cAl$.  }
    \label{fig-ae_ab}
  \end{center}
\end{figure}

The LEP extracted values of $\cAb$ and $\cAc$ are in agreement with
the SLD measurements, but somewhat lower than the Standard Model
predictions (0.935 and 0.668, respectively, essentially independent of
$\Mt$ and $\MH$).  The combination of LEP and SLD of $\cAb$ is 2.5
sigma below the Standard Model, while $\cAc$ agrees well with the
expectation.  This is mainly because the $\cAb$ value, deduced from the
measured $\Afbzb$ and the combined $\cAl$, is significantly lower than
both the Standard Model and the direct measurement of $\cAb$, this can
also be seen in Figure~\ref{fig-ae_ab}.

\boldmath
\section{The Effective Vector and Axial-Vector Coupling Constants}
\label{sec-GAGV}
\unboldmath

The partial widths of the $\Zzero$ into leptons and the lepton
forward-backward asymmetries (Section~\ref{sec-LS}), the $\tau$
polarisation and the $\tau$ polarisation asymmetry
(Section~\ref{sec-TP}) are combined to determine the effective
vector and axial-vector couplings for $\rm e$, $\mu$ and $\tau$. The
asymmetries (Equations~(\ref{eqn-apol}) and~(\ref{eqn-taupol}))
determine the ratio $\gvl/\gal$ (Equation~(\ref{eqn-cAf})),
while the leptonic partial widths determine the sum of the squares of
the couplings:
\begin{eqnarray}
\label{eqn-Gll}
\Gll & = &
{{\GF\MZ^3}\over{6\pi\sqrt 2}}
(\gvl^{2}+\gal^{2})(1+\delta^{QED}_\ell)\,,
\end{eqnarray}
where $\delta^{QED}_\ell=3q^2_\ell\alpha(\MZ^2)/(4\pi)$, with $q_\ell$
denoting the electric charge of the lepton, accounts for final state
photonic corrections. Corrections due to lepton masses, neglected in
Equation~\ref{eqn-Gll}, are taken into account for the results
presented below.

\begin{figure}[htbp]
\begin{center}
  \mbox{\includegraphics[width=0.9\linewidth]{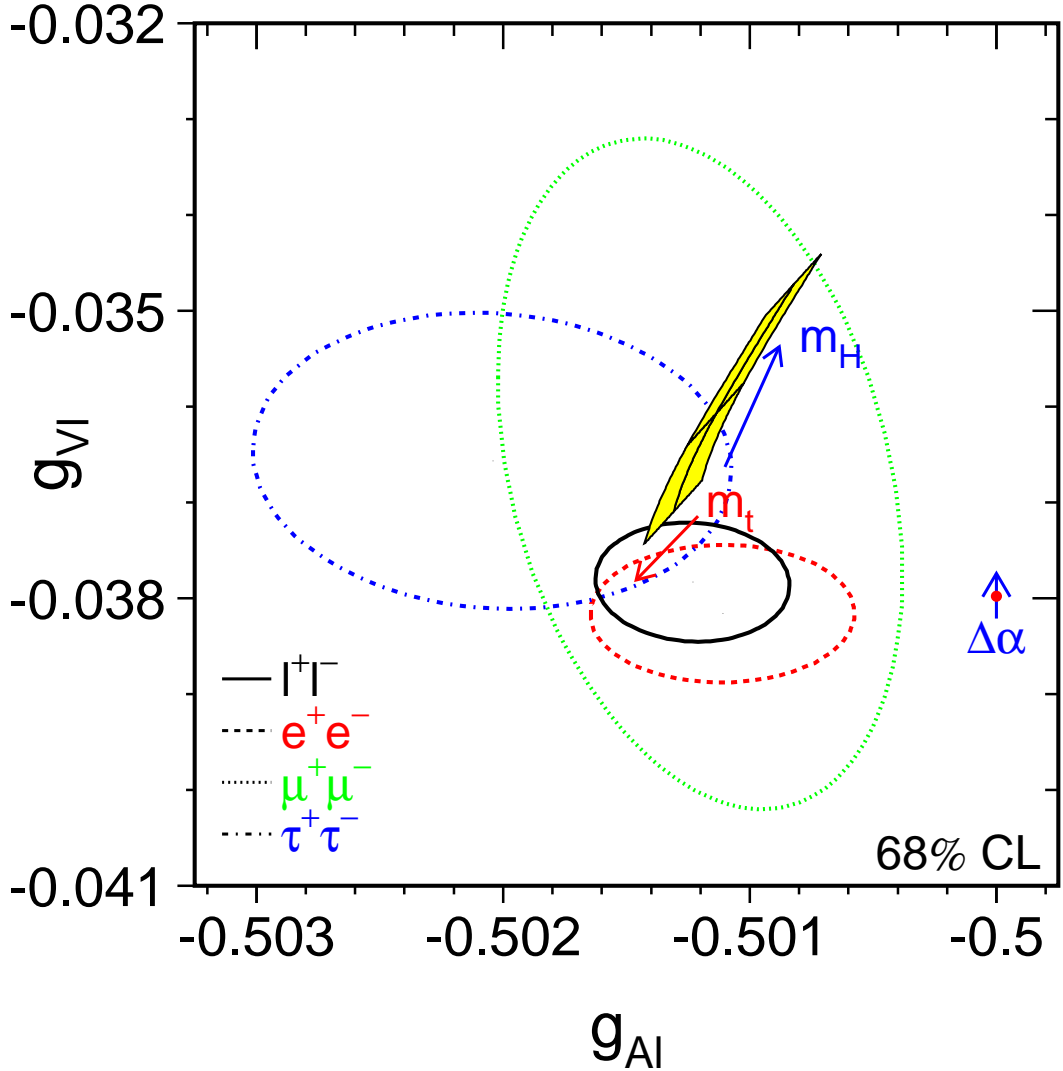}}
\end{center}
\caption[]{
  Contours of 68\% probability in the ($\gvl$,$\gal$) plane from LEP
  and SLD measurements.  The solid contour results from a fit to the
  LEP and SLD results assuming lepton universality. The shaded region
  corresponds to the Standard Model prediction for $\Mt = 174.3 \pm
  5.1$~\GeV{} and $\MH=300^{+700}_{-186}~\GeV$. The arrows point in
  the direction of increasing values of $\Mt$ and $\MH$.  Varying the
  hadronic vacuum polarisation by $\dalhad=0.02761\pm0.00036$ yields
  an additional uncertainty on the Standard Model prediction indicated
  by the corresponding arrow.}
\label{fig-gagv}
\end{figure}

The averaged results for the effective lepton couplings are given in
Table~\ref{tab-coup} for both the LEP data alone as well as for the
LEP and SLD measurements.  Figure~\ref{fig-gagv} shows the 68\%
probability contours in the $\gal$-$\gvl$ plane for the individual
lepton species. The signs of $\gal$ and $\gvl$ are based on the
convention $\gae < 0$. With this convention the signs of the couplings
of all charged leptons follow from LEP data alone.  The measured
ratios of the $\rm e$, $\mu$ and $\tau$ couplings provide a test of
lepton universality and are shown in Table~\ref{tab-coup}.  All values
are consistent with lepton universality.  The combined results
assuming universality are also given in the table and are shown as a
solid contour in Figure~\ref{fig-gagv}.

The neutrino couplings to the $\Zzero$ can be derived from the
measured value of the invisible width of the $\Zzero$, $\Ginv$ (see
Table~\ref{tab-widths}), attributing it exclusively to the decay into
three identical neutrino generations ($\Ginv=3\Gnn$) and assuming
$\gan\equiv\gvn\equiv\gn$.  The relative sign of $\gn$ is chosen to be
in agreement with neutrino scattering data\cite{ref:CHARMIIgn},
resulting in $\gn = +0.50068\pm 0.00075$.

In addition, the couplings analysis is extended to include also the
heavy-flavour measurements as presented in Section~\ref{sec-HFSUM}.
Assuming neutral-current lepton universality, the effective coupling
constants are determined jointly for leptons as well as for b and c
quarks. QCD corrections, modifying Equation~\ref{eqn-Gll}, are taken
from the Standard Model, as is also done to obtain the quark pole
asymmetries, see Section~\ref{sec:asycorrections}.

\begin{figure}[htbp]
\begin{center}
  \mbox{\includegraphics[width=0.6\linewidth]{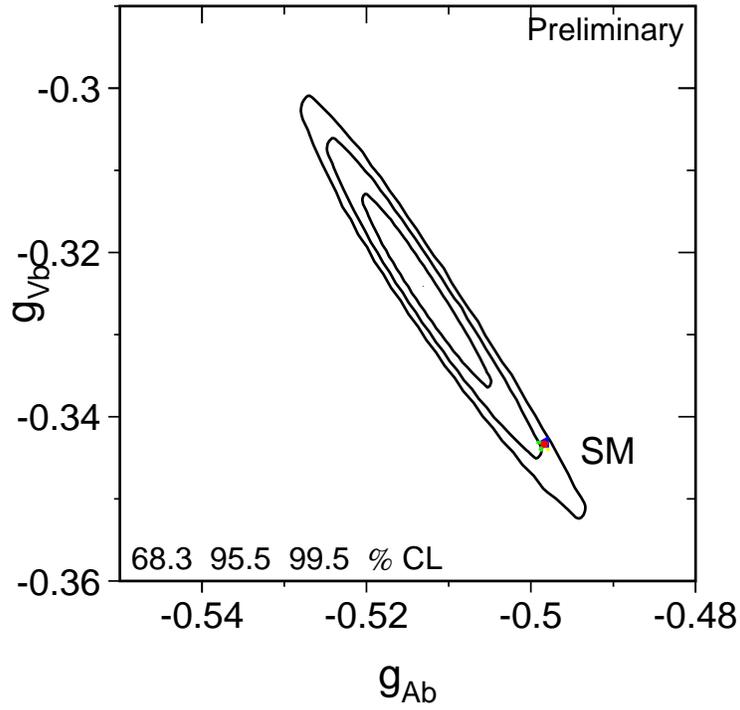}}\\
  \mbox{\includegraphics[width=0.6\linewidth]{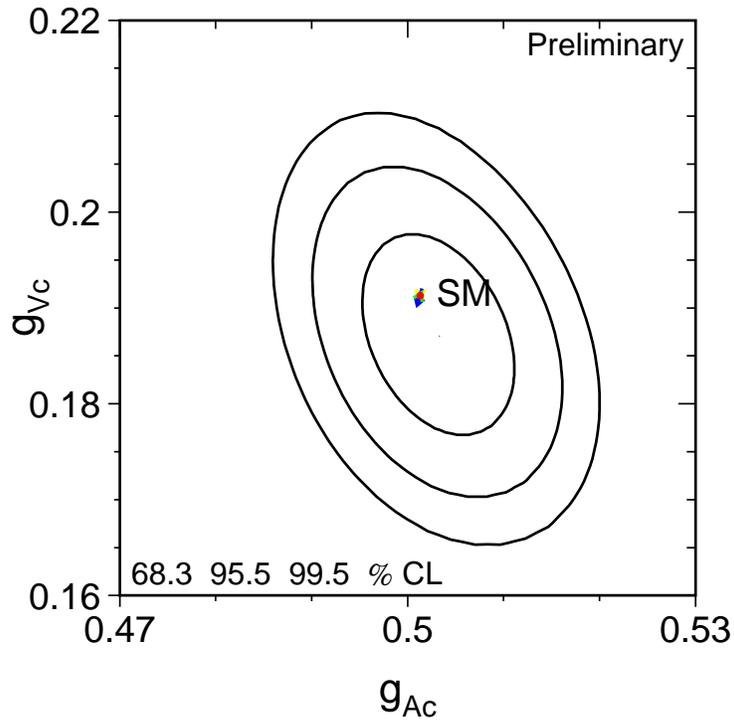}}
\end{center}
\caption[]{
  Contours of 68.3, 95.5 and 99.5\% probability in the ($\gvq$,$\gaq$)
  plane from LEP and SLD measurements for b and c quarks and assuming
  lepton universality. The dot corresponds to the Standard Model
  prediction for $\Mt = 174.3 \pm 5.1$~\GeV{},
  $\MH=300^{+700}_{-186}~\GeV$ and $\dalhad=0.02761\pm0.00036$. }
\label{fig-gaqgvq}
\end{figure}

The results are also reported in Table~\ref{tab-coup} and shown in
Figure~\ref{fig-gaqgvq}.  The deviation of the b-quark couplings from
the Standard Model expectation is mainly caused by the combined value
of $\cAb$ being low as discussed in Section~\ref{sec-AF} and shown in
Figure~\ref{fig-ae_ab}.

\begin{table}[htbp]
\renewcommand{\arraystretch}{1.15}
\begin{center}
\begin{tabular}{|l||c|c|}
\hline
&\multicolumn{2}{|c|}{Without Lepton Universality:} \\
 & LEP & LEP+SLD\\
\hline
\hline
$\gae$    & $-0.50112 \pm 0.00035$ & $-0.50111 \pm 0.00035$ \\
$\gamu$   & $-0.50115 \pm 0.00056$ & $-0.50120 \pm 0.00054$ \\
$\gatau$  & $-0.50204 \pm 0.00064$ & $-0.50204 \pm 0.00064$ \\
$\gve$    & $-0.0378  \pm 0.0011 $ & $-0.03816 \pm 0.00047$ \\
$\gvmu$   & $-0.0376  \pm 0.0031 $ & $-0.0367  \pm 0.0023 $ \\
$\gvtau$  & $-0.0368  \pm 0.0011 $ & $-0.0366  \pm 0.0010 $ \\
\hline
\hline
&\multicolumn{2}{|c|}{Ratios of couplings:} \\
 & LEP & LEP+SLD\\
\hline
\hline
$\gamu/\gae$ & $1.0001\pm0.0014$  &$1.0002\pm0.0014$\\
$\gatau/\gae$& $1.0018\pm0.0015$  &$1.0019\pm0.0015$\\
$\gvmu/\gve$ & $0.995\pm0.096$    &$0.962\pm0.063$\\
$\gvtau/\gve$& $0.972\pm0.041$    &$0.958\pm0.029$\\
\hline
\hline
&\multicolumn{2}{|c|}{With Lepton Universality:   } \\
 & LEP & LEP+SLD\\
\hline
\hline
$\gal$    & $-0.50126 \pm 0.00026$& $-0.50123 \pm 0.00026$\\
$\gvl$    & $-0.03736 \pm 0.00066$& $-0.03783 \pm 0.00041$\\
\hline
$\gn$     & $+0.50068 \pm 0.00075$& $+0.50068 \pm 0.00075$\\
\hline
\hline
&\multicolumn{2}{|c|}{With Lepton Universality   } \\
&\multicolumn{2}{|c|}{and Heavy Flavour Results: } \\
 & LEP & LEP+SLD\\
\hline
\hline
$\gal$    & $-0.50126 \pm 0.00026$ & $-0.50125 \pm 0.00026$ \\
$\gab$    & $-0.5152  \pm 0.0080 $ & $-0.5132  \pm 0.0051 $ \\
$\gac$    & $+0.5015  \pm 0.0081 $ & $+0.5033  \pm 0.0052 $ \\
$\gvl$    & $-0.03736 \pm 0.00066$ & $-0.03755 \pm 0.00037$ \\
$\gvb$    & $-0.321   \pm 0.012  $ & $-0.3241  \pm 0.0077 $ \\
$\gvc$    & $+0.178   \pm 0.011  $ & $+0.1871  \pm 0.0069 $ \\
\hline
\end{tabular}
\end{center}
\caption[]{%
  Results for the effective vector and axial-vector couplings derived
  from the LEP data and the combined LEP and SLD data without and with 
  the assumption of lepton universality. Note that the results, in
  particular for b quarks, are highly correlated.}
\label{tab-coup}
\end{table}

\clearpage

\boldmath
\section{The Leptonic Effective Electroweak Mixing Angle $\swsqeffl$}
\label{sec-SW}
\unboldmath

The asymmetry measurements from LEP 
and SLD can be combined into a single
parameter, the effective electroweak mixing angle, $\swsqeffl$,
defined as:
\begin{eqnarray}
\label{eqn-sw}
\swsqeffl & \equiv &
\frac{1}{4}\left(1-\frac{\gvl}{\gal}\right)\,,
\end{eqnarray}
without making strong model-specific assumptions.

\begin{figure}[p]
  \begin{center}
    \leavevmode
    \includegraphics[width=0.9\linewidth]{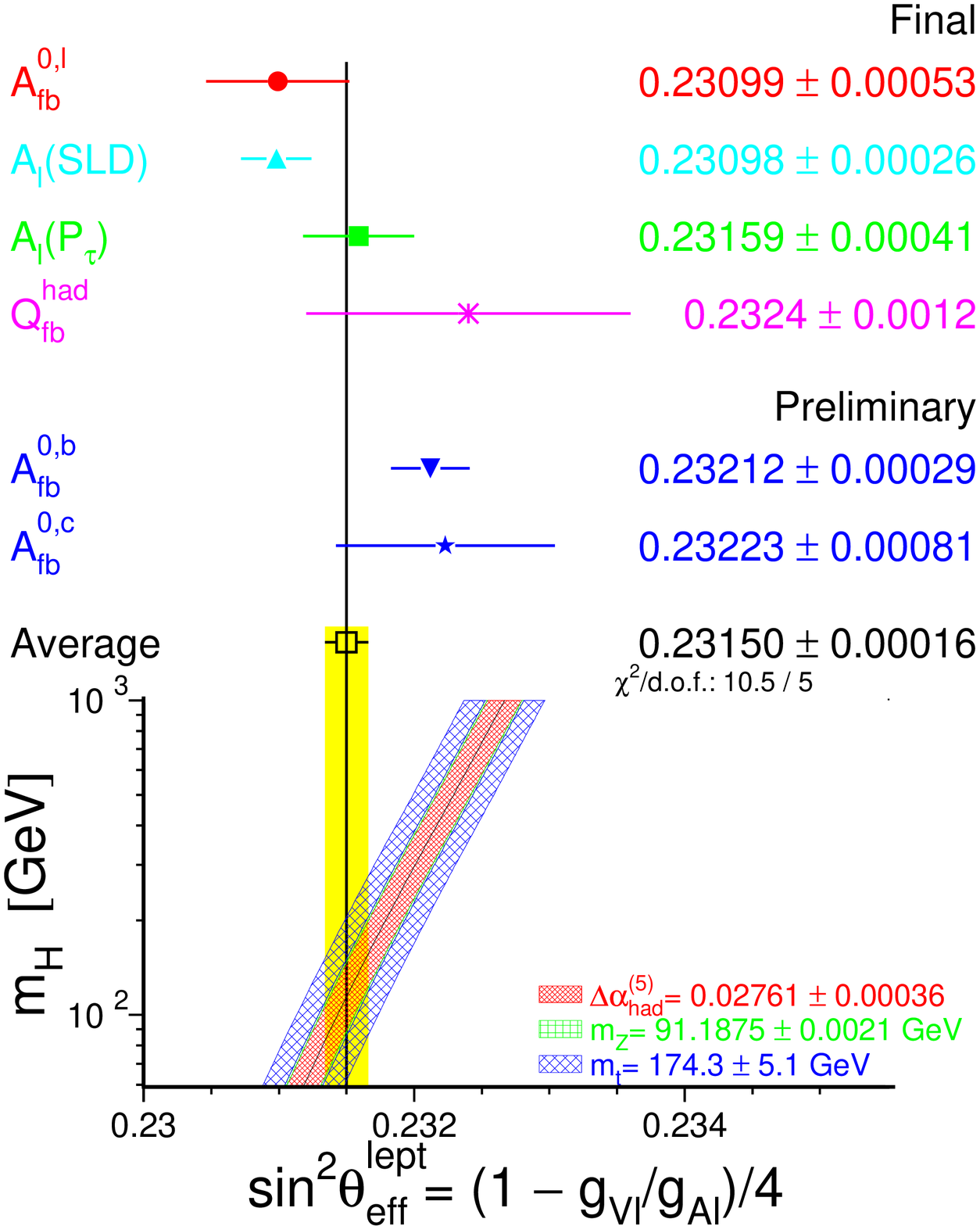}
  \end{center}
  \caption[]{
    Comparison of several determinations of $\swsqeffl$ from 
    asymmetries.  In the average, the small correlation between
    $\Afbzb$ and $\Afbzc$ is included.
    Also shown is the prediction of the Standard Model
    as a function of $\MH$.  The width of the Standard Model band is
    due to the uncertainties in
    $\Delta\alpha_{\mathrm{had}}^{(5)}(\MZ^2)$ (see Chapter~\ref{sec-MSM}),
    $\MZ$ and $\Mt$.
    The total width of the band is the linear sum of these effects.}
  \label{fig-swsq}
\end{figure}

For a combined average of $\swsqeffl$ from $\Afbzl$, $\cAt$ and $\cAe$
only the assumption of lepton universality, already inherent in the
definition of $\swsqeffl$, is needed.  Also the value derived from the
measurements of $\cAl$ from SLD is given.  We also include the
hadronic forward-backward asymmetries, assuming the difference between
$\swsqefff$ for quarks and leptons to be given by the Standard Model.
This is justified within the Standard Model as the hadronic
asymmetries $\Afbzb$ and $\Afbzc$ have a reduced sensitivity to the
small non-universal corrections specific to the quark vertex.  The
results of these determinations of $\swsqeffl$ and their combination
are shown in Table~\ref{tab-swsq} and in Figure~\ref{fig-swsq}.  The
combinations based on the leptonic results plus $\cAl$(SLD) and on the
hadronic forward-backward asymmetries differ by 2.9 standard
deviations, caused by the two most precise measurements of
$\swsqeffl$, $\cAl$ (SLD) dominated by $\ALRz$, and $\Afbzb$ (LEP),
likewise differing by 2.9 standard deviations.
This is the same effect as discussed already in sections~\ref{sec-AF}
and~\ref{sec-GAGV} and shown in Figures~\ref{fig-ae_ab}
and~\ref{fig-gaqgvq}: the deviation in $\cAb$ as extracted from
$\Afbzb$ discussed above is reflected in the value of $\swsqeffl$
extracted from $\Afbzb$ in this analysis.

\begin{table}[htbp]
\renewcommand{\arraystretch}{1.25}
\begin{center}
\begin{tabular}{|l||c|c|c|c|}
\hline
     & $\swsqeffl$&\mco{Average by Group}&Cumulative &       \\
     &            &\mco{of Observations} &Average    &$\chi^2$/d.o.f.\\
\hline
\hline
$\Afbzl$         & $0.23099\pm 0.00053$ &&& \\
$\cAl~(\ptau)$   & $0.23159\pm 0.00041$ &$0.23137\pm0.00033$
                                 &                   &0.8/1\\
\hline
$\cAl$ (SLD)     &$0.23098\pm0.00026$ &                   
                                 &$0.23113\pm0.00021$&1.6/2\\
\hline
$\Afbzb$  & $0.23212\pm 0.00029$ &&& \\
$\Afbzc$  & $0.23223\pm 0.00081$ &&& \\
$\avQfb$  & $0.2324 \pm 0.0012 $ &$0.23214\pm0.00027$
                                 &$0.23150\pm0.00016$&10.5/5\\
\hline
\end{tabular}\end{center}
\caption[]{
  Determinations of $\swsqeffl$ from asymmetries.  
  The second
  column lists the $\swsqeffl$ values derived from the quantities
  listed in the first column. The third column contains the averages
  of these numbers by groups of observations, where the groups are
  separated by the horizontal lines. The fourth column shows the
  cumulative averages. The $\chi^2$ per degree of freedom for the
  cumulative averages is also given. The averages are performed
  including the small correlation between $\Afbzb$ and $\Afbzc$.
  The average of all six results has a probability of 7.0\%.}
\label{tab-swsq}
\end{table}

\boldmath
\chapter{Constraints on the Standard Model}
\label{sec-MSM}
\unboldmath

\updates{Updated preliminary and published measurements as discussed
  in the previous chapters are taken into account, as well as new
  corrections in the measurement of atomic parity violation in
  Caesim.}

\section{Introduction}

The precise electroweak measurements performed at LEP and SLC and
elsewhere can be used to check the validity of the Standard Model and,
within its framework, to infer valuable information about its
fundamental parameters. The accuracy of the measurements makes them
sensitive to the mass of the top quark $\Mt$, and to the mass of the
Higgs boson $\MH$ through loop corrections. While the leading $\Mt$
dependence is quadratic, the leading $\MH$ dependence is logarithmic.
Therefore, the inferred constraints on $\MH$ are much weaker than
those on $\Mt$. 

\section{Measurements}

The LEP and SLD measurements used are summarised in
Table~\ref{tab-SMIN}. Also shown are the results of the Standard Model
fit to all results.

The final results on the W-boson mass by UA2\cite{bib-UA2MW} and
CDF\cite{bib-CDFMW1,bib-CDFMW2} and D\O\cite{bib-D0MW} in Run-I, and
the W-boson width by CDF\cite{bib-CDFGW} and D\O\cite{bib-D0GW} in
Run-I were recently combined based on a detailed treatment of common
systematic uncertainties. The results are\cite{bib-MWGWAVE-02}: $\MW =
80454\pm59~\MeV$, $\GW = 2115\pm106~\MeV$, with a correlation of
$-17.4\%$.  Combining these results with the updated preliminary LEP-2
measurements as presented in Chapter~\ref{sec-MW}, the new preliminary
world averages used in the following analyses are:
\begin{eqnarray}
\MW & = & 80.426 \pm 0.034~\GeV\\
\GW & = &  2.139 \pm 0.069~\GeV\,
\end{eqnarray}
with a correlation of $-6.7\%$. 

For the mass of the top quark, $\Mt$, the results from
CDF\cite{bib-topCDF} and D\O\cite{bib-topD0} are
combined~\citen{bib-Tevatop}, with the result $\Mt=174.3\pm51.~\GeV$.
In addition, the final result of the NuTeV collaboration on
neutrino-nucleon neutral to charged current cross section
ratios~\cite{bib-NuTeV-final}, and the measurements of atomic parity
violation in caesium\cite{QWCs:exp:1, QWCs:exp:2}, with the numerical
result\cite{QWCs:theo:2003} taken from a recent revised analysis of
QED radiative corrections applied to the raw measurement, are included
in some of the analyses shown below.

Although the $\nu{\cal N}$ result is quoted in terms of
$\swsq=1-\MW^2/\MZ^2=0.2277\pm0.0016$, radiative corrections result in
small $\Mt$ and $\MH$ dependences\footnote{The formula used is
  $\delta\sin^2\theta_W = -0.00022 \frac{\Mt^2 -
    (175\GeV)^2}{(50\GeV)^2} + 0.00032 \ln(\frac{\MH}{150\GeV}).$ See
  Reference \citen{bib-NuTeV-final} for details.}  that are included
in the fit.  Note that the NuTeV result in terms of the on-shell
electroweak mixing angle is 2.9 standard deviations higher
than the expectation.

An additional input parameter, not shown in the table, is the Fermi
constant $G_F$, determined from the $\mu$ lifetime, $G_F = 1.16637(1)
\cdot 10^{-5} \GeV^{-2}$\cite{bib-Gmu}.  The relative error of $G_F$
is comparable to that of $\MZ$; both errors have negligible effects on
the fit results.

\begin{table}[p]
\begin{center}
\renewcommand{\arraystretch}{1.10}
\begin{tabular}{|ll||r|r|r|r|}
\hline
 && \mcc{Measurement with}  &\mcc{Systematic} & \mcc{Standard} & \mcc{Pull} \\
 && \mcc{Total Error}       &\mcc{Error}      & \mcc{Model fit}&            \\
\hline
\hline
&&&&& \\[-3mm]
& $\Delta\alpha^{(5)}_{\mathrm{had}}(\MZ^2)$\cite{bib-BP01}
                & $0.02761 \pm 0.00036$ & 0.00035 &0.02767& $-0.2$ \\
&&&&& \\[-3mm]
\hline
a) & \underline{LEP}     &&&& \\
   & line-shape and      &&&& \\
   & lepton asymmetries: &&&& \\
&$\MZ$ [\GeV{}] & $91.1875\pm0.0021\pz$
                & ${}^{(a)}$0.0017$\pz$ &91.1875$\pz$ & $ 0.0$ \\
&$\GZ$ [\GeV{}] & $2.4952 \pm0.0023\pz$
                & ${}^{(a)}$0.0012$\pz$ & 2.4960$\pz$ & $-0.4$ \\
&$\shad$ [nb]   & $41.540 \pm0.037\pzz$ 
                & ${}^{(b)}$0.028$\pzz$ &41.478$\pzz$ & $ 1.7$ \\
&$\RZ$          & $20.767 \pm0.025\pzz$ 
                & ${}^{(b)}$0.007$\pzz$ &20.742$\pzz$ & $ 1.0$ \\
&$\Afbzl$       & $0.0171 \pm0.0010\pz$ 
                & ${}^{(b)}$0.0003\pz & 0.0164\pz     & $ 0.8$ \\
&+ correlation matrix Table~\ref{tab-zparavg} &&&& \\
&                                             &&&& \\[-3mm]
&$\tau$ polarisation:                         &&&& \\
&$\cAl~(\ptau)$ & $0.1465\pm 0.0033\pz$ 
                & 0.0016$\pz$ & 0.1477$\pz$ & $-0.4$ \\
                      &                       &&&& \\[-3mm]
&$\qq$ charge asymmetry:                      &&&& \\
&$\swsqeffl(\Qfbhad)$
                & $0.2324\pm0.0012\pz$ 
                & 0.0010$\pz$ & 0.2314$\pz$ & $ 0.8$ \\
&                                             &&&& \\[-3mm]
\hline
b) & \underline{SLD}\cite{ref:sld-s99} &&&& \\
&$\cAl$ (SLD)   & $0.1513\pm 0.0021\pz$ 
                & 0.0010$\pz$ & 0.1478$\pz$ & $ 1.7$ \\
&&&&& \\[-3mm]
\hline
c) & \underline{LEP and SLD Heavy Flavour} &&&& \\
&$\Rbz{}$        & $0.21638\pm0.00066$  
                 & 0.00049     & 0.21579     & $ 0.9$ \\
&$\Rcz{}$        & $0.1720\pm0.0030\pz$
                 & 0.0020$\pz$ & 0.1723$\pz$ & $-0.1$ \\
&$\Afbzb{}$      & $0.0997\pm0.0016\pz$
                 & 0.0007$\pz$ & 0.1036$\pz$ & $-2.4$ \\
&$\Afbzc{}$      & $0.0706\pm0.0035\pz$
                 & 0.0017$\pz$ & 0.0740$\pz$ & $-1.0$ \\
&$\cAb$          & $0.925\pm 0.020\pzz$
                 & 0.014$\pzz$ & 0.935$\pzz$ & $-0.5$ \\
&$\cAc$          & $0.670\pm 0.026\pzz$
                 & 0.016$\pzz$ & 0.668$\pzz$ & $ 0.1$ \\
&+ correlation matrix Table~\ref{tab:14parcor} &&&& \\
&                                              &&&& \\[-3mm]
\hline
d) & \underline{Additional} &&&& \\
&$\MW$ [\GeV{}] ($\pp$ and LEP-2)
& $80.426 \pm 0.034\pzz$ &      $\pzz$   & 80.385$\pzz$ & $ 1.2$ \\
&$\GW$ [\GeV{}] ($\pp$ and LEP-2)
& $ 2.139 \pm 0.069\pzz$ &      $\pzz$   &  2.093$\pzz$ & $ 0.7$ \\
&$\swsq$ ($\nu{\cal N}$\cite{bib-NuTeV-final})
& $0.2277\pm0.0016\pz$   & 0.0009$\pz$   & 0.2229$\pz$  & $ 2.9$ \\
&$\Mt$ [\GeV{}] ($\pp$\cite{bib-Tevatop})
& $174.3\pm 5.1\pzz\pzz$ & 4.0$\pzz\pzz$ & 174.3$\pzz\pzz$ & $ 0.0$ \\
&$\QWCs$~\cite{QWCs:theo:2003}
& $-72.84\pm 0.46\pzz\pz$& 0.36$\pz\pzz$ & $-72.90\pz\pzz$ & $ 0.1$ \\
\hline
\end{tabular}\end{center}
\caption[]{
  Summary of measurements included in the combined analysis of
  Standard Model parameters. Section~a) summarises LEP averages,
  Section~b) SLD results ($\swsqeffl$ includes $\ALR$ and the
  polarised lepton asymmetries), Section~c) the LEP and SLD heavy
  flavour results and Section~d) electroweak measurements from $\pp$
  colliders and $\nu$N scattering.  The total errors in column 2
  include the systematic errors listed in column 3.  Although the systematic
  errors include both correlated and uncorrelated sources, the determination 
  of the systematic part of each error is approximate.  The $\SM$
  results in column~4 and the pulls (difference between measurement
  and fit in units of the total measurement error) in column~5 are
  derived from the Standard Model fit including all data
  (Table~\ref{tab-BIGFIT}, column~5) with the Higgs mass treated as a
  free parameter.\\
  $^{(a)}$\small{The systematic errors on $\MZ$ and $\GZ$ contain the
    errors arising from the uncertainties in the LEP energy only.}\\
  $^{(b)}$\small{Only common systematic errors are indicated.}\\
}
\label{tab-SMIN}
\end{table}

\section{Theoretical and Parametric Uncertainties}

Detailed studies of the theoretical uncertainties in the Standard
Model predictions due to missing higher-order electroweak corrections
and their interplay with QCD corrections are carried out by the
working group on `Precision calculations for the $\Zzero$
resonance'\cite{bib-PCLI}, and more recently in~\cite{bib-PCP99}.
Theoretical uncertainties are evaluated by comparing different but,
within our present knowledge, equivalent treatments of aspects such as
resummation techniques, momentum transfer scales for vertex
corrections and factorisation schemes.  The effects of these
theoretical uncertainties are reduced by the inclusion of higher-order
corrections\cite{bib-twoloop,bib-QCDEW} in the electroweak
libraries\cite{bib-SMNEW}.

The recently calculated complete fermionic two-loop corrections on
$\MW$~\cite{FHWW-f2l-MW} are currently only used in the determination
of the theoretical uncertainty.  Their effect on $\MW$ is small
compared to the current experimental uncertainty on $\MW$. However,
the naive propagation of this new $\MW$ to
$\swsqeffl=\kappa(1-\MW^2/\MZ^2)$, keeping the electroweak form-factor
$\kappa$ unmodified, shows a more visible effect as $\swsqeffl$ is
measured very precisely.  Thus the corresponding calculations for
$\swsqeffl$ (or $\kappa$) and for the partial Z widths are urgently
needed; in particular since partial cancellations of these new
corrections in the product $\kappa(1-\MW^2/\MZ^2)=\swsqeffl$ could be
expected~\cite{BGPWprivate}.

The use of the QCD corrections\cite{bib-QCDEW} increases the value of
$\alfmz$ by 0.001, as expected.  The effects of missing higher-order
QCD corrections on $\alfmz$ covers missing higher-order electroweak
corrections and uncertainties in the interplay of electroweak and QCD
corrections and is estimated to be at least 0.002~\cite{bib-SMALFAS}.
A discussion of theoretical uncertainties in the determination of
$\alfas$ can be found in References~\citen{bib-PCLI}
and~\citen{bib-SMALFAS}.  The determination of the size of remaining
theoretical uncertainties is under continued study.

The theoretical errors discussed above are not included in the results
presented in Table~\ref{tab-BIGFIT}.  At present the impact of
theoretical uncertainties on the determination of $\SM$ parameters
from the precise electroweak measurements is small compared to the
error due to the uncertainty in the value of $\alpha(\MZ^2)$, which is
included in the results.

The uncertainty in $\alpha(\MZ^2)$ arises from the contribution of
light quarks to the photon vacuum polarisation
($\Delta\alpha_{\mathrm{had}}^{(5)}(\MZ^2)$):
\begin{equation}
\alpha(\MZ^2) = \frac{\alpha(0)}%
   {1 - \Delta\alpha_\ell(\MZ^2) -
   \Delta\alpha_{\mathrm{had}}^{(5)}(\MZ^2) -
   \Delta\alpha_{\mathrm{top}}(\MZ^2)} \,,
\end{equation}
where $\alpha(0)=1/137.036$.  The top contribution, $-0.00007(1)$,
depends on the mass of the top quark, and is therefore determined
inside the electroweak libraries\cite{bib-SMNEW}.  The leptonic
contribution is calculated to third order\cite{bib-alphalept} to be
$0.03150$, with negligible uncertainty.

For the hadronic contribution, we no longer use the value $0.02804 \pm
0.00065$\cite{bib-JEG2}, but rather the new evaluation
$0.02761\pm0.0036$~\cite{bib-BP01} which takes into account the
recently published results on electron-positron annihilations into
hadrons at low centre-of-mass energies by the BES
collaboration~\cite{BES_01}.  This reduced uncertainty still causes an
error of 0.00013 on the $\SM$ prediction of $\swsqeffl$, and errors of
0.2~\GeV{} and 0.1 on the fitted values of $\Mt$ and $\log(\MH)$,
included in the results presented below.  The effect on the $\SM$
prediction for $\Gll$ is negligible.  The $\alfmz$ values for the
$\SM$ fits presented here are stable against a variation of
$\alpha(\MZ^2)$ in the interval quoted.

There are also several evaluations of
$\Delta\alpha^{(5)}_{\mathrm{had}}(\MZ^2)$%
\cite{bib-Swartz,bib-Zeppe,bib-Alemany,bib-Davier,bib-alphaKuhn,bib-Erler,bib-ADMartin,bib-jeger99,bib-TY0102}
which are more theory-driven.  One of the most recent of these
(Reference \citen{bib-TY0102}) also includes the new results from BES,
yielding $0.02747\pm0.00012$.  To show the effects of the uncertainty
of $\alpha(\MZ^2)$, we also use this evaluation of the hadronic vacuum
polarisation.  Note that all these evaluations obtain values for
$\Delta\alpha^{(5)}_{\mathrm{had}}(\MZ^2)$ consistently lower than -
but still in agreement with - the old value of $0.02804 \pm 0.00065$.

\section{Selected Results}

Figure~\ref{fig-gllsef} shows a comparison of the leptonic partial
width from LEP (Table~\ref{tab-widths}) and the effective electroweak
mixing angle from asymmetries measured at LEP and SLD
(Table~\ref{tab-swsq}), with the Standard Model. Good agreement with
the $\SM$ prediction is observed.  The point with the arrow indicates
the prediction if among the electroweak radiative corrections only the
photon vacuum polarisation is included, which shows that LEP+SLD data
are sensitive to non-trivial electroweak corrections.  Note that the
error due to the uncertainty on $\alpha(\MZ^2)$ (shown as the length
of the arrow) is not much smaller than the experimental error on
$\swsqeffl$ from LEP and SLD.  This underlines the continued
importance of a precise measurement of
$\sigma(\mathrm{e^+e^-\rightarrow hadrons})$ at low centre-of-mass
energies.

Of the measurements given in Table~\ref{tab-SMIN}, $\RZ$ is one of the
most sensitive to QCD corrections.  For $\MZ=91.1875$~\GeV{}, and
imposing $\Mt=174.3\pm5.1$~\GeV{} as a constraint,
$\alfas=0.1224\pm0.0038$ is obtained.  Alternatively, $\sll$ (see
Table~\ref{tab-widths}) which has higher sensitivity to QCD
corrections and less dependence on $\MH$ yields:
$\alfas=0.1180\pm0.0030$.  Typical errors arising from the variation
of $\MH$ between $100~\GeV$ and $200~\GeV$ are of the order of
$0.001$, somewhat smaller for $\sll$.  These results on $\alfas$, as
well as those reported in the next section, are in very good agreement
with recently determined world averages ($\alfmz=0.118 \pm
0.002$\cite{common_bib:pdg2000}, or $\alfmz=0.1178 \pm 0.0033$ based
solely on NNLO QCD results excluding the LEP lineshape results and
accounting for correlated errors\cite{Siggi-Bethke-alpha-s}).

\section{Standard Model Analyses}

In the following, several different Standard Model fits to the data
reported in Table~\ref{tab-BIGFIT} are discussed.  The $\chi^2$
minimisation is performed with the program MINUIT~\cite{MINUIT}, and
the predictions are calculated with TOPAZ0~\cite{ref:TOPAZ0} and
ZFITTER~\cite{ref:ZFITTER}.  The somewhat increased $\chi^2$/d.o.f.{}
for all of these fits is caused by the same effect as discussed in the
previous chapter, namely the large dispersion in the values of the
leptonic effective electroweak mixing angle measured through the
various asymmetries.  For the analyses presented here, this dispersion
is interpreted as a fluctuation in one or more of the input
measurements, and thus we neither modify nor exclude any of them.  A
further drastic increase in $\chi^2$/d.o.f.{} is observed when the
NuTeV result on $\swsq$ is included in the analysis.

To test the agreement between the LEP data and the Standard Model, a
fit to the LEP data (including the $\LEPII$ $\MW$ and $\GW$
determinations) leaving the top quark mass and the Higgs mass as free
parameters is performed.  The result is shown in
Table~\ref{tab-BIGFIT}, column~1.  This fit shows that the LEP data
predicts the top mass in good agreement with the direct measurements.
In addition, the data prefer an intermediate Higgs-boson mass, albeit
with very large errors.  The strongly asymmetric errors on $\MH$ are
due to the fact that to first order, the radiative corrections in the
Standard Model are proportional to $\log(\MH)$.

\begin{figure}[htbp]
\begin{center}
  \leavevmode \includegraphics[width=0.9\linewidth]{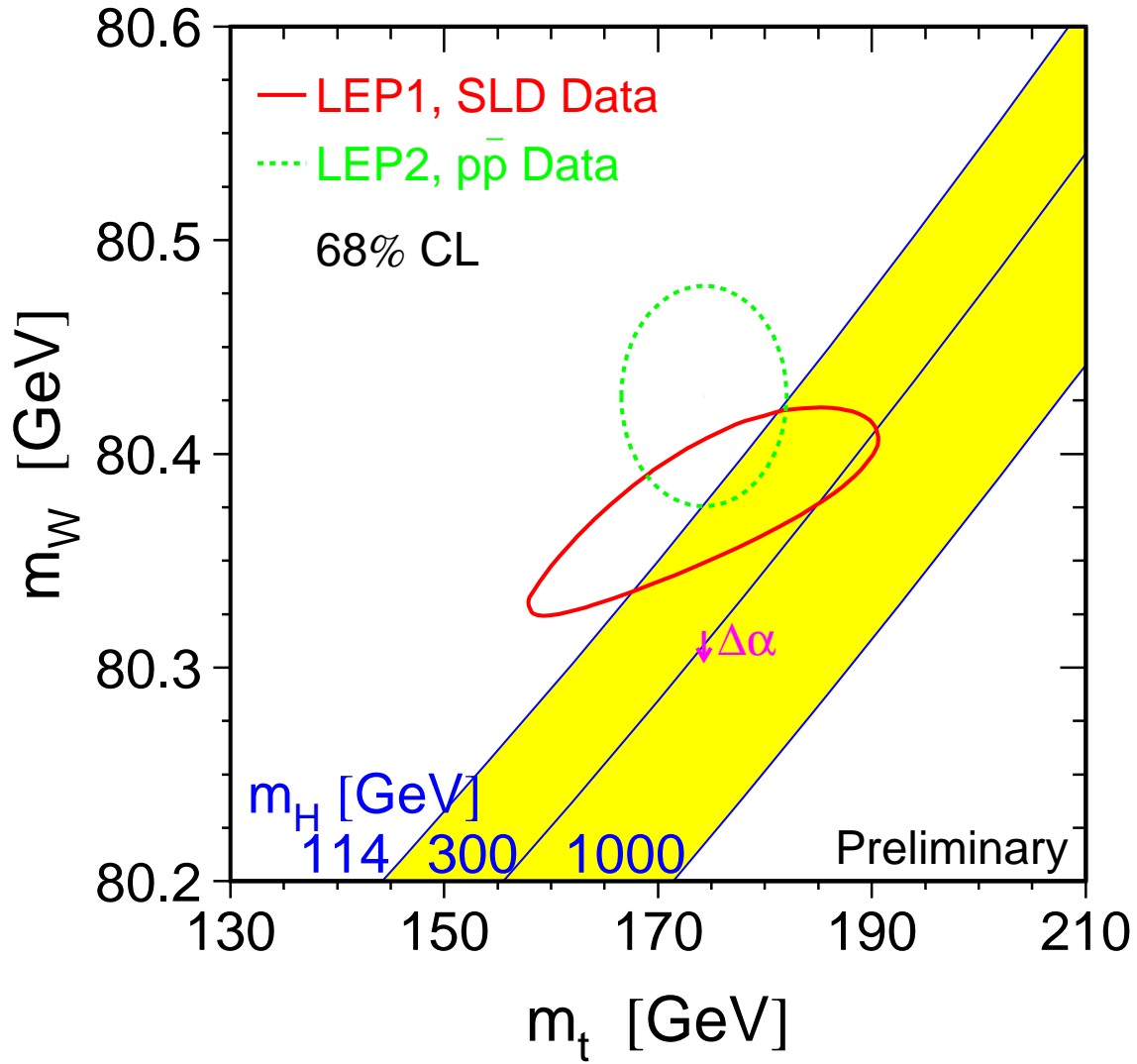}
\caption[]{
  The comparison of the indirect measurements of $\MW$ and $\Mt$
  ($\LEPI$+ SLD data) (solid contour) and the direct
  measurements ($\pp$ colliders and $\LEPII$ data) (dashed contour).  In both
  cases the 68\% CL contours are plotted.  Also shown is the Standard
  Model relationship for the masses as a function of the Higgs mass. The
  arrow labelled $\Delta\alpha$ shows the variation of this relation if
  $\alpha(\MZ^2)$ is changed by one standard deviation. This variation
  gives an additional uncertainty to the Standard Model band 
  shown in the figure.}
\label{fig:mtmW}
\end{center}
\end{figure}

The data can also be used within the Standard Model to determine the
top quark and W masses indirectly, which can be compared to the direct
measurements performed at the $\pp$ colliders and \LEPII.  In the
second fit, all LEP and SLD results in Table~\ref{tab-SMIN}, except
the measurements of $\MW$ and $\GW$, are used.  The results are shown
in column~2 of Table~\ref{tab-BIGFIT}.  The indirect measurements of
$\MW$ and $\Mt$ from this data sample are shown in
Figure~\ref{fig:mtmW}, compared with the direct measurements. Also
shown are the Standard Model predictions for Higgs masses between 114
and 1000~\GeV.  As can be seen in the figure, the indirect and direct
measurements of $\MW$ and $\Mt$ are now in good agreement, and both sets
prefer a low value of the Higgs mass.

\begin{figure}[htbp]
\begin{center}
  \mbox{\includegraphics[width=0.9\linewidth]{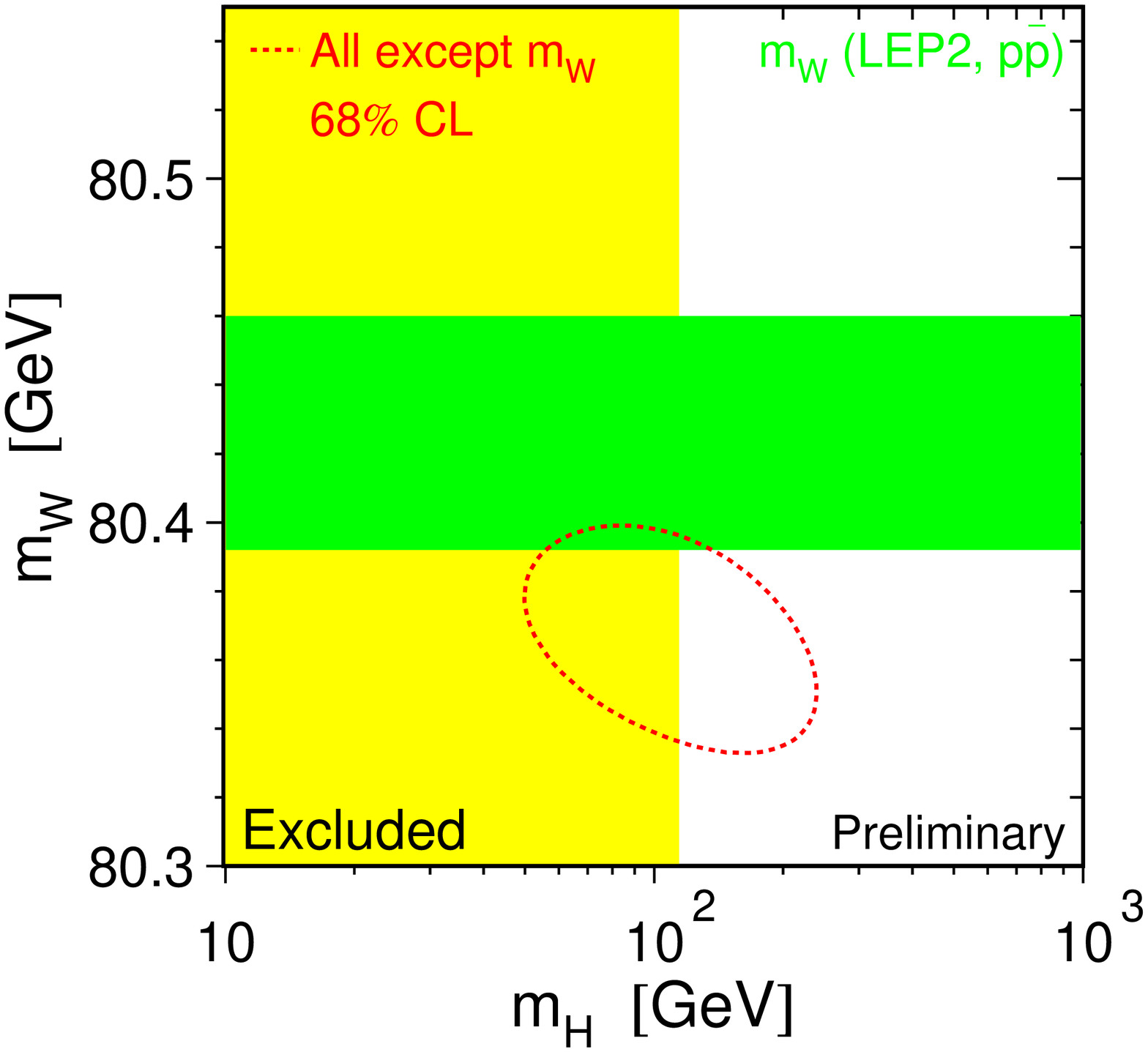}}
\end{center}
\vspace*{-0.6cm}
\caption[]{
  The 68\% confidence level contour in $\MW$ and $\MH$ for the fit to
  all data except the direct measurement of $\MW$, indicated by the
  shaded horizontal band of $\pm1$ sigma width.  The vertical band
  shows the 95\% CL exclusion limit on $\MH$ from the direct search.
  }
\label{fig-mhmw}
\end{figure}

\begin{figure}[htbp]
\begin{center}
  \mbox{\includegraphics[width=0.9\linewidth]{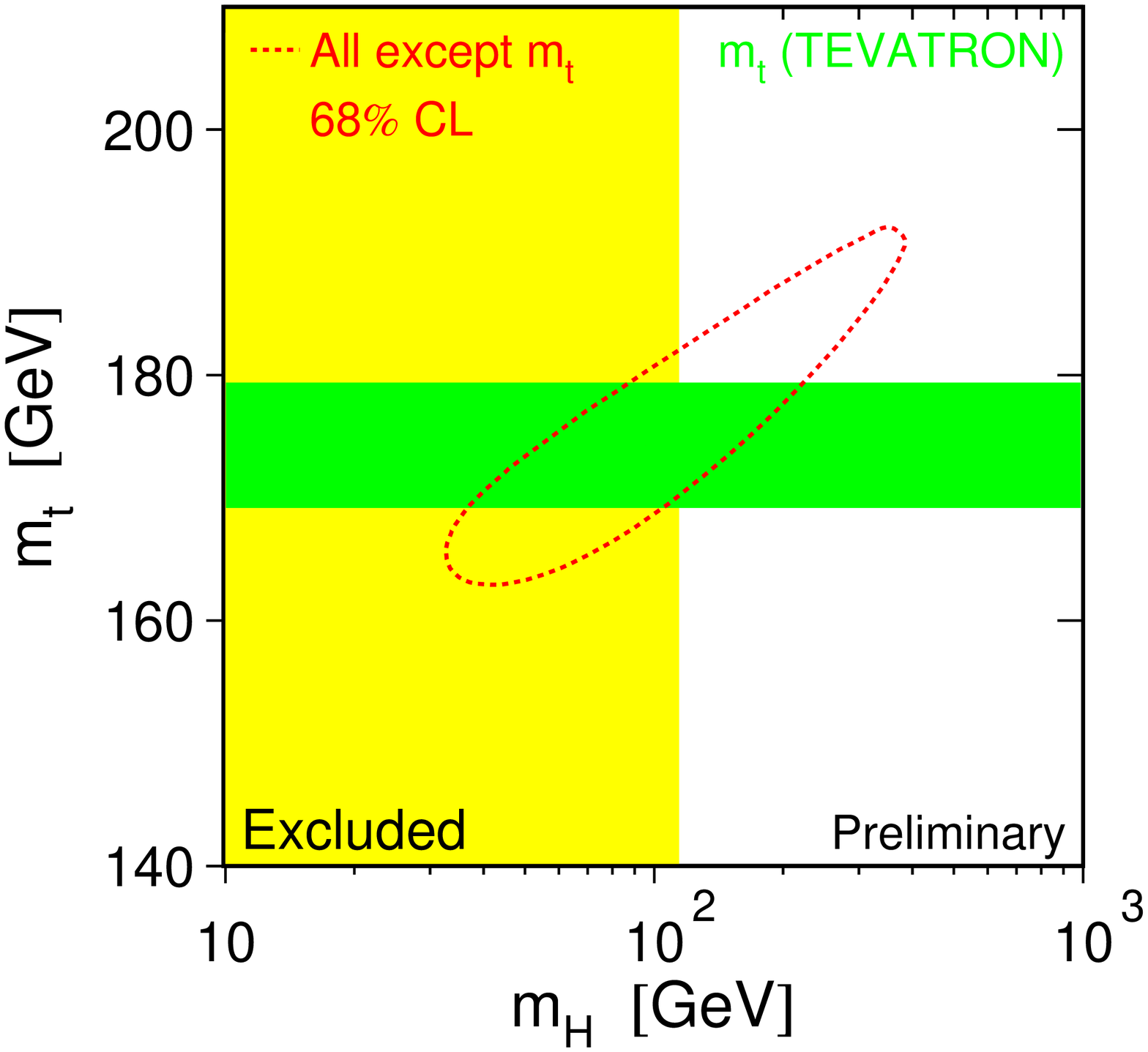}}
\end{center}
\vspace*{-0.6cm}
\caption[]{
  The 68\% confidence level contour in $\Mt$ and $\MH$ for the fit to
  all data except the direct measurement of $\Mt$, indicated by the
  shaded horizontal band of $\pm1$ sigma width.  The vertical band
  shows the 95\% CL exclusion limit on $\MH$ from the direct search.
  }
\label{fig-mhmt}
\end{figure}

For the third fit, the direct $\Mt$ measurement is used to obtain the
best indirect determination of $\MW$.  The result is shown in column~3
of Table~\ref{tab-BIGFIT} and in Figure~\ref{fig-mhmw}.  Also here,
the indirect determination of W boson mass $80.378\pm0.023$ \GeV\ is
in good agreement with the combination of direct measurements from \LEPII\ 
and $\pp$ colliders of $\MW= 80.426\pm0.034$ \GeV.  For the next fit,
(column~4 of Table~\ref{tab-BIGFIT} and Figure~\ref{fig-mhmt}), the
direct $\MW$ and $\GW$ measurements from LEP and $\pp$ colliders are
included to obtain $\Mt= 179^{+11}_{-9}$ \GeV, in very good agreement
with the direct measurement of $\Mt = 174.3\pm5.1$ \GeV. Compared to
the second fit, the error on $\log\MH$ increases due to effects from
higher-order terms.

\begin{figure}[htbp]
\begin{center}
  \mbox{\includegraphics[width=0.9\linewidth]{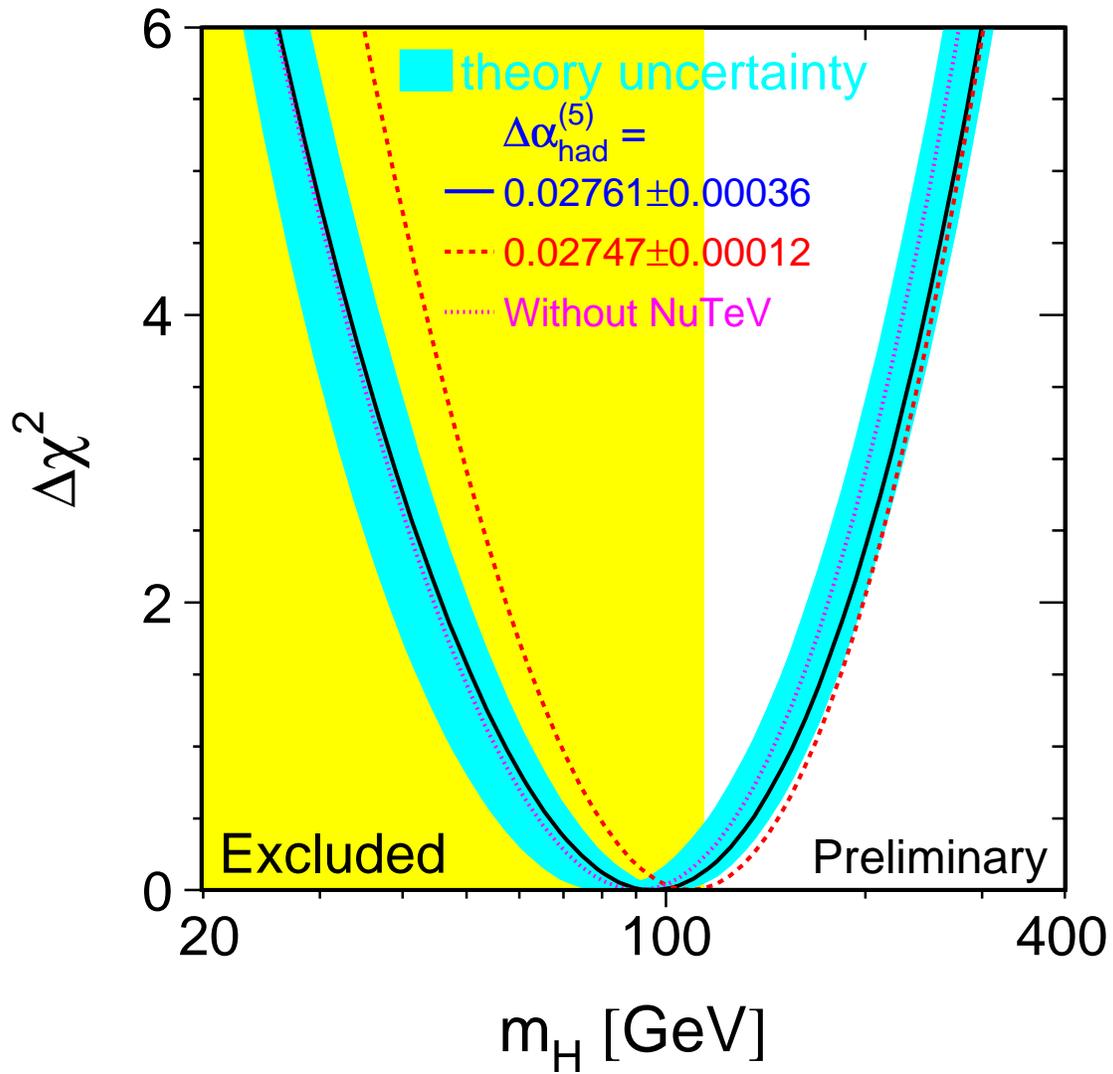}}
\end{center}
\vspace*{-0.6cm}
\caption[]{%
  $\Delta\chi^{2}=\chi^2-\chi^2_{min}$ {\it vs.} $\MH$ curve.  The
  line is the result of the fit using all data (last column of
  Table~\protect\ref{tab-BIGFIT}); the band represents an estimate of
  the theoretical error due to missing higher order corrections.  The
  vertical band shows the 95\% CL exclusion limit on $\MH$ from the
  direct search.  The dashed curve is the result obtained using the
  evaluation of $\Delta\alpha^{(5)}_{\mathrm{had}}(\MZ^2)$ from
  Reference~\citen{bib-TY0102}. }
\label{fig-chiex}
\end{figure}

Finally, the best constraints on $\MH$ are obtained when all data are
used in the fit.  The results of this fit are shown in column~6 of
Table~\ref{tab-BIGFIT} and Figure~\ref{fig-chiex}.  While the
$\chi^2$/d.o.f.{} increases by 8.7 units for the additional degree of
freedom due to the NuTeV result, the fit results themselves are rather
stable when excluding this measurement, as shown in column~5.  In
Figure~\ref{fig-chiex} the observed value of $\Delta\chi^2 \equiv
\chi^2 - \chi^2_{\mathrm{min}}$ as a function of $\MH$ is plotted for
the fit including all data.  The solid curve is the result using
ZFITTER, and corresponds to the last column of Table~\ref{tab-BIGFIT}.
The shaded band represents the uncertainty due to uncalculated
higher-order corrections, as estimated by TOPAZ0 and ZFITTER.
The 95\% confidence level upper limit on $\MH$ (taking the band into
account) is 219 \GeV.  The 95\% C.L. lower limit on $\MH$ of
114.4~\GeV{} obtained from direct searches\cite{ref:LEP-HIGGS} is not
used in the determination of this limit.  Also shown is the result
(dashed curve) obtained when using
$\Delta\alpha^{(5)}_{\mathrm{had}}(\MZ^2)$ of Reference
\citen{bib-TY0102}.

\begin{figure}[p]
\begin{center}
  \mbox{\includegraphics[width=0.9\linewidth]{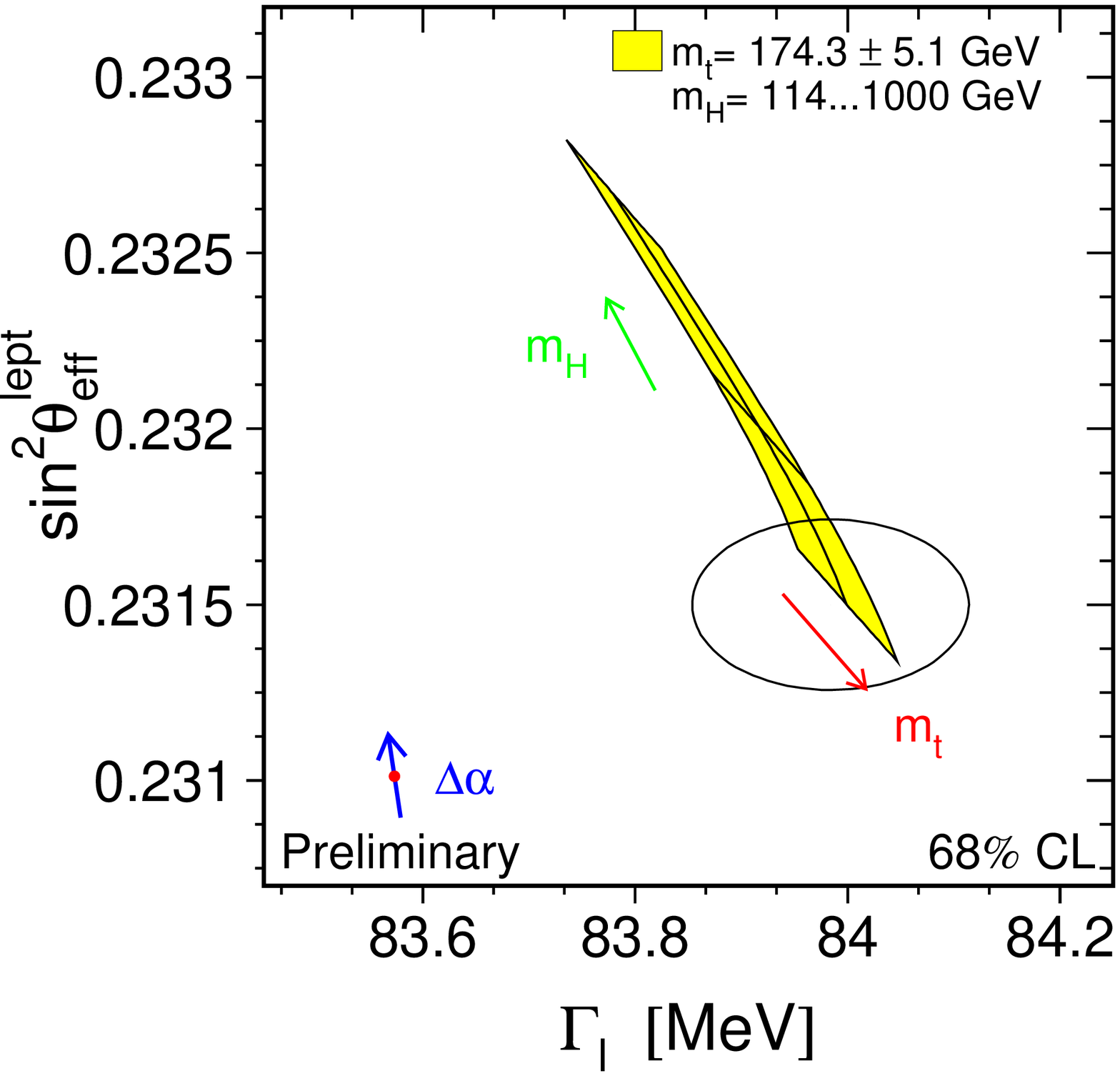}}
\end{center}
\caption[]{%
  $\LEPI$+SLD measurements of $\swsqeffl$ (Table~\ref{tab-swsq}) and
  $\Gll$ (Table~\ref{tab-widths}) and the Standard Model prediction.
  The point shows the predictions if among the electroweak radiative
  corrections only the photon vacuum polarisation is included. The
  corresponding arrow shows variation of this prediction if
  $\alpha(\MZ^2)$ is changed by one standard deviation. This variation
  gives an additional uncertainty to the Standard Model prediction
  shown in the figure.  }
\label{fig-gllsef}
\end{figure}

\begin{figure}[p]
\vspace*{-2.0cm}
\begin{center}
  \mbox{\includegraphics[height=21cm]{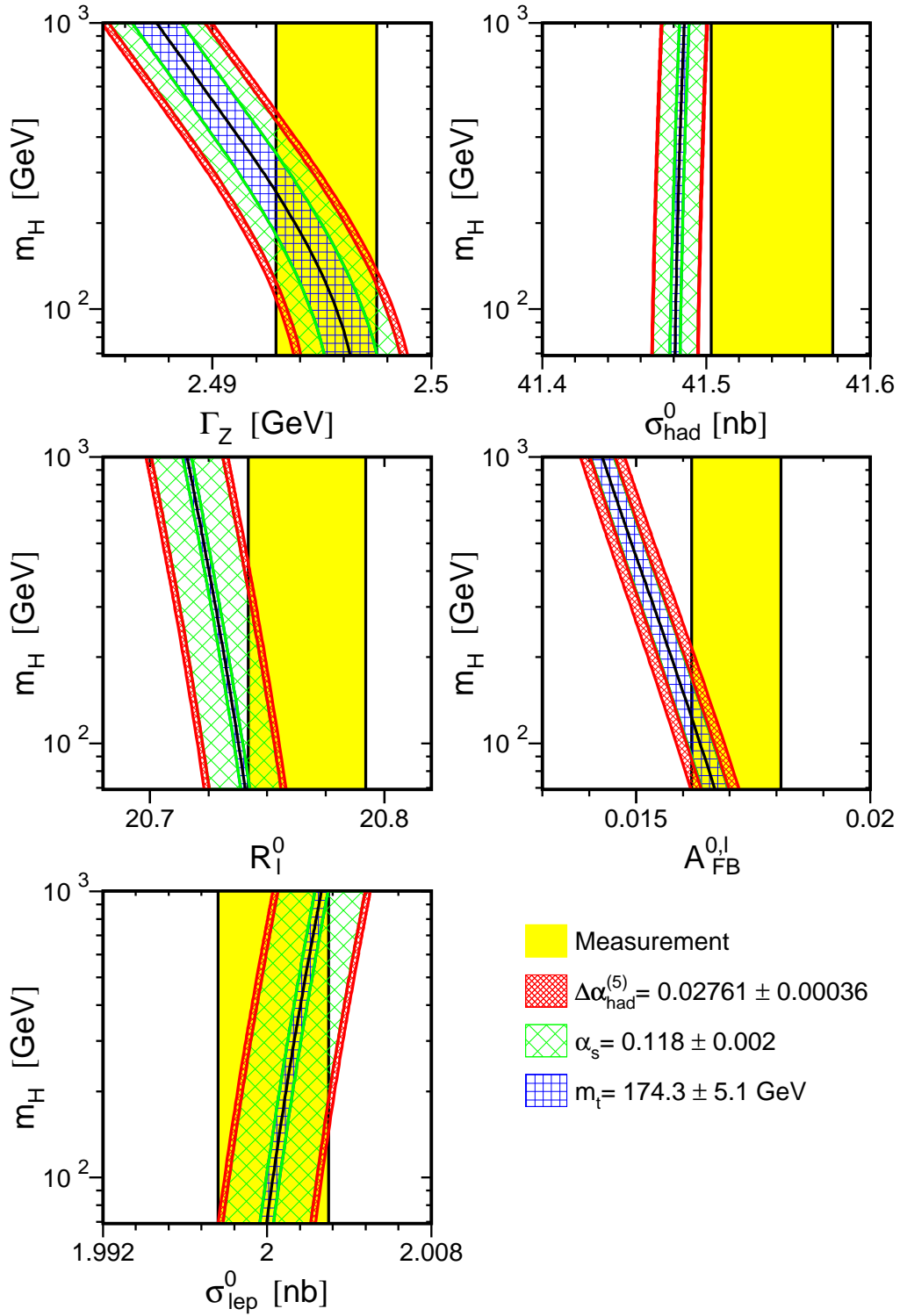}}
\end{center}
\vspace*{-0.6cm}
\caption[]{%
  Comparison of $\LEPI$ measurements with the Standard Model
  prediction as a function of $\MH$.  The measurement with its error
  is shown as the vertical band.  The width of the Standard Model band
  is due to the uncertainties in
  $\Delta\alpha^{(5)}_{\mathrm{had}}(\MZ^2)$, $\alfmz$ and $\Mt$.  The
  total width of the band is the linear sum of these effects.  }
\label{fig-higgs1}
\end{figure}

\begin{figure}[p]
\vspace*{-2.0cm}
\begin{center}
  \mbox{\includegraphics[height=21cm]{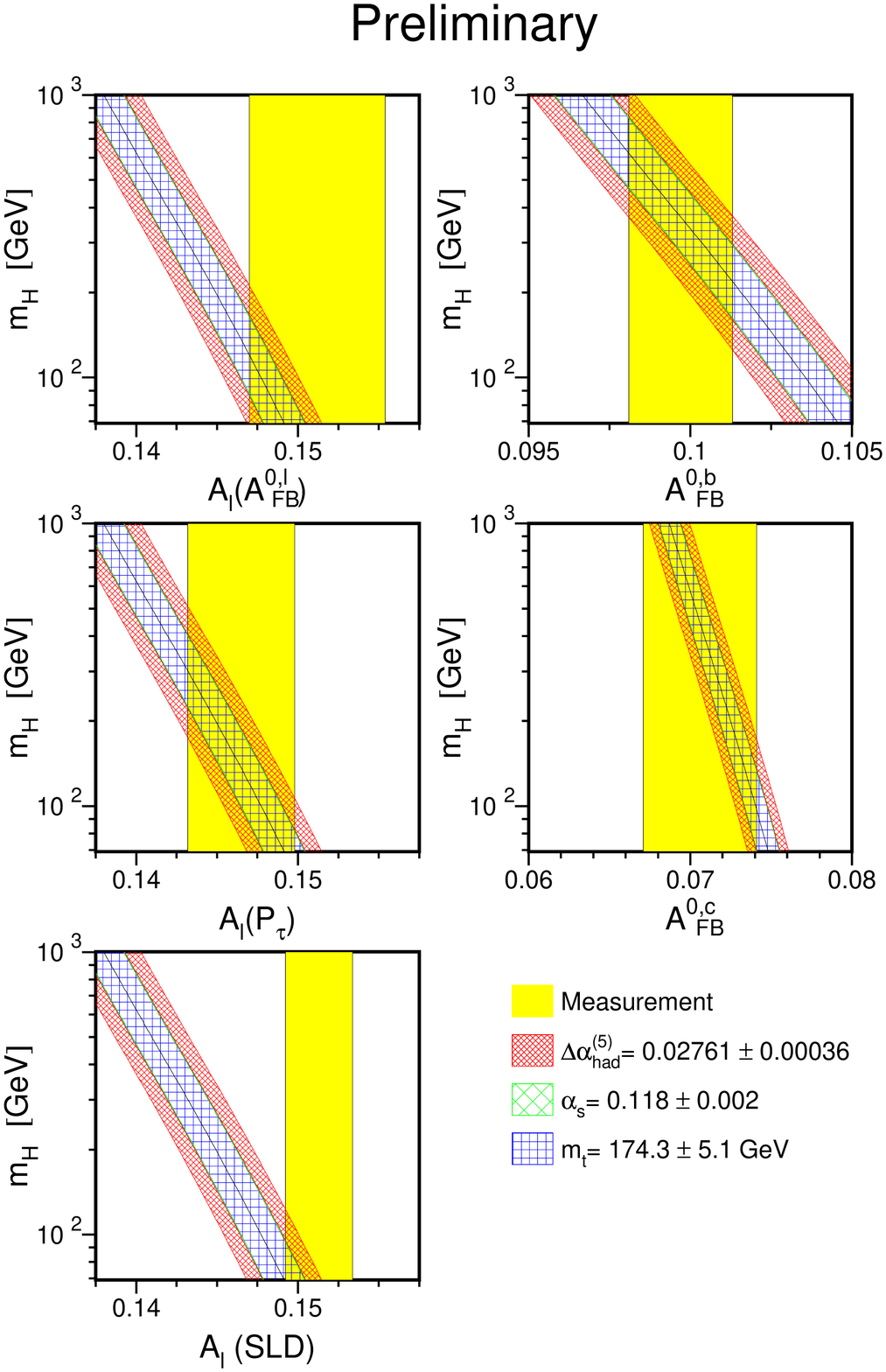}}
\end{center}
\vspace*{-0.6cm}
\caption[]{%
  Comparison of $\LEPI$ measurements with the Standard Model
  prediction as a function of $\MH$.  The measurement with its error
  is shown as the vertical band.  The width of the Standard Model band
  is due to the uncertainties in
  $\Delta\alpha^{(5)}_{\mathrm{had}}(\MZ^2)$, $\alfmz$ and $\Mt$.  The
  total width of the band is the linear sum of these effects.  Also
  shown is the comparison of the SLD measurement of $\cAl$, dominated
  by $\ALRz$, with the Standard Model. }
\label{fig-higgs2}
\end{figure}

\begin{figure}[p]
\vspace*{-2.0cm}
\begin{center}
  \mbox{\includegraphics[height=21cm]{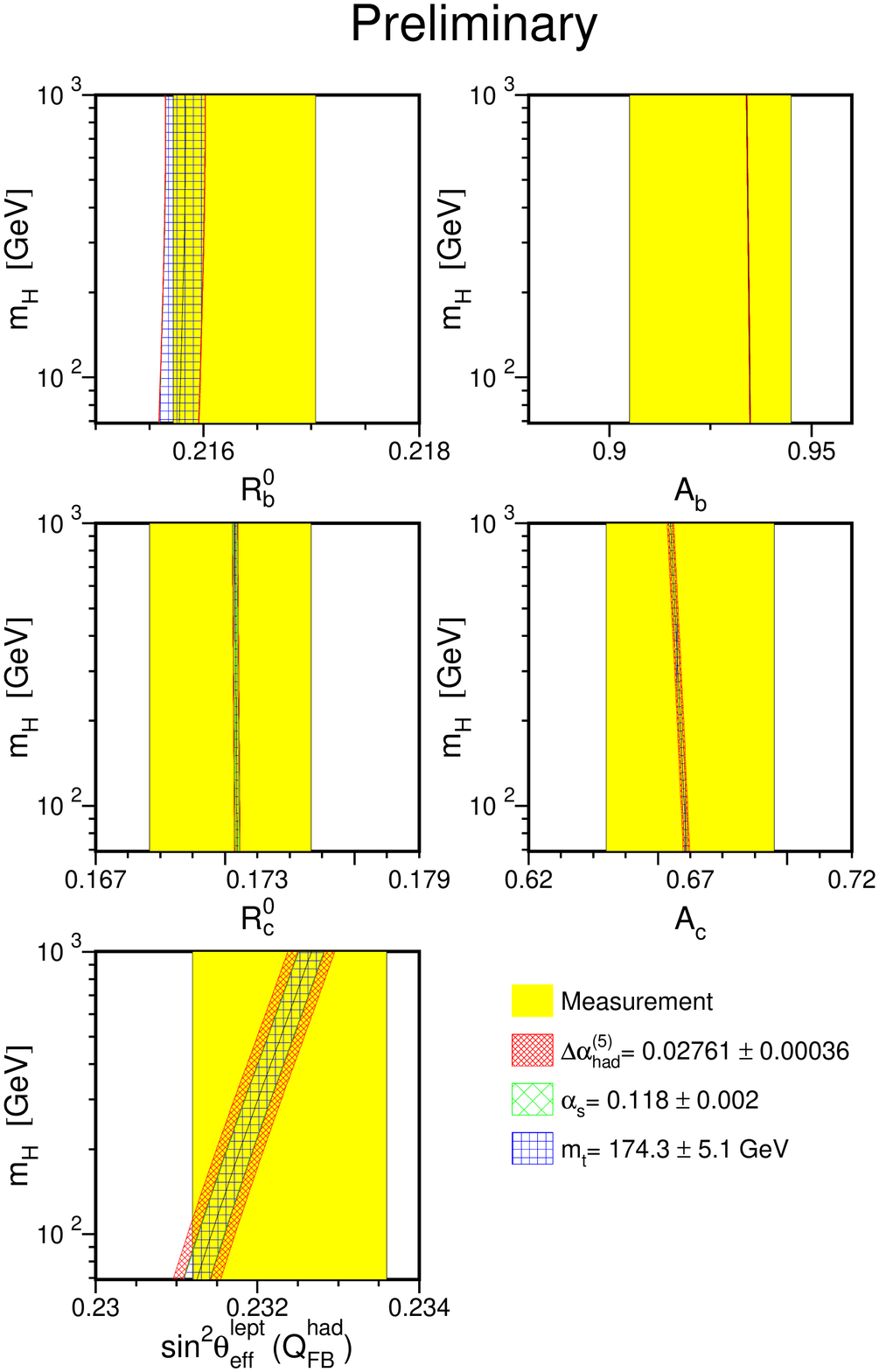}}
\end{center}
\vspace*{-0.6cm}
\caption[]{%
  Comparison of $\LEPI$ and SLD heavy-flavour measurements with the
  Standard Model prediction as a function of $\MH$.  The measurement
  with its error is shown as the vertical band.  The width of the
  Standard Model band is due to the uncertainties in
  $\Delta\alpha^{(5)}_{\mathrm{had}}(\MZ^2)$, $\alfmz$ and $\Mt$.  The
  total width of the band is the linear sum of these effects. Also
  shown is the comparison of the $\LEPI$ measurement of the inclusive
  hadronic charge asymmetry $\Qfbhad$ with the Standard Model. }
\label{fig-higgs3}
\end{figure}

\begin{figure}[p]
\vspace*{-2.0cm}
\begin{center}
  \mbox{\includegraphics[height=21cm]{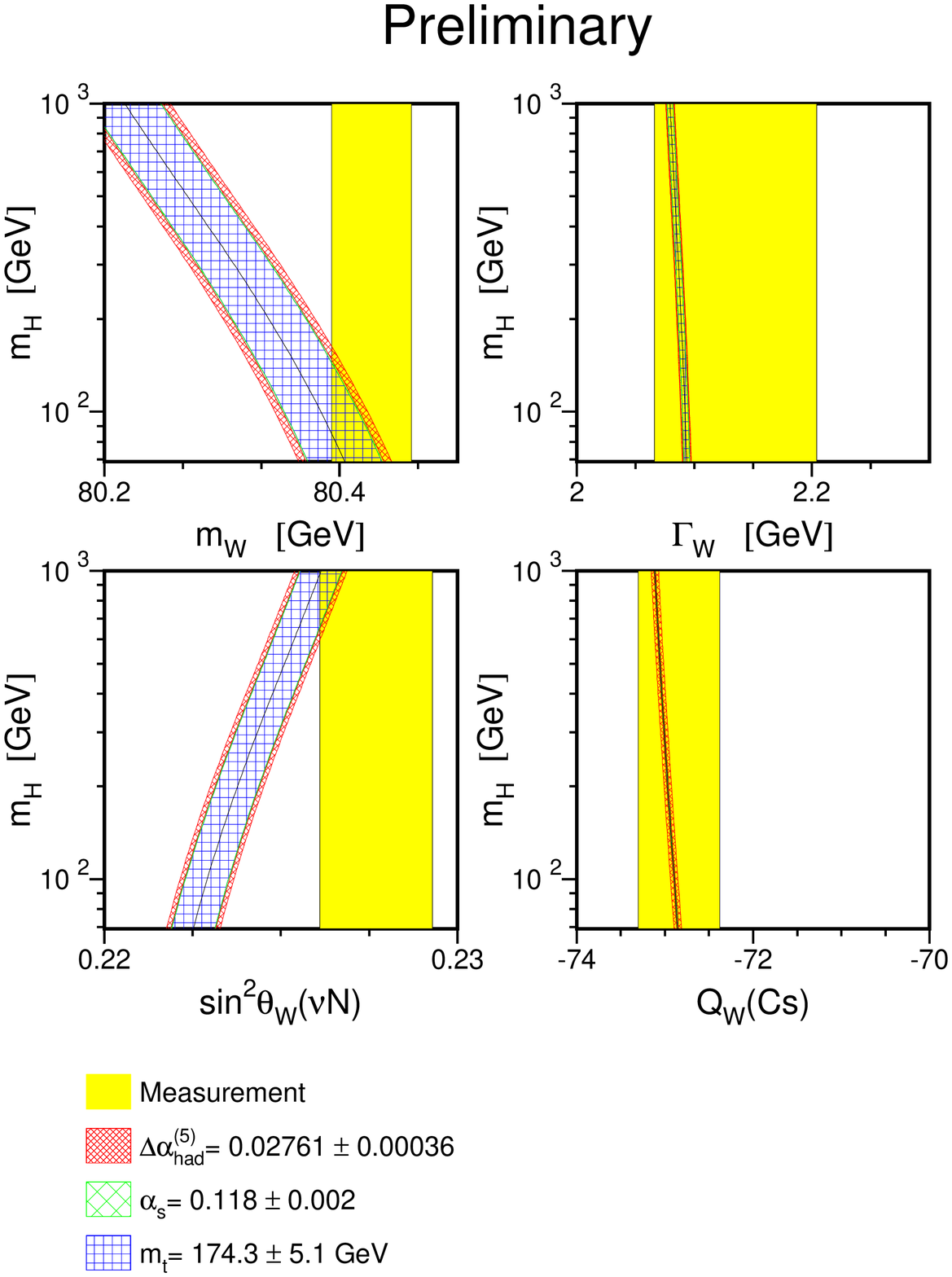}}
\end{center}
\vspace*{-0.6cm}
\caption[]{%
  Comparison of $\MW$ and $\GW$ measured at LEP-2 and $\pp$ colliders,
  of $\swsq$ measured by NuTeV and of APV in caesium with the Standard
  Model prediction as a function of $\MH$.  The measurement with its
  error is shown as the vertical band.  The width of the Standard
  Model band is due to the uncertainties in
  $\Delta\alpha^{(5)}_{\mathrm{had}}(\MZ^2)$, $\alfmz$ and $\Mt$.  The
  total width of the band is the linear sum of these effects.  }
\label{fig-higgs4}
\end{figure}

\begin{figure}[p]
\vspace*{-2.0cm}
\begin{center}
  \mbox{\includegraphics[height=21cm]{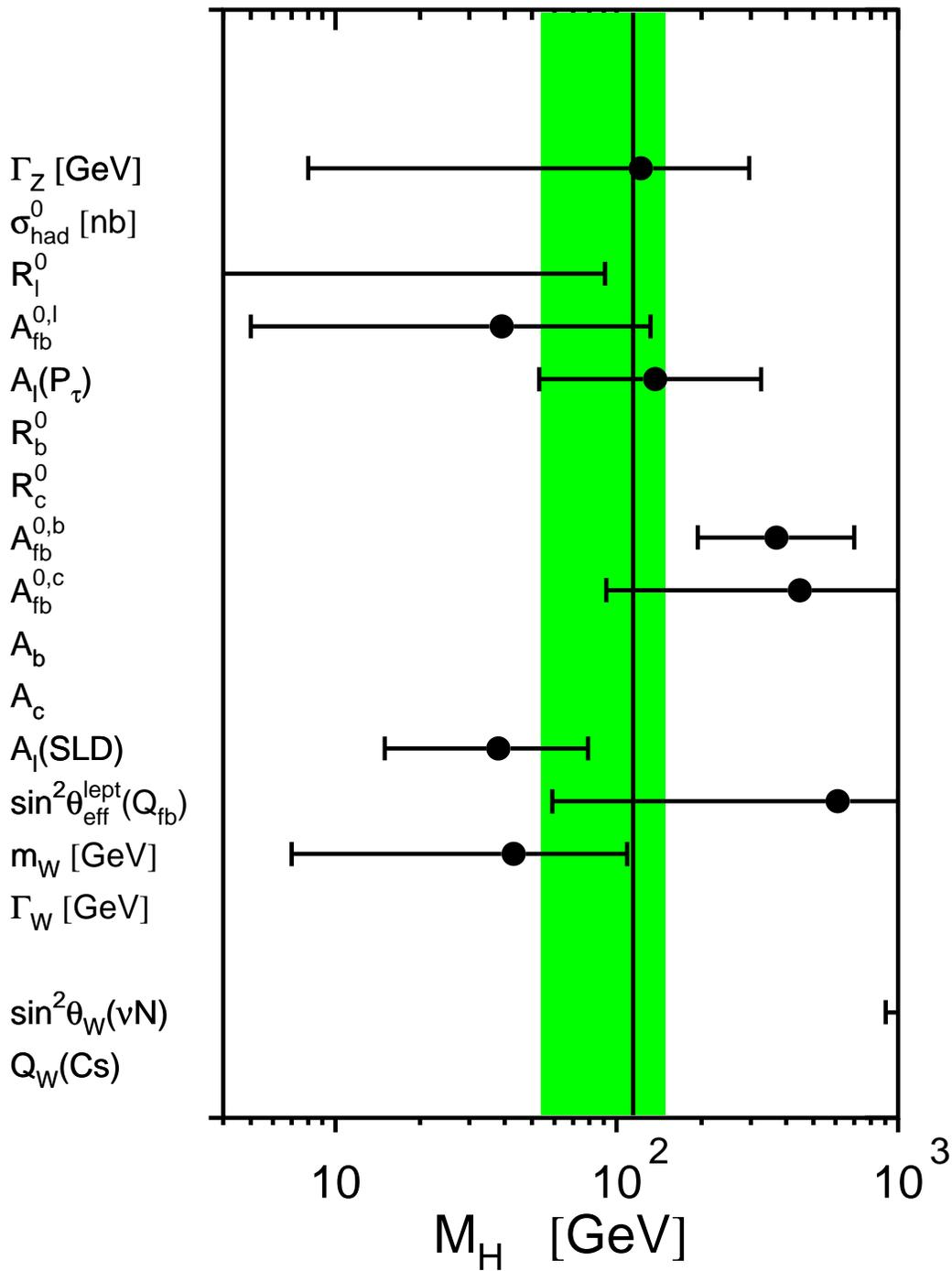}}
\end{center}
\vspace*{-0.6cm}
\caption[]{Constraints on the mass of the Higgs boson from each
  pseudo-observable. The Higgs-boson mass and its 68\% CL uncertainty
  is obtained from a five-parameter Standard Model fit to the
  observable, constraining
  $\Delta\alpha^{(5)}_{\mathrm{had}}(\MZ^2)=0.02761\pm0.00036$,
  $\alfmz=0.118\pm0.002$, $\MZ=91.1875\pm0.0021~\GeV$ and
  $\Mt=174.3\pm5.1~\GeV$. Because of these four common constraints the
  resulting Higgs-boson mass values cannot be combined.  The shaded
  band denotes the overall constraint on the mass of the Higgs boson
  derived from all pseudo-observables including the above four
  Standard Model parameters as reported in the last column of 
  Table~\ref{tab-BIGFIT}.
  }
\label{fig-higgs-obs}
\end{figure}

In Figures~\ref{fig-higgs1} to~\ref{fig-higgs4} the sensitivity of the
LEP and SLD measurements to the Higgs mass is shown.  Besides the
measurement of the W mass, the most sensitive measurements are the
asymmetries, \ie, $\swsqeffl$.  A reduced uncertainty for the value of
$\alpha(\MZ^2)$ would therefore result in an improved constraint on
$\log\MH$ and thus $\MH$, as already shown in Figures~\ref{fig-gllsef}
and \ref{fig-chiex}. Given the constraints on the other four Standard
Model input parameters, each observable is equivalent to a constraint
on the mass of the Standard Model Higgs boson. The constraints on the
mass of the Standard Model Higgs boson resulting from each observable
are compared in Figure~\ref{fig-higgs-obs}.  For vey low Higgs-masses,
these constraints are qualitative only as the effects of real
Higgs-strahlung, neither included in the experimental analyses nor in
the SM calculations of expectations, may then become sizeable.

\vfill

\begin{table}[htbp]
\renewcommand{\arraystretch}{1.5}
  \begin{center}
 \begin{sideways}
 \begin{minipage}[b]{\textheight}
 \begin{center}
    \leavevmode
    \begin{tabular}{|c||c|c|c|c|c|c|}
\hline
  &    - 1 -              &     - 2 -          &      - 3 -            &    - 4 -              &     - 5 -          &      - 6 -    \\
  & LEP including         & all Z-pole         & all Z-pole data       & all Z-pole data       &    all data        & all data      \\[-3mm]
  & $\LEPII$ $\MW$, $\GW$ & data               &    plus   $\Mt$       & plus $\MW$, $\GW$     &    except NuTeV    &               \\
\hline
\hline
$\Mt$\hfill[\GeV] 
& $181^{+13}_{-11}$      & $172^{+12}_{-9}$      & $173.8^{+4.7}_{-4.6}$ & $179^{+11}_{- 9}$      & $175.3^{+4.4}_{-4.3}$ & $174.3^{+4.5}_{-4.4}$  \\
$\MH$\hfill[\GeV] 
& $212^{+338}_{-125}$    & $89^{+122}_{-45}$     & $103^{+65}_{-41}$   & $117^{+163}_{-63}$     & $91^{+55}_{-36}$      & $ 96^{+60}_{-38}$   \\
$\log(\MH/\GeV)$  
& $2.33^{+0.41}_{-0.38}$ & $1.95^{+0.38}_{-0.31}$ &  $2.01^{+0.21}_{-0.22}$ & $2.07^{+0.38}_{-0.33}$ & $1.96^{+0.20}_{-0.22}$ &  $1.98^{+0.21}_{-0.22}$  \\
$\alfmz$          
& $0.1200\pm 0.0030$     & $0.1187\pm 0.0027$ & $0.1188\pm0.0027$   & $0.1187\pm 0.0027$     & $0.1185\pm 0.0027$ & $0.1186\pm0.0027$  \\
\hline
$\chi^2$/d.o.f.{} ($P$)
& $11.0/9~(27\%)$        & $14.7/10~(14\%)$   & $14.8/11~(19\%)$     & $16.5/12~(17\%)$       & $16.7/14~(28\%)$   & $25.4/15~(4.5\%)$  \\
\hline
\hline
$\swsqeffl$ & $\pz0.23165$& $\pz0.23147$& $\pz0.23147$            & $\pz0.23140$& $\pz0.23138$& $\pz0.23143$ \\[-1mm]
            & $\pm0.00018$& $\pm0.00016$& $\pm0.00016$            & $\pm0.00015$& $\pm0.00014$& $\pm0.00014$ \\
$\swsq$     & $\pz0.22311$& $\pz0.22313$& $\pz0.22302$            & $\pz0.22261$& $\pz0.22272$& $\pz0.22289$ \\[-1mm]
            & $\pm0.00053$& $\pm0.00064$& $\pm0.00044$            & $\pm0.00046$& $\pm0.00036$& $\pm0.00036$ \\
$\MW$\hfill[\GeV]
& $80.374\pm0.027$ & $80.373\pm0.032$   & $80.378\pm0.023$   & $80.400\pm0.023$ & $80.394\pm0.019$   & $80.385\pm0.019$     \\
\hline
    \end{tabular}
  \end{center}
    \caption[]{
      Results of the fits to: (1) LEP data alone, (2) all Z-pole data
      (LEP-1 and SLD), (3) all Z-pole data plus direct $\Mt$
      determinations, (4) all Z-pole data plus direct $\MW$ and direct
      $\GW$ determinations, (5) all data (including APV) except NuTeV,
      and (6) all data.  As the sensitivity to $\MH$ is logarithmic,
      both $\MH$ as well as $\log(\MH/\GeV)$ are quoted.  The bottom
      part of the table lists derived results for $\swsqeffl$, $\swsq$
      and $\MW$.  See text for a discussion of theoretical errors not
      included in the errors above.  }
    \label{tab-BIGFIT}
\end{minipage}
 \end{sideways}
 \end{center}
\renewcommand{\arraystretch}{1.0}
\end{table}
\vfill

\clearpage


\boldmath
\chapter{Conclusions}
\label{sec-Conc}
\unboldmath

The combination of the many precise electroweak results yields
stringent constraints on the Standard Model.  In addition, the results
are sensitive to the Higgs mass.  Most measurements agree well with
the predictions.  The spread in values of the various determinations
of the effective electroweak mixing angle is somewhat larger than expected.
Within the Standard Model analysis, this seems to be caused by the
measurement of the forward-backward asymmetry in b-quark production,
showing the largest pull of all Z-pole measurements w.r.t. the
Standard Model expectation.  The final result of the NuTeV collaboration on
the electroweak mixing angle differs more, by close to three standard
deviations from the Standard Model expectation calculated based on all
other precision electroweak measurements.

The LEP and SLD experiments wish to stress that this report reflects a
preliminary status of their results at the time of the 2003 summer
conferences.  A definitive statement on these results must wait for
publication by each collaboration.

\boldmath
\section*{Prospects for the Future}
\label{sec-Future}
\unboldmath

Most of the measurements from data taken at or near the Z resonance,
both at LEP as well as at SLC, that are presented in this report are
either final or are being finalised.  The main improvements will
therefore take place in the high energy data, with more than
700~pb$^{-1}$ per experiment.  The measurements of $\MW$ are likely to
reach a precision not too far from the uncertainty on the prediction
obtained via the radiative corrections of the \Zzero{} data, providing
a further important test of the Standard Model.  In the measurement of
the triple and quartic electroweak gauge boson self couplings, the
analysis of the complete $\LEPII$ statistics, together with the
increased sensitivity at higher beam energies, will lead to an
improvement in the current precision.
\section*{Acknowledgements}

We would like to thank the CERN accelerator divisions for the
efficient operation of the LEP accelerator, the precise information on
the absolute energy scale and their close cooperation with the four
experiments.  The SLD collaboration would like to thank the SLAC
accelerator department for the efficient operation of the SLC
accelerator.  We would also like to thank members of the CDF, D\O{}
and NuTeV collaborations for making results available to
us in advance of the conferences and for useful discussions concerning
their combination.  Finally, the results of the section on Standard
Model constraints would not be possible without the close
collaboration of many theorists.

\clearpage

\begin{appendix}

\input{s03_hftab}
\input{s03_hffit}

\input{4f_app}

\input{cr_app}

\end{appendix}

\clearpage

\bibliographystyle{lep2unsrt}
\bibliography{s03_ew,s03_hf,common,gg,ff,smat,fsi,be,4f_s03,gc,mw}

\vfill

\section*{Links to LEP results on the World Wide Web}
The physics notes describing the preliminary results of the four LEP
experiments submitted to the 2003 summer conferences, as well as
additional documentation from the LEP electroweak working
group are available on the World Wide Web at: \\[2mm]
\begin{tabular}{ll}
  ALEPH:   &
  {\tt http://alephwww.cern.ch/ALPUB/oldconf/oldconf\_03.html}\\
  DELPHI:  &
  {\tt http://delphiwww.cern.ch/\~{}pubxx/delsec/conferences/summer03/}\\
  L3:      &
  {\tt http://l3.web.cern.ch/l3/conferences/Aachen2003/}\\
  OPAL:    &
  {\tt http://opal.web.cern.ch/Opal/pubs/eps2003/abstr.html}\\     
  LEP-EWWG: &
  {\tt http://cern.ch/LEPEWWG/}\\
\end{tabular}

\end{document}

%% file: s03_hf.tex

\updates{DELPHI has finalised their b-asymmetry using jet- and vertex charge
tagging. OPAL has published a new analysis of the b- and c-asymmetry with 
leptons.}

\section{Introduction}

The relevant quantities in the heavy quark sector at \LEPI/SLD which are
currently determined by the combination procedure are:
\begin{itemize}
\item The ratios
  of the b and c quark partial widths of the Z to its total hadronic
  partial width: $\Rbz \equiv \Gbb / \Ghad$ and $\Rcz \equiv \Gcc /
  \Ghad$. (The symbols \Rb, \Rc{} are used to denote the
  experimentally measured ratios of event rates or cross sections.)
\item The forward-backward asymmetries, \Abb{} and \Acc.
\item The final state coupling parameters $\cAb,\,\cAc$ obtained from the
  left-right-forward-backward asymmetry at SLD.
\item The semileptonic branching ratios, $\Brbl$, $\Brbclp$ and $\Brcl$, and
  the average time-integrated $\Bzero\Bzerob$ mixing parameter, $\chiM$. 
  These are often determined at the same time or with similar methods
  as the asymmetries.
  Including them in the combination greatly reduces the errors.
  For example $\chiM$ parameterises the probability that a b-quark decays
  into a negative lepton which is the charge tagging efficiency in the
  asymmetry analyses. For this reason the errors coming from the mixture of
  different lepton sources in $\bb$ events cancel largely in the asymmetries
  if they are analyses together with $\chiM$.
\item The probability that a c quark produces a $\Dplus$, $\Ds$, 
 $\Dstarp$ meson\footnote{%
   Actually the product $\PcDst$ is fitted because this quantity is
   needed and measured by the LEP experiments.}  or a charmed baryon.
 The probability that a c quark fragments into a $\Dzero$ is
 calculated from the constraint that the probabilities for the weakly
 decaying charmed hadrons add up to one.  
\end{itemize}
A full description of the averaging procedure is published in \cite{ref:lephf};
the main motivations for the procedure are outlined here.  
Several analyses measure
more than one parameter simultaneously, for example the asymmetry measurements
with leptons or D mesons.
Some of the measurements of electroweak parameters depend explicitly
on the values of other parameters, for example \Rb{} depends on \Rc.
The common tagging and analysis techniques lead to common sources of
systematic uncertainty, in particular for the double-tag measurements
of \Rb.  The starting point for the combination is to ensure that all
the analyses use a common set of assumptions for input parameters
which give rise to systematic uncertainties.  
The input parameters are updated
and extended \cite{ref:lephfnew} to accommodate new analyses
and more recent measurements.  The correlations and interdependencies
of the input measurements are then taken into account in a $\chi^2$
minimisation which results in the combined electroweak parameters and
their correlation matrix.

\section{Summary of Measurements and Averaging Procedure}

All measurements are presented by the LEP and SLD collaborations in
a consistent manner for the purpose of combination.
The tables prepared by the experiments include a detailed breakdown of
the systematic error of each measurement and its dependence on other
electroweak parameters. Where necessary, the experiments apply small
corrections to their results in order to use agreed values and ranges
for the input parameters to calculate systematic errors.  The
measurements, corrected where necessary, are summarised in
Appendix~\ref{app-HF-tab} in Tables~\ref{tab:Rbinp}--\ref{tab:RcPcDstinp},
where the statistical and systematic errors are quoted separately.
The correlated systematic entries are from physics sources shared with one or
more other results in the tables and are derived from the full
breakdown of common systematic uncertainties. The uncorrelated
systematic entries come from the remaining sources.


\subsection{Averaging Procedure}

A $\chi^2$ minimisation procedure is used to derive the values of the
heavy-flavour electroweak parameters,  following the procedure
described in 
Reference~\citen{ref:lephf}.  The full statistical and systematic
covariance matrix for all measurements is calculated.  This
correlation matrix takes into account correlations between different measurements
of one experiment and between different experiments.  The
explicit dependence of each measurement on the other parameters is
also accounted for.  

Since c-quark events form the main background in the \Rb{} analyses,
the value of \Rb{} depends on the value of \Rc. If \Rb{} and \Rc{} were
measured in the same analysis, this would be reflected in the correlation
matrix for the results.  However the analyses do not determine \Rb\ 
and \Rc\ simultaneously but instead measure \Rb\ for an assumed value
of \Rc. In this case the dependence is parameterised as
\begin{eqnarray}
 \Rb & = & 
 \Rb^{\rm{meas}} + a(\Rc) \frac {(\Rc - \Rc^{\rm{used}} )} {\Rc}.
\label{eq:rbrc}
\end{eqnarray}
In this expression, $\Rb^{\rm{meas}}$ is the result of the analysis
assuming a value of $\Rc = \Rc^{\rm{used}}$. The values of
$\Rc^{\rm{used}}$ and the coefficients $a(\Rc)$ are given in
Table~\ref{tab:Rbinp} where appropriate. The dependence of all other
measurements on other electroweak parameters is treated in the same
way, with coefficients $a(x)$ describing the dependence on parameter
$x$.

\subsection{Partial Width Measurements}

The measurements of \Rb{} and \Rc{} fall into two categories. In the
first, called a single-tag measurement, a method to select b or c
events is devised, and the number of tagged events is counted. This
number must then be corrected for backgrounds from other flavours and
for the tagging efficiency to calculate the true fraction of hadronic
\Zzero{} decays of that flavour. The dominant systematic errors come
from understanding the branching ratios and detection efficiencies
which give the overall tagging efficiency. For the second technique,
called a double-tag measurement, each event is divided into two
hemispheres.  With $N_t$ being the number of tagged hemispheres,
$N_{tt}$ the number of events with both hemispheres tagged and
$N_{\rm{had}}$ the total number of hadronic \Zzero{} decays one has
\begin{eqnarray}
   \frac{N_t}{2N_{\rm{had}}} &=& \effb \Rb
                        + \effc  \Rc +
                        \effuds ( 1 - \Rb - \Rc ) ,\\
   \frac{N_{tt}}{N_{\rm{had}}} &=& \Cb \effb^2 \Rb
                +    \Cc \effc^2 \Rc +
                          {\cal C}_{\mathrm{uds}} \effuds^2 ( 1 - \Rb - \Rc ) ,
\end{eqnarray}
where $\effb$, $\effc$ and $\effuds$ are the tagging efficiencies per
hemisphere for b, c and light-quark events, and $\Cq \ne 1$ accounts
for the fact that the tagging efficiencies between the hemispheres may
be correlated.  In the case of \Rb{} one has $\effb\gg\effc\gg\effuds$,
$\Cb \approx 1$. The correlations for the other flavours can be
neglected. These equations can be solved to give \Rb{} and $\effb$.
Neglecting the c and uds backgrounds and the correlations, they are
approximately given by
\begin{eqnarray}
\effb &\approx& 2 N_{tt} / N_t  , \\
\Rb &\approx& N_t^2 / (4N_{tt}N_{\rm{had}}).
\end{eqnarray}
The double-tagging method has the advantage that the b tagging
efficiency is derived 
from the data, reducing the systematic
error. The residual background of other flavours in the sample, and
the evaluation of the correlation between the tagging efficiencies in
the two hemispheres of the event are the main sources of systematic
uncertainty in such an analysis.

In the standard approach each hemisphere is simply tagged as b or
non-b. This method can be enhanced by using more tags. All additional 
efficiencies can be determined from the data, reducing the statistical 
uncertainties without adding new systematic uncertainties.

Small corrections must be applied to the results to obtain the partial
width ratios \Rbz{} and \Rcz{} from the cross section ratios \Rb{} and \Rc{}.
These corrections depend slightly on the 
invariant mass cutoff of the simulations used by the experiments;
they are applied by the collaborations before the combination.

The partial width measurements included are:
\begin{itemize}
\item Lifetime (and lepton) double-tag measurements for \Rb{} from
  ALEPH\cite{ref:alife}, DELPHI\cite{ref:drb}, L3\cite{ref:lrbmixed},
  OPAL\cite{ref:omixed} and SLD\cite{ref:SLD_R_B}.  These are the most
  precise determinations of \Rb.
  Since they completely dominate the combined result, no other \Rb{}
  measurements are used at present.
  The basic features of the double-tag technique are discussed above.
  In the ALEPH, DELPHI, OPAL and SLD measurements the charm rejection is
  enhanced by using the invariant mass information. DELPHI, OPAL and SLD
  also add kinematic information from the particles at the
  secondary vertex.
  The ALEPH and DELPHI measurements make use of several different
  tags, which reduces largely the statistical error. This allows
  to cut harder in the primary b-tag so that, due to the higher
  b-purity, also the systematic uncertainties are reduced.
\item Analyses with D/$\Dstarpm$ mesons to measure \Rc{} from
  ALEPH, DELPHI and OPAL.
  All measurements are constructed in such a way that no assumptions about
  charm fragmentation are necessary as these are determined from the
  \LEPI\ data.  The
  available measurements can be divided into three groups:
\begin{itemize}
\item inclusive/exclusive double tag (ALEPH\cite{ref:arcd}, 
  DELPHI\cite{ref:drcd,ref:drcc}, OPAL\cite{ref:orcd}): In a first
  step $\Dstarpm$ mesons are reconstructed in the decay channel
  ${\rm D}^{*+} \rightarrow \pi^+ \Dzero$
  using several decay channels of the $\Dzero$
  and their production rate is measured\footnote{
    If not explicitely mentioned charge conjugate states are always included},
  which depends on the product
  $\Rc \times \PcDst$.  This sample of $\cc$ (and $\bb$) events is
  then used to measure $\PcDst$ using a slow pion tag in the opposite
  hemisphere.  In the ALEPH measurement only \Rc{} is given and
  no explicit $\PcDst$ is available. 
\item exclusive double tag (ALEPH\cite{ref:arcd}): 
  This analysis uses exclusively
  reconstructed $\Dstarp$, $\Dzero$ and $\Dplus$ mesons in different
  decay channels. It has lower statistics but better purity than the
  inclusive analyses.
\item reconstruction of all weakly decaying charmed states
  (ALEPH\cite{ref:arcc},  DELPHI\cite{ref:drcc}, OPAL\cite{ref:orcc}): 
  These analyses make the assumption that the production fractions
  of $\Dzero$, $\Dplus$, $\Ds$ and $\Lc$ 
  in c-quark jets of $\cc$ events add up to one with small corrections
  due to unmeasured charmed strange baryons.
  This is a single tag measurement, relying only on knowing
  the decay branching ratios of the charm hadrons.  
  These analyses are also used to measure the c hadron production
  ratios which are needed for the \Rb{} analyses.  
\end{itemize}
\item A lifetime plus mass double tag from SLD to measure
  \Rc\cite{ref:SLD_R_C}.  This analysis uses the same tagging
  algorithm as the SLD \Rb{} analysis, but with the neural net tuned to
  tag charm. Although the
  charm tag has a purity of about 84\%, most of the background is from
  b which can be measured with high precision from the b/c mixed tag rate.
\item A measurement of \Rc{} using single leptons assuming $\Brcl$ from
  ALEPH \cite{ref:arcd}.
\end{itemize}
To avoid effects from nonlinearities in the fit, for the inclusive/exclusive
single/double tag and for the charm-counting analyses, the products
\RcPcDst, \RcfDz, \RcfDp, \RcfDs{} and \RcfLc that are actually
measured in the analyses are directly used as inputs to the fit.
The measurements of the production rates of weakly decaying charmed
hadrons, especially \RcfDs{} and \RcfLc{} have substantial errors due
to the uncertainties in the branching ratios of the decay mode used. 
These errors are relative so that the absolute errors are smaller
when the measurements fluctuate downwards, leading to a potential bias 
towards lower averages.
To avoid this bias, for the production rates of weakly decaying charmed 
hadrons the logarithm of the production rates instead of the rates themselves
are input to the fit. For \RcfDz{} and \RcfDp{} the difference between
the results using the logarithm or the value itself is negligible. For
\RcfDs{} and \RcfLc{} the difference in the extracted value of $\Rc$ 
is about one
tenth of a standard deviation.

\subsection{Asymmetry Measurements}
\label{sec:asycorrections}
All b and c asymmetries given by the experiments correspond to full
acceptance.

The QCD corrections to the forward-backward asymmetries depend
strongly on the experimental analyses.  For this reason the numbers
given by the collaborations are also corrected for QCD effects. A
detailed description of the procedure can be found
in \cite{ref:afbqcd} with updates reported in \cite{ref:lephfnew}.

For the heavy-flavour combinations described in this chapter, the LEP peak and
off-peak asymmetries are corrected to $ \sqrt {s} = 91.26$ \GeV{}
using the predicted dependence from ZFITTER\cite{ref:ZFITTER}. The
slope of the asymmetry around $\MZ$ depends only on the axial coupling
and the charge of the initial and final state fermions and is thus
independent of the value of the asymmetry itself, i.e., the effective
electroweak mixing angle.

After calculating the overall averages, the quark pole asymmetries
$\Afbzq$, defined in terms of effective couplings, are derived from
the measured asymmetries by applying corrections as listed in
Table~\ref{tab:aqqcor}. These corrections are due to the energy shift
from 91.26 \GeV{} to $\MZ$, initial state radiation, $\gamma$ exchange
and $\gamma$-$\Zzero$ interference.  A very small correction due to
the nonzero value of the b quark mass is included in the last
correction.  All corrections are calculated using ZFITTER.

\begin{table}[bhtb]
\begin{center}
\begin{tabular}{|l||l|l|}
\hline
Source   & $\delta A_{\mathrm{FB}}^{\mathrm{b}}$
         & $\delta A_{\mathrm{FB}}^{\mathrm{c}}$ \\
\hline
\hline
$\sqrt{s} = \MZ $       & $ -0.0013 $  & $ -0.0034$  \\
QED corrections         & $ +0.0041 $  & $ +0.0104$  \\
$\gamma$, $\gamma$-$\Zzero$, mass & $ -0.0003 $  & $ -0.0008$  \\
\hline
\hline
Total                   & $ +0.0025 $  & $ +0.0062$  \\
\hline
\end{tabular}
\end{center}
\caption[]{%
  Corrections to be applied to the quark asymmetries as 
   $A_{\mathrm{FB}}^0 = A_{\mathrm{FB}}^{\mathrm{meas}}
  + \delta A_{\mathrm{FB}}$.}
\label{tab:aqqcor}
\end{table}

The SLD left-right-forward-backward asymmetries are also corrected for all
radiative effects and are directly presented in terms of $\cAb$ and $\cAc$.

The measurements used are:
\begin{itemize}
\item Measurements of \Abb{} and \Acc{} using leptons from 
  ALEPH\cite{ref:alasy}, DELPHI\cite{ref:dlasy}, L3\cite{ref:llasy} and
  OPAL\cite{ref:olasy}.  
  These analyses measure either \Abb{} only or \Abb{} and \Acc{} 
  from a fit to the lepton
  spectra. In the case of OPAL the lepton information is combined
  with hadronic variables in a neural net. DELPHI uses in addition lifetime
  information and jet-charge in the hemisphere opposite to the lepton to
  separate the different lepton sources.
  Some asymmetry analyses also measure $\chiM$ within the same analysis.
\item Measurements of \Abb{} based on lifetime tagged events with a
  hemisphere charge measurement from ALEPH\cite{ref:ajet}, 
  DELPHI\cite{ref:dnnasy}, L3\cite{ref:ljet} and OPAL\cite{ref:ojet}.
  These measurements dominate the combined result. 
  The DELPHI measurement using jetcharge only \cite{ref:djasy} is no longer
  used in the combination because of the large correlation with the
  measurement combining the jetcharge with vertex related variables in a 
  neural net \cite{ref:dnnasy}.
\item Analyses with D mesons to measure \Acc{} from
  ALEPH\cite{ref:adsac} or \Acc{} and \Abb{} from
  DELPHI\cite{ref:ddasy} and OPAL\cite{ref:odsac}.
\item Measurements of \cAb{} and \cAc{} from SLD.
  These results include measurements using 
  lepton \cite{ref:SLD_AQL}, 
  D meson \cite{ref:SLD_ACD} and 
  vertex mass plus hemisphere charge \cite{ref:SLD_ABJ} 
  tags, which have similar sources of
  systematic errors as the LEP asymmetry measurements. 
  SLD also uses vertex mass for bottom or charm tagging in conjunction
  with a kaon tag or a vertex charge tag for both $\cAb$ and $\cAc$ 
  measurements \cite{ref:SLD_ABK,ref:SLD_vtxasy,ref:SLD_ACV}.
\end{itemize}
Since all asymmetry measurements use the full event sample the analyses using
different techniques from the same collaboration are statistically correlated.
These correlations are evaluated by the experiments and included in the
combination procedure. The correlations between the b-asymmetry measurements
with jetcharge and with leptons range between 6\% and 30\%.

\subsection{Other Measurements}

The measurements of the charmed hadron fractions $\PcDst$, $\fDp$, $\fDs$
and $\fcb$ are included in the \Rc{} measurements and are described there.

ALEPH\cite{ref:abl}, DELPHI\cite{ref:dbl}, L3\cite{ref:lbl,ref:lrbmixed} and 
OPAL\cite{ref:obl} measure $\Brbl$, $\Brbclp$ and $\chiM$ or a subset of them
from a sample of leptons opposite to a b-tagged hemisphere and from a
double lepton sample. 
DELPHI\cite{ref:drcd} and OPAL\cite{ref:ocl} measure $\Brcl$ from a sample
opposite to a high energy $\Dstarpm$.
%
\section{Results}\label{sec-HFSUM}

\begin{figure}[htbp]
\vspace*{-0.6cm}
\begin{center}
  \mbox{\includegraphics[width=0.9\linewidth]{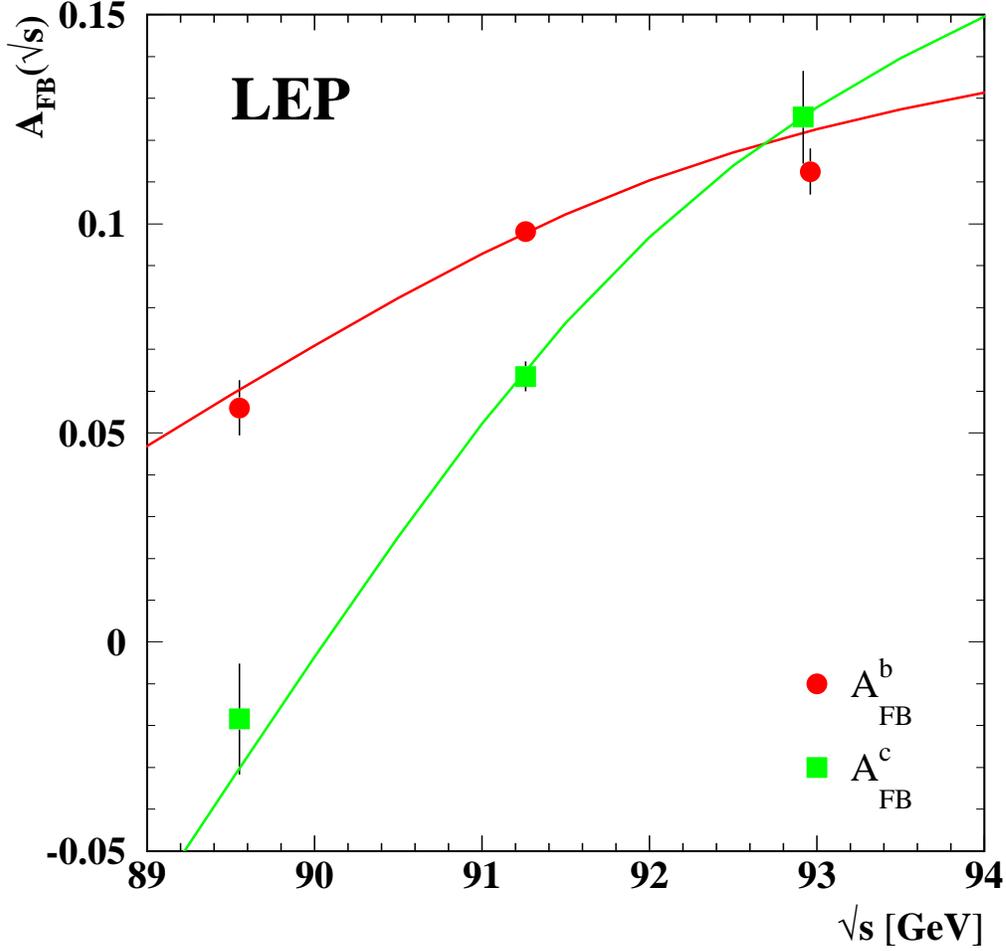}}
\end{center}
\caption[]{%
  Measured asymmetries for b and c quark final states as a function of
  the centre-of-mass energy. The Standard-Model expectations are shown
  as the lines calculated for $\Mt=175~\GeV$ and $\MH=300~\GeV$.  }
\label{fig-afbene}
\end{figure}

In a first fit the asymmetry measurements on peak, above peak and
below peak are corrected to three common centre-of-mass energies and
are then  combined at each energy point. The results of
this fit, including the SLD results, are given in
Appendix~\ref{app-HF-fit}.  The dependence of the average asymmetries on
centre-of-mass energy agrees with the prediction of the Standard
Model, as shown in Figure~\ref{fig-afbene}.
A second fit is made to derive the pole asymmetries $\Afbzq$ from the
measured quark asymmetries, in which all the off-peak asymmetry
measurements are corrected to the peak energy before combining. This fit
determines a total of 14 parameters:
the two partial widths, two LEP asymmetries, 
two coupling parameters from SLD,
three semileptonic branching ratios, the average mixing parameter and the
probabilities for c quark to fragment into a $\Dplus$, a $\Ds$, a
$\Dstarp$, or a charmed baryon.
If the SLD measurements are excluded from the fit there are 12 parameters to
be determined.
Results for
the non-electroweak parameters are independent of the
treatment of the off-peak asymmetries and the SLD data.

\subsection{Results of the 12-Parameter Fit to the LEP Data}
\label{sec-HFSUM-LEP}

Using the full averaging procedure gives the following combined
results for the electroweak parameters:
\begin{eqnarray}
  \label{eqn-hf4}
  \Rbz    &=& 0.21643 \pm  0.00073  \\
  \Rcz    &=& 0.1690  \pm  0.0047   \nonumber \\
  \Afbzb  &=& 0.0998  \pm  0.0016   \nonumber \\
  \Afbzc  &=& 0.0702  \pm  0.0036   \,,\nonumber
\end{eqnarray}
where all corrections to the asymmetries and partial widths are
applied.  The $\chi^2/$d.o.f.{} is $49/(96-12)$. The corresponding
correlation matrix is given in Table~\ref{tab:12parcor}.

\begin{table}[htbp]
\begin{center}
\begin{tabular}{|l||rrrr|}
\hline
&\makebox[1.2cm]{\Rbz}
&\makebox[1.2cm]{\Rcz}
&\makebox[1.2cm]{$\Afbzb$}
&\makebox[1.2cm]{$\Afbzc$}\\
\hline
\hline
\Rbz      & $  1.00$&$ -0.17$&$ -0.08$&$  0.07$ \\
\Rcz      & $ -0.17$&$  1.00$&$  0.08$&$ -0.06$ \\
$\Afbzb$  & $ -0.08$&$  0.08$&$  1.00$&$  0.14$ \\
$\Afbzc$  & $  0.07$&$ -0.06$&$  0.14$&$  1.00$ \\
\hline
\end{tabular}
\end{center}
\caption[]{
  The correlation matrix for the four electroweak parameters from the
  12-parameter fit.}
\label{tab:12parcor}
\end{table}

\subsection{Results of the 14-Parameter Fit to LEP and SLD Data}
\label{sec-HFSUM-LEP-SLD}

Including the SLD results for \Rb, \Rc, \cAb{} and \cAc{} into the fit the
following results are obtained:
\begin{eqnarray}
  \label{eqn-hf6}
  \Rbz    &=&  0.21638 \pm 0.00065  \,,\\
  \Rcz    &=&  0.1720  \pm 0.0030   \nonumber\,,\\
  \Afbzb  &=&  0.0997  \pm 0.0016   \nonumber\,,\\
  \Afbzc  &=&  0.0706  \pm 0.0035   \nonumber\,,\\
  \cAb    &=&  0.925   \pm 0.020    \nonumber\,,\\
  \cAc    &=&  0.670   \pm 0.026    \nonumber\,,
\end{eqnarray}
with a $\chi^2/$d.o.f.{} of $53/(105-14)$. The corresponding
correlation matrix is given in Table~\ref{tab:14parcor}
and the largest errors for the electroweak parameters are listed in Table
\ref{tab:hferrbk}.
If only statistical errors are used in the fit the $\chi^2/$d.o.f.{} is 
$91/(105-14)$. This indicates that the data fluctuate towards a very small 
$\chi^2$ and that probably systematic uncertainties are estimated in a 
conservative way.

\begin{figure}[htbp]
\vspace*{-0.6cm}
\begin{center}
  \mbox{\includegraphics[width=0.9\linewidth]{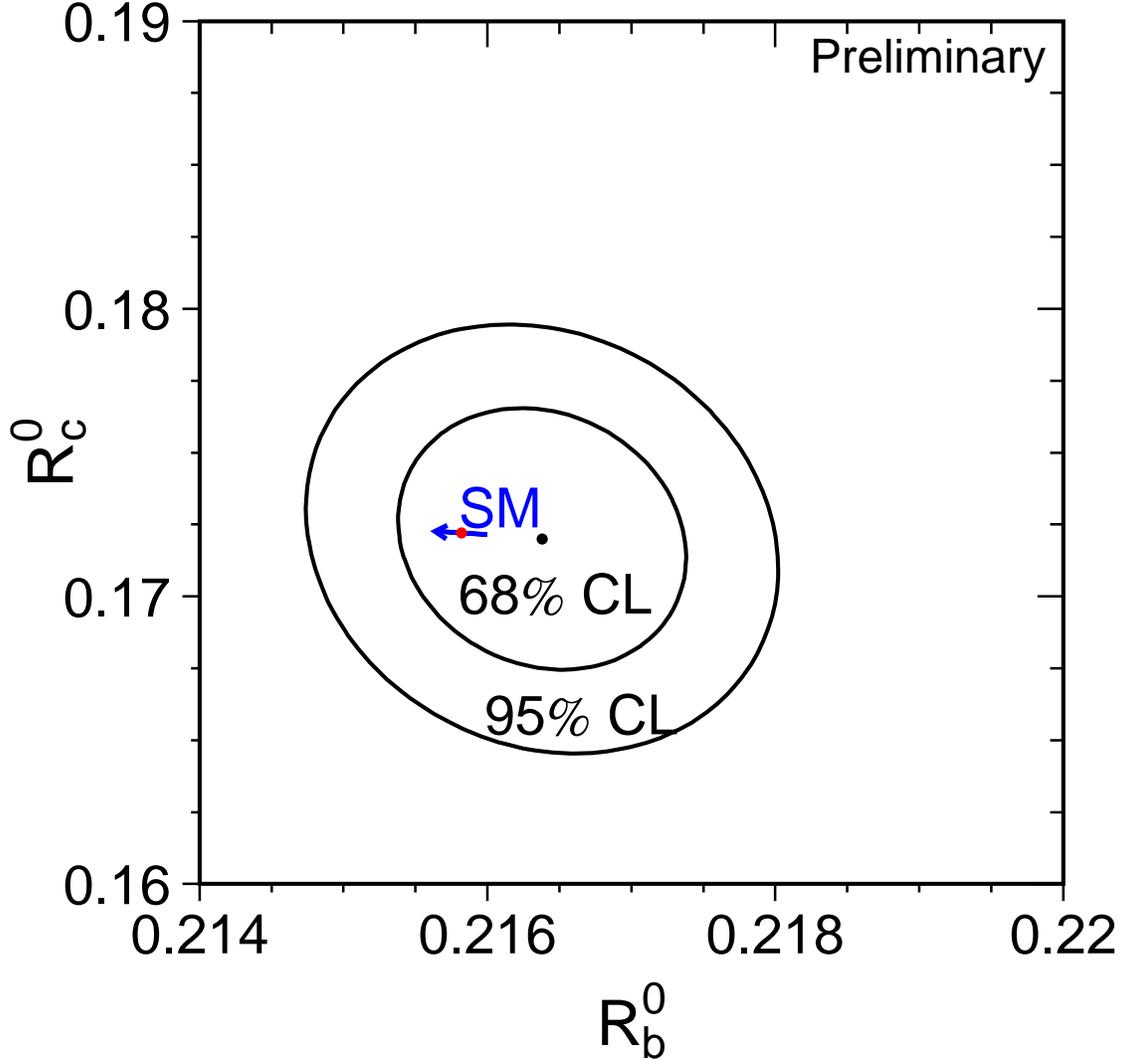}}
\end{center}
\caption[]{%
  Contours in the ($\Rbz$,$\Rcz$) plane derived from the LEP+SLD
  data, corresponding to 68\% and 95\% confidence levels assuming
  Gaussian systematic errors. The Standard Model prediction for
  $\Mt=174.3 \pm 5.1$~\GeV{} is also shown. The arrow points in the
  direction of increasing values of $\Mt$.  }
\label{fig-RbRc}
\end{figure}

In deriving
these results the parameters $\cAb$ and $\cAc$ are treated as
independent of the forward-backward asymmetries $\Afbzb$ and $\Afbzc$
(but see Section~\ref{sec-AF} for a joint analysis). In
Figure~\ref{fig-RbRc} the results for $\Rbz$ and $\Rcz$ are shown
compared with the Standard Model expectation.

\begin{table}[htbp]
\begin{center}
\begin{tabular}{|l||rrrrrr|}
\hline
&\makebox[1.2cm]{\Rbz}
&\makebox[1.2cm]{\Rcz}
&\makebox[1.2cm]{$\Afbzb$}
&\makebox[1.2cm]{$\Afbzc$}
&\makebox[0.9cm]{\cAb}
&\makebox[0.9cm]{\cAc}\\
\hline
\hline
\Rbz     & $  1.00$&$ -0.14$&$ -0.10$&$  0.07$&$ -0.07$&$  0.05$ \\
\Rcz     & $ -0.14$&$  1.00$&$  0.03$&$ -0.06$&$  0.03$&$ -0.05$ \\  
$\Afbzb$ & $ -0.10$&$  0.03$&$  1.00$&$  0.15$&$  0.03$&$ -0.01$ \\
$\Afbzc$ & $  0.07$&$ -0.06$&$  0.15$&$  1.00$&$ -0.02$&$  0.04$ \\
\cAb     & $ -0.07$&$  0.03$&$  0.03$&$ -0.02$&$  1.00$&$  0.13$ \\
\cAc     & $  0.05$&$ -0.05$&$ -0.01$&$  0.04$&$  0.13$&$  1.00$ \\
\hline
\end{tabular}
\end{center}
\caption[]{
  The correlation matrix for the six electroweak parameters from the
  14-parameter fit.  }
\label{tab:14parcor}
\end{table}

\begin{table}[htbp]
\begin{center}
\begin{tabular}{|c|c|c|c|c|c|c|}
\hline
&\makebox[1.2cm]{\Rbz}
&\makebox[1.2cm]{\Rcz}
&\makebox[1.2cm]{$\Afbzb$}
&\makebox[1.2cm]{$\Afbzc$}
&\makebox[0.9cm]{\cAb}
&\makebox[0.9cm]{\cAc}\\
 & $(10^{-3})$ & $(10^{-3})$ & $(10^{-3})$ & $(10^{-3})$ 
 & $(10^{-2})$ & $(10^{-2})$ \\
\hline
statistics & 
$0.43$ & $2.3$ & $1.5$ & $3.0$ & $1.5$ & $2.1$ \\
internal systematics &
$0.28$ & $1.4$ & $0.6$ & $1.4$ & $1.4$ & $1.6$ \\
QCD effects &
$0.18$ & $0.1$ & $0.4$ & $0.1$ & $0.3$ & $0.2$ \\
semil. B,D decay model &
$0.01$ & $0.1$ & $0.1$ & $0.5$ & $0.1$ & $0.1$ \\
BR(D $\rightarrow$ neut.)&
$0.14$ & $0.3$ & $0$   & $0$   & $0$   & $0$ \\
D decay multiplicity &
$0.13$ & $0.3$ & $0  $ & $0.2$ & $0$   & $0$ \\
BR(D$^+ \rightarrow$ K$^- \pi^+ \pi^+) $&
$0.08$ & $0.2$ & $0  $ & $0.1$ & $0$   & $0  $ \\
BR($\Ds \rightarrow \phi \pi^+) $&
$0.02$ & $0.5$ & $0  $ & $0.1$  & $0$   & $0$ \\
BR($\Lambda_{\mathrm{c}} \rightarrow $p K$^- \pi^+) $&
$0.06$ & $0.5$ & $0  $ & $0.1$ & $0$   & $0  $ \\
D lifetimes&
$0.06$ & $0.1$ & $0  $ & $0.2$ & $0$   & $0$ \\
gluon splitting &
$0.22$ & $0.1$ & $0.1$ & $0.2$ & $0.1$ & $0.1$ \\
c fragmentation &
$0.10$ & $0.2$ & $0.1$ & $0.1$ & $0.1$ & $0.1$ \\
light quarks&
$0.07$ & $0.2$ & $0  $ & $0  $ & $0$   & $0  $ \\
beam polarisation&
$0$    & $0$   & $0$   & $0$   & $0.5$ & $0.4$ \\
\hline
total &
$0.65$ & $3.1$ & $1.6$ & $3.5$ & $2.0$ & $2.6$ \\
\hline
\end{tabular}
\end{center}
\caption[]{
The dominant error sources for the electroweak parameters from the 14-parameter
fit.
}
\label{tab:hferrbk}
\end{table}

Amongst the non-electroweak observables the B semileptonic branching
fraction ($\Brbl \, = \, 0.1071 \pm 0.0022$) is of special interest. 
The dominant error source on this quantity is the dependence on the 
semileptonic decay models $\bl$, $\cl$ with 
\begin{equation}
\Delta \Brbl_{\bl-\rm{modelling}}  = 0.0012.
\end{equation}
Extensive studies have been made to understand the size of this error.
Amongst the electroweak quantities the quark asymmetries with leptons
depend also on the assumptions on the decay model while the
asymmetries using other methods usually do not. The fit implicitly
requires that the different methods give consistent results. This
effectively constrains the decay model and thus reduces the error from
this source in the fit result for $\Brbl$.

To get a conservative estimate of the modelling error in $\Brbl$ the
fit is repeated removing all asymmetry measurements. The result
of this fit is
\begin{equation}
\Brbl \, = \, 0.1065 \pm 0.0023
\end{equation}
with
\begin{equation}
\Delta \Brbl_{\bl-\rm{modelling}} = 0.0014.
\end{equation}


%% file: gg.tex
\section{Introduction}
The reaction $\eeggga$ provides a clean test of QED at LEP energies
and is well suited to detect the presence of non-standard physics.
The differential QED cross-section at the Born level in the
relativistic limit is given by \cite{ref:QED,ref:radcor}:
\begin{equation}
\xb = \frac{\alpha^2}{s} 
\frac{1+\cos^2\theta}{1 -\cos^2\theta} \; .
\end{equation}
Since the two final state particles are identical the polar angle
$\theta$ is defined such that $\ct > 0$. Various models with 
deviations from this cross-section will be discussed in section \ref{gg:sec:fit}.
Results on the $\ge$2-photon  final state using the high energy data 
collected by the four LEP collaborations are reported by the individual
experiments \cite{gg:ref:LEPGG}.
Here the results of the LEP working group %
dedicated to the combination of the $\eeggga$ measurements
are reported.  Results are given for the averaged total cross-section
and for global fits to the differential cross-sections.

\section{Event Selection}
This channel is very clean and the event selection, which is similar
for all experiments, is based on the presence of at least two
energetic clusters in the electromagnetic calorimeters.  A minimum
energy is required, typically $(E_1+ E_2)/\sqrt{s}$ larger than 0.3 to
0.6, where $E_1$ and $E_2$ are the energies of the two most energetic
photons.  In order to remove $\ee$ events, charged tracks are in
general not allowed except when they can be associated to a photon
conversion in one hemisphere.

The polar angle is defined in order to minimise effects due to 
initial state radiation as
\[
\ct =\left.\left| \sin (\frac{\theta_1 - \theta_2}{2}) \right| 
        \right/ \sin (\frac{\theta_1 + \theta_2}{2}) \; ,   \] 
where $\theta_1$ and $\theta_2$ are the polar angles of the two most energetic photons.
The acceptance in polar angle is in the range of 0.90 to 0.96 on 
$|\ct|$, depending on the experiment.

With these criteria, the selection efficiencies are in the range of
68\% to 98\% and the residual background (from $\ee$ events and
from $\eetautau$ with $\tau^{\pm} \rightarrow\rm e^{\pm}\nu
\bar{\nu}$) is very small, 0.1\% to 1\%.  Detailed descriptions of
the event selections performed by the four collaborations can be found
in \cite{gg:ref:LEPGG}.

\section{Total cross-section}

\begin{figure}[t]
   \begin{center} \mbox{
          \epsfxsize=16.0cm
           \epsffile{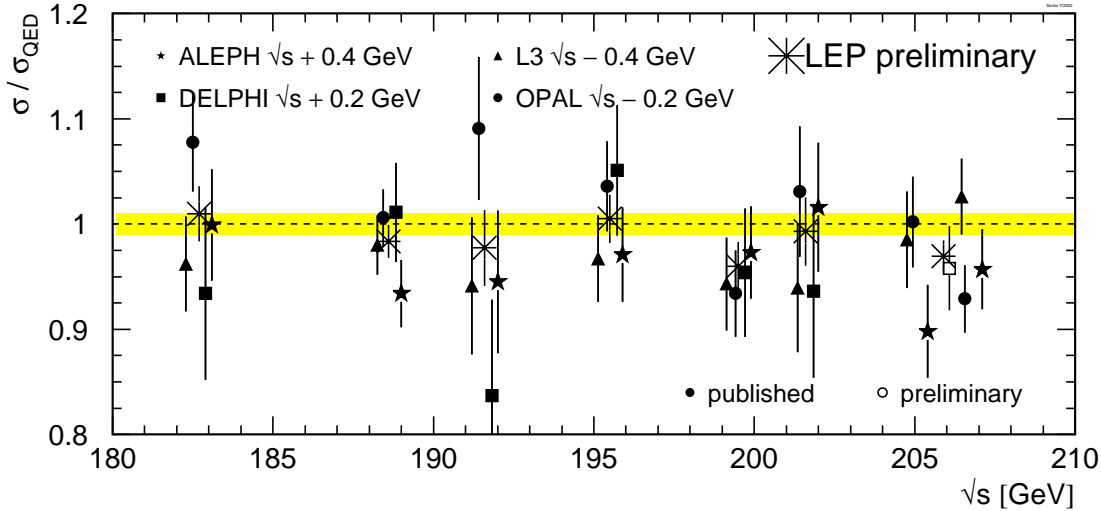}
           } \end{center}
\vspace{-1cm}
\caption{Cross-section ratios 
$r = \sigma_{\rm meas} / \sigma_{\rm QED}$ at different energies.
The measurements of the single experiments are displaced by 
$\pm$ 200 or 400 \MeV\ from the actual energy for clarity. Filled symbols
indicate published results, open symbols stand for preliminary numbers.
The average over the experiments at each energy is shown as a star. 
Measurements between 203 and 209 \GeV\ are averaged to one energy point. 
The theoretical error is not included in the experimental errors 
but is represented as the shaded band.
}
\label{gg:fig:xsn}
\end{figure}

The total cross-sections are combined using a $\chi^2$ minimisation.
For simplicity, given the different angular acceptances,
the ratios of the measured cross-sections relative to the QED 
expectation, \mbox{$r = \sigma_{\rm meas} / \sigma_{\rm QED}$},
are averaged. Figure \ref{gg:fig:xsn} shows the measured ratios $r_{i,k}$ 
of the experiments $i$ at energies $k$ with their statistical
and systematic errors %
added in quadrature. There are no significant sources of experimental 
systematic errors that are correlated between experiments. The theoretical error on 
the QED prediction, which is fully correlated between energies and experiments
is taken into account after the combination.

Denoting with $\Delta$ the vector of residuals between the measurements
and the expected ratios, three different averages are performed:
\begin{enumerate}
\item per energy $k=1,\ldots,7$: $\Delta_{i,k} = r_{i,k} - x_k$ 
\item per experiment $i=1,\ldots,4$: $\Delta_{i,k} = r_{i,k} - y_i$ 
\item global value:  $\Delta_{i,k} = r_{i,k} - z$ 
\end{enumerate}
The seven fit parameters per energy $x_k$ are shown in Figure 
\ref{gg:fig:xsn} as LEP combined cross-sections. They are correlated
with correlation coefficients ranging from 5\% to 20\%. 
The four fit-parameters per experiment $y_i$ are uncorrelated
between each other, the results are given in Table \ref{gg:tab:xsn}
together with the single global fit parameter $z$.

No significant deviations from the QED expectations are found.
The global ratio is below unity by 1.8 standard deviations not 
accounting for the error on the radiative corrections.
This theory error can be assumed to be about 10\% of the applied
radiative correction and hence depends on the selection. 
For this combination it is assumed to be 1\% which is of 
same size as the experimental error (1.0\%).

\begin{table}[hbt]
\begin{center}
\begin{tabular}{|l|r@{$\pm$}l|}\hline
Experiment & \multicolumn{2}{c|}{cross-section ratio} \\\hline\hline
ALEPH  & 0.953 & 0.024 \\
DELPHI & 0.976 & 0.032 \\
L3     & 0.978 & 0.018 \\
OPAL   & 0.999 & 0.016 \\ \hline
global & 0.982 & 0.010 \\ \hline
\end{tabular}
\caption[]{Cross-section ratios 
$r = \sigma_{\rm meas} / \sigma_{\rm QED}$ for the four LEP experiments
averaged over all energies and the global average over all experiments
and energies. The error includes the statistical and experimental
systematic error but no error from theory.
}
\label{gg:tab:xsn}
\end{center}
\end{table}

\begin{figure}[btp]
   \begin{center} 
   \mbox{\epsfxsize=8.0cm\epsffile{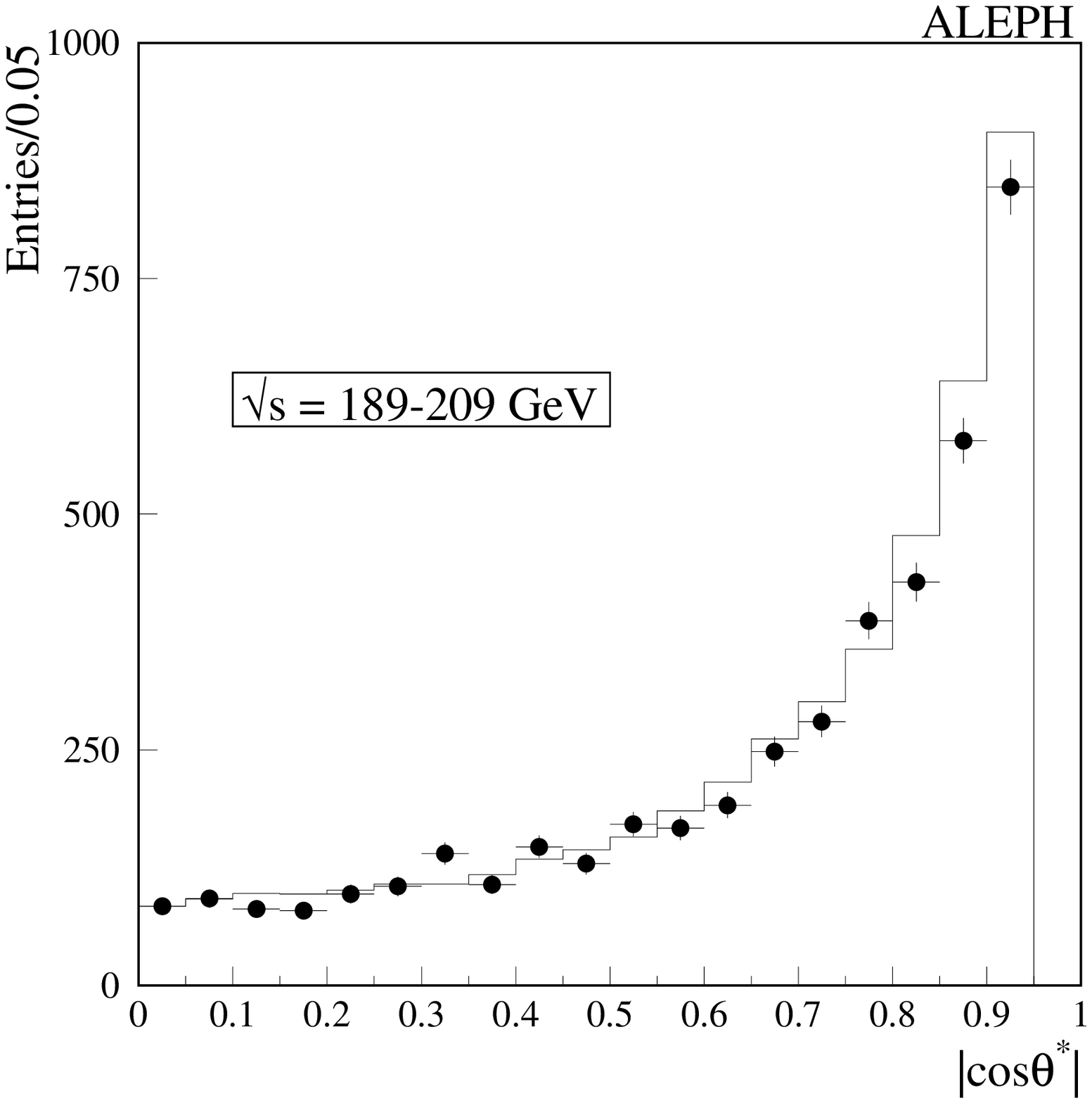}}
   \raisebox{1.5cm}{\epsfxsize=8.0cm\epsffile{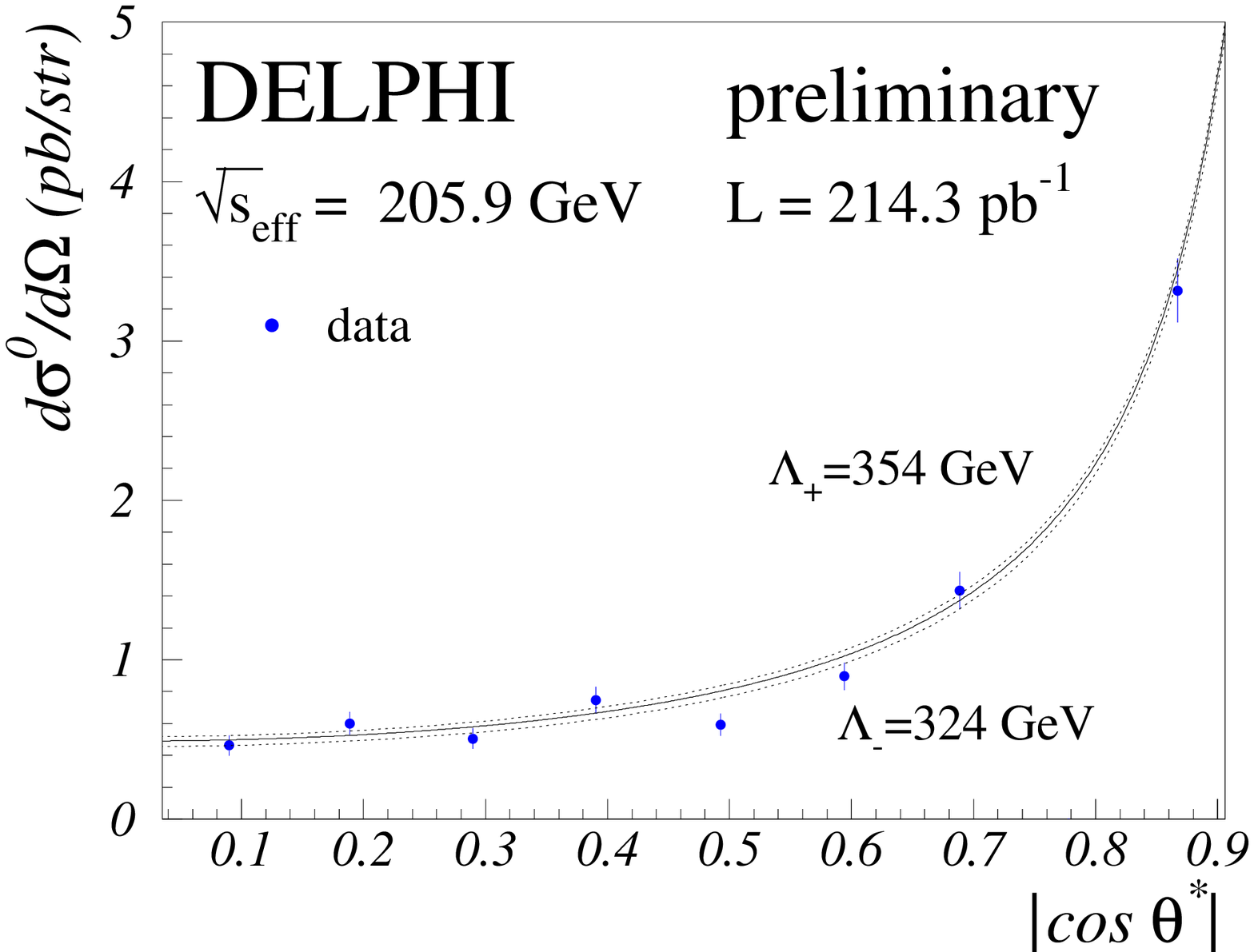}}\\
   \mbox{\epsfxsize=8.0cm\epsffile{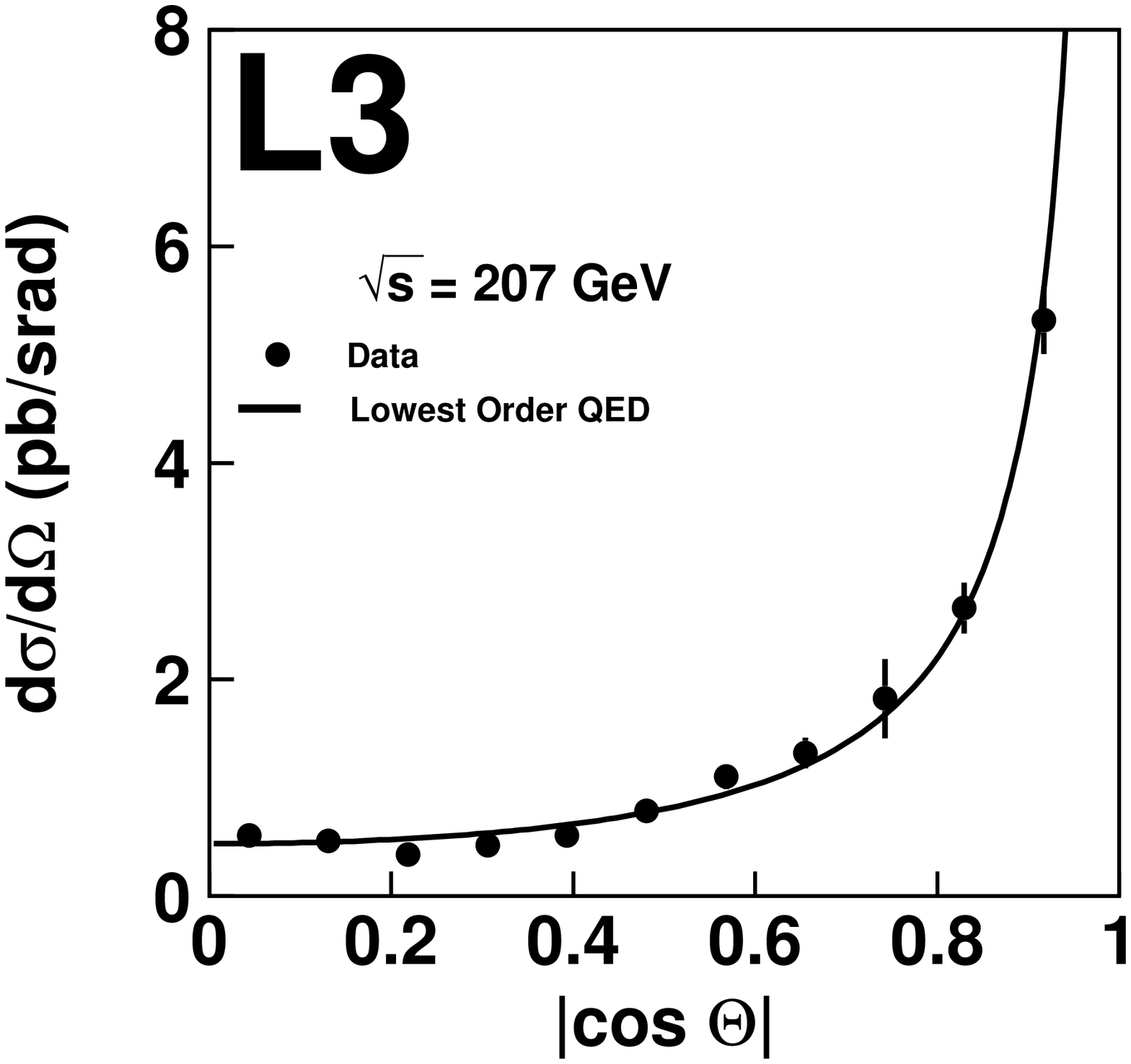}}
   \mbox{\epsfxsize=8.0cm\epsffile{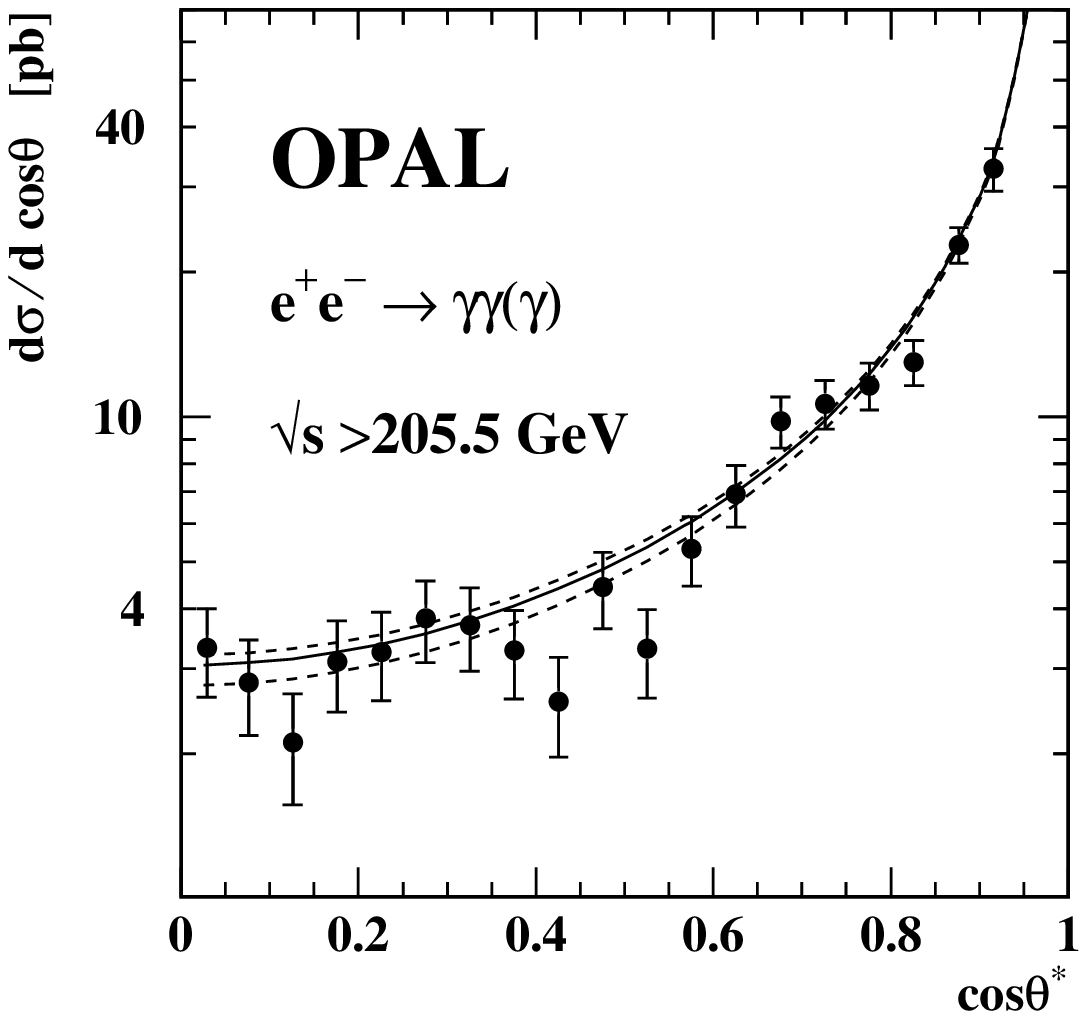}}
   \end{center}
\caption{Examples for angular distributions of the four LEP experiments.
Points are the data and the curves are the QED prediction (solid) and
the individual fit results for $\Lpm$ (dashed). ALEPH shows the
uncorrected number of observed events, the expectation is presented as
histogram. 
}
\label{gg:fig:ADLO}
\end{figure}

\section{Global fit to the differential cross-sections}
\label{gg:sec:fit}

\begin{table}[tb]
\begin{center}
\begin{tabular}{|l|c|c|c|c|c|} \hline
  & \multicolumn{2}{c|}{data used} & \multicolumn{2}{c|}{sys. error $[ \% ]$}&$\left | \rm{cos} \theta \right |$ \\ 
  & published & preliminary & experimental & theory & \\\hline
ALEPH  & 189 -- 207 &  --  & 2 & 1& 0.95 \\
DELPHI & 189 -- 202 & 206 & 2.5 & 1 & 0.90 \\
L3     & 183 -- 207 &  --  & 2.1 & 1 & 0.96 \\
OPAL   & 183 -- 207 &  --  & 0.6 -- 2.9 & 1 & 0.93 \\\hline
\end{tabular}
\caption[]{ The data samples used for the global fit to the
  differential cross-sections, the systematic errors, the assumed
  error on the theory and the polar angle acceptance for the LEP
  experiments.}
\label{gg:tab:stat} 
\end{center} 
\end{table}

The global fit is based on angular distributions at energies between
183 and 207 \GeV\ from the individual experiments. As an example,
angular distributions from each experiment are shown in
Figure~\ref{gg:fig:ADLO}. Combined differential cross-sections are not
available yet, since they need a common binning of the distributions.
All four experiments give results including the whole year 2000 
data-taking. Apart from the 2000 DELPHI data all inputs are final, 
as shown in Table~\ref{gg:tab:stat}.  
The systematic errors arise from the luminosity
evaluation (including theory uncertainty on the small-angle Bhabha
cross-section computation), from the selection efficiency and the
background evaluations and from radiative corrections. The last
contribution, owing to the fact that the available $\eeggga$
cross-section calculation is based on $\cal O$$(\alpha^3)$ code,
is assumed to be 1\% and is considered correlated among energies and experiments.

Various model predictions
are fitted to these angular distributions taking into account the
experimental systematic error correlated between energies for each
experiment and the error on the theory.
A binned log likelihood fit is performed
with one free parameter for the model and five fit parameters
used to keep the normalisation free within the systematic errors
of the theory and the four experiments. Additional fit parameters are
needed to accommodate the angular dependent systematic errors of OPAL.

The following models of new physics are considered. 
The simplest ansatz is a short-range exponential deviation from the 
Coulomb field parameterised by cut-off 
parameters $\Lpm$~\cite{gg:ref:drell,gg:ref:low}. 
This leads to a differential cross-section of the form
\begin{equation}
\xl   =  \xb \pm \frac{\alpha^2 \pi s}{\Lambda_\pm^4}(1+\cos^2{\theta}) \; .
\label{gg:lambda}
\end{equation}

New effects can also be introduced in effective Lagrangian theory
\cite{gg:ref:eboli}. Here dimension-6 terms lead to anomalous 
$\rm ee\gamma$ couplings. The resulting deviations in the differential 
cross-section are similar in form to those given in 
Equation~\ref{gg:lambda}, but with a slightly different definition of the
parameter: $\Lambda_6^4 = \frac{2}{\alpha}\Lambda_+^4$.
While for the ad hoc included cut-off parameters $\Lpm$ both signs are
allowed the physics motivated parameter $\Lambda_6$ occurs only with the
positive sign.
Dimension 7 and 8 Lagrangians introduce $\rm ee\gamma\gamma$ contact
interactions and result in an angle-independent term added to the Born
cross-section:
\begin{equation}
\xq  =  \xb + \frac{s^2}{16}\frac{1}{\Lambda'{}^6} \; .
\end{equation}
The associated parameters are given by 
$\Lambda_7 = \Lambda'$ and $\Lambda_8^4 = m_{\rm e} {\Lambda'}^3$ for
dimension~7 and dimension~8 couplings, respectively.
The subscript refers to the dimension of the Lagrangian.

Instead of an ordinary electron, an excited electron $\rm e^\ast$
with mass $\mestar$
could be exchanged in the $t$-channel \cite{gg:ref:low,gg:ref:estar}. 
In the most general case $\rm \rm e^\ast e \gamma$ couplings would lead
to a large anomalous magnetic moment of the electron 
\cite{gg:ref:g2_brodsky}. 
This effect can be avoided by a chiral magnetic coupling of the form~\cite{gg:ref:boudjema:1993}:
\begin{equation}
{\cal L}_{\rm e^\ast e \gamma} = 
\frac{1}{2\Lambda} \bar{e^\ast} \sigma^{\mu\nu}
\left[ g f \frac{\tau}{2}W_{\mu\nu} + g' f' \frac{Y}{2} B_{\mu\nu}
\right] e_L + \mbox{h.c.} \; ,
\end{equation}
where $\tau$ are the Pauli matrices and $Y$ is the hypercharge.
The parameters of the model are the compositeness scale $\Lambda$
and the weight factors $f$ and $f'$ associated to the gauge fields 
$W$ and $B$ with Standard Model couplings $g$ and $g'$.
For the process $\eeggga$, 
the following cross-section results~\cite{gg:ref:vachon}: 
\begin{eqnarray}
\xe & = & \xb  \\
 & + & \frac{\alpha^2 \pi}{2}\frac{f_\gamma^4}{\Lambda^4}\mestar^2 \left[
\frac{p^4}{(p^2-\mestar^2)^2} + \frac{q^4}{(q^2-\mestar^2)^2} +
\frac{\frac{1}{2} s^2 \sin^2\theta}{(p^2-\mestar^2)(q^2-\mestar^2)} \right]
\; , \nonumber \end{eqnarray}
with $f_\gamma = -\frac{1}{2}(f+f')$, $p^2=-\frac{s}{2}(1-\ct)$ and 
$q^2=-\frac{s}{2}(1+\ct)$. 
Effects vanish in the case of $f = -f'$. The cross-section does not
depend on the sign of $f_\gamma$.

Theories of quantum gravity in extra spatial dimensions could solve the 
hierarchy problem because gravitons would be allowed to travel in 
more than 3+1 space-time dimensions \cite{gg:ref:ad}. 
While in these models the Planck mass $M_D$
in $D=n+4$ dimensions is chosen to be of electroweak scale the usual
Planck mass $M_{\rm Pl}$ in four dimensions would be
\begin{equation} M_{\rm Pl}^2 = R^n M_D^{n+2} \; ,\end{equation}
where $R$ is the compactification radius of the additional dimensions.
Since gravitons couple to the energy-momentum tensor, their
interaction with photons is as weak as with fermions. However, the huge
number of Kaluza-Klein excitation modes in the extra dimensions may 
give rise to
observable effects. These effects depend on the scale $M_s (\sim M_D)$ 
which may be as low as ${\cal O}(\rm TeV)$. Model dependencies
are absorbed in the parameter $\lambda$ which cannot be explicitly
calculated without knowledge of the full theory, the sign is undetermined. 
The parameter $\lambda$ is expected to be 
of ${\cal O}(1)$ and for this analysis it is assumed that $\lambda = \pm 1$. 
The expected differential cross-section is given by \cite{gg:ref:ad}:
\begin{equation}
\xg = \xb - {\alpha s} \; \frac{\lambda}{M_s^4}\;(1+\cos^2{\theta})
    + \frac{s^3}{8 \pi} \;  \frac{\lambda^2}{M_s^8} \;(1-\cos^4{\theta})
    \; .
\end{equation}

\section{Fit Results}

Where possible the fit parameters are chosen such that the likelihood
function is approximately Gaussian. The preliminary results of the
fits to the differential cross-sections are given in
Table~\ref{gg:tab:results}.  No significant deviations with respect to
the QED expectations are found (all the parameters are compatible with
zero) and therefore 95\% confidence level limits are obtained by
renormalising the probability distribution of the fit parameter to the
physically allowed region. 
The asymmetric limits $x_{95}^{\pm}$ 
on the fitting parameter are obtained by:
\begin{equation} \frac{\int^{x_{95}^+}_0 \Gamma(x,\mu ,\sigma ) dx}
         {\int^{\infty   }_0 \Gamma(x,\mu ,\sigma ) dx} = 0.95 \; 
    \hspace{7mm}\mbox{and}\hspace{7mm}
    \frac{\int_{x_{95}^-}^0 \Gamma(x,\mu ,\sigma ) dx}
         {\int_{-\infty  }^0 \Gamma(x,\mu ,\sigma ) dx} = 0.95 \; , 
         \label{limeq}\end{equation}
where $\Gamma$ is a Gaussian with the central value and error of the fit
result denoted by $\mu$ and $\sigma$, respectively. This is equivalent
to the integration of a Gaussian probability function as a
function of the fit parameter. The 95 \% CL limits on the model parameters 
are derived from the limits on the 
fit parameters, e.g. the limit on $\Lambda_+$ is obtained as 
$[x_{95}^+(\Lambda^{-4}_{\pm})]^{-1/4}$.

\begin{figure}[hbtp]
\vspace*{-1cm}
   \begin{center}\mbox{
          \epsfxsize=15.0cm
           \epsffile{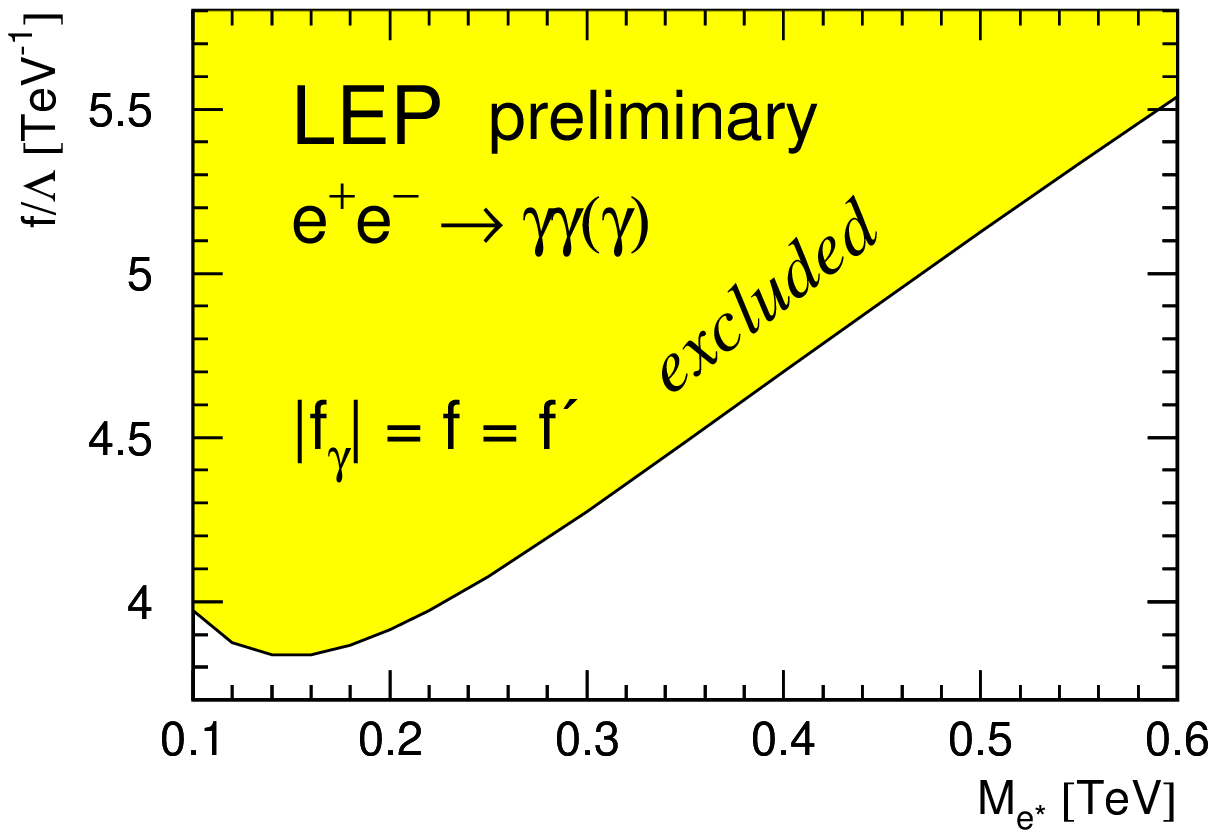}
           } \end{center}
\vspace{-1cm}
\caption{95\% CL limits on the coupling $f_\gamma/\Lambda$ of an excited
electron as a function of $\mestar$.
In the case of $f=f'$ it follows that $|f_\gamma| = f$.
It is assumed that $\Lambda=\mestar$.}
\label{gg:fig:estar}
\vspace*{-2cm}
   \begin{center}\mbox{
          \epsfxsize=15.0cm
           \epsffile{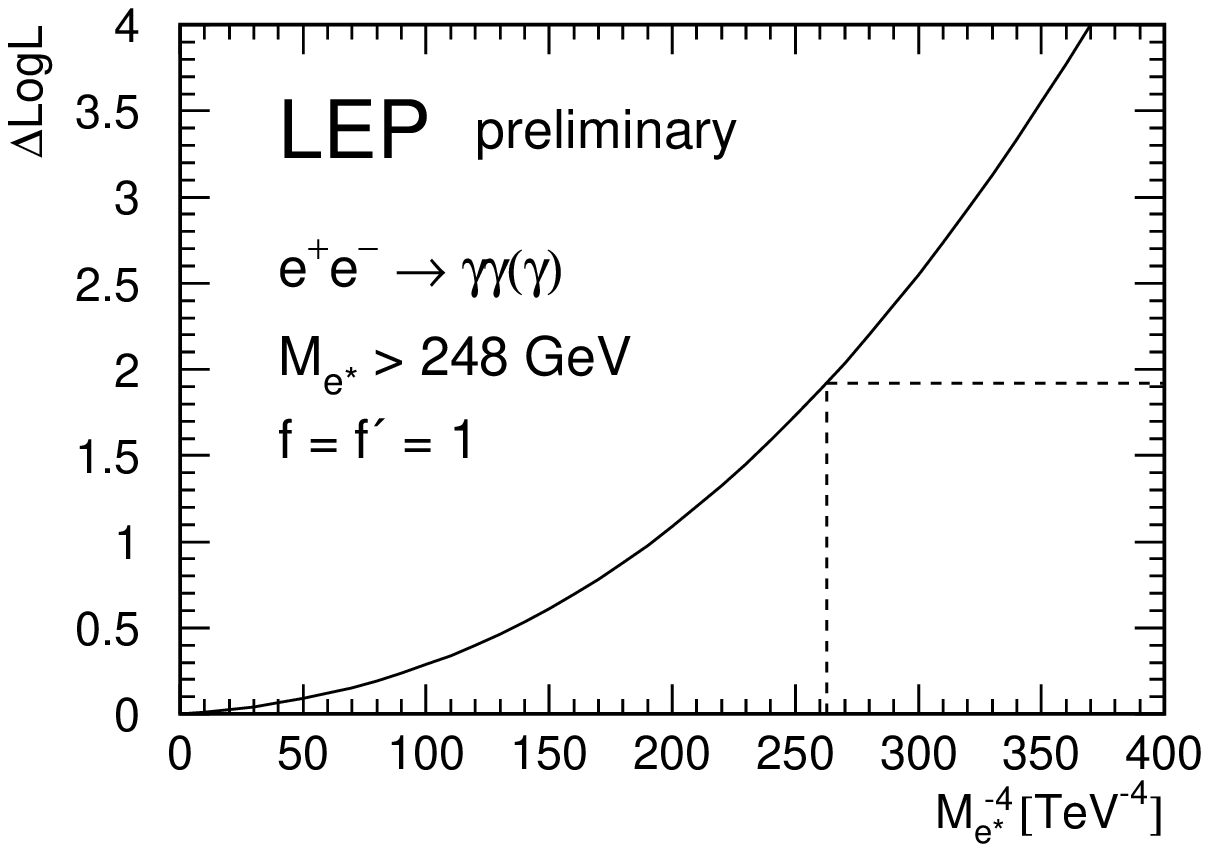}
           } \end{center}
\vspace{-1cm}
\caption{Log likelihood difference 
$\Delta\mbox{LogL} = -\ln{\cal L}+\ln{\cal L}_{\rm max}$
as a function of $\mestar^{-4}$. The coupling is fixed at $f = f' = 1$. 
The value corresponding to $\Delta\mbox{LogL} = 1.92$ is $\mestar$ = 248 \GeV.}
\label{gg:fig:ll}
\end{figure}

The only model with more than one free model parameter is the search 
for excited electrons. In this case only one out of the two parameters 
$f_\gamma$ and $\mestar$ is determined while the other is fixed.
It is assumed that $\Lambda=\mestar$. For limits on the coupling
$f_\gamma/\Lambda$ 
a scan over $\mestar$ is performed. The fit result at 
$\mestar = 200 \mbox{GeV}$ is included in Table~\ref{gg:tab:results},
limits for all masses are presented in Figure~\ref{gg:fig:estar}. 
For the determination of the excited electron mass 
the fit cannot be expressed in terms of a linear fit parameter.
For $|f_\gamma| =1$ the curve of the negative log likelihood, 
$\Delta\mbox{LogL}$, as a function of $\mestar$ is shown
in Figure \ref{gg:fig:ll}. The value corresponding to 
$\Delta\mbox{LogL} = 1.92$ is \mbox{$\mestar$ = 248 \GeV}.

\begin{table}[htb]
\begin{center}
\renewcommand{\arraystretch}{1.5}
\begin{tabular}{|c|c|r@{ }l|}\hline
Fit parameter & Fit result &
\multicolumn{2}{c|}{95\%\ CL limit [\GeV]}\\  \hline
 &  & $\Lambda_+ >$ &  392 \\
 \raisebox{2.2ex}[-2.2ex]{$\Lpm^{-4}$} &
 \raisebox{2.2ex}[-2.2ex]{$
 \left(-12.5{+25.1 \atop -24.7}\right)\cdot 10^{-12}$ \GeV$^{-4}$}
 & $\Lambda_- > $&  364  \\\hline
$\Lambda_7^{-6}$ & $ \left(-0.91{+1.81 \atop -1.78}\right)\cdot
                                  10^{-18}$ \GeV$^{-6}$
& $\Lambda_7 > $&  831   \\ \hline
\multicolumn{2}{|c|}{derived from $\Lambda_+$}& $\Lambda_6 > $&  1595   \\
\multicolumn{2}{|c|}{derived from $\Lambda_7$}& $\Lambda_8 > $&  23.3   \\\hline
 &  & $\lambda = +1$: $M_s >$ &  933  \\
 \raisebox{2.2ex}[-2.2ex]{$\lambda/M_s^4$} &
 \raisebox{2.2ex}[-2.2ex]{ $ \left(0.29{+0.57 \atop -0.58}\right)\cdot
                                         10^{-12}$ \GeV$^{-4} $ }
 & $\lambda = -1$: $M_s >$&  1010 \\ \hline
 $f_\gamma^4 (\mestar=200 \rm \GeV)$ & 
  $0.037{+0.202 \atop -0.198 }$ & 
 \multicolumn{2}{c|}{$f_\gamma/\Lambda <  3.9 \mbox{ \TeV}^{-1}$} \\ \hline
\end{tabular}
\caption[]{ The preliminary combined fit parameters 
and the 95$\%$  confidence level limits for the four LEP experiments.}
\label{gg:tab:results} 
\end{center} 
\end{table}

\section{Conclusion}
The LEP collaborations study the $\eeggga$ channel up to the highest
available centre-of-mass energies. The total cross-section results are
combined in terms of the ratios with respect to the QED expectations.
No deviations are found. The differential cross-sections are fit
following different parametrisations from models predicting deviations
from QED. No evidence for deviations is found and therefore combined
95\% confidence level limits are given.

%% file: ff.tex
\section{Introduction}

\begin{table}[p]
 \begin{center}
 \begin{tabular}{|c|c|c|c|}
  \hline
   Year & Nominal Energy & Actual Energy & Luminosity \\
        &     $\GeV$     &    $\GeV$     &  pb$^{-1}$ \\
  \hline
  \hline
   1995 &      130       &    130.2      & $\sim 3  $ \\
        &      136       &    136.2      & $\sim 3  $ \\
  \cline{2-4}
        &  $133^{\ast}$ &     133.2      & $\sim 6  $ \\
  \hline
   1996 &      161       &    161.3      & $\sim 10 $ \\
        &      172       &    172.1      & $\sim 10 $ \\
  \cline{2-4}
        &  $167^{\ast}$ &     166.6      & $\sim 20 $ \\
  \hline
   1997 &      130       &    130.2      & $\sim 2  $ \\
        &      136       &    136.2      & $\sim 2  $ \\
        &      183       &    182.7      & $\sim 50 $ \\
  \hline
   1998 &      189       &    188.6      & $\sim 170$ \\
  \hline
   1999 &      192       &    191.6      & $\sim 30 $ \\
        &      196       &    195.5      & $\sim 80 $ \\
        &      200       &    199.5      & $\sim 80 $ \\
        &      202       &    201.6      & $\sim 40 $ \\
  \hline
   2000 &      205       &    204.9      & $\sim 80 $ \\
        &      207       &    206.7      & $\sim140 $ \\
  \hline
 \end{tabular}
 \end{center}
 \caption{The nominal and actual centre-of-mass energies for data
          collected during $\LEPII$ operation in each year. The approximate
          average luminosity analysed per experiment at each energy is also
          shown. Values marked with 
          a $^{\ast}$ are average energies for 1995 and 1996 used 
          for heavy flavour results. The data taken at nominal energies of
          130 GeV and 136 GeV in 1995 and 1997 are combined by most 
          experiments.}
 \label{ff:tab:ecms}
\end{table}

During the $\LEPII$ program LEP delivered collisions
at energies from $\sim 130$ $\GeV$ to $\sim 209$ $\GeV$. The 4 LEP experiments
have made measurements on the $\eeff$ process over this range of energies,
and a preliminary combination of these data is discussed in this note.
 
In the years 1995 through 1999 LEP delivered luminosity at a number of 
distinct centre-of-mass energy points. In 2000 most of the luminosity
was delivered close to 2 distinct energies, but there was also
a significant fraction of the luminosity delivered in, more-or-less, a 
continuum of energies. To facilitate the combination of the data,
the 4 LEP experiments all divided the data they collected in 2000 
into two energy bins: from 202.5 to 205.5 $\GeV$; and 205.5 $\GeV$ and above.
The nominal and actual centre-of-mass energies to which the LEP data are
averaged for each year are given in Table~\ref{ff:tab:ecms}.

A number of measurements on the process $\eeff$ exist and are combined.
The preliminary averages of cross-section and forward-backward asymmetry
measurements are discussed in Section \ref{ff:sec-ave-xsc-afb}.
The results presented in this section update those presented 
in~\cite{bib-EWEP-02}.
Complete results of the combinations are available on the 
web page~\cite{ff:ref:ffbar_web}.
In Section~\ref{ff:sec-dsdc} a preliminary average of the differential
cross-sections measurements, $\dsdc$, for the channels $\eeee$,
$\eemumu$ and $\eetautau$ is presented. 
In Section~\ref{ff:sec-hvflv} a preliminary combination of the
heavy flavour results $\Rb$, $\Rc$, $\Abb$ and $\Acc$ from $\LEPII$ is 
presented. In Section~\ref{ff:sec-interp} the combined results are interpreted
in terms of contact interactions and the exchange of $\Zprime$ bosons, the 
exchange of leptoquarks or squarks and the exchange of gravitons in large
extra dimensions. The results are summarised in section~\ref{ff:sec-conc}.

\section{Averages for Cross-sections and Asymmetries}
\label{ff:sec-ave-xsc-afb}

In this section the results of the preliminary combination of
cross-sections and asymmetries are given.
The individual experiments' analyses of cross-sections and forward-backward
asymmetries are discussed in~\cite{ff:ref:expts}. 

Cross-section results are combined for the $\eeqq$, $\eemumu$ and $\eetautau$ 
channels, forward-backward asymmetry measurements are combined for
the $\mumu$ and $\tautau$ final states. The averages are made for the
samples of events with high effective centre-of-mass energies, $\sqrt{\spr}$. 
\begin{figure}[tp]
 \begin{center}
 \mbox{\epsfig{file=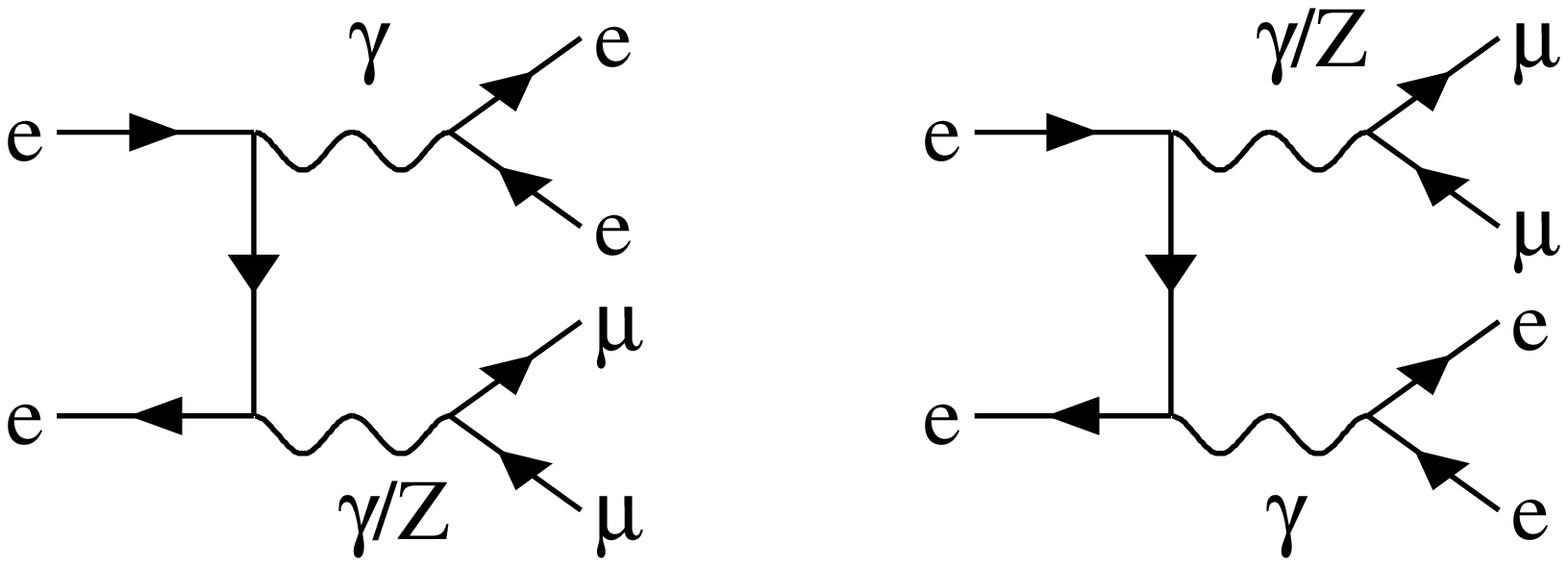,width=14cm}}
 \end{center}
 \caption{Diagrams leading to the the production of initial state non-singlet
          electron-positron pairs in $\eemumu$, which are considered as signal
          in the common signal definition.}
\label{ff:fig:isnspairs}
\end{figure}
Individual experiments have their own \ff\ signal definitions; corrections are
applied to bring the measurements to a common signal definitions:
\begin{itemize}
\item $\sqrt{\spr}$ is taken to be the mass of the
      $s$-channel propagator, with the $\ff$ signal being defined by the cut 
      $\sqrt{\spr/s} > 0.85$. 

\item ISR-FSR photon interference is subtracted to 
      render the propagator mass unambiguous.

\item Results are given for the full $4\pi$ angular acceptance. 

\item Initial state non-singlet diagrams \cite{ff:ref:lepffwrkshp}, 
      see for example Figure~\ref{ff:fig:isnspairs},
      which lead to events containing additional fermions pairs are considered
      as part of the two fermion signal. In such events, the additional 
      fermion pairs are typically lost down the beampipe of the experiments, 
      such that the visible event topologies are usually similar to a 
      difermion events with photons radiated from the initial state.
\end{itemize}
The corrected measurement of a cross-section or a forward backward asymmetry, 
$\mathrm{M_{LEP}}$, corresponding to the
common signal definition,  is computed from the
experimental measurement $\mathrm{M_{exp}}$,
\begin{eqnarray}
{\mathrm{M_{LEP}}} = {\mathrm{M_{exp}}} + ({\mathrm{P_{LEP}}} -
                                              {\mathrm{P_{exp}}}),
\end{eqnarray}
\noindent
where $\mathrm{P_{exp}}$ is the prediction for the measurement obtained 
for the experiments signal definition and $\mathrm{P_{LEP}}$ is the
prediction for the common signal definition. The predictions are computed with
ZFITTER~\cite{ff:ref:ZFITTER}.

In choosing a common signal definition there is a tension between the need to
have a definition which is practical to implement in event generators and
semi-analytical calculations, one which comes close to describing the 
underlying hard processes and one which most closely matches what is actually
measured in experiments. Different signal definitions represent different 
balances between these needs. To illustrate how different choices 
would effect the quoted results a second signal definition is 
studied by calculating different predictions using ZFITTER:
\begin{itemize}
 \item For dilepton events, $\sqrt{\spr}$ is taken to be 
       the bare invariant mass of the outgoing difermion pair (\ie,
       the invariant mass excluding all radiated photons). 
       
 \item For hadronic events, it is taken to be the mass of the $s$-channel 
       propagator. 

 \item In both cases, ISR-FSR photon interference is included and the signal
       is defined by the cut $\sqrt{\spr/s} > 0.85$. When calculating the 
       contribution to the hadronic cross-section due to ISR-FSR interference,
       since the propagator mass is ill-defined, it is replaced by the bare 
       $\qq$ mass.
\end{itemize}
The definition of the hadronic cross-section is close to that used to
define the signal for the heavy quark measurements given in 
Section~\ref{ff:sec-hvflv}.

Theoretical uncertainties associated with the Standard Model predictions 
for each of the measurements are not included during the averaging procedure,
but must be included when assessing the compatibility of the data with 
theoretical predictions.
The theoretical uncertainties on the Standard Model predictions amount to
$0.26\%$ on $\sigma(\qq)$, $0.4\%$ on $\sigma(\mumu)$ and
$\sigma(\tautau)$, $2\%$ on $\sigma(\ee)$, 
and 0.004 on the leptonic forward-backward 
asymmetries~\cite{ff:ref:lepffwrkshp}.

The average is performed using the best linear unbiased estimator
technique (BLUE)~\cite{common_bib:BLUE}, which is equivalent to a $\chi^{2}$ 
minimisation. All data from nominal centre-of-mass 
energies of 130--207 GeV are averaged at the same time.

Particular care is taken to ensure that the correlations between the 
hadronic cross-sections are reasonably estimated. 
The errors are broken down into 5 
categories, with the ensuing correlations accounted for in the combinations:
\begin{itemize}

\item[1)] The statistical uncertainty plus uncorrelated systematic 
uncertainties, combined in quadrature.

\item[2)] The systematic uncertainty for the final state X which is 
fully correlated between energy points for that experiment.

\item[3)] The systematic uncertainty for experiment Y which is fully 
correlated  between different final states for this energy point.

\item[4)] The systematic uncertainty for the final state X which is 
fully correlated between energy points  and between different experiments.

\item[5)] The systematic uncertainty which is fully correlated between 
energy points and between different experiments for all final states.
\end{itemize}
Uncertainties in the hadronic cross-sections arising from fragmentation
models and modelling of ISR are treated as fully correlated between 
experiments. Despite some differences between the models used and the 
methods of evaluating the errors in the different experiments, 
there are significant common elements in the estimation of these sources 
of uncertainty. 

New, preliminary, results from ALEPH are included in the average. The 
updated ALEPH measurements use a lower cut on the effective
centre-of-mass energy, which makes the signal definition of 
ALEPH closer to the combined LEP signal definition.

Table~\ref{ff:tab:xsafbres} gives the averaged cross-sections
and forward-backward asymmetries for all energies.
The differences in the results obtained when using predictions
of ZFITTER for the second signal definition are also given.
The differences are significant when compared to the precision obtained
from averaging together the measurements at all energies.
The $\chi^{2}$ per degree of freedom for the average of the $\LEPII$ $\ff$ 
data is $160/180$. Most correlations are rather small, with the largest 
components at any given pair of energies being between the hadronic 
cross-sections. The other off-diagonal terms in the correlation 
matrix are smaller than $10\%$. The correlation matrix between the 
averaged hadronic cross-sections at different centre-of-mass energies 
is given in Table~\ref{ff:tab:hadcorrel}.

\begin{figure}[p]
 \begin{center}
 \mbox{\epsfig{file=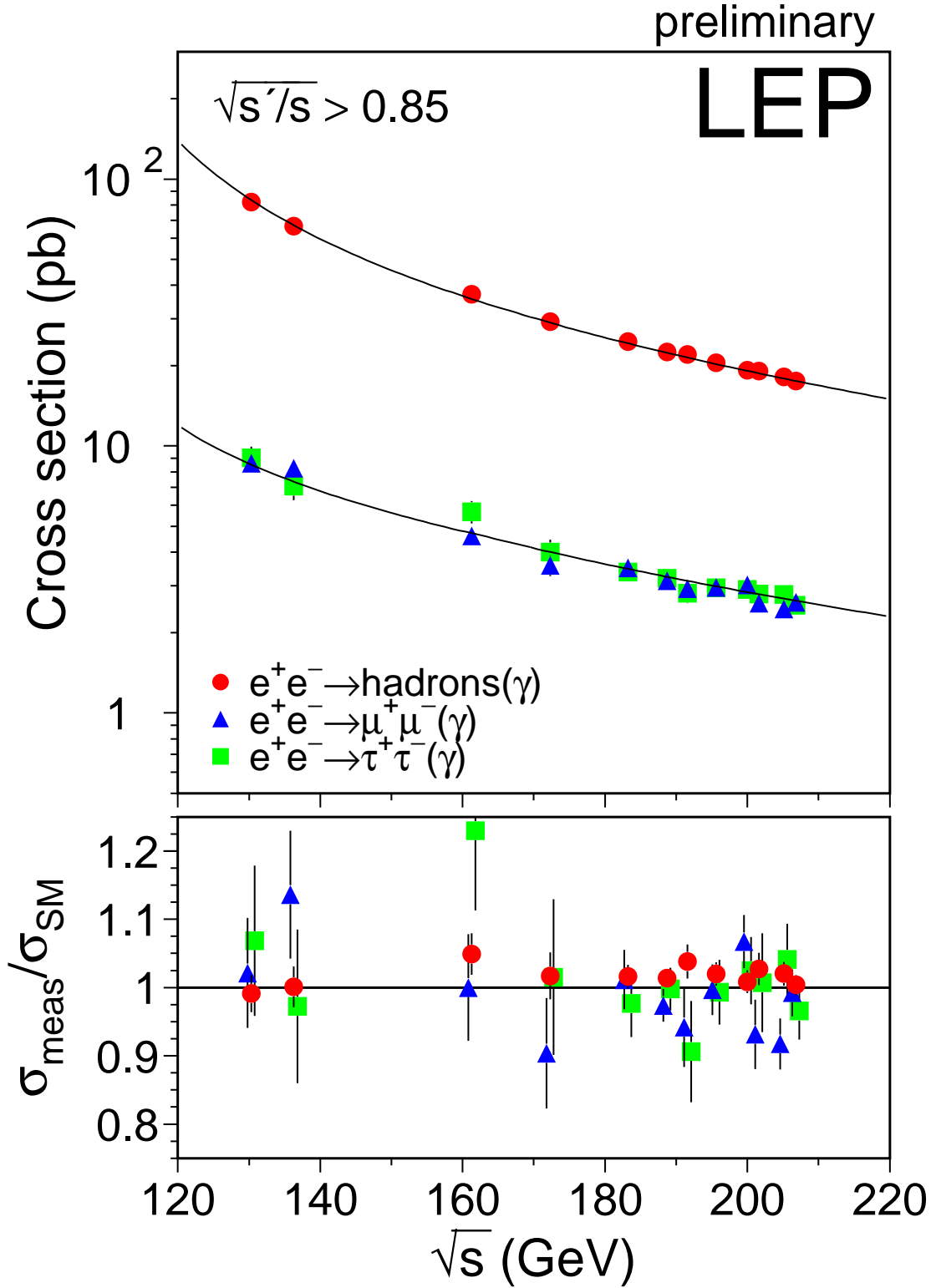,width=15cm}}
 \end{center}
 \caption{Preliminary combined LEP results on the cross-sections for 
          $\qq$, $\mumu$ and $\tautau$ final states, as a function of 
          centre-of-mass energy. The expectations of the SM, 
          computed with ZFITTER~\capcite{ff:ref:ZFITTER}, are shown as curves.
          The lower plot shows the ratio of the data divided by the SM.}
\label{ff:fig-xs_lep}
\end{figure}

\begin{figure}[p]
 \begin{center}
 \mbox{\epsfig{file=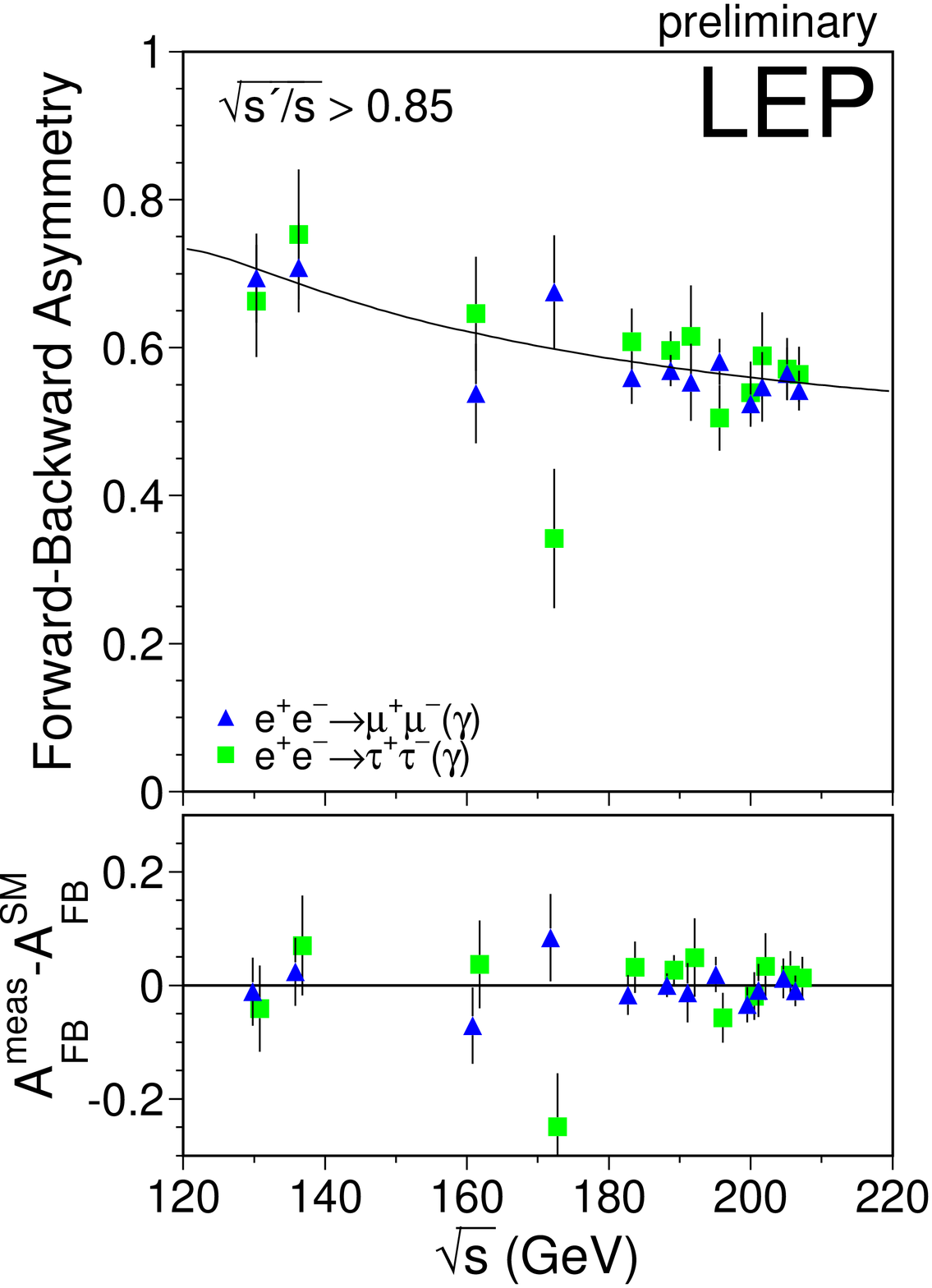,width=15cm}}
 \end{center}
 \caption{Preliminary combined LEP results on the forward-backward 
          asymmetry for $\mumu$  and $\tautau$ final states as a function of 
          centre-of-mass energy. The expectations of the SM 
          computed  with ZFITTER~\capcite{ff:ref:ZFITTER}, are shown as 
          curves. The lower plot shows differences between the data 
          and the SM.}
 \label{ff:fig-afb_lep}
\end{figure}

Figures~\ref{ff:fig-xs_lep} and~\ref{ff:fig-afb_lep} show the LEP 
averaged cross-sections and asymmetries, respectively, as a 
function of the centre-of-mass energy, together with the SM predictions. 
There is good agreement between the SM expectations and the measurements of the
individual experiments and the combined averages.
The cross-sections for hadronic final states at most of the energy points 
are somewhat above the SM expectations. Taking into account the correlations
between the data points and also taking into account the theoretical error
on the SM predictions, 
the ratio of the measured cross-sections to the SM expectations, averaged over 
all energies, is approximately a $1.7$ standard deviation excess. 
It is concluded that there is no significant evidence in the results of the
combinations for physics beyond the SM in the process $\eeff$.

\begin{table}[p]
 \begin{center}
 \begin{turn}{90}
 \begin{tabular}{cc}
 \begin{tabular}{|c|c|r@{$\pm$}l|c|c|}
 \hline
 $\sqrt{s}$ &          &
 \multicolumn{2}{c|}{Average} &
                    &
                               \\
 ($\GeV$) & Quantity   &
 \multicolumn{2}{|c|}{value}  &
 \multicolumn{1}{|c|}{SM} &
 \multicolumn{1}{|c|}{$\Delta$}  \\
 \hline\hline
  130 & $\sigma(q\overline{q})$                      & 82.1   &  2.2   & 82.8   & -0.3   \\
  130 & $\sigma(\mu^{+}\mu^{-})$                     &  8.62  &  0.68  &  8.44  & -0.33  \\
  130 & $\sigma(\tau^{+}\tau^{-})$                   &  9.02  &  0.93  &  8.44  & -0.11  \\
  130 & $\mathrm{A_{FB}}(\mu^{+}\mu^{-})$            &  0.694 &  0.060 &  0.705 &  0.012 \\
  130 & $\mathrm{A_{FB}}(\tau^{+}\tau^{-})$          &  0.663 &  0.076 &  0.704 &  0.012 \\
 \hline
  136 & $\sigma(q\overline{q})$                      & 66.7   &  2.0   & 66.6   & -0.2   \\
  136 & $\sigma(\mu^{+}\mu^{-})$                     &  8.27  &  0.67  &  7.28  & -0.28  \\
  136 & $\sigma(\tau^{+}\tau^{-})$                   &  7.078 &  0.820 &  7.279 & -0.091 \\
  136 & $\mathrm{A_{FB}}(\mu^{+}\mu^{-})$            &  0.708 &  0.060 &  0.684 &  0.013 \\
  136 & $\mathrm{A_{FB}}(\tau^{+}\tau^{-})$          &  0.753 &  0.088 &  0.683 &  0.014 \\
 \hline
  161 & $\sigma(q\overline{q})$                      & 37.0   &  1.1   & 35.2   & -0.1   \\
  161 & $\sigma(\mu^{+}\mu^{-})$                     &  4.61  &  0.36  &  4.61  & -0.18  \\
  161 & $\sigma(\tau^{+}\tau^{-})$                   &  5.67  &  0.54  &  4.61  & -0.06  \\
  161 & $\mathrm{A_{FB}}(\mu^{+}\mu^{-})$            &  0.538 &  0.067 &  0.609 &  0.017 \\
  161 & $\mathrm{A_{FB}}(\tau^{+}\tau^{-})$          &  0.646 &  0.077 &  0.609 &  0.016 \\
 \hline
  172 & $\sigma(q\overline{q})$                      & 29.23  &  0.99  & 28.74  & -0.12  \\
  172 & $\sigma(\mu^{+}\mu^{-})$                     &  3.57  &  0.32  &  3.95  & -0.16  \\
  172 & $\sigma(\tau^{+}\tau^{-})$                   &  4.01  &  0.45  &  3.95  & -0.05  \\
  172 & $\mathrm{A_{FB}}(\mu^{+}\mu^{-})$            &  0.675 &  0.077 &  0.591 &  0.018 \\
  172 & $\mathrm{A_{FB}}(\tau^{+}\tau^{-})$          &  0.342 &  0.094 &  0.591 &  0.017 \\
 \hline
  183 & $\sigma(q\overline{q})$                      & 24.59  &  0.42  & 24.20  & -0.11  \\
  183 & $\sigma(\mu^{+}\mu^{-})$                     &  3.49  &  0.15  &  3.45  & -0.14  \\
  183 & $\sigma(\tau^{+}\tau^{-})$                   &  3.37  &  0.17  &  3.45  & -0.05  \\
  183 & $\mathrm{A_{FB}}(\mu^{+}\mu^{-})$            &  0.559 &  0.035 &  0.576 &  0.018 \\
  183 & $\mathrm{A_{FB}}(\tau^{+}\tau^{-})$          &  0.608 &  0.045 &  0.576 &  0.018 \\
 \hline
  189 & $\sigma(q\overline{q})$                      & 22.47  &  0.24  & 22.156 & -0.101 \\
  189 & $\sigma(\mu^{+}\mu^{-})$                     &  3.123 &  0.076 &  3.207 & -0.131 \\
  189 & $\sigma(\tau^{+}\tau^{-})$                   &  3.20  &  0.10  &  3.20  & -0.048 \\
  189 & $\mathrm{A_{FB}}(\mu^{+}\mu^{-})$            &  0.569 &  0.021 &  0.569 &  0.019 \\
  189 & $\mathrm{A_{FB}}(\tau^{+}\tau^{-})$          &  0.596 &  0.026 &  0.569 &  0.018 \\
 \hline
 \end{tabular}
 &
 \begin{tabular}{|c|c|r@{$\pm$}l|c|c|}
 \hline
 $\sqrt{s}$ &          &
 \multicolumn{2}{c|}{Average} &
                    &
                               \\
 ($\GeV$) & Quantity   &
 \multicolumn{2}{|c|}{value}  &
 \multicolumn{1}{|c|}{SM} &
 \multicolumn{1}{|c|}{$\Delta$}  \\
 \hline\hline
  192 & $\sigma(q\overline{q})$                      & 22.05  &  0.53  & 21.24  & -0.10  \\
  192 & $\sigma(\mu^{+}\mu^{-})$                     &  2.92  &  0.18  &  3.10  & -0.13  \\
  192 & $\sigma(\tau^{+}\tau^{-})$                   &  2.81  &  0.23  &  3.10  & -0.05  \\
  192 & $\mathrm{A_{FB}}(\mu^{+}\mu^{-})$            &  0.553 &  0.051 &  0.566 &  0.019 \\
  192 & $\mathrm{A_{FB}}(\tau^{+}\tau^{-})$          &  0.615 &  0.069 &  0.566 &  0.019 \\
 \hline
  196 & $\sigma(q\overline{q})$                      & 20.53  &  0.34  & 20.13  & -0.09  \\
  196 & $\sigma(\mu^{+}\mu^{-})$                     &  2.94  &  0.11  &  2.96  & -0.12  \\
  196 & $\sigma(\tau^{+}\tau^{-})$                   &  2.94  &  0.14  &  2.96  & -0.05  \\
  196 & $\mathrm{A_{FB}}(\mu^{+}\mu^{-})$            &  0.581 &  0.031 &  0.562 &  0.019 \\
  196 & $\mathrm{A_{FB}}(\tau^{+}\tau^{-})$          &  0.505 &  0.044 &  0.562 &  0.019 \\
 \hline
  200 & $\sigma(q\overline{q})$                      & 19.25  &  0.32  & 19.09  & -0.09  \\
  200 & $\sigma(\mu^{+}\mu^{-})$                     &  3.02  &  0.11  &  2.83  & -0.12  \\
  200 & $\sigma(\tau^{+}\tau^{-})$                   &  2.90  &  0.14  &  2.83  & -0.04  \\
  200 & $\mathrm{A_{FB}}(\mu^{+}\mu^{-})$            &  0.524 &  0.031 &  0.558 &  0.019 \\
  200 & $\mathrm{A_{FB}}(\tau^{+}\tau^{-})$          &  0.539 &  0.042 &  0.558 &  0.019 \\
 \hline
  202 & $\sigma(q\overline{q})$                      & 19.07  &  0.44  & 18.57  & -0.09  \\
  202 & $\sigma(\mu^{+}\mu^{-})$                     &  2.58  &  0.14  &  2.77  & -0.12  \\
  202 & $\sigma(\tau^{+}\tau^{-})$                   &  2.79  &  0.20  &  2.77  & -0.04  \\
  202 & $\mathrm{A_{FB}}(\mu^{+}\mu^{-})$            &  0.547 &  0.047 &  0.556 &  0.020 \\
  202 & $\mathrm{A_{FB}}(\tau^{+}\tau^{-})$          &  0.589 &  0.059 &  0.556 &  0.019 \\
 \hline
  205 & $\sigma(q\overline{q})$                      & 18.17  &  0.31  & 17.81  & -0.09  \\
  205 & $\sigma(\mu^{+}\mu^{-})$                     &  2.45  &  0.10  &  2.67  & -0.11  \\
  205 & $\sigma(\tau^{+}\tau^{-})$                   &  2.78  &  0.14  &  2.67  & -0.042 \\
  205 & $\mathrm{A_{FB}}(\mu^{+}\mu^{-})$            &  0.565 &  0.035 &  0.553 &  0.020 \\
  205 & $\mathrm{A_{FB}}(\tau^{+}\tau^{-})$          &  0.571 &  0.042 &  0.553 &  0.019 \\
 \hline
  207 & $\sigma(q\overline{q})$                      & 17.49  &  0.26  & 17.42  & -0.08  \\
  207 & $\sigma(\mu^{+}\mu^{-})$                     &  2.595 &  0.088 &  2.623 & -0.111 \\
  207 & $\sigma(\tau^{+}\tau^{-})$                   &  2.53  &  0.11  &  2.62  & -0.04  \\
  207 & $\mathrm{A_{FB}}(\mu^{+}\mu^{-})$            &  0.542 &  0.027 &  0.552 &  0.020 \\
  207 & $\mathrm{A_{FB}}(\tau^{+}\tau^{-})$          &  0.564 &  0.037 &  0.551 &  0.019 \\
 \hline
 \end{tabular}
 \end{tabular}
 \end{turn}
 \end{center}
\caption{Preliminary combined LEP results for $\eeff$, with cross
 section quoted in units of picobarn.
 All the results correspond to the first signal definition. The Standard Model
 predictions are from ZFITTER \capcite{ff:ref:ZFITTER}.
 The difference, $\Delta$, in the predictions of ZFITTER for 
 second definition relative to the first are given in the final column.
 The quoted uncertainties do not include the theoretical 
 uncertainties on the corrections discussed in the text.}
\label{ff:tab:xsafbres}
\end{table}
\begin{table}[p]
 \vskip 3cm
 \begin{center}
 \begin{turn}{90}
 \begin{tabular}{|c|c|c|c|c|c|c|c|c|c|c|c|c|}
 \hline
 $\begin{array}[b]{c}\roots \\ (\GeV) \end{array}$
       & 130    & 136    & 161    & 172    & 183    & 189    & 192    & 196    & 200    & 202    & 205    & 207    \\
 \hline\hline
 130 &  1.000 &  0.071 &  0.080 &  0.072 &  0.114 &  0.146 &  0.077 &  0.105 &  0.120 &  0.086 &  0.117 &  0.138 \\
 136 &  0.071 &  1.000 &  0.075 &  0.067 &  0.106 &  0.135 &  0.071 &  0.097 &  0.110 &  0.079 &  0.109 &  0.128 \\
 161 &  0.080 &  0.075 &  1.000 &  0.077 &  0.120 &  0.153 &  0.080 &  0.110 &  0.125 &  0.090 &  0.124 &  0.145 \\
 172 &  0.072 &  0.067 &  0.077 &  1.000 &  0.108 &  0.137 &  0.072 &  0.099 &  0.112 &  0.081 &  0.111 &  0.130 \\
 183 &  0.114 &  0.106 &  0.120 &  0.108 &  1.000 &  0.223 &  0.117 &  0.158 &  0.182 &  0.129 &  0.176 &  0.208 \\
 189 &  0.146 &  0.135 &  0.153 &  0.137 &  0.223 &  1.000 &  0.151 &  0.206 &  0.235 &  0.168 &  0.226 &  0.268 \\
 192 &  0.077 &  0.071 &  0.080 &  0.072 &  0.117 &  0.151 &  1.000 &  0.109 &  0.126 &  0.090 &  0.118 &  0.138 \\
 196 &  0.105 &  0.097 &  0.110 &  0.099 &  0.158 &  0.206 &  0.109 &  1.000 &  0.169 &  0.122 &  0.162 &  0.190 \\
 200 &  0.120 &  0.110 &  0.125 &  0.112 &  0.182 &  0.235 &  0.126 &  0.169 &  1.000 &  0.140 &  0.184 &  0.215 \\
 202 &  0.086 &  0.079 &  0.090 &  0.081 &  0.129 &  0.168 &  0.090 &  0.122 &  0.140 &  1.000 &  0.132 &  0.153 \\
 205 &  0.117 &  0.109 &  0.124 &  0.111 &  0.176 &  0.226 &  0.118 &  0.162 &  0.184 &  0.132 &  1.000 &  0.213 \\
 207 &  0.138 &  0.128 &  0.145 &  0.130 &  0.208 &  0.268 &  0.138 &  0.190 &  0.215 &  0.153 &  0.213 &  1.000 \\
 \hline
 \end{tabular}
 \end{turn}
 \end{center}
\caption{The correlation coefficients between averaged hadronic cross-sections
         at different energies.}
\label{ff:tab:hadcorrel}
 \vskip 5cm
\end{table}

\clearpage

\section{Averages for Differential Cross-sections}
\label{ff:sec-dsdc}

\subsection{{\boldmath{\ee}} final state}
\label{ff:sec-dsdc-ee}

The LEP experiments have measured the differential cross-section, $\dsdc$, 
for the $\eeee$ channel.%
A preliminary combination of these results is made by performing 
a $\chi^{2}$ fit to the measured differential cross-sections,
using the statistical errors as given by the experiments. In contrast
to the muon and tau  channels (Section~\ref{ff:sec-dsdc-mm-tt})
the higher statistics makes the use of expected statistical errors unnecessary.
The combination includes data from 189 $\GeV$ to 207 $\GeV$ from
all experiments but DELPHI. 
The data used in the 
combination are summarised in Table~\ref{ff:tab:ee_inputs}. 

Each experiment's data are binned according to an agreed common definition,
which takes into account the large forward peak of Bhabha scattering:
\begin{itemize}
 \item 10 bins for $\cos\theta$ between $0.0$  and $0.90$ and
 \item  5 bins for $\cos\theta$ between $-0.90$ and $0.0$
\end{itemize}
at each energy. The scattering angle, $\theta$, is
the angle of the negative lepton with respect to the incoming electron 
direction in the lab coordinate system. 
The outer acceptances of the most 
forward and most backward bins for which the experiments present
their data are different. The ranges in $\cos\theta$ of 
the individual experiments and the average are given in 
Table~\ref{ff:tab:acptee}. Except for the binning, each experiment uses their 
own signal definition, for example different experiments have different
acollinearity cuts to select events.
The signal definition used for the LEP average 
corresponds to an acollinearity cut of $\rm 10^{\circ}$. The experimental
measurements are corrected to the common signal definition following the
procedure described in Section~\ref{ff:sec-ave-xsc-afb}. The theoretical 
predictions are taken from the Monte Carlo event generator 
BHWIDE~\cite{ff:ref:BHWIDE}.

Correlated systematic errors between different experiments, energies and bins
at the same energy, arising from uncertainties on the overall normalisation,
and from migration of events between forward and backward bins with the same
absolute value of $\cos\theta$ due to uncertainties in the corrections for
charge confusion, were considered in the averaging procedure.

\begin{figure}[p]
 \begin{center}
  \epsfig{file=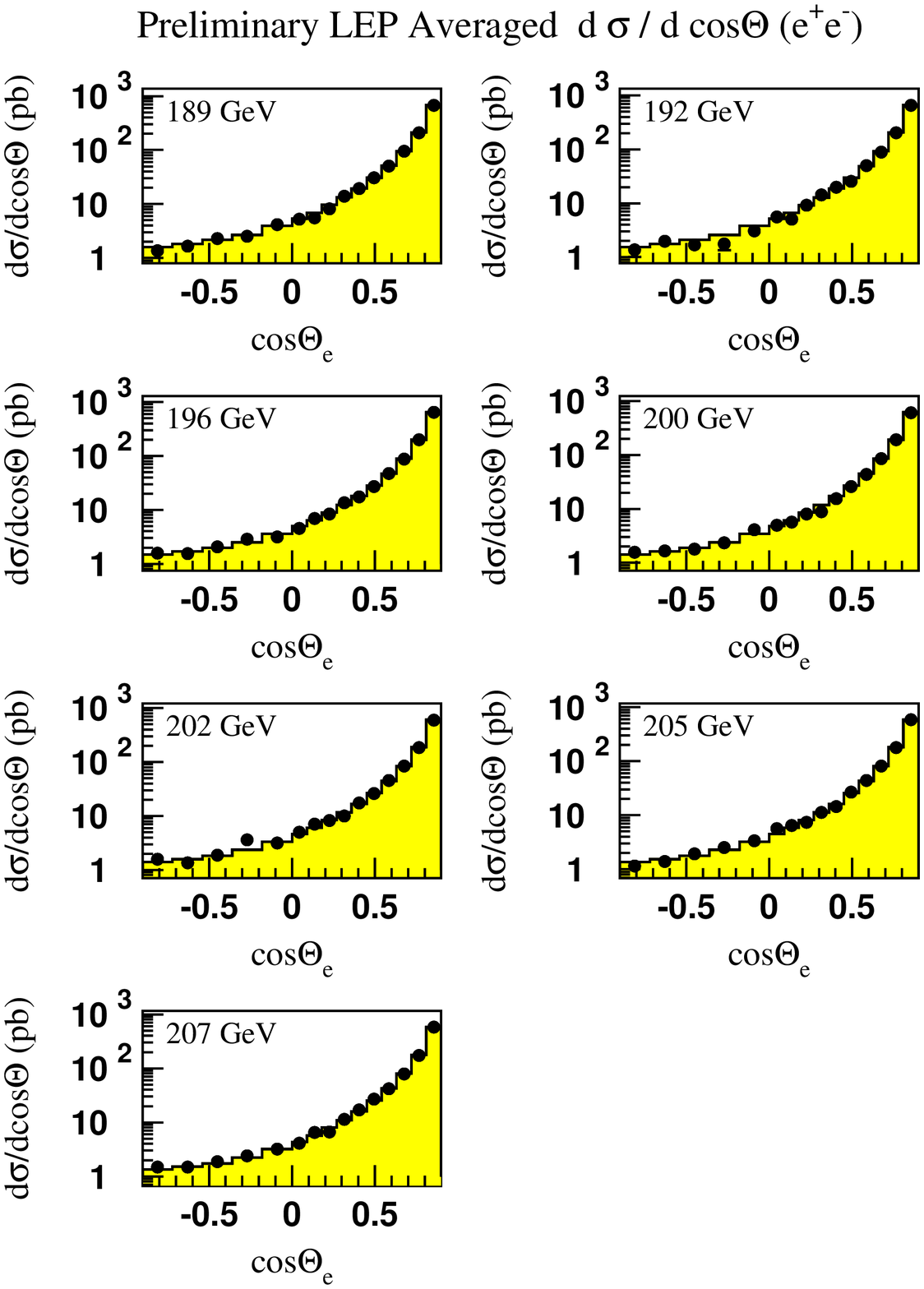,width=0.98\textwidth}
 \end{center}
 \caption{LEP averaged differential cross-sections for $\eeee$ at
          energies of 189--207 $\GeV$. The SM
          predictions, shown as solid histograms, are computed with
          BHWIDE~\capcite{ff:ref:BHWIDE}.}
 \label{ff:fig:dsdc-res-ee}
 \vskip 2cm 
\end{figure}

An average for all energies between 189--207 $\GeV$ is performed.
The results of the averages are shown in Figure~\ref{ff:fig:dsdc-res-ee}.
The $\chi^{2}$ per degree of freedom for the average is $190.8/189$.

The correlations between bins in the average are well below $5\%$ of the total
error on the averages in each bin for most of the cases, and exceed $10\%$ for
the most forward bin for the energy points with the highest accumulated
statistics.
The agreement between the averaged data and the predictions from the Monte
Carlo generator BHWIDE is good.
\subsection{{\boldmath{\mumu}}and {\boldmath{\tautau}} final states}
\label{ff:sec-dsdc-mm-tt}

The LEP experiments have measured the differential cross-section, $\dsdc$, 
for the $\eemumu$ and $\eetautau$ channels for samples of events 
with high effective centre-of-mass energy, $\sqrt{s'/s}>0.85$. 
A preliminary combination of these results is
made using the BLUE technique. The statistical error associated with
each measurement is taken as the expected statistical error on the 
differential cross-section, computed from the expected number of events 
in each bin for each experiment. Using a Monte Carlo simulation it has 
been shown that this method provides a good approximation to the exact 
likelihood method based on Poisson statistics~\cite{ff:ref:lepff-osaka}.

The combination includes data from 183 $\GeV$ to 207 $\GeV$, but not all 
experiments
provided data at all energies. 
The data used in the combination are summarised in
Table~\ref{ff:tab:inputs}.

Each experiment's data are binned in 10 bins of $\cos\theta$ at each 
energy, using their own signal definition. The scattering angle, $\theta$, is
the angle of the negative lepton with respect to the incoming electron 
direction in the lab coordinate system. The outer acceptances of the most 
forward and most backward bins for which the four experiments present 
their data are different. This was accounted for as part of the correction to 
a common signal definition. The ranges in $\cos\theta$ for the measurements of 
the individual experiments and the average are given in 
Table~\ref{ff:tab:acpt}. The signal definition used corresponded 
to the first definition given in Section~\ref{ff:sec-ave-xsc-afb}.

Correlated systematic errors between different experiments, channels and 
energies, arising from uncertainties on the overall normalisation are 
considered in the averaging procedure.
All data from all energies are combined in a single fit to obtain
averages at each centre-of-mass energy yielding the full covariance matrix 
between the different measurements at all energies.

\begin{figure}[p]
 \begin{center}
  \epsfig{file=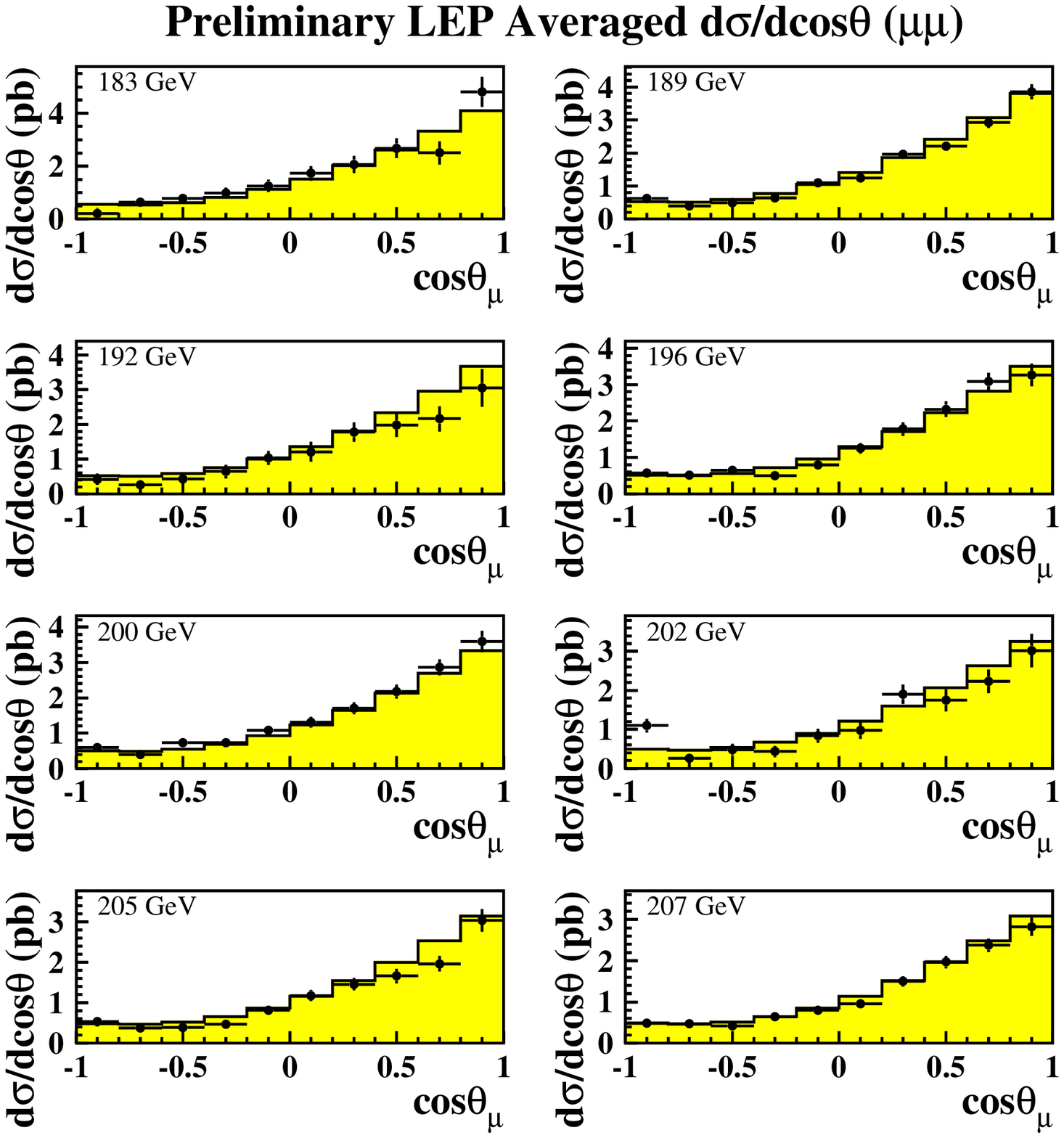,width=0.98\textwidth}
 \end{center}
 \caption{LEP averaged differential cross-sections for $\eemumu$ at
          energies of 183--207 $\GeV$. The SM
          predictions, shown as solid histograms, are computed with
          ZFITTER~\capcite{ff:ref:ZFITTER}.}
 \label{ff:fig:dsdc-res-mm}
 \vskip 2cm 
\end{figure}

\begin{figure}[p]
 \begin{center}
  \epsfig{file=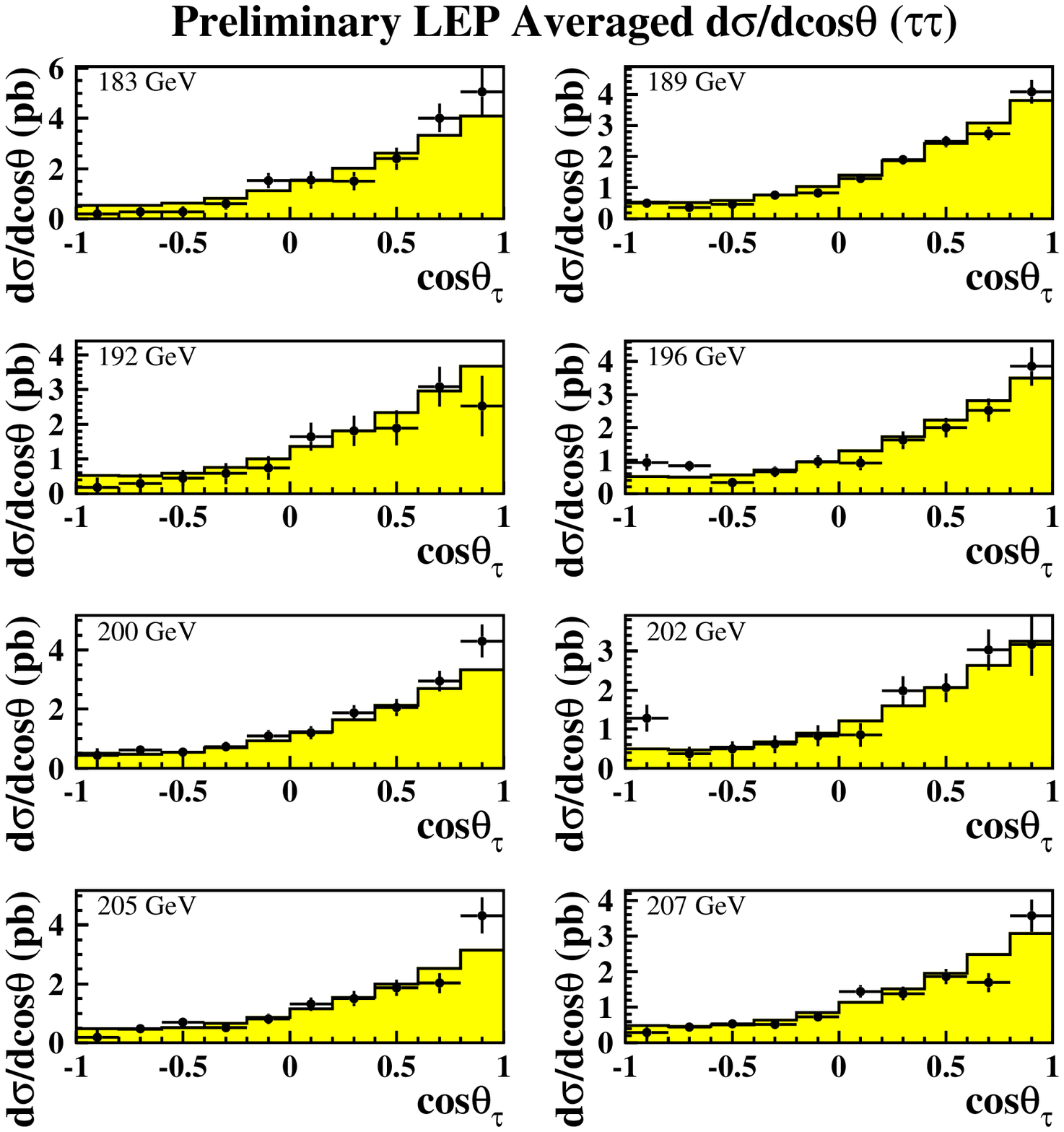,width=0.98\textwidth}
 \end{center}
 \caption{LEP averaged differential cross-sections for $\eetautau$ at 
          energies of 183--207 $\GeV$. The SM
          predictions, shown as solid histograms, are computed with
          ZFITTER~\capcite{ff:ref:ZFITTER}.}
 \label{ff:fig:dsdc-res-tt}
 \vskip 2cm 
\end{figure}

The results of the averages are shown in Figures~\ref{ff:fig:dsdc-res-mm} 
and~\ref{ff:fig:dsdc-res-tt}. 
The correlations between bins in the average are less that 
$2\%$ of the total error on the averages in each bin.
Overall the agreement between the averaged data and the predictions
is reasonable, with a $\chi^{2}$ of $200$ for $160$ degrees of freedom. 
At 202 $\GeV$ the measured differential cross-sections in the most backward 
bins, $-1.00 < \cos\theta < 0.8$, for both muon and tau final states are 
above the predictions. The data at 202 $\GeV$ suffer
from rather low delivered luminosity, with less than 4 events
expected in each experiment in each channel in this backward 
$\cos\theta$ bin. The agreement between the data
and the predictions in the same $\cos\theta$ bin is more consistent at 
higher energies. 

\begin{table}[htbp]
 \begin{center}
 \begin{tabular}{|l|cccc|}
 \hline
                     & \multicolumn{4}{|c|}{$\eeee$}             \\
 \cline{2-5}
  $\sqrt{s}$($\GeV$) &      A   &      D   &      L   &      O   \\
 \hline 
  189                & {\sc{P}} & {\sc{-}} & {\sc{P}} & {\sc{F}} \\
 \hline
  192--202           & {\sc{P}} & {\sc{-}} & {\sc{P}} & {\sc{P}} \\
 \hline
  205--207           & {\sc{P}} & {\sc{-}} & {\sc{P}} & {\sc{P}} \\
 \hline
 \end{tabular}
 \end{center}
 \caption{Differential cross-section data provided by the LEP 
          collaborations (ALEPH, DELPHI, L3 and OPAL) for $\eeee$.
          Data indicated with {\sc{F}} are final, published data.
          Data marked with {\sc{P}} are preliminary. 
          Data marked with a {\sc{-}} were not available for combination.}
 \label{ff:tab:ee_inputs}
\end{table}
\begin{table}[htbp]
 \begin{center}
 \begin{tabular}{|l|c|c|}
  \hline
  Experiment                       & $\cos\theta_{min}$ & $\cos\theta_{max}$ \\
  \hline
  \hline 
   ALEPH  ($\sqrt{s'/s}>0.85$)     &    $-0.90$         &     $0.90$         \\
   L3     (acol. $<\ 25^{\circ}$)  &    $-0.72$         &     $0.72$         \\
   OPAL   (acol. $<\ 10^{\circ}$)  &    $-0.90$         &     $0.90$         \\
  \hline
  \hline
   Average (acol. $<\ 10^{\circ}$) &    $-0.90$         &     $0.90$         \\
  \hline
 \end{tabular}
 \end{center}
 \caption{The acceptances for which experimental data are presented 
          for the $\eeee$ channel
          and the acceptance for the LEP average.}
 \label{ff:tab:acptee}
\end{table}
\begin{table}[htbp]
 \begin{center}
 \begin{tabular}{|l|cccc|cccc|}
 \hline
                     & \multicolumn{4}{|c|}{$\eemumu$}           
                     & \multicolumn{4}{|c|}{$\eetautau$}         \\
 \cline{2-9}
  $\sqrt{s}$($\GeV$) &      A   &      D   &      L   &      O   
                     &      A   &      D   &      L   &      O   \\
 \hline 
 \hline
  183                & {\sc{-}} & {\sc{F}} & {\sc{-}} & {\sc{F}} 
                     & {\sc{-}} & {\sc{F}} & {\sc{-}} & {\sc{F}} \\
 \hline
  189                & {\sc{P}} & {\sc{F}} & {\sc{F}} & {\sc{F}}  
                     & {\sc{P}} & {\sc{F}} & {\sc{F}} & {\sc{F}} \\
 \hline
  192--202           & {\sc{P}} & {\sc{P}} & {\sc{P}} & {\sc{P}} 
                     & {\sc{P}} & {\sc{P}} & {\sc{-}} & {\sc{P}} \\
 \hline
  205--207           & {\sc{P}} & {\sc{P}} & {\sc{P}} & {\sc{P}} 
                     & {\sc{P}} & {\sc{P}} & {\sc{-}} & {\sc{P}} \\
 \hline
 \end{tabular}
 \end{center}
 \caption{Differential cross-section data provided by the LEP 
          collaborations (ALEPH, DELPHI, L3 and OPAL) for $\eemumu$ and 
          $\eetautau$ combination at different centre-of-mass energies. 
          Data indicated with {\sc{F}} are final, published data. 
          Data marked with {\sc{P}} are preliminary. 
          Data marked with a {\sc{-}} were not available for combination.}
 \label{ff:tab:inputs}
\end{table}
\begin{table}[htbp]
 \begin{center}
 \begin{tabular}{|l|c|c|}
  \hline
  Experiment                   & $\cos\theta_{min}$ & $\cos\theta_{max}$ \\
  \hline
  \hline 
   ALEPH                       &    $-0.95$         &     $0.95$         \\
   DELPHI ($\eemumu$ 183)      &    $-0.94$         &     $0.94$         \\
   DELPHI ($\eemumu$ 189--207) &    $-0.97$         &     $0.97$         \\
   DELPHI ($\eetautau$)        &    $-0.96$         &     $0.96$         \\
   L3                          &    $-0.90$         &     $0.90$         \\
   OPAL                        &    $-1.00$         &     $1.00$         \\
  \hline
  \hline
   Average                     &    $-1.00$         &     $1.00$         \\
  \hline
 \end{tabular}
 \end{center}
 \caption{The acceptances for which experimental data are presented 
          and the acceptance for the LEP average.
          For DELPHI the acceptance is shown for the different channels and 
          for the muons for different centre of mass energies. For all other
          experiments the acceptance is the same for muon and tau-lepton 
          channels and for all energies provided.}
 \label{ff:tab:acpt}
\end{table}

\clearpage

\section{Averages for Heavy Flavour Measurements}
\label{ff:sec-hvflv}

This section presents a preliminary combination of both 
published~\cite{ff:ref:hfpublished}
and preliminary~\cite{ff:ref:hfpreliminary} measurements of the 
ratios cross section ratios $R_{\mathrm{q}}$ defined as  
${\mathrm{\frac{\sigma_{q \overline {q} } }{\sigma_{had}}}}$
for b and c production, $\Rb$ and $\Rc$,  
and the forward-backward asymmetries, $\Abb$ and 
$\Acc$, from the LEP collaborations at centre-of-mass 
energies in the range of 130 $\GeV$ to 207 $\GeV$. 
Table~\ref{ff:tab:hfinput} summarises all the inputs that have been combined so far.

A common signal definition is defined for all the measurements, requiring:
\begin{list}{$\bullet$}{\setlength{\itemsep}{0ex}
                        \setlength{\parsep}{0ex}
                        \setlength{\topsep}{0ex}}
 \item{an effective centre-of-mass energy $\sqrt{s^{\prime}} > 0.85 \sqrt{s}$}
 \item{no subtraction of ISR and FSR photon interference contribution and}
 \item{extrapolation to full angular acceptance.}
\end{list}
Systematic errors are divided into three categories: uncorrelated errors, 
errors correlated between the measurements of each experiment, and 
errors common to all experiments. 

Due to the fact that $\Rc$ measurements are only provided by a single 
experiment and are strongly correlated with $\Rb$ measurements, 
it was decided to fit the b sector and c sector separately, 
the other flavour's measurements being fixed to their Standard Model 
predictions. 
In addition, these fitted values are used to set limits upon physics beyond 
the Standard Model, such as contact term interactions, in which only one 
quark flavour is assumed to be effected by the new physics during each fit,
therefore this averaging method is consistent with the interpretations.

Full details concerning the combination procedure
can be found in~\cite{ff:ref:hfconfnote}. 

\begin{figure}[p]
\begin{center}
\mbox{\epsfig{file=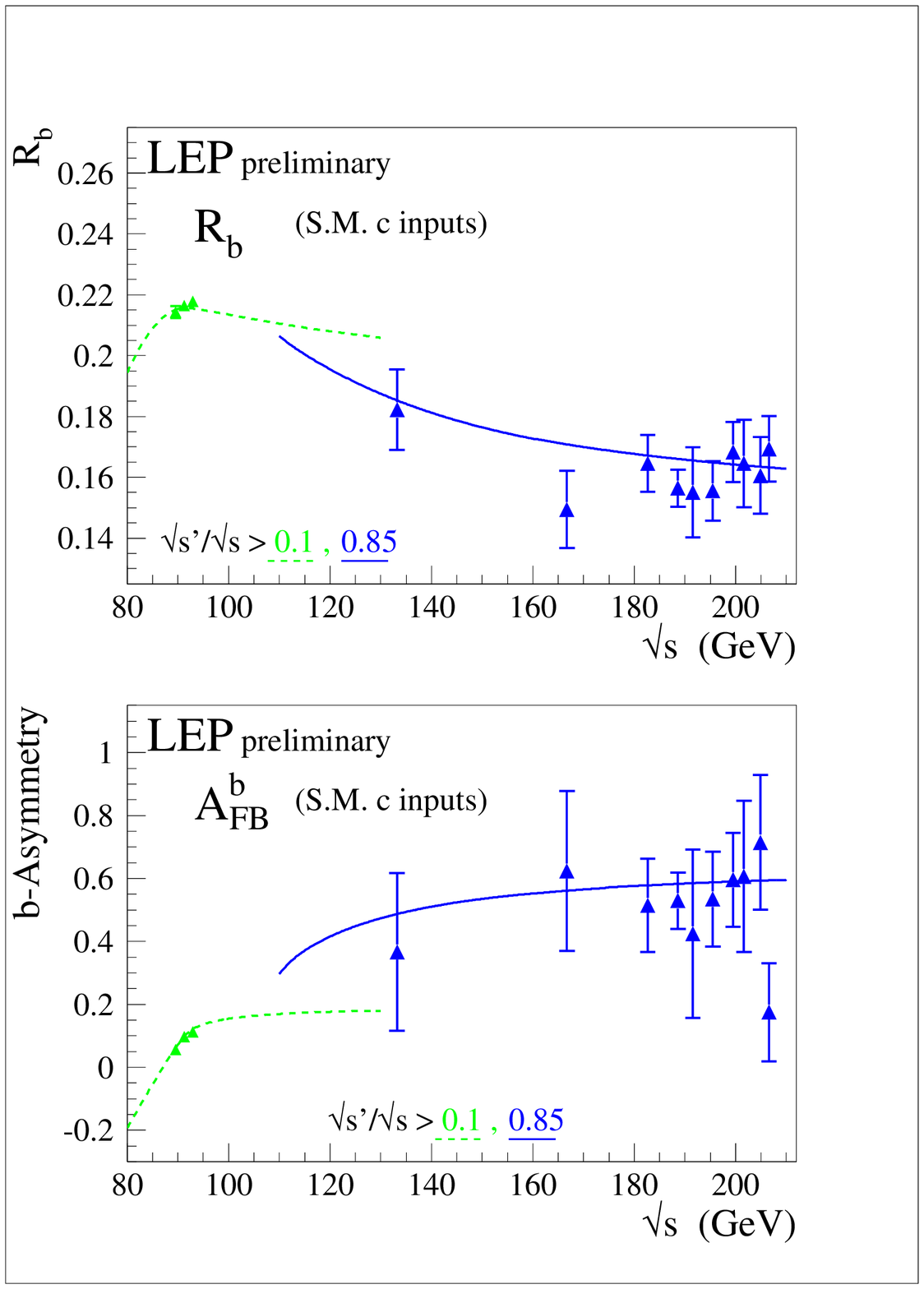,height=20cm}}
\end{center}
\caption{Preliminary combined LEP measurements of $\Rb$ and $\Abb$. 
  Solid lines represent the Standard Model prediction for the high
  $\sqrt{s'}$ selection used at $\LEPII$ and dotted lines the inclusive
  prediction used at $\LEPI$. Both are computed with
  ZFITTER\capcite{ff:ref:hfzfit}. The $\LEPI$ measurements have been
  taken from \capcite{ff:ref:hflep1-99}.}
\label{ff:fig:hfbres}
\end{figure}

\begin{figure}[p]
\begin{center}
\mbox{\epsfig{file=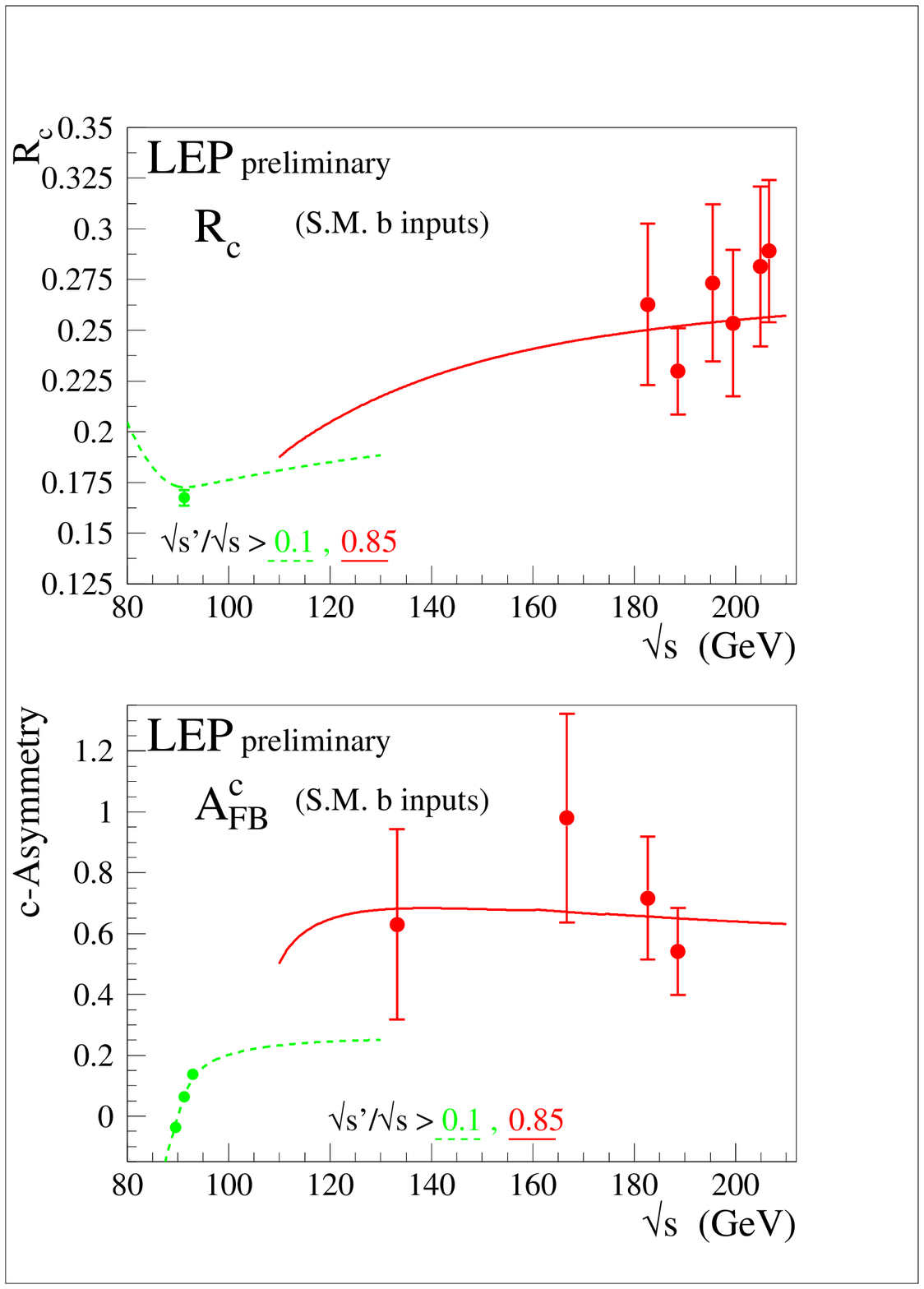,height=20cm}}
\end{center}
\caption{Preliminary combined LEP measurements of $\Rc$ and $\Acc$. 
  Solid lines represent the Standard Model prediction for the high
  $\sqrt{s'}$ selection used at $\LEPII$ and dotted lines the
  inclusive prediction used at $\LEPI$.  Both are computed with
  ZFITTER~\capcite{ff:ref:hfzfit}. The $\LEPI$ measurements have been
  taken from~\capcite{ff:ref:hflep1-99}.}
\label{ff:fig:hfcres} 
\end{figure}

The results of the combination are presented in Table~\ref{ff:tab:hfbresults} 
and Table~\ref{ff:tab:hfcresults} 
and in Figures~\ref{ff:fig:hfbres} and~\ref{ff:fig:hfcres}. 
The results for both b and c sector are in agreement with the Standard 
Model predictions of ZFITTER. 
The averaged discrepancies with respect to the Standard Model predictions 
is -2.08 $\sigma$ for $\Rb$, +0.30 $\sigma$ for $\Rc$, -1.56  $\sigma$ for $\Abb$ and -0.24 $\sigma $ for $\Acc$. 
A list of the error contributions from the combination at 189~$\GeV$ is shown 
in Table~\ref{ff:tab:hferror}.

\begin{table}[htbp]
\begin{center}
\begin{tabular}{|l|cccc|cccc|cccc|cccc|}
\hline 
 $\sqrt{s}$ ($\GeV$)
            & \multicolumn{4}{|c|}{$\Rb$}
            & \multicolumn{4}{|c|}{$\Rc$}
            & \multicolumn{4}{|c|}{$\Abb$}
            & \multicolumn{4}{|c|}{$\Acc$} \\
\cline{2-17}
            & A & D & L & O & A & D & L & O & A & D & L & O & A & D & L & O \\
\hline\hline
133         & {\sc{F}} & {\sc{F}} & {\sc{F}} & {\sc{F}}
            & {\sc{-}} & {\sc{-}} & {\sc{-}} & {\sc{-}}  
            & {\sc{-}} & {\sc{F}} & {\sc{-}} & {\sc{F}}  
            & {\sc{-}} & {\sc{F}} & {\sc{-}} & {\sc{F}} \\
\hline
167         & {\sc{F}} & {\sc{F}} & {\sc{F}} & {\sc{F}}
            & {\sc{-}} & {\sc{-}} & {\sc{-}} & {\sc{-}} 
            & {\sc{-}} & {\sc{F}} & {\sc{-}} & {\sc{F}} 
            & {\sc{-}} & {\sc{F}} & {\sc{-}} & {\sc{F}}   \\
\hline 
183         & {\sc{F}} & {\sc{P}} & {\sc{F}} & {\sc{F}}
            & {\sc{F}} & {\sc{-}} & {\sc{-}} & {\sc{-}}  
            & {\sc{F}} & {\sc{-}} & {\sc{-}} & {\sc{F}} 
            & {\sc{P}} & {\sc{-}} & {\sc{-}} & {\sc{F}}  \\
\hline 
189         & {\sc{P}} & {\sc{P}} & {\sc{F}} & {\sc{F}}
            & {\sc{P}} & {\sc{-}} & {\sc{-}} & {\sc{-}}  
            & {\sc{P}} & {\sc{P}} & {\sc{F}} & {\sc{F}} 
            & {\sc{P}} & {\sc{-}} & {\sc{-}} & {\sc{F}} \\
\hline 
192 to 202  & {\sc{P}} & {\sc{P}} & {\sc{P}} & {\sc{-}} 
            & {\sc{P*}} & {\sc{-}} & {\sc{-}} & {\sc{-}} 
            & {\sc{P}} & {\sc{P}} & {\sc{-}} & {\sc{-}} 
            & {\sc{-}} & {\sc{-}} & {\sc{-}} & {\sc{-}} \\ 
\hline 
205 and 207 & {\sc{-}} & {\sc{P}} & {\sc{P}} & {\sc{-}} 
            & {\sc{P}} & {\sc{-}} & {\sc{-}} & {\sc{-}} 
            & {\sc{P}} & {\sc{P}} & {\sc{-}} & {\sc{-}} 
            & {\sc{-}} & {\sc{-}} & {\sc{-}} & {\sc{-}} \\
\hline
\end{tabular}
\end{center}
\caption{Data provided by the ALEPH, DELPHI, L3, OPAL collaborations 
         for combination at different centre-of-mass energies. 
         Data indicated with {\sc{F}} are final, published data. 
         Data marked with {\sc{P}} are preliminary and for data marked 
         with {\sc{P*}}, not all energies are supplied.
         Data marked with a {\sc{-}} were not supplied for combination.}
\label{ff:tab:hfinput} 
\end{table}
\begin{table}[htbp]
\begin{center}
\begin{tabular}{|l|c|c|}
\hline 
$\sqrt{s}$ ($\GeV$) & $\Rb$
                    & $\Abb$ \\
\hline\hline
133      & 0.1822 $\pm$ 0.0132 & 0.367 $\pm$ 0.251 \\
         & (0.1867)            & (0.504)           \\
\hline
167      & 0.1494 $\pm$ 0.0127 & 0.624 $\pm$ 0.254 \\
         & (0.1727)            & (0.572)           \\
\hline 
183      & 0.1646 $\pm$ 0.0094 & 0.515 $\pm$ 0.149 \\
         & (0.1692)            & (0.588)           \\
\hline 
189      & 0.1565 $\pm$ 0.0061 & 0.529 $\pm$ 0.089 \\
         & (0.1681)            &  (0.593)          \\
\hline
192      & 0.1551 $\pm$ 0.0149 & 0.424 $\pm$ 0.267 \\
         & (0.1676)            & (0.595)           \\
\hline 
196      & 0.1556 $\pm$ 0.0097 & 0.535 $\pm$ 0.151 \\
         & (0.1670)            & (0.598)           \\
\hline
200      & 0.1683 $\pm$ 0.0099 & 0.596 $\pm$ 0.149 \\
         & (0.1664)            & (0.600)           \\
\hline
202      & 0.1646 $\pm$ 0.0144 & 0.607 $\pm$ 0.241 \\
         & (0.1661)            & (0.601)           \\
\hline
205      & 0.1606 $\pm$ 0.0126 & 0.715 $\pm$ 0.214 \\
         & (0.1657)            & (0.603)           \\
\hline
207      & 0.1694 $\pm$ 0.0107 & 0.175 $\pm$ 0.156 \\
         & (0.1654)            & (0.604)           \\
\hline
\end{tabular}
\end{center}
\caption[]{Combined results on $\Rb $ and $\Abb$. Quoted errors 
represent the statistical and systematic errors added in quadrature. 
For comparison, the Standard Model predictions computed with 
ZFITTER~\capcite{ff:ref:hfzfit} are given in parentheses. }
\label{ff:tab:hfbresults}
\end{table}
\begin{table}[htbp]
\begin{center}
\begin{tabular}{|l|c|c|}
\hline 
$\sqrt{s}$ ($\GeV$) & $\Rc$
                    & $\Acc$ \\
\hline\hline
133      & -   & 0.630 $\pm$ 0.313 \\
         &     & (0.684)           \\
\hline
167      & -   & 0.980 $\pm$ 0.343 \\
         &     & (0.677)           \\
\hline 
183      & 0.2628 $\pm$ 0.0397  & 0.717 $\pm$ 0.201 \\
         & (0.2472)    & (0.663)           \\
\hline 
189      & 0.2298 $\pm$ 0.0213  & 0.542 $\pm$ 0.143 \\
         & (0.2490)    & (0.656)           \\
\hline
196      & 0.2734 $\pm$ 0.0387 & - \\
         & (0.2508)            &   \\
\hline
200      & 0.2535 $\pm$ 0.0360 & - \\
         & (0.2518)            &   \\
\hline
205      & 0.2816 $\pm$ 0.0394 & - \\
         & (0.2530)            &   \\
\hline
207      & 0.2890 $\pm$ 0.0350 & - \\
         & (0.2533)            &   \\
\hline
\end{tabular}
\end{center}
\caption{Combined results on $\Rc$ and $\Acc$. Quoted errors 
represent the statistical and systematic errors added in quadrature. 
For comparison, the Standard Model predictions computed with 
ZFITTER~\capcite{ff:ref:hfzfit} are given in parentheses. }
\label{ff:tab:hfcresults}
\end{table}
\begin{table}[htbp]
\begin{center}
\begin{tabular}{|l|c|c||c|c|}
\hline 
Error list & $\Rb$ (189 $\GeV$) 
           & $\Abb$ (189 $\GeV$) 
           & $\Rc$ (189 $\GeV$) 
           & $\Acc$ (189 $\GeV$)  \\

\hline\hline
statistics    & 0.0057  & 0.084 & 0.0169 & 0.119 \\ 
\hline 
internal syst & 0.0020  & 0.025 & 0.0109 & 0.042 \\
common syst   & 0.0007  & 0.011 & 0.0072 & 0.069 \\
total syst    & 0.0021  & 0.027 & 0.0130 & 0.081 \\ 
\hline 
total error   & 0.0061  & 0.089 & 0.0213 & 0.143 \\ 
\hline 
\end{tabular}
\end{center}
\caption{Error breakdown at 189 $\GeV$.}
\label{ff:tab:hferror} 
\end{table}
\section{Interpretation}
\label{ff:sec-interp}

The combined measurements presented above are interpreted in a variety
of models.
The cross-section and asymmetry results are used to place limits 
on contact interactions between leptons and quarks and, using 
the results on heavy flavour production, on contact interaction between 
electrons and $b$ and $c$ quarks specifically.
Limits on the mass of a possible additional heavy neutral boson, $\Zprime$,
are obtained for a variety of models.
Using the combined differential cross-sections for \ee\ final states, 
limits on contact interactions in the $\eeee$ channel and limits on the 
scale of gravity in models with large extra-dimensions are presented.
Limits are also derived on the masses of leptoquarks - assuming
a coupling of electromagnetic strength. 
In all cases the Born level predictions for the physics beyond the Standard 
Model have been corrected to take into account QED radiation.

\subsection{Contact Interactions}
\label{ff:sec-cntc}

The averages of cross-sections and forward-backward asymmetries for 
muon-pair and tau-lepton pair and the cross-sections for $\qq$ 
final states are used to search for 
contact interactions between fermions. 

Following~\cite{ff:ref:ELPthr}, contact interactions are parameterised 
by an effective Lagrangian, $\cal{L}_{\mathrm{eff}}$, which is added to the 
Standard Model Lagrangian and has the form:
\begin{eqnarray}
 \mbox{$\cal{L}$}_{\mathrm{eff}} = 
                        \frac{g^{2}}{(1+\delta)\Lambda^{2}} 
                          \sum_{i,j=L,R} \eta_{ij} 
                           \overline{e}_{i} \gamma_{\mu} e_{i}
                            \overline{f}_{j} \gamma^{\mu} f_{j},
\end{eqnarray}
where $g^{2}/{4\pi}$ is taken to be 1 by convention, $\delta=1 (0)$ for 
$f=e ~(f \neq e)$, $\eta_{ij}=\pm 1$ or $0$ for different interaction types,
$\Lambda$ is the scale of the contact interactions,
$e_{i}$ and $f_{j}$ are left or right-handed spinors. 
By assuming different helicity coupling between the initial 
state and final state currents, a set of different models can be defined
from this Lagrangian~\cite{ff:ref:Kroha}, with either
constructive ($+$) or destructive ($-$) interference between the 
Standard Model process and the contact interactions. The models and 
corresponding choices of $\eta_{ij}$ are given in Table~\ref{ff:tab:cntcdef}.
The models LL$^{\pm}$, RR$^{\pm}$, VV$^{\pm}$, AA$^{\pm}$, LR$^{\pm}$, 
RL$^{\pm}$, V0$^{\pm}$, A0$^{\pm}$ are considered here since 
these models lead to large deviations in $\eeff$ at LEP II.
The corresponding energies scales for the models with constructive
or destructive interference are denoted by $\Lambda^{+}$ and $\Lambda^{-}$
respectively.

For leptonic final states 4 different fits are made
\begin{itemize}
 \item individual fits to contact interactions in $\eemumu$ and $\eetautau$
       using the measured cross-sections and asymmetries,
 \item fits to $\eell$ (simultaneous fits to $\eemumu$ and $\eetautau$)
       again using the measured cross-sections and asymmetries,
 \item fits to $\eeee$, using the measured differential cross-sections. 
\end{itemize}
For the inclusive hadronic final states three different model 
assumptions are used to fit the total hadronic cross-section
\begin{itemize}
\item the contact interactions affect only one quark flavour of up-type 
      using the measured hadronic cross-sections,
\item the contact interactions affect only one quark flavour of down-type 
      using the measured hadronic cross-sections,
\item the contact interactions contribute to all quark final states with 
      the same strength. 
\end{itemize}

Limits on contact interactions between electrons and $b$ and $c$ quarks
are obtained using all the heavy flavour $\LEPII$ combined results 
from 133 $\GeV$ to 207 $\GeV$ given in Tables~\ref{ff:tab:hfbresults} 
and~\ref{ff:tab:hfcresults}.
For the purpose of fitting contact interaction models to the data, 
$\Rb$ and $\Rc$ are converted to cross-sections 
$\sigma_{\bb}$ and $\sigma_{\cc}$ using the averaged ${\qq}$ cross-section of 
section \ref{ff:sec-ave-xsc-afb} corresponding to the second signal 
definition.  
In the calculation of errors, the correlations between $\Rb$, $\Rc$ and 
$\sigma_{\qq}$ are assumed to be negligible.
These results are of particular interest since they are inaccessible
to ${\mathrm{p\bar{p}}}$ or ep colliders.

For the purpose of fitting contact interaction models to the data, 
the parameter $\epsilon=1/\Lambda^{2}$ is used, with
$\epsilon=0$ in the limit that there are no contact interactions. 
This parameter is allowed to take both positive and negative values in 
the fits. 
Theoretical uncertainties on the Standard Model predictions are taken 
from~\cite{ff:ref:lepffwrkshp}.

The values of $\epsilon$ extracted for each model are all compatible 
with the Standard Model expectation $\epsilon=0$, at the two standard 
deviation level. As expected, 
the errors on $\epsilon$ are typically a factor of two 
smaller than those obtained from a single LEP experiment with the same data
set. The fitted values of $\epsilon$ are converted into  
$95\%$ confidence level lower limits on $\Lambda$. 
The limits are obtained by integrating the likelihood function in 
$\epsilon$ over the physically allowed values\footnote{To be able to obtain 
confidence limits from the likelihood function in $\epsilon$ 
it is necessary to convert the likelihood to a probability density function 
for $\epsilon$; this is done by 
multiplying by a prior probability function. Simply integrating the 
likelihood over $\epsilon$ is equivalent to multiplying by a uniform 
prior probability function in $\epsilon$.}, 
$\epsilon \ge 0$ for each $\Lambda^{+}$ limit and $\epsilon \le 0$ for 
$\Lambda^{-}$ limits.

\begin{figure}[p]
 \begin{center}
  \begin{tabular}{cc}
  \epsfig{file=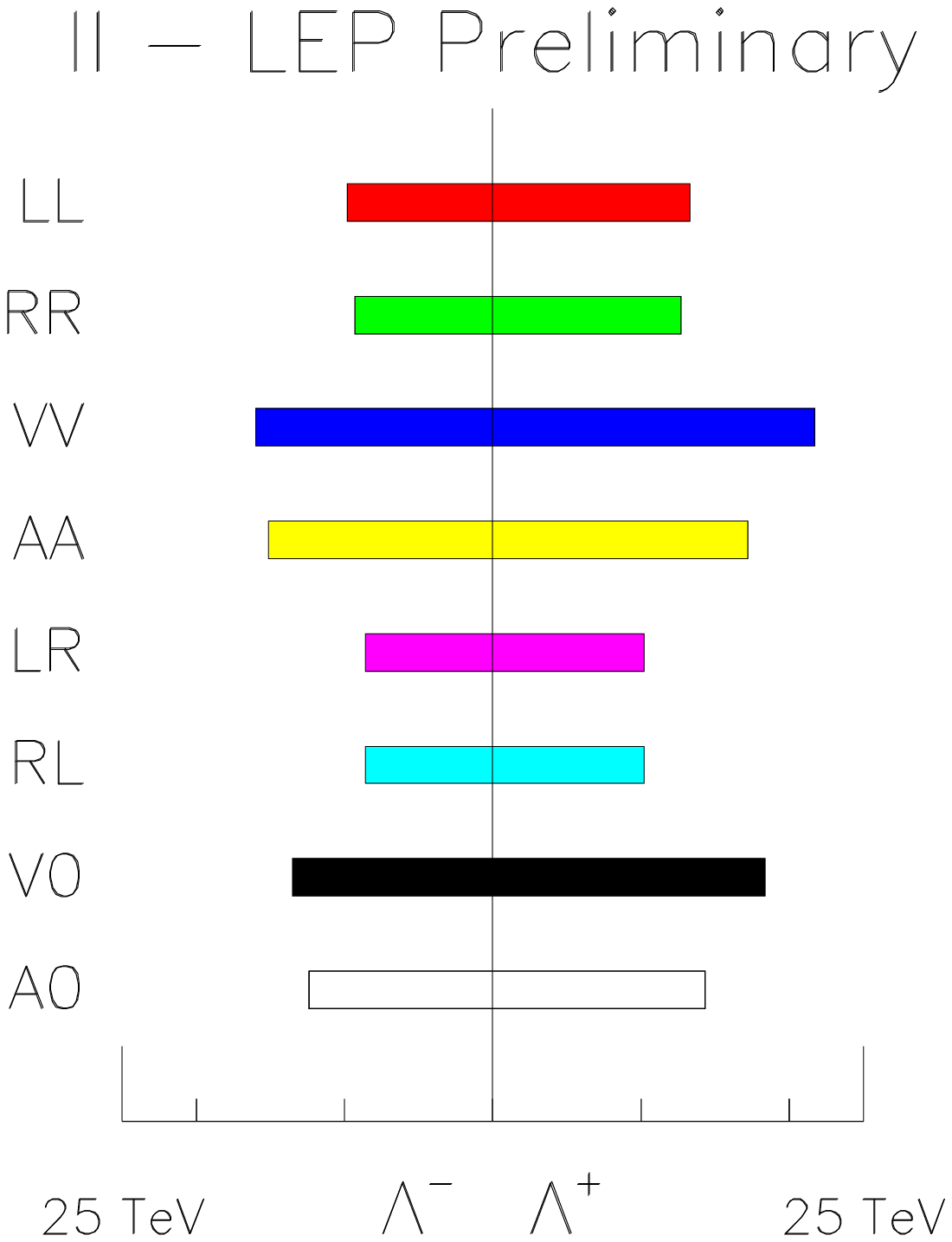,width=0.36\textwidth} &
  \epsfig{file=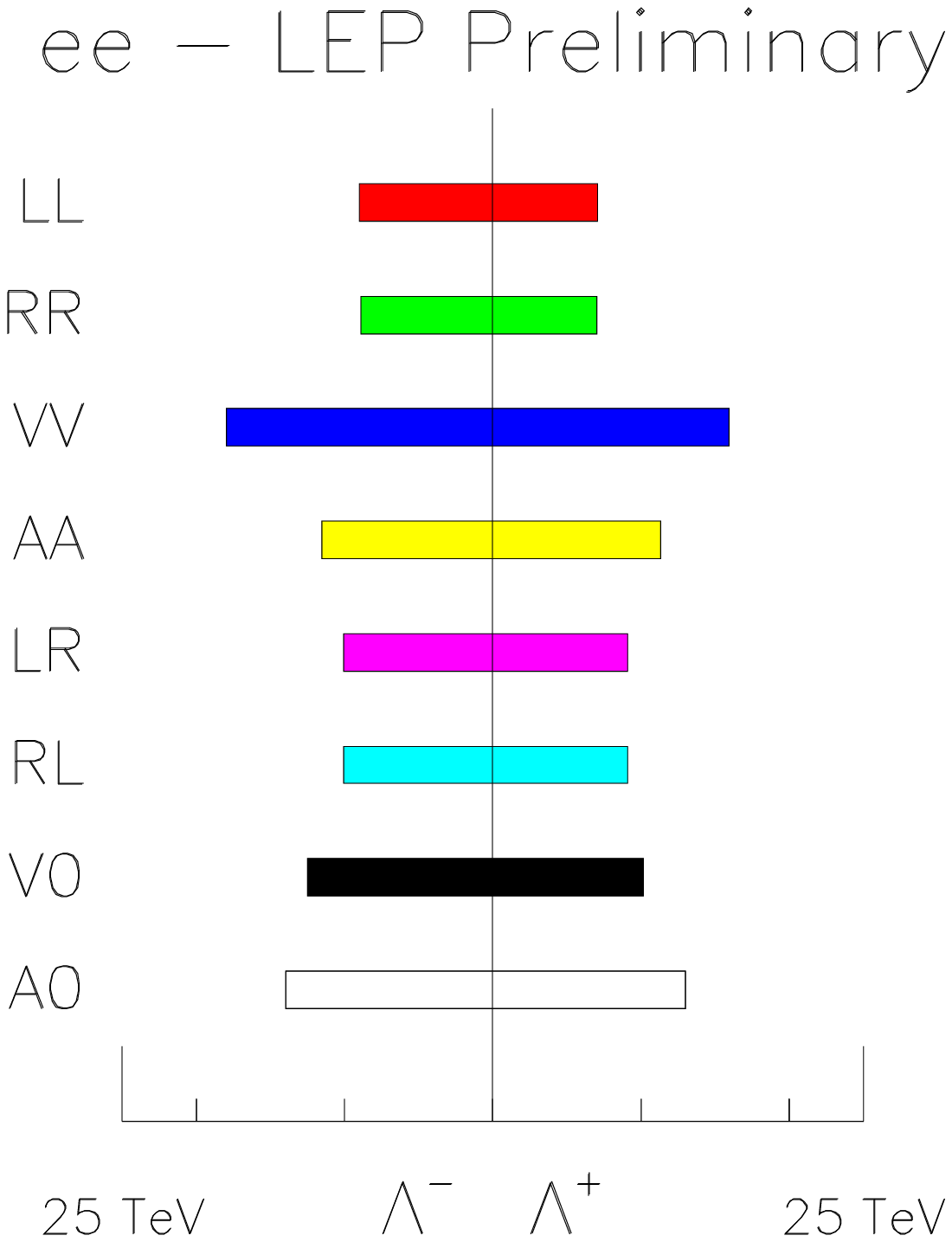,width=0.36\textwidth} \\
  \multicolumn{2}{c}
   {\epsfig{file=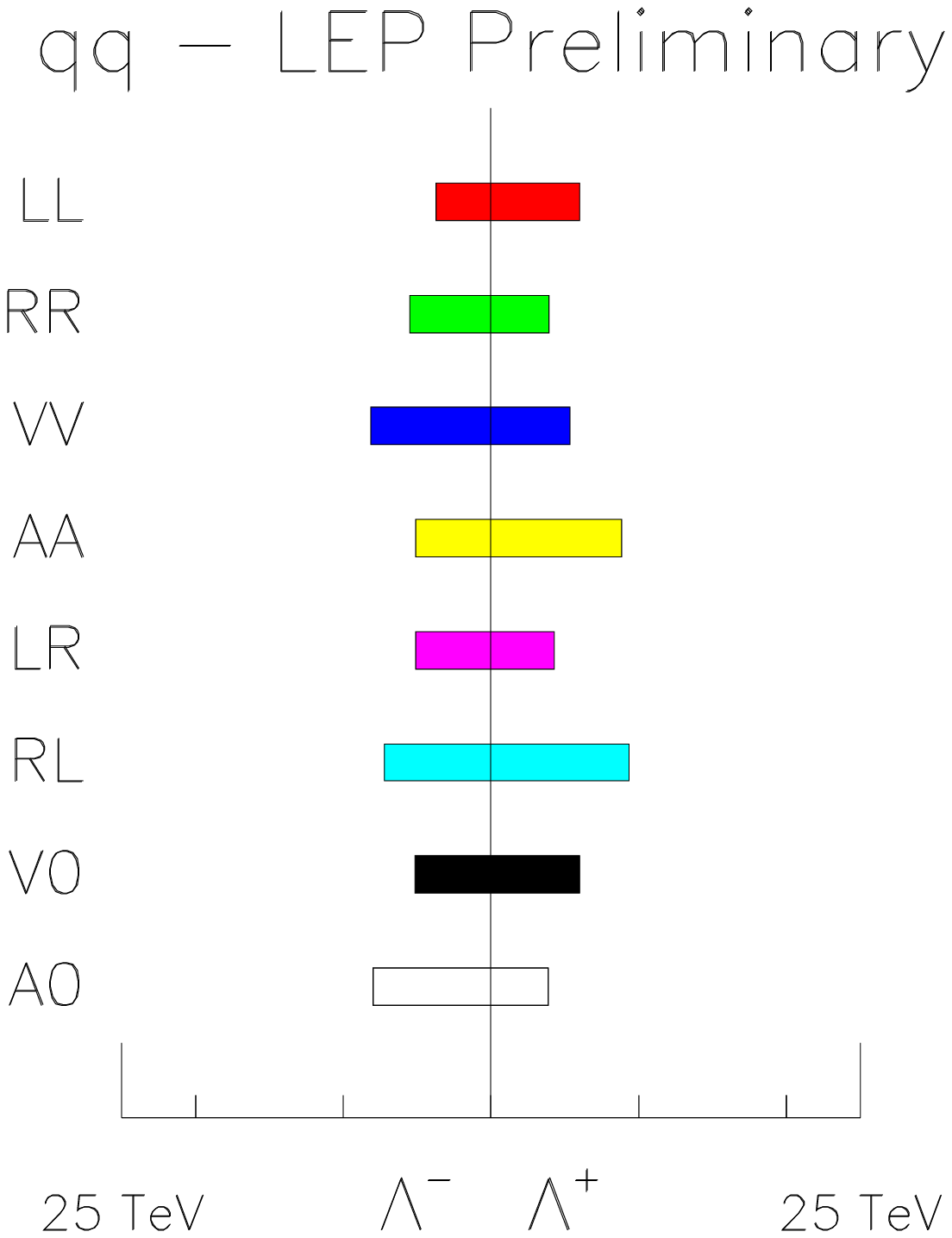,width=0.36\textwidth}} \\ 
  \epsfig{file=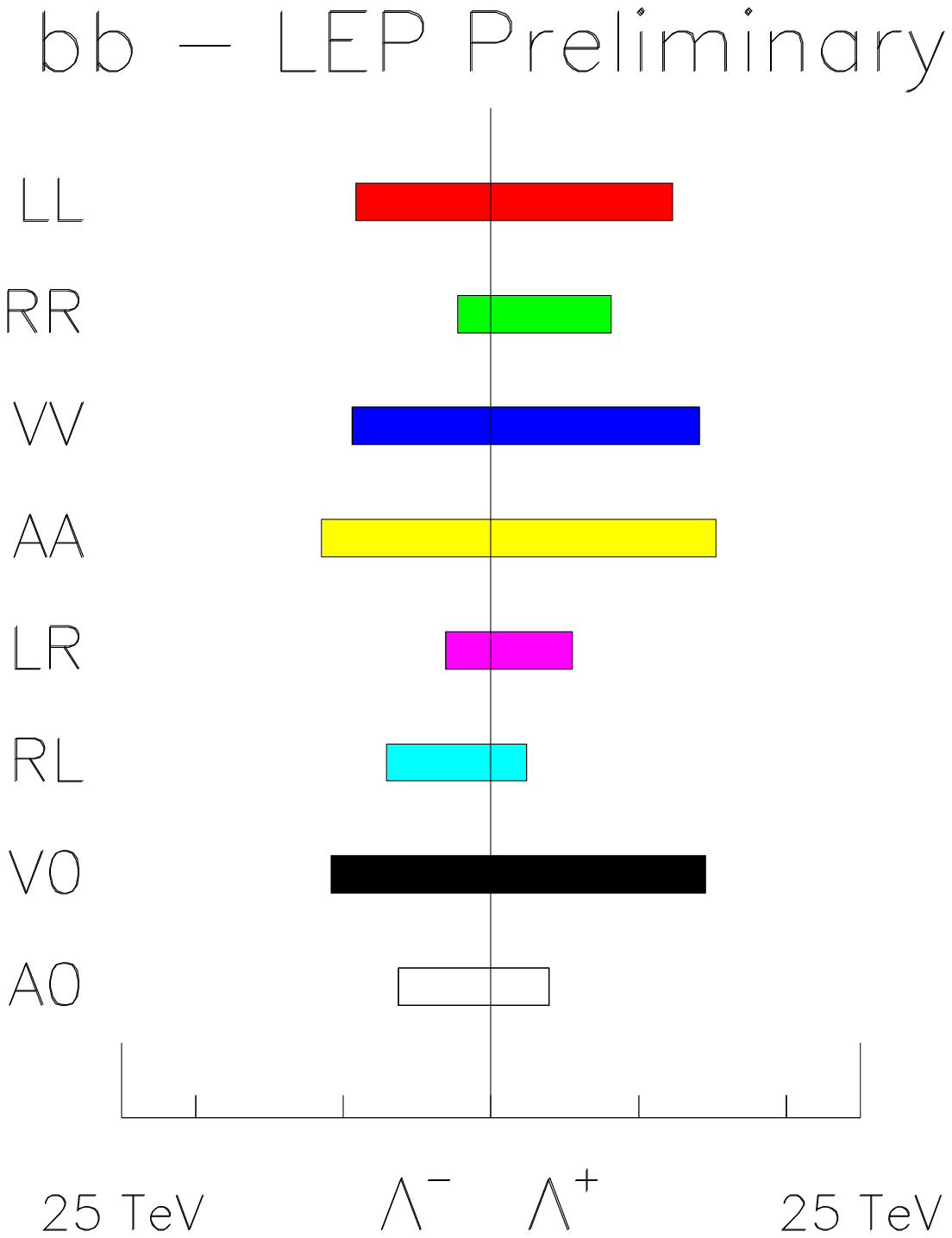,width=0.36\textwidth} &
  \epsfig{file=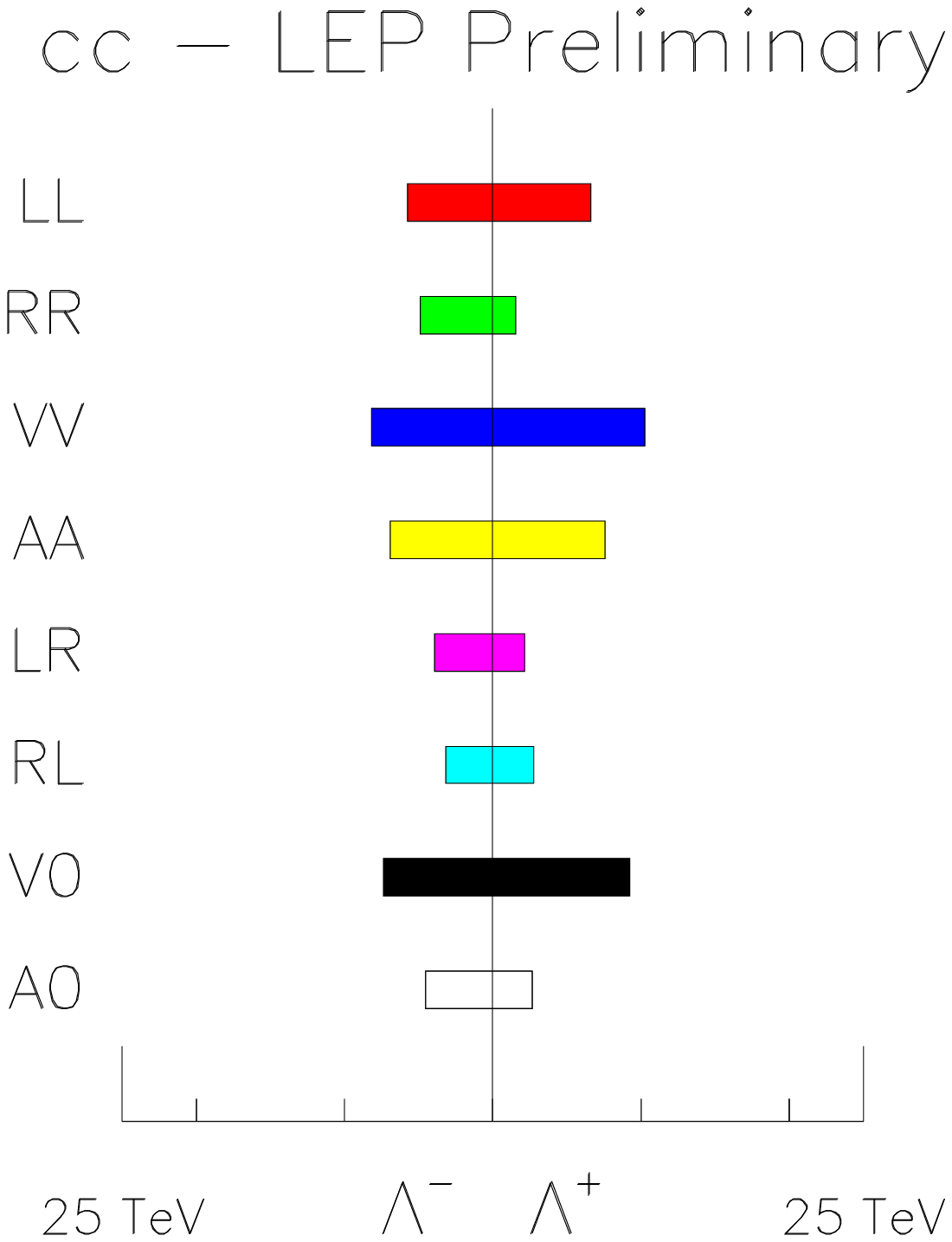,width=0.36\textwidth} \\
  \end{tabular}
 \end{center}
 \caption{The limits on $\Lambda$ for $\eell$ assuming
          universality in the contact interactions between 
          $\eell$ ($\ell \neq e$), for $\eeee$, for $\eeqq$ assuming 
          equal strength contact interactions for quarks and for
          $\eebb$ and $\eecc$.}
 \label{ff:fig:cntc}
\end{figure}

The fitted values of $\epsilon$ and their 68$\%$ confidence level 
uncertainties together with the 95$\%$ confidence level lower limit 
on ${\mathrm{\Lambda}}$ are shown in Table \ref{ff:tab:cntceps} for
the fits to $\eell$ ($\ell \neq e$), $\eeee$ , inclusive $\eeqq$, $\eebb$ 
and $\eecc$. Table \ref{ff:tab:cntclmb} shows only the limits 
obtained on the scale $\Lambda$ for other fits. The limits are shown 
graphically in Figure \ref{ff:fig:cntc}.

For the VV model with positive interference and assuming
electromagnetic coupling strength instead of $g^{2}/{4\pi} = 1$,
the scale $\Lambda$ obtained in the $\eeee$ channel is converted to 
an upper limit on the electron size:
\begin{eqnarray}
\mathrm{r_e < 1.4 \times 10^{-19} m}
\end{eqnarray}
Models with stronger couplings will make this upper limit even tighter.

\begin{table}[tp]
 \begin{center}
  \begin{tabular}{|c|c|c|c|c|}
   \hline
   Model      & $\eta_{LL}$ & $\eta_{RR}$ & $\eta_{LR}$ & $\eta_{RL}$ \\
   \hline\hline
   LL$^{\pm}$ &   $\pm 1$   &      0      &      0      &      0      \\
   \hline
   RR$^{\pm}$ &      0      &   $\pm 1$   &      0      &      0      \\
   \hline
   VV$^{\pm}$ &   $\pm 1$   &   $\pm 1$   &   $\pm 1$   &   $\pm 1$   \\
   \hline
   AA$^{\pm}$ &   $\pm 1$   &   $\pm 1$   &   $\mp 1$   &   $\mp 1$   \\
   \hline
   LR$^{\pm}$ &      0      &      0      &   $\pm 1$   &      0      \\
   \hline
   RL$^{\pm}$ &      0      &      0      &      0      &   $\pm 1$   \\
   \hline
   V0$^{\pm}$ &   $\pm 1$   &   $\pm 1$   &      0      &      0      \\
   \hline
   A0$^{\pm}$ &      0      &      0      &  $\pm 1$    &   $\pm 1$   \\
   \hline
  \end{tabular}
 \end{center}
 \caption{Choices of $\eta_{ij}$ for different contact interaction models}
 \label{ff:tab:cntcdef}.
\end{table}
\begin{table} 
 \renewcommand{\arraystretch}{1.2}
  \begin{center}
  \begin{tabular}{cc}

\begin{tabular}{|c||c|cc|}
\hline
\multicolumn{4}{|c|}{$\eell$} \\
\hline 
\hline
       & $\epsilon$   & $\Lambda^{-}$ & $\Lambda^{+}$ \\
 Model & (TeV$^{-2}$) &     (TeV)     &     (TeV)     \\
\hline 
\hline    
~~~~LL~~~~ &  -0.0044$^{+0.0035}_{-0.0035}$ &  9.8  & 13.3 \\
\hline
    RR     &  -0.0049$^{+0.0039}_{-0.0039}$ &  9.3  & 12.7 \\
\hline                                                     
    VV     &  -0.0016$^{+0.0013}_{-0.0014}$ & 16.0  & 21.7  \\
\hline
    AA     &  -0.0013$^{+0.0017}_{-0.0017}$ & 15.1  & 17.2  \\
\hline                                                     
    LR     &  -0.0036$^{+0.0052}_{-0.0054}$ &  8.6  & 10.2 \\
\hline
    RL     &  -0.0036$^{+0.0052}_{-0.0054}$ &  8.6  & 10.2 \\
\hline                                                     
    V0     &  -0.0023$^{+0.0018}_{-0.0018}$ & 13.5  & 18.4  \\
\hline
    A0     &  -0.0018$^{+0.0026}_{-0.0026}$ & 12.4  & 14.3 \\
\hline
\end{tabular}
&
\begin{tabular}{|c||c|cc|}
\hline
\multicolumn{4}{|c|}{$\eeee$} \\
\hline 
\hline
       & $\epsilon$   & $\Lambda^{-}$ & $\Lambda^{+}$ \\
 Model & (TeV$^{-2}$) &      (TeV)    &     (TeV)     \\
\hline    
\hline    
~~~~LL~~~~ &  ~0.0049$^{+0.0084}_{-0.0084}$ & 9.0  & 7.1  \\
\hline
    RR     &  ~0.0056$^{+0.0082}_{-0.0092}$ & 8.9  & 7.0  \\
\hline                                                     
    VV     &  ~0.0004$^{+0.0022}_{-0.0016}$ &18.0  &15.9  \\
\hline
    AA     &  ~0.0009$^{+0.0041}_{-0.0039}$ &11.5  &11.3  \\
\hline                                                     
    LR     &  ~0.0008$^{+0.0064}_{-0.0052}$ &10.0  & 9.1  \\
\hline
    RL     &  ~0.0008$^{+0.0064}_{-0.0052}$ &10.0  & 9.1  \\ 
\hline                                                     
    V0     &  ~0.0028$^{+0.0038}_{-0.0045}$ &12.5  &10.2  \\
\hline
    A0     &  -0.0008$^{+0.0028}_{-0.0030}$ &14.0  &13.0  \\
\hline
\end{tabular}
\\

\\
\multicolumn{2}{c}{
\begin{tabular}{|c||c|cc|}
\hline
\multicolumn{4}{|c|}{$\eeqq$} \\
\hline 
\hline
       & $\epsilon$   & $\Lambda^{-}$ & $\Lambda^{+}$ \\
 Model & (TeV$^{-2}$) &      (TeV)    &     (TeV)     \\
\hline    
\hline    
~~~~LL~~~~ &  ~0.0152$^{+0.0064}_{-0.0076}$ & 3.7  & 6.0  \\
\hline
    RR     &  -0.0208$^{+0.0103}_{-0.0082}$ & 5.5  & 3.9  \\
\hline                                                     
    VV     &  -0.0096$^{+0.0051}_{-0.0037}$ & 8.1  & 5.3  \\
\hline
    AA     &  ~0.0068$^{+0.0033}_{-0.0034}$ & 5.1  & 8.8  \\
\hline                                                     
    LR     &  -0.0308$^{+0.0172}_{-0.0055}$ & 5.1  & 4.3  \\
\hline
    RL     &  -0.0108$^{+0.0057}_{-0.0054}$ & 7.2  & 9.3  \\ 
\hline                                                     
    V0     &  ~0.0174$^{+0.0057}_{-0.0074}$ & 5.1  & 6.0  \\
\hline
    A0     &  -0.0092$^{+0.0049}_{-0.0041}$ & 8.0  & 3.9  \\
\hline
\end{tabular}
}
\\

\\
\begin{tabular}{|c|c|cc|}
\hline
\multicolumn{4}{|c|}{$\eebb$} \\
\hline
\hline
       & $\epsilon$   & $\Lambda^{-}$ & $\Lambda^{+}$ \\
 Model & (TeV$^{-2}$) &      (TeV)    &     (TeV)     \\
\hline
\hline
~~~~LL~~~~ & -0.0038$^{+ 0.0044}_{- 0.0047}$ &    9.1 &   12.3 \\
\hline
    RR     & -0.1729$^{+ 0.1584}_{- 0.0162}$ &    2.2 &    8.1 \\
\hline
    VV     & -0.0040$^{+ 0.0039}_{- 0.0041}$ &    9.4 &   14.1 \\
\hline
    AA     & -0.0022$^{+ 0.0029}_{- 0.0031}$ &   11.5 &   15.3 \\
\hline
    LR     & -0.0620$^{+ 0.0692}_{- 0.0313}$ &    3.1 &    5.5 \\
\hline
    RL     &  0.0180$^{+ 0.1442}_{- 0.0249}$ &    7.0 &    2.4 \\
\hline
    V0     & -0.0028$^{+ 0.0032}_{- 0.0033}$ &   10.8 &   14.5 \\
\hline
    A0     &  0.0375$^{+ 0.0193}_{- 0.0379}$ &    6.3 &    3.9 \\
\hline
\end{tabular}
&
\begin{tabular}{|c|c|cc|}
\hline
\multicolumn{4}{|c|}{$\eecc$} \\
\hline
\hline
       & $\epsilon$   & $\Lambda^{-}$ & $\Lambda^{+}$ \\
 Model & (TeV$^{-2}$) &      (TeV)    &     (TeV)     \\
\hline
\hline
~~~~LL~~~~ & -0.0091$^{+ 0.0126}_{- 0.0126}$ &    5.7 &    6.6 \\
\hline
    RR     &  0.3544$^{+ 0.0476}_{- 0.3746}$ &    4.9 &    1.5 \\
\hline
    VV     & -0.0047$^{+ 0.0057}_{- 0.0060}$ &    8.2 &   10.3 \\
\hline
    AA     & -0.0059$^{+ 0.0095}_{- 0.0090}$ &    6.9 &    7.6 \\
\hline
    LR     &  0.1386$^{+ 0.0555}_{- 0.1649}$ &    3.9 &    2.1 \\
\hline
    RL     &  0.0106$^{+ 0.0848}_{- 0.0757}$ &    3.1 &    2.8 \\
\hline
    V0     & -0.0058$^{+ 0.0075}_{- 0.0071}$ &    7.4 &    9.2 \\
\hline
    A0     &  0.0662$^{+ 0.0564}_{- 0.0905}$ &    4.5 &    2.7 \\
\hline
\end{tabular}

\end{tabular}
   \caption{The fitted values of $\epsilon$ and the derived 95\% confidence 
            level lower limits on the parameter $\Lambda$
            of contact interaction derived from fits to lepton-pair
            cross-sections and asymmetries and from fits to hadronic 
            cross-sections. The limits $\Lambda_+$ and $\Lambda_-$
            given in TeV correspond to the upper and lower signs of the 
            parameters $\eta_{ij}$ in Table \ref{ff:tab:cntcdef}.
            For $\leptlept$ ($\ell \neq e$) the couplings to $\mumu$ and 
            $\tautau$ are a assumed to be universal and for inclusive 
            $\qq$ final states 
            all quarks are assumed to experience contact interactions 
            with the same strength.}
  \label{ff:tab:cntceps}
  \end{center} 
\end{table}
\begin{table}[tp]
 \renewcommand{\arraystretch}{1.1}
  \begin{center}
  \begin{tabular}{c}
    \begin{tabular}{|c||cc|cc|}
\hline
\multicolumn{5}{|c|}{leptons} \\
\hline
\hline       
           &\multicolumn{2}{|c|}{$\mu^+\mu^-$}
           &\multicolumn{2}{|c|}{$\tau^+ \tau^-$} \\
 Model     &~~$\Lambda_-$~~
                 &~~$\Lambda_+$~~
                       &~~$\Lambda_-$~~
                            &~~$\Lambda_+$~~ \\
\hline \hline  
~~~~LL~~~~ & 8.5  & 12.5 & 9.1  & 8.6  \\
    RR     & 8.1  & 11.9 & 8.7  & 8.2  \\
\hline                                                     
    VV     & 14.3 & 19.7 & 14.2 & 14.5 \\
    AA     & 12.7 & 16.4 & 14.0 & 11.3 \\
\hline                                                     
    LR     & 7.9  & 8.9  & 2.2  & 7.9  \\ 
    RL     & 7.9  & 8.9  & 2.2  & 7.9  \\
\hline                                                     
    V0     & 11.7 & 17.2 & 12.7 & 11.8 \\
    A0     & 11.5 & 12.4 & 9.8  & 10.8 \\
\hline
    \end{tabular}
\\

\\
    \begin{tabular}{|c||cc|cc|}
\hline
\multicolumn{5}{|c|}{hadrons} \\
\hline
\hline
           &\multicolumn{2}{|c|}{up-type}
           &\multicolumn{2}{|c|}{down-type} \\
 Model     &~~$\Lambda_-$~~
                 &~~$\Lambda_+$~~
                       &~~$\Lambda_-$~~
                             &~~$\Lambda_+$~~ \\
\hline \hline  
~~~~LL~~~~ & 6.7  & 10.2 & 10.6 & 6.0  \\
    RR     & 5.7  & 8.3  & 2.2  & 4.3  \\
\hline                   
    VV     & 9.6  & 14.3 & 11.4 & 7.0  \\
    AA     & 8.0  & 11.5 & 13.3 & 7.7  \\
\hline
    LR     & 4.2  & 2.3  & 2.7  & 3.5  \\
    RL     & 3.5  & 2.8  & 4.2  & 2.4  \\
\hline                                                     
    V0     & 8.7  & 13.4 & 12.5 & 7.1  \\
    A0     & 4.9  & 2.8  & 4.2  & 3.3  \\
\hline
    \end{tabular}
   \end{tabular}
   \caption{The 95\% confidence level lower limits on the parameter 
            $\Lambda$
            of contact interaction derived from fits to lepton-pair 
            cross-sections and asymmetries and from fits to hadronic 
            cross-sections. The limits $\Lambda_+$ and $\Lambda_-$
            given in TeV correspond to the upper and lower signs of the 
            parameters $\eta_{ij}$ in Table \ref{ff:tab:cntcdef}.
            For hadrons the limits for up-type and down-type quarks
            are derived assuming a single up or down type quark undergoes
            contact interactions.}
   \label{ff:tab:cntclmb}
  \end{center}
\end{table}
\subsection{Models with $\mathbf{\Zprime}$ Bosons}
\label{ff:sec-zprime}

The combined hadronic and leptonic cross-sections and the leptonic 
forward-backward asymmetries are used to fit the data to models including 
an additional, heavy, neutral boson, $\Zprime$.

Fits are made to $\MZp$, the mass of a $\Zprime$ for models 
resulting from an E$_6$ GUT and L-R symmetric models~\cite{ff:ref:zprime-thry}
and for the Sequential Standard Model (SSM)~\cite{ff:ref:sqsm}, which proposes the 
existence of a $\Zprime$ with exactly the same coupling to fermions as 
the standard Z. $\LEPII$ data alone does not significantly constrain
the mixing angle between the Z and $\Zprime$ fields, $\thtzzp$.
However results from a single experiment, in which $\LEPI$ data is used in the 
fit, show that the mixing is consistent with zero (see for 
example~\cite{ff:ref:lep1zprime}). So for these fits $\thtzzp$ was fixed to 
zero.

\begin{figure}[tp]
  \begin{flushleft}
   \begin{tabular}{ll}  
    \mbox{\epsfig{file=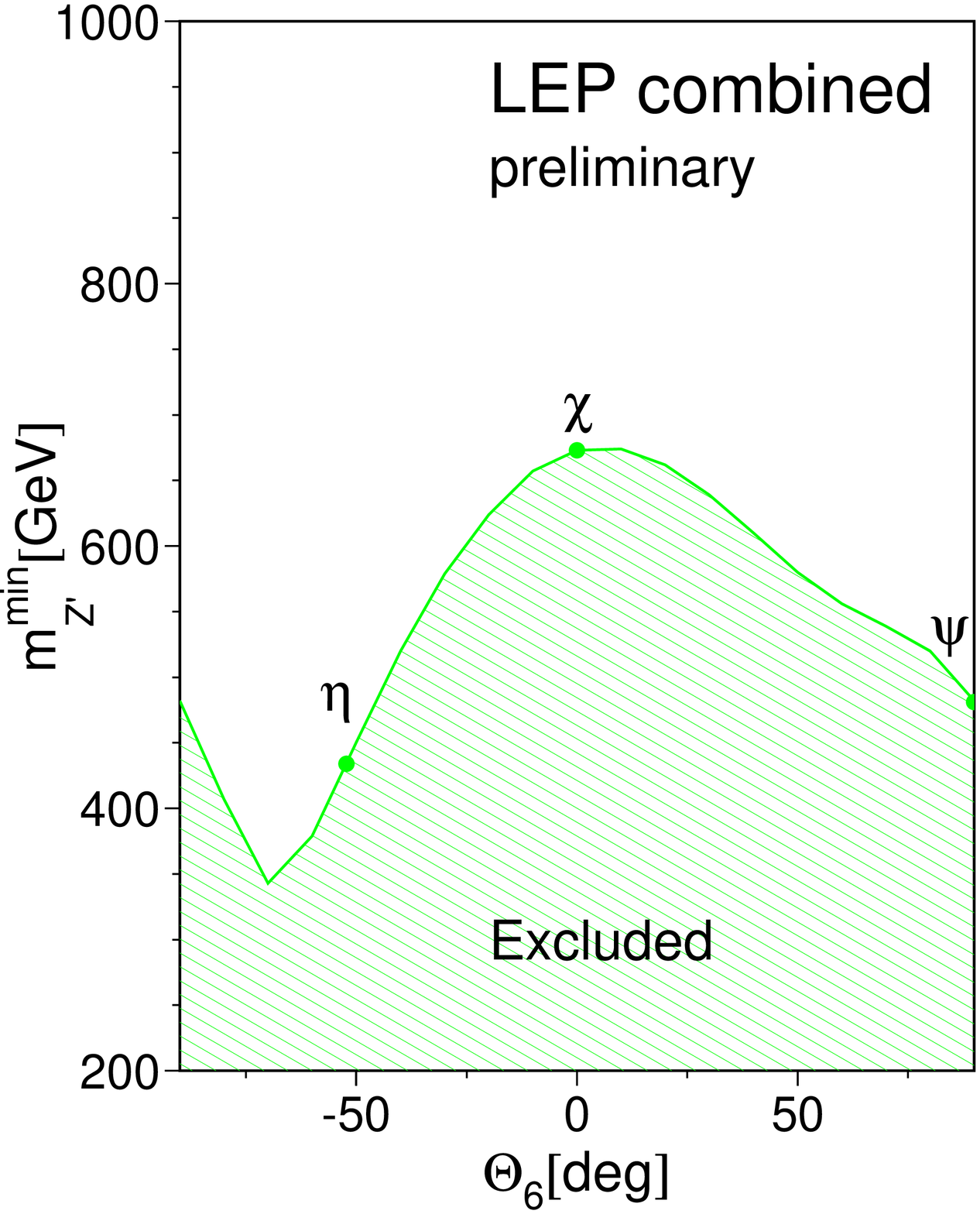,width=0.50\textwidth}}
    \mbox{\epsfig{file=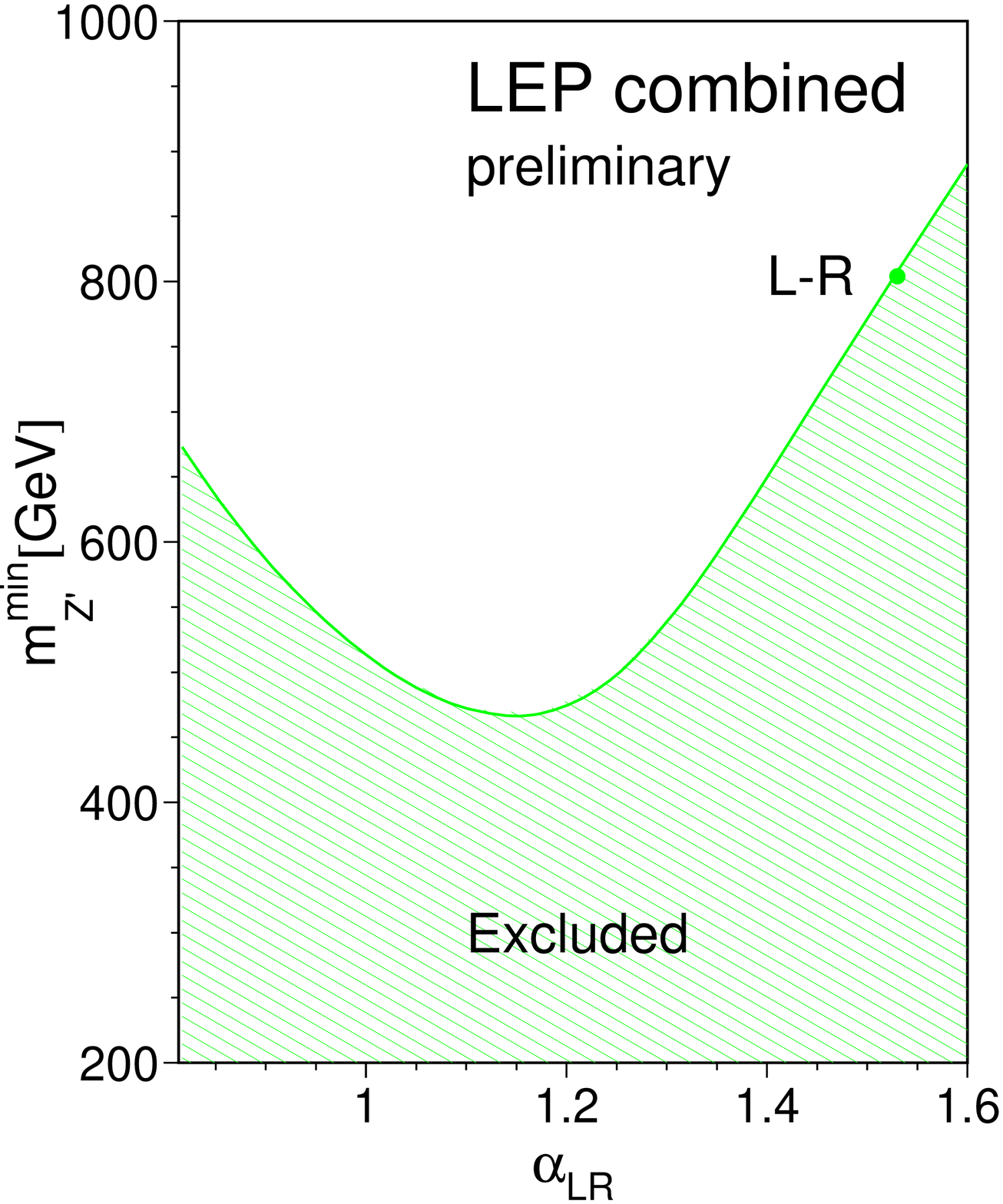,width=0.50\textwidth}}
   \end{tabular}
  \caption{The 95\% confidence level limits on $\MZp$ as a function of 
           the model parameter $\theta_6$ for E$_6$ models and 
           $\alpha_{\mathrm{LR}}$ for left-right models. 
           The Z-$\Zprime$ mixing is fixed, $\thtzzp=0$.}  
   \label{ff:fig:zp_e6-lr}
  \end{flushleft}
\end{figure}

No significant evidence is found for the existence of a $\Zprime$ boson
in any of the models. 
The procedure to find limits on the Z$'$ mass corresponds to that in case 
of  contact interactions: for large masses the exchange of a Z$'$ can be 
approximated by contact terms, $\Lambda \propto \MZp$.
The lower limits on the Z$'$ mass are shown in Figure \ref{ff:fig:zp_e6-lr} 
varying the parameters $\theta_6$ for the E$_6$ models and  
$\alpha_{\mathrm{LR}}$ for the left-right models. 
The results for the specific models 
$\chi,~\psi~,\eta$ ($\theta_6=0,~\pi/2,~- \arctan \sqrt{5/3}$), 
L-R ($\alpha_{\mathrm{LR}}$=1.53) and SSM are shown in 
Table~\ref{ff:tab:zprime_mass_lim}.

\begin{table}[tp]
  \renewcommand{\arraystretch}{1.15}
  \begin{center}
    \begin{tabular}{|c||c|c|c|c|c|}
\hline
Z$'$ model                   & $\chi$ & $\psi$ & $\eta$ & L-R   & SSM  \\
  \hline  \hline 
M$_{Z'}^{limit}$ (GeV/c$^2$) & 673    & 481    & 434    & 804   & 1787 \\
 \hline
 \end{tabular}
 \caption{The 95\% confidence level lower limits on the $\Zprime$ mass for
 $\chi,~\psi,~\eta$, L-R and SSM models.}
 \label{ff:tab:zprime_mass_lim}
 \end{center} 
\end{table}
\subsection{Leptoquarks and R-parity violating squarks }
\label{ff:sec-lq-sq}

Leptoquarks (LQ) would mediate quark-lepton transitions. 
Following the notations in  Reference~\cite{ff:ref:lq-thry,ff:ref:lq-squ}, 
scalar leptoquarks, $S_I$, and vector leptoquarks,$V_I$ are indicated 
based on spin and isospin $I$. Leptoquarks with the same Isospin but 
with different hypercharges are distinguished by an additional tilde. 
See Reference~\citen{ff:ref:lq-squ} for further details.
They carry fermion numbers, $F=L+3B$. 
It is assumed that leptoquark couplings to quark-lepton 
pairs preserve baryon- and lepton-number. 
The couplings $g_L,~g_R$, are labelled according to the chirality 
of the lepton.        

\SBL{1/2} and \SL{0} leptoquarks are equivalent to up-type anti-squarks and 
down-type squarks, respectively. Limits in terms of the leptoquark coupling 
are  then exactly equivalent to limits on $\mathrm \lambda_{1jk}$ in the 
Lagrangian ${\mathrm  \lambda_{1jk}L_{1}Q_{j}\bar D_{k}}$. 

At LEP, the exchange of a leptoquark can modify the hadronic
cross-sections and asymmetries, as described at the Born level by the equations
given in Reference~\citen{ff:ref:lq-squ}. Using the LEP combined measurements 
of hadronic cross-sections, and the measurements of heavy quark production, 
$\Rb$, $\Rc$, $\Abb$ and $\Acc$, upper limits can be set on the leptoquark's 
coupling $g$ as a function of its mass \MLQ\ for leptoquarks coupling electrons to first, second and third generation quarks.
For convenience, one type of leptoquark is assumed to be much lighter than 
the others. Furthermore, experimental constraints on the product $g_L  g_R$ 
allow the study leptoquarks assuming either only $g_L \neq 0$
or $g_R \neq 0$. Limits are then denoted by either (L) for leptoquarks coupling
to left handed leptons or (R) for leptoquarks coupling to right handed leptons.

In the processes $\eeuu$ and $\eedd$ first generation leptoquarks could be 
exchanged in $u$- or $t$-channel (F=2 or  F=0) which would lead to a change 
of the hadronic cross-section.
In the processes $\eecc$ and $\eebb$ the exchange of leptoquarks with 
cross-generational couplings can alter the \qq\ angular distribution, 
especially at low polar angle. 
The reported measurements on heavy quark production have been extrapolated 
to $4\pi$ acceptance, using SM predictions, from the measurements performed 
in restricted angular ranges, corresponding to the acceptance of the
vertex-detector in each experiment.
Therefore, when fitting limits on leptoquarks' coupling to the 2nd or 3rd 
generation of quarks, the LEP combined results for b and c sector are 
extrapolated back to an angular range of $\left| \cos \theta \right| < 0.85$
using ZFITTER predictions.  

The following measurements are used to constrain different types of leptoquarks
\begin{itemize}  
\item For leptoquarks coupling electrons to 1$^{\mathrm{st}}$ generation 
      quarks, all LEP combined hadronic cross-sections at centre-of-mass 
      energies from 130 GeV to 207 GeV are used

\item For leptoquarks coupling electrons to 2$^{\mathrm{nd}}$ generation 
      quarks, $\sigma_{\cc}$ is calculated from $\Rc$ and the hadronic 
      cross-section at the energy points where $\Rc$ is 
      measured. The measurements of $\sigma_{\cc}$ and $\Acc$ are then 
      extrapolated back to $\left| \cos \theta \right| < 0.85$.
      Since measurements in the c-sector are scarce and originate from, 
      at most, 2 experiments, hadronic cross-sections, extrapolated down to 
      $\left| \cos \theta \right| < 0.85$ are also used in the fit, with an 
      average $10\%$ correlated errors. 

\item For leptoquarks coupling electrons to 3$^{\mathrm{rd}}$ generation 
      quarks, only ${\mathrm \sigma_{b\bar b}}$ and \Abb, extrapolated 
      back to a $\left| \cos \theta \right| < 0.85$ are used.
\end{itemize}

\begin{figure}[tp]
\begin{center}
\mbox{\epsfig{file=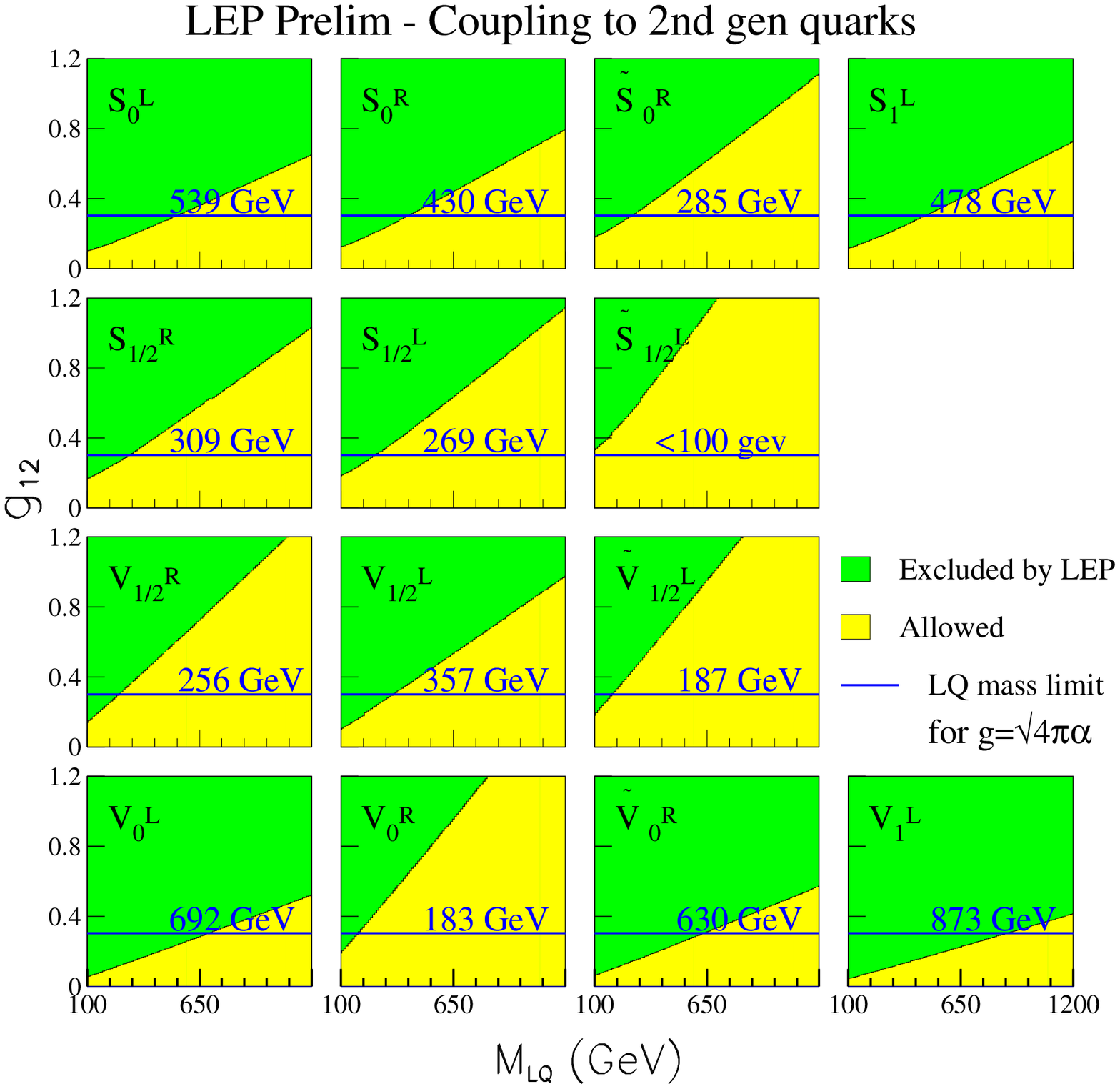,width=15cm}}
\caption{95\% confidence level limit on the coupling 
         of leptoquarks to 2nd generation of quarks.}  
\label{ff:fig:lq-2nd} 
\end{center}
\end{figure}

\begin{figure}[tp]
\begin{center}
\begin{tabular}{c}
\mbox{\epsfig{file=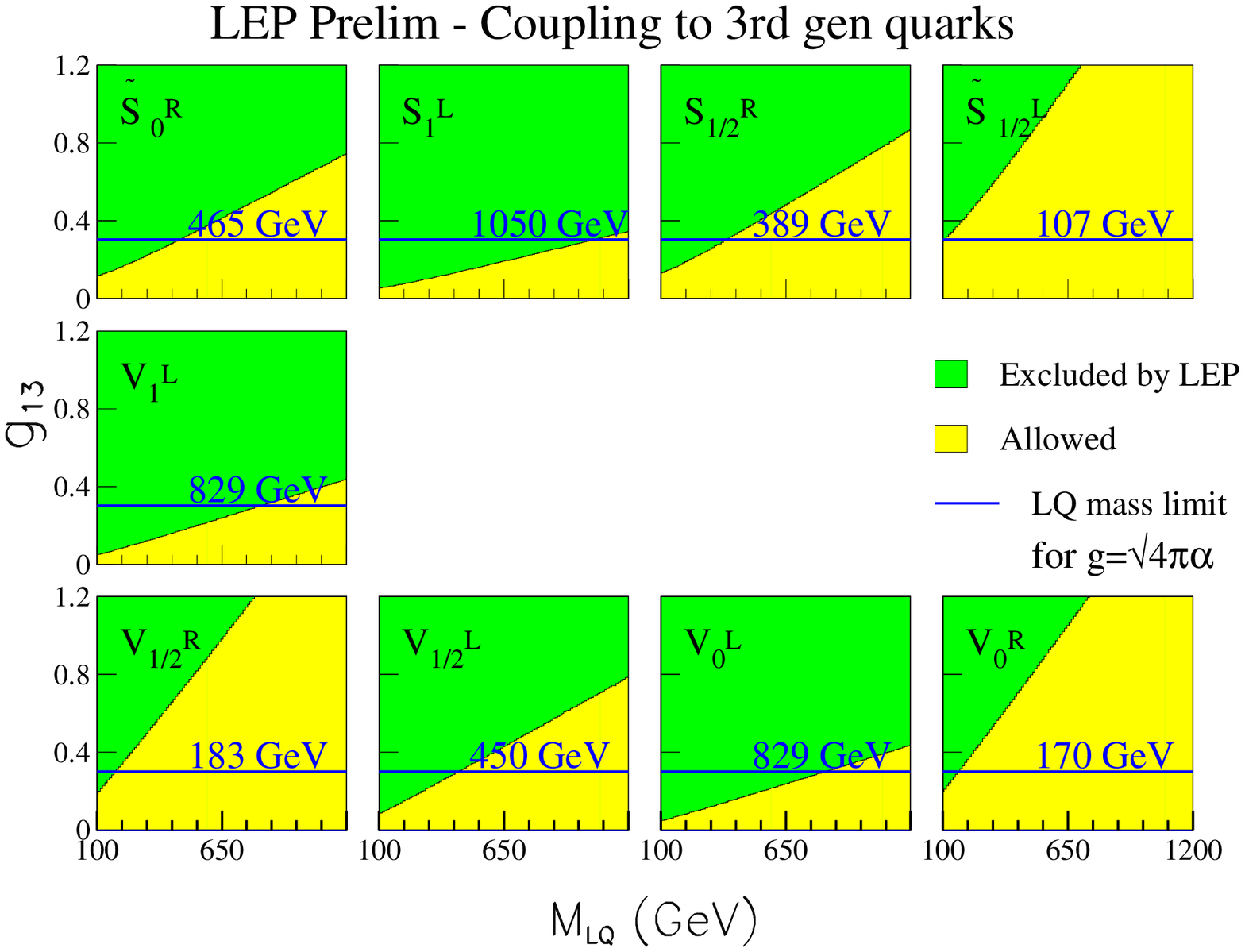,width=15cm}}
\end{tabular}
\caption{95\% confidence level limit on the coupling 
         of leptoquarks to 3rd generation of quarks.}  
\label{ff:fig:lq-3rd} 
\end{center}
\end{figure}

The 95$\%$ confidence level lower limits on masses $\MLQ$ are derived 
assuming a coupling of electromagnetic strength, 
$g = \sqrt{4\pi \alpha_{em}}$, where $\alpha_{em}$ is the fine structure 
constant. The results are summarised in    
Table~\ref{ff:tab:lq-mass}. These results complement the leptoquark searches 
at HERA~\cite{ff:ref:lq-h1,ff:ref:lq-zeus} and the 
Tevatron~\cite{ff:ref:lq-tevatron}.
Figures~\ref{ff:fig:lq-2nd} and \ref{ff:fig:lq-3rd} give the 95\% confidence 
level limits on the 
coupling as a function of the leptoquark mass for leptoquarks coupling 
electrons to the second and third generations of quarks.

\begin{table}[tp]                                                              
  \renewcommand{\arraystretch}{1.35}
 \begin{center}
\begin{tabular}{|c|c c|c|c c|c|c|}
\hline
 \multicolumn{8}{|c|}{Limit on scalar LQ mass (GeV/$c^{2}$)} \\
\hline\hline
 & \SL{0} & \SR{0} & \SBR{0} & \SL{\lqhalf} & \SR{\lqhalf} & \SBL{\lqhalf} & \SL{1} \\
\hline\hline
$LQ_{1st}$ &  655  &  520  &   202   &   178  &   232  &   -  &  361 \\
\hline
$LQ_{2nd}$ &  539  &  430  &   285   &   269  &   309  &   -  &  478 \\
\hline
$LQ_{3rd}$ &  NA  &  NA   &   465   &   NA  &   389  &   107  &  1050 \\
\hline
\multicolumn{8}{c}{\null}\\

\hline
 \multicolumn{8}{|c|}{Limit on vector LQ mass (GeV/$c^{2}$)} \\
\hline\hline
  & \VL{0} & \VR{0} & \VBR{0} & \VL{\lqhalf} & \VR{\lqhalf} & \VBL{\lqhalf} & \VL{1} \\
\hline\hline
$ LQ_{1st}$ &  917  &   165  &   489   &   303  &   227  &   176   &   659  \\
\hline
$ LQ_{2nd}$ &  692  &   183  &   630   &   357  &   256  &   187   &   873  \\
\hline
$ LQ_{3rd}$ &  829   &   170  &   NA   &   451  &   183  &   NA   &   829

  \\
\hline
\end{tabular}
\caption{$95\%$ confidence level lower limits on the LQ mass for leptoquarks 
         coupling between electrons and
         the first, second and third generation of quarks.
         A dash indicates that no limit can be  set and N.A denotes 
         leptoquarks coupling only to top quarks and hence not visible at LEP.}
\label{ff:tab:lq-mass}
\end{center} 
\end{table}                                                                  
\subsection{Low Scale Gravity in Large Extra Dimensions}
\label{ff:sec-grav}

The averaged differential cross-sections for $\eeee$ are used to search for 
the effects of graviton exchange in large extra dimensions.

A new approach to the solution of the hierarchy problem has
been proposed in~\cite{ff:ref:ADD,ff:ref:ADD2,ff:ref:ADD3}, which brings close
the electroweak scale $\rm m_{EW} \sim 1\; TeV$ and the
Planck scale $\rm M_{Pl} = \frac{1}{\sqrt{G_N}} \sim 10^{15}\; TeV$.
In this framework the effective 4 dimensional $\rm M_{Pl}$ is
connected to a new $\rm M_{Pl(4+n)}$ scale in a (4+n) dimensional
theory:
\begin{eqnarray}
\mathrm{M_{Pl}^2 \sim M_{Pl(4+n)}^{2+n} R^n},
\end{eqnarray}
where there are n extra compact spatial dimensions of radius
$\rm \sim R$.

In the production of fermion- or boson-pairs in $\ee$ collisions this class of
models can be manifested through virtual effects due to the exchange of
gravitons (Kaluza-Klein excitations). As discussed in~\cite{ff:ref:Hewett,ff:ref:Rizzo,ff:ref:Giudice,ff:ref:Lykken,ff:ref:Shrock},
the exchange of spin-2 gravitons modifies in a unique way the differential 
cross-sections for fermion pairs, providing clear signatures. These models 
introduce an effective scale (ultraviolet cut-off).
Adopting the notation from~\cite{ff:ref:Hewett} 
the gravitational mass scale is called $\mathrm{M_H}$. 
The cut-off scale is supposed to be of the order of the
fundamental gravity scale in 4+n dimensions.

The parameter $\varepsilon_{H}$ is defined as
\begin{eqnarray}
 \varepsilon_{H} = \frac{\lambda}{\mathrm{M_H^4}},
\end{eqnarray}
where the coefficient $\rm \lambda$ is of $\rm \mathcal{O}(1)$ and can not be
calculated explicitly without knowledge of the full quantum gravity
theory. In the following analysis we will assume that
$\rm \lambda = \pm 1$ in order to study both the cases of positive
and negative interference.
To compute the deviations from the Standard Model due to virtual graviton
exchange the calculations~\cite{ff:ref:Giudice,ff:ref:Rizzo} were used.

Theoretical uncertainties on the Standard Model predictions are taken 
from~\cite{ff:ref:lepffwrkshp}. The full correlation matrix of the 
differential cross-sections, obtained in our averaging procedure, is
used in the fits. This is an improvement compared to previous combined analyses
of published or preliminary LEP data on Bhabha scattering, performed before
this detailed information was available (see 
e.g.~\cite{ff:ref:Bourilkov:1999,ff:ref:Bourilkov:2000,ff:ref:Bourilkov:2001}).

\begin{figure}[p]
 \begin{center}
  \epsfig{file=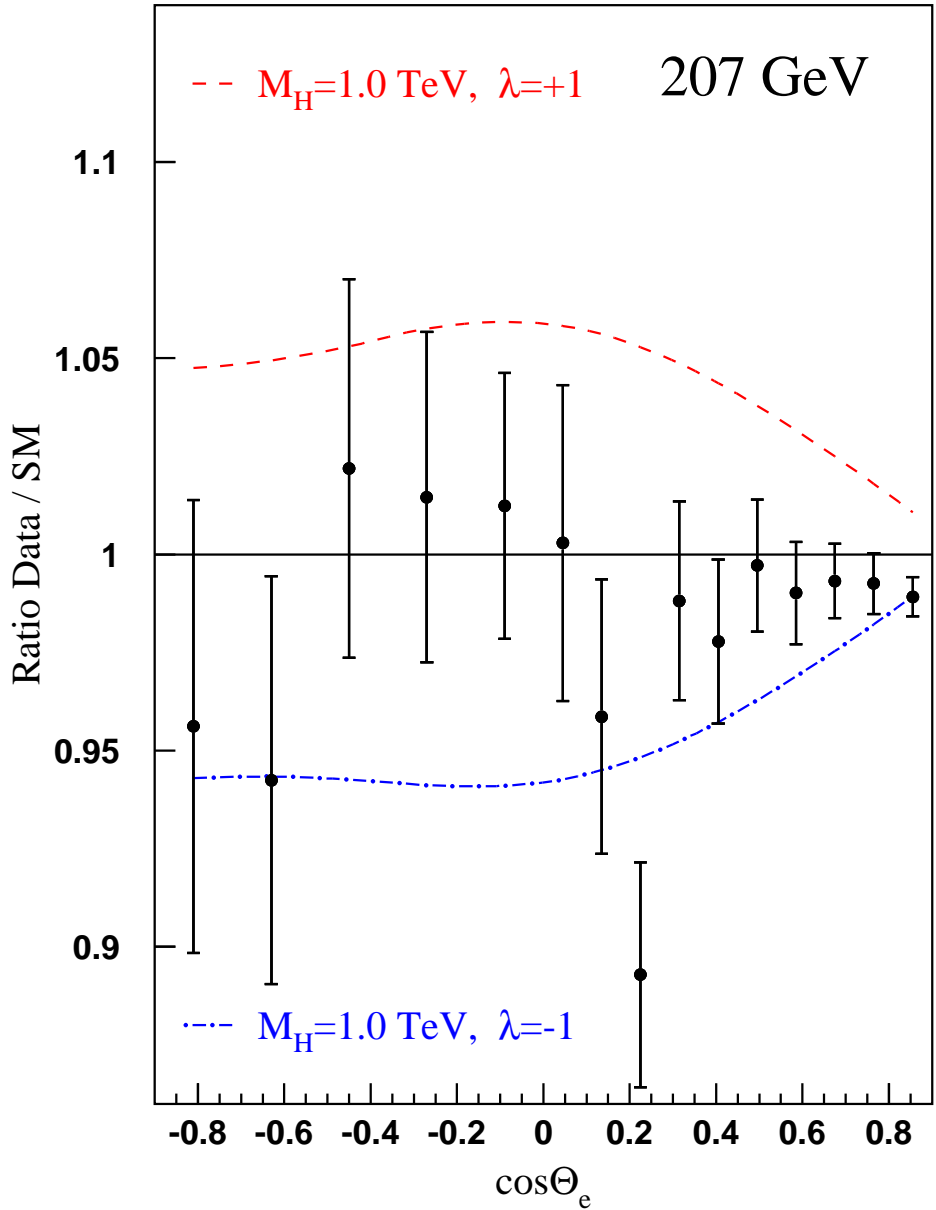,width=0.8\textwidth}
 \end{center}
 \caption{Ratio of the LEP averaged differential cross-section for $\eeee$
          at energy of 207 $\GeV$ compared to the SM prediction. The effects
          expected from virtual graviton exchange are also shown.}
 \label{ff:fig:dsdc-ee-207-lsg}
 \vskip 2cm 
\end{figure}

The extracted value of $\varepsilon_{H}$ is compatible 
with the Standard Model expectation $\varepsilon_{H}=0$.
The errors on $\varepsilon_{H}$ are $\sim 1.5$
smaller than those obtained from a single LEP experiment with the same data
set. The fitted value of $\varepsilon_{H}$ is converted into  
$95\%$ confidence level lower limits on $\mathrm{M_H}$
by integrating the likelihood function over the 
physically allowed values, $\varepsilon_{H} \ge 0$ for $\lambda = +1$ and 
$\varepsilon_{H} \le 0$ for $\lambda = -1$ giving:
\begin{eqnarray}
\mathrm{M_H} & > & 1.20~\TeV\qquad\mathrm{for}~\lambda = +1\,, \\
\mathrm{M_H} & > & 1.09~\TeV\qquad\mathrm{for}~\lambda = -1\,.
\end{eqnarray}
An example of our analysis for the highest energy point is
shown in Figure~\ref{ff:fig:dsdc-ee-207-lsg}.

The interference of virtual graviton exchange amplitudes with both 
t-channel and s-channel Bhabha scattering amplitudes makes this the
most sensitive search channel at LEP. The results obtained here would not
be strictly valid if the luminosity measurements of the LEP experiments,
based on the very same process, are also significantly affected by graviton
exchange.
As shown in~\cite{ff:ref:Bourilkov:1999}, the effect on the cross-section
in the luminosity angular range is so small that it can safely be neglected
in this analysis.

\section{Summary}
\label{ff:sec-conc}

A preliminary combination of the $\LEPII$ $\eeff$ cross-sections (for hadron, 
muon and tau-lepton final states) and
forward-backward asymmetries (for muon and tau final states) 
from LEP running at energies from 130~$\GeV$  to 207~$\GeV$ has been made. 
The results from the four LEP experiments are in good 
agreement with each other. 
The averages for all energies are shown given in Table~\ref{ff:tab:xsafbres}.
Overall the data agree 
with the Standard Model predictions of ZFITTER, although the combined hadronic 
cross-sections are on average $1.7$ standard deviations above the predictions.
Further information is available at~\cite{ff:ref:ffbar_web}.

Preliminary differential cross-sections, $\dsdc$, for $\eeee$, $\eemumu$ and
$\eetautau$ were combined. Results are shown in 
Figures~\ref{ff:fig:dsdc-res-ee}, \ref{ff:fig:dsdc-res-mm} 
and~\ref{ff:fig:dsdc-res-tt}.

A preliminary average of results on heavy flavour production at $\LEPII$ 
has also been made for measurements of $\Rb$, $\Rc$, $\Abb$
and $\Acc$, using results from LEP centre-of-mass energies
from 130 to 207 $\GeV$. Results are given in Tables~\ref{ff:tab:hfbresults}
and~\ref{ff:tab:hfcresults}  and 
shown graphically in Figures~\ref{ff:fig:hfbres} and~\ref{ff:fig:hfcres}.
The results are in good agreement with the predictions of the SM. 

The preliminary averaged cross-section and forward-backward asymmetry results 
together with the combined results on heavy flavour production have been
interpreted in a  variety of models. 
Limits on the scale of contact interactions between leptons and quarks 
and in $\eeee$ and also 
between electrons and specifically $\bb$ and $\cc$ final states have been 
determined.
A full set of limits are given in Tables~\ref{ff:tab:cntceps} and
\ref{ff:tab:cntclmb}.
The $\LEPII$ averaged cross-sections have been used to obtain lower limits 
on the mass of a possible $\Zprime$ boson 
in different models. Limits range from $340$ to $1787$ $\GeV/c^2$ depending
on the model. 
Limits on the masses of leptoquarks have been derived from the hadronic 
cross-sections. The limits range from $101$ to $1036$ $\GeV/c^2$ depending on 
the type of leptoquark.
Limits on the scale of gravity in models with large extra dimensions have been 
obtained from combined differential cross-sections for $\eeee$; for
positive interference between the new physics and the Standard model the limit
is $1.20$ TeV and for negative interference $1.09$ TeV.

%% file: smat.tex
\section{Introduction}

The S-Matrix ansatz provides a coherent way of describing
LEP measurements of the cross-section and forward-backward 
asymmetries in $s$-channel $\eeff$ processes at centre-of-mass energies 
around the $\Zzero$ resonance, from the
{\LEPI} program, and the measurements at centre-of-mass 
energies from  130 -- 207~GeV from the {\LEPII} program.

Compared with the standard 5 and 9 parameter descriptions of the
measurements at the $\Zzero$~\cite{smat:ref:5and9}, the S-Matrix formalism
includes an extra 3 parameters (assuming lepton universality) or 7 parameters
(without lepton universality) which explicitly determine the
contributions to the cross-sections and forward-backward asymmetries
of the interference between the exchange of a $\Zzero$ and a photon.
The {\LEPI} data alone cannot tightly constrain these interference 
terms, in particular the interference term for hadronic cross-sections, 
since their contributions are small around the $\Zzero$ resonance and change 
sign at the pole.
Due to strong correlations between the size of the hadronic interference 
term and the mass of the $\Zzero$, this leads to a larger error on the 
fitted mass of the $\Zzero$ compared to the standard 5 and 9 parameter fits, 
where the hadronic interference term is fixed to the value predicted in the
Standard Model. Including the {\LEPII} data 
leads to a significant improvement in the constraints on the interference terms
and a corresponding reduction in the uncertainty on the mass of the 
$\Zzero$. This results in a measurement of {$\MZ$} which is almost as 
sensitive as the standard results, 
but without constraining the interference to the Standard Model prediction.
This chapter describes the first, preliminary, combination of data from the
full data sets of the 4 LEP experiments, to obtain a LEP combined results 
on the parameters of the S-Matrix ansatz. These results update those
of a previous combination~\cite{smat:ref:smat1997} which was based on
preliminary {\LEPI} data and only partial statistics from the full {\LEPII} 
data set.

Different strategies are used to combined the {\LEPI} and {\LEPII} data.
For {\LEPI} data, an average of the individual experiment's results
on the S-Matrix parameters is made. This approach is rather similar to
the method used to combine the results of the 5 and 9 parameter fits.
To include {\LEPII} data, a fit is made to LEP combined 
measurements of cross-sections and asymmetries above the $\Zzero$, 
taking into account the results of the {\LEPI} combination of S-Matrix
parameters.

In Section~\ref{smat:sec:ansatz} the parameters of the
S-Matrix ansatz are explained. In Sections~\ref{smat:sec:lep1} 
and~\ref{smat:sec:lep2} the average of the {\LEPI} data and the 
inclusion of the {\LEPII} data are described. The results are discussed in 
Section~\ref{smat:sec:discuss} and conclusions are drawn in 
Section~\ref{smat:sec:conc}.

\section{The S-Matrix Ansatz}
\label{smat:sec:ansatz}

The S-matrix
ansatz~\cite{smat:ref:bcms,*smat:ref:rgs,*smat:ref:arr,*smat:ref:tr}
is a rigorous approach to describe the cross-sections and
forward-backward asymmetries in the $s$-channel $\ee$ annihilations
under the assumption that the processes can be parameterised as the
exchange of a massless and a massive vector boson, in which the
couplings of the bosons including their interference are treated as
free parameters.

In this model, the cross-sections can be parametrised
 as follows:
\begin{equation}
\sigma^0_{tot, f}(s)=\frac{4}{3}\pi\alpha^2
\left[
      \frac{\gf^{tot}}{s}
     +\frac{\jf^{tot} (s-\MZbar^2) + \rf^{tot} \, s}
           {(s-\MZbar^2)^2 + \MZbar^2 \GZbar^2}
\right]
\,\,\,\mathrm{with}\,\,\mathrm{f=had,e,\mu,\tau}\,,
\label{smat:eqn:eq1}
\end{equation}
while the forward-backward asymmetries are given by:
\begin{equation}
A^0_\mathrm{fb, f}(s)=\pi\alpha^2
\left[
 \frac{\gf^{fb}}{s} +
 \frac{\jf^{fb} (s-\MZbar^2) + \rf^{fb} \, s}
           {(s-\MZbar^2)^2 + \MZbar^2 \GZbar^2}
\right]
                       / {\sigma^0_{\mathrm{tot, f}}(s)}\,,
\label{smat:eqn:eq2}
\end{equation}
where $\roots$ is the centre-of-mass energy.
The parameters $\rf$ and $\jf$ scale the $\Zzero$ exchange and the 
\mbox{$\Zzero-\gamma$} 
interference contributions to the total cross-section and forward-backward
asymmetries. The contribution $\gf$ of the pure $\gamma$ exchange was fixed to 
the value predicted by QED in all fits. Neither the hadronic charge
asymmetry, nor the flavour tagged quark forward-backward asymmetries are 
considered here, 
which leaves 16 free parameters to described the LEP data: 14 $\rf$ and
$\jf$ parameters and the mass and width of the massive $\Zzero$ resonance.
Applying the constraint of lepton universality reduces this to 8 
parameters.

In the Standard Model the $\Zzero$ exchange term, the $\Zzero-\gamma$ 
interference term and the photon exchange term are given in terms of the
fermion charges and their effective vector and axial couplings to the
$\Zzero$ by:
\begin{equation}
\begin{array}{l@{}l@{}l@{}l}
\displaystyle
\rtotf & = & \kappa^2
             \left[\gae^2+\gve^2\rule{0mm}{4mm}\right]
             \left[\gaf^2+\gvf^2\right]
            -2\kappa\,\gve\,\gvf C_{Im}         \\[5mm]
\jtotf & = & 2\kappa\,\gve\,\gvf \left(C_{Re}+C_{Im}\right) \\[5mm]
\gtotf & = & Q^2_{\mathrm{e}}Q^2_{\mathrm{f}}\left|F_A(\MZ)\right|^2 \\[5mm]
\rfbf  & = & 4\kappa^2\gae\,\gve\,\gaf\,\gvf
            -2\kappa\,\gae\,\gaf C_{Im}         \\[5mm]
\jfbf  & = & 2\kappa\,\gae\,\gaf \left(C_{Re}+C_{Im}\right) \\[5mm]
\gfbf  & = & 0 \,,
\end{array}
\end{equation}
with the following definitions:
\begin{equation}
\begin{array}{l}
\displaystyle
\kappa    = \dfrac{G_F\MZ^2}{2\sqrt{2\,}\pi\alpha} \approx 1.50\\[5mm]
C_{Im}    = \dfrac{\GZ}{\MZ}  \left.Q_{\mathrm{e}}Q_{\mathrm{f}}\right.
                                \mathrm{Im} \left\{F_A(\MZ)\right\} \\[5mm]
C_{Re}    =                   \left.Q_{\mathrm{e}}Q_{\mathrm{f}}\right.
                                \mathrm{Re} \left\{F_A(\MZ)\right\} \\[5mm]
F_A(\MZ)  = \dfrac{\alpha(\MZ)}{\alpha} \,,
\end{array}
\end{equation}
where $\alpha(\MZ)$ is the complex fine-structure constant, and 
$\alpha\equiv\alpha(0)$.
The photonic virtual and bremsstrahlung corrections are included through 
the convolution of Equations~\ref{smat:eqn:eq1} and~\ref{smat:eqn:eq2}
with radiator functions as in the 5 and 9 parameter fits.
The expressions of the S-Matrix parameters in terms of the
effective vector and axial-vector couplings given above 
neglect the imaginary parts of the effective couplings.

The usual definitions of the mass $\MZ$ and width $\GZ$ of a Breit-Wigner 
resonance are used, the width being $s$-dependent, such that:
\begin{equation}
\begin{array}{lllll@{}c@{}r}
  \MZ & \equiv  & \MZbar\sqrt{1+\GZbar^2/\MZbar^2} & \approx & \MZbar & + & 34.20~\MeV/c^2\phantom{\,,} \\[3mm]
  \GZ & \equiv  & \GZbar\sqrt{1+\GZbar^2/\MZbar^2} & \approx & \GZbar & + &  0.94~\MeV\,.\phantom{/c^2}
\end{array}
\label{smat:eqn:smat-ls}
\end{equation}

In the following fits, the predictions from the S-Matrix ansatz and
the QED convolution for cross-sections and asymmetries are made using
SMATASY~\cite{smat:ref:smatasy}, which in turn uses
ZFITTER~\cite{smat:ref:lep2xsafbave} to calculate the QED convolution
of the electroweak kernel.  In case of the $\ee$ final state,
$t$-channel and $s/t$ interference contributions are added to the
$s$-channel ansatz.

\section{LEP combination}
\label{smat:sec:comb}

In the following sections the combinations of the results from the
individual LEP experiments are described: firstly the {\LEPI}
combination, then the combination of both {\LEPI} and {\LEPII} data.
The results from these combinations are compared in
Section~\ref{smat:sec:discuss}.  Although all 16 parameters are
averaged during the combination, only results for the
parameters $\MZ$ and $\jtoth$ are reported here. Systematic studies
specific to the other parameters are ongoing.

\subsection{LEP-I combination}
\label{smat:sec:lep1}

Individual LEP experiments have their own determinations of the
16 S-Matrix parameters~\cite{smat:ref:aleph,smat:ref:delphi,smat:ref:l3lep1,smat:ref:opallep1} from {\LEPI} data alone, using the full {\LEPI} data sets.

These results are averaged using a multi-parameter BLUE technique
based on an extension of Reference~\citen{common_bib:BLUE}. Sources of 
systematic uncertainty correlated between the experiments have been 
investigated, using techniques described in~\cite{smat:ref:5and9} and are
accounted for in the averaging procedure and benefiting from the experience
gained in those combinations.

The parameters $\MZ$ and $\jtoth$ are the most sensitive of all 16
S-matrix parameters to the inclusion of the {\LEPII} data, and are
also the
most interesting ones in the context of the 5 and 9 parameter fits. For
these parameters the most significant source of systematic error which
is correlated between experiments comes from the uncertainty on the
$\ee$ collision energy as determined by models of the LEP RF system
and calibrations using the resonant depolarisation technique. These
errors amount to $\pm 3$ MeV on $\MZ$ and $\pm 0.16$ on $\jtoth$ with
a correlation coefficient of $-0.86$.  The LEP averaged values of
$\MZ$ and $\jtoth$ are given in Table~\ref{smat:tab:lepres}, together
with their correlation coefficient.  The $\chi^2/$D.O.F. for the
average of all 16 parameters is 62.0/48, corresponding to a
probability of $8\%$, which is acceptable.

\begin{table}[tpb]
\begin{center}
\renewcommand{\arraystretch}{1.2}
\begin{tabular}{|c|r@{$\pm$}l|r@{$\pm$}l|c|}
\hline
 & \multicolumn{2}{c|}{ $\MZ$ [GeV] }& \multicolumn{2}{c|}{ $\jtoth$}
 & correlation \\ \hline\hline
{\LEPI} only  & 91.1925 & 0.0059 &  -0.084 & 0.324  & -0.935 \\
{\LEPI} \& {\LEPII} & 91.1869 & 0.0023 &   0.277 & 0.065  & -0.461 \\ \hline
\end{tabular}
\caption{Averaged {\LEPI} and {\LEPII} S-Matrix results for $\MZ$ and $\jtoth$.}
\label{smat:tab:lepres}
\end{center}
\end{table}
\subsection{{\LEPI} and {\LEPII} combination}
\label{smat:sec:lep2}

Some experiments have determined S-Matrix parameters using 
their {\LEPI} and {\LEPII} measured cross-sections and forward-backward
asymmetries~\cite{smat:ref:aleph,smat:ref:delphi,smat:ref:l3lep2,smat:ref:opallep2}. To do a full LEP combination
would require each experiment to provide S-Matrix results and would require 
an analysis of the correlated systematic errors on each measured parameter.

However, preliminary combinations of the measurements of forward-backward 
asymmetries and cross-sections from all 4 LEP experiments, for the 
full~{\LEPII} period, have already been made~\cite{smat:ref:lep2xsafbave} and 
correlations between these measurements have been estimated. The combination
procedure averages measurements of cross-sections and asymmetry for those
events with reduced centre-of-mass energies, $\rootsp$, close to the actual 
centre-of-mass energy of the $\ee$ beams, $\roots$, removing those
events which are less sensitive to the $\Zzero-\gamma$ interference where, 
predominantly, initial state radiation reduces the centre-of-mass
energy to close to the mass of the $\Zzero$. The only significant correlations
are those between hadronic cross-section measurements at different energies,
which are around 20--40\%, depending on energies.

The predictions from SMATASY are fitted to the combined 
{\LEPII} cross-section and forward-backward asymmetry 
measurements~\cite{smat:ref:lep2xsafbave}.
The signal definition 1 of Reference~\citen{smat:ref:lep2xsafbave} is 
used for the data and for the predictions of SMATASY. 
Theoretical uncertainties on the S-Matrix predictions for the {\LEPII}
results and on the corrections of the LEP II data to the common signal 
defintion are taken to be the same as for the 
Standard Model predictions of ZFITTER~\cite{smat:ref:lep2xsafbave} which
are dominated by uncertainties in the QED convolution. These 
amount to a relative uncertainty of $0.26\%$ on the hadronic cross-sections, 
fully correlated between all {\LEPII} energies.

The fit also uses as inputs the averaged {\LEPI} S-Matrix
parameters and covariance matrix. These inputs effectively constrain those
parameters, such as $\MZ$, which are not accurately determined by 
{\LEPII} data.
There are no significant correlations between the {\LEPI} and {\LEPII} inputs.

The LEP averaged values of $\MZ$ and $\jtoth$ for both {\LEPI} and {\LEPII}
data are given in Table~\ref{smat:tab:lepres}, together with their
correlation coefficient. 
The $\chi^2/$D.O.F. for the average of all 16 parameters 
is 64.4/60, corresponding to a probability of $33\%$, which is good.

\subsection{Discussion}
\label{smat:sec:discuss}

In the {\LEPI} combination the measured values of the Z boson mass
$\MZ = 91.1925 \pm 0.0059$~GeV agrees well with the results of the standard 
9 parameter fit ($91.1876 \pm 0.0021$~GeV) albeit with a significantly
larger error, resulting from the correlation with the large uncertainty 
on $\jtoth$ which is then the dominant source of uncertainty on $\MZ$ in the
S-Matrix fits.
The measured value of $\jtoth = -0.084 \pm 0.324 $ , also agrees with the 
prediction of the Standard Model ($0.2201^{+0.0032}_{-0.0137}$). 

\begin{figure}[p]
 \begin{center}
  \mbox{\epsfig{file=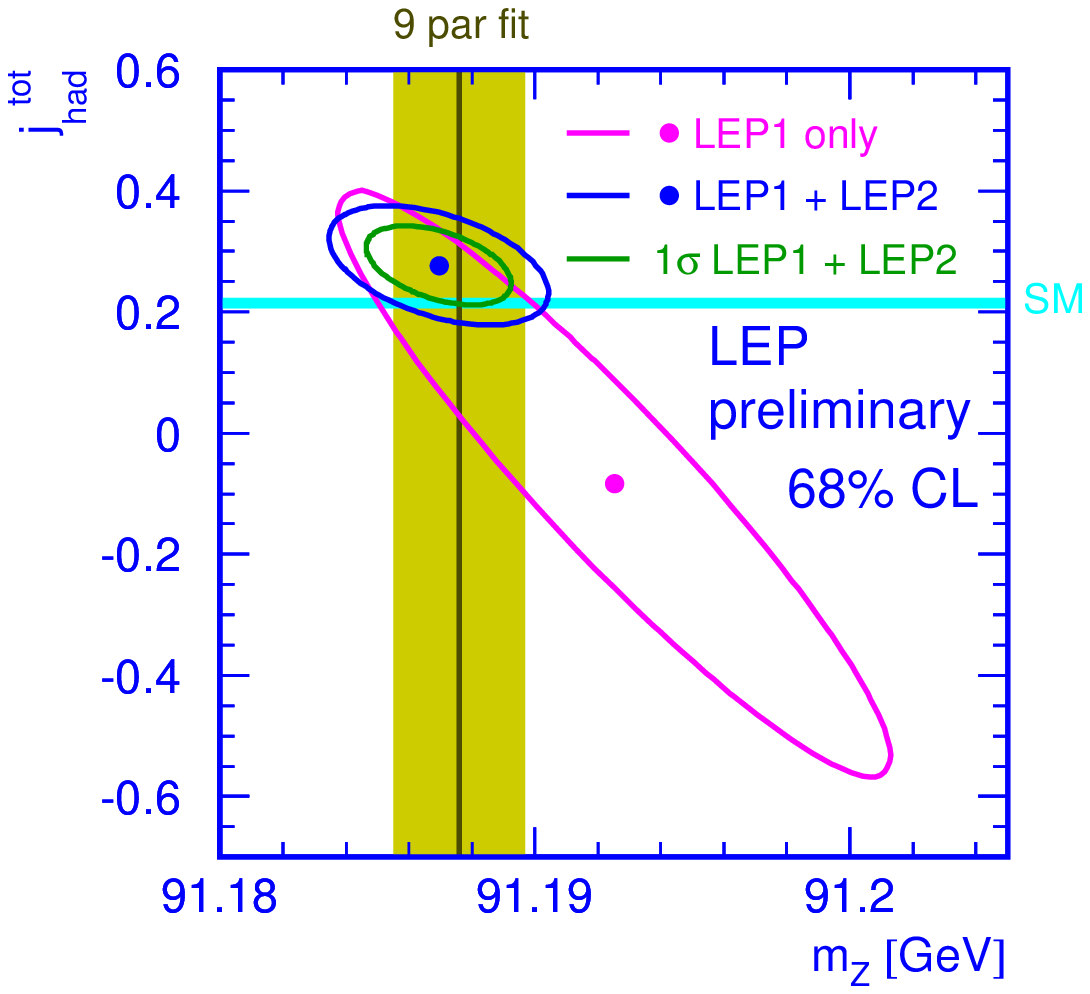,width=\textwidth}}
  \caption{Error ellipses for $\MZ$ and $\jtoth$ for {\LEPI} (at 39\% and
    68\%) and the combination of {\LEPI} and {\LEPII} (at 68\%).}
  \label{smat:fig:mzjtoth}
 \end{center}
\end{figure}

Including the {\LEPII} data brings a significant improvement in the
uncertainty on the size of the interference between $\Zzero$ and photon
exchange compared to {\LEPI} data alone.  The measured value 
$\jtoth = 0.277 \pm 0.065$, agrees well with the values predicted from the 
Standard Model. Correspondingly, the uncertainty
on the the mass of the $\Zzero$ in this ansatz, $2.3$ MeV, is close to 
the precision obtained from {\LEPI} data alone using the standard 9 parameter
fit, $2.1$ MeV. The slightly larger error is due to the uncertainty on
$\jtoth$ which amounts to 0.9~MeV. 
The measured value, $\MZ = 91.1869 \pm 0.0023$~GeV, agrees with that 
obtained from the standard 9 parameter fits.
The results are summarised in Figure~\ref{smat:fig:mzjtoth}.

The good agreement found between the values of $\MZ$ and $\jtoth$ and their
expectations provide a validation of the approach taken in the standard 5 
and 9 parameter fits, in which the size of the interference between $\Zzero$
boson and photon exchange in the hadronic cross-sections was fixed to the
Standard Model expectation.

The precision on $\jtoth$ is slightly better than that obtained by the VENUS 
collaboration~\cite{smat:ref:venus} of $\pm 0.08$, which was obtained using
preliminary results from {\LEPI} and their own 
measurements of the hadronic cross-section below the $\Zzero$ resonance.
The measurement of the hadronic cross-sections from VENUS~\cite{smat:ref:venus}
and TOPAZ~\cite{smat:ref:topaz} could be included in the future to give a 
further reduction in the uncertainty on $\jtoth$.

Work is in progress to understand those sources of systematic error,
correlated between experiments, which are significant
for the remaining S-Matrix parameter that have not been presented here. 
In particular, for $\jtote$ and $\jfbe$, it is important to understand the 
errors resulting from $t$-channel contributions to the $\eeee$ process.
These errors have only limited impact on the standard 5 and 9 parameter fits.

\section{Conclusion}
\label{smat:sec:conc}

Results for the S-Matrix parameter $\MZ$ and $\jtoth$ have been presented 
for {\LEPI} data alone and for a fit using the full data sets for
{\LEPI}  and {\LEPII} from all 4 LEP experiments. 
Inclusion of {\LEPII} data brings a significant improvement 
in the determination of $\jtoth$, the fitted value $0.277 \pm 0.065$, agrees 
well with the values predicted from the Standard Model. As a result
in the improvement of the uncertainty in $\jtoth$, the uncertainty on the 
fitted value of $\MZ$ approaches that of the standard 5 and 9 parameter
fits and the measured value $\MZ = 91.1869 \pm 0.0023$~GeV is compatible
with that from the standard fits.

%% file: 4f_s03.tex
\section{Introduction}
\label{4f_sec:introduction}

This chapter summarises what has been updated in the combination of published 
and preliminary
results of the four LEP experiments on four-fermion cross-sections 
for the Summer 2003 Conferences.
If not stated otherwise, all presented results use the full LEP2 data 
sample at \CoM\ energies up to 209 GeV, supersede the results presented
at the Summer 2002 Conferences~\cite{4f_bib:4f_s02} and have to be 
considered as preliminary.

The \CoM\ energies and the corresponding integrated luminosities are provided
by the experiments and are the same used for previous 
conferences. The LEP energy value in each point (or group
of points) is the luminosity-weighted average of those values. 

Cross-section results from different experiments are combined 
by $\chi^2$ minimisation using the Best Linear Unbiased Estimate method 
described in Ref.~\cite{common_bib:BLUE}, properly taking into account the correlations 
between the systematic uncertainties.

The detailed inputs from the experiments and the resulting LEP
combined values, with the full breakdown of systematic errors 
is described in Appendix~\ref{4f_sec:appendix}.
Experimental results are compared with recent theoretical predictions,
many of which were developed in the framework 
of the LEP2 \MC\ workshop~\cite{4f_bib:fourfrep}. 

\section{W-pair production cross-section}
\label{4f_sec:WWxsec}

All experiments have published final results 
on the W-pair ({\sc CC03}~\cite{4f_bib:fourfrep}) 
production cross-section and W branching ratios for \CoM\ energies 
from~161 to~189 GeV~\cite{4f_bib:adloww161,4f_bib:adloww172,
4f_bib:aleww183,4f_bib:delww183,4f_bib:ltrww183,4f_bib:opaww183,
4f_bib:aleww189,4f_bib:delww189,4f_bib:ltrww189,4f_bib:opaww189}.
DELPHI has contributed final results to the Summer Conferences 
2003~\cite{4f_bib:delww}
also at higher energies, whereas ALEPH, L3 and OPAL 
measurements at $\roots=192$--207~GeV still have to be considered as
preliminary~\cite{4f_bib:alewwsc01,4f_bib:ltrwwsc02,4f_bib:opawwsc01}.

Because of the DELPHI update for the Summer 2003 Conferences, new 
LEP averages of the cross-section measurements at the eight \CoM\ energies between 
183 and 207 GeV have been computed, using the same grouping of the systematic 
errors consolidated in previous combinations~\cite{4f_bib:4f_s02}.

The detailed inputs used for the combinations are given in 
Appendix~\ref{4f_sec:appendix}.

The measured statistical errors are used for the combination; 
after building the full 32$\times$32 covariance matrix for the measurements,
the $\chi^2$ minimisation fit is performed by matrix algebra,
as described in Ref.~\cite{4f_bib:valassi},
and is cross-checked using Minuit~\cite{MINUIT}.

The results from each experiment 
for the W-pair production cross-section
are shown in Table~\ref{4f_tab:wwxsec}, 
together with the LEP combination at each energy. 
All measurements assume Standard Model values for the W decay branching fractions.
The results for \CoM\ energies between 183 and 207 GeV,
for which new LEP averages have been computed,
supersede the ones presented in~\cite{4f_bib:4f_s02}.
For completeness, 
the measurements at 161 and 172~GeV are also listed in the table.

\renewcommand{\arraystretch}{1.2}
\begin{table}[hbtp]
\begin{center}
\hspace*{-0.5cm}
\begin{tabular}{|c|c|c|c|c|c|r|} 
\hline
\roots & \multicolumn{5}{|c|}{WW cross-section (pb)} 
       & \multicolumn{1}{|c|}{$\chi^2/\textrm{d.o.f.}$} \\
\cline{2-6} 
(GeV)      & \Aleph\                & \Delphi\               &
             \Ltre\                 & \Opal\                 &
             LEP                    &                        \\
\hline
161.3      & $\phz4.23\pm0.75^*$    & 
             $\phz3.67^{\phz+\phz0.99\phz*}_{\phz-\phz0.87}$& 
             $\phz2.89^{\phz+\phz0.82\phz*}_{\phz-\phz0.71}$& 
             $\phz3.62^{\phz+\phz0.94\phz*}_{\phz-\phz0.84}$& 
             $\phz3.69\pm0.45\phs^*$  & 
             $\left\} \hspace*{2mm} \phz1.3\phz/\phz3 \right.$ \\
172.1      & $11.7\phz\pm1.3\phz^*$ & $11.6\phz\pm1.4\phz^*$ &
             $12.3\phz\pm1.4\phz^*$ & $12.3\phz\pm1.3\phz^*$ &
             $12.0\phz\pm0.7\phz\phs^*$ & 
             $\left\} \hspace*{2mm} \phz0.22/\phz3 \right.$ \\
182.7      & $15.57\pm0.68^*$       & $16.07\pm0.70^*$       &
             $16.53\pm0.72^*$       & $15.43\pm0.66^*$       &
             $15.81\pm0.37\phs^*$     & 
             \multirow{8}{20.3mm}{$
               \hspace*{-0.3mm}
               \left\}
                 \begin{array}[h]{rr}
                   &\multirow{8}{8mm}{\hspace*{-4.2mm}26.3/24}\\
                   &\\ &\\ &\\ &\\ &\\ &\\ &\\  
                 \end{array}
               \right.
               $}\\
188.6      & $15.71\pm0.38^*$       & $16.09\pm0.42^*$       &
             $16.24\pm0.43^*$       & $16.30\pm0.39^*$       &
             $16.04\pm0.22\phs^*$   & \\
191.6      & $17.23\pm0.91\phs$     & $16.64\pm1.00^*$     &
             $16.39\pm0.93\phs$     & $16.60\pm0.99\phs$     &
             $16.68\pm0.49\phs$     & \\
195.5      & $17.00\pm0.57\phs$     & $17.04\pm0.60^*$     &
             $16.67\pm0.60\phs$     & $18.59\pm0.75\phs$     &
             $17.16\pm0.33\phs$     & \\
199.5      & $16.98\pm0.56\phs$     & $17.39\pm0.57^*$     &
             $16.94\pm0.62\phs$     & $16.32\pm0.67\phs$     &
             $16.89\pm0.32\phs$     & \\
201.6      & $16.16\pm0.76\phs$     & $17.37\pm0.82^*$     &
             $16.95\pm0.88\phs$     & $18.48\pm0.92\phs$     &
             $17.11\pm0.43\phs$     & \\
204.9      & $16.57\pm0.55\phs$     & $17.56\pm0.59^*$     &
             $17.35\pm0.64\phs$     & $15.97\pm0.64\phs$     &
             $16.80\pm0.32\phs$     & \\
206.6      & $17.32\pm0.45\phs$     & $16.35\pm0.47^*$     &
             $17.96\pm0.51\phs$     & $17.77\pm0.57\phs$     &
             $17.28\pm0.27\phs$     & \\
\hline
\end{tabular}
\caption{%
W-pair production cross-section from the four LEP
experiments and combined values at all recorded \CoM\ energies.
All results are preliminary, with the exception of those indicated by $^*$. 
The measurements between 183 and 207 GeV
have been combined in one global fit, taking into account 
inter-experiment as well as inter-energy correlations of systematic errors.
The results for the combined LEP W-pair production cross-section 
at 161 and 172~GeV are taken 
from~\protect\cite{4f_bib:lepewwg97,4f_bib:lepewwg98} respectively.}
\label{4f_tab:wwxsec}
\end{center}
\vspace*{-4mm}
\end{table}
\renewcommand{\arraystretch}{1.}

\begin{figure}[p]
\centering
\epsfig{figure=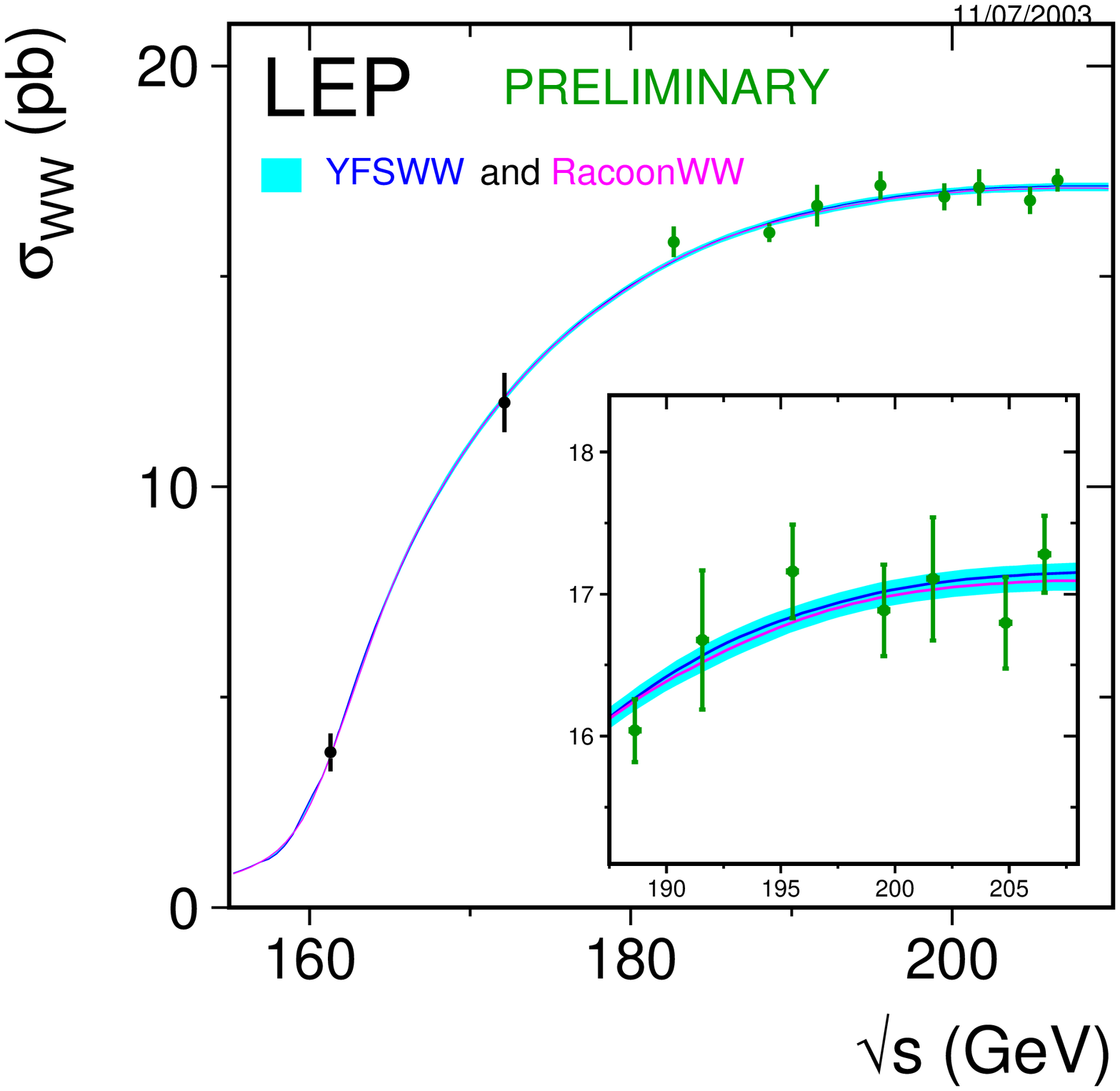,width=0.9\textwidth}
\caption{%
  Measurements of the W-pair production cross-section,
  compared to the predictions 
  of \RacoonWW~\protect\cite{4f_bib:racoonww} 
  and \YFSWW~\protect\cite{4f_bib:yfsww}. 
  The shaded area represents the uncertainty on the theoretical predictions,
  estimated in $\pm$2\% for $\roots\!<\!170$ GeV
  and ranging from 0.7 to 0.4\% above 170~GeV.
}
\label{4f_fig:sww_vs_sqrts}
\end{figure}

Figure~\ref{4f_fig:sww_vs_sqrts} shows the combined LEP W-pair cross-section 
measured as a function of the \CoM\ energy.
The experimental points are compared with the theoretical calculations 
from \YFSWW~\cite{4f_bib:yfsww} and \RacoonWW~\cite{4f_bib:racoonww} 
between 155 and 215 GeV for $\Mw=80.35$~GeV.
The two codes have been extensively compared and 
agree at a level better than 0.5\% 
at the LEP2 energies~\cite{4f_bib:fourfrep}.
The calculations above 170 GeV, 
based for the two programs on the so-called leading pole~(LPA) 
or double pole approximations~(DPA)~\cite{4f_bib:dpa}, 
have theoretical uncertainties 
decreasing from 0.7\% at 170 GeV
to about 0.4\% at \CoM\ energies larger than 200 GeV,
while in the threshold region, where the codes are run in Improved Born
Approximation, a larger theoretical uncertainty of 2\% is 
assigned~\cite{4f_bib:dpaerr}.
This theoretical uncertainty is represented by
the blue band in Figure~\ref{4f_fig:sww_vs_sqrts}. 
An error of 50 MeV on the W mass would translate 
into additional errors of 0.1\% (3.0\%) 
on the cross-section predictions at 200 GeV (161 GeV, respectively).
All results, up to the highest \CoM\ energies, 
are in agreement with the considered theoretical predictions.

The agreement between the measured W-pair cross-section,
$\sww^\mathrm{meas}$,
and its expectation according to a given theoretical model,
$\sww^\mathrm{theo}$,
can be expressed quantitatively in terms of their ratio
\begin{equation}
\rww = \frac{\sww^\mathrm{meas}}{\sww^\mathrm{theo}} ,
\end{equation}
averaged over the measurements performed by the four experiments 
at different energies in the LEP2 region.
The above procedure has been used to compare the measurements
at the eight energies between 183 and 207 GeV to the predictions 
of \Gentle~\cite{4f_bib:gentle}, \KoralW~\cite{4f_bib:koralw}, 
\YFSWW~\cite{4f_bib:yfsww} and \RacoonWW~\cite{4f_bib:racoonww}.
The measurements at 161 and 172 GeV
have not been used in the combination 
because they were performed using data samples of low statistics
and because of the high sensitivity of the cross-section 
to the value of the W mass at these energies.

The combination of the ratio $\rww$ is performed
using as input from the four experiments the 32 cross-sections
measured at each of the eight energies.
These are then converted into 32 ratios by dividing them
by the considered theoretical predictions,
listed in Appendix~\ref{4f_sec:appendix}.
The full 32$\times$32 covariance matrix for the ratios
is built taking into account the same sources
of systematic errors used for the combination 
of the W-pair cross-sections at these energies.

The small statistical errors 
on the theoretical predictions at the various energies,
taken as fully correlated for the four experiments
and uncorrelated between different energies, are also translated into errors 
on the individual measurements of $\rww$.
The theoretical errors on the predictions,
due to the physical and technical precision of the generators used, 
are not propagated to the individual ratios but are used when comparing 
the combined values of $\rww$ to unity.
For each of the four models considered,
two fits are performed:
in the first, eight values of $\rww$ at the different energies are extracted,
averaged over the four experiments;
in the second, only one value of $\rww$ is determined,
representing the global agreement of measured and predicted cross-sections
over the whole energy range.

\renewcommand{\arraystretch}{1.2}
\begin{table}[bhtp]
\vspace*{-0mm}
\begin{center}
\hspace*{-0.3cm}
\begin{tabular}{|c|c|c|} 
\hline
\roots (GeV) & $\rww^{\footnotesize\YFSWW}$ & $\rww^{\footnotesize\RacoonWW}$ \\
\hline
182.7      & $1.030\pm0.024$ & $1.029\pm0.024$ \\
188.6      & $0.986\pm0.014$ & $0.987\pm0.014$ \\
191.6      & $1.007\pm0.030$ & $1.010\pm0.030$ \\
195.5      & $1.019\pm0.020$ & $1.021\pm0.020$ \\
199.5      & $0.992\pm0.019$ & $0.995\pm0.019$ \\
201.6      & $1.002\pm0.025$ & $1.004\pm0.026$ \\
204.9      & $0.981\pm0.019$ & $0.984\pm0.019$ \\
206.6      & $1.008\pm0.016$ & $1.011\pm0.016$ \\
\hline
$\chi^2$/d.o.f & 26.3/24         & 26.3/24        \\
\hline
\hline
Average        & $0.997\pm0.010$ & $0.999\pm0.010$ \\
\hline
$\chi^2$/d.o.f & 32.7/31         & 32.7/31        \\
\hline
\end{tabular}
\caption{%
Ratios of LEP combined W-pair cross-section measurements
to the expectations according to 
\YFSWW~\protect\cite{4f_bib:yfsww} and 
\RacoonWW~\protect\cite{4f_bib:racoonww}.
For each of the two models,
two fits are performed,
one to the LEP combined values 
of $\rww$ at the eight energies between 183 and 207~GeV,
and another to the LEP combined average of $\rww$ over all energies.
The results of the fits are given in the table
together with the resulting $\chi^2$.
Both fits take into account inter-experiment 
as well as inter-energy correlations of systematic errors.
}
\label{4f_tab:wwratio}
\end{center}
\vspace*{-6mm}
\end{table}
\renewcommand{\arraystretch}{1.}

\begin{figure}[tp]
\centering
\vspace*{-0.5truecm}
\mbox{
  \fbox{\epsfig{figure=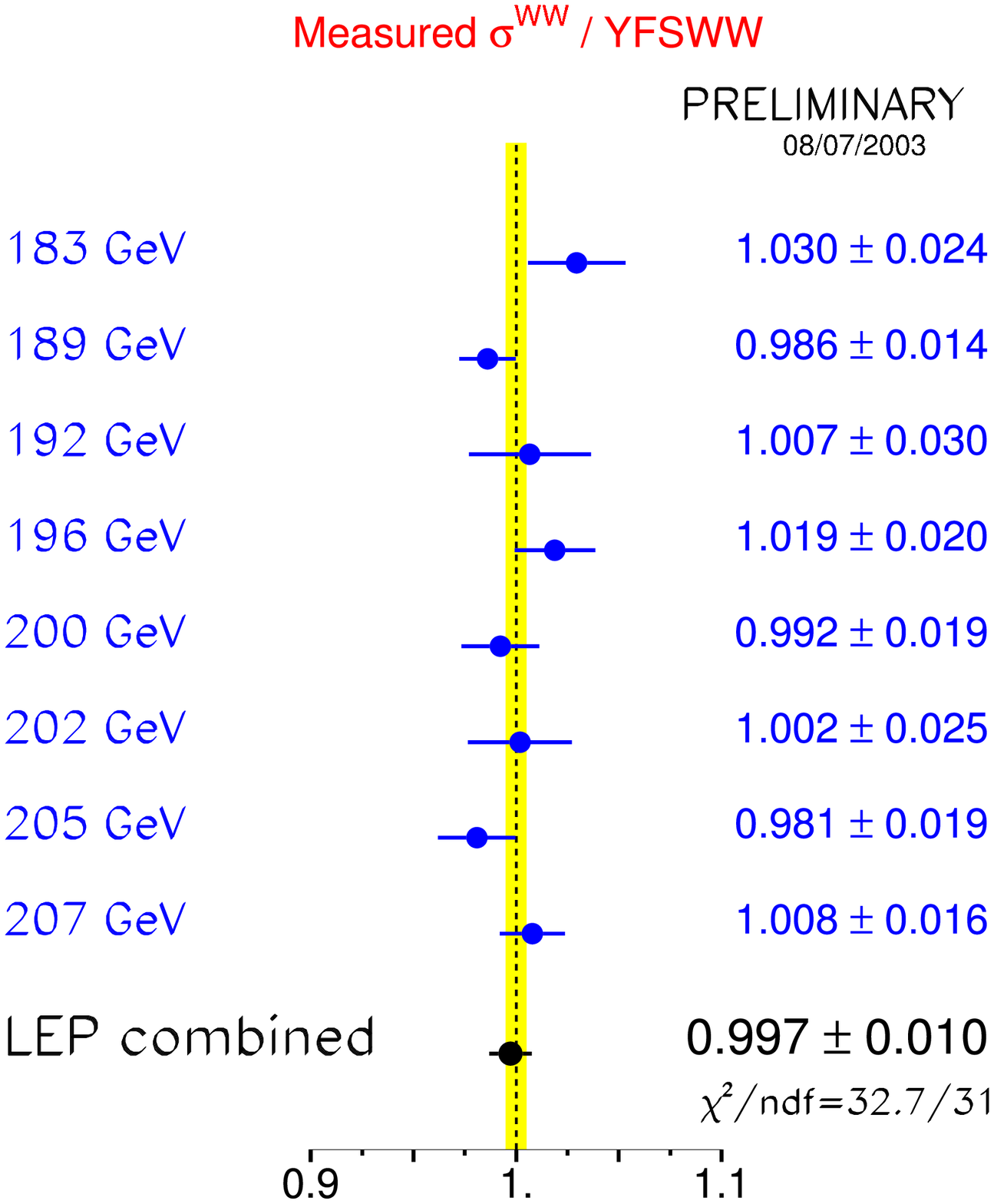,width=0.45\textwidth}}
  \hspace*{0.04\textwidth}
  \fbox{\epsfig{figure=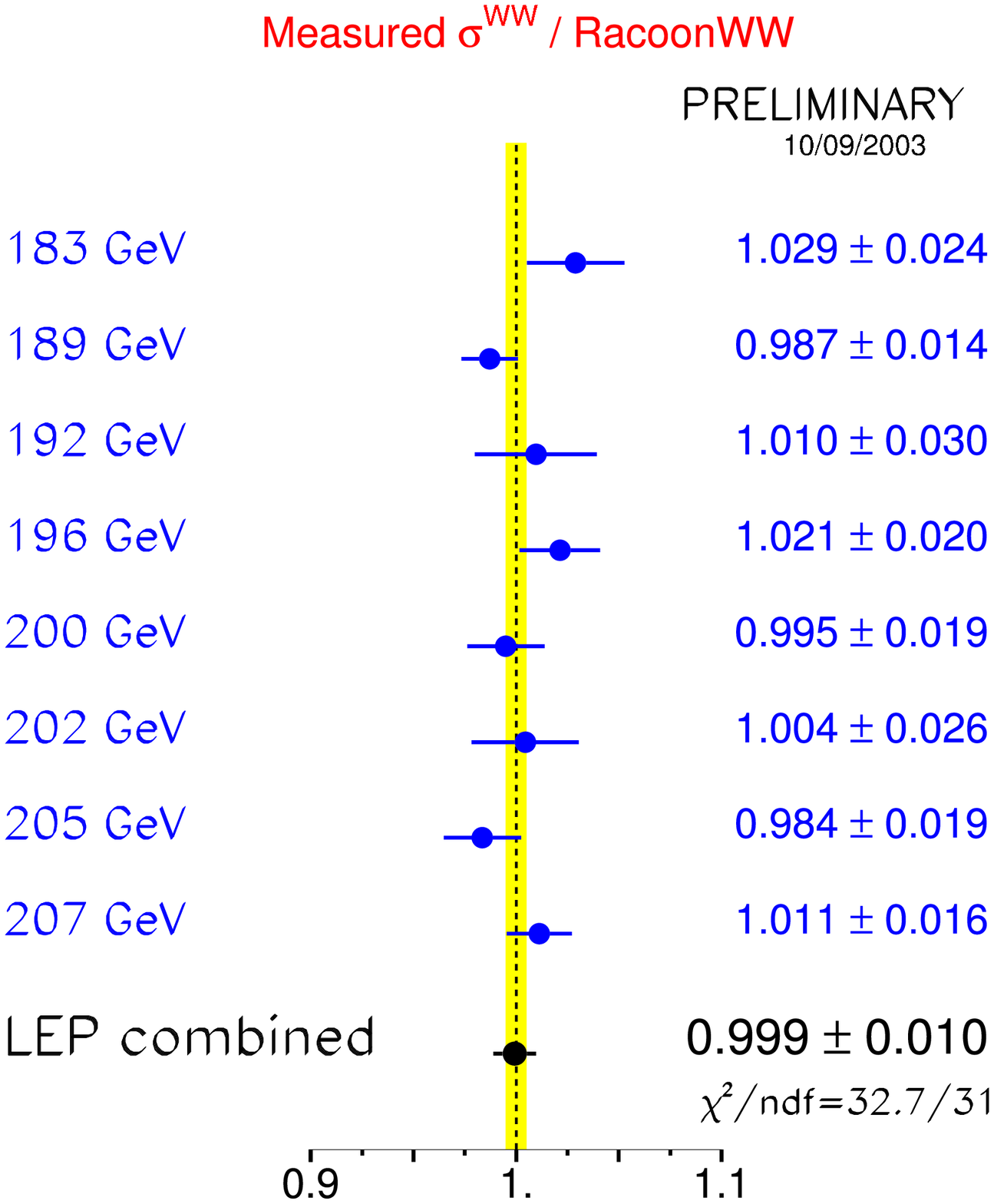,width=0.45\textwidth}}
  }
\vspace*{-0.5truecm}
\caption{%
  Ratios of LEP combined W-pair cross-section measurements
  to the expectations according to 
  \YFSWW~\protect\cite{4f_bib:yfsww} and 
  \RacoonWW~\protect\cite{4f_bib:racoonww}
  The yellow bands represent constant relative errors 
  of 0.5\% on the two 
  cross-section predictions.
}
\label{4f_fig:rww}
\end{figure} 

The results of the two fits to $\rww$
for \YFSWW\ and \RacoonWW\ are given in Table~\ref{4f_tab:wwratio}.
As already qualitatively noted from Figure~\ref{4f_fig:sww_vs_sqrts},
the LEP measurements of the W-pair cross-section above threshold
are in very good agreement to the predictions.
In contrast, the predictions from \Gentle\ and \KoralW\ are
more than 2\% too high with respect to the measurements; the equivalent
values of $\rww$ in those cases are, respectively, $0.973\pm0.010$
and $0.978\pm0.010$.
The main differences between these two sets of predictions
come from non-leading $\oa$ electroweak radiative corrections
to the W-pair production process and non-factorisable corrections,
which are included 
(in the LPA/DPA approximation~\cite{4f_bib:dpa})
in both \YFSWW\ and \RacoonWW, but not in \Gentle\ and \KoralW.
The data clearly prefer the computations which
more precisely include $\oa$ radiative corrections.

The results of the fits for \YFSWW\ and \RacoonWW are also shown 
in Figure~\ref{4f_fig:rww}, where relative errors of 0.5\% on the 
cross-section predictions have been assumed.
For simplicity in the figure the energy dependence of the theory error
on the W-pair cross-section has been neglected.

\section{W branching ratios}
\label{4f_sec:WWBR}

From the partial cross-sections WW$\rightarrow\mathrm{4f}$ measured by the 
four experiments at all energies above 161~GeV, the W decay branching fractions 
\mbox{$\mathcal{B}(\mathrm{W}\rightarrow\textrm{f}\overline{\textrm{f}}')$}
are determined, with and without the assumption of lepton universality.

The two combinations performed, with and without the assumption of lepton 
universality, use as inputs from the experiments the three leptonic branching 
fractions, with their systematic and observed statistical errors
and their correlation matrices.
In the fit with lepton universality, the branching fraction to hadrons is 
determined from that to leptons by constraining the sum to unity.
The part of the systematic error correlated between experiments is properly
accounted for when building the full covariance matrix.

\begin{figure}[tp]
\centering
\vspace*{-0.5truecm}
\mbox{
  \fbox{\epsfig{figure=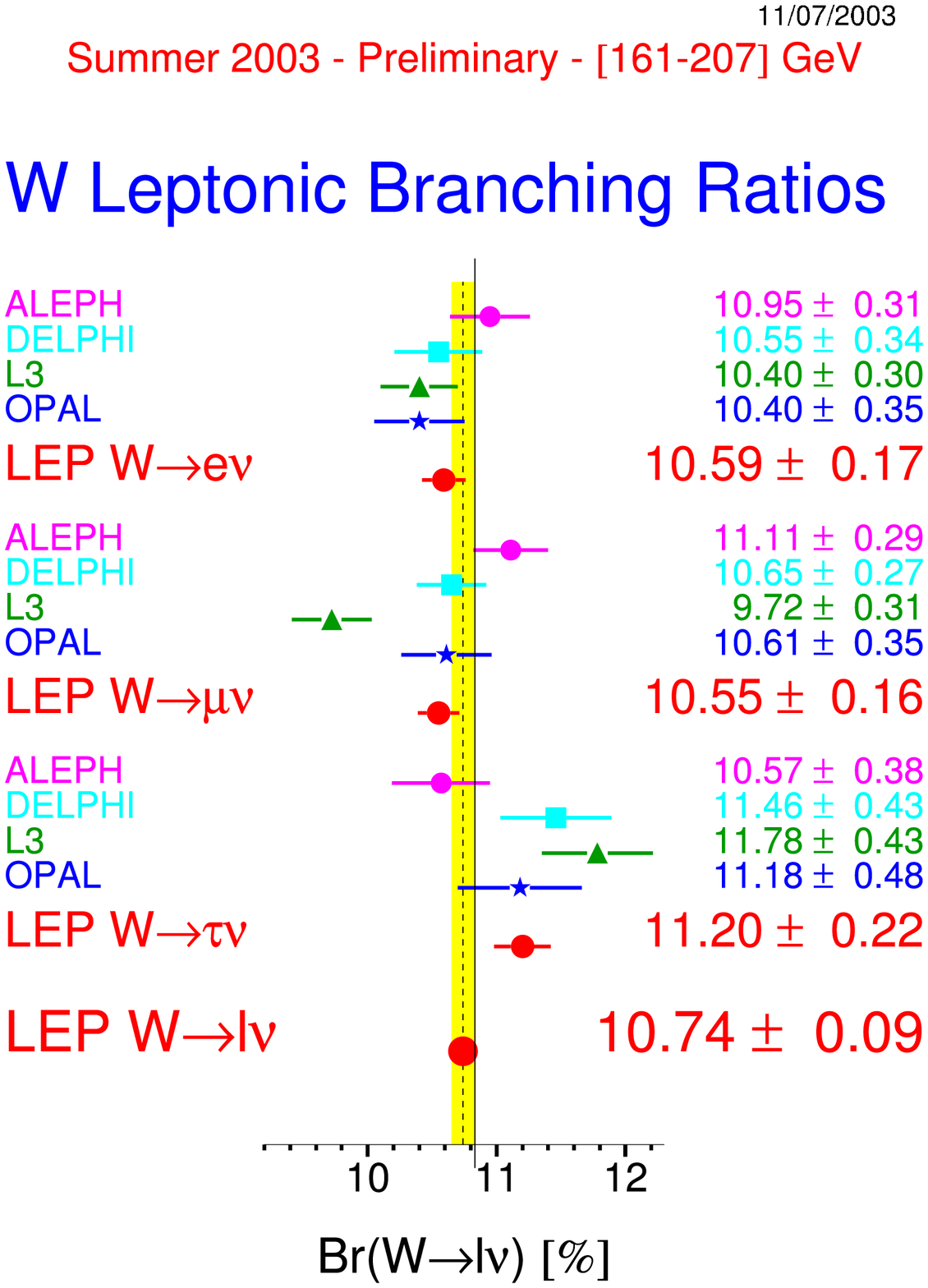,width=0.45\textwidth}}
  \hspace*{0.04\textwidth}
  \fbox{\epsfig{figure=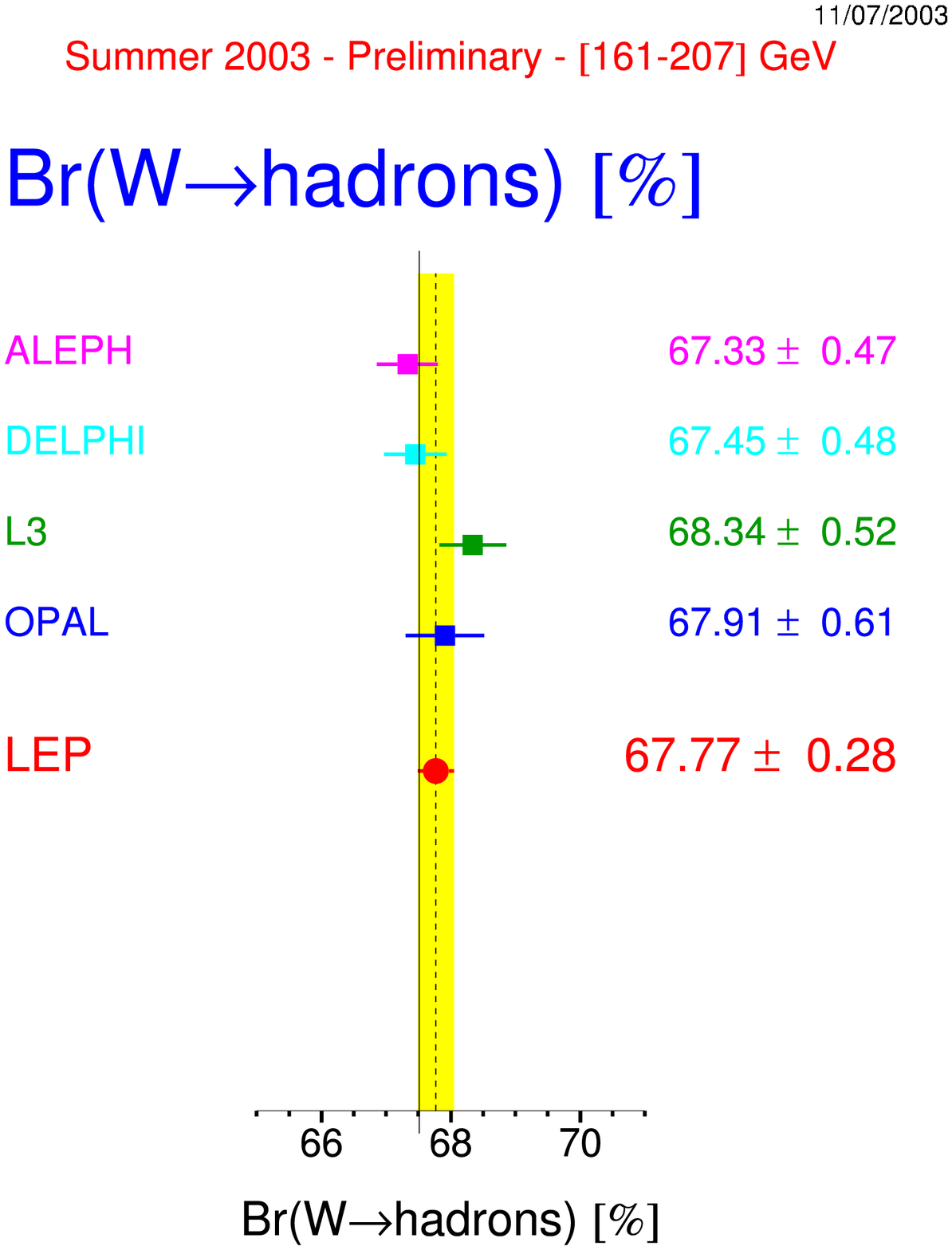,width=0.45\textwidth}}
  }
\vspace*{-0.5truecm}
\caption{%
  Leptonic and hadronic W branching fractions, as measured by 
  the experiments, and the LEP combined values according to the
  procedures described in the text.}
\label{4f_fig:brw}
\end{figure}

The detailed inputs used for the combinations are given in 
Appendix~\ref{4f_sec:appendix}.
The results from each experiment are given in Table~\ref{tab:wwbra} 
together with the result of the LEP combination. The same results are 
shown in Figure~\ref{4f_fig:brw}.

\renewcommand{\arraystretch}{1.2}
\begin{table}[hbtp]
\begin{center}
\begin{tabular}{|c|c|c|c|c|} 
\cline{2-5}
\multicolumn{1}{c|}{$\quad$} & \multicolumn{3}{|c|}{Lepton} & Lepton \\
\multicolumn{1}{c|}{$\quad$} & \multicolumn{3}{|c|}{non--universality} & universality \\
\hline
Experiment 
         & \wwbr(\Wtoenu) & \wwbr(\Wtomnu) & \wwbr(\Wtotnu)  
         & \wwbr({\mbox{$\mathrm{W}\rightarrow\mathrm{hadrons}$}}) \\
         & [\%] & [\%] & [\%] & [\%]  \\
\hline
\Aleph\  & $10.95\pm0.31$ & $11.11\pm0.29$ & $10.57\pm0.38$ & $67.33\pm0.47$ \\
\Delphi\ & $10.55\pm0.34^*$ & $10.65\pm0.27^*$ & $11.46\pm0.43^*$ & $67.45\pm0.48^*$ \\
\Ltre\   & $10.40\pm0.30$ 
                       & $\phz9.72\pm0.31$ & $11.78\pm0.43$ & $68.34\pm0.52$ \\
\Opal\   & $10.40\pm0.35$ & $10.61\pm0.35$ & $11.18\pm0.48$ & $67.91\pm0.61$ \\
\hline
LEP      & $10.59\pm0.17$ & $10.55\pm0.16$ & $11.20\pm0.22$ & $67.77\pm0.28$ \\
\hline
$\chi^2/\textrm{d.o.f.}$ & \multicolumn{3}{|c|}{15.3/9} & 20.8/11 \\
\hline
\end{tabular}
\caption{
  Summary of W branching fractions derived from W-pair production 
  cross sections measurements up to 207 GeV \CoM\ energy. All results
  are preliminary with the exception of those indicated by $^*$. } 
\label{tab:wwbra} 
\end{center}
\end{table}
\renewcommand{\arraystretch}{1.}

The results of the fit which does not make use of the lepton universality
assumption show a negative correlation of 19.6\% (14.8\%) between the 
\Wtotnu\  and \Wtoenu\  (\Wtomnu)  branching fractions, while between the
electron and muon decay channels there is a positive correlation of 9.2\%.
The two-by-two comparison of these branching fractions 
allows to test lepton universality 
in the decay of on--shell W bosons at the level of 2.9\%:
\begin{eqnarray*}
\wwbr\mathrm{(\Wtomnu)} \, / \, \wwbr\mathrm{(\Wtoenu)} \,
& = & 0.997 \pm 0.021 \, ,\\
\wwbr\mathrm{(\Wtotnu)} \; / \, \wwbr\mathrm{(\Wtoenu)} \,
& = & 1.058 \pm 0.029 \, ,\\
\wwbr\mathrm{(\Wtotnu)} \, / \, \wwbr\mathrm{(\Wtomnu)} 
& = & 1.061 \pm 0.028 \, .
\end{eqnarray*}
The branching fractions are all consistent with each other within the errors.

Assuming lepton universality,
the measured hadronic branching fraction 
is $67.77\pm0.18\mathrm{(stat.)}\pm0.22\mathrm{(syst.)}\%$ 
and the leptonic one 
is $10.74\pm0.06\mathrm{(stat.)}\pm0.07\mathrm{(syst.)}\%$.
These results are consistent with their Standard Model expectations,
of 67.51\% and 10.83\% respectively.
The systematic error receives equal contributions 
from the correlated and uncorrelated sources.

Within the Standard Model, 
the branching fractions of the W boson depend on the six matrix elements 
$|\mathrm{V}_{\mathrm{qq'}}|$ of the Cabibbo--Kobayashi--Maskawa (CKM) 
quark mixing matrix not involving the top quark. 
In terms of these matrix elements, 
the leptonic branching fraction of the W boson 
$\mathcal{B}(\Wtolnu)$ is given by
\begin{equation*}
  \frac{1}{\mathcal{B}(\Wtolnu)}\quad = \quad 3 
  \Bigg\{ 1 + 
          \bigg[ 1 + \frac{\alpha_s(\mathrm{M}^2_{\mathrm{W}})}{\pi} 
          \bigg] 
          \sum_{\tiny\begin{array}{c}i=(u,c),\\j=(d,s,b)\\\end{array}}
          |\mathrm{V}_{ij}|^2 
  \Bigg\},
\end{equation*} 
where $\alpha_s(\mathrm{M}^2_{\mathrm{W}})$ is the strong coupling
constant. 
Taking $\alpha_s(\mathrm{M}^2_{\mathrm{W}})=0.119\pm0.002$~\cite{bib:pdg02},
and using the experimental knowledge of the sum
$|\mathrm{V}_{ud}|^2+|\mathrm{V}_{us}|^2+|\mathrm{V}_{ub}|^2+
 |\mathrm{V}_{cd}|^2+|\mathrm{V}_{cb}|^2=1.0476\pm0.0074$~\cite{bib:pdg02}, 
the above result can be interpreted as a measurement of $|\mathrm{V}_{cs}|$ 
which is the least well determined of these matrix elements:
\begin{equation*}
  |\mathrm{V}_{cs}|\quad=\quad0.989\,\pm\,0.014.
\end{equation*}
The error includes 
a $\pm0.0006$ contribution from the uncertainty on $\alpha_s$
and a $\pm0.004$ contribution from the uncertainties 
on the other CKM matrix elements,
the largest of which is that on $|\mathrm{V}_{cd}|$.
These contributions are negligible
in the error on this determination of $|\mathrm{V}_{cs}|$,
which is dominated by the $\pm0.013$ experimental error 
from the measurement of the W branching fractions. The value of $|\mathrm{V}_{cs}|$
is in agreement with unity.

\clearpage

\section{Z-pair production cross-section}
\label{4f_sec:ZZxsec}

With respect to previous combinations\cite{bib-EWEP-02}, final results
results of the measurements of the Z-pair production cross-section,
defined as the {\sc NC02}~\cite{4f_bib:fourfrep} contribution to
four-fermion cross-sections, are now available from
DELPHI~\cite{4f_bib:delzz}, L3~\cite{4f_bib:ltrzz183a,
  4f_bib:ltrzz183b,4f_bib:ltrzz189,4f_bib:ltrzz1999,4f_bib:ltrzzsc03}
and OPAL~\cite{4f_bib:opazz189,4f_bib:opazz}.  ALEPH published final
results at 183 and 189 GeV~\cite{4f_bib:alezz189} and contributed
preliminary results for all other energies up to 207
GeV~\cite{4f_bib:alezzsc01}.

\renewcommand{\arraystretch}{1.2}
\begin{table}[bp]
\begin{center}
\begin{tabular}{|c|c|c|c|c|c|c|} 
\hline
\roots & \multicolumn{5}{|c|}{ZZ cross-section (pb)} & \\
\cline{2-6} 
(GeV) & \Aleph\ & \Delphi\ & \Ltre\ & \Opal\ & LEP & $\chi^2/\textrm{d.o.f.}$ \\ 
\hline
182.7 & 
$0.11^{\phz+\phz0.16\phz*}_{\phz-\phz0.12}$ & 
$0.35^{\phz+\phz0.20\phz*}_{\phz-\phz0.15}$ & 
$0.31\pm0.17^*$ & 
$0.12^{\phz+\phz0.20\phz*}_{\phz-\phz0.18}$ &
$0.22\pm0.08\phs^*$ & 
             \multirow{8}{20.3mm}{$
               \hspace*{-0.3mm}
               \left\}
                 \begin{array}[h]{rr}
                   &\multirow{8}{8mm}{\hspace*{-4.2mm}16.1/24}\\
                   &\\ &\\ &\\ &\\ &\\ &\\ &\\  
                 \end{array}
               \right.
               $}\\
188.6 & 
$0.67^{\phz+\phz0.14\phz*}_{\phz-\phz0.13}$ & 
$0.52^{\phz+\phz0.12\phz*}_{\phz-\phz0.11}$ & 
$0.73\pm0.15\phs^*$ & 
$0.80^{\phz+\phz0.15\phz*}_{\phz-\phz0.14}$ &
$0.66\pm0.07\phs^*$ &  \\
191.6 & 
$0.53^{\phz+\phz0.34}_{\phz-\phz0.27}\phzs$ &
$0.63^{\phz+\phz0.36*}_{\phz-\phz0.30}\phzs$ &
$0.29\pm0.22^*$ & 
$1.29^{\phz+\phz0.48*}_{\phz-\phz0.41}\phzs$ &
$0.65\pm0.17\phs$ &  \\
195.5 & 
$0.69^{\phz+\phz0.23}_{\phz-\phz0.20}\phzs$ & 
$1.05^{\phz+\phz0.25*}_{\phz-\phz0.22}\phzs$ & 
$1.18\pm0.26^*$ & 
$1.13^{\phz+\phz0.27*}_{\phz-\phz0.25}\phzs$ &
$0.99\pm0.12\phs$ &  \\
199.5 & 
$0.70^{\phz+\phz0.22}_{\phz-\phz0.20}\phzs$ & 
$0.75^{\phz+\phz0.20*}_{\phz-\phz0.18}\phzs$ & 
$1.25\pm0.27^*$ & 
$1.05^{\phz+\phz0.26*}_{\phz-\phz0.23}\phzs$ &
$0.90\pm0.12\phs$ &  \\
201.6 & 
$0.70^{\phz+\phz0.33}_{\phz-\phz0.28}\phzs$ & 
$0.85^{\phz+\phz0.33*}_{\phz-\phz0.28}\phzs$ & 
$0.95\pm0.39^*$ & 
$0.79^{\phz+\phz0.36*}_{\phz-\phz0.30}\phzs$ &
$0.81\pm0.17\phs$ &  \\
204.9 & 
$1.21^{\phz+\phz0.26}_{\phz-\phz0.23}\phzs$ & 
$1.03^{\phz+\phz0.23*}_{\phz-\phz0.20}\phzs$ & 
$0.77^{\phz+\phz0.21}_{\phz-\phz0.19}\phzs$ & 
$1.07^{\phz+\phz0.28*}_{\phz-\phz0.25}\phzs$ &
$0.98\pm0.13\phs$ &  \\
206.6 & 
$1.01^{\phz+\phz0.19}_{\phz-\phz0.17}\phzs$ & 
$0.96^{\phz+\phz0.16*}_{\phz-\phz0.15}\phzs$ & 
$1.09^{\phz+\phz0.18}_{\phz-\phz0.17}\phzs$ & 
$0.97^{\phz+\phz0.20*}_{\phz-\phz0.19}\phzs$ &
$0.99\pm0.09\phs$ & \\
\hline
\end{tabular}
\caption{%
Z-pair production cross-sections from the four LEP
experiments and combined values 
for the eight energies between 183 and 207~GeV.
All results are preliminary with the exception of those indicated by $^*$.}
\label{4f_tab:zzxsec}
\end{center}
\end{table}
\renewcommand{\arraystretch}{1.}

\begin{figure}[p]
\centering
\epsfig{figure=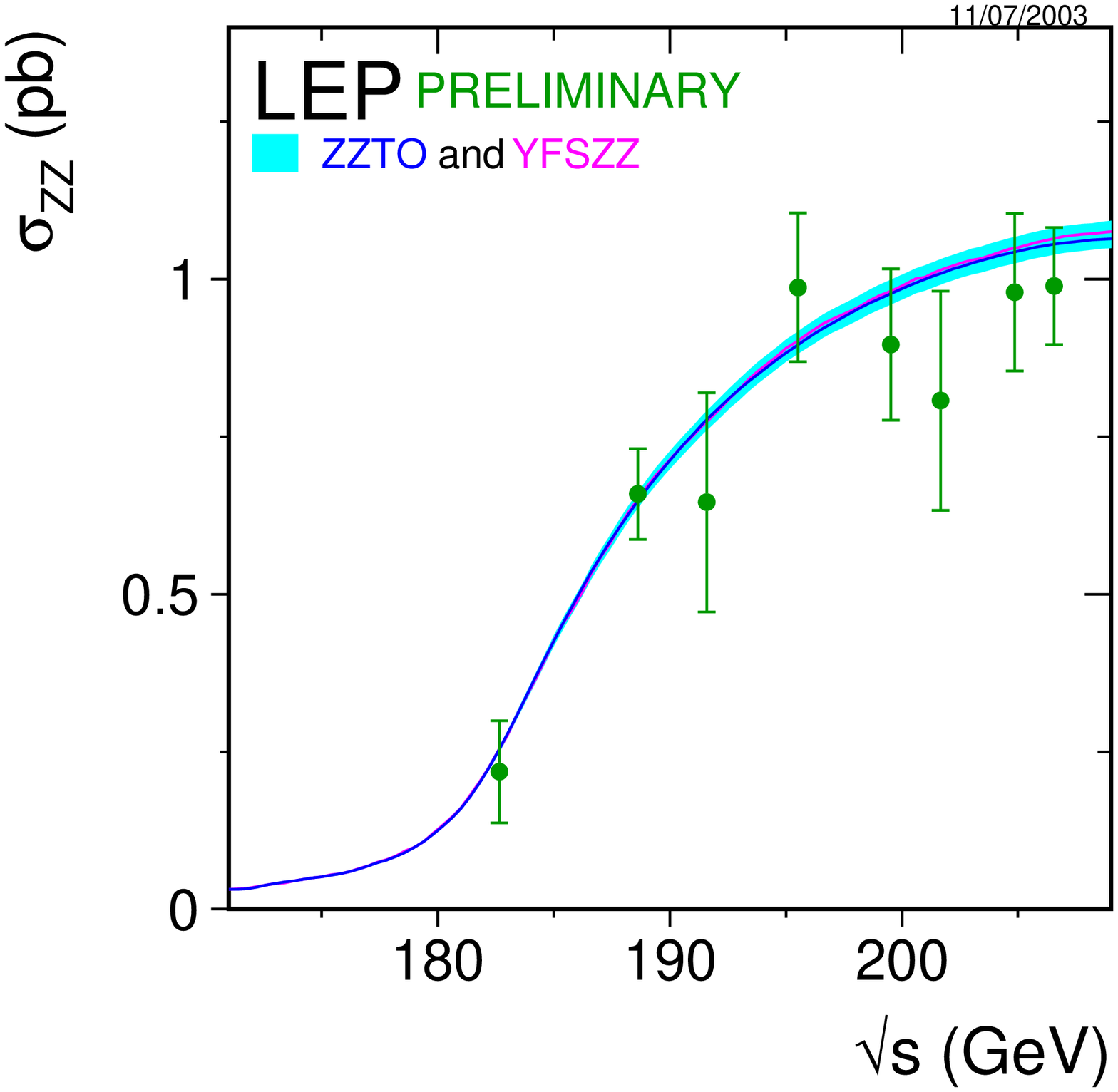,width=0.9\textwidth}
\caption{%
  Measurements of the Z-pair production cross-section,
  compared to the predictions 
  of \YFSZZ~\protect\cite{4f_bib:yfszz} and \ZZTO~\protect\cite{4f_bib:zzto}. 
  The shaded area represent the $\pm2$\% uncertainty 
  on the predictions.
}
\label{4f_fig:szz_vs_sqrts}
\end{figure}

\begin{table}[bhp]
\begin{center}
\begin{tabular}{|c|c|c|} 
\hline
\roots (GeV)      & $\rzz^{\footnotesize\ZZTO}$ 
           & $\rzz^{\footnotesize\YFSZZ}$ \\
\hline
182.7             & $0.857\pm0.320$ & $0.857\pm0.320$  \\
188.6             & $1.017\pm0.113$ & $1.007\pm0.111$  \\
191.6             & $0.831\pm0.225$ & $0.826\pm0.224$  \\
195.5             & $1.100\pm0.133$ & $1.100\pm0.133$  \\
199.5             & $0.915\pm0.125$ & $0.912\pm0.124$  \\
201.6             & $0.799\pm0.174$ & $0.795\pm0.173$  \\
204.9             & $0.937\pm0.121$ & $0.931\pm0.120$  \\
206.6             & $0.937\pm0.091$ & $0.928\pm0.090$  \\
\hline
$\chi^2$/d.o.f    & 16.1/24         & 16.1/24         \\
\hline
\hline
Average           & $0.952\pm0.052$ & $0.945\pm0.052$  \\
\hline
$\chi^2$/d.o.f    & 19.1/31         & 19.1/31        \\
\hline
\end{tabular}
\caption{%
Ratios of LEP combined Z-pair cross-section measurements
to the expectations according to 
\ZZTO~\protect\cite{4f_bib:zzto} and \YFSZZ~\protect\cite{4f_bib:yfszz}.
The results of the combined fits are given in the table together with 
the resulting $\chi^2$.
Both fits take into account inter-experiment 
as well as inter-energy correlations of systematic errors.
}
\label{4f_tab:zzratio}
\end{center}
\end{table}
\renewcommand{\arraystretch}{1.}

\begin{figure}[tp]
\centering
\vspace*{-0.5truecm}
\mbox{
  \fbox{\epsfig{figure=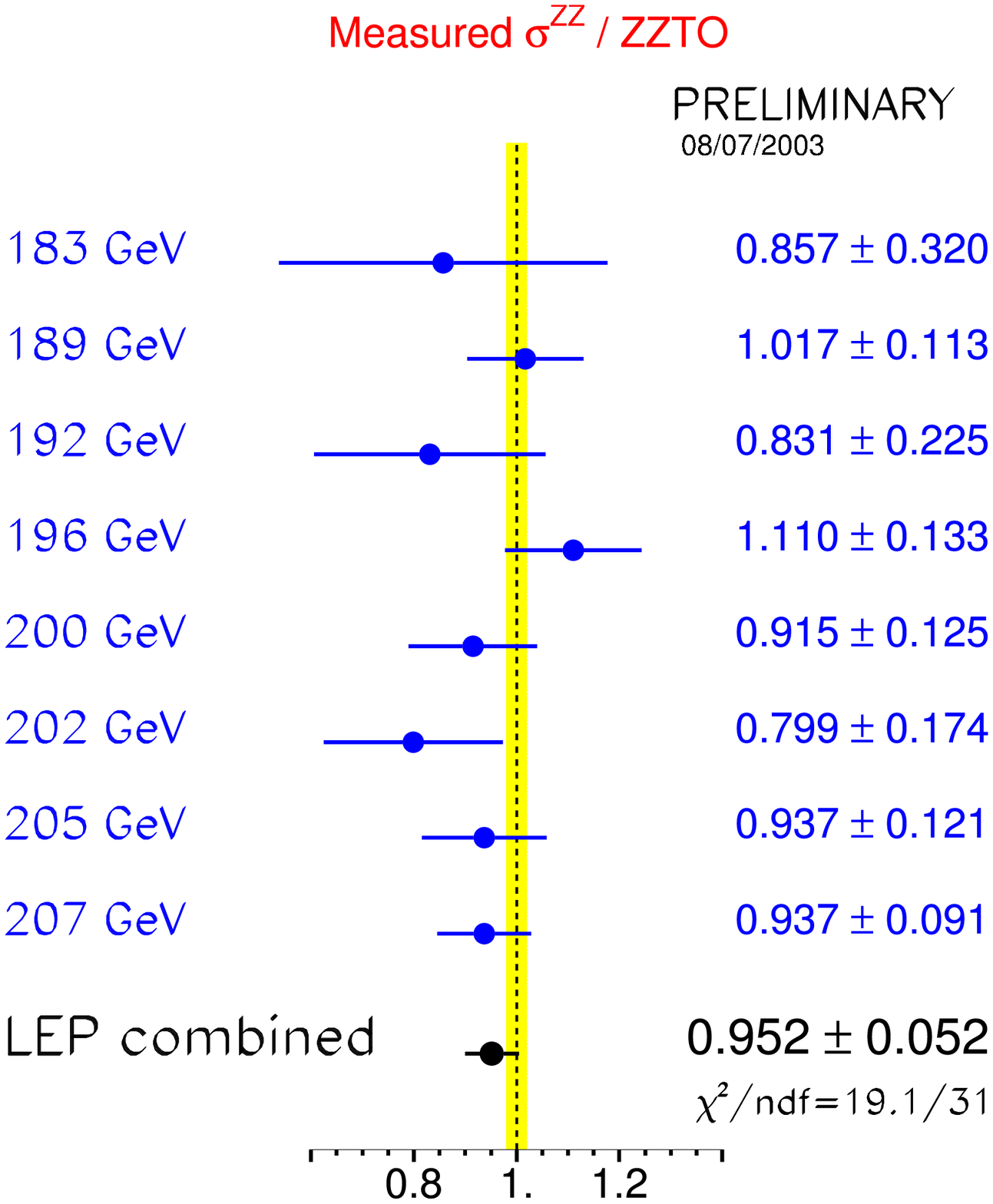,width=0.45\textwidth}}
  \hspace*{0.04\textwidth}
  \fbox{\epsfig{figure=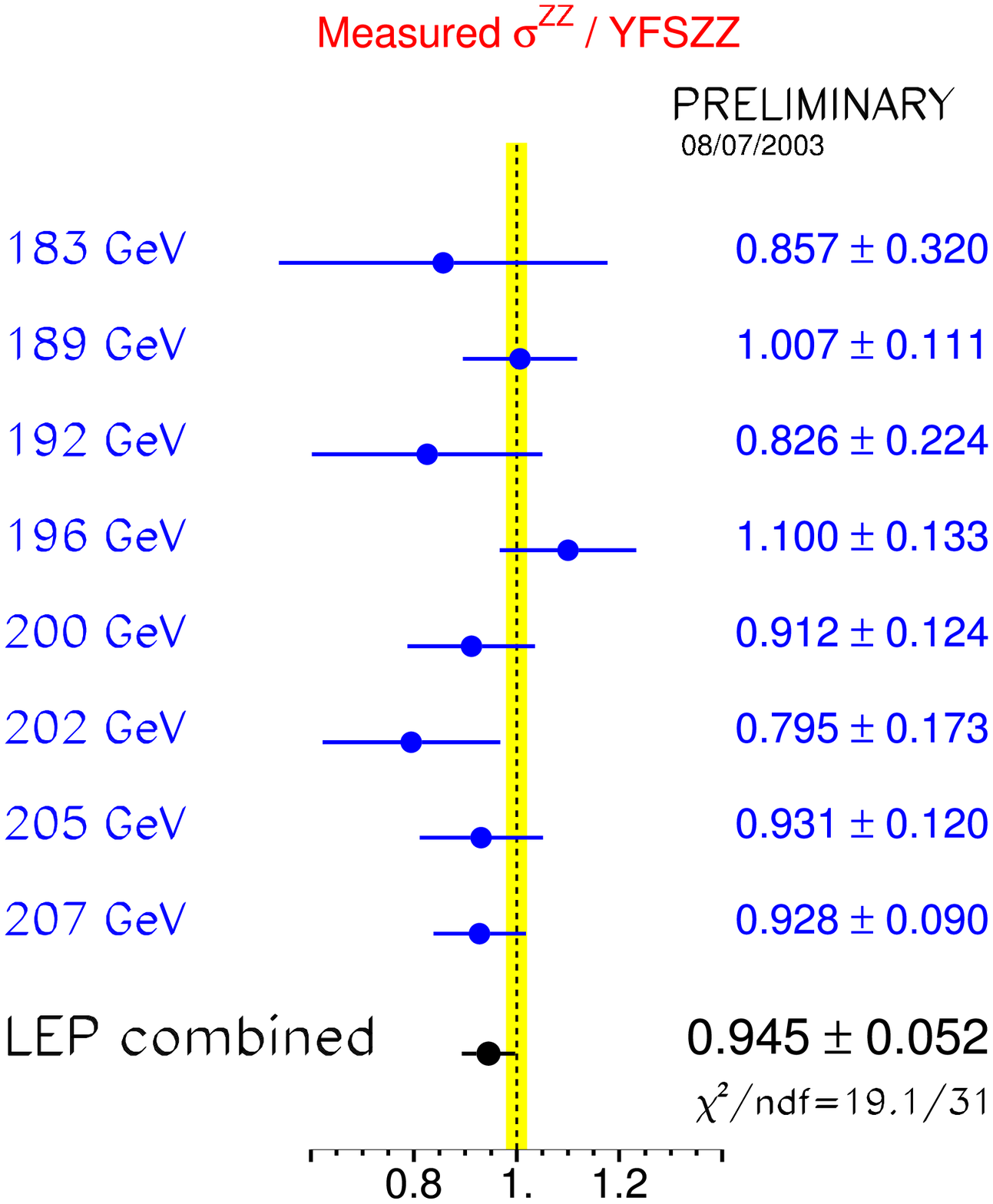,width=0.45\textwidth}}
  }
\vspace*{-0.5truecm}
\caption{%
  Ratios of LEP combined Z-pair cross-section measurements
  to the expectations according to 
  \ZZTO~\protect\cite{4f_bib:zzto} and 
  \YFSZZ~\protect\cite{4f_bib:yfszz}
  The yellow bands represent constant relative errors 
  of 2\% on the two cross-section predictions.
}
\label{4f_fig:rzz}
\end{figure} 

The combination of results is performed with the same technique used
for the WW cross-section. The symmetrized expected statistical error 
of each analysis is used, to avoid biases due to the limited number of 
selected events. 
All the cross-sections used for the combination and presented in 
Table~\ref{4f_tab:zzxsec} are determined by the experiments 
using the frequentist approach, i.e. without assuming any prior for 
the value of the cross-section itself.
This is now changed with respect to the 2002 Summer Conferences
combination, where the \Delphi\ cross-sections determined in a 
Bayesian approach were used.   

The measurements are shown in Figure~\ref{4f_fig:szz_vs_sqrts} 
as a function of the LEP \CoM\ energy,
where they are compared to the \YFSZZ~\cite{4f_bib:yfszz} and
\ZZTO~\cite{4f_bib:zzto} predictions.
Both these calculations have an estimated 
uncertainty of $\pm2\%$~\cite{4f_bib:fourfrep}. 
The data do not show any significant deviation 
from the theoretical expectations.

In analogy with the W-pair cross-section, a value for $\rzz$ can 
also be determined: its definition and the procedure of the combination
follows the one described for $\rww$.
The data are compared with the \YFSZZ\ and \ZZTO\ predictions; 
Table~\ref{4f_tab:zzratio} reports the numerical values of $\rzz$ in energy 
and combined, whereas figure~\ref{4f_fig:rzz} show them in comparison
to unity, where the $\pm$2\% error on the theoretical ZZ cross-section is 
shown as a yellow band. The experimental accuracy on the combined 
value of $\rzz$ is about 5\%.

The theory predictions, the details of the experimental inputs
with the the breakdown of the error contributions and the LEP combined 
values of the total cross-sections and the ratios to theory are
reported in Appendix~\ref{4f_sec:appendix}. 

\clearpage

\section{Single-Z production cross-section}
\label{4f_sec:zeexsec}

The LEP combination of the single-Z production cross-section
has been updated using preliminary \Aleph~\cite{4f_bib:alezeesc03},
and \Delphi~\cite{4f_bib:delswsc03} inputs to the Summer 2003 Conferences
and final \Ltre~\cite{4f_bib:ltrzee} results.
With respect to the last Summer Conferences \Aleph\ provided inputs
in the hadronic channel as well and \Ltre\ measurements are now final.
Single-Z production at LEP2 is studied considering only the 
$eeq\bar{q}$, $ee\mu\mu$ final states with the following phase space cuts 
and assuming one visible electron:
\mbox{$m_{q\bar{q}}(m_{\mu\mu})>60$~GeV/c$^2$},
\mbox{$\theta_\mathrm{e^+}<12^{\circ}$},
\mbox{$12^{\circ}<\theta_\mathrm{e^-}<120^{\circ}$} and 
\mbox{$E_\mathrm{e^-}>$3~GeV}, with obvious notation and where the angle is
defined with respect to the beam pipe, with the positron direction
being along $+z$ 
and the electron direction being along $-z$. 
Corresponding cuts are imposed when the positron is visible:
\mbox{$\theta_\mathrm{e^-}>168^{\circ}$},
\mbox{$60^{\circ}<\theta_\mathrm{e^+}<168^{\circ}$} and 
\mbox{$E_\mathrm{e^+}>$3~GeV}.

\renewcommand{\arraystretch}{1.2}
\begin{table}[htb]
\begin{center}
\hspace*{-0.0cm}
\begin{tabular}{|c|c|c|c|c|c|c|} 
\hline
\roots & \multicolumn{5}{|c|}{Single-Z hadronic cross-section (pb)} 
       & \\
\cline{2-6} 
(GeV) & \Aleph\ & \Delphi\ & \Ltre\ & \Opal\ & LEP & $\chi^2/\textrm{d.o.f.}$ \\
\hline
182.7 & $0.25^{\phz+\phz0.20\phz}_{\phz-\phz0.15}$ & $0.56^{\phz+\phz0.28\phz}_{\phz-\phz0.23}$ &
        $0.51^{\phz+\phz0.19\phz*}_{\phz-\phz0.16}$ & --- &
$0.44\pm0.11\phs$ &  
             \multirow{8}{20.3mm}{$
               \hspace*{-0.3mm}
               \left\}
                 \begin{array}[h]{rr}
                   &\multirow{8}{8mm}{\hspace*{-4.2mm}12.4/16}\\
                   &\\ &\\ &\\ &\\ &\\ &\\ &\\  
                 \end{array}
               \right.
               $}\\

188.6 & $0.40^{\phz+\phz0.12}_{\phz-\phz0.11}\phs$ & 
        $0.65^{\phz+\phz0.16}_{\phz-\phz0.15}\phs$ &
        $0.55^{\phz+\phz0.11\phz*}_{\phz-\phz0.10}$ & --- &
$0.52\pm0.07\phs$ & \\

191.6 & $0.58^{\phz+\phz0.37}_{\phz-\phz0.27}\phs$ & 
        $0.63^{\phz+\phz0.40}_{\phz-\phz0.30}\phs$ &
        $0.60^{\phz+\phz0.27*}_{\phz-\phz0.22}\phs$ & --- &
$0.60\pm0.15\phs$ & \\

195.5 & $0.68^{\phz+\phz0.22}_{\phz-\phz0.18}\phs$ & 
        $0.66^{\phz+\phz0.22}_{\phz-\phz0.19}\phs$ &
        $0.40^{\phz+\phz0.13*}_{\phz-\phz0.11}\phs$ & --- &
$0.53\pm0.10\phs$ & \\

199.5 & $0.57^{\phz+\phz0.19}_{\phz-\phz0.17}\phs$ & 
        $0.57^{\phz+\phz0.20}_{\phz-\phz0.17}\phs$ &
        $0.33^{\phz+\phz0.13*}_{\phz-\phz0.11}\phs$ & --- &
$0.47\pm0.10\phs$ & \\

201.6 & $0.82^{\phz+\phz0.33}_{\phz-\phz0.26}\phs$ & 
        $0.19^{\phz+\phz0.21}_{\phz-\phz0.16}\phs$ &
        $0.81^{\phz+\phz0.27*}_{\phz-\phz0.23}\phs$ & --- &
$0.65\pm0.13\phs$ & \\

204.9 & $0.41^{\phz+\phz0.16}_{\phz-\phz0.13}\phs$ & 
        $0.37^{\phz+\phz0.18}_{\phz-\phz0.15}\phs$ & 
        $0.56^{\phz+\phz0.16*}_{\phz-\phz0.14}\phs$ & --- &
$0.46\pm0.10\phs$ & \\

206.6 & $0.66^{\phz+\phz0.16}_{\phz-\phz0.14}\phs$ & 
        $0.68^{\phz+\phz0.16}_{\phz-\phz0.14}\phs$ & 
        $0.59^{\phz+\phz0.12*}_{\phz-\phz0.11}\phs$ & --- &
$0.63\pm0.08\phs$ & \\
\hline
\end{tabular}
\end{center}
\caption{%
  Single-Z hadronic production cross-section from the four LEP
  experiments and combined values for the eight energies between 
  183 and 207~GeV. All results are preliminary with the exception 
  of those indicated by $^*$.}
\label{4f_tab:szxsecqq}
\vspace*{-0.1cm}
\end{table}
\renewcommand{\arraystretch}{1.}

\renewcommand{\arraystretch}{1.2}
\begin{table}[htb]
\begin{center}
\hspace*{-0.0cm}
\begin{tabular}{|c|c|c|c|c|c|} 
\hline
  & \multicolumn{5}{|c|}{Single-Z cross-section into muons(pb)} \\
\cline{2-6} 
    & \Aleph\ & \Delphi\ & \Ltre\ & \Opal\ & LEP  \\
\hline
Av. \roots (GeV) & 196.67 & 197.10 & --- & --- & 196.88   \\
$\sigma_{Zee\rightarrow \mu\mu ee}$ 
& $0.057\pm0.014\phs$ &  $0.070^{\phz+\phz0.023\phz}_{\phz-\phz0.019}$ & --- & --- &
$0.063\pm0.011\phs$  \\
\hline
\end{tabular}
\end{center}
\caption{%
  Preliminary energy averaged single-Z production cross-section into muons 
  from the four LEP experiments and combined values .}
\label{4f_tab:szxsecmm}
\vspace*{-0.1cm}
\end{table}
\renewcommand{\arraystretch}{1.}

\begin{figure}[p]
\centering
\epsfig{figure=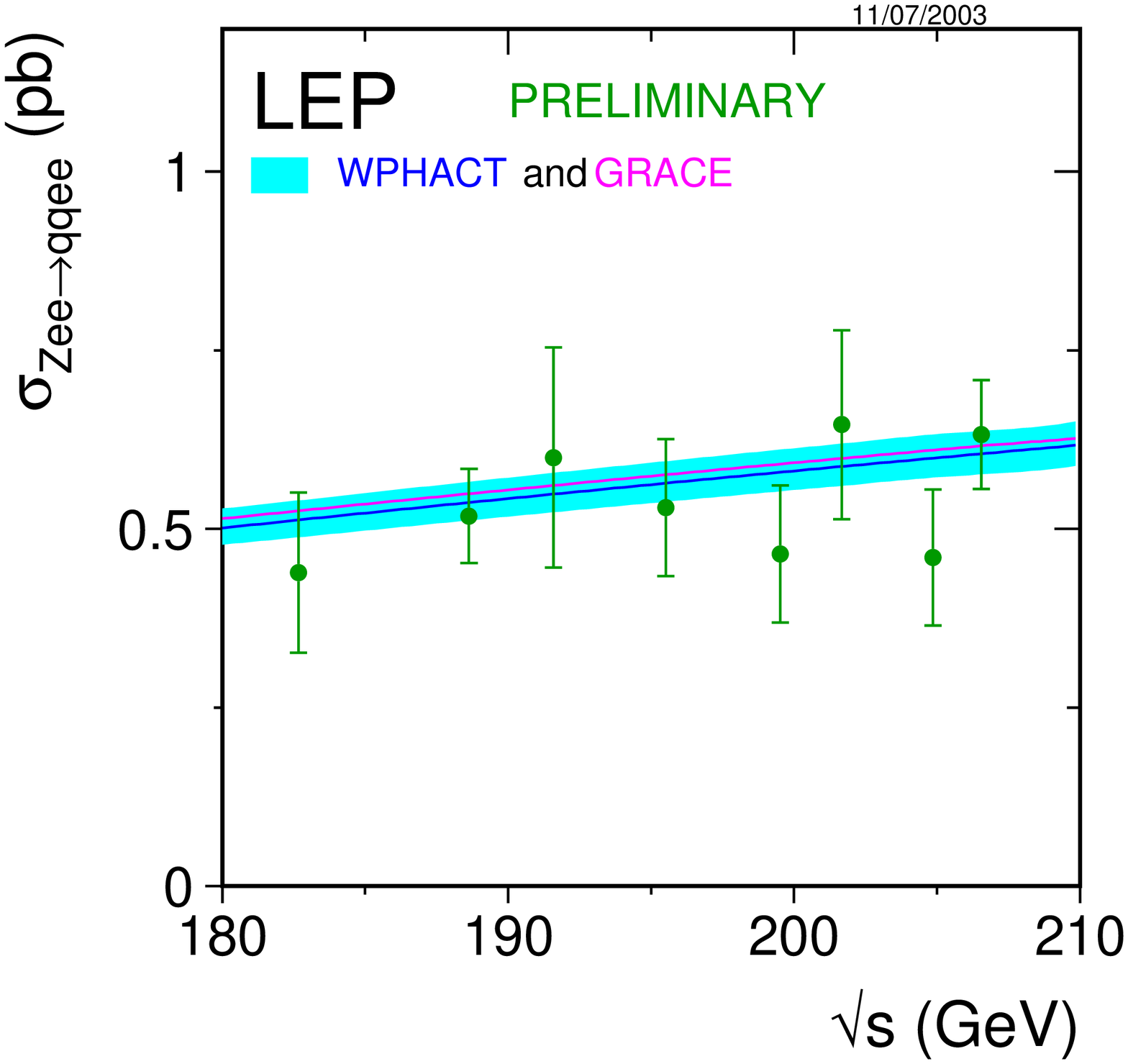,width=0.9\textwidth}
\caption{%
  Measurements of the single-Z hadronic production cross-section,
  compared to the predictions of \WPHACT\ and \Grace. 
  The shaded area represents the $\pm5$\% uncertainty 
  on the predictions.
}
\label{4f_fig:szee}
\end{figure}

\begin{figure}[tp]
\centering
\vspace*{-0.5truecm}
\mbox{
  \fbox{\epsfig{figure=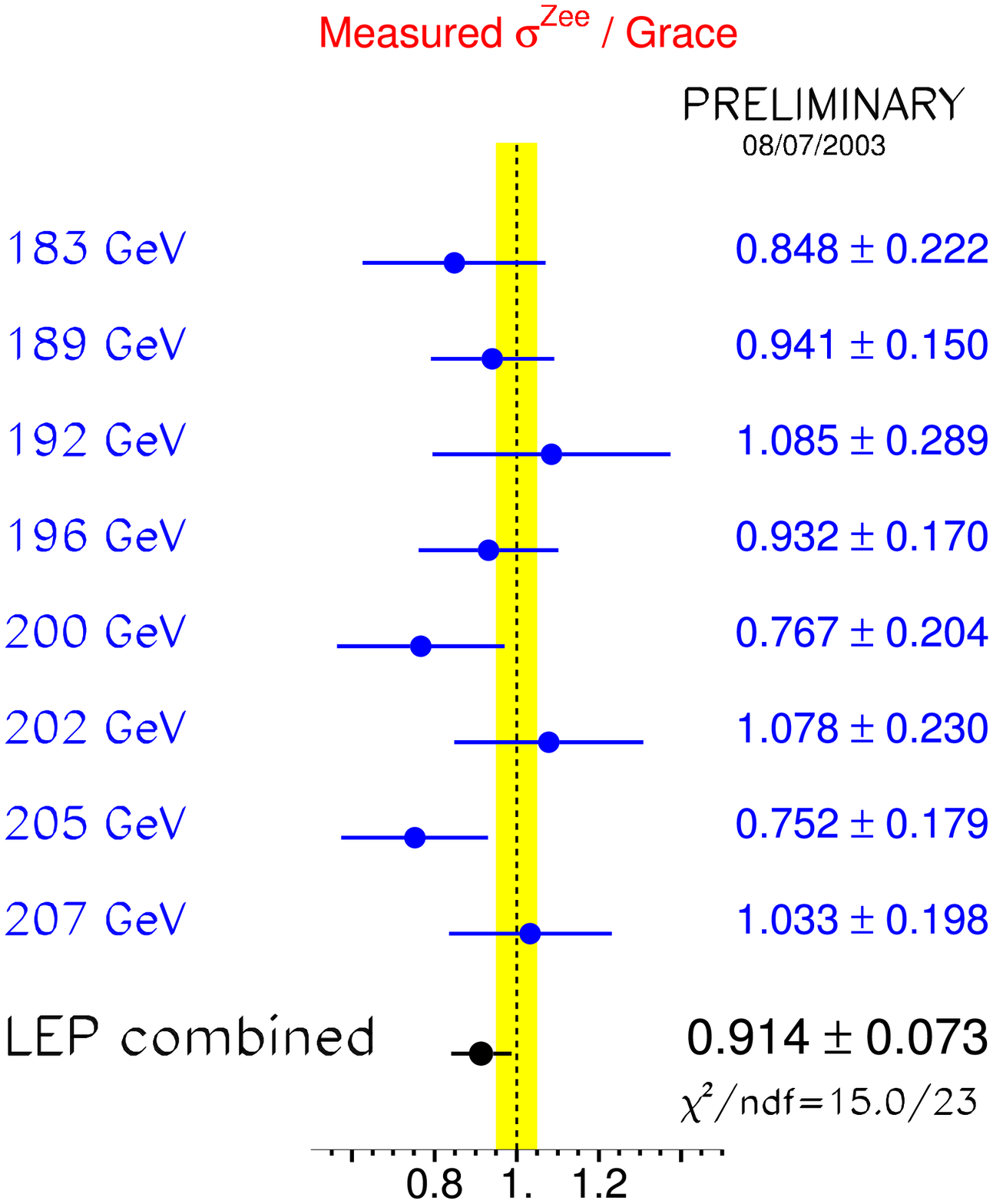,width=0.45\textwidth}}
  \hspace*{0.04\textwidth}
  \fbox{\epsfig{figure=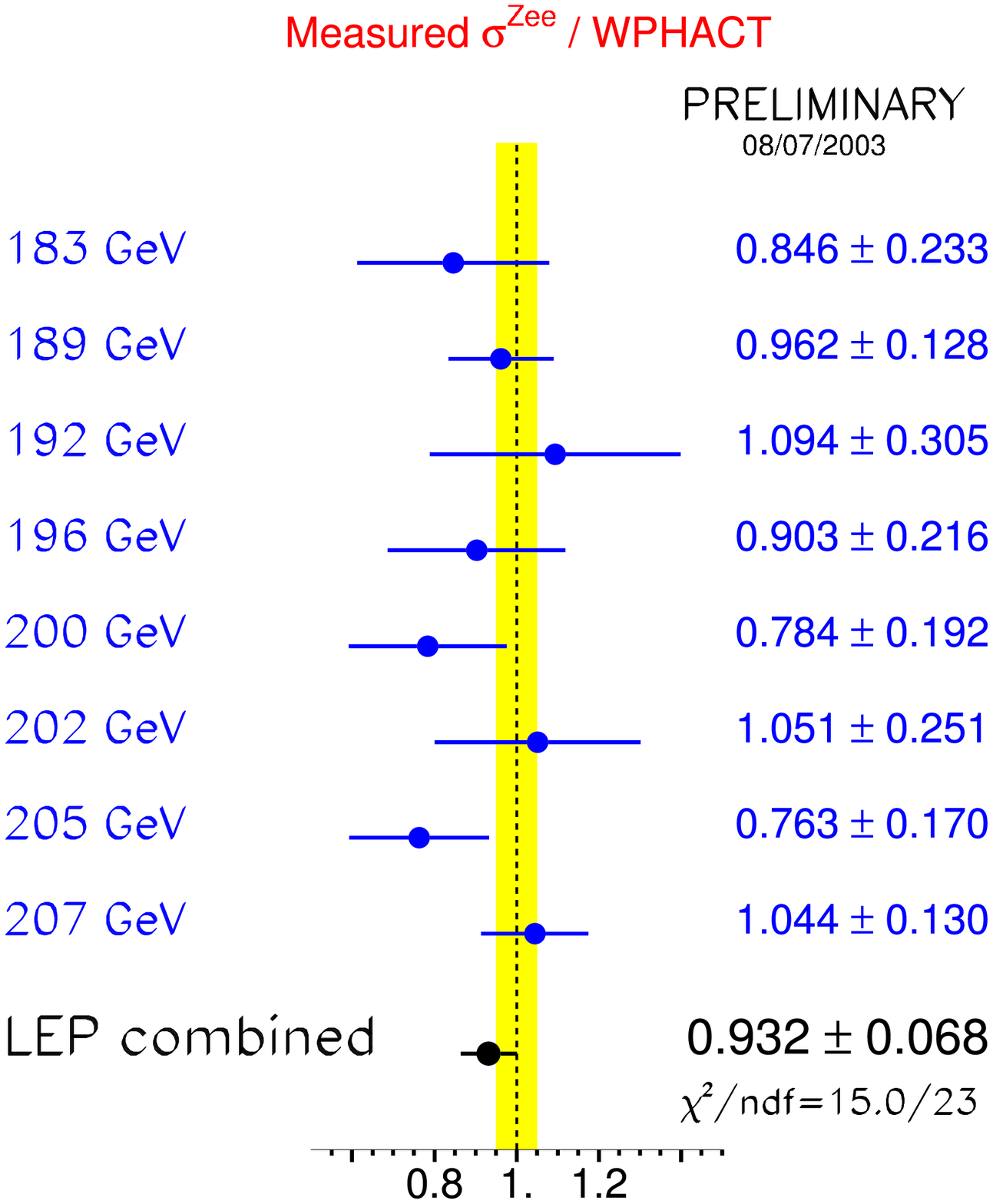,width=0.45\textwidth}}
  }
\vspace*{-0.5truecm}
\caption{%
  Ratios of LEP combined single-Z hadronic cross-section measurements
  to the expectations according to 
  \Grace~\protect\cite{4f_bib:grace} and 
  \WPHACT~\protect\cite{4f_bib:wphact}.
  The yellow bands represent constant relative errors 
  of 5\% on the two cross-section predictions.
}
\label{4f_fig:rzee}
\end{figure}

Tables~\ref{4f_tab:szxsecqq} and~\ref{4f_tab:szxsecmm} synthesize the inputs
by the experiments and the corresponding LEP combinations in the hadronic and
muon channel, respectively. 
The $ee\mu\mu$ cross-section is already combined in energy by the individual
experiments to increase the statistics of the data.
The combination accounts for energy and experiment
correlation of the systematic errors. 
The results in the hadronic channel are compared with the \WPHACT\ and \Grace\
predictions as a function of the \CoM\ energy and shown in figure~\ref{4f_fig:szee}.
Table~\ref{4f_tab:zeeratio} and figure~\ref{4f_fig:rzee} show the preliminary
values of the ratio between measured and expected cross-sections at the various
energy points and the combined value; the testing accuracy of the combined value 
is about 7\% with three experiments contributing in the average.

The detailed breakdown of the inputs of the experiments with the split up of the 
systematic contribution according to the correlations for the single-Z cross-section
and its ratio to theory can be found in Appendix~\ref{4f_sec:appendix}.

\begin{table}[bhtp]
\vspace*{-0mm}
\begin{center}
\hspace*{-0.3cm}
\begin{tabular}{|c|c|c|} 
\hline
\roots (GeV) & $\rzee^{\footnotesize\Grace}$ & $\rzee^{\footnotesize\WPHACT}$ \\
\hline
182.7             & $0.848\pm0.222$ & $0.846\pm0.233$  \\
188.6             & $0.941\pm0.150$ & $0.962\pm0.128$  \\
191.6             & $1.085\pm0.289$ & $1.094\pm0.305$  \\
195.5             & $0.932\pm0.170$ & $0.903\pm0.216$  \\
199.5             & $0.767\pm0.204$ & $0.784\pm0.192$  \\
201.6             & $1.078\pm0.230$ & $1.051\pm0.251$  \\
204.9             & $0.752\pm0.179$ & $0.763\pm0.170$  \\
206.6             & $1.033\pm0.198$ & $1.044\pm0.130$  \\
\hline
$\chi^2$/d.o.f    &  12.2/16          & 12.2/16         \\
\hline
\hline
Average           & $0.914\pm0.073$ & $0.932\pm0.068$  \\
\hline
$\chi^2$/d.o.f    & 15.0/23         & 15.0/23        \\
\hline
\end{tabular}
\caption{%
  Ratios of LEP combined single-Z hadronic cross-section measurements
  to the expectations according to \Grace~\protect\cite{4f_bib:grace}
  and \WPHACT~\protect\cite{4f_bib:wphact}.  The resulting averages
  over energies are also given.  The averages take into account
  inter-experiment as well as inter-energy correlations of systematic
  errors.  }
\label{4f_tab:zeeratio}
\end{center}
\vspace*{-6mm}
\end{table}
\renewcommand{\arraystretch}{1.}

\clearpage

\section{WW$\gamma$ production cross-section}
\label{4f_sec:wwgxsec}

A LEP combination of the WW$\gamma$ production cross-section
has been performed using final \Delphi~\cite{4f_bib:delwwg},
\Ltre~\cite{4f_bib:ltrwwg} and \Opal~\cite{4f_bib:opawwg}
inputs to the Summer 2003 Conferences.
The signal is defined as the part of the WW$\gamma$ process with the following 
cuts to the photon: 
\mbox{$E_{\gamma}>$5~GeV},
\mbox{$|\cos\theta_{\gamma}|<$0.95}, 
\mbox{$|\cos\theta_{\gamma,f}|<$0.90} and
\mbox{$m_W-2\Gamma_W<m_{ff'}<m_W+2\Gamma_W$}
where $\theta_{\gamma,f}$ is the angle between the photon and the closest 
charged fermion and $m_{ff'}$ is the invariant mass of fermions from the Ws.

In order to increase the statistics the LEP combination is performed in energy
intervals rather than at each energy point; they are defined according to the 
LEP2 running period where more statistics was accumulated.
The luminosity weighted \CoM\ per interval is determined in each experiment
and then combined to obtain the corresponding value in the combination.
Table~\ref{4f_tab:wwgxsec} reports those energies and the cross-sections
measured by the experiments, together with the combined LEP values.
With respect to the 2002 Summer Conferences \Opal\ inputs are now used, 
whereas \Delphi\ and \Ltre\ cross-sections are rescaled to better match the 
LEP agreed signal definition. 

\renewcommand{\arraystretch}{1.2}
\begin{table}[hbt]
\begin{center}
\hspace*{-0.0cm}
\begin{tabular}{|c|c|c|c|c|c|} 
\hline
\roots & \multicolumn{5}{|c|}{WW$\gamma$ cross-section (pb)}  \\
\cline{2-6} 
(GeV) & \Aleph\ & \Delphi\ & \Ltre\ & \Opal\ & LEP  \\
\hline
188.6 & --- & $0.05\pm0.08\phs$ & $0.20\pm0.09\phs$ & $0.16\pm0.04\phs$ & $0.15\pm0.03\phs$ \\
194.4 & --- & $0.17\pm0.12\phs$ & $0.17\pm0.10\phs$ & $0.17\pm0.06\phs$ & $0.17\pm0.05\phs$ \\
200.2 & --- & $0.34\pm0.12\phs$ & $0.43\pm0.13\phs$ & $0.21\pm0.06\phs$ & $0.27\pm0.05\phs$ \\
206.1 & --- & $0.18\pm0.08\phs$ & $0.13\pm0.08\phs$ & $0.30\pm0.05\phs$ & $0.24\pm0.04\phs$ \\
\hline
\end{tabular}
\end{center}
\vspace*{-0.3cm}
\caption{%
  WW$\gamma$ production cross-section from the four LEP experiments and combined values for 
  the four energy bins. All results are final.}
\label{4f_tab:wwgxsec}
\end{table}
\renewcommand{\arraystretch}{1.}

\begin{figure}[p]
\centering
\epsfig{figure=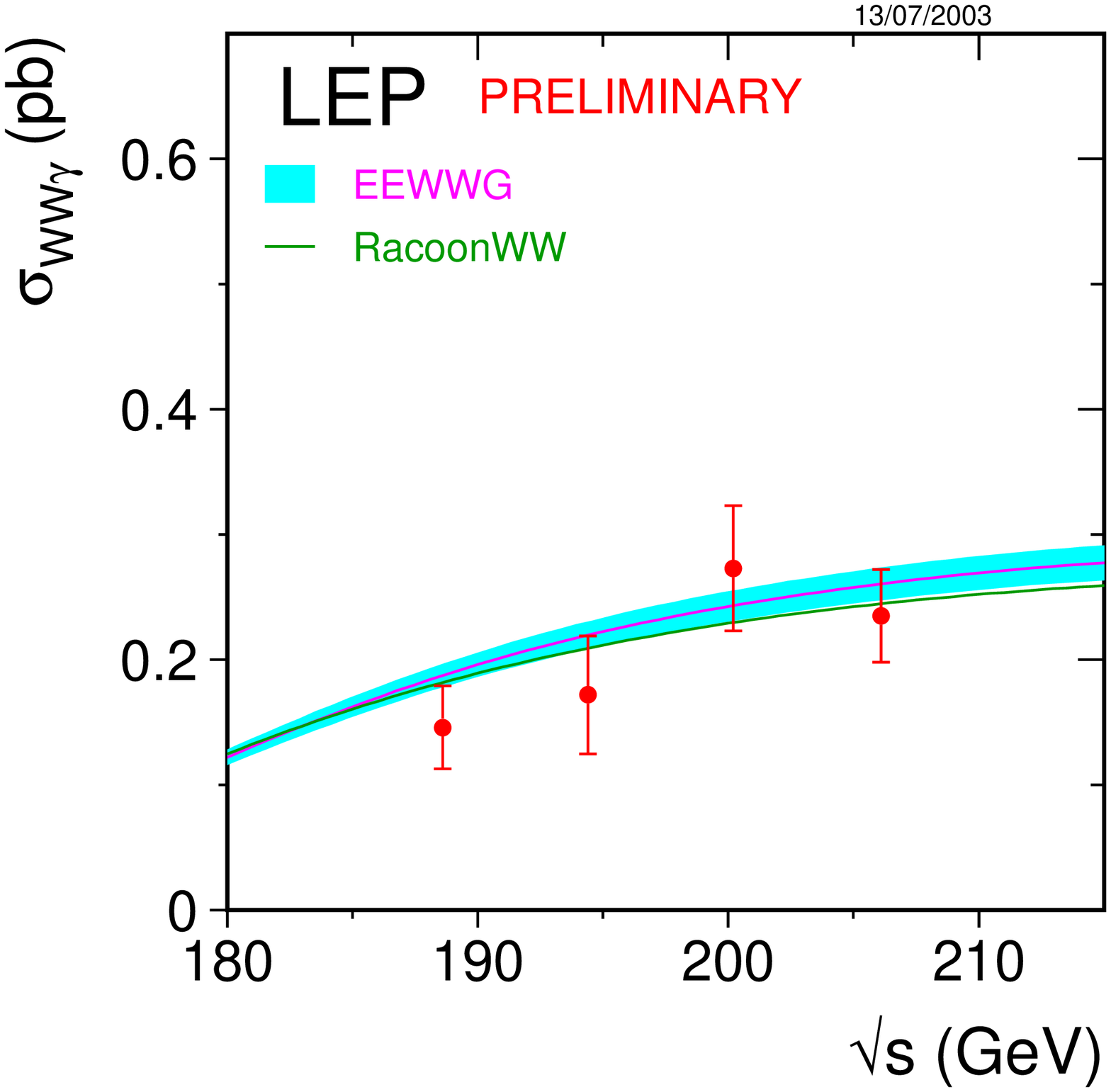,width=0.9\textwidth}
\caption{%
  Measurements of the WW$\gamma$ production cross-section,
  compared to the predictions of \EEWWG~\protect\cite{4f_bib:eewwg} and
  \RacoonWW~\protect\cite{4f_bib:racoonww}. 
  The shaded area in the \EEWWG\ curve represents the $\pm5$\% uncertainty 
  on the predictions.
}
\label{4f_fig:swwg}
\end{figure}

Figure~\ref{4f_fig:swwg} shows the combined data points compared with the
cross-section prediction by \EEWWG~\cite{4f_bib:eewwg} and by 
\RacoonWW. The \RacoonWW\ is shown in the figure without any theory error band.

\section{Summary}
\label{4f_sec:summary}
The updated LEP combinations of the W-pair, Z-pair, single-Z, WW$\gamma$ cross-sections
and the W branching ratios, based on data collected up to 209 GeV by the four LEP experiments,
have been presented. 
All measurements agree with the theoretical predictions.

This note still reflects a preliminary status of the analyses 
at the time of the Summer 2003 Conferences.
A definitive statement on these results and the ones not updated for these Conferences
must wait for publication by each collaboration.
Further work on the possibility of providing a LEP combination of W angular differential
distributions and other cross-sections in the neutral current sector 
(Z$\gamma^*$, Z$\gamma\gamma$) are ongoing.

%% file: gc.tex
\section{Introduction}
\label{sec:gc_introduction}

The measurement of  gauge boson couplings and the
search for possible anomalous contributions due to the effects of new
physics beyond the Standard Model are among the principal physics
aims at \LEPII~\cite{gc_bib:LEP2YR}.
Combined preliminary measurements of triple gauge boson
couplings are presented here. Results from W-pair production are
combined in single and two-parameter fits, including updated results from
ALEPH, L3 and OPAL as well as an improved treatment of the main systematic
effect in our previous combination, the uncertainty in the $O(\alpha_{em})$
correction.     
An updated combination of quartic gauge coupling (QGC) results for the \ZZgg\
vertex is also presented, including data from ALEPH, L3 and OPAL.
The combination of QGCs associated with the \WWgg\ vertex, including the sign
convention as reported 
in~\cite{gc_bib:Montagna:2001ej,gc_bib:denner} and the reweighting based
on~\cite{gc_bib:Montagna:2001ej} is foreseen for our next report. 
The combination of neutral TGCs measured in ZZ production (f-couplings) has
been updated, including new results from L3 and OPAL.
The combinations for neutral TGCs accessible through ${\rm Z} \gamma$
production (h-couplings) reported in 2001 still remain valid ~\cite{gc_bib:budapest01}.

The W-pair production process, $\mathrm{e^+e^-\rightarrow\WW}$,
involves charged triple gauge boson vertices between the $\WW$ and
the Z or photon.  During \LEPII\ operation, about 10,000 W-pair
events were collected by each experiment.  
Single W ($\enw$) and single photon ($\nng$) production at LEP are also
sensitive to the $\WWg$ vertex. Results from these channels are also included
in the combination for some experiments; the individual references should be
consulted for details.

For the charged TGCs, Monte Carlo calculations
(RacoonWW~\cite{common_bib:racoonww} and
YFSWW~\cite{common_bib:yfsww}) incorporating an improved treatment of
$O(\alpha_{em})$ corrections to the WW production have become our standard by
now. The corrections affect the measurements of the charged
TGCs in W-pair production. 
Results, some of them preliminary, including these
$O(\alpha_{em})$ corrections have been submitted from all four LEP
collaborations ALEPH~\cite{gc_bib:ALEPH-cTGC}, DELPHI~\cite{gc_bib:DELPHI-cTGC}, L3~\cite{gc_bib:L3-cTGC} and OPAL~\cite{gc_bib:OPAL-cTGC3}. 
LEP combinations are made for the charged TGC measurements in single- and
two-parameter fits. 

At centre-of-mass energies exceeding twice the Z boson mass, pair
production of Z bosons is kinematically allowed. Here, one searches
for the possible existence of triple vertices involving only neutral
electroweak gauge bosons. Such vertices could also contribute to
Z$\gamma$ production.  In contrast to triple gauge boson vertices with
two charged gauge bosons, purely neutral gauge boson vertices do not
occur in the Standard Model of electroweak interactions.

Within the Standard Model, quartic electroweak gauge boson vertices
with at least two charged gauge bosons exist. In $\ee$ collisions at
\LepII\ centre-of-mass energies, the $\WWZg$ and $\WWgg$ vertices
contribute to $\WWg$ and $\nngg$ production in $s$-channel and
$t$-channel, respectively.  The effect of the Standard Model quartic
electroweak vertices is below the sensitivity of \LepII.  
Quartic gauge boson vertices with only neutral bosons, like the
\ZZgg\ vertex, do not exist in the Standard Model. However, anomalous QGCs
associated with this vertex are studied at LEP.

Anomalous quartic vertices are searched for in the production of
$\WWg$, $\nngg$ and $\Zgg$ final states. The couplings related to the \ZZgg\
and \WWgg\ vertices are assumed to be different~\cite{gc_bib:QGC-Belanger},
and are therefore treated separately.
In this report, we only combine the results for the anomalous couplings
associated with the \ZZgg\ vertex. The combination of the \WWgg\ vertex
couplings is foreseen for the near future.

\subsection{Charged Triple Gauge Boson Couplings}
\label{sec:gc_cTGCs}

The parametrisation of the charged triple gauge boson vertices is
described in References~\cite{gc_bib:GAEMERS,gc_bib:Hagiwara1987vm,
gc_bib:HAGIWARA,gc_bib:BILENKY,gc_bib:KUSS,
gc_bib:PAPADOPOULOSCP,gc_bib:LEP2YR}.  
The most general Lorentz invariant
Lagrangian which describes the triple gauge boson interaction has
fourteen independent complex couplings, seven describing the
WW$\gamma$ vertex and seven describing the WWZ vertex.  Assuming
electromagnetic gauge invariance as well as C and P conservation, the
number of independent TGCs reduces to five.  A common set is \{$\gz,
\kz, \kg, \lz$, $\lg$\} where $\gz = \kz = \kg = 1$ and $\lz = \lg =
0$ in the \SM.  The parameters proposed in~\cite{gc_bib:LEP2YR} and used by
the LEP experiments are $\gz$, $\lg$ and $\kg$ with the gauge constraints:
\begin{eqnarray}
\kz & = & \gz - (\kg - 1) \twsq \,, \\
\lz & = & \lg \,,
\end{eqnarray}                               
where $\theta_W$ is the weak mixing angle.  The
couplings are considered as real, with the imaginary parts fixed to
zero. In contrast to previous LEP
combinations~\cite{gc_bib:moriond01,gc_bib:budapest01}, we are quoting
the measured coupling values themselves and not their deviation from the
Standard Model. 

Note that the photonic couplings $\lg$ and $\kg$ are related to the
magnetic and electric properties of the W-boson. One can write the
lowest order terms for a multipole expansion describing the W-$\gamma$
interaction as a function of $\lg$ and $\kg$. For the magnetic dipole
moment $\mu_{W}$ and the electric quadrupole moment $q_{W}$ one
obtains $e(1+\kappa_{\gamma}+\lambda_{\gamma})/2\MW$ and
$-e(\kappa_{\gamma}-\lambda_{\gamma})/\MW^2$, respectively.

The inclusion of $O(\alpha_{em})$ corrections in the Monte Carlo
calculations has a considerable effect on the charged TGC measurement. Both the
total cross-section and the differential distributions are affected. The  
cross-section is reduced by 1-2\% (depending on the energy). Amongst
the differential distributions, the effects are naturally more complex. The
polar W$^-$ production angle carries most of the information on the TGC
parameters; its 
shape is modified to be more forwardly peaked. In a fit to data, the
$O(\alpha_{em})$ effect
manifests itself as a negative shift of the obtained TGC values with a
magnitude of typically -0.015 for \lg\ and \gz\, and -0.04 for \kg.

\subsection{Neutral Triple Gauge Boson Couplings}
\label{sec:gc_nTGCs}

There are two classes of Lorentz invariant structures associated with
neutral TGC vertices which preserve $U(1)_{em}$ and Bose symmetry, as
described in~\cite{gc_bib:Hagiwara1987vm,gc_bib:Gounaris2000tb}.

The first class refers to anomalous Z$\gamma\gamma^*$ and $\rm Z\gamma
\rm Z^*$ couplings which are accessible at LEP in the process
$\mathrm{e^{+} e^{-}} \rightarrow {\rm Z} \gamma$. The parametrisation
contains eight couplings: $h_i^{V}$ with $i=1,...,4$ and $V=\gamma$,Z.
The superscript $\gamma$ refers to Z$\gamma\gamma^*$ couplings and
superscript Z refers to $\rm Z\gamma \rm Z^*$ couplings.  The photon
and the Z boson in the final state are considered as on-shell
particles, while the third boson at the vertex, the $s$-channel
internal propagator, is off shell.  The couplings $h_{1}^{V}$ and
$h_{2}^{V}$ are CP-odd while $h_{3}^{V}$ and $h_{4}^{V}$ are CP-even.

The second class refers to anomalous $\rm{ZZ}\gamma^*$ and
$\rm{ZZZ}^*$ couplings which are accessible at \LEPII\ in the process
$\mathrm{e^{+} e^{-}} \rightarrow$ ZZ.  This anomalous vertex is
parametrised in terms of four couplings: $f_{i}^{V}$ with $i=4,5$ and
$V=\gamma$,Z.  The superscript $\gamma$ refers to ZZ$\gamma^*$
couplings and the superscript Z refers to $\rm{ZZZ}^*$ couplings,
respectively.  Both Z bosons in the final state are assumed to be
on-shell, while the third boson at the triple vertex, the $s$-channel
internal propagator, is off-shell.
The couplings $f_{4}^{V}$ are CP-odd whereas $f_{5}^{V}$ are CP-even.

The $h_i^{V}$ and $f_{i}^{V}$ couplings are assumed to be real and they
vanish at tree level in the Standard Model.

\subsection{Quartic Gauge Boson Couplings}
\label{sec:gc_QGCs}
The couplings associated with the two QGC vertices \WWgg\ and \ZZgg\ are
assumed to be different, and are by convention treated as separate couplings
at LEP. In this report, we only combine QGCs related to the \ZZgg\ vertex.
The contribution of such anomalous quartic gauge boson couplings is described by two coupling parameters \acl\ and \azl,
which are zero in the Standard
Model~\cite{gc_bib:QGC-BelBou,4f_bib:eewwg}.  
Events from $\nngg$ and $\Zgg$ final states can originate from the \ZZgg\
vertex and are therefore used to study anomalous QGCs.

\section{Measurements}
\label{sec:gc_data}

The combined results presented here are obtained from charged and neutral
electroweak gauge boson coupling measurements, and from quartic gauge boson
couplings measurements as discussed above.  
The individual references should be consulted for details about the data
samples used. 

The charged TGC analyses of ALEPH, DELPHI, L3 and OPAL use data collected
at \LEPII\ up to centre-of-mass energies of 209~\GeV. These
analyses use different
channels, typically the semileptonic and fully hadronic W-pair
decays~\cite{gc_bib:ALEPH-cTGC,gc_bib:DELPHI-cTGC,
  gc_bib:L3-cTGC,gc_bib:OPAL-cTGC3}. 
The full data set is analysed by ALEPH, L3 and OPAL, whereas DELPHI presently
uses all data at 189 $\GeV$ and above. 
Anomalous
TGCs affect both the total production cross-section and the shape of
the differential cross-section as a function of the polar W$^-$
production angle.  The relative contributions of each helicity state
of the W bosons are also changed, which in turn affects the
distributions of their decay products.  The analyses presented by each
experiment make use of different combinations of each of these
quantities.  In general, however, all analyses use at least the
expected variations of the total production cross-section and the
W$^-$ production angle. Results from $\enw$ and $\nng$ production are
included by some 
experiments.  Single W production is particularly sensitive to \kg,
thus providing information complementary to that from W-pair
production.

The $h$-coupling analyses of ALEPH, DELPHI and L3 use data
collected up to centre-of-mass energies of 209~\GeV. The
OPAL measurements so far use the data at 189~\GeV.  The results of the
$f$-couplings are obtained from the whole data set above the
ZZ-production threshold by all of the experiments.  The experiments
already pre-combine different processes and final states for each of
the couplings.  For the neutral TGCs, the analyses use measurements of
the total cross sections of Z$\gamma$ and ZZ production and the
differential distributions: the $h_i^V$
couplings~\cite{gc_bib:ALEPH-nTGC,gc_bib:DELPHI-nTGC,gc_bib:L3-hTGC,
  gc_bib:OPAL-hTGC} and the $f_i^V$
couplings~\cite{gc_bib:ALEPH-nTGC,gc_bib:DELPHI-nTGC,gc_bib:L3-fTGC,
  gc_bib:OPAL-fTGC} are determined.

The combination of quartic gauge boson couplings associated with
the \ZZgg\ vertex is at present based on analyses of
ALEPH~\cite{gc_bib:ALEPH-QGC}, L3~\cite{gc_bib:L3-QGC} 
and OPAL~\cite{gc_bib:OPAL-QGC}.
The L3 analysis uses data from the \qqgg\
final state all at centre-of-mass energies above the Z resonance, from 130
GeV to 207 GeV. 
Both ALEPH and OPAL analyse the $\nngg$ final state, with ALEPH using data
from centre-of-mass energies ranging from 183 GeV to 209 GeV, and OPAL from
189 GeV to 209 GeV.  

\section{Combination Procedure}
\label{sec:gc_combination}

The combination is based on the individual likelihood functions from the four
LEP experiments.
Each experiment provides the negative log likelihood, $\LL$, as a
function of the coupling parameters to be combined.  
The single-parameter analyses are performed fixing 
all other parameters to their Standard Model values.  The
two-parameter analyses are performed setting the remaining
parameters to their Standard Model values. For the charged TGCs, the
gauge constraints listed in Section~\ref{sec:gc_cTGCs} are always
enforced.

The $\LL$ functions from each experiment include statistical as well
as those systematic uncertainties which are considered as
uncorrelated between experiments.  For both single- and
multi-parameter combinations, the individual $\LL$ functions are
combined.  It is necessary to use the $\LL$ functions directly in the
combination, since in some cases they are not parabolic, and hence it is not
possible to properly combine the results by simply taking weighted
averages of the measurements.

The main contributions to the systematic uncertainties that are
uncorrelated between experiments arise from detector effects,
background in the selected signal samples, limited Monte Carlo
statistics and the fitting method.  Their importance varies for each
experiment and the individual references should be consulted for
details.

In the neutral TGC sector, the systematic uncertainties arising from the
theoretical cross 
section prediction in Z$\gamma$-production ($\simeq 1\%$ in the
$\qq\gamma$- and $\simeq 2\%$ in the $\nng$ channel) are treated as
correlated.
For ZZ production, the uncertainty on the theoretical cross section
prediction is small compared to the statistical accuracy and therefore
is neglected.  Smaller sources of correlated systematic uncertainties,
such as those arising from the LEP beam energy, are for simplicity
treated as uncorrelated.

The combination procedure for neutral TGCs, where the relative
systematic uncertainties are small, is unchanged with respect to the
previous LEP combinations of electroweak gauge boson
couplings~\cite{gc_bib:moriond01,gc_bib:budapest01}. 
The correlated systematic uncertainties in the $h$-coupling analyses
are taken into account by scaling the combined log-likelihood
functions by the squared ratio of the sum of statistical and
uncorrelated systematic uncertainty over the total uncertainty
including all correlated uncertainties.  For the general case of
non-Gaussian probability density functions, this treatment of the
correlated errors is only an approximation; it also neglects
correlations in the systematic uncertainties between the parameters in
multi-parameter analyses.

In the charged TGC sector, systematic uncertainties considered
correlated between the experiments are the theoretical cross
section prediction ($0.5\%$ for W-pair production and $5\%$ for single W
production), hadronisation effects, the final
state interactions, namely Bose-Einstein correlations and colour reconnection,
and the uncertainty in the radiative corrections themselves. 
The latter was the dominant systematic error in our previous combination,
where we used a conservative estimate, the full effect from applying the
$O(\alpha_{em})$ corrections. 
New preliminary analyses on the subject are now available from several LEP
experiments~\cite{gc_bib:ALEPH-cTGC}, based on comparisons 
of fully simulated events using two different leading-pole approximation
schemes (LPA-A and LPA-B)~\cite{gc_bib:LPA_A-B}.
In addition, the availability of comparisons of both generators
incorporating $O(\alpha_{em})$ corrections (RacoonWW and
YFSWW~\cite{common_bib:racoonww,common_bib:yfsww}) makes 
it now possible to perform a more realistic estimation of this effect. 
In general, the TGC shift
measured in the comparison of the two generators is found to be larger than
the effect from the different LPA schemes.
This improved
estimation, whilst still being conservative, reduces the systematic
uncertainty from $O(\alpha_{em})$ corrections by about a
third for $\gz$ and $\lg$ and roughly halves it for $\kg$, compared to the
full $O(\alpha_{em})$ correction applied previously. The application of this
reduced systematic error renders the charged TGC measurements statistics
dominated. 

In case of the charged TGCs, the systematic uncertainties considered
correlated between the experiments amount to 58\% of the combined
statistical and uncorrelated uncertainties for $\lg$ and $\gz$, while for
$\kg$ it is 68\%.  This means that the measurements of $\lg$, $\gz$ and $\kg$
are now clearly limited by statistics.
An improved combination
procedure~\cite{gc_bib:Alcaraz} is used for the charged TGCs.  
This procedure allows the combination of statistical and correlated
systematic uncertainties, independently of the analysis method chosen by
the individual experiments. 

The combination of charged TGCs uses the likelihood curves and
correlated systematic errors submitted by each of the four experiments. 
The procedure is based on the introduction of an additional free parameter to
take into account the systematic uncertainties, which are treated as shifts on
the fitted TGC value, and are assumed to have a Gaussian distribution. 
A simultaneous minimisation of both parameters (TGC
and systematic error) is performed to the log-likelihood function. 

In detail, the combination proceeds in the following way: the set of
measurements from the LEP experiments
ALEPH, DELPHI, OPAL and L3 is given with statistical plus uncorrelated
systematic uncertainties in terms of likelihood curves:  
$-\log{\mathcal L}^A_{stat}(x)$,
$-\log{\mathcal L}^D_{stat}(x)$ 
$-\log{\mathcal L}^L_{stat}(x)$ 
and $-\log{\mathcal L}^O_{stat}(x)$, 
respectively, where $x$ is the coupling parameter in question. 
Also given are the shifts for each of the five totally correlated sources
of uncertainty mentioned above; each source $S$ is 
leading to systematic errors $\sigma^S_A$, $\sigma^S_D$, $\sigma^S_L$ and
$\sigma^S_O$.

Additional parameters $\Delta^S$ are included in order to take into 
account a Gaussian distribution for each of the systematic uncertainties.
The procedure then consists in minimising the function:
\noindent
\begin{eqnarray}
-\log {\mathcal L}_{total} = 
\sum_{E=A,D,L,O} \log {\mathcal L}^E_{stat} 
(x-\sum_{S=DPA,\sigma_{WW},HAD,BE,CR}(\sigma^S_E \Delta^S))
 + \sum_{S} {\displaystyle \frac{(\Delta^S)^{2}}{2}} \\ \nonumber
\end{eqnarray}

\noindent
where $x$ and $\Delta_S$ are the free parameters, and the sums run over the
four experiments and the five systematic errors.
The resulting uncertainty on $x$ will take into account all sources 
of uncertainty, yielding a measurement of the coupling with the error
representing statistical and systematic sources.
The projection of the minima of the log-likelihood as a function of $x$ 
gives the combined log-likelihood curve including statistical and
systematic uncertainties. 
The advantage over the scaling method used previously is that it
treats systematic uncertainties that are correlated between the experiments
correctly, while not forcing the averaging of these systematic
uncertainties into one global LEP systematics scaling factor. In other words,
the (statistical) precision of each experiment now gets reduced by its own
correlated systematic errors, instead of an averaged LEP systematic
error. 
The method has been cross-checked against the scaling method, and was found
to give comparable results. 
The inclusion of
the systematic uncertainties lead to small differences as expected by the
improved treatment of correlated systematic errors, a similar behaviour as
seen in Monte Carlo comparisons of these two combinations methods
~\cite{gc_bib:renaud}. Furthermore, it was shown that the minimisation-based
combination method used for the charged TGCs agrees with the method
based on optimal observables, where systematic effects are included directly
in the mean values of the optimal observables (see~\cite{gc_bib:renaud}), 
for any realistic ratio of statistical and systematic uncertainties. 
Further details on the improved combination method can be found
in~\cite{gc_bib:Alcaraz}. 

In the combination of the QGCs, the influence of correlated systematic
uncertainties is considered negligible compared to the statistical error,
arising from the small number of selected events. Therefore, the QGCs are
combined by adding the log-likelihood curves from the single experiments. 

For all single- and multi-parameter results quoted in numerical form,
the one standard deviation uncertainties (68\% confidence level) are
obtained by taking the coupling values for which $\Delta\LL=+0.5$
above the minimum.  The 95\% confidence level (C.L.)  limits are given
by the coupling values for which $\Delta\LL=+1.92$ above the minimum.
Note that in the case of the neutral TGCs, double minima structures
appear in the negative log-likelihood curves.  For multi-parameter
analyses, the two dimensional 68\%~C.L.~contour curves for any pair of
couplings are obtained by requiring $\Delta\LL=+1.15$, while for the
95\% C.L.~contour curves $\Delta\LL=+3.0$ is required.  Since the
results on the different parameters and parameter sets are obtained
from the same data sets, they cannot be combined.

\section{Results}

We present results from the four LEP experiments on the various
electroweak gauge boson couplings, and their combination.
The charged TGC combination has been updated with the inclusion of recent
results from ALEPH, L3 and OPAL.
The neutral TGC results include an update of the $f_i^V$ combinations,
whilst the $h_i^V$ combinations remain unchanged since
our last note~\cite{gc_bib:budapest01}.  The results
quoted for each individual experiment are calculated using the methods
described in Section~\ref{sec:gc_combination}.  Therefore they may differ
slightly from those reported in the individual references, as the experiments
in general use other methods to combine the data from different channels, and
to include systematic uncertainties. 
In particular for the charged couplings, experiments using a
combination method based on optimal observables (ALEPH, OPAL) 
obtain results with small differences
compared to the values given by our combination technique. These small
differences have been studied in Monte Carlo tests and are well
understood~\cite{gc_bib:renaud}.  
For the $h$-coupling result from OPAL and DELPHI, a slightly
modified estimate of the systematic uncertainty due to the theoretical
cross section prediction is responsible for slightly different limits
compared to the published results.

\subsection{Charged Triple Gauge Boson Couplings}

The individual analyses and results of the experiments for the charged couplings are described in~\cite{gc_bib:ALEPH-cTGC,gc_bib:DELPHI-cTGC,
gc_bib:L3-cTGC,gc_bib:OPAL-cTGC3}.

\subsubsection*{Single-Parameter Analyses}
\begin{figure}[htbp]
\begin{center}
\includegraphics[width=\linewidth]{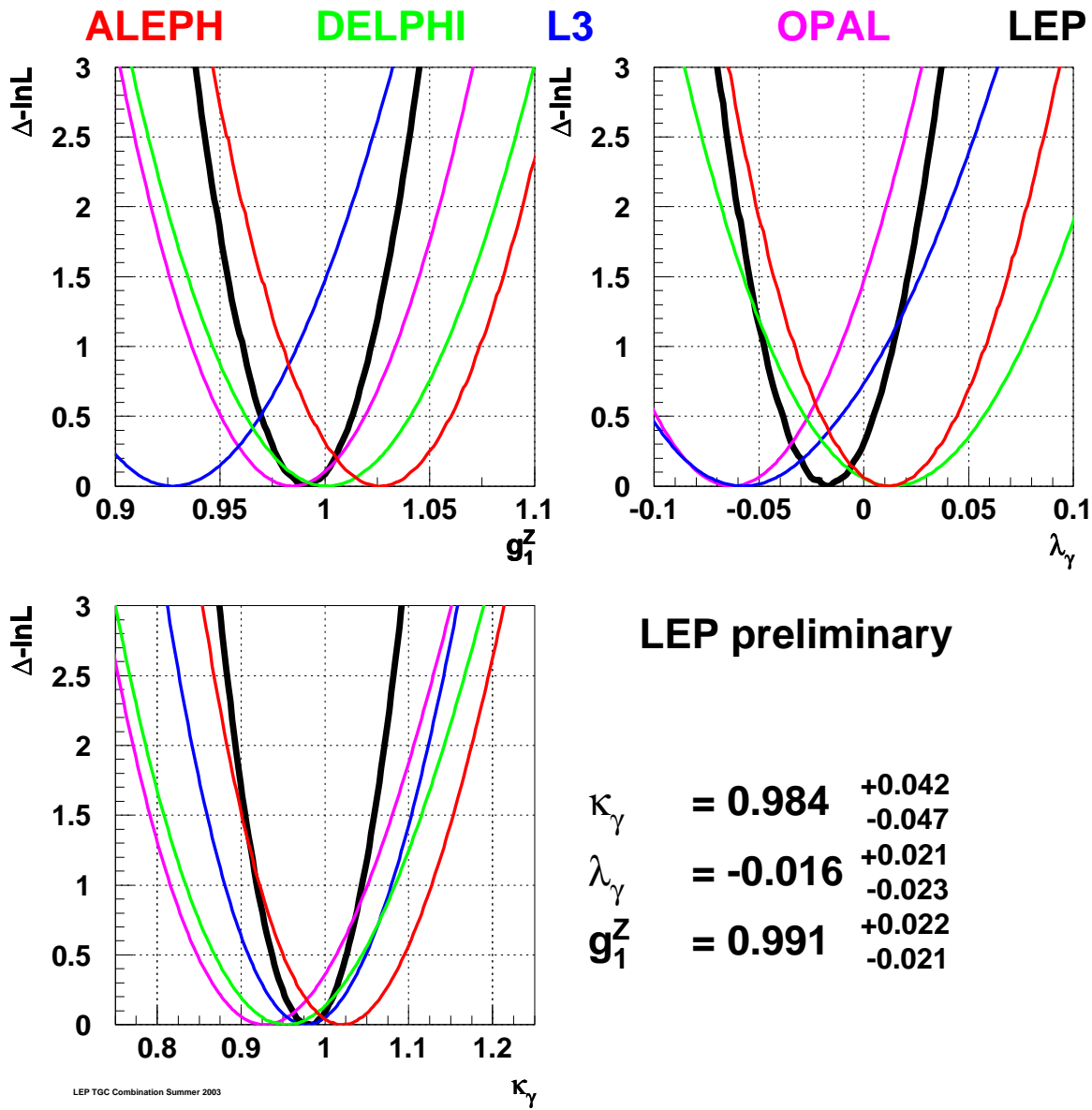}
\caption[]{
  The $\LL$ curves of the four experiments (thin lines) and the LEP
  combined curve (thick line) for the three charged TGCs $\gz$,
  $\kg$ and $\lg$.  In each case, the minimal value is subtracted.  }
\label{fig:cTGC-1}
\end{center}
\end{figure}

The results of single-parameter fits from each experiment are shown in
Table~\ref{tab:cTGC-1-ADLO}, where the errors include both statistical
and systematic effects. The individual $\LL$ curves and their sum are shown in
Figure~\ref{fig:cTGC-1}.  The results of the combination are given in
Table~\ref{tab:cTGC-1-LEP}. A list of the systematic errors treated as fully
correlated between the LEP experiments, and their shift on the combined fit
result are given in Table ~\ref{tab:cTGC-syst}.

\subsubsection*{Two-Parameter Analyses}
\begin{figure}[htbp]
\begin{center}
\includegraphics[width=\linewidth]{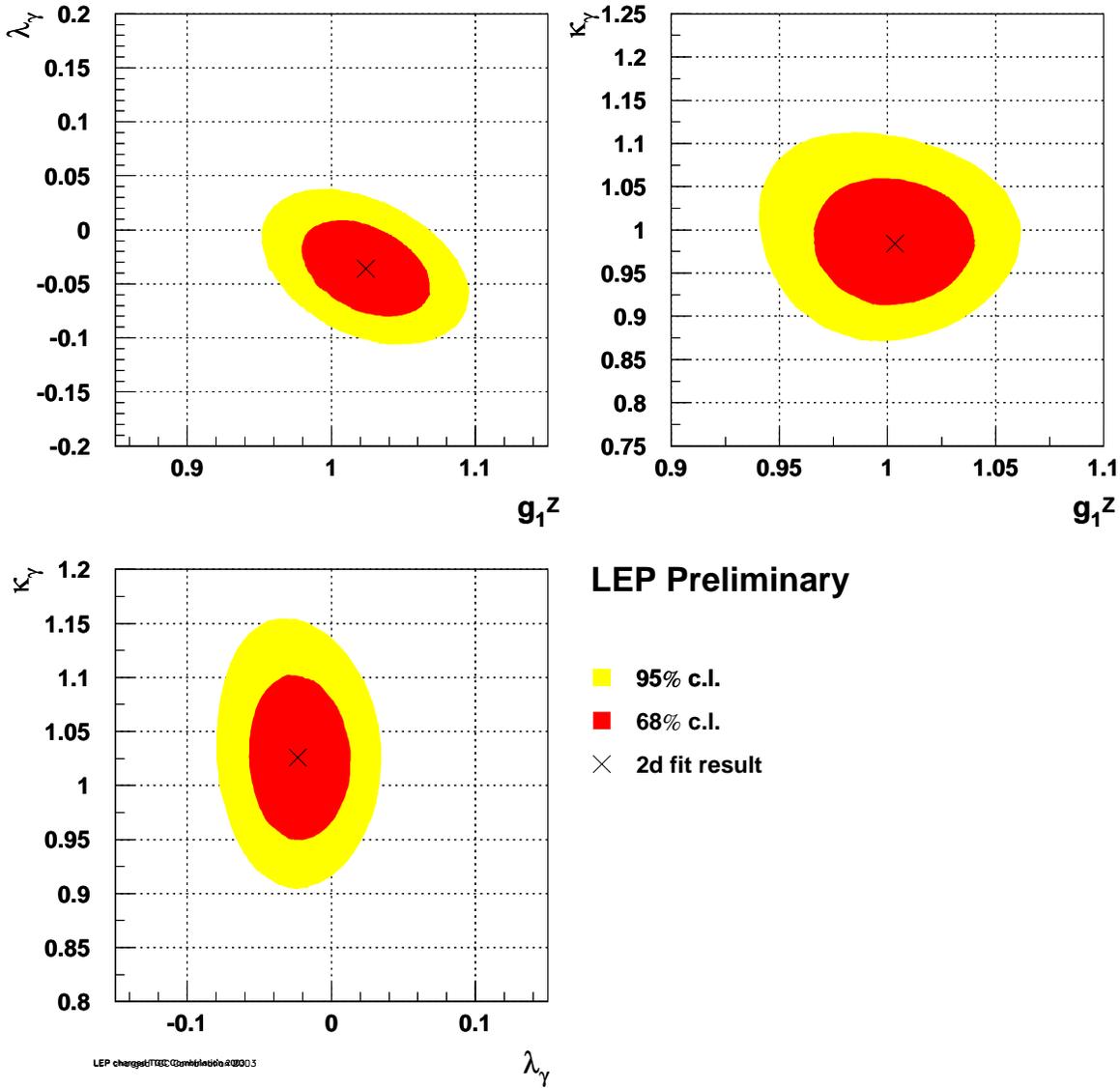}
\caption[]{
The 68\% and 95\% confidence level contours for the three two-parameter fits
to the charged TGCs $\gz$-$\lg$, $\gz$-$\kg$ and $\lg$-$\kg$. 
The fitted coupling
value is indicated with a cross; the Standard Model value for each fit is in
the centre of the grid. The contours include the contribution from systematic
uncertainties.}
 \label{fig:cTGC-2}
\end{center}
\end{figure}

Contours at 68\% and 95\% confidence level for the combined two-parameter
fits are shown in Figure~\ref{fig:cTGC-2}. The numerical results of the
combination are given in Table~\ref{tab:cTGC-2D-LEP}. The errors include both statistical and systematic effects. 

\begin{table}[htbp]
\begin{center}
\renewcommand{\arraystretch}{1.3}
\begin{tabular}{|l||r|r|r|r|} 
\hline
Parameter  & ALEPH   & DELPHI  &  L3   & OPAL  \\
\hline
\hline
\gz       & $1.026\apm{0.034}{0.033}$ & $1.002\apm{0.038}{0.040}$ 
           & $0.928\apm{0.042}{0.041}$ & $0.985\apm{0.035}{0.034}$ \\ 
\hline
\kg       & $1.022\apm{0.073}{0.072}$ & $0.955\apm{0.090}{0.086}$  
           & $0.922\apm{0.071}{0.069}$ & $0.929\apm{0.085}{0.081}$ \\  
\hline
\lg        & $0.012\apm{0.033}{0.032}$ & $0.014\apm{0.044}{0.042}$  
           & $-0.058\apm{0.047}{0.044}$ & $-0.063\apm{0.036}{0.036}$ \\
\hline
\end{tabular}
\caption[]{The measured central values and one standard deviation errors
  obtained by the four LEP experiments.  In each case the parameter
  listed is varied while the remaining two are fixed to their Standard Model
  values. Both
  statistical and systematic errors are included. The values given here
  differ slightly from the ones quoted in the individual contributions from
  the four LEP experiments, as a different combination method is used. See
  text in section \ref{sec:gc_combination} for details. }
\label{tab:cTGC-1-ADLO}
\end{center}
\end{table}

\begin{table}[htbp]
\begin{center}
\renewcommand{\arraystretch}{1.3}
\begin{tabular}{|l||r|c|} 
\hline
Parameter  & 68\% C.L.   & 95\% C.L.      \\
\hline
\hline
$\gz$     & $0.991\apm{0.022}{0.021} $  & [$0.949,~~1.034$]  \\ 
\hline
$\kg$     & $0.984\apm{0.042}{0.047}$  & [$0.895,~~1.069$]  \\ 
\hline
$\lg$     & $-0.016\apm{0.021}{0.023}$  & [$-0.059,~~0.026$]  \\ 
\hline
\end{tabular}
\caption[]{ The combined 68\% C.L. errors and 95\% C.L. intervals
  obtained combining the results from the four LEP experiments.  In
  each case the parameter listed is varied while the other two are
  fixed to their Standard Model values.  Both statistical and systematic
  errors are included.  }
 \label{tab:cTGC-1-LEP}
\end{center}
\end{table}

\begin{table}[htbp]
\begin{center}
\renewcommand{\arraystretch}{1.3}
\begin{tabular}{|l||r|r|r|} 
\hline
Source  & \gz  & \lg   & \kg  \\
\hline
\hline
$O(\alpha_{em})$ correction  & 0.010 & 0.010  & 0.020 \\ 
$\sigma_{WW}$ prediction   & 0.003  & 0.005 & 0.014 \\ 
Hadronisation   & 0.004 & 0.002 & 0.004 \\ 
Bose-Einstein Correlation   & 0.005  & 0.004  & 0.009 \\ 
Colour Reconnection   & 0.005 & 0.004 & 0.010 \\
$\sigma_{single W}$ prediction  & - & - &  0.011 \\ 
\hline
\end{tabular}
\caption[]{The systematic uncertainties considered correlated between the LEP
  experiments in the charged TGC combination and their effect on the combined
  fit results.}
\label{tab:cTGC-syst}
\end{center}
\end{table}

\begin{table}[htbp]
\begin{center}
\begin{tabular}{|l||r|r|rr|} \hline
Parameter   & 68\% C.L.  & 95\% C.L.   & \multicolumn{2}{|c|}{Correlations}  \\
\hline \hline
\gz       & $1.004\apm{0.024}{0.025}$ & [$+0.954,~~+1.050$]
  & 1.00 & +0.11 \\  
\kg       & $0.984\apm{0.049}{0.049}$ & [$+0.894,~~+1.084$]  
 & +0.11 & 1.00 \\ 
\hline
\gz       & $1.024\apm{0.029}{0.029}$ & [$+0.966,~~+1.081$] & 1.00
  & -0.40  \\  
\lg       & $-0.036\apm{0.029}{0.029}$ & [$-0.093,~~+0.022$] 
 & -0.40  & 1.00 \\ 
\hline
\kg        & $1.026\apm{0.048}{0.051}$ & [$+0.928,~~+1.127$]
 & 1.00  & +0.21 \\ 
\lg        & $-0.024\apm{0.025}{0.021}$ & [$-0.068,~~+0.023$] 
  & +0.21 & 1.00 \\ 
\hline
\end{tabular}
\end{center}
\caption{ The measured central values, one standard deviation errors and
  limits at 95\% confidence level, 
    obtained by combining the four LEP experiments for the 
    two-parameter fits of the charged TGC parameters.
    Since the shape of the log-likelihood
  is not parabolic, there is some ambiguity in the definition of the
  correlation coefficients and the values quoted here are approximate.
    The listed parameters are varied while the
    remaining one is fixed to its Standard Model value.
    Both statistical and systematic errors are included.
  }
\label{tab:cTGC-2D-LEP}
\end{table}

\clearpage

\clearpage

\subsection{Neutral Triple Gauge Boson Couplings in Z\boldmath$\gamma$ 
Production}

The individual analyses and results of the experiments for the $h$-couplings 
are described in~\cite{gc_bib:ALEPH-nTGC,gc_bib:DELPHI-nTGC,
gc_bib:L3-hTGC,gc_bib:OPAL-hTGC}.

\subsubsection*{Single-Parameter Analyses}
\begin{figure}[p]
\begin{center}
\includegraphics[width=\linewidth]{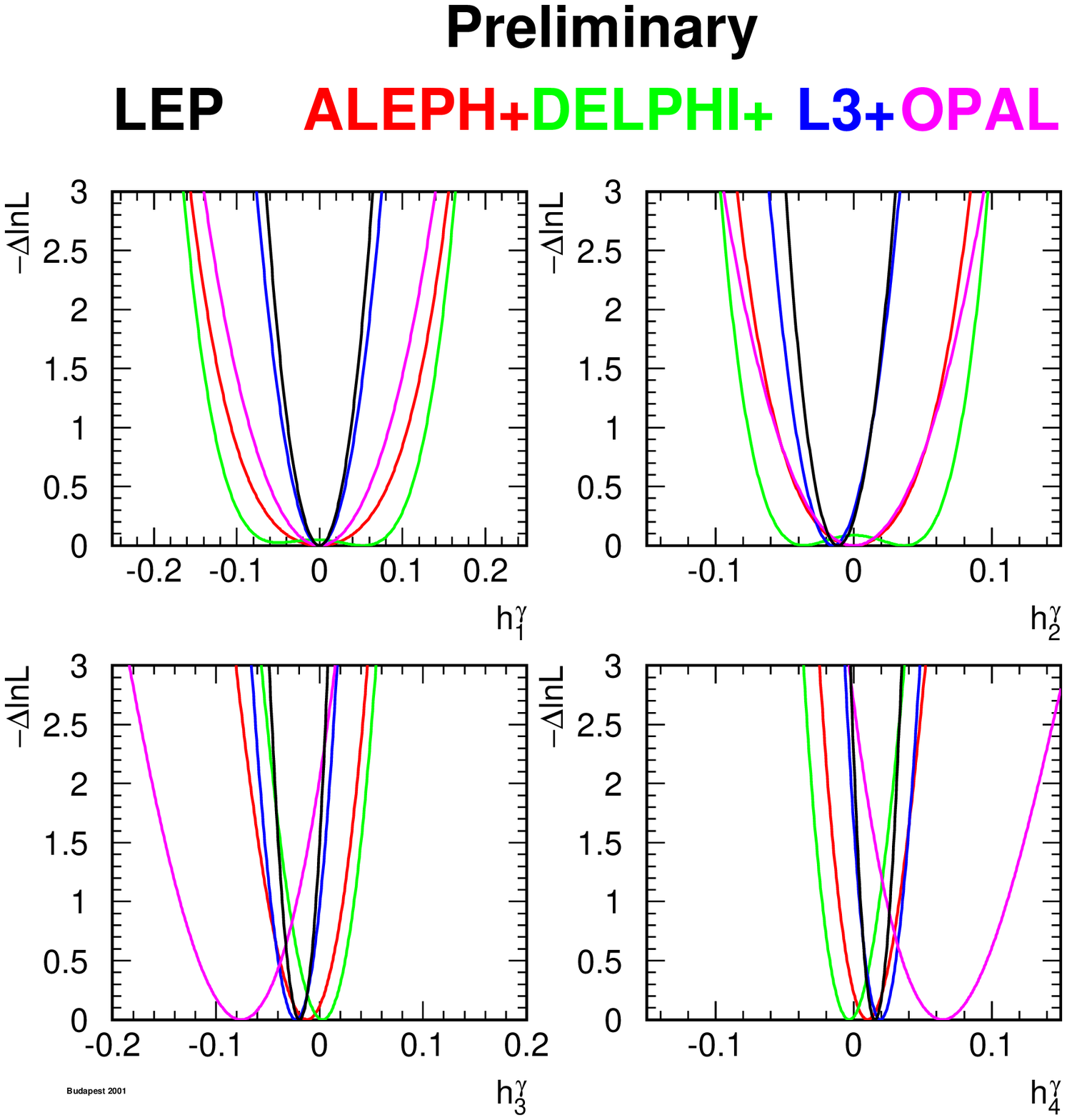}
\caption[]{
  The $\LL$ curves of the four experiments, and the LEP combined curve
  for the four neutral TGCs $h_i^\gamma,~i=1,2,3,4$. In each case, the
  minimal value is subtracted.  }
\label{fig:gc_hgTGC-1}
\end{center}
\end{figure}

\begin{figure}[p]
\begin{center}
\includegraphics[width=\linewidth]{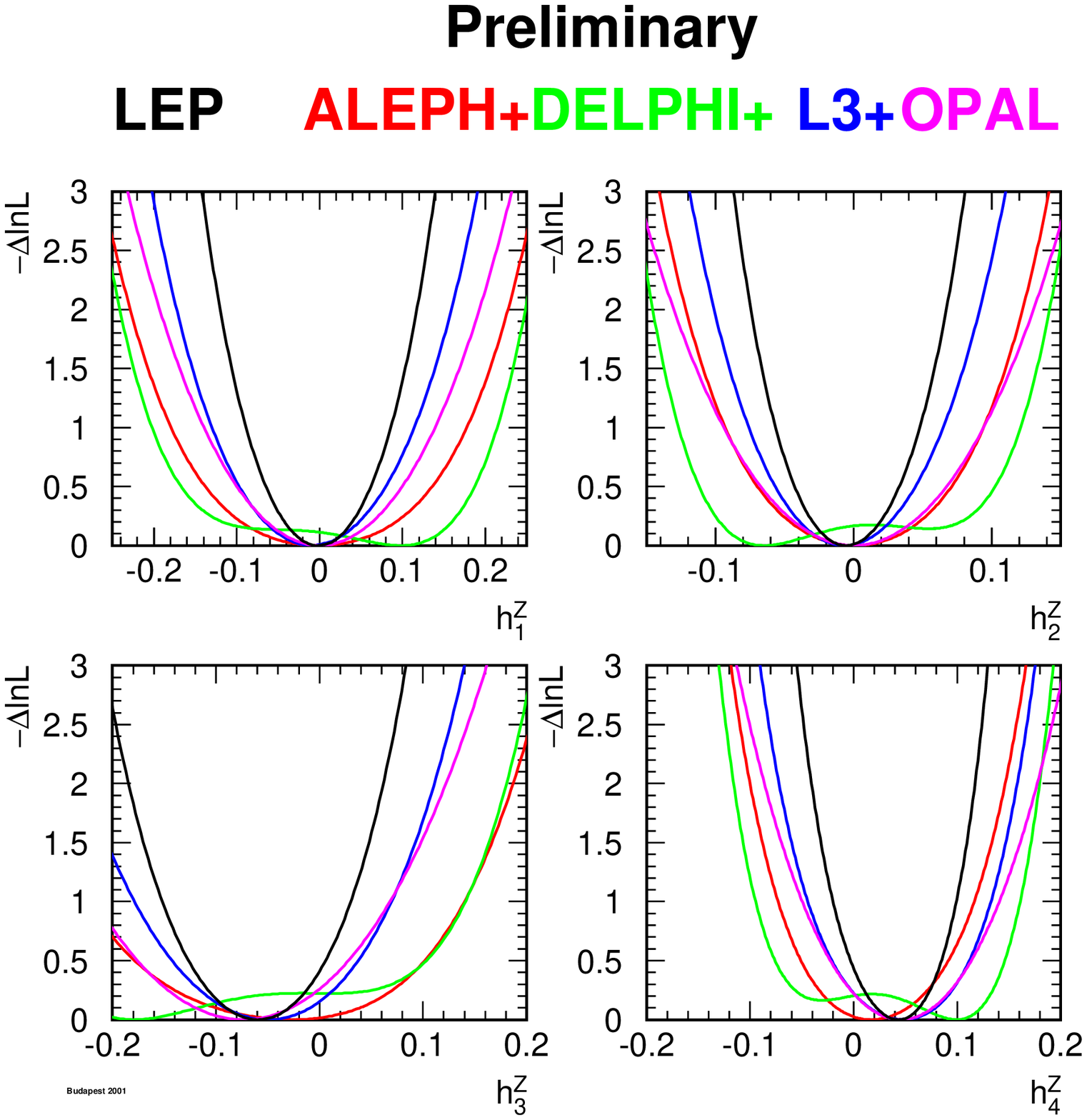}
\caption[]{
  The $\LL$ curves of the four experiments, and the LEP combined curve
  for the four neutral TGCs $h_i^Z,~i=1,2,3,4$.  In each case, the
  minimal value is subtracted.  }
\label{fig:gc_hzTGC-1}
\end{center}
\end{figure}

The results for each experiment are shown in
Table~\ref{tab:gc_hTGC-1-ADLO}, where the errors include both
statistical and systematic uncertainties.  The individual $\LL$ curves
and their sum are shown in Figures~\ref{fig:gc_hgTGC-1}
and~\ref{fig:gc_hzTGC-1}.  The results of the combination are given in
Table~\ref{tab:gc_hTGC-1-LEP}.  From
Figures~\ref{fig:gc_hgTGC-1} and \ref{fig:gc_hzTGC-1} it is clear that the
sensitivity of the L3 analysis~\cite{gc_bib:L3-hTGC} is the highest
amongst the LEP experiments. This is partially due to the use of a larger
phase space region, which increases the statistics by about a factor two,
and partially due to additional information from using an optimal-observable
technique.

\subsubsection*{Two-Parameter Analyses}

\begin{figure}[p]
\begin{center}
\includegraphics[width=0.49\linewidth]{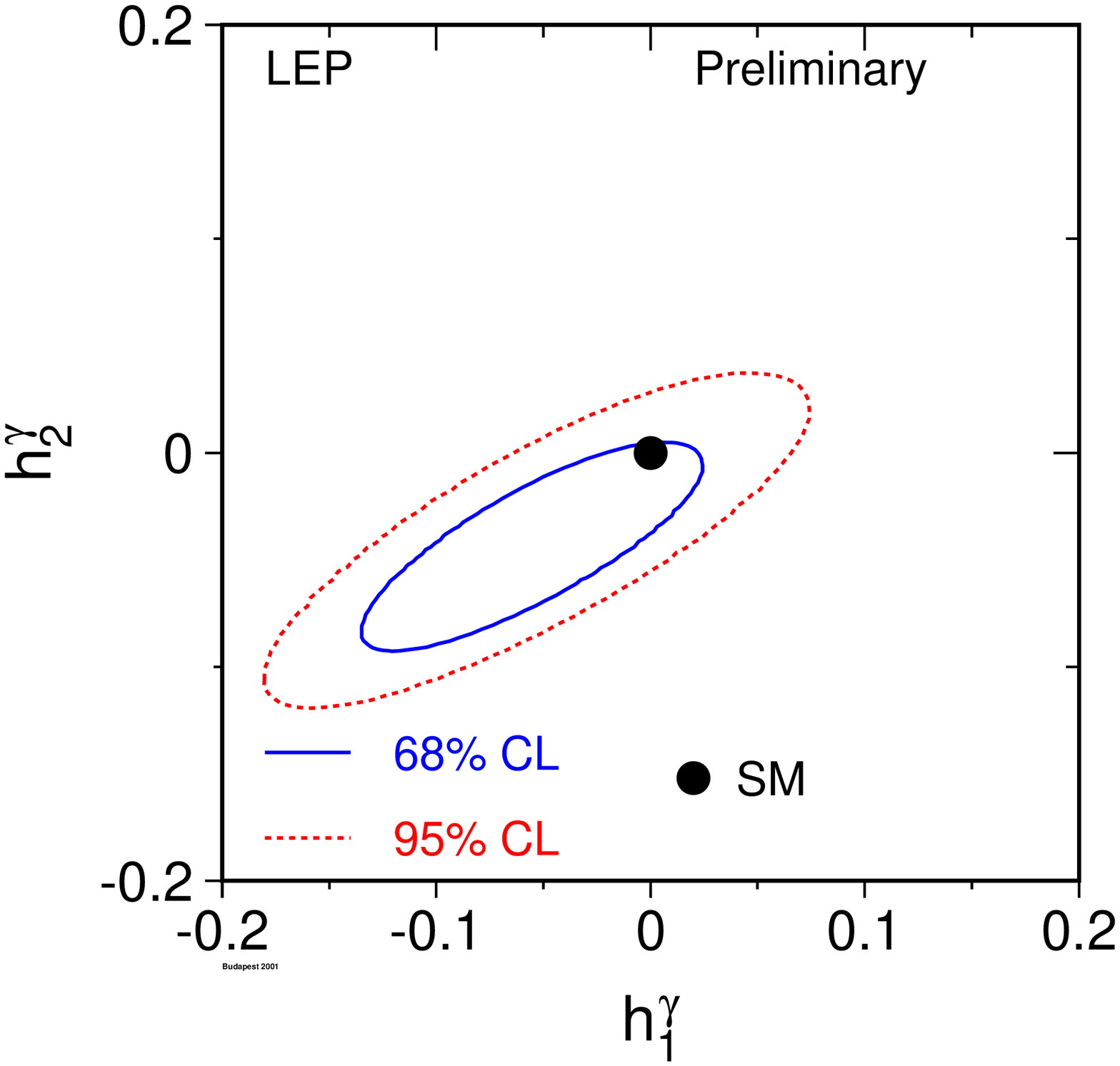}\hfill
\includegraphics[width=0.49\linewidth]{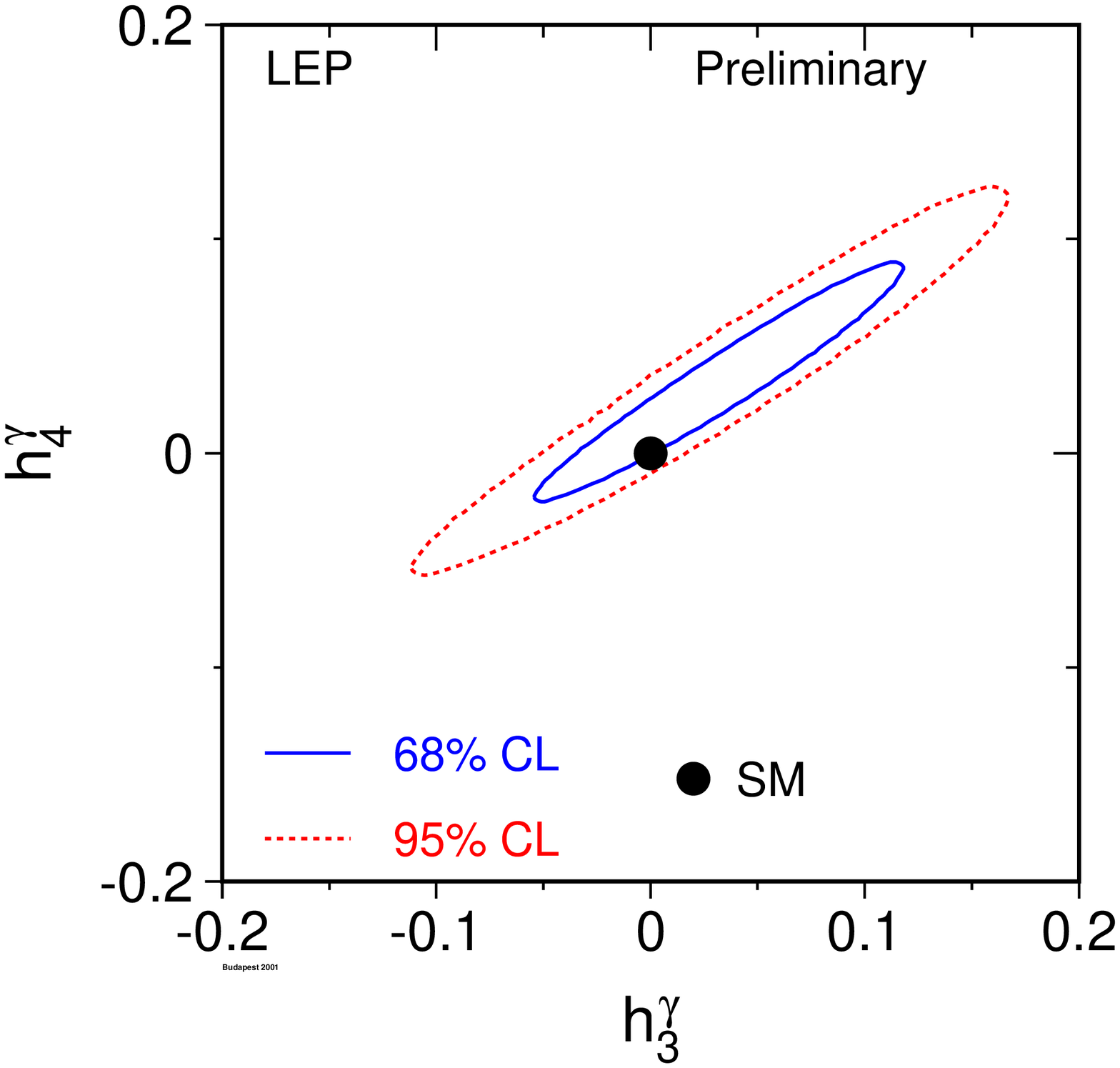}\\
\includegraphics[width=0.49\linewidth]{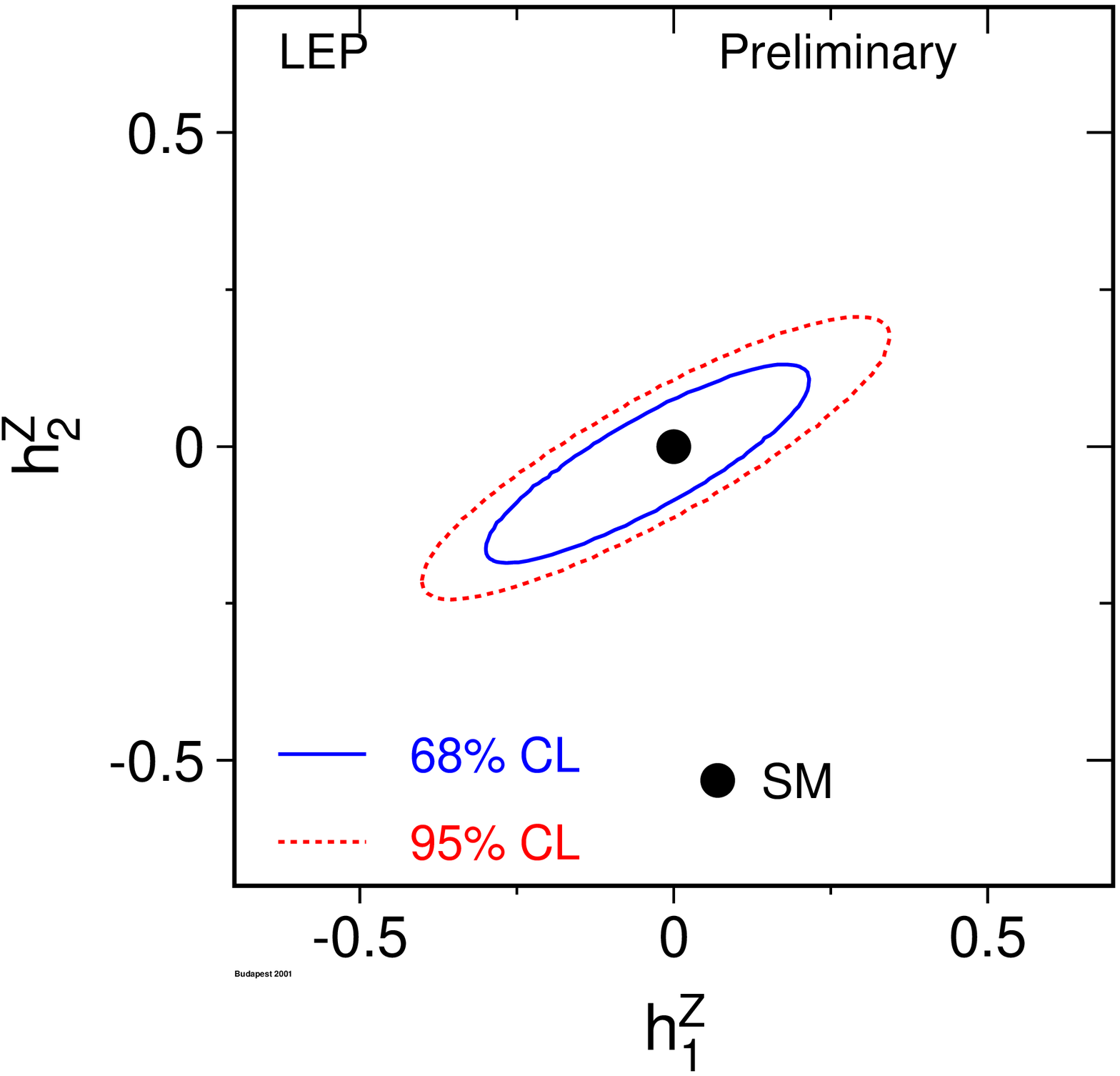}\hfill
\includegraphics[width=0.49\linewidth]{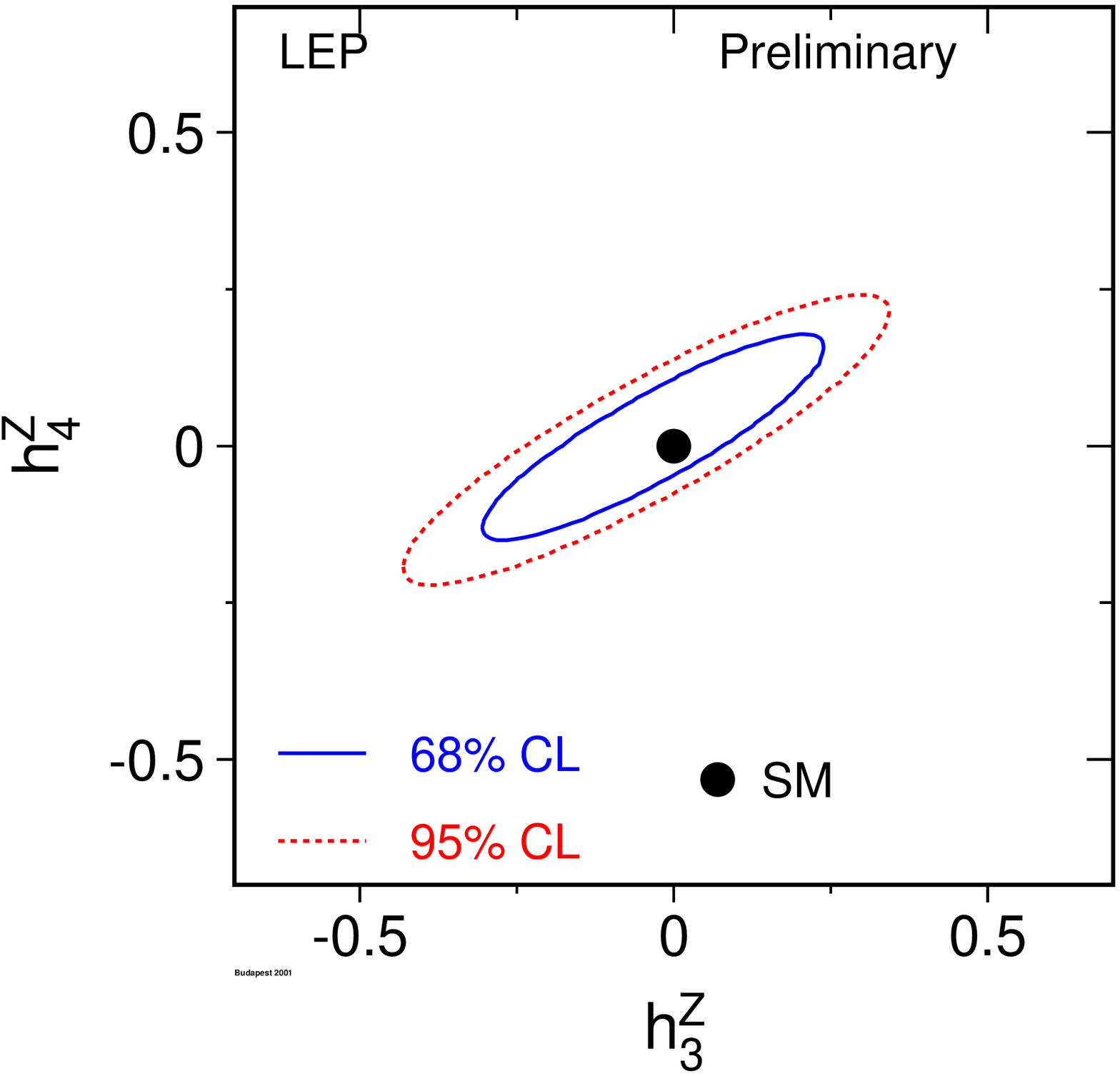}
\caption[]{
  Contour curves of 68\% C.L. and 95\% C.L. in the planes
  $(h_1^\gamma,h_2^\gamma)$, $(h_3^\gamma,h_4^\gamma)$,
  $(h_1^Z,h_2^Z)$ and $(h_3^Z,h_4^Z)$ showing the LEP combined result.
  }
\label{fig:gc_hTGC-2}
\end{center}
\end{figure}

The results for each experiment are shown in
Table~\ref{tab:gc_hTGC-2-ADLO}, where the errors include both
statistical and systematic uncertainties.  The 68\% C.L. and 95\% C.L.
contour curves resulting from the combinations of the two-dimensional
likelihood curves are shown in Figure~\ref{fig:gc_hTGC-2}.  The LEP
average values are given in Table~\ref{tab:gc_hTGC-2-LEP}.

\begin{table}[htbp]
\begin{center}
\renewcommand{\arraystretch}{1.3}
\begin{tabular}{|l||r|r|r|r|} 
\hline
Parameter  & ALEPH & DELPHI  &  L3  & OPAL \\
\hline
\hline
$h_1^\gamma$ & [$-0.14,~~+0.14$]  & [$-0.15,~~+0.15$]   & [$-0.06,~~+0.06$]   & [$-0.13,~~+0.13$] \\
\hline                             
$h_2^\gamma$ & [$-0.07,~~+0.07$]  & [$-0.09,~~+0.09$]   & [$-0.053,~~+0.024$] & [$-0.089,~~+0.089$] \\
\hline                             
$h_3^\gamma$ & [$-0.069,~~+0.037$]& [$-0.047,~~+0.047$] & [$-0.062,~~-0.014$] & [$-0.16,~~+0.00$] \\
\hline                             
$h_4^\gamma$ & [$-0.020,~~+0.045$]& [$-0.032,~~+0.030$] & [$-0.004,~~+0.045$] & [$+0.01,~~+0.13$] \\
\hline                             
$h_1^Z$      & [$-0.23,~~+0.23$]  & [$-0.24,~~+0.25$]   & [$-0.17,~~+0.16$]   & [$-0.22,~~+0.22$] \\
\hline                             
$h_2^Z$      & [$-0.12,~~+0.12$]  & [$-0.14,~~+0.14$]   & [$-0.10,~~+0.09$]   & [$-0.15,~~+0.15$] \\
\hline                             
$h_3^Z$      & [$-0.28,~~+0.19$]  & [$-0.32,~~+0.18$]   & [$-0.23,~~+0.11$]   & [$-0.29,~~+0.14$] \\
\hline                             
$h_4^Z$      & [$-0.10,~~+0.15$]  & [$-0.12,~~+0.18$]   & [$-0.08,~~+0.16$]   & [$-0.09,~~+0.19$] \\
\hline
\end{tabular}
\caption[]{The 95\% C.L. intervals ($\Delta\LL=1.92$) measured by
  the ALEPH, DELPHI, L3 and OPAL.  In each case the parameter listed is varied
  while the remaining ones are fixed to their Standard Model values.
  Both statistical and systematic uncertainties are included.  }
\label{tab:gc_hTGC-1-ADLO}
\end{center}
\end{table}

\begin{table}[htbp]
\begin{center}
\renewcommand{\arraystretch}{1.3}
\begin{tabular}{|l||c|} 
\hline
Parameter     & 95\% C.L.      \\
\hline
\hline
$h_1^\gamma$  & [$-0.056,~~+0.055$]  \\ 
\hline
$h_2^\gamma$  & [$-0.045,~~+0.025$]  \\ 
\hline
$h_3^\gamma$  & [$-0.049,~~-0.008$]  \\ 
\hline
$h_4^\gamma$  & [$-0.002,~~+0.034$]  \\ 
\hline
$h_1^Z$       & [$-0.13,~~+0.13$]  \\ 
\hline
$h_2^Z$       & [$-0.078,~~+0.071$]  \\ 
\hline
$h_3^Z$       & [$-0.20,~~+0.07$]  \\ 
\hline
$h_4^Z$       & [$-0.05,~~+0.12$]  \\ 
\hline
\end{tabular}
\caption[]{ The 95\% C.L. intervals ($\Delta\LL=1.92$) obtained
  combining the results from the four experiments.  In each case the
  parameter listed is varied while the remaining ones are fixed to
  their Standard Model values.  Both statistical and systematic
  uncertainties are included.  }
 \label{tab:gc_hTGC-1-LEP}
\end{center}
\end{table}

\begin{table}[htbp]
\begin{center}
\renewcommand{\arraystretch}{1.3}
\begin{tabular}{|l||r|r|r|} 
\hline
Parameter  & ALEPH & DELPHI  &  L3  \\
\hline
\hline
$h_1^\gamma$ & [$-0.32,~~+0.32$] & [$-0.28,~~+0.28$] & [$-0.17,~~+0.04$] \\
$h_2^\gamma$ & [$-0.18,~~+0.18$] & [$-0.17,~~+0.18$] & [$-0.12,~~+0.02$] \\
\hline                                               
$h_3^\gamma$ & [$-0.17,~~+0.38$] & [$-0.48,~~+0.20$] & [$-0.09,~~+0.13$] \\
$h_4^\gamma$ & [$-0.08,~~+0.29$] & [$-0.08,~~+0.15$] & [$-0.04,~~+0.11$] \\
\hline                                               
$h_1^Z$      & [$-0.54,~~+0.54$] & [$-0.45,~~+0.46$] & [$-0.48,~~+0.33$] \\
$h_2^Z$      & [$-0.29,~~+0.30$] & [$-0.29,~~+0.29$] & [$-0.30,~~+0.22$] \\
\hline
$h_3^Z$      & [$-0.58,~~+0.52$] & [$-0.57,~~+0.38$] & [$-0.43,~~+0.39$] \\
$h_4^Z$      & [$-0.29,~~+0.31$] & [$-0.31,~~+0.28$] & [$-0.23,~~+0.28$] \\
\hline
\end{tabular}
\caption[]{The 95\% C.L. intervals ($\Delta\LL=1.92$) measured  by
  ALEPH, DELPHI and L3.  In each case the two parameters listed are varied
  while the remaining ones are fixed to their Standard Model values.
  Both statistical and systematic uncertainties are included.  }
\label{tab:gc_hTGC-2-ADLO}
\end{center}
\end{table}

\begin{table}[htbp]
\begin{center}
\renewcommand{\arraystretch}{1.3}
\begin{tabular}{|l||c|rr|} 
\hline
Parameter  & 95\% C.L. & \multicolumn{2}{|c|}{Correlations} \\
\hline
\hline
$h_1^\gamma$  & [$-0.16,~~+0.05$]    & $ 1.00$ & $+0.79$ \\ 
$h_2^\gamma$  & [$-0.11,~~+0.02$]    & $+0.79$ & $ 1.00$ \\ 
\hline
$h_3^\gamma$  & [$-0.08,~~+0.14$]    & $ 1.00$ & $+0.97$ \\ 
$h_4^\gamma$  & [$-0.04,~~+0.11$]    & $+0.97$ & $ 1.00$ \\ 
\hline
$h_1^Z$       & [$-0.35,~~+0.28$]    & $ 1.00$ & $+0.77$ \\ 
$h_2^Z$       & [$-0.21,~~+0.17$]    & $+0.77$ & $ 1.00$ \\ 
\hline
$h_3^Z$       & [$-0.37,~~+0.29$]    & $ 1.00$ & $+0.76$ \\ 
$h_4^Z$       & [$-0.19,~~+0.21$]    & $+0.76$ & $ 1.00$ \\ 
\hline
\end{tabular}
\caption[]{ The 95\% C.L. intervals ($\Delta\LL=1.92$) obtained
  combining the results from ALEPH, DELPHI and L3.  In each case the two
  parameters listed are varied while the remaining ones are fixed to
  their Standard Model values.  Both statistical and systematic
  uncertainties are included.  Since the shape of the log-likelihood
  is not parabolic, there is some ambiguity in the definition of the
  correlation coefficients and the values quoted here are approximate.
  }
 \label{tab:gc_hTGC-2-LEP}
\end{center}
\end{table}

\clearpage

\clearpage

\subsection{Neutral Triple Gauge Boson Couplings in ZZ Production}

The individual analyses and results of the experiments for the $f$-couplings 
are described in~\cite{gc_bib:ALEPH-nTGC,gc_bib:DELPHI-nTGC,gc_bib:L3-fTGC,gc_bib:OPAL-fTGC}.

\subsubsection*{Single-Parameter Analyses}

\begin{figure}[p]
\begin{center}
\includegraphics[width=\linewidth]{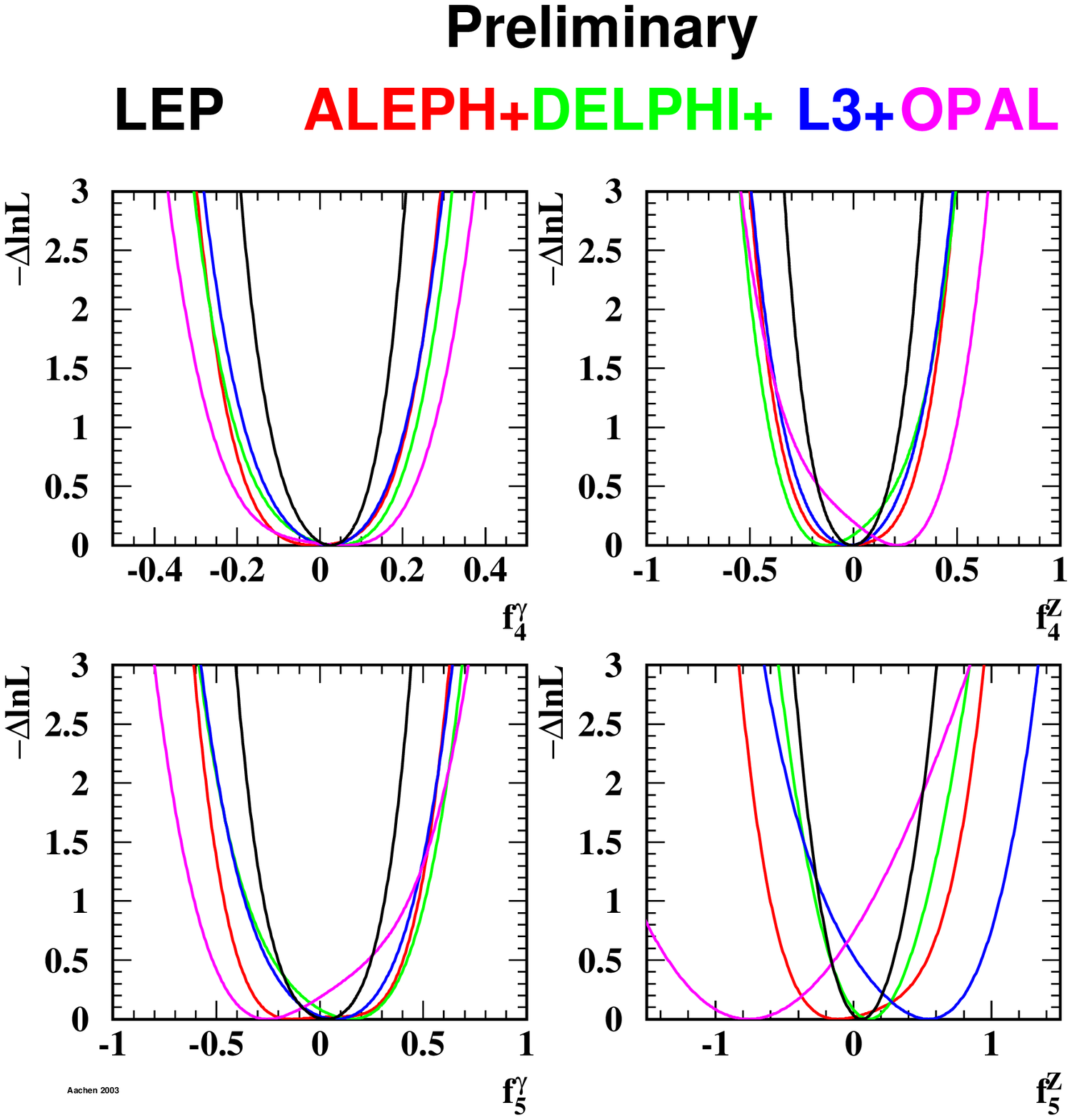}
\caption[]{
  The $\LL$ curves of the four experiments, and the LEP combined curve
  for the four neutral TGCs $f_i^V,~V=\gamma,Z,~i=4,5$.  In each case,
  the minimal value is subtracted.  }
\label{fig:gc_fTGC-1}
\end{center}
\end{figure}

The results for each experiment are shown in
Table~\ref{tab:gc_fTGC-1-ADLO}, where the errors include both statistical
and systematic uncertainties.  The individual $\LL$ curves and their sum are
shown in Figure~\ref{fig:gc_fTGC-1}.  The results of the combination are
given in Table~\ref{tab:gc_fTGC-1-LEP}.

\subsubsection*{Two-Parameter Analyses}

\begin{figure}[p]
\begin{center}
\includegraphics[width=0.55\linewidth]{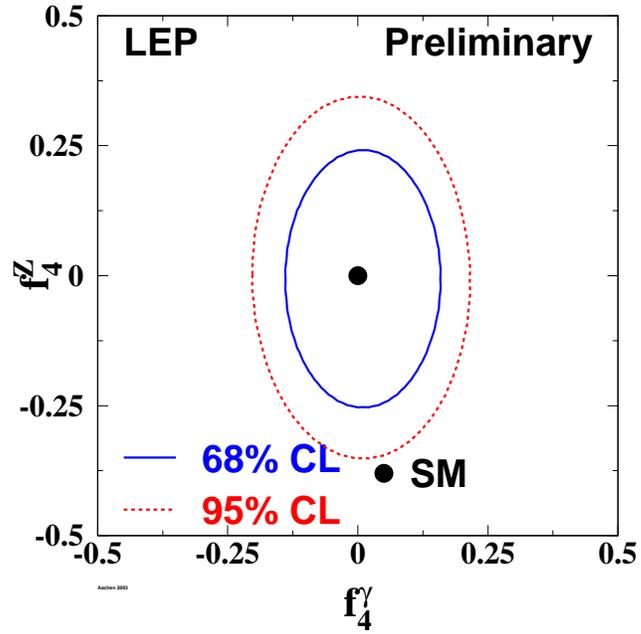}\\
\includegraphics[width=0.55\linewidth]{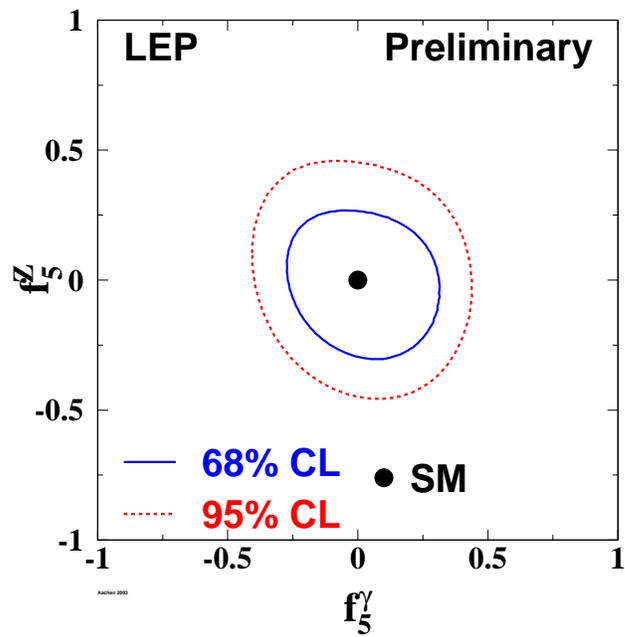}
\caption[]{
  Contour curves of 68\% C.L. and 95\% C.L. in the plane
  $(f_4^\gamma,f_4^Z)$ and $(f_5^\gamma,f_5^Z)$ showing the LEP
  combined result.  }
\label{fig:gc_fTGC-2}
\end{center}
\end{figure}

The results from each experiment are shown in
Table~\ref{tab:gc_fTGC-2-ADLO}, where the errors include both
statistical and systematic uncertainties.  The 68\% C.L. and 95\% C.L.
contour curves resulting from the combinations of the two-dimensional
likelihood curves are shown in Figure~\ref{fig:gc_fTGC-2}.  The LEP
average values are given in Table~\ref{tab:gc_fTGC-2-LEP}.

\begin{table}[htbp]
\begin{center}
\renewcommand{\arraystretch}{1.3}
\begin{tabular}{|l||r|r|r|r|} 
\hline
Parameter  & ALEPH & DELPHI  &  L3   & OPAL  \\
\hline
\hline
$f_4^\gamma$ & [$-0.26,~~+0.26$] & [$-0.26,~~+0.28$] & [$-0.28,~~+0.28$] & [$-0.32,~~+0.33$] \\ 
\hline
$f_4^Z$      & [$-0.44,~~+0.43$] & [$-0.49,~~+0.42$] & [$-0.48,~~+0.46$] & [$-0.45,~~+0.58$] \\ 
\hline
$f_5^\gamma$ & [$-0.54,~~+0.56$] & [$-0.48,~~+0.61$] & [$-0.39,~~+0.47$] & [$-0.71,~~+0.59$] \\ 
\hline
$f_5^Z$      & [$-0.73,~~+0.83$] & [$-0.42,~~+0.69$] & [$-0.35,~~+1.03$] & [$-0.94,~~+0.25$] \\ 
\hline
\end{tabular}
\caption[]{The 95\% C.L. intervals ($\Delta\LL=1.92$) measured by
  ALEPH, DELPHI, L3 and OPAL.  In each case the parameter listed is varied
  while the remaining ones are fixed to their Standard Model values.
  Both statistical and systematic uncertainties are included.  }
\label{tab:gc_fTGC-1-ADLO}
\end{center}
\end{table}

\begin{table}[htbp]
\begin{center}
\renewcommand{\arraystretch}{1.3}
\begin{tabular}{|l||c|} 
\hline
Parameter     & 95\% C.L.     \\
\hline
\hline
$f_4^\gamma$  & [$-0.17,~~+0.19$]  \\ 
\hline
$f_4^Z$       & [$-0.30,~~+0.30$]  \\ 
\hline
$f_5^\gamma$  & [$-0.32,~~+0.36$]  \\ 
\hline
$f_5^Z$       & [$-0.34,~~+0.38$]  \\ 
\hline
\end{tabular}
\caption[]{ The 95\% C.L. intervals ($\Delta\LL=1.92$) obtained
  combining the results from all four experiments.  In each case the
  parameter listed is varied while the remaining ones are fixed to
  their Standard Model values.  Both statistical and systematic
  uncertainties are included.  }
 \label{tab:gc_fTGC-1-LEP}
\end{center}
\end{table}

\begin{table}[htbp]
\begin{center}
\renewcommand{\arraystretch}{1.3}
\begin{tabular}{|l||r|r|r|r|} 
\hline
Parameter  & ALEPH & DELPHI  &  L3   & OPAL  \\

\hline
\hline
$f_4^\gamma$ & [$-0.26,~~+0.26$] & [$-0.26,~~+0.28$]& [$-0.28,~~+0.28$] & [$-0.32,~~+0.33$] \\ 
$f_4^Z$      & [$-0.44,~~+0.43$] & [$-0.49,~~+0.42$]& [$-0.48,~~+0.46$] & [$-0.47,~~+0.58$]  \\ 
\hline                                              
$f_5^\gamma$ & [$-0.52,~~+0.53$] & [$-0.52,~~+0.61$]& [$-0.52,~~+0.62$] & [$-0.67,~~+0.62$] \\ 
$f_5^Z$      & [$-0.77,~~+0.86$] & [$-0.44,~~+0.69$]& [$-0.47,~~+1.39$] & [$-0.95,~~+0.33$] \\ 
\hline
\end{tabular}
\caption[]{The 95\% C.L. intervals ($\Delta\LL=1.92$) measured by
  ALEPH, DELPHI, L3 and OPAL.  In each case the two parameters listed are
  varied while the remaining ones are fixed to their Standard Model
  values.  Both statistical and systematic uncertainties are included.
  }
\label{tab:gc_fTGC-2-ADLO}
\end{center}
\end{table}

\begin{table}[htbp]
\begin{center}
\renewcommand{\arraystretch}{1.3}
\begin{tabular}{|l||c|rr|} 
\hline
Parameter     & 95\% C.L. & \multicolumn{2}{|c|}{Correlations} \\
\hline
\hline
$f_4^\gamma$  &[$-0.17,~~+0.19$] & $ 1.00$ & $ 0.07$\\ 
$f_4^Z$       &[$-0.30,~~+0.29$] & $ 0.07$ & $ 1.00$\\ 
\hline
$f_5^\gamma$  &[$-0.34,~~+0.38$] & $ 1.00$ & $-0.17$\\ 
$f_5^Z$       &[$-0.38,~~+0.36$] & $-0.17$ & $ 1.00$\\ 
\hline
\end{tabular}
\caption[]{ The 95\% C.L. intervals ($\Delta\LL=1.92$) obtained
  combining the results from all four experiments.  In each case the
  two parameters listed are varied while the remaining ones are fixed
  to their Standard Model values.  Both statistical and systematic
  uncertainties are included. Since the shape of the log-likelihood is
  not parabolic, there is some ambiguity in the definition of the
  correlation coefficients and the values quoted here are approximate.
  }
 \label{tab:gc_fTGC-2-LEP}
\end{center}
\end{table}

\clearpage

\clearpage

\subsection{Quartic Gauge Boson Couplings}
\begin{figure}[htbp]
\begin{center}
\includegraphics[width=\linewidth]{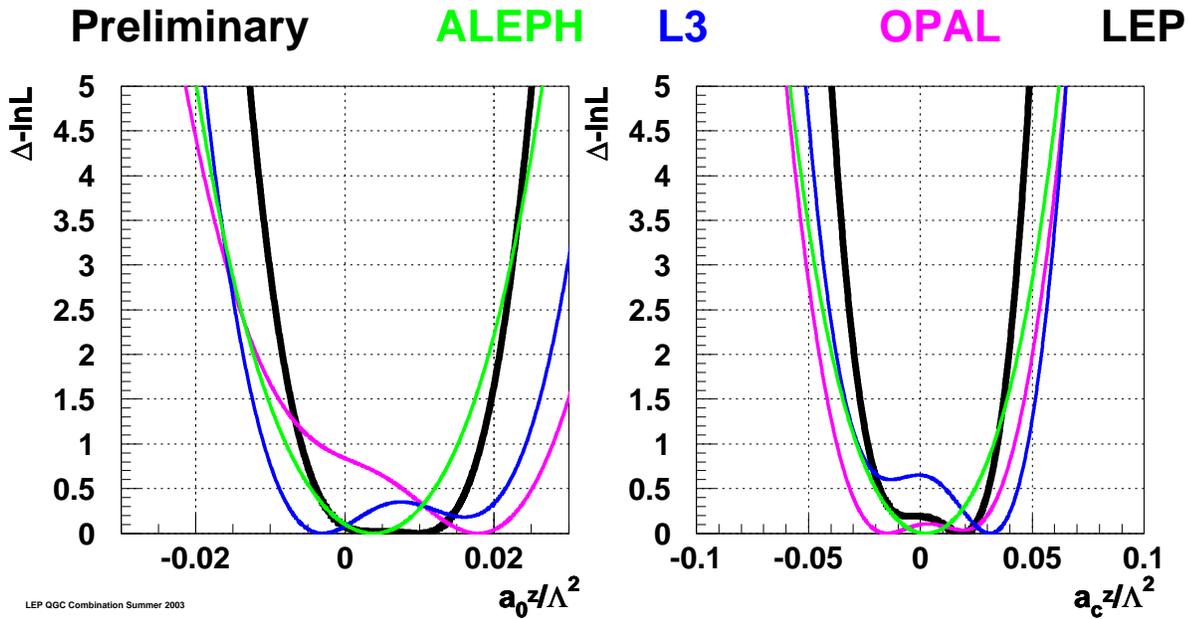}
\caption[]{
  The $\LL$ curves of L3 and OPAL (thin lines) and the 
  combined curve (thick line) for the QGCs \acl\ and \azl, associated with
  the \ZZgg\  vertex.  In each case, the minimal value is subtracted.  } 
\label{fig:gc_QGC-1}
\end{center}
\end{figure}

The individual numerical results from the experiments participating in the
combination, and the combined result are shown in Table~\ref{tab:gc_QGCs}. 
The corresponding $\LL$ curves are shown in
Figure~\ref{fig:gc_QGC-1}. The errors include both statistical and
systematic uncertainties. 

\begin{table}[ht]
\begin{center}
\begin{tabular}{|l||c|c|c|c|} \hline
Parameter & ALEPH  & L3  & OPAL   & Combined  \\
\hline \hline
\acl  &  [$-0.041,~~+0.044$]  &  [$-0.037,~~+0.054$] & 
         [$-0.045,~~+0.050$]  &  [$-0.029,~~+0.039$]  \\  
\azl  &  [$-0.012,~~+0.019$]  &  [$-0.014,~~+0.027$] & 
         [$-0.012,~~+0.031$]  &  [$-0.008,~~+0.021$] \\ 
\hline
\end{tabular}
\end{center}
\caption{ The limits for the QGCs \acl\ and \azl\ associated
    with the \ZZgg\ vertex at 95\% confidence level for ALEPH, L3 and OPAL, 
    and the LEP result obtained by combining them. 
    Both statistical and systematic errors are included.
  }
\label{tab:gc_QGCs}
\end{table}

\section*{Conclusions}
Combinations of charged and neutral triple gauge boson couplings, as well as
quartic gauge boson couplings associated with the \ZZgg\ vertex were made,
based on results from the four LEP experiments ALEPH, DELPHI, L3 and OPAL.
No significant deviation
from the Standard Model prediction is seen for any of the electroweak
gauge boson couplings studied. 
With the LEP-combined charged TGC results,
the existence of triple gauge boson couplings among the electroweak
gauge bosons is experimentally verified.
As an example, these data allow
the Kaluza-Klein theory~\cite{gc_bib:klein}, in which $\kg = -2$, to be
excluded completely~\cite{gc_bib:maiani}.

%% file: fsi_cr.tex
\section{Introduction}
 
 In \WWtoqqqq\ events, the products of the two (colour singlet) W
 decays in general have a significant space-time overlap as the
 separation of their decay vertices, $\tau_W \sim 1/\Gamma_W\approx
 0.1$~fm, is small compared to characteristic hadronic distance scales
 of $\sim 1$~fm.  Colour reconnection, also known as colour
 rearrangement (CR), was first introduced in\cite{bib:cr:GPZ} and
 refers to a reorganisation of the colour flow between the two W
 bosons.  A precedent is set for such effects by colour suppressed B
 meson decays, \eg\ $B \rightarrow J/\psi K$, where there is
 ``cross-talk'' between the two original colour singlets,
 $\bar{\mathrm c}$+s and
 c+spectator\cite{bib:cr:GPZ,bib:cr:SK_MODELS}.
 
 QCD interference effects between the colour singlets in \WW\ decays
 during the perturbative phase are expected to be small, affecting the
 W mass by $\sim (\frac{\alfas}{\pi N_{\mathrm{colours}}})^2 \Gamma_W$
 $\sim \cal{O}(\mathrm{1~\MeV})$ \cite{bib:cr:SK_MODELS}. In contrast,
 non-perturbative effects involving soft gluons with energies less
 than \GW\ may be significant, with effects on \MW\ $\sim
 \cal{O}(\mathrm{10~\MeV})$.  To estimate the impact of this phenomenon
 a variety of phenomenological models have been developed
 \cite{bib:cr:SK_MODELS,bib:cr:ARIADNECR_MODEL,HERWIG6,bib:cr:GH_MODEL,bib:cr:EG_MODEL,bib:cr:RATHSMAN},
 some of which are compared with data in this note.
 
 Many observables have been considered in the search for an
 experimental signature of colour reconnection.  The inclusive
 properties of events such as the mean charged particle multiplicity,
 distributions of thrust, rapidity, transverse momentum and
 $\ln(1/x_p)$ are found to have limited
 sensitivity\cite{bib:cr:OPAL_CR,bib:cr:DELPHI_CR,bib:cr:ALEPH_CR,bib:cr:L3_CR}.
 The effects of CR are predicted to be numerically larger in these
 observables when only higher mass hadrons such as kaons and protons
 are considered\cite{bib:cr:SK_HEAVYHAD}.  However, experimental
 investigations\cite{bib:cr:DELPHI_CR,bib:cr:OPAL_HEAVYHAD} find no
 significant gain in sensitivity due to the low production rate of
 such species in W decays and the finite size of the data sample.
 
 More recently, in analogy with the ``string effect'' analysis in
 3-jet $\eeqq g$
 events\cite{bib:cr:JADE_STRING2,*bib:cr:JADE_STRING3,%
 *bib:cr:JADE_STRING4,*bib:cr:JADE_STRING5,*bib:cr:TPC2GAM_STRING1,%
 *bib:cr:TPC2GAM_STRING2,*bib:cr:TASSO_STRING1}, the so-called
 ``particle flow'' method
 \cite{bib:cr:pflow1,bib:cr:pflow2,bib:cr:OXFORD_WS} has been
 investigated by all LEP collaborations
 \cite{bib:cr:ALEPH_PF,bib:cr:DELPHI_PF,bib:cr:L3_PF,bib:cr:OPAL_PF}.
 In this, pairs of jets in \WWtoqqqq\ events are associated with the
 decay of a W, after which four jet-jet regions are chosen: two corresponding
 to jets sharing the same W parent (intra-W), and two in which the
 parents differ (inter-W).  As there is a two-fold ambiguity in the
 assignment of inter-W regions, the configuration having the smaller sum
 of inter-W angles is chosen.
 
 Particles are projected onto the planes defined by these jet pairs
 and the particle density constructed as a function of $\phi$, the
 projected angle relative to one jet in each plane.  To account for
 the variation in the opening angles, $\phi_0$, of the jet-jet pairs
 defining each plane, the particle
 densities in $\phi$ are constructed as functions of normalised
 angles, $\phi_r=\phi/\phi_0$, by a simple rescaling of the projected
 angles for each particle, event by event.  Particles having
 projected angles $\phi$ smaller than $\phi_0$ in at least one of the
 four planes are considered further.  This gives particle densities,
 $\frac{1}{\Nevt}\dndphir$, in four regions with $\phi_r$ in the range
 0--1, and where $n$ and \Nevt\ are the number of particles and
 events, respectively.
 
 As particle density reflects the colour flow in an event, CR models
 predict a change in the relative particle densities between inter-W
 and intra-W regions.  On average, colour reconnection is expected to
 affect the particle densities of both inter-W regions in the same way
 and so they are added together, as are the two intra-W regions.  The
 observable used to quantify such changes, \Rn, is defined:
\begin{equation}
 \Rn =
  \frac{\frac{1}{\Nevt}\int^{0.8}_{0.2} \dndphir (\mathrm{intra-W}) \dphir}
       {\frac{1}{\Nevt}\int^{0.8}_{0.2} \dndphir (\mathrm{inter-W}) \dphir} \,.
 \label{fsi:cr:eq:Rn}
\end{equation}
 As the effects of CR are expected to be enhanced for low momentum
 particles far from the jet axis, the range of integration excludes
 jet cores ($\phi_r\approx 0$ and $\phi_r\approx 1$).  The precise
 upper and lower limits are optimised by model studies of predicted
 sensitivity.
 
 Each LEP experiment has developed its own variation on this analysis,
 differing primarily in the selection of \WWtoqqqq\ events.  In
 \Ltre\cite{bib:cr:L3_PF} and \Delphi\cite{bib:cr:DELPHI_PF}, events
 are selected in a very particular configuration (``topological
 selection'') by imposing restrictions on the jet-jet angles and on
 the jet resolution parameter for the three- to four-jet transition
 (Durham or LUCLUS schemes).  This selects events which are more planar
 than those in the inclusive \WWtoqqqq\ sample and the association between
 jet pairs and W's is given by the relative angular separation of the
 jets.  The overall efficiency for selecting events is $\sim15$\%.
 The \Aleph\cite{bib:cr:ALEPH_PF} and \Opal\cite{bib:cr:OPAL_PF} event
 selections are based on their W mass analyses.  Assignment of pairs
 of jets to W's also follows that used in measuring \MW, using either
 a 4-jet matrix element \cite{ALEPH-MW} or a multivariate algorithm
 \cite{OPAL-MW}.  These latter selections have much higher
 efficiencies, varying from 45\% to 90\%, but lead to samples of events
 having a less planar topology and hence a more complicated colour
 flow. ALEPH also uses the topological selection for consistency
 checks.
 
 The data are corrected bin-by-bin for background contamination in the
 inter-W and intra-W regions separately.  The possibility of CR
 effects existing in background processes, such as \ZZtoqqqq, is
 neglected.  Since the data are not corrected for the effects of event
 selection, momentum resolution and finite acceptance, the values of
 \Rn\ measured by the experiments cannot be compared directly with one
 another.  However, it is possible to perform a relative comparison by
 using a common sample of Monte Carlo events, processed using the
 detector simulation program of each experiment.

\section{Combination Procedure}
 
 The measured values of \Rn\ can be compared after they have been
 normalised using a common sample of events, processed using the
 detector simulation and particle flow analysis of each experiment.  A
 variable, $r$, is constructed:
\begin{equation}
          r = \frac{\Rndata}{\Rnnocr} \,,
\end{equation}
 where \Rndata\ and \Rnnocr\ are the values of \Rn\ measured by each
 experiment in data and in a common sample of events without CR.  In
 the absence of CR, all experiments should find $r$ consistent with
 unity.  The default no-CR sample used for this normalisation consists
 of \eeWW\ events produced using the \KoralW\cite{bib:fsi:KORALW} event
 generator and hadronised using either the \Jetset\cite{JETSET},
 \Ariadne\cite{ARIADNE} or \Herwig\cite{HERWIG6} model depending on
 the colour reconnection model being tested.  Input from experiments used to
 perform the combination is given in terms of \Rn\ and detailed in
 Appendix~\ref{fsi:cr:app:inputs}.

 \subsection{Weights}
 
 The statistical precision of \Rn\ measured by the experiments does
 not reflect directly the sensitivity to CR, for example the
 measurements of \Aleph\ and \Opal\ have efficiencies several times
 larger than the topological selections of \Ltre\ and \Delphi, yet
 only yield comparable sensitivity.  The relative sensitivity of the
 experiments may also be model dependent.  Therefore, results are
 averaged using model dependent weights, \ie\ 
\begin{equation}
  w_i = \frac{(\Rni - \Rninocr)^2}
             {\sigsqRn(\mathrm{stat.}) + \sigsqRn (\mathrm{syst.})}
             \,,
 \label{fsi:cr:eq:weight}
\end{equation}
 where \Rni\ and \Rninocr\ represent the \Rn\ values for CR model $i$
 and its corresponding no-CR scenario, and \sigsqRn\ are the total
 statistical and systematic uncertainties.  To test models, \Rn\ 
 values using common samples are provided by experiments for each of
 the following models:
 \begin{enumerate}
  \item \SKI, 100\% reconnected (\KoralW\ $+$ \Jetset),
  \item \Ariadne-II, inter-W reconnection rate about 22\% (\KoralW\ $+$ \Ariadne),
  \item \Herwig\ CR, reconnected fraction $\frac{1}{9}$ (\KoralW\ $+$
  \Herwig).
 \end{enumerate}
 Samples in parentheses are the corresponding no-CR scenarios used to
 define $w_i$.  In each case, $\KoralW$ is used to generate the events
 at least up to the four-fermion level. 
 These special Monte Carlo samples (called ``Cetraro''
 samples) have been generated with the ALEPH tuned parameters,
 obtained with hadronic Z decays, and have been processed through the
 detector simulation of each experiment.

 \subsection{Combination of centre-of-mass energies}
 
 The common files required to perform the combination are only
 available at a single centre-of-mass energy ($E_{\mathrm{cm}}$) of
 188.6~GeV.  The data from the experiments can only therefore be
 combined at this energy.  The procedure adopted to combine all LEP
 data is summarised below.
 
 \Rn\ is measured in each experiment at each centre-of-mass energy, in
 both data and Monte Carlo.
 The predicted variation of \Rn\ with centre-of-mass energy is
 determined separately by each experiment using its own samples of 
 simulated \eeWW\ events, with hadronisation performed using the no-CR
 \Jetset\ model.  This variation is parametrised by fitting
 a polynomial to these simulated \Rn.  The \Rn\ measured in data are
 subsequently extrapolated to the reference energy of 189~GeV using
 this function, and the weighted average of the rescaled values in
 each experiment is used as input to the combination.

\section{Systematics}

 The sources of potential systematic uncertainty identified are
 separated into those which are correlated between experiments and
 those which are not.  For correlated sources,
 the component correlated between all experiments is assigned as the
 smallest uncertainty found in any single experiment, with the
 quadrature remainder treated as an uncorrelated contribution.
 Preliminary estimates of the dominant systematics on \Rn\ are given
 in Appendix~\ref{fsi:cr:app:inputs} for each experiment, and
 described below.

 \subsection{Hadronisation}
 
 This is assigned by comparison of the single sample of \WW\ events
 generated using \KoralW, and hadronised with three different models,
 \ie\ \Jetset, \Herwig\ and \Ariadne.  The systematic is assigned as
 the spread of the \Rn\ values obtained when using the various models
 given in Appendix~\ref{fsi:cr:app:inputs}.  This is treated as a
 correlated uncertainty.

 \subsection{Bose-Einstein Correlations}
 
 Although a recent analysis by \Delphi\ reports the observation of
 inter-W Bose-Einstein correlation (BEC) in \WWtoqqqq\ events with a
 significance of 2.9 standard deviations for like-sign pairs and 1.9
 standard deviations for unlike-sign pairs~\cite{be:DELPHI03},
 analyses by other
 collaborations\cite{bib:cr:ALEPH_BEC,be:L302,bib:cr:OPAL_BEC} find no
 significant evidence for such effects, see also chapter~\ref{sec-BE}.
 Therefore, BEC effects are only considered within each W separately.
 The estimated uncertainty is assigned, using common MC samples, as
 the difference in \Rn\ between an intra-W BEC sample and the
 corresponding no-BEC sample.  This is treated as correlated between
 experiments.

 \subsection{Background}
 
 Background is dominated by the \eeqq\ process, with a smaller
 contribution from \ZZtoqqqq\ diagrams.  As no common background
 samples exist, apart from dedicated ones for BEC analyses, experiment
 specific samples are used.  The uncertainty is defined as the
 difference in the \Rn\ value relative to that obtained using the
 default background model and assumed cross-sections in each
 experiment.

  \subsubsection{\eeqq}
  
 The systematic is separated into two components, one accounting for
 the shape of the background, the other for the uncertainty in the
 value of the background cross-section, $\sigma(\eeqq)$.
 
 Uncertainty in the shape is estimated by comparing hadronisation
 models.  Experiments typically have large samples simulated using
 2-fermion event generators hadronised with various models.  This
 uncertainty is assigned as $\pm\frac{1}{2}$ of the largest difference
 between any pair of hadronisation models and treated as uncorrelated
 between experiments.
 
 The second uncertainty arises due to the accuracy of the
 experimentally measured cross-sections.  The systematic is assigned
 as the larger of the deviations in \Rn\ caused when $\sigma(\eeqq)$
 is varied by $\pm 10$\% from its default value.  This variation was
 based on the conclusions of a study comparing four-jet data
 with models\cite{bib:LEP2_MCWS}, and is significantly larger than the
 $\sim 1$\% uncertainty in the inclusive \eeqq\ ($\sqrt{s'/s}>0.85$)
 cross-section measured by the LEP2 2-fermion group.  It is treated as
 correlated between experiments.

 \subsubsection{\ZZtoqqqq}
 
 Similarly to the \eeqq\ case, this background cross-section is varied
 by $\pm 15$\%.  For comparison, the uncertainty on $\sigma(\ZZ)$
 measured by the LEP2 4-fermion group is $\sim 11$\% at
 $\sqrt{s}\simeq 189$~\GeV.  It is treated as correlated between
 experiments.

 \subsubsection{\WWtoqqlv}
 
 Semi-leptonic WW decays which are incorrectly identified as
 \WWtoqqqq\ events are the third main category of background, and its
 contribution is very small.  The fraction of \WWtoqqlv\ events
 present in the sample used for the particle flow analysis varies in
 the range 0.04--2.2\% between the experiments.  The uncertainty in
 this background consists of hadronisation effects and also
 uncertainty in the cross-section.  As this source is a very small
 background relative to those discussed above, and the effect of
 either varying the cross-section by its measured uncertainty or of
 changing the hadronisation model do not change the measured \Rn\ 
 significantly, this source is neglected.

 \subsection{Detector Effects}
 
 The data are not corrected for the effects of finite resolution or
 acceptance.  Various studies have been carried out, e.g. by analysing
 \WWtoqqlv\ events in the same way as \WWtoqqqq\ events in order to
 validate the method and the choice of energy flow objects used to
 measure the particle yields between jets~\cite{bib:cr:L3_PF}.  To
 take into account the effects of detector resolution and acceptance,
 ALEPH, L3 and OPAL have studied the impact of changing the object
 definition entering the particle flow distributions and have assigned
 a systematic error from the difference in the measured \Rn.

 \subsection{Centre-of-mass energy dependence}
 
 As there may be model dependence in the parametrised energy
 dependence, the second order polynomial used to perform the
 extrapolation to the reference energy of 189~GeV is usually
 determined using several different models, with and without colour
 reconnection.  DELPHI, L3 and OPAL use differences relative to the
 default no-CR model to assign a systematic uncertainty while ALEPH
 takes the spread of the results obtained with all the models with and
 without CR which have been used.  This error is assumed to be
 uncorrelated between experiments.

 \subsection{Weighting function}
 
 The weighting function of Equation~\ref{fsi:cr:eq:weight} could
 justifiably be modified such that only the uncorrelated components of
 the systematic uncertainty appear in the denominator.  To accommodate
 this, the average is performed using both variants of the weighting
 function.  This has an insignificant effect on the consistency
 between data and model under test, e.g. for \SKI\ the result is
 changed by 0.02 standard deviations, and this effect is therefore neglected.

\section{Combined Results}
 
 Experiments provide their results in the form of \Rn\ (or changes to
 \Rn) at a reference centre-of-mass energy of 189~\GeV\ by scaling
 results obtained at various energies using the predicted energy
 dependence of their own no-CR MC samples. This avoids having to
 generate common samples at multiple centre-of-mass energies.
 
 The detailed results from all experiments are included
 in Appendix~\ref{fsi:cr:app:inputs}.  These consist of preliminary
 results, taken from the publicly available
 notes\cite{bib:cr:ALEPH_PF,bib:cr:DELPHI_PF,bib:cr:L3_PF,bib:cr:OPAL_PF},
 and additional information from analysis of Monte Carlo samples.  The
 averaging procedure itself is carried out by each of the experiments
 and good agreement is obtained.
 
\begin{figure}[tbhp]
 \centerline{\epsfig{file=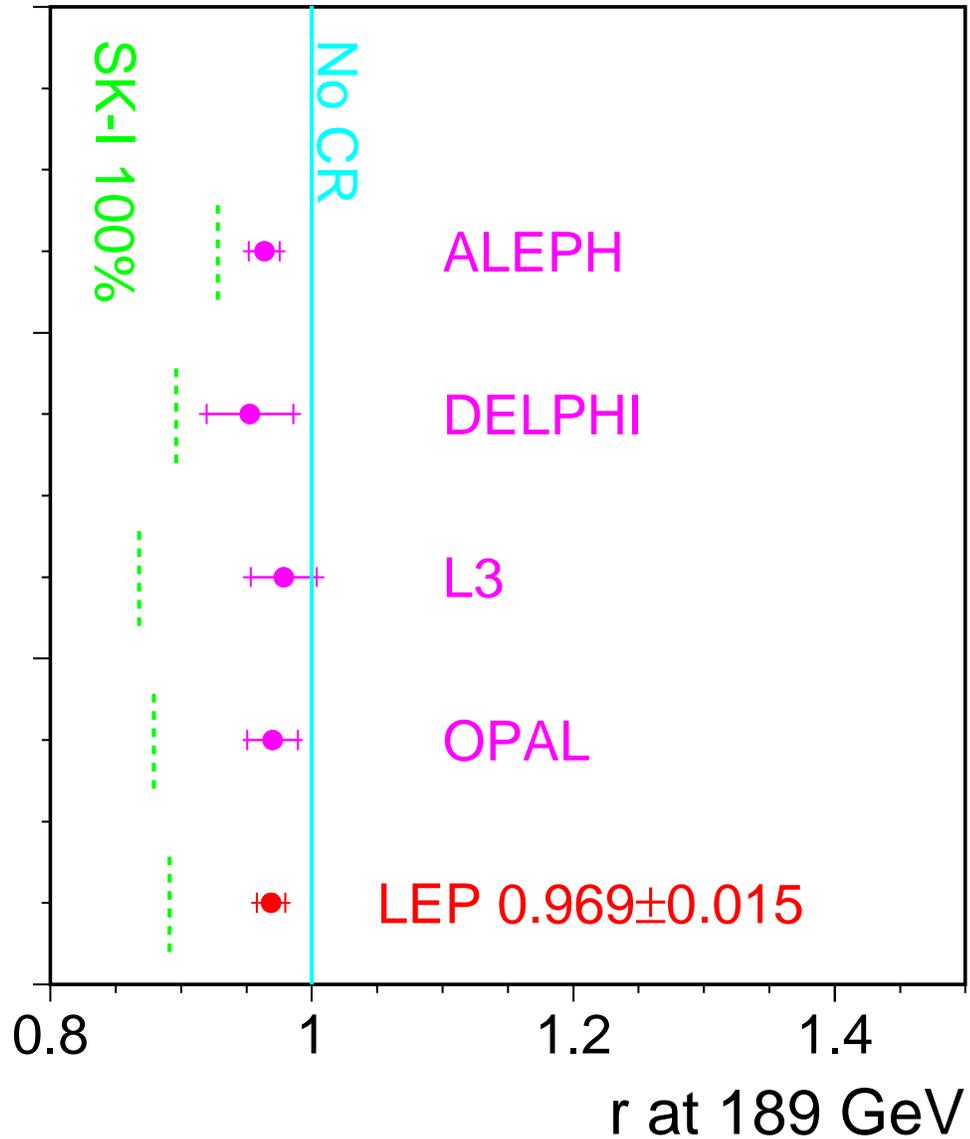,width=0.9\textwidth}}
 \caption[Particle flow combination, \SKI\ model.]  {Preliminary
   particle flow results using all data, combined to test the limiting
   case of the \SKI\ model in which more than 99.9\% of the events are
   colour reconnected. The error bars correspond to the total error
   with the inner part showing the statistical uncertainty.  The
   predicted values of $r$ for this CR model are indicated separately 
   for the analysis of each experiment by dashed lines.}
 \label{fsi:cr:fig:SKI_comb}
\end{figure}

 An example of this averaging to test an extreme scenario of the \SKI\ 
 CR model (full reconnection) is given in
 Appendix~\ref{fsi:cr:app:results}.  The average obtained in this case
 is:
\begin{eqnarray}
    r(data)    &  = & 0.969 \pm 0.011 (\mathrm{stat.})
                \pm 0.009 (\mathrm{syst.~corr.})
                \pm 0.006 (\mathrm{syst.~uncorr.}) \,, \\
   r (\mathrm{\SKI}\ 100\%) & = & 0.8909  \,.
 \end{eqnarray}
 The measurements of each experiment and this combined result are
 shown in Figure~\ref{fsi:cr:fig:SKI_comb}.  As the sensitivity of the
 analysis is different for each experiment, the value of $r$ predicted
 by the \SKI\ model is indicated separately for each experiment by a
 dashed line in the figure.  Thus the data disagree with the extreme
 scenario of this particular model at a level of 5.2 standard
 deviations. The data from the four experiments are consistent with
 each other and tend to prefer an intermediate colour reconnection
 scenario rather than the no colour reconnection one at the level of 2.2
 standard deviations in the \SKI\ framework.
 
 \subsection{Parameter space in \SKI\ model}
 
\begin{figure}[tbhp]
 \centerline{\epsfig{file=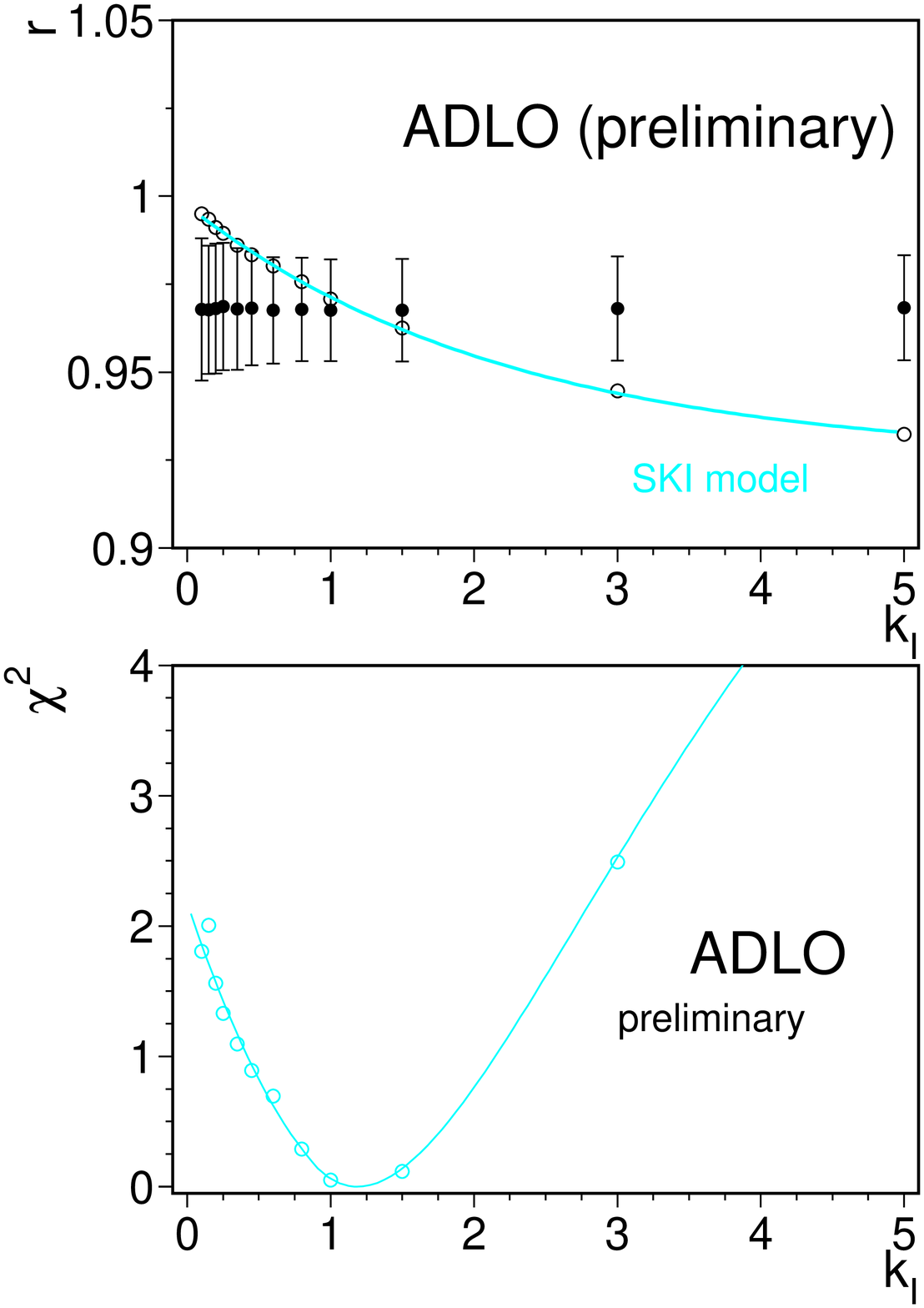,width=0.8\textwidth}}
 \caption[Constraints on \SKI\ model parameter, \kI.] {Comparison of the
   LEP average $r$ values with the \SKI\ model prediction obtained as
   a function of the \kI\ parameter.  The comparisons are performed
   after extrapolation of data to the reference centre-of-mass energy
   of 189~GeV.  In the upper plot, the solid line is the result of
   fitting a function of the form $r(k_{I}) = p_{1} (1-exp(-p_{2}
   k_{I}))+p_{3}$ to the MC predictions.  The lower plot shows the
   corresponding $\chi^{2}$ curve obtained from this comparison.  The
   best agreement between the model and the data is obtained when
   $\kI = 1.18$.}
 \label{fsi:cr:fig:ki_scan}
\end{figure}

\begin{figure}[tbhp]
 \centerline{\epsfig{file=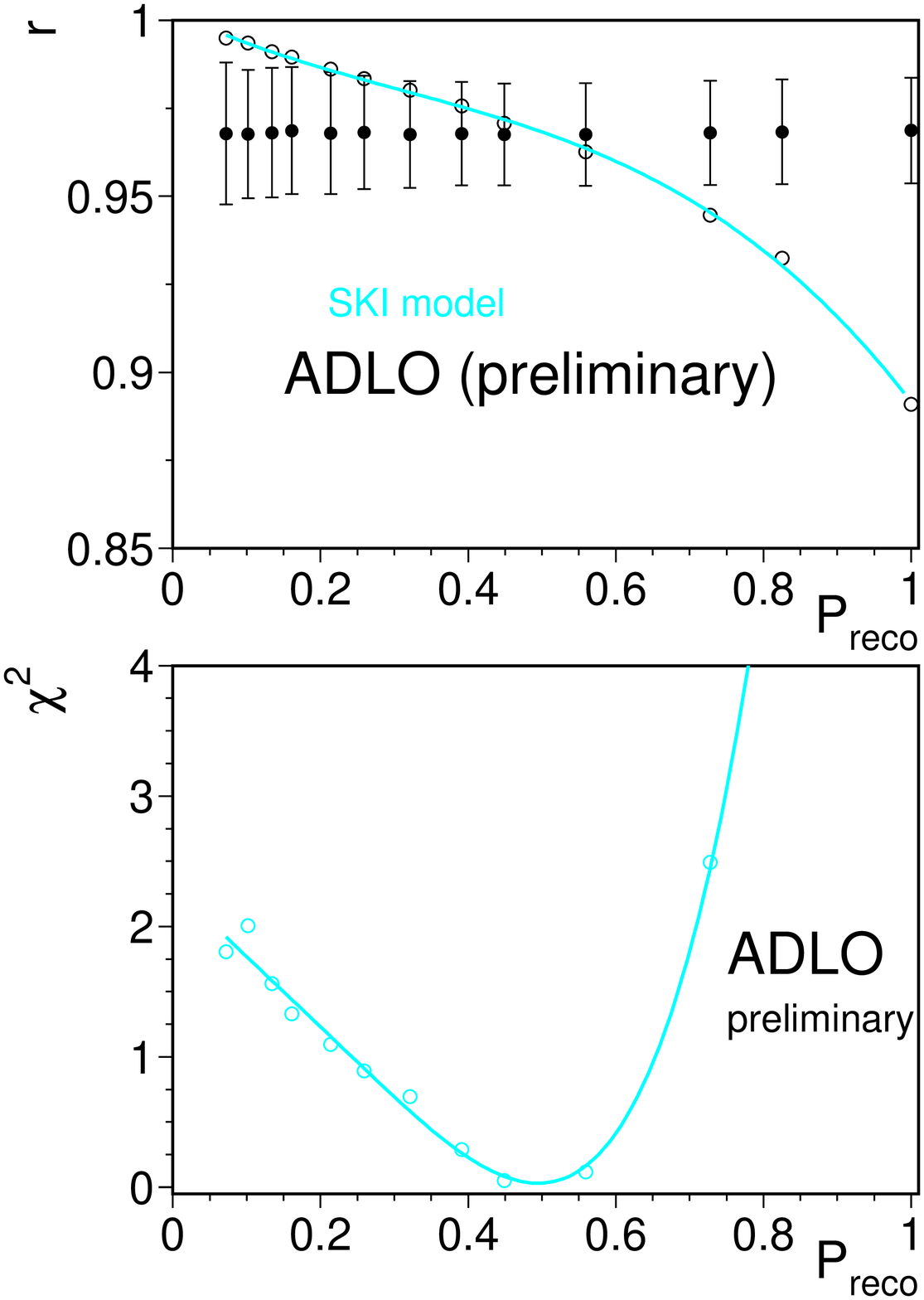,width=0.8\textwidth}}
 \caption[Constraints on \SKI\ reconnection probability.] {Comparison of the
   LEP average $r$ values with the \SKI\ model prediction obtained as
   a function of the reconnection probability.  In the upper plot, the
   solid line is the result of fitting a third order polynomial
   function to the MC predictions.  The lower plot shows a $\chi^{2}$
   curve obtained from this comparison using all LEP data at the
   reference centre-of-mass energy of 189 GeV.  The best agreement
   between the model and the data is obtained when 49\% of events are
   reconnected in this model.}
\label{fsi:cr:fig:preco_scan}
\end{figure}

 In the \SKI\ model, the reconnection probability is governed by an
 arbitrary, free parameter, \kI.  By comparing the data with model
 predictions evaluated at a variety of \kI\ values, it is possible to
 determine the reconnection probability that is most consistent with
 data, which can in turn be used to estimate the corresponding bias in
 the measured \MW.  By repeating the averaging procedure using model
 inputs for the set of \kI\ values given in
 Table~\ref{fsi:cr:tab:cetraro}, including a re-evaluation of the
 weights for each value of \kI, it is found that the data prefer a
 value of $\kI =1.18$ as shown in Figure~\ref{fsi:cr:fig:ki_scan}. The
 68\% confidence level lower and upper limits are 0.39 and 2.13
 respectively.  The LEP averages in $r$ obtained for the different \kI\
 values are summarised in Table~\ref{fsi:cr:tab:average}.  They
 correspond to a preferred reconnection probability of 49\% in this model at
 189 GeV as illustrated in Figure~\ref{fsi:cr:fig:preco_scan}.

 The small variations observed in the LEP average value of $r$ and its
 corresponding error as a function of $k_{I}$ (or $P_{reco}$) are
 essentially due to changes in the relative weighting of the
 experiments.

 \subsection{\Ariadne\ and \Herwig\ models}
\begin{figure}[tbhp]
   \centerline{\epsfig{file=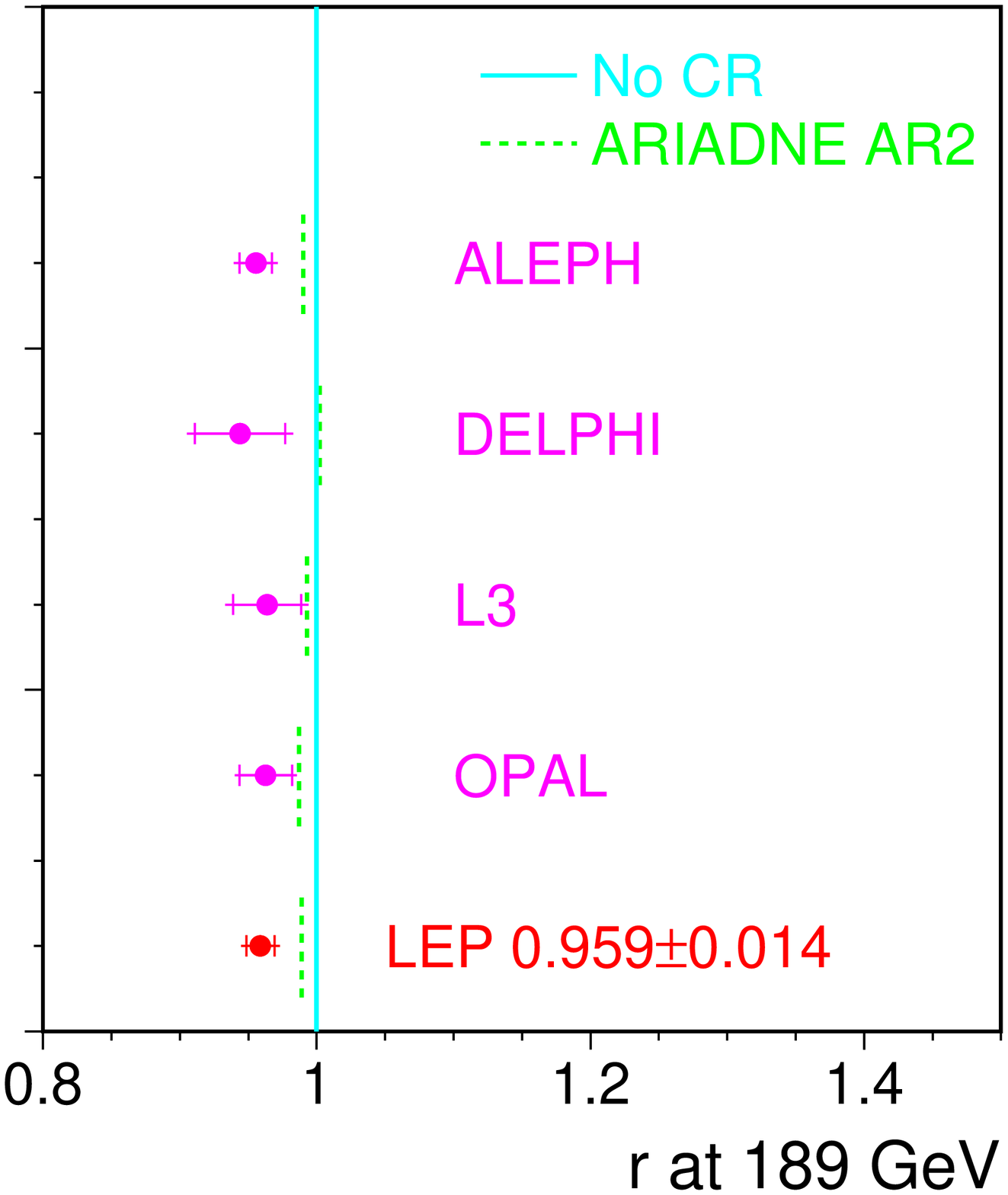,width=0.5\textwidth}
     \epsfig{file=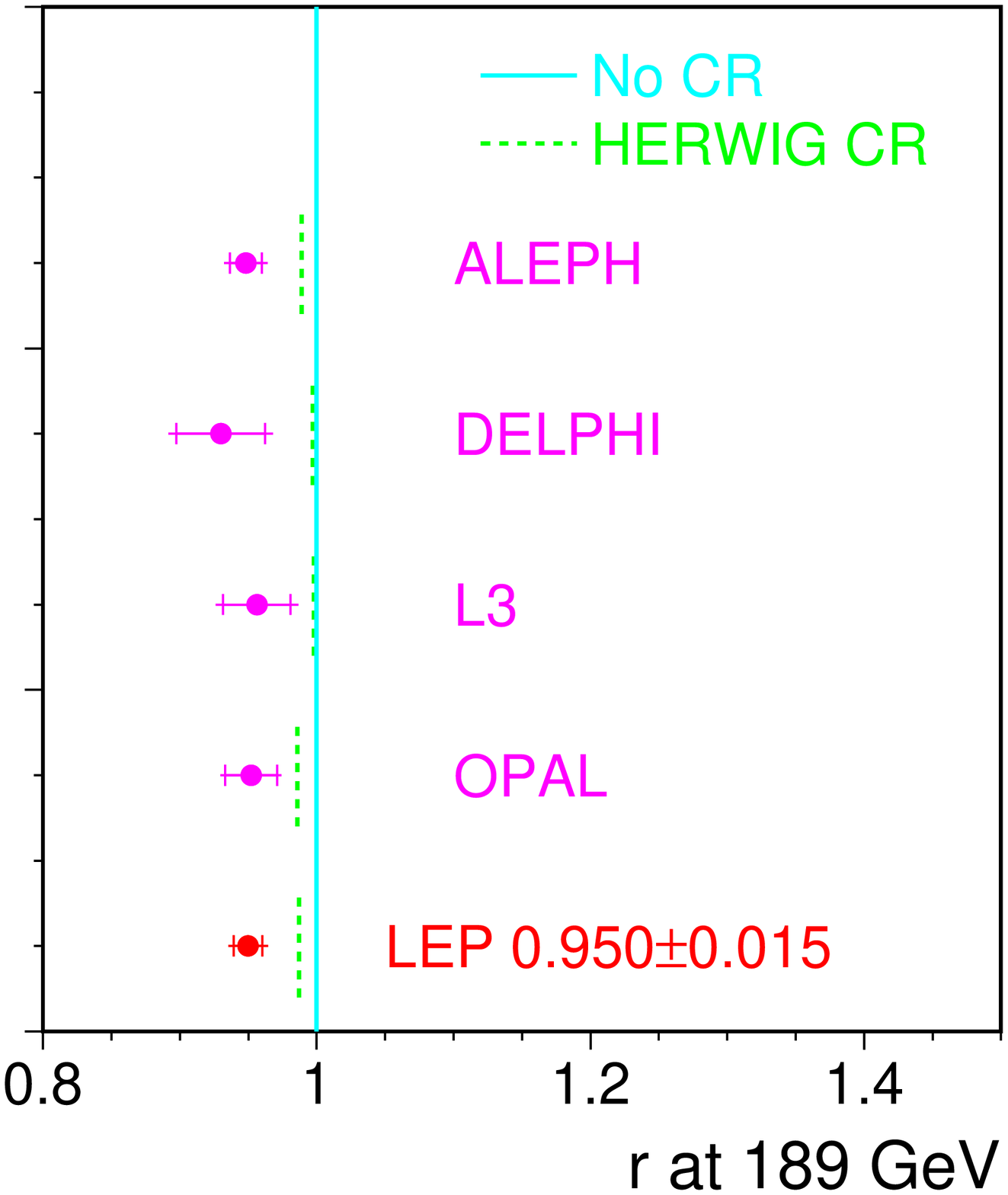,width=0.5\textwidth}}
  \caption[Particle flow combination, \ARII\ and \Herwig\ CR models.]
  {Preliminary particle flow results using all data, combined to test
    the \Ariadne\ and \Herwig\ colour reconnection models, based on
    the predicted sensitivity.  The predicted values of $r$ for this
    CR model are indicated separately for the analysis of each
    experiment by dashed lines.}
 \label{fsi:cr:fig:arhw}
 \end{figure}

 The combination procedure has been applied to common samples of
 \Ariadne\ and \Herwig\ Monte Carlo models. The \Rn\ average values
 obtained with these models based on their respective predicted
 sensitivity are summarised in Table~\ref{fsi:cr:tab:average2}.  The
 four experiments have observed a weak sensitivity to these colour
 reconnected samples with the particle flow analysis, as can be seen
 from Figure~\ref{fsi:cr:fig:arhw}.
 
\section{Summary}
 
 A first, preliminary combination of the LEP particle flow results is
 presented, using the entire LEP2 data sample.  The data disfavour by
 5.2 standard deviations an extreme version of the \SKI\ model in
 which colour reconnection has been forced to occur in essentially all
 events.  The combination procedure has been generalised to the \SKI\ 
 model as a function of its variable reconnection probability. The
 combined data are described best by the model where 49\% of events
 at 189 GeV are reconnected, corresponding to $\kI =1.18$.  The LEP
 data, averaged using weights corresponding to $\kI=1.0$, \ie\ closest
 to the optimal fit, do not exclude the no colour reconnection
 hypothesis, deviating from it by 2.2 standard deviations.  A 68\%
 confidence level range has been determined for \kI\ and corresponds
 to [0.39,2.13].
 
 For both the \Ariadne\ and \Herwig\ models, which do not contain
 adjustable colour reconnection parameters, differences between the
 results of the colour reconnected and the no-CR scenarios are small
 and do not allow the particle flow analysis to discriminate between
 them.  To test consistency between data and the no-CR models, the
 data are averaged using weights where the factor accounting for
 predicted sensitivity to a given CR model has been set to unity.  The
 \Rn\ values obtained with the no colour reconnection \Herwig\ and
 \Ariadne\ models, using the common Cetraro samples, differ from the
 measured data value by 3.7 and 3.1 standard deviations.
 
 The observed deviations of the \Rn\ values from all no colour reconnection
 models may indicate a possible systematic effect in the description
 of particle flow for 4-jet events.  Independent studies of particle
 flow in WW semileptonic events as well as other CR-oriented analyses
 are required to investigate this.

%% file: be.tex
\section{Introduction}

The LEP experiments have measured the strength of particle
correlations between two hadronic systems obtained from W-pair decay
occuring close in space-time at \LEPII.  The work presented in this
chapter is focused on so-called Bose-Einstein (BE) correlations, i.e.,
the enhanced probability of production of pairs (multiplets) of
identical mesons close together in phase space. The effect is readily
observed in particle physics, in particular in hadronic decays of the
Z boson, and is qualitatively understood as a result of
quantum-mechanical interference originating from the symmetry of
the amplitude of the particle production process under exchange of
identical mesons.

The presence of correlations between hadrons coming from the decay of
a $\WW$ pair, in particular those between hadrons originating from
different Ws, can affect the direct reconstruction of the mass of the
initial W bosons.  The measurement of the strength of these
correlations can be used for the estimation of the systematic
uncertainty of the W mass measurement.
  
\section{Method} 
 
The principal method~\cite{be:chekanov}, called ``mixing method'',
used in this measurement is based on the direct comparison of
2-particle spectra of genuine hadronic WW events and of mixed WW
events.  The latter are constructed by mixing the hadronic parts of
two semileptonic WW events (first used in ~\cite{be:DELPHI97}). Such a
reference sample has the advantage of reproducing the correlations
between particles belonging to the same W, while the particles from
different Ws are uncorrelated by construction.
  
This method gives a model-independent estimate of the interplay
between the two hadronic systems, for which BE correlations and also
colour reconnection are considered as dominant sources. The
possibility of establishing the strength of inter-W correlations in a
model-independent way is rather unique; most correlations do carry an
inherent model dependence on the reference sample. In the present
measurement, the model dependence is limited to the background
subtraction.

\section{Distributions}    

The two-particle correlations are evaluated using two-particle
densities defined in terms of the 4-momentum transfer
$Q=\sqrt{-(p_{1}-p_{2})^{2}}$, where $p_1,p_2$ are the 4-momenta of
the two particles:
\begin{equation}
\rho_{2}(Q)=\frac{1}{N_{ev}}\frac{dn_{pairs}}{dQ}
\end{equation}
Here $n_{pairs}$ stands for number of like-sign (unlike-sign)
2-particle permutations.\footnote{For historical reasons, the number
  of particle permutations rather than combinations is used in
  formulas.  For the same reason, a factor 2 appears in front of
  $\rho_{2}^{mix}$ in eq.~\ref{eq:fsi:rho-mix}. The experimental
  statistical errors are, however, based on the number of particle
  pairs, i.e., 2-particle combinations.}  In the case of two
stochastically independent hadronically decaying Ws the two-particle
inclusive density is given by:
\begin{equation}
 \rho_{2}^{WW}~=~\rho_{2}^{W^{+}}+\rho_{2}^{W^{-}}+2 \rho_{2}^{mix},
\label{eq:fsi:rho-mix}
\end{equation}
where $\rho_{2}^{mix}$ can be expressed via single-particle inclusive
density $\rho_1(p)$ as:
\begin{equation}
\rho_2^{mix}(Q)~=~\int d^{4}p_{1}d^{4}p_{2}\rho^{W^{+}}(p_1)\rho^{W^{-}}(p_2)\delta(Q^{2}+(p_{1}-p_{2})^{2})\delta(p_1^2-m_{\pi}^2)\delta(p_2^2-m_{\pi}^2).
\end{equation}
Assuming further that:
\begin{equation}
\rho_{2}^{W^{+}}(Q)~=~\rho_{2}^{W^{-}}(Q)~=~\rho_{2}^{W}(Q),
\end{equation}
we obtain:
\begin{equation}
\rho_{2}^{WW}(Q)~=~2\rho_{2}^{W}(Q)+2\rho_2^{mix}(Q).
\end{equation}
In the mixing method, we obtain $\rho_2^{mix}$ by combining two
hadronic W systems from two different semileptonic WW events.  The
direct search for inter-W BE correlations is done using the difference
of 2-particle densities:
\begin{equation}
\Delta \rho(Q) ~=~ \rho_{2}^{WW}(Q)-2\rho_{2}^{W}(Q)-2\rho_2^{mix}(Q),
\label{eq:be:dr}
\end{equation}
or, alternatively, their ratio:
\begin{equation}
    D(Q)~=~\frac{\rho_{2}^{WW}(Q)}{2\rho_{2}^{W}(Q)+2\rho^{mix}(Q)}
        ~=~ 1 + \frac{ \Delta \rho(Q) }{2\rho_{2}^{W}(Q)+2\rho^{mix}(Q)} .
\label{eq:be:D}
\end{equation}
In case of $\Delta\rho(Q)$, we look for a deviation from 0, while in
case of $D(Q)$, inter-W BE correlations would manifest themselves by
deviation from 1.  The event mixing procedure may introduce artificial
distortions, or may not fully account for some detector effects or for
correlations other than BE correlations, causing a deviation of
$\Delta\rho(Q)$ from zero or D from unity for data as well as Monte
Carlo without inter-W BE correlations.  These possible effects are
reduced by using the double ratio or the double difference:
\begin{equation}
    D'(Q)~=~\frac{D(Q)_{data}}{D(Q)_{MC,no inter}}\hspace{0.2cm} ,\hspace{1cm}
\Delta \rho'(Q) ~=~\Delta \rho(Q)_{data} - \Delta \rho(Q)_{MC,no inter}
\hspace{0.2cm} ,
\end{equation}
where $D(Q)_{MC,no inter}$ and  $\Delta \rho(Q)_{MC,no inter}$ are derived 
from a MC without inter-W BE correlations.

In addition to the mixing method, ALEPH \cite{be:ALEPH00} also uses
the double ratio of like-sign pairs ($N_{\pi}^{++,--}(Q)$) and
unlike-sign pairs $N_{\pi}^{+-}(Q)$ corrected with Monte-Carlo
simulations not including BE effects:
\begin{equation}
R^*(Q) ~=~ \left.\left(\frac{ N_{\pi}^{++,--}(Q) }
{ N_{\pi}^{+-}(Q) } \right)^{data}\right/
\left(\frac{ N_{\pi}^{++,--}(Q) }
{ N_{\pi}^{+-}(Q) }\right)^{MC}_{noBE}.
\end{equation}

\section{Results}

\begin{figure}[htbp]
\begin{center}
\epsfig{file=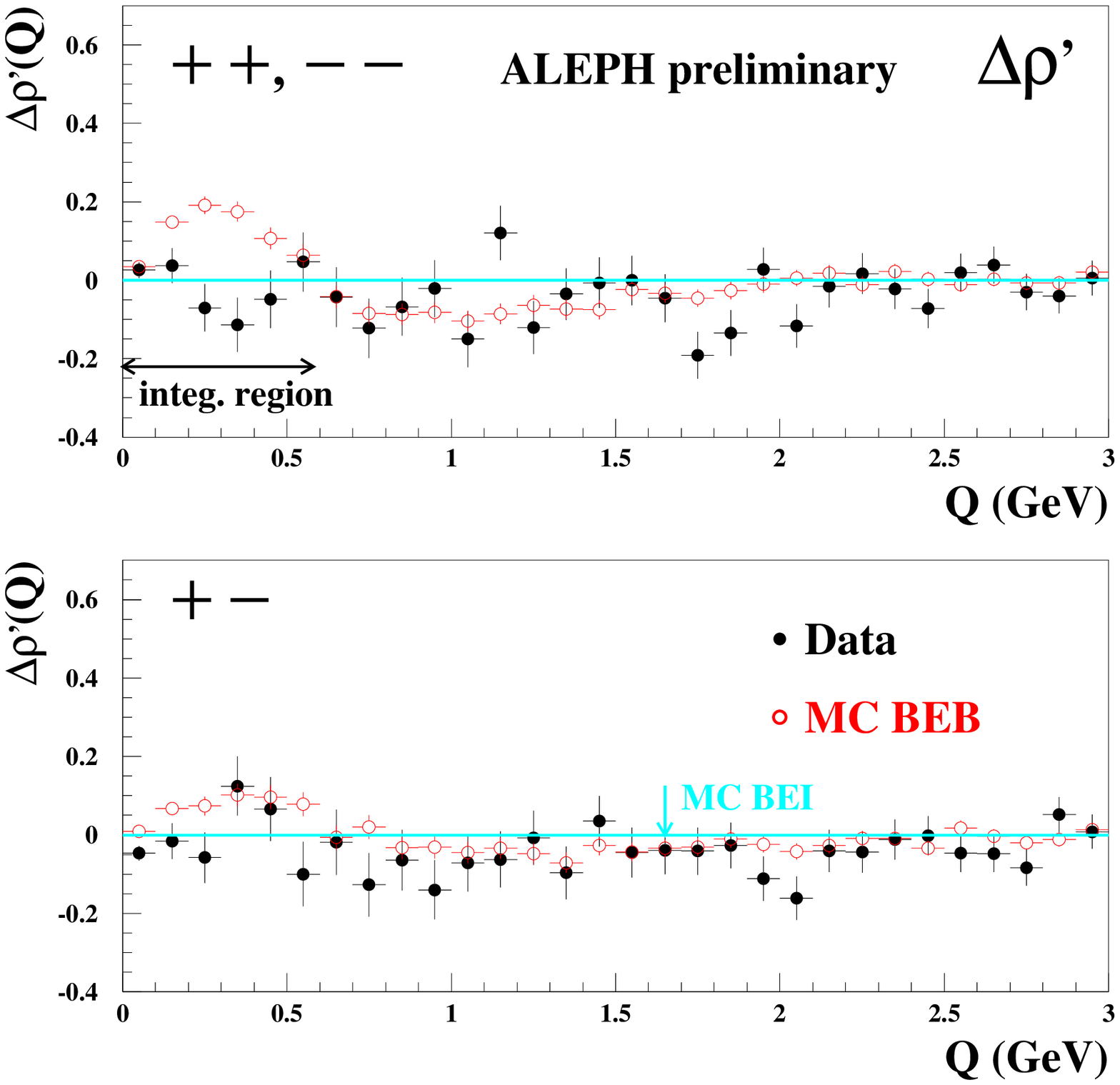,width=0.9\linewidth}
\end{center}
\caption[]{Distribution of the quantity $\Delta\rho'$ for like- and
  unlike-sign pairs as a function of
  $Q$ as measured by the ALEPH collaboration. }
\label{be:fig:aleph}
\end{figure}

\begin{figure}[htbp]
\begin{center}
\epsfig{file=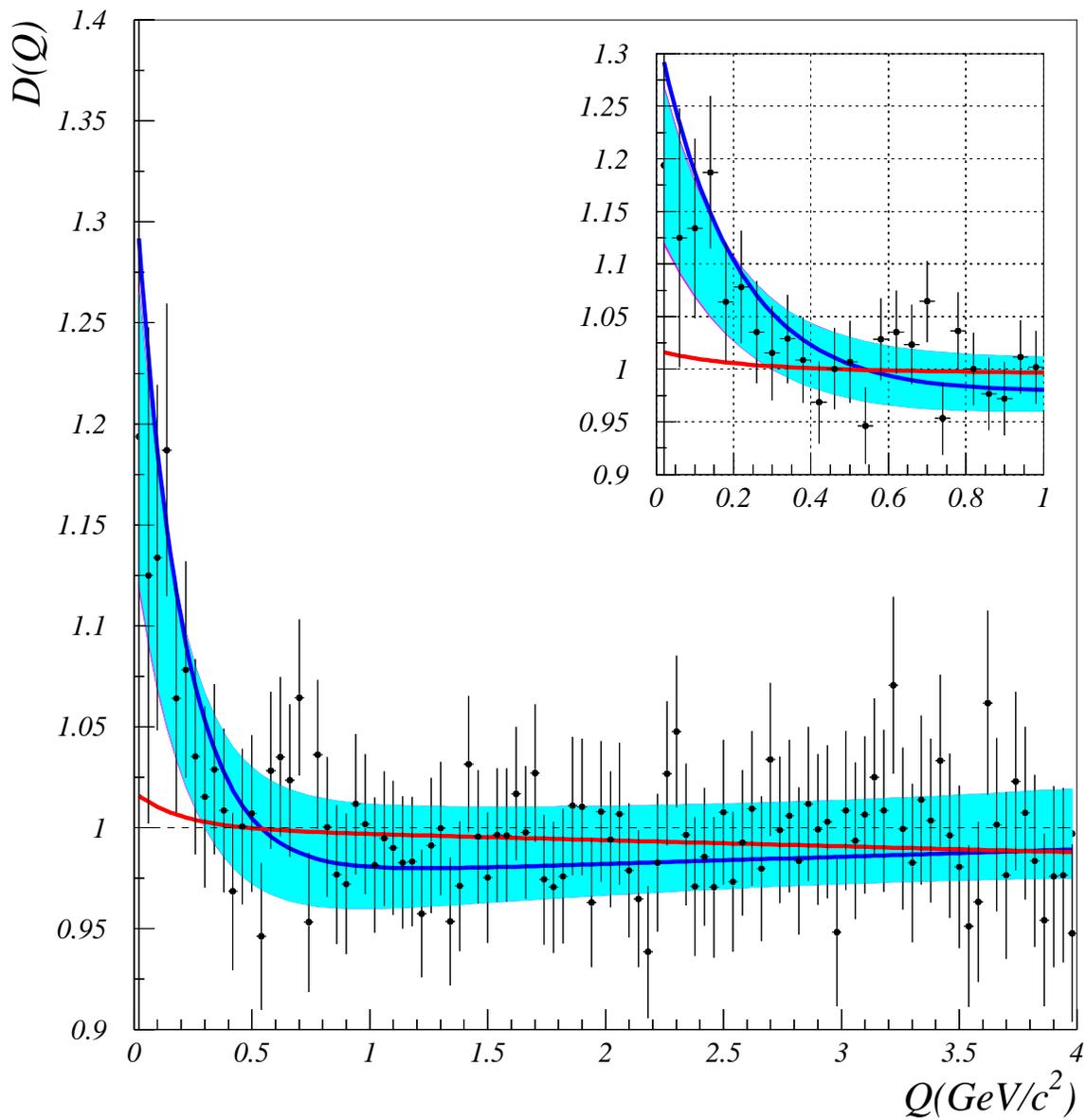,width=\linewidth}
\end{center}
\caption[]{Distributions of the quantity D for like-sign pairs 
as a function of $Q$ as measured by the DELPHI
  collaboration.The shadowed region shows the fit results. }
\label{be:fig:delphi}
\end{figure}

\begin{figure}[htbp]
\begin{center}
\epsfig{file=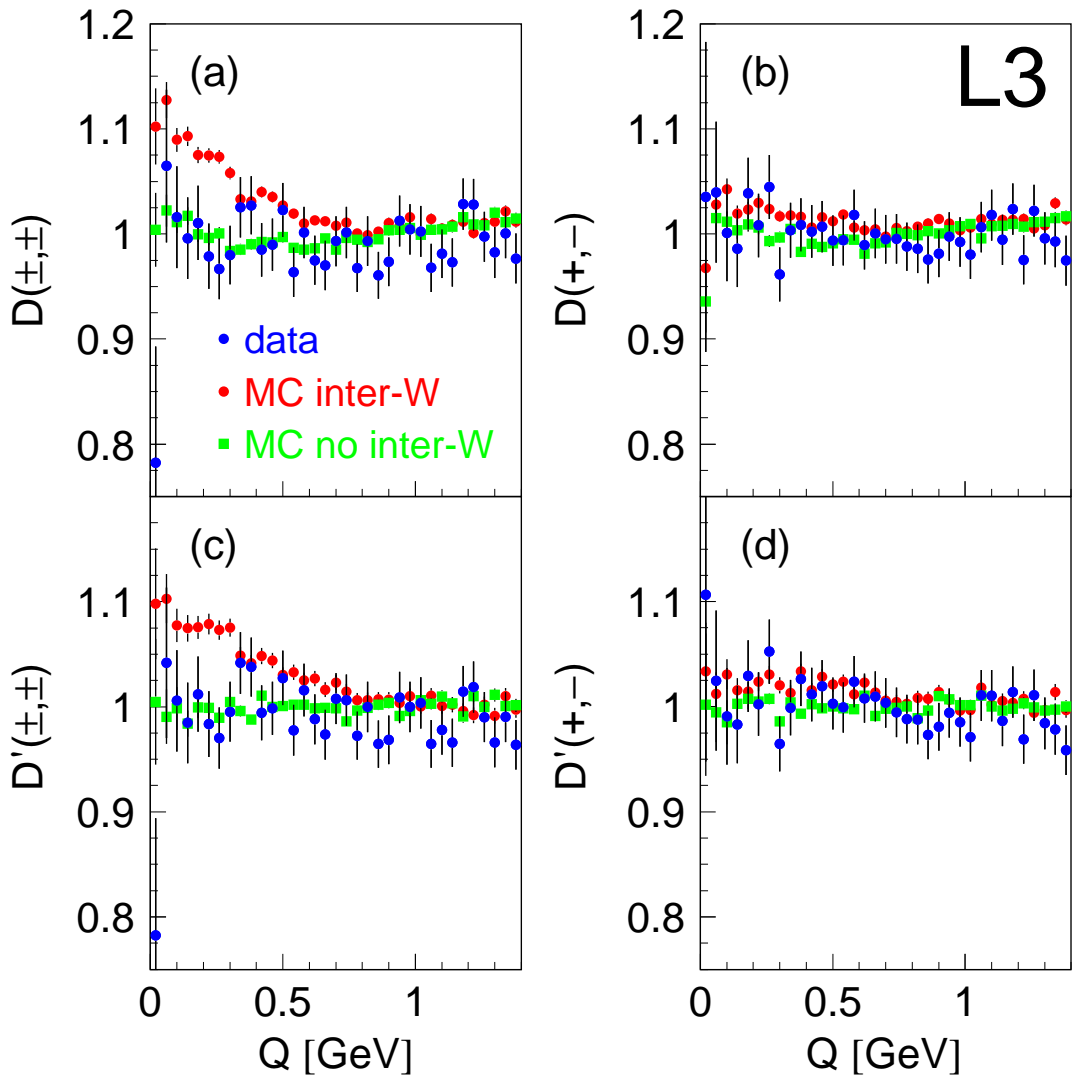,width=\linewidth}
\end{center}
\caption[]{Distributions of the quantity D and D' for like- and
  unlike-sign pairs as a function of $Q$ as measured by the L3
  collaboration.}
\label{be:fig:l3}
\end{figure}

\begin{figure}[htbp]
\begin{center}
\epsfig{file=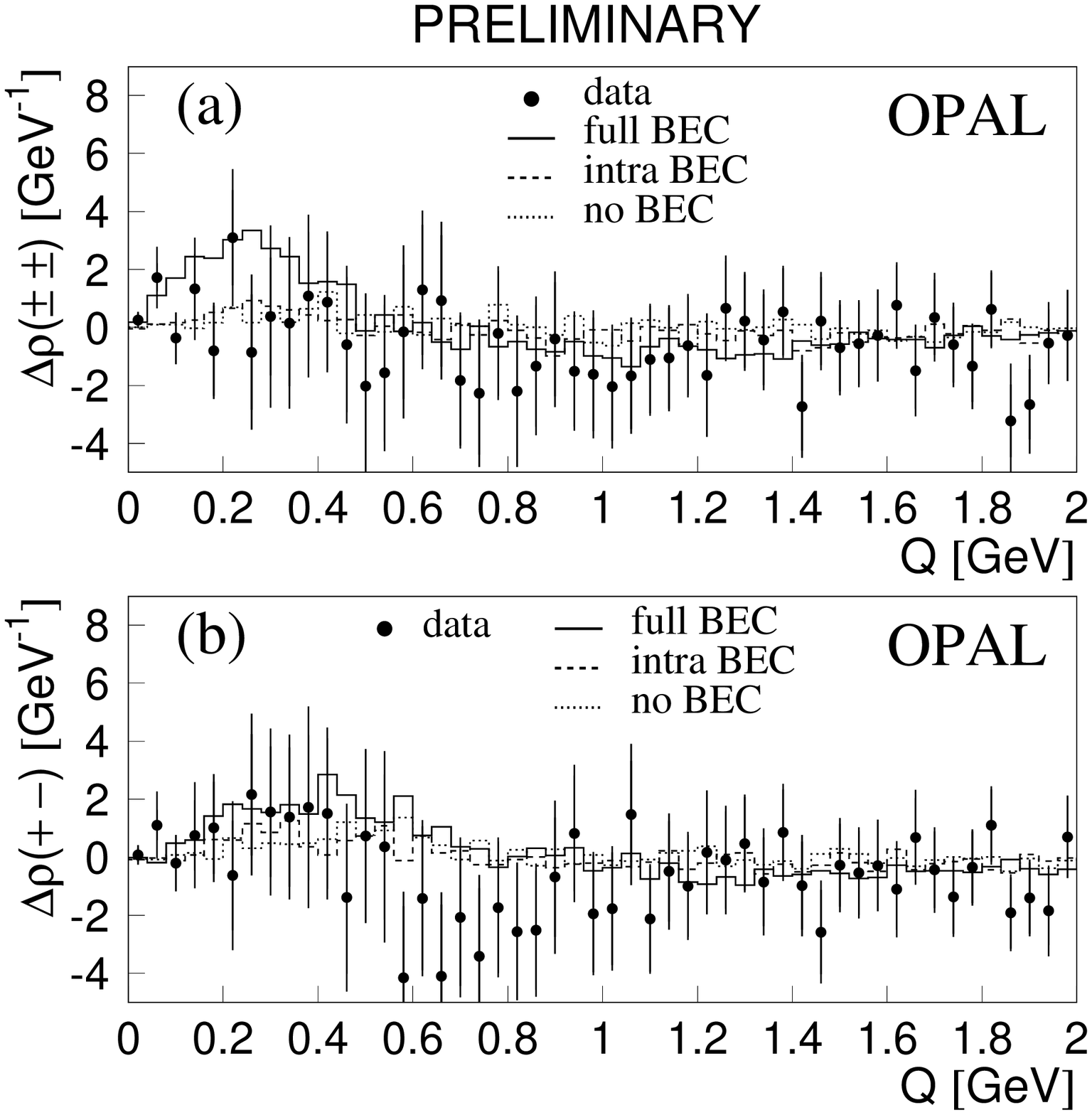,width=0.9\linewidth}
\end{center}
\caption[]{Distribution of the quantity $\Delta\rho$ for like- and
  unlike-sign pairs as a function of
  $Q$ as measured by the OPAL collaboration. }
\label{be:fig:opal}
\end{figure}

Four LEP experiment have submitted results applying the mixing method
to the full 
LEP2 data sample. 
As examples, the distributions of  $\Delta\rho$ measured by
ALEPH~\cite{be:ALEPH03}, D measured by 
DELPHI~\cite{be:DELPHI03}, D and D' measured by L3~\cite{be:L302}
and $\Delta\rho$ measured by OPAL~\cite{be:OPAL03}
are shown in Figures~\ref{be:fig:aleph}, \ref{be:fig:delphi}, 
\ref{be:fig:l3} and~\ref{be:fig:opal}, respectively.
In addition ALEPH have submitted results using $R^*(Q)$ variable
 based on data collected at
centre-of-mass energies up to $189~\GeV$ \cite{be:ALEPH00}.

A simple combination procedure is available through a $\chi^2$ average
of the numerical results of each experiment with respect to a specific
BE model under study, here based on comparisons with various (tuned)
versions of the LUBOEI
model~\cite{be:PYTHIA57,be:LoSj,be:ALEPH03,be:DELPHI03, be:L302, be:OPAL03}.  
The tuning is
performed by adjusting the parameters of the model to reproduce
correlations in samples of Z$^0$ and semileptonic W decays, and
applying identical parameters to the modelling of inter-W correlations
(so-called ``full BE'' scenario). In this way the tuning of each
experiment takes into account detector systematics in track
measurements of different experiments. 

An important advantage of the combination procedure used here is that
it allows the combination of results obtained using different
analyses.  
The combination procedure assumes a linear dependence of the observed
size of BE correlations on various estimators used to analyse the
different distributions.  It is also verified that there is a linear
dependence between the measured W mass shift and the values of these
estimators~\cite{be:lep-be}.  The estimators are: 
the integral of the $\Delta\rho(Q)$ distribution (ALEPH, L3, OPAL);
 the parameter $\Lambda$ when fitting the 
function $N(1+\delta
Q)(1+\Lambda\exp(-k^2Q^2))$ to the $D'(Q)$ distribution, with $N$ 
fixed to unity (L3), or $\delta$
fixed to zero and $k$ fixed to the value
obtained from a fit to the full BE sample (ALEPH); 
 the parameter $\Lambda$ when fitting the 
function $N(1+\delta
Q)(1+\Lambda\exp(-RQ))$ to the $D(Q)$ distribution, with $R$ fixed to the value
obtained from a fit to the full BE sample (DELPHI, L3);  
and finally the integral of the
term describing the BE correlation part, $\int
\lambda\exp(-\sigma^2Q^2)$, when fitting the function
$\kappa(1+\epsilon Q)(1+\lambda\exp(-\sigma^2Q^2))$ to the $R^*(Q)$
distribution (ALEPH).

The size of the correlations for like-sign pairs
of particles measured in terms of these estimators is compared with
the values expected in the model with and without inter-W correlations
in Table~\ref{be:table:bei}. 
Table~\ref{be:table:rel} summarizes the normalized fractions of the
model seen.
Note that DELPHI also finds a 1.9 standard deviation effect for pairs
of unlike-sign particles from different W bosons\cite{be:DELPHI03},
similar to the prediction of the LUBOEI model with full strength
correlations.
 
\begin{figure}[htbp]
   \begin{center}
     \mbox{\hspace*{-1.0cm}\epsfig{file=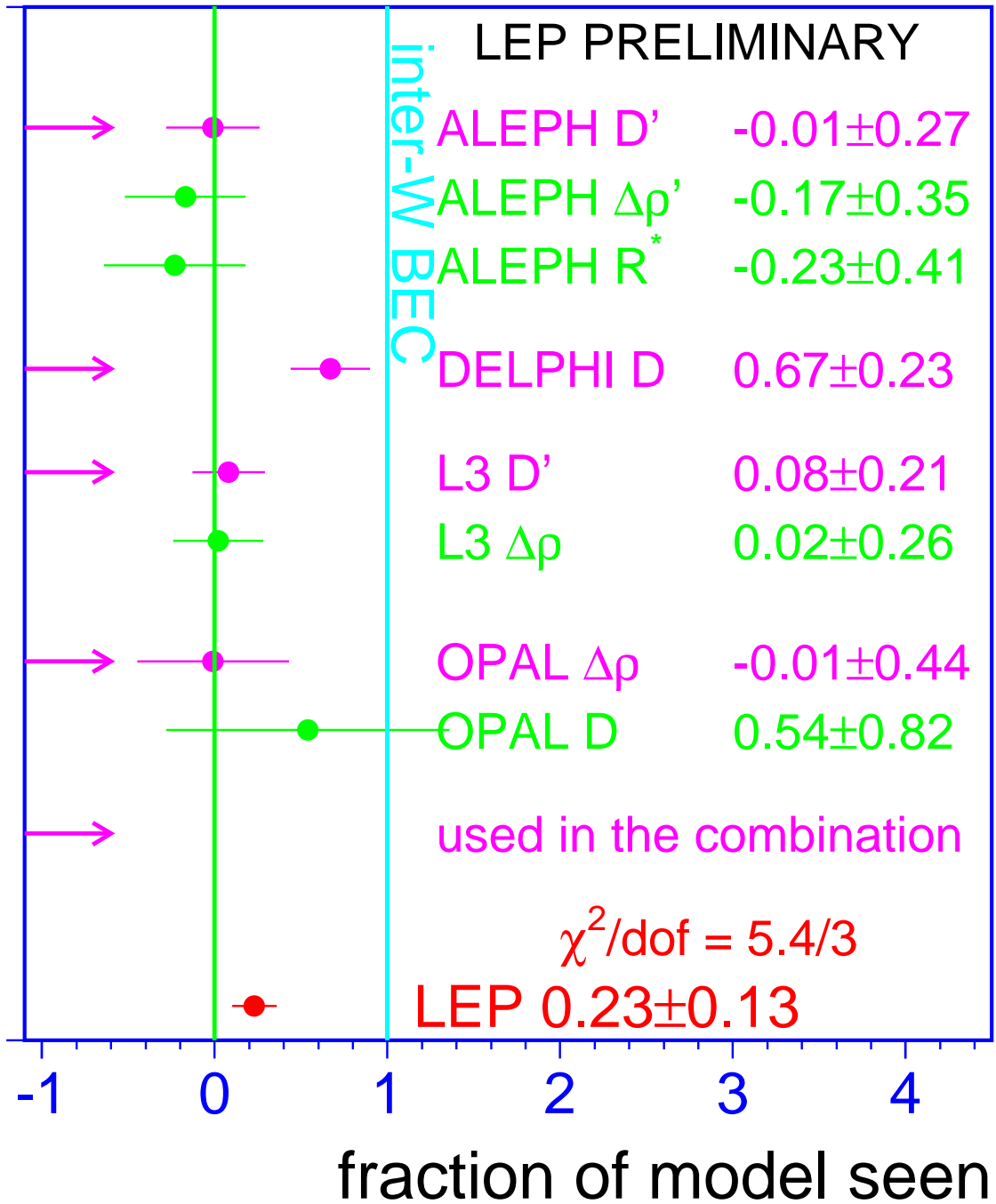,width=0.9\linewidth}}
   \end{center}
 \caption[]{%
   $\chi^2$ combination of the measured size of correlations expressed
   as the relative fraction of the model with inter-W correlations. }
 \label{be:chi-comb}
\end{figure}

For the combination of the above
measurements one has to take into account correlations between
them. Correlations between results of the same experiment are
strong and are not available. It is however found, for example, 
that taking reasonable value of these correlations 
and combining three ALEPH measurements, one obtaines the normalized 
fractions of the
model seen very close to the one of the most precise measurement.
Therefore, for simplicity, the combination of the most precise
measurements of each experiment is made here: 
D' from ALEPH, D from Delphi, D' from L3 and $\Delta\rho$ from OPAL.
In this combination
only the uncertainties in the understanding of the
background contribution in the data are treated as correlated between
experiments (denoted as ``corr. syst.''  in Table~\ref{be:table:bei}).
The combination via a MINUIT fit gives:
\begin{equation}
\frac{\mathrm{data - model(noBE)}}{\mathrm{model(fullBE)-model(noBE)}}
 ~ = ~ 0.23 \pm 0.13 \,, %
\end{equation}  
where ``noBE'' includes correlations between decay products of each W,
but not the ones between decay products of different Ws and ``fullBE''
includes all the correlations.  A $\chi^2$/dof=5.4/3 of the
fit is observed.  The measurements and their average are shown in
Figure~\ref{be:chi-comb}. The measurements used in the combination are
marked with arrow.

\begin{table}[ht]
\begin{center}
 \begin{tabular}{|l|l|l|l|l|l|c|}
 \hline
  & data--noBE &  stat. & syst. & corr. syst. &fullBE--noBE & Ref. \\ \hline
ALEPH (fit to D')  &  $-$0.001 & 0.015 & 0.014 & 0.002 &  0.077 &\cite{be:ALEPH03} \\
ALEPH (integral of $\Delta\rho$)  &  $-$0.124 & 0.148 & 0.200 & 0.001 &  0.720 &\cite{be:ALEPH03} \\  
ALEPH (fit to $R^*$)  &  $-$0.004 & 0.0062 & 0.0036 & negligible &  0.0177 &\cite{be:ALEPH00} \\ 
DELPHI (fit to $D$) & $+$0.241 & 0.075 & 0.038 & 0.017 & 0.36 &\cite{be:DELPHI03} \\
L3 (fit to $D'$)  & $+$0.008 & 0.018 & 0.012 & 0.004 &  0.103 & \cite{be:L302} \\
L3 (integral of $\Delta\rho$) &   $+$0.03 & 0.33 &  0.15 &  0.055 & 1.38 & \cite{be:L302} \\ 
OPAL (integral of $\Delta\rho$) &   $-$0.01 & 0.27 &  0.21 &  negligible & 0.77 & \cite{be:OPAL03} \\
OPAL (fit to D) &   $+$0.069 & 0.105 &  0.069 &  0.010 & 0.139 & \cite{be:OPAL03} \\
\hline 
\end{tabular}
\caption[]{
  An overview of the input values for the $\chi^2$ combination: the
  difference between the measured correlations and the model without
  inter-W correlations (data--noBE), the corresponding statistical
  (stat.)~and total systematic (syst.)~errors, the correlated
  systematic error contribution (corr.~syst.), and the difference
  between ``full BE'' and ``no BE'' scenario.  }
\label{be:table:bei}
\end{center}
\end{table}

\begin{table}[ht]
\begin{center}     
 \begin{tabular}{|l|c|c|c|}
 \hline
  & fraction of the model &  stat. & syst.  \\ \hline
ALEPH (fit to D') &  $-$0.01 & 0.19 & 0.18  \\ 
ALEPH (integral of $\Delta\rho$) &  $-$0.17 & 0.21 & 0.28  \\ 
ALEPH (fit to $R^*$) &  $-$0.23 & 0.35 & 0.20  \\ 
DELPHI (fit to $D$) &   $+$0.67 & 0.21 & 0.11  \\
L3  (fit to $D'$) &   $+$0.08 & 0.17 & 0.12  \\
L3 (integral of $\Delta\rho$) &   $+$0.02 & 0.24 & 0.11  \\ 
OPAL (integral of $\Delta\rho$) &   $-$0.01 & 0.35 & 0.27  \\
OPAL  (fit to $D$) &   $+$0.50 & 0.76 & 0.50  \\
\hline 
 \end{tabular}
\caption[]{
  The measured size of correlations expressed as the relative fraction
  of the model with inter-W correlations.}
\label{be:table:rel}
\end{center}
\end{table}

The result of the $\chi^2$ combination of the measurements 
can be translated into a 68\% confidence level upper limit on
the shift of the W mass measurements due to the BE correlations
between particles from different Ws, $\Delta\MW$, assuming a linear
dependence of $\Delta\MW$ on the size of the correlation.  For the
specific BE model investigated, LUBOEI, a shift of $-35~\MeV$ in the W
mass is obtained at full BE correlation strength\cite{be:LEPW}. Thus
the preliminary 68\% CL upper limit on the magnitude of the mass shift
within the LUBOEI model is:
\begin{eqnarray}
 |\Delta\MW| & = & (0.23+0.13) \cdot 35~\MeV
             ~ = ~ 13~\MeV~~ (+1~\sigma~\mathrm{limit})\,.
\end{eqnarray}

%% file: mw.tex
\section{Introduction}

The W boson mass and width results presented in this chapter are obtained from
data recorded over a range of centre-of-mass energies,
$\roots=161-209$~\GeV, during the 1996-2000 operation of the LEP
collider. The results reported by the ALEPH, DELPHI and L3
collaborations include an analysis of the year 2000 data, and have an
integrated luminosity per experiment of about $700$~\ipb. The OPAL
collaboration has analysed the data up to and including 1999 and has
an integrated luminosity of approximately $450$~\ipb. The ALEPH result does not include an analysis of the small amount of data (about $10$~\ipb) collected in 1996 at a centre-of-mass energy of 172 GeV.

The results on the W mass and width quoted below correspond to a 
definition based on a Breit-Wigner denominator with an $s$-dependent width,
$|(s-\Mw^2) + i s \Gw/\Mw|$. 

\section{W Mass Measurements}

Since 1996 the LEP $\epem$ collider has been operating above the
threshold for $\WW$ pair production. Initially, 10~\ipb\ of data
were recorded close to the $\WW$ pair production threshold. At this
energy the $\WW$ cross section is sensitive to the W boson mass, $\Mw$.
Table \ref{mw:tab:wmass_threshold} summarises the W mass results from the four 
LEP collaborations based 
on these data~\cite{common_bib:adloww161}.
\begin{table}[htbp]
 \begin{center}
  \begin{tabular}{|r|c|}\hline
     \multicolumn{2}{|c|}{THRESHOLD ANALYSIS~\cite{common_bib:adloww161}} \\
Experiment &   \Mw(threshold)/\GeVm     \\ \hline
   ALEPH    & $80.14\pm0.35$  \\ 
   DELPHI   & $80.40\pm0.45$  \\
   L3       & $80.80^{+0.48}_{-0.42}$  \\
   OPAL     & $80.40^{+0.46}_{-0.43}$  \\ \hline
\end{tabular}
 \caption{W mass measurements from the $\WW$ threshold cross section 
          at $\roots=161$~\GeV. The errors
          include statistical and systematic contributions.}
 \label{mw:tab:wmass_threshold}
\end{center}
\end{table} 

Subsequently LEP has operated at energies significantly above the
$\WW$ threshold, where the $\epem\rightarrow\WW$ cross section has
little sensitivity to $\Mw$. For these higher energy data $\Mw$ is
measured through the direct reconstruction of the W boson's invariant
mass from the observed jets and leptons.  Table
\ref{mw:tab:wmass_experiments} summarises the W mass results presented
individually by the four LEP experiments using the direct
reconstruction method.  The combined values of $\Mw$ from each
collaboration take into account the correlated systematic
uncertainties between the decay channels and between the different
years of data taking. In addition to the combined numbers, each
experiment presents mass measurements from $\WWqqln$ and $\WWqqqq$
channels separately.  The DELPHI and OPAL collaborations provide results from independent fits to the data in the $\qqln$ and $\qqqq$
decay channels separately and hence account for correlations between
years but do not need to include correlations between the two channels. The
$\qqln$ and $\qqqq$ results quoted by the ALEPH and L3 collaborations
are obtained from a simultaneous fit to all data which, in addition to
other correlations, takes into account the correlated systematic
uncertainties between the two channels. The L3 result is unchanged
when determined through separate fits.  The 
systematic uncertainties in the $\WWqqqq$ channel show
a large variation between experiments; this is caused by
differing estimates of the possible effects of Colour Reconnection
(CR) and Bose-Einstein Correlations (BEC), discussed below.
The systematic errors in the $\WWqqln$ channel are dominated by
uncertainties from hadronisation, with estimates ranging from
15 to 30~$\MeVm$.

The results presented in this note differ from those in the previous
combination \cite{bib-EWEP-02} due to revised measurements from the
ALEPH Collaboration~\cite{mw:bib:ALEPHRevised}; otherwise the results
are identical. The ALEPH measurements have been revised due to a
change in their event reconstruction algorithm. This change makes the
analysis less sensitive to detector simulation inaccuracies which were
not taken into account in their previous preliminary result.

\begin{table}[htbp]
 \begin{center}
  \begin{tabular}{|r|c|c||c|}\hline
    \multicolumn{1}{|c|}{ } & \multicolumn{3}{c|}{DIRECT RECONSTRUCTION } \\
           & \WWqqln         & \WWqqqq         & Combined        \\   
Experiment & \Mw/\GeVm        & \Mw/\GeVm        & \Mw/\GeVm        \\ \hline
     ALEPH \cite{mw:bib:ALEPHRevised}
           & $80.375 \pm 0.062$ & $80.431 \pm 0.117$ & $80.385 \pm 0.058$ \\ 
     DELPHI \cite{common_bib:delww172,mw:bib:D-mw183,mw:bib:D-mw189,mw:bib:D-mw20X}  
           & $80.414 \pm 0.089$ & $80.374 \pm 0.119$ & $80.402 \pm 0.075$ \\ 
     L3 \cite{common_bib:ltrww172,mw:bib:L-mw183,mw:bib:L-mw189,mw:bib:L-mw19X,mw:bib:L-mw20X}      
           & $80.314 \pm 0.087$ & $80.485 \pm 0.127$ & $80.367 \pm 0.078$ \\ 
     OPAL\cite{common_bib:opaww172,mw:bib:O-mw183,mw:bib:O-mw189,mw:bib:O-mw19X,mw:bib:O-mwlvlv}
           & $80.516 \pm 0.073$ & $80.407 \pm 0.120$ & $80.495 \pm 0.067$ \\ \hline
\end{tabular}
 \caption{Preliminary W mass measurements from direct reconstruction
         ($\roots=172-209$~\GeV). Results are given for the
         semi-leptonic, fully-hadronic channels and the combined value.
           The $\WWqqln$ results from the OPAL collaboration include mass information from 
         the $\WWlnln$ channel. The results given here differ from those in the publications of the individual experiments as they have been recalculated imposing common FSI uncertainties.}
 \label{mw:tab:wmass_experiments}
\end{center}
\end{table}

\section{Combination Procedure}
 
A combined LEP W mass measurement is obtained from the results
of the four experiments. In order to perform a reliable combination of
the measurements, a more detailed input than that given in
Table~\ref{mw:tab:wmass_experiments} is required.  Each experiment
provided a W mass measurement for both the $\WWqqln$ and $\WWqqqq$
channels for each of the data taking years (1996-2000) that it had
analysed. In addition to the four threshold measurements a total of 36
direct reconstruction measurements are supplied: DELPHI
provided 10 measurements (1996-2000), L3 gave 8 measurements
(1996-2000) having already combined the 1996 and 1997 results, ALEPH provided 8 measurements (1997-2000) and OPAL also gave 8 measurements (1996-1999). The $\WWlnln$ channel is also
analysed by the OPAL(1997-1999)
collaboration; the lower precision results obtained from this channel
are combined with the $\WWqqln$ channel mass
determinations.

Subdividing the results by data-taking years enables a proper
treatment of the correlated systematic uncertainty from the LEP beam
energy and other dependences on the centre-of-mass energy or
data-taking period.  A detailed breakdown of the sources of systematic
uncertainty are provided for each result and the correlations
specified. The inter-year, inter-channel and inter-experiment
correlations are included in the combination. The main sources of
correlated systematic errors are: colour reconnection, Bose-Einstein
correlations, hadronisation, the LEP beam energy, and uncertainties
from initial and final state radiation. The full correlation matrix
for the LEP beam energy is employed\cite{mw:bib:energy}.
The combination is performed and the evaluation of the components of
the total error assessed using the Best Linear Unbiased Estimate
(BLUE) technique, see Reference~\citen{common_bib:BLUE}.

\begin{figure}[tbp]
\begin{center}
 \mbox{\epsfxsize=9.5cm\epsffile{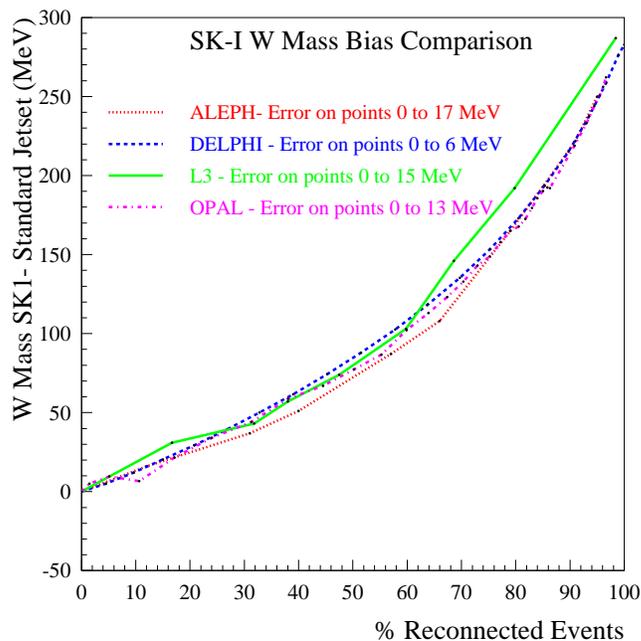}}
\caption{W mass bias obtained in the SK-I model of colour reconnection
  relative to a simulation without colour reconnection as a function
  of the fraction of events reconnected for the fully-hadronic decay channel
  at a centre of mass energy of 189 \GeV .  The
  analyses of the four LEP experiments show similar sensitivity to
  this effect. The points connected by the lines have correlated
  uncertainties increasing to the right in the range indicated.}
 \label{mw:fig:sk1}
\end{center}
\end{figure}

A preliminary study of colour reconnection has been made by the LEP
experiments using the particle flow method \cite{mw:bib:CRcomb} on a
sample of fully-hadronic WW events, see chapter~\ref{sec-CR}.  These
results are interpreted in terms of the reconnection parameter $k_i$
of the SK-I model \cite{mw:bib:ski} and yield a $68\%$ confidence
level range of:
\begin{eqnarray}
 0.39 < k_i < 2.13 \,.
\end{eqnarray}
The method was found to be insensitive to the \HERWIG\ and \ARIADNE-II
models of colour reconnection.

\begin{figure}[tbp]
\begin{center}
 \mbox{\epsfxsize=9.5cm\epsffile{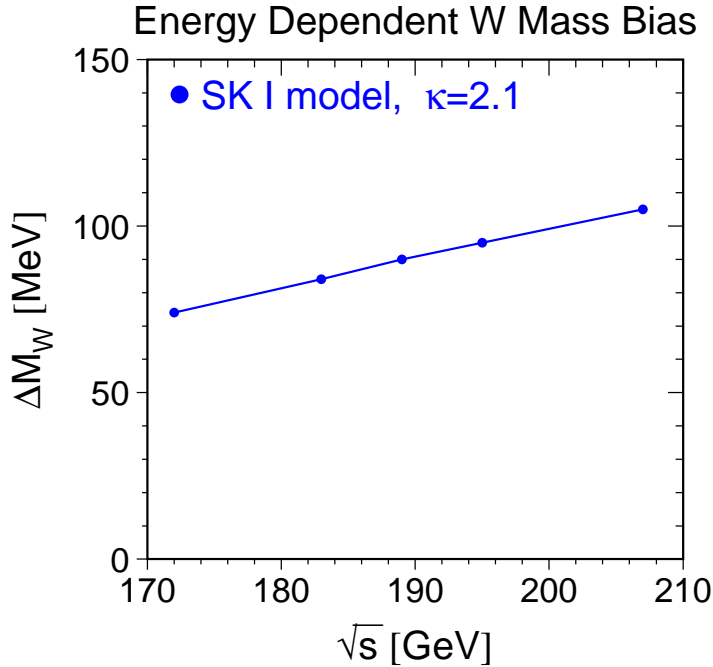}}
\caption{The values used in the W Mass combination for the uncertainty due to colour reconnection are shown as a function of the centre of mass energy. These values were obtained from a linear fit to simulation results obtained with the SK1 model of colour reconnection at $k_i = 2.13$. }
 \label{mw:fig:sk1cm}
\end{center}
\end{figure}

Studies of simulation samples have demonstrated that the four
experiments are equally sensitive to colour reconnection effects, {\em
  i.e.} when looking at the same CR model similar biases are seen by
all experiments. This is shown in Figure \ref{mw:fig:sk1} for the SKI
model as a function of the fraction of reconnected events.  For this
reason a common value for all experiments of the CR systematic
uncertainty is used in the combination.

For this combination, no offset has been applied to the central value
of \Mw\ due to colour reconnection effects and a symmetric systematic
error has been imposed. The $\Mw$ error is set from a linear
extrapolation of simulation results obtained at $k_i = 2.13$, the
values used in the combination ere: $74~\MeVm$ shift for the 1996
data at a centre-of-mass energy of $172~\GeV$, $84~\MeVm$ for 1997 at
$183~\GeV$, $90~\MeV$ for 1998 at $189~\GeV$, $95~\MeV$ for 1999 at
$195~\GeV$ and $105~\MeVm$ for 2000 at $207~\GeV$, they are shown in
Figure \ref{mw:fig:sk1cm}. Previous \Mw\ combinations have relied upon
theoretical expectations of colour reconnection effects, in which
there is considerable uncertainty. This new data driven approach
achieves a more robust uncertainty estimate at the expense of a
significantly increased colour reconnection uncertainty. The
\ARIADNE-II and \HERWIG\ models of colour reconnection have also been
studied and the W Mass shift was found to be lower than that from SK1
with $k_i = 2.13$ used for the combination.
 
\label{mw:sec:cr}

For Bose-Einstein correlations, a similar test has been made of the
respective experimental sensitivities with the LUBOEI
\cite{mw:bib:LUBOEI} model: the experiments observed compatible mass
shifts. A common value of the systematic uncertainty from BEC of
35~\MeVm\ is assumed from studies of the LUBOEI model. This value may
be compared with recent direct measurements from LEP of this effect, 
Chapter~\ref{sec-BE}, where the observed
Bose-Einstein effect was of smaller magnitude than in the LUBOEI
model, see chapter~\ref{sec-BE}. Hence, the currently assigned
35~\MeVm\ uncertainty is considered a conservative estimate.

\section{LEP Combined W Boson Mass }

The combined W mass from direct reconstruction is
\begin{eqnarray}
        \Mw(\mathrm{direct}) = 80.412\pm0.029(\mathrm{stat.})\pm0.031(\mathrm{syst.})~\GeVm,
\end{eqnarray}
with a $\chi^2$/d.o.f. of 28.2/33, corresponding to a $\chi^2$ probability 
of 70\%.
The weight of the fully-hadronic channel in the combined fit is
0.10. This reduced weight is a consequence of the relatively large
size of the current estimates of the systematic errors from 
CR and BEC. Table \ref{mw:tab:errors} gives a breakdown of the contribution
to the total error of the various sources of systematic errors. The largest
contribution to the systematic error comes from hadronisation uncertainties,
which are conservatively treated as correlated between the two channels, between experiments and between years. In the absence of systematic effects the current LEP statistical precision on $\Mw$ would be $21$~$\MeVm$: the statistical error contribution in the LEP combination is larger than this (29~$\MeVm$) due to the significantly reduced weight of the fully-hadronic channel.
\begin{table}[tbp]
 \begin{center}
  \begin{tabular}{|l|r|r||r|}\hline
       Source  &  \multicolumn{3}{|c|}{Systematic Error on \Mw\ ($\MeVm$)}  \\  
                             &  \qqln & \qqqq  & Combined  \\ \hline   
 ISR/FSR                     &  8 &  8 &  8 \\
 Hadronisation               & 19 & 18 & 18 \\
 Detector Systematics        & 14 & 10 & 14 \\
 LEP Beam Energy             & 17 & 17 & 17 \\
 Colour Reconnection         & $-$& 90 &  9 \\
 Bose-Einstein Correlations  & $-$& 35 &  3 \\
 Other                       &  4 &  5 &  4 \\ \hline
 Total Systematic            & 31 & 101 & 31 \\ \hline
 Statistical                 & 32 & 35 & 29 \\ \hline\hline
 Total                       & 44 & 107 & 43 \\ \hline
 & & & \\
 Statistical in absence of Systematics  & 32 & 28 & 21 \\ \hline

\end{tabular}
 \caption{Error decomposition for the combined LEP W mass results. 
          Detector systematics include uncertainties
          in the jet and lepton energy scales and resolution. The `Other'
          category refers to errors, all of which are uncorrelated
          between experiments, arising from: simulation statistics,
          background estimation, four-fermion treatment, fitting method 
          and event selection. The error decomposition 
          in the $\qqln$ and $\qqqq$
          channels refers to the independent fits to the results from 
          the two channels separately.}
 \label{mw:tab:errors}
\end{center}
\end{table}

In addition to the above results, the W boson mass is measured at
LEP from the 10~\ipb\ per experiment of
data recorded at threshold for W pair production:
\begin{eqnarray}
      {\Mw(\mathrm{threshold}) = 
  80.40\pm0.20(\mathrm{stat.})\pm
          0.07(\mathrm{syst.})\pm0.03(\mathrm{E_{beam}})~\GeVm}.
\end{eqnarray}
When the threshold measurements are combined with the much more precise results obtained from direct reconstruction one achieves a W mass measurement of 
\begin{eqnarray}
           \Mw = 80.412\pm0.029(\mathrm{stat.})\pm0.031(\mathrm{syst.}) \GeVm.
\end{eqnarray}
The LEP beam energy uncertainty is the only correlated systematic error source 
between the threshold and direct reconstruction measurements. 
The threshold measurements have a weight of only $0.03$ in the combined fit.
This LEP combined result is compared with the  results (threshold and direct reconstruction combined) of the four LEP experiments in Figure \ref{mw:fig:mwgw}.

\section{Consistency Checks}

The difference between the combined W boson mass measurements
obtained from the fully-hadronic and semi-leptonic channels,
$\Delta\Mw(\qqqq-\qqln)$, is determined:
\begin{eqnarray*}
 \Delta\Mw(\qqqq-\qqln) =  +22\pm43~\MeVm.    
\end{eqnarray*}

A significant non-zero value for $\Delta\Mw$ could indicate that CR and BEC
effects are biasing the value of \Mw\ determined from \WWqqqq\ events.
Since $\Delta\Mw$ is primarily of interest as a check of the possible
effects of final state interactions, the errors from CR and BEC are
set to zero in its determination. The result is obtained from a fit
where the imposed correlations are the same as those for the results
given in the previous sections. This result is almost unchanged if the
systematic part of the error on $\Mw$ from hadronisation effects is
considered as uncorrelated between channels, although the uncertainty
increases by 16\%: $\Delta\Mw=19\pm50~\MeVm$. 

\begin{figure}[hbt]
\begin{center}
 \mbox{\epsfxsize=8.25cm\epsffile{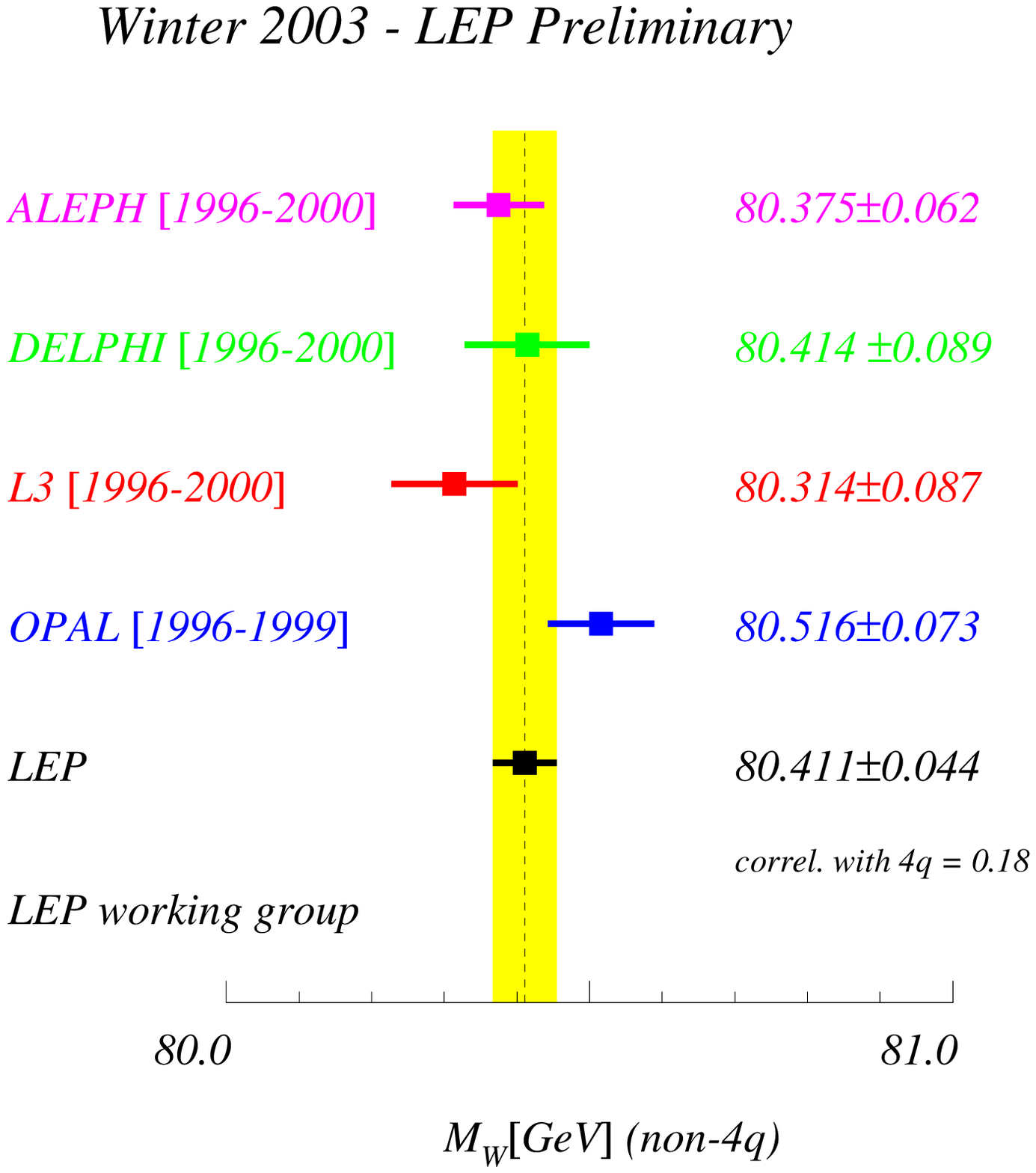} \newline
       \epsfxsize=8.25cm\epsffile{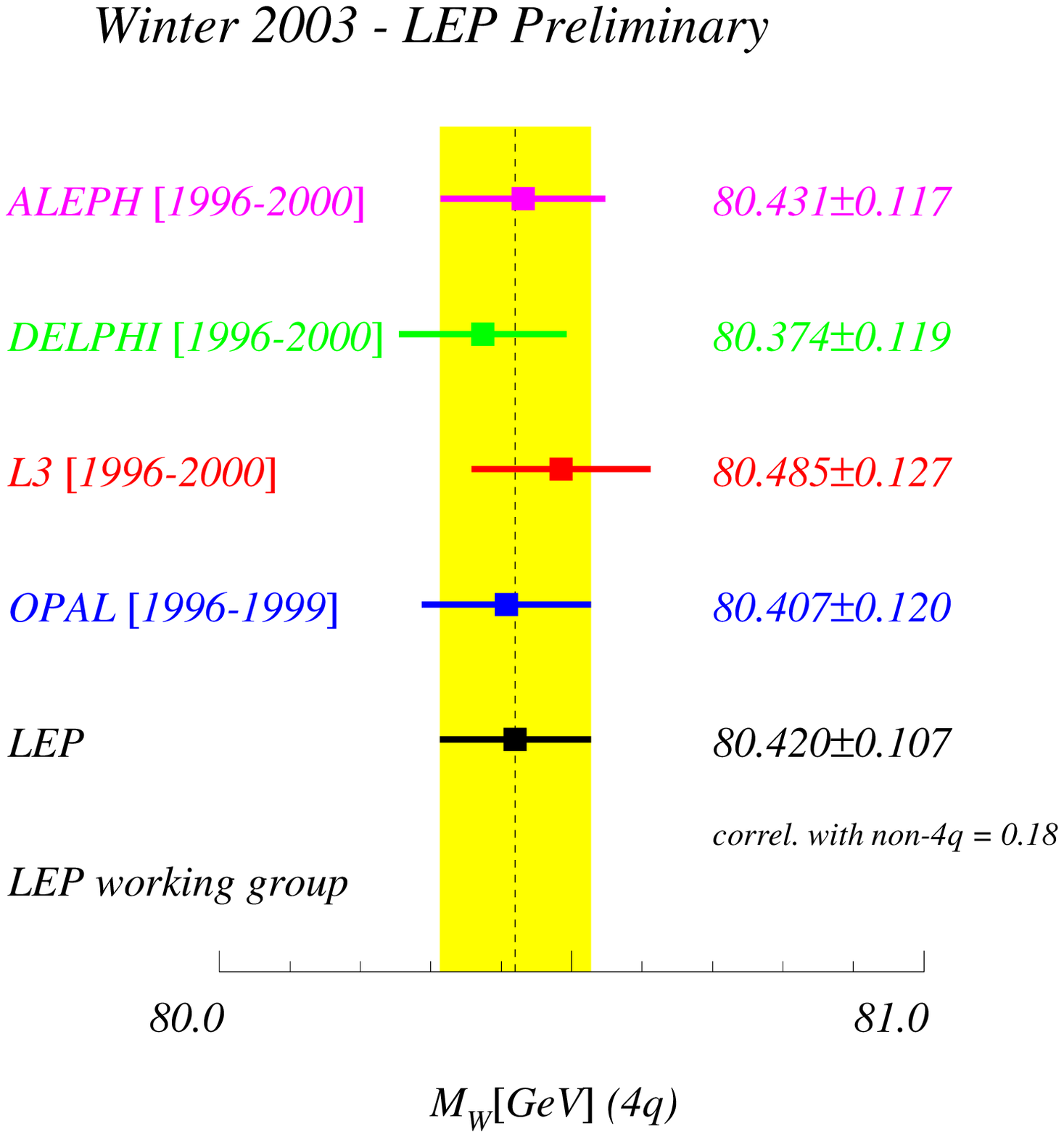}} 
\vspace*{-0.5cm}
\caption{\label{mw:fig-qqlnqqqq} 
          The W mass measurements
          from the $\WWqqln$ (left) and $\WWqqqq$ (right) channels 
          obtained by the four LEP collaborations compared to the 
          combined value. The combined values take into account 
          correlations between experiments, years and the two channels.
          In the LEP combination of 
          the $\qqqq$ results common values (see text) for the CR and BEC
          errors are used. 
          The ALEPH and L3 $\qqln$ and $\qqqq$ results are
          correlated since they are obtained from a fit to both channels 
          taking into account inter-channel correlations. The $\Mw$ values from the experiments have been recalculated 
          for this plot including the common LEP CR and BEC errors.}
 \end{center}
\end{figure}

The masses from the two channels obtained from this fit with the BEC and CR errors now included are:
\begin{eqnarray*}
\Mw(\WWqqln) = 80.411\pm0.032(\mathrm{stat.})\pm0.030(\mathrm{syst.})~\GeVm,\\
\Mw(\WWqqqq) = 80.420\pm0.035(\mathrm{stat.})\pm0.101(\mathrm{syst.})~\GeVm.  
\end{eqnarray*}
These two results are correlated and have a correlation coefficient of 
0.18. The value of $\chi^2$/d.o.f is 28.2/32, corresponding to a
$\chi^2$ probability of 66$\%$.  
These results and the correlation between them
can be used to combine the two measurements or to form the mass 
difference. The LEP combined results from the two channels 
are compared with those quoted by the individual experiments in 
Figure \ref{mw:fig-qqlnqqqq}, where the common CR and BEC errors have been imposed. 

Experimentally, separate $\Mw$ measurements are obtained from the
$\WWqqln$ and $\WWqqqq$ channels for each of the years of data. 
The combination using only the $\qqlv$ measurements yields:
\begin{eqnarray*}
 \Mwindep(\WWqqln) = 80.413\pm0.032(\mathrm{stat.})\pm0.031(\mathrm{syst.})~\GeVm. 
\end{eqnarray*}
The  systematic error is dominated by
hadronisation uncertainties  ($\pm19$~$\MeVm$) and the 
uncertainty in the LEP beam energy  ($\pm17$~$\MeVm$).
The combination using only the $\qqqq$ measurements gives:
\begin{eqnarray*}
 \Mwindep(\WWqqqq) = 80.411\pm0.035(\mathrm{stat.})\pm0.107(\mathrm{syst.})~\GeVm.  
\end{eqnarray*}
where the dominant contributions to the systematic error are from 
CR ($\pm90$~$\MeVm$) and BEC ($\pm35$~$\MeVm$).

\section{LEP Combined W Boson Width}

\begin{figure}[hbt]
\begin{center}
 \mbox{\epsfxsize=8.25cm\epsffile{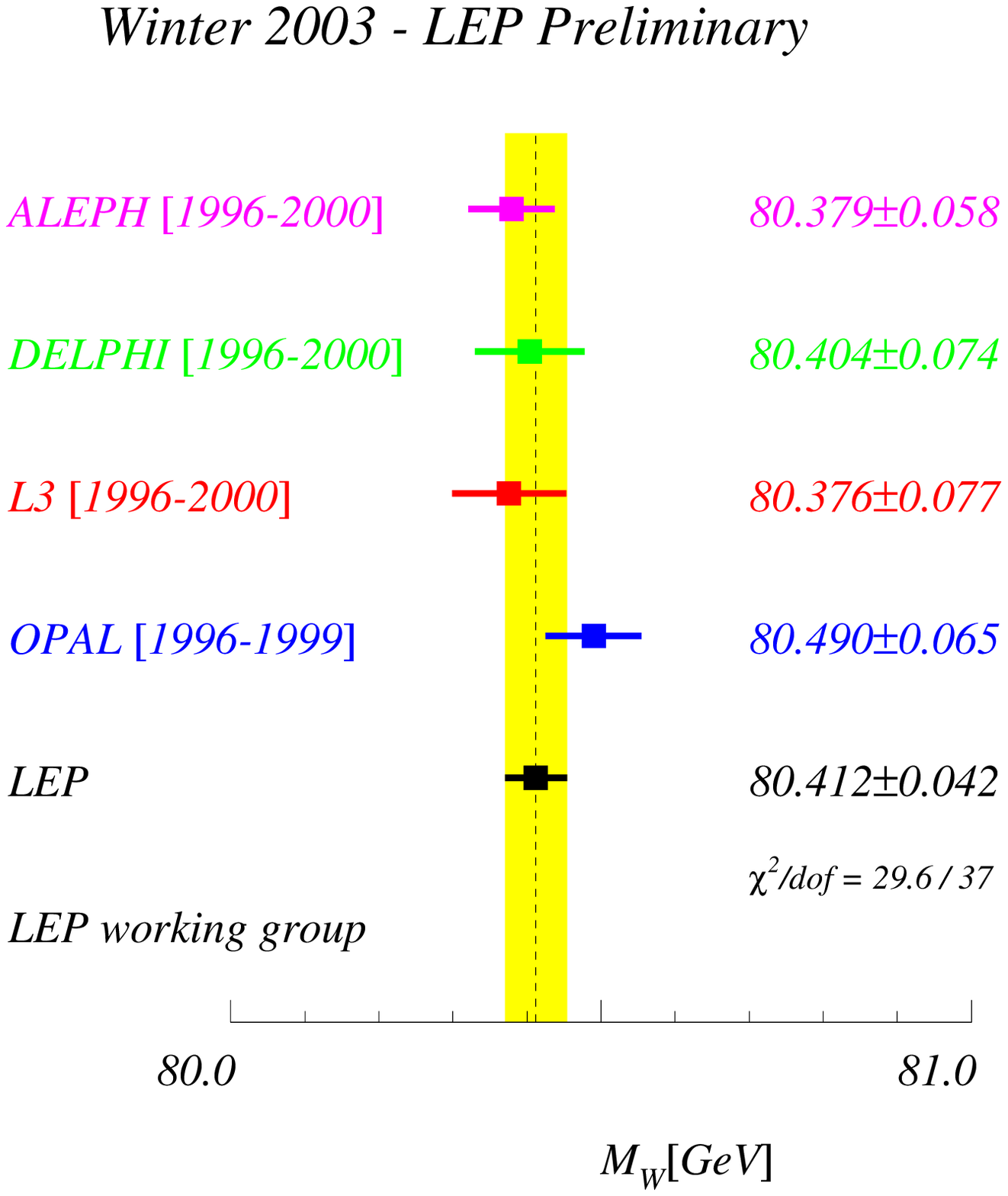}
       \epsfxsize=8.25cm\epsffile{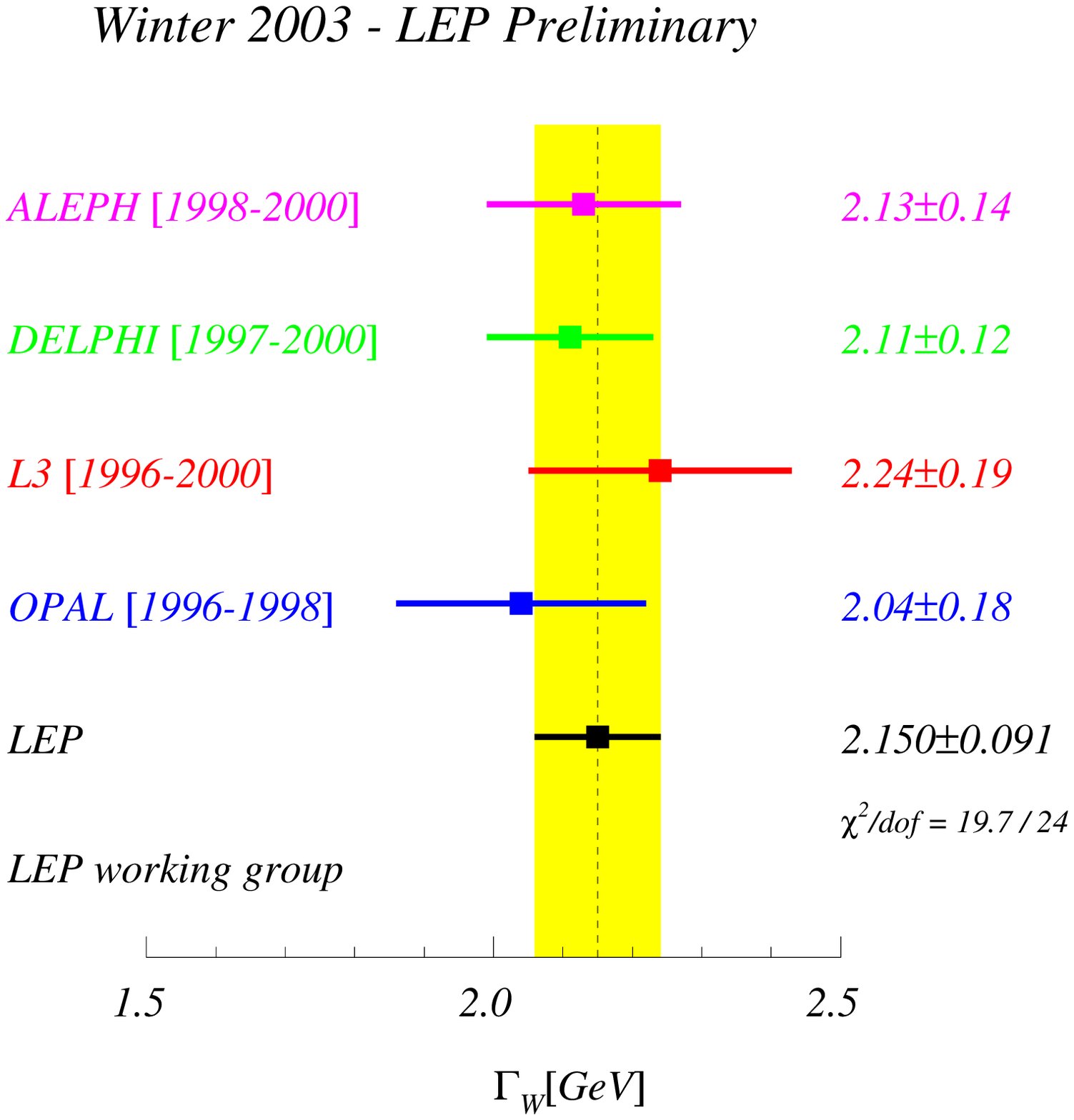}} 
\vskip-0.5cm
\caption{\label{mw:fig:mwgw} 
          The combined results for the measurements of the
          W mass (left) and W width (right) compared to the results  
          obtained by the four LEP collaborations. The combined
          values take into account correlations between experiments
          and years and hence, in general, do not give the same central 
          value as a simple average. In the LEP combination of 
          the $\qqqq$ results common values (see text) for the CR and BEC
          errors are used. The individual and combined $\Mw$ results 
          include the measurements from
          the threshold cross section. The $\Mw$ values from the experiments 
          have been recalculated for this plot including the common LEP 
          CR and BEC errors.}
 \end{center}
\end{figure}

The method of direct reconstruction is also well suited to the
direct measurement of the width of the W boson. The results of the four 
LEP experiments are shown in Table \ref{mw:tab:wwidth_experiments}
and in Figure \ref{mw:fig:mwgw}.
\begin{table}[htbp]
 \begin{center}
  \begin{tabular}{|c|c|}\hline
  Experiment & \Gw\ (\GeVm)        \\ \hline
   ALEPH    & $2.13\pm0.11\pm0.09$ \\ 
   DELPHI   & $2.11\pm0.10\pm0.07$ \\
   L3       & $2.24\pm0.11\pm0.15$ \\
   OPAL     & $2.04\pm0.16\pm0.09$ \\ \hline
\end{tabular}
 \caption{Preliminary W width measurements ($\roots=172-209$~\GeV) 
         from the individual experiments. The first error is statistical
         and the second systematic.}
 \label{mw:tab:wwidth_experiments}
\end{center}
\end{table}

Each experiment provided a W width measurement for both $\WWqqln$ and
$\WWqqqq$ channels for each of the data taking years (1996-2000) that
it has analysed. A total of 25 measurements are supplied: ALEPH
provided 3 $\WWqqqq$ results (1998-2000) and two $\WWqqln$ results
(1998-1999), DELPHI 8 measurements (1997-2000), L3 8 measurements
(1996-2000) having already combined the 1996 and 1997 results and OPAL
provided 4 measurements (1996-1998) where for the first two years the
$\WWqqln$ and $\WWqqqq$ results are already combined.

A common colour reconnection error of 65 $\MeVm$ and a common
Bose-Einstein correlation error of 35 $\MeVm$ are used in the
combination.  These common errors were determined such that the same 
error was obtained on \Gw\ as when using the BEC/CR errors supplied by the experiments. The change in the value of the width is only 2 \MeVm.  The BEC and CR values supplied by the experiments were based on  studies of phenomenological models of these effects, the uncertainty has not yet been determined from the particle flow  measurements of colour reconnection. 

A simultaneous fit to the results of the four LEP collaborations is
performed in the same way as for the $\Mw$ measurement. Correlated
systematic uncertainties are taken into account and the combination gives: 
\begin{eqnarray}
      \Gw = 2.150\pm0.068(\mathrm{stat.})\pm0.060
                                     (\mathrm{syst.})~\GeVm,
\end{eqnarray}
with a $\chi^2$/d.o.f. of 19.7/24, corresponding to a
$\chi^2$ probability of 71$\%$.

\section{Summary}

The results of the four LEP experiments on the mass and width of the W
boson are combined taking into account correlated systematic
uncertainties, giving:
\begin{eqnarray*}
       \Mw & = & 80.412\pm0.042~\GeVm, \\
       \Gw & = &  2.150\pm0.091~\GeVm.
\end{eqnarray*}
The statistical correlation between mass and width is small and 
neglected.  Their correlation due to common systematic effects is 
under study.

%% file: s03_hftab.tex
\chapter{The Measurements used in the Heavy Flavour Averages}

\label{app-HF-tab}


\newcommand{\SLDrb}{ref:SLD_R_B}
\newcommand{\SLDrc}{ref:SLD_R_C}
\newcommand{\alasy}{ref:alasy}
\newcommand{\dlasy}{ref:dlasy}
\newcommand{\llasy}{ref:llasy}
\newcommand{\SLDacbl}{ref:SLD_AQL}
\newcommand{\SLDabj}{ref:SLD_ABJ}
\newcommand{\SLDabk}{ref:SLD_ABK}

In the following 20 tables the results used in the combination are listed.
In each case an indication of the dataset used and the type of analysis is
given.
Preliminary results are indicated by the symbol ``*''. 
The values of centre-of-mass energy are given where
relevant.  In each table, following the number quoted in the referenced
publication , the value of each measurement really used in the average
is given followed by the statistical error, the internal systematic, 
the systematic error common to more than one measurement,  
the effect of a $\pm 1 \sigma$ change in all the other averaged parameters 
on the value used in the average for this measurement, 
the total systematic error, and the total error. 

Contributions to the common systematic error quoted here are from
any physics source that is potentially common between the different
experiments. Detector systematics that are common between different analyses
of the same experiment are considered internal.
\begin{table}[p]
\begin{center}
\begin{tabular}{|l||c|c|c|c|c|}
 \hline
 & \mca{1}{ALEPH} & \mca{1}{DELPHI} & \mca{1}{L3} & \mca{1}{OPAL} & \mca{1}{SLD} \\ \hline 
 & 92-95&92-95&94-95&92-95&93-98* \\ 
 & multi&multi&multi&multi&multi \\  
 & \tmcite{ref:alife}&\tmcite{ref:drb}&\tmcite{ref:lrbmixed}&\tmcite{ref:omixed}&\tmcite{\SLDrb} \\ \hline \hline 
Published \Rbz{} & 0.2159&0.21634&0.2174&0.2178&   \\ \hline \hline
Used \Rbz{} & 0.2158&0.21644&0.2167&0.2176&0.2163 \\ \hline \hline
Statistical & 0.0009&0.00067&0.0013&0.0011&0.0009 \\ \hline 
Internal Systematic & 0.0007&0.00038&0.0014&0.0009&0.0006 \\ 
Common Systematic & 0.0006&0.00039&0.0018&0.0008&0.0005 \\  
Other Param. Sys. & 0.0001&0.00014&0.0010&0.0004&0.0002 \\ \hline 
Total Systematic & 0.0009&0.00056&0.0025&0.0012&0.0008 \\ \hline \hline
Total Error & 0.0013&0.00087&0.0028&0.0017&0.0012 \\ \hline 
\end{tabular}
\end{center}
\caption{The measurements of \Rbz{}. 
  All measurements use a lifetime tag enhanced by other features like 
  invariant mass cuts or high $p_T$ leptons. 
 }
\label{tab:Rbinp}
\end{table}
\begin{table}[p]
\begin{center}
\begin{sideways}
\begin{minipage}[b]{\textheight}
\begin{center}
\begin{tabular}{|l||c|c|c|c|c|c|c|c|}
 \hline
 & \mca{3}{ALEPH} & \mca{2}{DELPHI} & \mca{2}{OPAL} & \mca{1}{SLD} \\ \hline 
 & 91-95&91-95&92-95&92-95&92-95&91-94&90-95&93-97* \\ 
 & $D$-meson&$c$-count&lepton&$c$-count&$D$-meson&$c$-count&$D$-meson&$D$-meson \\  
 &   & (result) &  & (result) & (result) & (result) & (result) &   \\  
 & \tmcite{ref:arcd}&\tmcite{ref:arcc}&\tmcite{ref:arcd}&\tmcite{ref:drcc}&\tmcite{ref:drcd,ref:drcc}&\tmcite{ref:orcc}&\tmcite{ref:orcd}&\tmcite{\SLDrc} \\ \hline \hline 
Published \Rcz{} & 0.1689&0.1738&0.1675&0.1692&0.1610&0.167&0.180&   \\ \hline \hline
Used \Rcz{} & 0.1682&0.1735&0.1684&0.1693&0.1610&0.164&0.177&0.1740 \\ \hline \hline
Statistical & 0.0082&0.0051&0.0062&0.0050&0.0104&0.012&0.010&0.0031 \\ \hline 
Internal Systematic & 0.0077&0.0057&0.0042&0.0050&0.0064&0.013&0.010&0.0014 \\ 
Common Systematic & 0.0029&0.0094&0.0042&0.0077&0.0061&0.010&0.006&0.0015 \\  
Other Param. Sys. & 0.0000&0.0000&0.0053&0.0000&0.0008&0.000&0.000&0.0002 \\ \hline 
Total Systematic & 0.0082&0.0110&0.0080&0.0092&0.0088&0.016&0.012&0.0021 \\ \hline \hline
Total Error & 0.0116&0.0122&0.0101&0.0105&0.0136&0.020&0.015&0.0037 \\ \hline 
\end{tabular}
\end{center}
\caption{The measurements of \Rcz{}. 
 ``$c$-count'' denotes the determination of \Rcz{} from the sum of production rates of weakly decaying charmed hadrons. 
 ``$D$-meson'' denotes any single/double tag 
 analysis using exclusive and/or inclusive  $D$ meson reconstruction. 
 The columns with the mention ``(result)'' are not directly used in the  
 global average, only the corresponding measurements 
 (\PcDst, \RcfDp, \RcfDs, \RcfLc, \RcfDz and \RcPcDst) are included.  }
\label{tab:Rcinp}
\end{minipage}
\end{sideways}
\end{center}
\end{table}
\begin{table}[p]
\begin{center}
\begin{sideways}
\begin{minipage}[b]{\textheight}
\begin{center}
\begin{tabular}{|l||c|c|c|c|c|c|c|c|c|c|c|}
 \hline
 & \mca{4}{ALEPH} & \mca{3}{DELPHI} & \mca{1}{L3} & \mca{3}{OPAL} \\ \hline 
 & 91-95 & 91-95 & 91-95&91-95&91-95&92-95&92-00*&90-95&91-00&90-00&90-95 \\ 
 & lepton & lepton & lepton&jet&lepton&$D$-meson&multi&lepton&jet&lepton&$D$-meson \\  
 & \tmcite{\alasy} & \tmcite{\alasy} & \tmcite{\alasy}&\tmcite{ref:ajet}&\tmcite{\dlasy}&\tmcite{ref:ddasy}&\tmcite{ref:dnnasy}&\tmcite{\llasy}&\tmcite{ref:ojet}&\tmcite{ref:olasy}&\tmcite{ref:odsac} \\ \hline \hline 
\roots\ (\GeV) & 88.38 & 89.38 & 90.21&89.47&89.434&89.434&89.449&89.50&89.50&89.51&89.49 \\ \hline 
Published \Abl & -13.10 & 5.47 & -0.42&4.36&6.6&5.67&6.37&6.11&5.82&4.70&-8.6 \\ \hline \hline
Used \Abl & \mca{3}{5.21} &4.59&6.4&4.81&6.64&6.29&5.99&5.24&-4.8 \\ \hline \hline
Statistical & \mca{3}{1.78} &1.19&2.2&7.31&1.43&2.93&1.53&1.77&10.4 \\ \hline 
Internal Systematic & \mca{3}{0.08} &0.04&0.2&0.68&0.21&0.31&0.09&0.11&2.0 \\ 
Common Systematic & \mca{3}{0.10} &0.01&0.1&0.21&0.02&0.17&0.04&0.05&1.1 \\  
Other Param. Sys. & \mca{3}{0.15} &0.14&0.2&0.75&0.02&0.07&0.09&0.16&0.8 \\ \hline 
Total Systematic & \mca{3}{0.19} &0.15&0.3&1.04&0.21&0.36&0.14&0.20&2.4 \\ \hline \hline
Total Error & \mca{3}{1.79} &1.20&2.2&7.39&1.45&2.95&1.54&1.78&10.7 \\ \hline 
\end{tabular}
\end{center}
\caption{The measurements of \Abl. 
The "Used" values are quoted at $\sqrt{s} = 89.55 \, \GeV$. All numbers are given in \%. 
 }
\label{tab:Ablinp}
\end{minipage}
\end{sideways}
\end{center}
\end{table}
\begin{table}[p]
\begin{center}
\begin{tabular}{|l||c|c|c|c|c|c|c|c|}
 \hline
 & \mca{4}{ALEPH} & \mca{2}{DELPHI} & \mca{2}{OPAL} \\ \hline 
 & 91-95 & 91-95 & 91-95&91-95&91-95&92-95&90-00&90-95 \\ 
 & lepton & lepton & lepton&$D$-meson&lepton&$D$-meson&lepton&$D$-meson \\  
 & \tmcite{\alasy} & \tmcite{\alasy} & \tmcite{\alasy}&\tmcite{ref:adsac}&\tmcite{\dlasy}&\tmcite{ref:ddasy}&\tmcite{ref:olasy}&\tmcite{ref:odsac} \\ \hline \hline 
\roots\ (\GeV) & 88.38 & 89.38   & 90.21&89.37&89.434&89.434&89.51&89.49 \\ \hline 
Published \Acl & -12.4 & -2.28 & -0.34&-1.0&3.0&-4.96&-6.83&3.9 \\ \hline \hline
Used \Acl & \mca{3}{-1.50} &0.2&3.4&-4.35&-6.23&2.5 \\ \hline \hline
Statistical & \mca{3}{2.44} &4.3&3.4&3.55&2.48&4.9 \\ \hline 
Internal Systematic & \mca{3}{0.17} &0.9&0.3&0.34&0.89&0.8 \\ 
Common Systematic & \mca{3}{0.07} &0.1&0.1&0.11&0.09&0.3 \\  
Other Param. Sys. & \mca{3}{0.14} &0.2&0.2&0.10&0.25&0.1 \\ \hline 
Total Systematic & \mca{3}{0.23} &0.9&0.4&0.37&0.93&0.8 \\ \hline \hline
Total Error & \mca{3}{2.45} &4.4&3.5&3.57&2.65&5.0 \\ \hline 
\end{tabular}
\end{center}
\caption{The measurements of \Acl. 
The "Used" values are quoted at $\sqrt{s} = 89.55 \, \GeV$. All numbers are given in \%. 
 }
\label{tab:Aclinp}
\end{table}
\begin{table}[p]
\begin{center}
\begin{sideways}
\begin{minipage}[b]{\textheight}
\begin{center}
\begin{tabular}{|l||c|c|c|c|c|c|c|c|c|c|}
 \hline
 & \mca{2}{ALEPH} & \mca{3}{DELPHI} & \mca{2}{L3} & \mca{3}{OPAL} \\ \hline 
 & 91-95&91-95&91-95&92-95&92-00*&91-95&90-95&91-00&90-00&90-95 \\ 
 & lepton&jet&lepton&$D$-meson&multi&jet&lepton&jet&lepton&$D$-meson \\  
 & \tmcite{\alasy}&\tmcite{ref:ajet}&\tmcite{\dlasy}&\tmcite{ref:ddasy}&\tmcite{ref:dnnasy}&\tmcite{ref:ljet}&\tmcite{\llasy}&\tmcite{ref:ojet}&\tmcite{ref:olasy}&\tmcite{ref:odsac} \\ \hline \hline 
\roots\ (\GeV) & 91.21&91.23&91.26&91.235&91.231&91.24&91.26&91.26&91.25&91.24 \\ \hline 
Published \Abp & 9.52&10.00&10.04&7.62&9.58&9.31&9.80&9.77&9.716&9.4 \\ \hline \hline
Used \Abp & 9.98&10.03&10.15&7.86&9.67&9.27&9.65&9.71&9.767&9.7 \\ \hline \hline
Statistical & 0.40&0.27&0.55&1.90&0.32&1.01&0.65&0.36&0.398&2.6 \\ \hline 
Internal Systematic & 0.07&0.10&0.17&0.53&0.15&0.51&0.27&0.15&0.073&2.1 \\ 
Common Systematic & 0.10&0.02&0.16&0.57&0.04&0.21&0.16&0.08&0.133&0.3 \\  
Other Param. Sys. & 0.12&0.05&0.10&0.18&0.03&0.07&0.12&0.05&0.098&0.2 \\ \hline 
Total Systematic & 0.17&0.12&0.25&0.80&0.15&0.56&0.33&0.18&0.180&2.1 \\ \hline \hline
Total Error & 0.44&0.29&0.60&2.06&0.35&1.15&0.73&0.40&0.437&3.4 \\ \hline 
\end{tabular}
\end{center}
\caption{The measurements of \Abp. 
The "Used" values are quoted at $\sqrt{s} = 91.26 \, \GeV$. All numbers are given in \%. 
 }
\label{tab:Abpinp}
\end{minipage}
\end{sideways}
\end{center}
\end{table}
\begin{table}[p]
\begin{center}
\begin{tabular}{|l||c|c|c|c|c|c|c|}
 \hline
 & \mca{2}{ALEPH} & \mca{2}{DELPHI} & \mca{1}{L3} & \mca{2}{OPAL} \\ \hline 
 & 91-95&91-95&91-95&92-95&90-95&90-00&90-95 \\ 
 & lepton&$D$-meson&lepton&$D$-meson&lepton&lepton&$D$-meson \\  
 & \tmcite{\alasy}&\tmcite{ref:adsac}&\tmcite{\dlasy}&\tmcite{ref:ddasy}&\tmcite{\llasy}&\tmcite{ref:olasy}&\tmcite{ref:odsac} \\ \hline \hline 
\roots\ (\GeV) & 91.21&91.22&91.26&91.235&91.24&91.25&91.24 \\ \hline 
Published \Acp & 6.45&6.3&6.31&6.59&7.84&5.683&6.3 \\ \hline \hline
Used \Acp & 6.63&6.3&6.25&6.49&8.18&5.701&6.5 \\ \hline \hline
Statistical & 0.56&0.9&0.92&0.93&3.03&0.539&1.2 \\ \hline 
Internal Systematic & 0.24&0.2&0.52&0.26&1.71&0.192&0.5 \\ 
Common Systematic & 0.22&0.2&0.26&0.07&0.58&0.221&0.3 \\  
Other Param. Sys. & 0.20&0.0&0.21&0.03&0.73&0.202&0.0 \\ \hline 
Total Systematic & 0.38&0.3&0.61&0.27&1.95&0.356&0.6 \\ \hline \hline
Total Error & 0.68&0.9&1.10&0.97&3.60&0.645&1.3 \\ \hline 
\end{tabular}
\end{center}
\caption{The measurements of \Acp. 
The "Used" values are quoted at $\sqrt{s} = 91.26 \, \GeV$. All numbers are given in \%. 
 }
\label{tab:Acpinp}
\end{table}
\begin{table}[p]
\begin{center}
\begin{sideways}
\begin{minipage}[b]{\textheight}
\begin{center}
\begin{tabular}{|l||c|c|c|c|c|c|c|c|c|c|c|}
 \hline
 & \mca{4}{ALEPH} & \mca{3}{DELPHI} & \mca{1}{L3} & \mca{3}{OPAL} \\ \hline 
 & 91-95 & 91-95 & 91-95&91-95&91-95&92-95&92-00*&90-95&91-00&90-00&90-95 \\ 
 & lepton & lepton & lepton&jet&lepton&$D$-meson&multi&lepton&jet&lepton&$D$-meson \\  
 & \tmcite{\alasy} & \tmcite{\alasy} & \tmcite{\alasy}&\tmcite{ref:ajet}&\tmcite{\dlasy}&\tmcite{ref:ddasy}&\tmcite{ref:dnnasy}&\tmcite{\llasy}&\tmcite{ref:ojet}&\tmcite{ref:olasy}&\tmcite{ref:odsac} \\ \hline \hline 
\roots\ (\GeV) & 92.05 & 92.94 & 93.90&92.95&92.990&92.990&92.990&93.10&92.91&92.95&92.95 \\ \hline 
Published \Abh & 11.10 & 10.43 & 13.77&11.72&10.9&8.82&10.41&13.71&12.21&10.31&-2.1 \\ \hline \hline
Used \Abh & \mca{3}{11.11} &11.69&11.4&8.59&10.40&13.70&12.24&10.06&-0.2 \\ \hline \hline
Statistical & \mca{3}{1.43} &0.98&1.8&6.16&1.16&2.39&1.23&1.47&8.7 \\ \hline 
Internal Systematic & \mca{3}{0.19} &0.11&0.1&0.88&0.27&0.30&0.16&0.09&2.0 \\ 
Common Systematic & \mca{3}{0.16} &0.02&0.1&0.48&0.03&0.19&0.12&0.17&1.2 \\  
Other Param. Sys. & \mca{3}{0.19} &0.12&0.2&0.54&0.04&0.14&0.10&0.17&0.7 \\ \hline 
Total Systematic & \mca{3}{0.31} &0.16&0.3&1.14&0.28&0.38&0.23&0.26&2.4 \\ \hline \hline
Total Error & \mca{3}{1.46} &0.99&1.8&6.27&1.19&2.42&1.25&1.50&9.0 \\ \hline 
\end{tabular}
\end{center}
\caption{The measurements of \Abh. 
The "Used" values are quoted at $\sqrt{s} = 92.94 \, \GeV$. All numbers are given in \%. 
 }
\label{tab:Abhinp}
\end{minipage}
\end{sideways}
\end{center}
\end{table}
\begin{table}[p]
\begin{center}
\begin{tabular}{|l||c|c|c|c|c|c|c|c|}
 \hline
 & \mca{4}{ALEPH} & \mca{2}{DELPHI} & \mca{2}{OPAL} \\ \hline 
 & 91-95 & 91-95 & 91-95&91-95&91-95&92-95&90-00&90-95 \\ 
 & lepton & lepton & lepton&$D$-meson&lepton&$D$-meson&lepton&$D$-meson \\  
 & \tmcite{\alasy} & \tmcite{\alasy} & \tmcite{\alasy}&\tmcite{ref:adsac}&\tmcite{\dlasy}&\tmcite{ref:ddasy}&\tmcite{ref:olasy}&\tmcite{ref:odsac} \\ \hline \hline 
\roots\ (\GeV) & 92.05 & 92.94 & 93.90  &92.96&92.990&92.990&92.95&92.95 \\ \hline 
Published \Ach & 10.58 & 11.93 & 12.09&11.0&10.8&11.80&14.59&15.8 \\ \hline \hline
Used \Ach & \mca{3}{11.94} &10.9&10.8&11.42&14.89&14.6 \\ \hline \hline
Statistical & \mca{3}{1.98} &3.3&2.8&3.09&2.02&4.0 \\ \hline 
Internal Systematic & \mca{3}{0.35} &0.7&0.4&0.55&0.50&0.7 \\ 
Common Systematic & \mca{3}{0.29} &0.1&0.2&0.09&0.24&0.5 \\  
Other Param. Sys. & \mca{3}{0.34} &0.2&0.3&0.09&0.43&0.1 \\ \hline 
Total Systematic & \mca{3}{0.57} &0.7&0.5&0.56&0.70&0.9 \\ \hline \hline
Total Error & \mca{3}{2.06} &3.4&2.8&3.15&2.14&4.1 \\ \hline 
\end{tabular}
\end{center}
\caption{The measurements of \Ach. 
The "Used" values are quoted at $\sqrt{s} = 92.94 \, \GeV$. All numbers are given in \%. 
 }
\label{tab:Achinp}
\end{table}
\begin{table}[p]
\begin{center}
\begin{tabular}{|l||c|c|c|c|}
 \hline
 & \mca{4}{SLD} \\ \hline 
 & 93-98&93-98&94-95&96-98* \\ 
 & lepton&jet&$K^{\pm}$&$K$+vertex \\  
 & \tmcite{\SLDacbl}&\tmcite{\SLDabj}&\tmcite{\SLDabk}&\tmcite{ref:SLD_vtxasy} \\ \hline \hline 
\roots\ (\GeV) & 91.28&91.28&91.28&91.28 \\ \hline 
Published \cAb & 0.919 & 0.907 &0.855 &   \\ \hline \hline
Used \cAb & 0.9387&0.9067&0.8551&0.9189 \\ \hline \hline
Statistical & 0.0296&0.0197&0.0880&0.0184 \\ \hline 
Internal Systematic & 0.0184&0.0235&0.1018&0.0179 \\ 
Common Systematic & 0.0079&0.0010&0.0065&0.0009 \\  
Other Param. Sys. & 0.0111&0.0008&0.0005&0.0023 \\ \hline 
Total Systematic & 0.0229&0.0235&0.1020&0.0181 \\ \hline \hline
Total Error & 0.0374&0.0307&0.1347&0.0258 \\ \hline 
\end{tabular}
\end{center}
\caption{The measurements of \cAb. 
 }
\label{tab:cAbinp}
\end{table}
\begin{table}[p]
\begin{center}
\begin{tabular}{|l||c|c|c|}
 \hline
 & \mca{3}{SLD} \\ \hline 
 & 93-98&93-98&96-98* \\ 
 & lepton&$D$-meson&$K$+vertex \\  
 & \tmcite{\SLDacbl}&\tmcite{ref:SLD_ACD}&\tmcite{ref:SLD_ACV} \\ \hline \hline 
\roots\ (\GeV) & 91.28&91.28&91.28 \\ \hline 
Published \cAc & 0.583 &0.688  &   \\ \hline \hline
Used \cAc & 0.5877&0.6889&0.6743 \\ \hline \hline
Statistical & 0.0552&0.0349&0.0290 \\ \hline 
Internal Systematic & 0.0452&0.0205&0.0240 \\ 
Common Systematic & 0.0211&0.0029&0.0016 \\  
Other Param. Sys. & 0.0171&0.0015&0.0021 \\ \hline 
Total Systematic & 0.0527&0.0207&0.0242 \\ \hline \hline
Total Error & 0.0763&0.0406&0.0378 \\ \hline 
\end{tabular}
\end{center}
\caption{The measurements of \cAc. 
 }
\label{tab:cAcinp}
\end{table}
\clearpage
\begin{table}[p]
\begin{center}
\begin{tabular}{|l||c|c|c|c|c|c|}
 \hline
 & \mca{1}{ALEPH} & \mca{1}{DELPHI} & \mca{2}{L3} & \mca{2}{OPAL} \\ \hline 
 & 91-95&94-95&94-95&92&92-95&92-95 \\ 
 & multi&multi&multi&multi&multi&multi \\  
 & \tmcite{ref:abl}&\tmcite{ref:dbl}&\tmcite{ref:lrbmixed}&\tmcite{ref:lbl}&\tmcite{ref:obl}&\tmcite{ref:obl} \\ \hline \hline 
Published \Brbl & 10.70&10.70&10.16&10.68&10.78&10.96 \\ \hline \hline
Used \Brbl & 10.74&10.70&10.26&10.81&\mca{2}{10.86}  \\ \hline \hline
Statistical & 0.10&0.14&0.09&0.11&\mca{2}{0.09}  \\ \hline 
Internal Systematic & 0.15&0.14&0.16&0.36&\mca{2}{0.21}  \\ 
Common Systematic & 0.23&0.43&0.31&0.22&\mca{2}{0.19}  \\  
Other Param. Sys. & 0.03&0.07&0.03&0.09&\mca{2}{0.02}  \\ \hline 
Total Systematic & 0.28&0.45&0.35&0.43&\mca{2}{0.29}  \\ \hline \hline
Total Error & 0.29&0.48&0.36&0.45&\mca{2}{0.30}  \\ \hline 
\end{tabular}
\end{center}
\caption{The measurements of \Brbl. 
 All numbers are given in \%. 
 }
\label{tab:Brblinp}
\end{table}
\begin{table}[p]
\begin{center}
\begin{tabular}{|l||c|c|c|c|}
 \hline
 & \mca{1}{ALEPH} & \mca{1}{DELPHI} & \mca{2}{OPAL} \\ \hline 
 & 91-95&94-95&92-95&92-95 \\ 
 & multi&multi&multi&multi \\  
 & \tmcite{ref:abl}&\tmcite{ref:dbl}&\tmcite{ref:obl}&\tmcite{ref:obl} \\ \hline \hline 
Published \Brbclp & 8.18&7.98&8.37&8.17 \\ \hline \hline
Used \Brbclp & 8.11&7.98&\mca{2}{8.42}  \\ \hline \hline
Statistical & 0.15&0.22&\mca{2}{0.15}  \\ \hline 
Internal Systematic & 0.18&0.16&\mca{2}{0.22}  \\ 
Common Systematic & 0.15&0.22&\mca{2}{0.32}  \\  
Other Param. Sys. & 0.05&0.04&\mca{2}{0.04}  \\ \hline 
Total Systematic & 0.24&0.27&\mca{2}{0.39}  \\ \hline \hline
Total Error & 0.29&0.35&\mca{2}{0.42}  \\ \hline 
\end{tabular}
\end{center}
\caption{The measurements of \Brbclp. 
 All numbers are given in \%. 
 }
\label{tab:Brbclpinp}
\end{table}
\begin{table}[p]
\begin{center}
\begin{tabular}{|l||c|c|}
 \hline
 & \mca{1}{DELPHI} & \mca{1}{OPAL} \\ \hline 
 & 92-95&90-95 \\ 
 & $D$+lepton&$D$+lepton \\  
 & \tmcite{ref:drcd}&\tmcite{ref:ocl} \\ \hline \hline 
Published $\Brcl$ & 9.58&9.5 \\ \hline \hline
Used $\Brcl$ & 9.67&9.6 \\ \hline \hline
Statistical & 0.42&0.6 \\ \hline 
Internal Systematic & 0.24&0.5 \\ 
Common Systematic & 0.13&0.4 \\  
Other Param. Sys. & 0.01&0.0 \\ \hline 
Total Systematic & 0.27&0.7 \\ \hline \hline
Total Error & 0.50&0.9 \\ \hline 
\end{tabular}
\end{center}
\caption{The measurements of $\Brcl$. 
 All numbers are given in \%. 
 }
\label{tab:Brclinp}
\end{table}
\begin{table}[p]
\begin{center}
\begin{tabular}{|l||c|c|c|c|}
 \hline
 & \mca{1}{ALEPH} & \mca{1}{DELPHI} & \mca{1}{L3} & \mca{1}{OPAL} \\ \hline 
 & 91-95&94-95&90-95&90-00 \\ 
 & multi&multi&lepton&lepton \\  
 & \tmcite{\alasy}&\tmcite{ref:dbl}&\tmcite{\llasy}&\tmcite{ref:olasy} \\ \hline \hline 
Published \chiM & 0.1246&0.127&0.1192&0.13121 \\ \hline \hline
Used \chiM & 0.1199&0.127&0.1199&0.13182 \\ \hline \hline
Statistical & 0.0049&0.013&0.0066&0.00463 \\ \hline 
Internal Systematic & 0.0021&0.005&0.0023&0.00149 \\ 
Common Systematic & 0.0040&0.003&0.0026&0.00369 \\  
Other Param. Sys. & 0.0012&0.001&0.0016&0.00163 \\ \hline 
Total Systematic & 0.0047&0.006&0.0038&0.00430 \\ \hline \hline
Total Error & 0.0068&0.014&0.0076&0.00632 \\ \hline 
\end{tabular}
\end{center}
\caption{The measurements of \chiM. 
 }
\label{tab:chiMinp}
\end{table}
\begin{table}[p]
\begin{center}
\begin{tabular}{|l||c|c|}
 \hline
 & \mca{1}{DELPHI} & \mca{1}{OPAL} \\ \hline 
 & 92-95&90-95 \\ 
 & $D$-meson&$D$-meson \\  
 & \tmcite{ref:drcd}&\tmcite{ref:orcd} \\ \hline \hline 
Published \PcDst & 0.174&0.15163 \\ \hline \hline
Used \PcDst & 0.174&0.15478 \\ \hline \hline
Statistical & 0.010&0.00384 \\ \hline 
Internal Systematic & 0.004&0.00452 \\ 
Common Systematic & 0.001&0.00499 \\  
Other Param. Sys. & 0.000&0.00207 \\ \hline 
Total Systematic & 0.004&0.00704 \\ \hline \hline
Total Error & 0.011&0.00802 \\ \hline 
\end{tabular}
\end{center}
\caption{The measurements of \PcDst. 
 }
\label{tab:PcDstinp}
\end{table}
\begin{table}[p]
\begin{center}
\begin{tabular}{|l||c|c|c|}
 \hline
 & \mca{1}{ALEPH} & \mca{1}{DELPHI} & \mca{1}{OPAL} \\ \hline 
 & 91-95&92-95&91-94 \\ 
 & $c$-count&$c$-count&$c$-count \\  
 & \tmcite{ref:arcc}&\tmcite{ref:drcc}&\tmcite{ref:orcc} \\ \hline \hline 
Published \RcfDp & 0.0409&0.03840&0.0393 \\ \hline \hline
Used \RcfDp & 0.0402&0.03859&0.0385 \\ \hline \hline
Statistical & 0.0014&0.00135&0.0056 \\ \hline 
Internal Systematic & 0.0012&0.00122&0.0026 \\ 
Common Systematic & 0.0029&0.00252&0.0028 \\  
Other Param. Sys. & 0.0012&0.00080&0.0015 \\ \hline 
Total Systematic & 0.0033&0.00291&0.0041 \\ \hline \hline
Total Error & 0.0036&0.00321&0.0069 \\ \hline 
\end{tabular}
\end{center}
\caption{The measurements of \RcfDp. 
 }
\label{tab:RcfDpinp}
\end{table}
\begin{table}[p]
\begin{center}
\begin{tabular}{|l||c|c|c|}
 \hline
 & \mca{1}{ALEPH} & \mca{1}{DELPHI} & \mca{1}{OPAL} \\ \hline 
 & 91-95&92-95&91-94 \\ 
 & $c$-count&$c$-count&$c$-count \\  
 & \tmcite{ref:arcc}&\tmcite{ref:drcc}&\tmcite{ref:orcc} \\ \hline \hline 
Published \RcfDs & 0.0199&0.02129&0.0161 \\ \hline \hline
Used \RcfDs & 0.0206&0.02134&0.0158 \\ \hline \hline
Statistical & 0.0036&0.00183&0.0048 \\ \hline 
Internal Systematic & 0.0011&0.00089&0.0007 \\ 
Common Systematic & 0.0047&0.00480&0.0037 \\  
Other Param. Sys. & 0.0003&0.00038&0.0006 \\ \hline 
Total Systematic & 0.0048&0.00489&0.0038 \\ \hline \hline
Total Error & 0.0060&0.00522&0.0061 \\ \hline 
\end{tabular}
\end{center}
\caption{The measurements of \RcfDs. 
 }
\label{tab:RcfDsinp}
\end{table}
\begin{table}[p]
\begin{center}
\begin{tabular}{|l||c|c|c|}
 \hline
 & \mca{1}{ALEPH} & \mca{1}{DELPHI} & \mca{1}{OPAL} \\ \hline 
 & 91-95&92-95&91-94 \\ 
 & $c$-count&$c$-count&$c$-count \\  
 & \tmcite{ref:arcc}&\tmcite{ref:drcc}&\tmcite{ref:orcc} \\ \hline \hline 
Published \RcfLc & 0.0169&0.01695&0.0107 \\ \hline \hline
Used \RcfLc & 0.0155&0.01701&0.0089 \\ \hline \hline
Statistical & 0.0017&0.00396&0.0065 \\ \hline 
Internal Systematic & 0.0005&0.00143&0.0008 \\ 
Common Systematic & 0.0038&0.00401&0.0028 \\  
Other Param. Sys. & 0.0004&0.00039&0.0005 \\ \hline 
Total Systematic & 0.0039&0.00428&0.0030 \\ \hline \hline
Total Error & 0.0042&0.00583&0.0072 \\ \hline 
\end{tabular}
\end{center}
\caption{The measurements of \RcfLc. 
 }
\label{tab:RcfLcinp}
\end{table}
\begin{table}[p]
\begin{center}
\begin{tabular}{|l||c|c|c|}
 \hline
 & \mca{1}{ALEPH} & \mca{1}{DELPHI} & \mca{1}{OPAL} \\ \hline 
 & 91-95&92-95&91-94 \\ 
 & $c$-count&$c$-count&$c$-count \\  
 & \tmcite{ref:arcc}&\tmcite{ref:drcc}&\tmcite{ref:orcc} \\ \hline \hline 
Published \RcfDz & 0.0961&0.09274&0.1016 \\ \hline \hline
Used \RcfDz & 0.0965&0.09292&0.1026 \\ \hline \hline
Statistical & 0.0031&0.00268&0.0079 \\ \hline 
Internal Systematic & 0.0036&0.00264&0.0033 \\ 
Common Systematic & 0.0042&0.00239&0.0038 \\  
Other Param. Sys. & 0.0018&0.00187&0.0016 \\ \hline 
Total Systematic & 0.0058&0.00402&0.0053 \\ \hline \hline
Total Error & 0.0066&0.00483&0.0095 \\ \hline 
\end{tabular}
\end{center}
\caption{The measurements of \RcfDz. 
 }
\label{tab:RcfDzinp}
\end{table}
\begin{table}[p]
\begin{center}
\begin{tabular}{|l||c|c|}
 \hline
 & \mca{1}{DELPHI} & \mca{1}{OPAL} \\ \hline 
 & 92-95&90-95 \\ 
 &  $D$-meson &$D$-meson \\  
 & \tmcite{ref:drcc}&\tmcite{ref:orcd} \\ \hline \hline 
Published \RcPcDst & 0.02829&0.027293 \\ \hline \hline
Used \RcPcDst & 0.02835&0.027102 \\ \hline \hline
Statistical & 0.00072&0.000472 \\ \hline 
Internal Systematic & 0.00082&0.000838 \\ 
Common Systematic & 0.00060&0.001022 \\  
Other Param. Sys. & 0.00087&0.000127 \\ \hline 
Total Systematic & 0.00133&0.001328 \\ \hline \hline
Total Error & 0.00151&0.001409 \\ \hline 
\end{tabular}
\end{center}
\caption{The measurements of \RcPcDst. 
 }
\label{tab:RcPcDstinp}
\end{table}

%% file: s03_hffit.tex
\chapter{Heavy-Flavour Fit including Off-Peak Asymmetries}\label{app-HF-fit}
The full 18 parameter fit to the LEP and SLD data gave the following results:
\begin{eqnarray*}
  \Rbz    &=& 0.21637   \pm  0.00066\\
  \Rcz    &=& 0.1721    \pm  0.0030 \\
  \Abl    &=& 0.0560    \pm  0.0066 \\
  \Acl    &=&-0.018     \pm  0.013  \\
  \Abp    &=& 0.0981    \pm  0.0017 \\
  \Acp    &=& 0.0635    \pm  0.0036 \\
  \Abh    &=& 0.1125    \pm  0.0054 \\
  \Ach    &=& 0.125     \pm  0.011  \\
  \cAb    &=& 0.925     \pm  0.020  \\
  \cAc    &=& 0.670     \pm  0.026  \\
  \Brbl   &=& 0.1070    \pm  0.0022 \\
  \Brbclp &=& 0.0802    \pm  0.0018 \\
  \Brcl   &=& 0.0971    \pm  0.0032 \\
  \chiM   &=& 0.1250    \pm  0.0039 \\
  \fDp    &=& 0.235     \pm  0.016  \\
  \fDs    &=& 0.124     \pm  0.026  \\
  \fcb    &=& 0.094     \pm  0.022  \\
  \PcDst  &=& 0.1623    \pm  0.0048 \,
\end{eqnarray*}
with a $\chi^2/$d.o.f.{} of  $48/(105-18)$. The corresponding correlation
matrix is given in Table~\ref{tab:18parcor}.
The energy for the  peak$-$2, peak and peak+2 results are respectively
89.55 \GeV{}, 91.26 \GeV{} and 92.94 \GeV.
Note that the asymmetry results shown here are not the pole
asymmetries shown in Section~\ref{sec-HFSUM-LEP-SLD}.
The non-electroweak parameters do not depend on the treatment of the 
asymmetries.

\begin{table}[p]
\begin{center}
\begin{sideways}
\begin{minipage}[b]{\textheight}
\begin{center}
\footnotesize
\begin{tabular}{|l||rrrrrrrrrrrrrrrrrr|}
\hline
&\makebox[0.45cm]{$1)$}
&\makebox[0.45cm]{$2)$}
&\makebox[0.45cm]{$3)$}
&\makebox[0.45cm]{$4)$}
&\makebox[0.45cm]{$5)$}
&\makebox[0.45cm]{$6)$}
&\makebox[0.45cm]{$7)$}
&\makebox[0.45cm]{$8)$}
&\makebox[0.45cm]{$9)$}
&\makebox[0.45cm]{$10)$}
&\makebox[0.45cm]{$11)$}
&\makebox[0.45cm]{$12)$}
&\makebox[0.45cm]{$13)$}
&\makebox[0.45cm]{$14)$}
&\makebox[0.45cm]{$15)$}
&\makebox[0.45cm]{$16)$}
&\makebox[0.45cm]{$17)$}
&\makebox[0.45cm]{$18)$}\\
&\makebox[0.45cm]{\Rb}
&\makebox[0.45cm]{\Rc}
&\makebox[0.45cm]{$\Abb$}
&\makebox[0.45cm]{$\Acc$}
&\makebox[0.45cm]{$\Abb$}
&\makebox[0.45cm]{$\Acc$}
&\makebox[0.45cm]{$\Abb$}
&\makebox[0.45cm]{$\Acc$}
&\makebox[0.45cm]{\cAb}
&\makebox[0.45cm]{\cAc}
&\makebox[0.45cm]{BR}
&\makebox[0.45cm]{BR}
&\makebox[0.45cm]{BR}
&\makebox[0.45cm]{\chiM}
&\makebox[0.45cm]{$\fDp$}
&\makebox[0.45cm]{$\fDs$}
&\makebox[0.45cm]{$f(c_{bar.})$}
&\makebox[0.55cm]{PcDst}\\
&
&
&\makebox[0.45cm]{$(-2)$}
&\makebox[0.45cm]{$(-2)$}
&\makebox[0.45cm]{(pk)}
&\makebox[0.45cm]{(pk)}
&\makebox[0.45cm]{$(+2)$}
&\makebox[0.45cm]{$(+2)$}
&
&
&\makebox[0.45cm]{$(1)$}
&\makebox[0.45cm]{$(2)$}
&\makebox[0.45cm]{$(3)$}
&
&
&
&
&\\
\hline\hline
 1) &$  1.00$&$ -0.15$&$ -0.02$&$  0.00$&$ -0.10$&$  0.07$&$ -0.04$&$  0.03$
    &$ -0.07$&$  0.05$&$ -0.07$&$ -0.03$&$ -0.01$&$  0.00$&$ -0.15$&$ -0.03$
    &$  0.12$&$  0.11$\\
 2) &$ -0.15$&$  1.00$&$  0.00$&$  0.01$&$  0.03$&$ -0.06$&$  0.01$&$ -0.03$
    &$  0.03$&$ -0.05$&$  0.04$&$ -0.01$&$ -0.29$&$  0.02$&$ -0.13$&$  0.17$
    &$  0.16$&$ -0.44$\\
 3) &$ -0.02$&$  0.00$&$  1.00$&$  0.13$&$  0.03$&$  0.00$&$  0.01$&$  0.00$
    &$  0.00$&$  0.00$&$  0.00$&$  0.00$&$  0.00$&$  0.01$&$  0.00$&$  0.00$
    &$  0.00$&$  0.00$\\
 4) &$  0.00$&$  0.01$&$  0.13$&$  1.00$&$  0.01$&$  0.02$&$  0.01$&$  0.01$
    &$  0.00$&$  0.00$&$  0.01$&$ -0.02$&$  0.02$&$  0.02$&$  0.00$&$  0.00$
    &$  0.00$&$  0.00$\\
 5) &$ -0.10$&$  0.03$&$  0.03$&$  0.01$&$  1.00$&$  0.16$&$  0.08$&$  0.02$
    &$  0.02$&$ -0.01$&$  0.00$&$ -0.05$&$  0.00$&$  0.11$&$  0.01$&$  0.00$
    &$ -0.01$&$ -0.01$\\
 6) &$  0.07$&$ -0.06$&$  0.00$&$  0.02$&$  0.16$&$  1.00$&$  0.02$&$  0.15$
    &$ -0.02$&$  0.04$&$  0.19$&$ -0.23$&$ -0.21$&$  0.08$&$ -0.04$&$ -0.02$
    &$  0.04$&$  0.03$\\
 7) &$ -0.04$&$  0.01$&$  0.01$&$  0.01$&$  0.08$&$  0.02$&$  1.00$&$  0.13$
    &$  0.01$&$  0.00$&$  0.00$&$ -0.03$&$  0.00$&$  0.03$&$  0.01$&$  0.00$
    &$  0.00$&$ -0.01$\\
 8) &$  0.03$&$ -0.03$&$  0.00$&$  0.01$&$  0.02$&$  0.15$&$  0.13$&$  1.00$
    &$ -0.01$&$  0.02$&$  0.07$&$ -0.08$&$ -0.14$&$  0.02$&$ -0.02$&$ -0.02$
    &$  0.02$&$  0.02$\\
 9) &$ -0.07$&$  0.03$&$  0.00$&$  0.00$&$  0.02$&$ -0.02$&$  0.01$&$ -0.01$
    &$  1.00$&$  0.13$&$ -0.02$&$  0.02$&$  0.03$&$  0.06$&$  0.00$&$  0.00$
    &$  0.00$&$ -0.02$\\
10) &$  0.05$&$ -0.05$&$  0.00$&$  0.00$&$ -0.01$&$  0.04$&$  0.00$&$  0.02$
    &$  0.13$&$  1.00$&$  0.02$&$ -0.04$&$ -0.03$&$  0.01$&$  0.00$&$  0.00$
    &$  0.00$&$  0.02$\\
11) &$ -0.07$&$  0.04$&$  0.00$&$  0.01$&$  0.00$&$  0.19$&$  0.00$&$  0.07$
    &$ -0.02$&$  0.02$&$  1.00$&$ -0.24$&$  0.00$&$  0.29$&$  0.04$&$  0.01$
    &$ -0.02$&$ -0.01$\\
12) &$ -0.03$&$ -0.01$&$  0.00$&$ -0.02$&$ -0.05$&$ -0.23$&$ -0.03$&$ -0.08$
    &$  0.02$&$ -0.04$&$ -0.24$&$  1.00$&$  0.10$&$ -0.23$&$  0.02$&$  0.00$
    &$ -0.01$&$  0.01$\\
13) &$ -0.01$&$ -0.29$&$  0.00$&$  0.02$&$  0.00$&$ -0.21$&$  0.00$&$ -0.14$
    &$  0.03$&$ -0.03$&$  0.00$&$  0.10$&$  1.00$&$  0.16$&$  0.00$&$ -0.02$
    &$ -0.01$&$  0.13$\\
14) &$  0.00$&$  0.02$&$  0.01$&$  0.02$&$  0.11$&$  0.08$&$  0.03$&$  0.02$
    &$  0.06$&$  0.01$&$  0.29$&$ -0.23$&$  0.16$&$  1.00$&$  0.02$&$  0.00$
    &$  0.00$&$  0.00$\\
15) &$ -0.15$&$ -0.13$&$  0.00$&$  0.00$&$  0.01$&$ -0.04$&$  0.01$&$ -0.02$
    &$  0.00$&$  0.00$&$  0.04$&$  0.02$&$  0.00$&$  0.02$&$  1.00$&$ -0.40$
    &$ -0.25$&$  0.10$\\
16) &$ -0.03$&$  0.17$&$  0.00$&$  0.00$&$  0.00$&$ -0.02$&$  0.00$&$ -0.02$
    &$  0.00$&$  0.00$&$  0.01$&$  0.00$&$ -0.02$&$  0.00$&$ -0.40$&$  1.00$
    &$ -0.48$&$ -0.09$\\
17) &$  0.12$&$  0.16$&$  0.00$&$  0.00$&$ -0.01$&$  0.04$&$  0.00$&$  0.02$
    &$  0.00$&$  0.00$&$ -0.02$&$ -0.01$&$ -0.01$&$  0.00$&$ -0.25$&$ -0.48$
    &$  1.00$&$ -0.14$\\
18) &$  0.11$&$ -0.44$&$  0.00$&$  0.00$&$ -0.01$&$  0.03$&$ -0.01$&$  0.02$
    &$ -0.02$&$  0.02$&$ -0.01$&$  0.01$&$  0.13$&$  0.00$&$  0.10$&$ -0.09$
    &$ -0.14$&$  1.00$\\
\hline
\end{tabular}
\normalsize
\end{center}
\caption[]{
  The correlation matrix for the set of the 18 heavy flavour
  parameters. BR(1), BR(2) and BR(3) denote $\Brbl$, $\Brbclp$ and $\Brcl$
  respectively, PcDst denotes $\PcDst$.  }
\label{tab:18parcor}
\end{minipage}
\end{sideways}
\end{center}
\end{table}

%% file: 4f_app.tex
\chapter{Detailed inputs and results on W-boson and four-fermion averages}
\label{4f_sec:appendix}

Tables~\ref{4f_tab:WWmeas}~-~\ref{4f_tab:rzeemeas}
give the details of the inputs and of the results
for the calculation of LEP averages
of the four-fermion cross-section and the corresponding cross-section ratios
For both inputs and results, whenever relevant,
the split up of the errors into their various components
is given in the table.

For each measurement, 
the Collaborations have privately 
provided
unpublished information which is necessary 
for the combination of LEP results,
such as the expected statistical error 
or the split up of the systematic uncertainty 
into its correlated and uncorrelated components.
Unless otherwise specified in the References,
all other inputs are taken from published papers 
and public notes submitted to conferences.

\begin{table}[hbtp]
\vspace*{-0.5cm}
\begin{center}
\begin{small}
\begin{tabular}{|c|ccccc|c|c|c|}
\cline{1-8}
\roots & & & {\scriptsize (LCEC)} & {\scriptsize (LUEU)} & 
{\scriptsize (LUEC)} & & &
\multicolumn{1}{|r}{$\quad$} \\
(GeV) & $\sww$ & 
$\Delta\sww^\mathrm{stat}$ &
$\Delta\sww^\mathrm{syst}$ &
$\Delta\sww^\mathrm{syst}$ &
$\Delta\sww^\mathrm{syst}$ &
$\Delta\sww^\mathrm{syst}$ &
$\Delta\sww$ & 
\multicolumn{1}{|r}{$\quad$} \\
\cline{1-8}
\multicolumn{8}{|c|}
{\Aleph~\cite{4f_bib:aleww183,4f_bib:aleww189,4f_bib:alewwsc01}} &
\multicolumn{1}{|r}{$\quad$} \\
\cline{1-8}
182.7 & 15.57 & $\pm$0.62 & $\pm$0.23 & $\pm$0.09 & $\pm$0.15 & $\pm$0.29& $\pm$0.68 & \multicolumn{1}{|r}{$\quad$} \\
188.6 & 15.71 & $\pm$0.34 & $\pm$0.11 & $\pm$0.09 & $\pm$0.11 & $\pm$0.18& $\pm$0.38 & \multicolumn{1}{|r}{$\quad$} \\
191.6 & 17.23 & $\pm$0.89 & $\pm$0.11 & $\pm$0.09 & $\pm$0.11 & $\pm$0.18& $\pm$0.91 & \multicolumn{1}{|r}{$\quad$} \\
195.5 & 17.00 & $\pm$0.54 & $\pm$0.11 & $\pm$0.09 & $\pm$0.11 & $\pm$0.18& $\pm$0.57 & \multicolumn{1}{|r}{$\quad$} \\
199.5 & 16.98 & $\pm$0.53 & $\pm$0.11 & $\pm$0.09 & $\pm$0.11 & $\pm$0.18& $\pm$0.56 & \multicolumn{1}{|r}{$\quad$} \\
201.6 & 16.16 & $\pm$0.74 & $\pm$0.11 & $\pm$0.09 & $\pm$0.11 & $\pm$0.18& $\pm$0.76 & \multicolumn{1}{|r}{$\quad$} \\
204.9 & 16.57 & $\pm$0.52 & $\pm$0.11 & $\pm$0.09 & $\pm$0.11 & $\pm$0.18& $\pm$0.55 & \multicolumn{1}{|r}{$\quad$} \\
206.6 & 17.32 & $\pm$0.41 & $\pm$0.11 & $\pm$0.09 & $\pm$0.11 & $\pm$0.18& $\pm$0.45 & \multicolumn{1}{|r}{$\quad$} \\
\cline{1-8}
\multicolumn{8}{|c|}
{\Delphi~\cite{4f_bib:delww183,4f_bib:delww189,4f_bib:delww}} &
\multicolumn{1}{|r}{$\quad$} \\
\cline{1-8}
182.7 & 16.07 & $\pm$0.68 & $\pm$0.09 & $\pm$0.09 & $\pm$0.08 & $\pm$0.15& $\pm$0.70 & \multicolumn{1}{|r}{$\quad$} \\
188.6 & 16.09 & $\pm$0.39 & $\pm$0.08 & $\pm$0.09 & $\pm$0.09 & $\pm$0.15& $\pm$0.42 & \multicolumn{1}{|r}{$\quad$} \\
191.6 & 16.64 & $\pm$0.99 & $\pm$0.09 & $\pm$0.10 & $\pm$0.09 & $\pm$0.16& $\pm$1.00 & \multicolumn{1}{|r}{$\quad$} \\
195.5 & 17.04 & $\pm$0.58 & $\pm$0.09 & $\pm$0.10 & $\pm$0.09 & $\pm$0.16& $\pm$0.60 & \multicolumn{1}{|r}{$\quad$} \\
199.5 & 17.39 & $\pm$0.55 & $\pm$0.09 & $\pm$0.10 & $\pm$0.09 & $\pm$0.16& $\pm$0.57 & \multicolumn{1}{|r}{$\quad$} \\
201.6 & 17.37 & $\pm$0.80 & $\pm$0.10 & $\pm$0.10 & $\pm$0.09 & $\pm$0.17& $\pm$0.82 & \multicolumn{1}{|r}{$\quad$} \\
204.9 & 17.56 & $\pm$0.57 & $\pm$0.10 & $\pm$0.10 & $\pm$0.09 & $\pm$0.17& $\pm$0.59 & \multicolumn{1}{|r}{$\quad$} \\
206.6 & 16.35 & $\pm$0.44 & $\pm$0.10 & $\pm$0.10 & $\pm$0.09 & $\pm$0.17& $\pm$0.47 & \multicolumn{1}{|r}{$\quad$} \\
\cline{1-8}
\multicolumn{8}{|c|}
{\Ltre~\cite{4f_bib:ltrww183,4f_bib:ltrww189,4f_bib:ltrwwsc02}} &
\multicolumn{1}{|r}{$\quad$} \\
\cline{1-8}
182.7 & 16.53 & $\pm$0.67 & $\pm$0.19 & $\pm$0.13 & $\pm$0.12 & $\pm$0.26& $\pm$0.72 & \multicolumn{1}{|r}{$\quad$} \\
188.6 & 16.24 & $\pm$0.37 & $\pm$0.18 & $\pm$0.06 & $\pm$0.12 & $\pm$0.22& $\pm$0.43 & \multicolumn{1}{|r}{$\quad$} \\
191.6 & 16.39 & $\pm$0.90 & $\pm$0.19 & $\pm$0.09 & $\pm$0.12 & $\pm$0.24& $\pm$0.93 & \multicolumn{1}{|r}{$\quad$} \\
195.5 & 16.67 & $\pm$0.55 & $\pm$0.19 & $\pm$0.09 & $\pm$0.12 & $\pm$0.24& $\pm$0.60 & \multicolumn{1}{|r}{$\quad$} \\
199.5 & 16.94 & $\pm$0.57 & $\pm$0.19 & $\pm$0.09 & $\pm$0.12 & $\pm$0.24& $\pm$0.62 & \multicolumn{1}{|r}{$\quad$} \\
201.6 & 16.95 & $\pm$0.85 & $\pm$0.19 & $\pm$0.09 & $\pm$0.12 & $\pm$0.24& $\pm$0.88 & \multicolumn{1}{|r}{$\quad$} \\
204.9 & 17.35 & $\pm$0.59 & $\pm$0.19 & $\pm$0.09 & $\pm$0.12 & $\pm$0.24& $\pm$0.64 & \multicolumn{1}{|r}{$\quad$} \\
206.6 & 17.96 & $\pm$0.45 & $\pm$0.19 & $\pm$0.09 & $\pm$0.12 & $\pm$0.24& $\pm$0.51 & \multicolumn{1}{|r}{$\quad$} \\
\cline{1-8}
\multicolumn{8}{|c|}
{\Opal~\cite{4f_bib:opaww183,4f_bib:opaww189,4f_bib:opawwsc01}} &
\multicolumn{1}{|r}{$\quad$} \\
\cline{1-8}
182.7 & 15.43 & $\pm$0.61 & $\pm$0.14 & $\pm$0.00 & $\pm$0.22 & $\pm$0.26& $\pm$0.66 & \multicolumn{1}{|r}{$\quad$} \\
188.6 & 16.30 & $\pm$0.35 & $\pm$0.11 & $\pm$0.12 & $\pm$0.07 & $\pm$0.18& $\pm$0.39 & \multicolumn{1}{|r}{$\quad$} \\
191.6 & 16.60 & $\pm$0.90 & $\pm$0.23 & $\pm$0.32 & $\pm$0.14 & $\pm$0.42& $\pm$0.99 & \multicolumn{1}{|r}{$\quad$} \\
195.5 & 18.59 & $\pm$0.61 & $\pm$0.23 & $\pm$0.34 & $\pm$0.14 & $\pm$0.43& $\pm$0.75 & \multicolumn{1}{|r}{$\quad$} \\
199.5 & 16.32 & $\pm$0.55 & $\pm$0.23 & $\pm$0.26 & $\pm$0.14 & $\pm$0.37& $\pm$0.67 & \multicolumn{1}{|r}{$\quad$} \\
201.6 & 18.48 & $\pm$0.82 & $\pm$0.23 & $\pm$0.33 & $\pm$0.14 & $\pm$0.42& $\pm$0.92 & \multicolumn{1}{|r}{$\quad$} \\
204.9 & 15.97 & $\pm$0.52 & $\pm$0.23 & $\pm$0.26 & $\pm$0.14 & $\pm$0.37& $\pm$0.64 & \multicolumn{1}{|r}{$\quad$} \\
206.6 & 17.77 & $\pm$0.42 & $\pm$0.23 & $\pm$0.28 & $\pm$0.14 & $\pm$0.38& $\pm$0.57 & \multicolumn{1}{|r}{$\quad$} \\
\cline{1-8}
\hline
\multicolumn{8}{|c|}{LEP Averages } & $\chi^2/\textrm{d.o.f.}$ \\
\hline
182.7 & 15.81 & $\pm$0.33 & $\pm$0.13 & $\pm$0.05 & $\pm$0.07 & $\pm$0.16& $\pm$0.37 & 
 \multirow{8}{20.3mm}{$
   \hspace*{-0.3mm}
   \left\}
     \begin{array}[h]{rr}
       &\multirow{8}{8mm}{\hspace*{-4.2mm}26.3/24}\\
       &\\ &\\ &\\ &\\ &\\ &\\ &\\  
     \end{array}
   \right.
   $}\\
188.6 & 16.04 & $\pm$0.19 & $\pm$0.10 & $\pm$0.05 & $\pm$0.05 & $\pm$0.12& $\pm$0.22 & \\
191.6 & 16.68 & $\pm$0.46 & $\pm$0.13 & $\pm$0.09 & $\pm$0.06 & $\pm$0.17& $\pm$0.49 & \\
195.5 & 17.16 & $\pm$0.29 & $\pm$0.12 & $\pm$0.07 & $\pm$0.06 & $\pm$0.15& $\pm$0.33 & \\
199.5 & 16.89 & $\pm$0.28 & $\pm$0.12 & $\pm$0.07 & $\pm$0.06 & $\pm$0.15& $\pm$0.32 & \\
201.6 & 17.11 & $\pm$0.40 & $\pm$0.12 & $\pm$0.08 & $\pm$0.06 & $\pm$0.16& $\pm$0.43 & \\
204.9 & 16.80 & $\pm$0.28 & $\pm$0.13 & $\pm$0.07 & $\pm$0.06 & $\pm$0.16& $\pm$0.32 & \\
206.6 & 17.28 & $\pm$0.23 & $\pm$0.12 & $\pm$0.07 & $\pm$0.06 & $\pm$0.15& $\pm$0.27 & \\
\hline
\end{tabular}
\end{small}
\caption[]{%
W-pair production cross-section (in pb) for different \CoM\ energies.
The first column contains the \CoM\ energy
and the second the measurements.
Observed statistical uncertainties are used in the fit
and are listed in the third column;
when asymmetric errors are quoted by the Collaborations,
the positive error is listed in the table and used in the fit.
The fourth, fifth and sixth columns contain
the components of the systematic errors,
as subdivided by the Collaborations into
LEP-correlated   energy-correlated   (LCEC),
LEP-uncorrelated energy-uncorrelated (LUEU),
LEP-uncorrelated energy-correlated   (LUEC).
The total systematic error is given in the seventh column,
the total error in the eighth.
For the LEP averages, the $\chi^2$ of the fit is also given
in the ninth column.}
\label{4f_tab:WWmeas} 
\end{center}
\end{table}

\begin{table}[hbtp]
\begin{center}
\hspace*{-0.3cm}
\renewcommand{\arraystretch}{1.2}
\begin{tabular}{|c|cccccccc|} 
\hline
\roots (GeV) 
      & 182.7 & 188.6 & 191.6 & 195.5 & 199.5 & 201.6 & 204.9 & 206.6 \\
\hline
182.7 & 1.000 & 0.240 & 0.138 & 0.198 & 0.206 & 0.152 & 0.209 & 0.243 \\
188.6 & 0.240 & 1.000 & 0.166 & 0.240 & 0.248 & 0.184 & 0.252 & 0.294 \\
191.6 & 0.138 & 0.166 & 1.000 & 0.137 & 0.142 & 0.105 & 0.144 & 0.167 \\
195.5 & 0.198 & 0.240 & 0.137 & 1.000 & 0.204 & 0.150 & 0.206 & 0.241 \\
199.5 & 0.206 & 0.248 & 0.142 & 0.204 & 1.000 & 0.156 & 0.214 & 0.249 \\
201.6 & 0.152 & 0.184 & 0.105 & 0.150 & 0.156 & 1.000 & 0.158 & 0.184 \\
204.9 & 0.209 & 0.252 & 0.144 & 0.206 & 0.214 & 0.158 & 1.000 & 0.253 \\
206.6 & 0.243 & 0.294 & 0.167 & 0.241 & 0.249 & 0.184 & 0.253 & 1.000 \\
\hline
\end{tabular}
\renewcommand{\arraystretch}{1.}
\caption[]{%
Correlation matrix for the LEP combined W-pair cross-sections
listed at the bottom of Table~\protect\ref{4f_tab:WWmeas}.
Correlations are all positive and range from 11\% to 29\%.}
\label{4f_tab:WWcorr} 
\end{center}
\end{table}

\begin{table}[hbtp]
\begin{center}
\hspace*{-0.3cm}
\renewcommand{\arraystretch}{1.2}
\begin{tabular}{|c|c|c|} 
\hline
\roots & \multicolumn{2}{|c|}{WW cross-section (pb)}                              \\
\cline{2-3} 
(GeV) & $\sww^{\footnotesize\YFSWW}$    
      & $\sww^{\footnotesize\RacoonWW}$ \\
\hline
182.7 & $15.361\pm0.005$ & $15.368\pm0.008$ \\
188.6 & $16.266\pm0.005$ & $16.249\pm0.011$ \\
191.6 & $16.568\pm0.006$ & $16.519\pm0.009$ \\
195.5 & $16.841\pm0.006$ & $16.801\pm0.009$ \\
199.5 & $17.017\pm0.007$ & $16.979\pm0.009$ \\
201.6 & $17.076\pm0.006$ & $17.032\pm0.009$ \\
204.9 & $17.128\pm0.006$ & $17.079\pm0.009$ \\
206.6 & $17.145\pm0.006$ & $17.087\pm0.009$ \\
\hline
\end{tabular}
\renewcommand{\arraystretch}{1.}
\caption[]{%
W-pair cross-section predictions (in pb) for different \CoM\ energies,
according to \YFSWW~\protect\cite{4f_bib:yfsww} and 
\RacoonWW~\protect\cite{4f_bib:racoonww},
for $\Mw=80.35$~GeV.
The errors listed in the table are only the statistical errors 
from the numerical integration of the cross-section.}
\label{4f_tab:WWtheo} 
\end{center}
\end{table}

\begin{table}[hbtp]
\begin{center}
\begin{small}
\begin{tabular}{|c|cccccc|c|c|}
\hline
\roots & & & {\scriptsize (LCEU)} & {\scriptsize (LCEC)} & 
{\scriptsize (LUEU)} & {\scriptsize (LUEC)} & & \\
(GeV) & $\rww$ & 
$\Delta\rww^\mathrm{stat}$ &
$\Delta\rww^\mathrm{syst}$ &
$\Delta\rww^\mathrm{syst}$ &
$\Delta\rww^\mathrm{syst}$ &
$\Delta\rww^\mathrm{syst}$ &
$\Delta\rww$ &
$\chi^2/\textrm{d.o.f.}$ \\
\hline
\hline

\multicolumn{9}{|c|}{\YFSWW~\cite{4f_bib:yfsww}}\\
\hline
182.7 & 1.030 & $\pm$0.021 & $\pm$0.000 & $\pm$0.009 & $\pm$0.003& $\pm$0.005 & $\pm$0.024&
\multirow{8}{20.3mm}{$
  \hspace*{-0.3mm}
  \left\}
    \begin{array}[h]{rr}
      &\multirow{8}{6mm}{\hspace*{-4.2mm}26.3/24}\\
      &\\ &\\ &\\ &\\ &\\ &\\ &\\  
    \end{array}
  \right.
  $}\\
188.6 & 0.986 & $\pm$0.011 & $\pm$0.000 & $\pm$0.006 & $\pm$0.003& $\pm$0.003 & $\pm$0.014&\\
191.6 & 1.007 & $\pm$0.028 & $\pm$0.000 & $\pm$0.008 & $\pm$0.005& $\pm$0.003 & $\pm$0.030&\\
195.5 & 1.019 & $\pm$0.017 & $\pm$0.000 & $\pm$0.007 & $\pm$0.004& $\pm$0.003 & $\pm$0.020&\\
199.5 & 0.992 & $\pm$0.017 & $\pm$0.000 & $\pm$0.007 & $\pm$0.004& $\pm$0.003 & $\pm$0.019&\\
201.6 & 1.002 & $\pm$0.024 & $\pm$0.000 & $\pm$0.007 & $\pm$0.004& $\pm$0.003 & $\pm$0.025&\\
204.9 & 0.981 & $\pm$0.016 & $\pm$0.000 & $\pm$0.007 & $\pm$0.004& $\pm$0.003 & $\pm$0.019&\\
206.6 & 1.008 & $\pm$0.013 & $\pm$0.000 & $\pm$0.007 & $\pm$0.004& $\pm$0.003 & $\pm$0.016&\\
\hline
Average & 
0.997 & $\pm$0.007 & $\pm$0.000 & $\pm$0.007 & $\pm$0.001& $\pm$0.003 & $\pm$0.010&
\hspace*{1.5mm}32.7/31\hspace*{-0.5mm}\\
\hline
\hline
\multicolumn{9}{|c|}{\RacoonWW~\cite{4f_bib:racoonww}}\\
\hline
182.7 & 1.029 & $\pm$0.021 & $\pm$0.001 & $\pm$0.009 & $\pm$0.003& $\pm$0.005 & $\pm$0.024&
\multirow{8}{20.3mm}{$
  \hspace*{-0.3mm}
  \left\}
    \begin{array}[h]{rr}
      &\multirow{8}{6mm}{\hspace*{-4.2mm}26.3/24}\\
      &\\ &\\ &\\ &\\ &\\ &\\ &\\  
    \end{array}
  \right.
  $}\\
188.6 & 0.987 & $\pm$0.011 & $\pm$0.001 & $\pm$0.006 & $\pm$0.003& $\pm$0.003 & $\pm$0.014&\\
191.6 & 1.010 & $\pm$0.028 & $\pm$0.001 & $\pm$0.008 & $\pm$0.005& $\pm$0.003 & $\pm$0.030&\\
195.5 & 1.021 & $\pm$0.017 & $\pm$0.001 & $\pm$0.007 & $\pm$0.004& $\pm$0.003 & $\pm$0.020&\\
199.5 & 0.995 & $\pm$0.017 & $\pm$0.001 & $\pm$0.007 & $\pm$0.004& $\pm$0.003 & $\pm$0.019&\\
201.6 & 1.004 & $\pm$0.024 & $\pm$0.001 & $\pm$0.007 & $\pm$0.005& $\pm$0.003 & $\pm$0.026&\\
204.9 & 0.984 & $\pm$0.017 & $\pm$0.001 & $\pm$0.007 & $\pm$0.004& $\pm$0.003 & $\pm$0.019&\\
206.6 & 1.011 & $\pm$0.013 & $\pm$0.001 & $\pm$0.007 & $\pm$0.004& $\pm$0.003 & $\pm$0.016&\\
\hline
Average & 
0.999 & $\pm$0.007 & $\pm$0.000 & $\pm$0.007 & $\pm$0.002& $\pm$0.003 & $\pm$0.010&
\hspace*{1.5mm}32.7/31\hspace*{-0.5mm}\\
\hline
\end{tabular}
\end{small}
\caption[]{%
Ratios of LEP combined W-pair cross-section measurements
to the expectations of the considered theoretical models,
for different \CoM\ energies and for all energies combined.
The first column contains the \CoM\ energy,
the second the combined ratios,
the third the statistical errors.
The fourth, fifth, sixth and seventh columns contain
the sources of systematic errors that are considered as 
LEP-correlated   energy-uncorrelated (LCEU),
LEP-correlated   energy-correlated   (LCEC),
LEP-uncorrelated energy-uncorrelated (LUEU),
LEP-uncorrelated energy-correlated   (LUEC).
The total error is given in the eighth column.
The only LCEU systematic sources considered 
are the statistical errors on the cross-section theoretical predictions,
while the LCEC, LUEU and LUEC sources are those coming from
the corresponding errors on the cross-section measurements.
For the LEP averages, the $\chi^2$ of the fit is also given
in the ninth column.}
\label{4f_tab:rWWmeas} 
\end{center}
\end{table}

 \renewcommand{\arraystretch}{1.2}
 \begin{table}[p]
 \begin{center}
 \begin{small}
 \hspace*{-0.0cm}
 \begin{tabular}{|l|cccc|c|c|c|}
 \cline{1-8}
 Decay & & & {\scriptsize (unc)} & {\scriptsize (cor)} & & & 
 3$\times$3 correlation \\
 channel & $\wwbr$ & 
 $\Delta\wwbr^\mathrm{stat}$ &
 $\Delta\wwbr^\mathrm{syst}$ &
 $\Delta\wwbr^\mathrm{syst}$ &
 $\Delta\wwbr^\mathrm{syst}$ &
 $\Delta\wwbr$ & 
 for $\Delta\wwbr$\\
 \cline{1-8}
 \multicolumn{8}{|c|}{\Aleph}\\
 \hline
 \BWtoenu & 
 10.95 & $\pm$0.27 & $\pm$0.15 & $\pm$0.04 & $\pm$0.16 & $\pm$0.31 &
 \multirow{3}{47mm}{\mbox{$\Biggl(\negthickspace\negthickspace$
                      \begin{tabular}{ccc}
                       \phm1.000 &    -0.048 &    -0.271 \\
                          -0.048 & \phm1.000 &    -0.253 \\
                          -0.271 &    -0.253 & \phm1.000 \\
                      \end{tabular}
                      $\negthickspace\negthickspace\Biggr)$} } \\
 \BWtomnu & 
 11.11 & $\pm$0.25 & $\pm$0.14 & $\pm$0.04 & $\pm$0.15 & $\pm$0.29 & \\
 \BWtotnu & 
 10.57 & $\pm$0.32 & $\pm$0.20 & $\pm$0.04 & $\pm$0.20 & $\pm$0.38 & \\
 \hline
 \multicolumn{8}{c}{}\\
 
 \cline{1-8}
 \multicolumn{8}{|c|}{\Delphi}\\
 \hline
 \BWtoenu & 
 10.55 & $\pm$0.31 & $\pm$0.13 & $\pm$0.05 & $\pm$0.14 & $\pm$0.34 &
 \multirow{3}{47mm}{\mbox{$\Biggl(\negthickspace\negthickspace$
                      \begin{tabular}{ccc}
                       \phm1.000 &     0.030 &    -0.340 \\
                           0.030 & \phm1.000 &    -0.170 \\
                          -0.340 &    -0.170 & \phm1.000 \\
                      \end{tabular}
                      $\negthickspace\negthickspace\Biggr)$} } \\
 \BWtomnu & 
 10.65 & $\pm$0.26 & $\pm$0.06 & $\pm$0.05 & $\pm$0.08 & $\pm$0.27 & \\
 \BWtotnu & 
 11.46 & $\pm$0.39 & $\pm$0.17 & $\pm$0.09 & $\pm$0.19 & $\pm$0.43 & \\
 \hline
 \multicolumn{8}{c}{}\\
 
 \cline{1-8}
 \multicolumn{8}{|c|}{\Ltre}\\
 \hline
 \BWtoenu & 
 10.40 & $\pm$0.26 & $\pm$0.13 & $\pm$0.06 & $\pm$0.14 & $\pm$0.30 & 
 \multirow{3}{47mm}{\mbox{$\Biggl(\negthickspace\negthickspace$
                      \begin{tabular}{ccc}
                       \phm1.000 &    -0.016 &    -0.279 \\
                          -0.016 & \phm1.000 &    -0.295 \\
                          -0.279 &    -0.295 & \phm1.000 \\
                      \end{tabular}
                      $\negthickspace\negthickspace\Biggr)$} } \\
 \BWtomnu & 
 \phz9.72 & $\pm$0.27 & $\pm$0.14 & $\pm$0.06 & $\pm$0.15 & $\pm$0.31 & \\
 \BWtotnu & 
 11.78 & $\pm$0.38 & $\pm$0.20 & $\pm$0.06 & $\pm$0.21 & $\pm$0.43 & \\
 \hline
 \multicolumn{8}{c}{}\\
 
 \cline{1-8}
 \multicolumn{8}{|c|}{\Opal}\\
 \hline
 \BWtoenu & 
 10.40 & $\pm$0.25 & $\pm$0.24 & $\pm$0.05 & $\pm$0.25 & $\pm$0.35 & 
 \multirow{3}{47mm}{\mbox{$\Biggl(\negthickspace\negthickspace$
                      \begin{tabular}{ccc}
                       \phm1.000 &     0.141 &    -0.179 \\
                           0.141 & \phm1.000 &    -0.174 \\
                          -0.179 &    -0.174 & \phm1.000 \\
                      \end{tabular}
                      $\negthickspace\negthickspace\Biggr)$} } \\
 \BWtomnu & 
 10.61 & $\pm$0.25 & $\pm$0.23 & $\pm$0.06 & $\pm$0.24 & $\pm$0.35 & \\
 \BWtotnu & 
 11.18 & $\pm$0.31 & $\pm$0.37 & $\pm$0.05 & $\pm$0.37 & $\pm$0.48 & \\
 \hline
 \multicolumn{8}{c}{}\\
 
 \cline{1-8}
 \multicolumn{7}{|c}{LEP Average (without lepton universality assumption)}
 &\multicolumn{1}{c|}{}\\
 \hline
 \BWtoenu & 
 10.59 & $\pm$0.14 & $\pm$0.08 & $\pm$0.05 & $\pm$0.10 & $\pm$0.17 &
 \multirow{3}{47mm}{\mbox{$\Biggl(\negthickspace\negthickspace$
                      \begin{tabular}{ccc}
                        \phm1.000 & \phm0.092 &    -0.196 \\
                        \phm0.092 & \phm1.000 &    -0.148 \\
                           -0.196 &    -0.148 & \phm1.000 \\
                      \end{tabular}
                      $\negthickspace\negthickspace\Biggr)$} } \\
 \BWtomnu & 
 10.55 & $\pm$0.13 & $\pm$0.07 & $\pm$0.05 & $\pm$0.09 & $\pm$0.16 & \\
 \BWtotnu & 
 11.20 & $\pm$0.18 & $\pm$0.11 & $\pm$0.06 & $\pm$0.13 & $\pm$0.22 & \\
 \hline
 $\chi^2/\textrm{d.o.f.}$ & \multicolumn{1}{|c|}{15.3/9} & 
 \multicolumn{6}{c}{}\\
 \cline{1-2} 
 \multicolumn{8}{c}{}\\
 
 \cline{1-7} 
 \multicolumn{7}{|c|}{LEP Average (with lepton universality assumption)}
 &\multicolumn{1}{c}{}\\
 \cline{1-7} 
 \BWtolnu & 
 10.74 & $\pm$0.06 & $\pm$0.05 & $\pm$0.05 & $\pm$0.07 & $\pm$0.09 & 
 \multicolumn{1}{c}{}\\
 {\mbox{$\mathcal{B}(\mathrm{W}\rightarrow\mathrm{had.})$}}  & 
 67.77 & $\pm$0.18 & $\pm$0.14 & $\pm$0.16 & $\pm$0.22 & $\pm$0.28 & 
 \multicolumn{1}{c}{}\\
 \cline{1-7} 
 $\chi^2/\textrm{d.o.f.}$ & \multicolumn{1}{|c|}{20.8/11} &
 \multicolumn{6}{c}{}\\
 \cline{1-2} 
 \end{tabular}
 \vspace*{0.5cm}
 
 \end{small}
 \caption[]{%
 W branching fraction measurements (in \%).
 The first column contains the decay channel, 
 the second the measurements,
 the third the statistical uncertainty.
 The fourth and fifth column list 
 the uncorrelated and correlated components
 of the systematic errors,
 as provided by the Collaborations.
 The total systematic error is given in the sixth column and
 the total error in the seventh. 
 Correlation matrices 
 for the three leptonic branching fractions 
 are given in the last column.}
 \label{4f_tab:Wbrmeas} 
 \end{center}
 \end{table}
 \renewcommand{\arraystretch}{1.}

\begin{table}[p]
\vspace*{-0.8cm}
\begin{center}
\begin{small}
\begin{tabular}{|c|ccccc|c|c|}
\cline{1-8}
\roots & & & {\scriptsize (LCEC)} & {\scriptsize (LUEU)} & 
{\scriptsize (LUEC)} & & 
\multicolumn{1}{|r}{$\quad$} \\
(GeV) & $\szz$ & 
$\Delta\szz^\mathrm{stat}$ &
$\Delta\szz^\mathrm{syst}$ &
$\Delta\szz^\mathrm{syst}$ &
$\Delta\szz^\mathrm{syst}$ &
$\Delta\szz$ & 
$\Delta\szz^\mathrm{stat\,(exp)}$ \\
\hline
\multicolumn{8}{|c|}
{\Aleph~\cite{4f_bib:alezz189,4f_bib:alezzsc01}} \\
\hline
182.7 & 0.11 & $^{+0.16}_{-0.11}$ & $\pm$0.01 & $\pm$0.03 & $\pm$0.03 & $^{+0.16}_{-0.12}$ & $\pm$0.14 \\
188.6 & 0.67 & $^{+0.13}_{-0.12}$ & $\pm$0.01 & $\pm$0.03 & $\pm$0.03 & $^{+0.14}_{-0.13}$ & $\pm$0.13 \\
191.6 & 0.53 & $^{+0.34}_{-0.27}$ & $\pm$0.01 & $\pm$0.01 & $\pm$0.01 & $^{+0.34}_{-0.27}$ & $\pm$0.33 \\
195.5 & 0.69 & $^{+0.23}_{-0.20}$ & $\pm$0.01 & $\pm$0.02 & $\pm$0.02 & $^{+0.23}_{-0.20}$ & $\pm$0.23 \\
199.5 & 0.70 & $^{+0.22}_{-0.20}$ & $\pm$0.01 & $\pm$0.02 & $\pm$0.02 & $^{+0.22}_{-0.20}$ & $\pm$0.23 \\
201.6 & 0.70 & $^{+0.33}_{-0.28}$ & $\pm$0.01 & $\pm$0.01 & $\pm$0.01 & $^{+0.33}_{-0.28}$ & $\pm$0.35 \\
204.9 & 1.21 & $^{+0.26}_{-0.23}$ & $\pm$0.01 & $\pm$0.02 & $\pm$0.02 & $^{+0.26}_{-0.23}$ & $\pm$0.27 \\
206.6 & 1.01 & $^{+0.19}_{-0.17}$ & $\pm$0.01 & $\pm$0.01 & $\pm$0.01 & $^{+0.19}_{-0.17}$ & $\pm$0.18 \\
\hline
\multicolumn{8}{|c|}
{\Delphi~\cite{4f_bib:delzz}} \\
\hline
182.7 & 0.35 & $^{+0.20}_{-0.15}$ & $\pm$0.01 & $\pm$0.00 & $\pm$0.02 & $^{+0.20}_{-0.15}$ & $\pm$0.16 \\
188.6 & 0.52 & $^{+0.12}_{-0.11}$ & $\pm$0.01 & $\pm$0.00 & $\pm$0.02 & $^{+0.12}_{-0.11}$ & $\pm$0.13 \\
191.6 & 0.63 & $^{+0.36}_{-0.30}$ & $\pm$0.01 & $\pm$0.01 & $\pm$0.02 & $^{+0.36}_{-0.30}$ & $\pm$0.35 \\
195.5 & 1.05 & $^{+0.25}_{-0.22}$ & $\pm$0.01 & $\pm$0.01 & $\pm$0.02 & $^{+0.25}_{-0.22}$ & $\pm$0.21 \\
199.5 & 0.75 & $^{+0.20}_{-0.18}$ & $\pm$0.01 & $\pm$0.01 & $\pm$0.01 & $^{+0.20}_{-0.18}$ & $\pm$0.21 \\
201.6 & 0.85 & $^{+0.33}_{-0.28}$ & $\pm$0.01 & $\pm$0.01 & $\pm$0.01 & $^{+0.33}_{-0.28}$ & $\pm$0.32 \\
204.9 & 1.03 & $^{+0.23}_{-0.20}$ & $\pm$0.02 & $\pm$0.01 & $\pm$0.01 & $^{+0.23}_{-0.20}$ & $\pm$0.23 \\
206.6 & 0.96 & $^{+0.16}_{-0.15}$ & $\pm$0.02 & $\pm$0.01 & $\pm$0.01 & $^{+0.16}_{-0.15}$ & $\pm$0.17 \\
\hline
\multicolumn{8}{|c|}
{\Ltre~\cite{4f_bib:ltrzz183a,4f_bib:ltrzz183b,4f_bib:ltrzz189,4f_bib:ltrzz1999,4f_bib:ltrzzsc03}} \\
\hline
182.7 & 0.31 & $\pm$0.16 & $\pm$0.05 & $\pm$0.00 & $\pm$0.01 & $\pm$0.17 & $\pm$0.16 \\
188.6 & 0.73 & $\pm$0.15 & $\pm$0.02 & $\pm$0.02 & $\pm$0.02 & $\pm$0.15 & $\pm$0.15 \\
191.6 & 0.29 & $\pm$0.22 & $\pm$0.01 & $\pm$0.01 & $\pm$0.02 & $\pm$0.22 & $\pm$0.34 \\
195.5 & 1.18 & $\pm$0.24 & $\pm$0.04 & $\pm$0.05 & $\pm$0.06 & $\pm$0.26 & $\pm$0.22 \\
199.5 & 1.25 & $\pm$0.25 & $\pm$0.04 & $\pm$0.05 & $\pm$0.07 & $\pm$0.27 & $\pm$0.24 \\
201.6 & 0.95 & $\pm$0.38 & $\pm$0.03 & $\pm$0.04 & $\pm$0.05 & $\pm$0.39 & $\pm$0.35 \\
204.9 & 0.77 & $^{+0.21}_{-0.19}$ & $\pm$0.01 & $\pm$0.01 & $\pm$0.04 & $^{+0.21}_{-0.19}$ & $\pm$0.22 \\
206.6 & 1.09 & $^{+0.17}_{-0.16}$ & $\pm$0.02 & $\pm$0.02 & $\pm$0.06 & $^{+0.18}_{-0.17}$ & $\pm$0.17 \\
\hline
\multicolumn{8}{|c|}
{\Opal~\cite{4f_bib:opazz189,4f_bib:opazz}}  \\
\hline
182.7 & 0.12 & $^{+0.20}_{-0.18}$ & $\pm$0.00 & $\pm$0.03 & $\pm$0.00 & $^{+0.20}_{-0.18}$ & $\pm$0.19 \\
188.6 & 0.80 & $^{+0.14}_{-0.13}$ & $\pm$0.01 & $\pm$0.05 & $\pm$0.03 & $^{+0.15}_{-0.14}$ & $\pm$0.14 \\
191.6 & 1.29 & $^{+0.47}_{-0.40}$ & $\pm$0.02 & $\pm$0.09 & $\pm$0.05 & $^{+0.48}_{-0.41}$ & $\pm$0.36 \\
195.5 & 1.13 & $^{+0.26}_{-0.24}$ & $\pm$0.02 & $\pm$0.06 & $\pm$0.05 & $^{+0.27}_{-0.25}$ & $\pm$0.25 \\
199.5 & 1.05 & $^{+0.25}_{-0.22}$ & $\pm$0.02 & $\pm$0.05 & $\pm$0.04 & $^{+0.26}_{-0.23}$ & $\pm$0.25 \\
201.6 & 0.79 & $^{+0.35}_{-0.29}$ & $\pm$0.02 & $\pm$0.05 & $\pm$0.03 & $^{+0.36}_{-0.30}$ & $\pm$0.37 \\
204.9 & 1.07 & $^{+0.27}_{-0.24}$ & $\pm$0.02 & $\pm$0.06 & $\pm$0.04 & $^{+0.28}_{-0.25}$ & $\pm$0.26 \\
206.6 & 0.97 & $^{+0.19}_{-0.18}$ & $\pm$0.02 & $\pm$0.05 & $\pm$0.04 & $^{+0.20}_{-0.19}$ & $\pm$0.20 \\
\hline
\multicolumn{7}{|c|}
{LEP} & $\chi^2/\textrm{d.o.f.}$ \\
\hline
182.7 & 0.22 & $\pm$0.08 & $\pm$0.02 & $\pm$0.01 & $\pm$0.01 & $\pm$0.08 & 
 \multirow{8}{20.3mm}{$
   \hspace*{-0.3mm}
   \left\}
     \begin{array}[h]{rr}
       &\multirow{8}{8mm}{\hspace*{-4.2mm}16.1/24}\\
       &\\ &\\ &\\ &\\ &\\ &\\ &\\  
     \end{array}
   \right.
   $}\\
188.6 & 0.66 & $\pm$0.07 & $\pm$0.01 & $\pm$0.01 & $\pm$0.01 & $\pm$0.07 & \\
191.6 & 0.65 & $\pm$0.17 & $\pm$0.01 & $\pm$0.02 & $\pm$0.01 & $\pm$0.17 & \\
195.5 & 0.99 & $\pm$0.11 & $\pm$0.02 & $\pm$0.02 & $\pm$0.02 & $\pm$0.12 & \\
199.5 & 0.90 & $\pm$0.12 & $\pm$0.02 & $\pm$0.02 & $\pm$0.02 & $\pm$0.12 & \\
201.6 & 0.81 & $\pm$0.17 & $\pm$0.02 & $\pm$0.02 & $\pm$0.01 & $\pm$0.17 & \\
204.9 & 0.98 & $\pm$0.12 & $\pm$0.01 & $\pm$0.01 & $\pm$0.02 & $\pm$0.13 & \\
206.6 & 0.99 & $\pm$0.09 & $\pm$0.02 & $\pm$0.01 & $\pm$0.02 & $\pm$0.09 & \\
\hline
\end{tabular}
\end{small}
\caption[]{%
Z-pair production cross-section (in pb) at different energies.
The first column contains the LEP \CoM\ energy,
the second the measurements and
the third the statistical uncertainty. 
The fourth, the fifth and the sixth columns list 
the different components of the systematic errors, 
as provided by the Collaborations.
The total error is given in the seventh column,
whereas the eighth column lists, for the four LEP measurements,
the symmetrized expected statistical error,
and for the LEP combined value,
the $\chi^2$ of the fit.}
\label{4f_tab:ZZmeas} 
\end{center}
\end{table}

\begin{table}[hbtp]
\begin{center}
\hspace*{-0.3cm}
\renewcommand{\arraystretch}{1.2}
\begin{tabular}{|c|c|c|} 
\hline
\roots & \multicolumn{2}{|c|}{ZZ cross-section (pb)}  \\
\cline{2-3} 
(GeV) & $\szz^{\footnotesize\YFSZZ}$    
      & $\szz^{\footnotesize\ZZTO}$ \\
\hline
182.7 & 0.254[1] & 0.25425[2] \\
188.6 & 0.655[2] & 0.64823[1] \\
191.6 & 0.782[2] & 0.77670[1] \\
195.5 & 0.897[3] & 0.89622[1] \\
199.5 & 0.981[2] & 0.97765[1] \\
201.6 & 1.015[1] & 1.00937[1] \\
204.9 & 1.050[1] & 1.04335[1] \\
206.6 & 1.066[1] & 1.05535[1] \\
\hline
\end{tabular}
\renewcommand{\arraystretch}{1.}
\caption[]{%
Z-pair cross-section predictions (in pb) interpolated at the data 
\CoM\ energies,according to the \YFSZZ~\protect\cite{4f_bib:yfszz} and 
\ZZTO~\protect\cite{4f_bib:zzto} predictions.
The numbers in brackets are the errors on the last digit and are coming
from the numerical integration of the cross-section only.}
\label{4f_tab:ZZtheo} 
\end{center}
\end{table}

\begin{table}[hbtp]
\begin{center}
\begin{small}
\begin{tabular}{|c|cccccc|c|c|}
\hline
\roots & & & {\scriptsize (LCEU)} & {\scriptsize (LCEC)} & 
{\scriptsize (LUEU)} & {\scriptsize (LUEC)} & & \\
(GeV) & $\rzz$ & 
$\Delta\rzz^\mathrm{stat}$ &
$\Delta\rzz^\mathrm{syst}$ &
$\Delta\rzz^\mathrm{syst}$ &
$\Delta\rzz^\mathrm{syst}$ &
$\Delta\rzz^\mathrm{syst}$ &
$\Delta\rzz$ &
$\chi^2/\textrm{d.o.f.}$ \\
\hline
\hline
\multicolumn{9}{|c|}{\YFSZZ~\cite{4f_bib:yfszz}}\\
\hline
182.7 & 0.857 & $\pm$0.307 & $\pm$0.018 & $\pm$0.068 & $\pm$0.041 & $\pm$0.040 & $\pm$0.320 &
\multirow{8}{20.3mm}{$
  \hspace*{-0.3mm}
  \left\}
    \begin{array}[h]{rr}
      &\multirow{8}{6mm}{\hspace*{-4.2mm}16.1/24}\\
      &\\ &\\ &\\ &\\ &\\ &\\ &\\  
    \end{array}
  \right.
  $}\\
188.6 & 1.007 & $\pm$0.104 & $\pm$0.020 & $\pm$0.019 & $\pm$0.022& $\pm$0.018 & $\pm$0.111&\\
191.6 & 0.826 & $\pm$0.220 & $\pm$0.017 & $\pm$0.014 & $\pm$0.025& $\pm$0.017 & $\pm$0.224&\\
195.5 & 1.100 & $\pm$0.127 & $\pm$0.022 & $\pm$0.021 & $\pm$0.019& $\pm$0.020 & $\pm$0.133&\\
199.5 & 0.912 & $\pm$0.119 & $\pm$0.019 & $\pm$0.018 & $\pm$0.016& $\pm$0.017 & $\pm$0.124&\\
201.6 & 0.795 & $\pm$0.170 & $\pm$0.016 & $\pm$0.017 & $\pm$0.015& $\pm$0.013 & $\pm$0.173&\\
204.9 & 0.931 & $\pm$0.116 & $\pm$0.019 & $\pm$0.014 & $\pm$0.013& $\pm$0.014 & $\pm$0.120&\\
206.6 & 0.928 & $\pm$0.085 & $\pm$0.019 & $\pm$0.014 & $\pm$0.010& $\pm$0.015 & $\pm$0.090&\\
\hline
Average & 
0.945 & $\pm$0.045 & $\pm$0.008 & $\pm$0.017 & $\pm$0.006& $\pm$0.016 & $\pm$0.052&
\hspace*{1.5mm}19.1/31\hspace*{-0.5mm}\\
\hline
\hline
\multicolumn{9}{|c|}{\ZZTO~\cite{4f_bib:zzto}}\\
\hline
182.7 & 0.857 & $\pm$0.307 & $\pm$0.018 & $\pm$0.068 & $\pm$0.041 & $\pm$0.040 & $\pm$0.320 &
\multirow{8}{20.3mm}{$
  \hspace*{-0.3mm}
  \left\}
    \begin{array}[h]{rr}
      &\multirow{8}{6mm}{\hspace*{-4.2mm}16.1/24}\\
      &\\ &\\ &\\ &\\ &\\ &\\ &\\  
    \end{array}
  \right.
  $}\\
188.6 & 1.017 & $\pm$0.105 & $\pm$0.021 & $\pm$0.019 & $\pm$0.022& $\pm$0.019 & $\pm$0.113&\\
191.6 & 0.831 & $\pm$0.222 & $\pm$0.017 & $\pm$0.014 & $\pm$0.025& $\pm$0.017 & $\pm$0.225&\\
195.5 & 1.100 & $\pm$0.127 & $\pm$0.022 & $\pm$0.021 & $\pm$0.019& $\pm$0.020 & $\pm$0.133&\\
199.5 & 0.915 & $\pm$0.120 & $\pm$0.019 & $\pm$0.018 & $\pm$0.016& $\pm$0.017 & $\pm$0.125&\\
201.6 & 0.799 & $\pm$0.171 & $\pm$0.016 & $\pm$0.017 & $\pm$0.015& $\pm$0.013 & $\pm$0.174&\\
204.9 & 0.937 & $\pm$0.117 & $\pm$0.019 & $\pm$0.014 & $\pm$0.013& $\pm$0.014 & $\pm$0.121&\\
206.6 & 0.937 & $\pm$0.085 & $\pm$0.019 & $\pm$0.014 & $\pm$0.011& $\pm$0.015 & $\pm$0.091&\\
\hline
Average & 
0.952 & $\pm$0.046 & $\pm$0.008 & $\pm$0.017 & $\pm$0.006& $\pm$0.016 & $\pm$0.052&
\hspace*{1.5mm}19.1/31\hspace*{-0.5mm}\\
\hline
\end{tabular}
\end{small}
\caption[]{%
Ratios of LEP combined Z-pair cross-section measurements
to the expectations, for different \CoM\ energies and for all energies combined.
The first column contains the \CoM\ energy,
the second the combined ratios,
the third the statistical errors.
The fourth, fifth, sixth and seventh columns contain
the sources of systematic errors that are considered as 
LEP-correlated   energy-uncorrelated (LCEU),
LEP-correlated   energy-correlated   (LCEC),
LEP-uncorrelated energy-uncorrelated (LUEU),
LEP-uncorrelated energy-correlated   (LUEC).
The total error is given in the eighth column.
The only LCEU systematic sources considered 
are the statistical errors on the cross-section theoretical predictions,
while the LCEC, LUEU and LUEC sources are those coming from
the corresponding errors on the cross-section measurements.
For the LEP averages, the $\chi^2$ of the fit is also given
in the ninth column.}
\label{4f_tab:rZZmeas} 
\end{center}
\end{table}

\begin{table}[p]
\renewcommand{\arraystretch}{1.2}
\vspace*{-0.0cm}
\begin{center}
\begin{small}
\begin{tabular}{|c|ccccc|c|c|}
\cline{1-8}
\roots & & & {\scriptsize (LCEC)} & {\scriptsize (LUEU)} & 
{\scriptsize (LUEC)} & & \\
(GeV) & $\szee$ & 
$\Delta\szee^\mathrm{stat}$ &
$\Delta\szee^\mathrm{syst}$ &
$\Delta\szee^\mathrm{syst}$ &
$\Delta\szee^\mathrm{syst}$ &
$\Delta\szee$ & 
$\Delta\szee^\mathrm{stat\,(exp)}$ \\
\hline
\multicolumn{8}{|c|}
{\Aleph~\cite{4f_bib:alezeesc03}} \\
\hline
182.7 & 0.25 & $^{+0.20}_{-0.15}$ & $\pm$0.00 & $\pm$0.02 & $\pm$0.00 & $^{+0.20}_{-0.15}$ & $\pm$0.20 \\
188.6 & 0.40 & $^{+0.12}_{-0.11}$ & $\pm$0.00 & $\pm$0.02 & $\pm$0.00 & $^{+0.12}_{-0.11}$ & $\pm$0.11 \\
191.6 & 0.58 & $^{+0.37}_{-0.27}$ & $\pm$0.00 & $\pm$0.03 & $\pm$0.00 & $^{+0.37}_{-0.27}$ & $\pm$0.30 \\
195.5 & 0.68 & $^{+0.22}_{-0.18}$ & $\pm$0.00 & $\pm$0.03 & $\pm$0.00 & $^{+0.22}_{-0.18}$ & $\pm$0.19 \\
199.5 & 0.57 & $^{+0.19}_{-0.17}$ & $\pm$0.00 & $\pm$0.03 & $\pm$0.00 & $^{+0.19}_{-0.17}$ & $\pm$0.17 \\
201.6 & 0.82 & $^{+0.33}_{-0.26}$ & $\pm$0.00 & $\pm$0.04 & $\pm$0.00 & $^{+0.33}_{-0.26}$ & $\pm$0.25 \\
204.9 & 0.41 & $^{+0.16}_{-0.13}$ & $\pm$0.00 & $\pm$0.02 & $\pm$0.00 & $^{+0.16}_{-0.13}$ & $\pm$0.16 \\
206.6 & 0.66 & $^{+0.16}_{-0.14}$ & $\pm$0.00 & $\pm$0.03 & $\pm$0.00 & $^{+0.16}_{-0.14}$ & $\pm$0.14 \\
\hline
\multicolumn{8}{|c|}
{\Delphi~\cite{4f_bib:delswsc03}} \\
\hline
182.7 & 0.56 & $^{+0.27}_{-0.22}$ & $\pm$0.01 & $\pm$0.06 & $\pm$0.02 & $^{+0.28}_{-0.23}$ & $\pm$0.24 \\
188.6 & 0.65 & $^{+0.15}_{-0.14}$ & $\pm$0.01 & $\pm$0.03 & $\pm$0.03 & $^{+0.16}_{-0.15}$ & $\pm$0.14 \\
191.6 & 0.63 & $^{+0.40}_{-0.30}$ & $\pm$0.01 & $\pm$0.03 & $\pm$0.03 & $^{+0.40}_{-0.30}$ & $\pm$0.33 \\
195.5 & 0.66 & $^{+0.22}_{-0.18}$ & $\pm$0.01 & $\pm$0.02 & $\pm$0.03 & $^{+0.22}_{-0.19}$ & $\pm$0.19 \\
199.5 & 0.57 & $^{+0.20}_{-0.17}$ & $\pm$0.01 & $\pm$0.02 & $\pm$0.02 & $^{+0.20}_{-0.17}$ & $\pm$0.18 \\
201.6 & 0.19 & $^{+0.21}_{-0.16}$ & $\pm$0.01 & $\pm$0.02 & $\pm$0.01 & $^{+0.21}_{-0.16}$ & $\pm$0.25 \\
204.9 & 0.37 & $^{+0.18}_{-0.15}$ & $\pm$0.01 & $\pm$0.02 & $\pm$0.02 & $^{+0.18}_{-0.15}$ & $\pm$0.19 \\
206.6 & 0.68 & $^{+0.16}_{-0.14}$ & $\pm$0.01 & $\pm$0.02 & $\pm$0.03 & $^{+0.16}_{-0.14}$ & $\pm$0.14 \\
\hline
\multicolumn{8}{|c|}
{\Ltre~\cite{4f_bib:ltrzee}} \\
\hline
182.7 & 0.51 & $^{+0.19}_{-0.16}$ & $\pm$0.02 & $\pm$0.01 & $\pm$0.03 & $^{+0.19}_{-0.16}$ & $\pm$0.16 \\
188.6 & 0.55 & $^{+0.10}_{-0.09}$ & $\pm$0.02 & $\pm$0.01 & $\pm$0.03 & $^{+0.11}_{-0.10}$ & $\pm$0.09 \\
191.6 & 0.60 & $^{+0.26}_{-0.21}$ & $\pm$0.01 & $\pm$0.01 & $\pm$0.03 & $^{+0.27}_{-0.22}$ & $\pm$0.21 \\
195.5 & 0.40 & $^{+0.13}_{-0.11}$ & $\pm$0.01 & $\pm$0.01 & $\pm$0.03 & $^{+0.13}_{-0.11}$ & $\pm$0.13 \\
199.5 & 0.33 & $^{+0.12}_{-0.10}$ & $\pm$0.01 & $\pm$0.01 & $\pm$0.03 & $^{+0.13}_{-0.11}$ & $\pm$0.14 \\
201.6 & 0.81 & $^{+0.27}_{-0.23}$ & $\pm$0.02 & $\pm$0.02 & $\pm$0.03 & $^{+0.27}_{-0.23}$ & $\pm$0.19 \\
204.9 & 0.56 & $^{+0.16}_{-0.14}$ & $\pm$0.01 & $\pm$0.01 & $\pm$0.03 & $^{+0.16}_{-0.14}$ & $\pm$0.14 \\
206.6 & 0.59 & $^{+0.12}_{-0.10}$ & $\pm$0.01 & $\pm$0.01 & $\pm$0.03 & $^{+0.12}_{-0.11}$ & $\pm$0.11 \\
\hline
\multicolumn{7}{|c|}
{LEP} & $\chi^2/\textrm{d.o.f.}$ \\
\hline
182.7 & 0.44 & $\pm$0.11 & $\pm$0.01 & $\pm$0.02 & $\pm$0.01 & $\pm$0.11 & 
 \multirow{8}{20.3mm}{$
   \hspace*{-0.3mm}
   \left\}
     \begin{array}[h]{rr}
       &\multirow{8}{8mm}{\hspace*{-4.2mm}12.4/16}\\
       &\\ &\\ &\\ &\\ &\\ &\\ &\\  
     \end{array}
   \right.
   $}\\
188.6 & 0.52 & $\pm$0.06 & $\pm$0.01 & $\pm$0.01 & $\pm$0.01 & $\pm$0.07 &  \\
191.6 & 0.60 & $\pm$0.15 & $\pm$0.01 & $\pm$0.02 & $\pm$0.01 & $\pm$0.15 &  \\
195.5 & 0.53 & $\pm$0.09 & $\pm$0.01 & $\pm$0.01 & $\pm$0.01 & $\pm$0.10 &  \\
199.5 & 0.47 & $\pm$0.09 & $\pm$0.01 & $\pm$0.02 & $\pm$0.01 & $\pm$0.10 &  \\
201.6 & 0.65 & $\pm$0.13 & $\pm$0.01 & $\pm$0.02 & $\pm$0.01 & $\pm$0.13 &  \\
204.9 & 0.46 & $\pm$0.09 & $\pm$0.01 & $\pm$0.01 & $\pm$0.01 & $\pm$0.10 &  \\
206.6 & 0.63 & $\pm$0.07 & $\pm$0.01 & $\pm$0.01 & $\pm$0.01 & $\pm$0.08 &  \\
\hline
\end{tabular}
\end{small}
\caption[]{%
Single-Z hadronic production cross-section (in pb) at different energies.
The first column contains the LEP \CoM\ energy,
and the second the measurements. 
The third column reports the statistical error, whereas in the fourth to the
sixth columns the different systematic uncertainties are listed.
The seventh column contains the total error and the eight lists,
for the four LEP measurements,
the symmetrized expected statistical error,
and for the LEP combined value,
the $\chi^2$ of the fit.}
\label{4f_tab:Zeemeas} 
\end{center}
\renewcommand{\arraystretch}{1.}
\end{table}

\begin{table}[hbtp]
\begin{center}
\hspace*{-0.3cm}
\renewcommand{\arraystretch}{1.2}
\begin{tabular}{|c|c|c|} 
\hline
\roots & \multicolumn{2}{|c|}{Zee cross-section (pb)}  \\
\cline{2-3} 
(GeV) & $\szee^{\footnotesize\WPHACT}$    
      & $\szee^{\footnotesize\Grace}$ \\
\hline
182.7 & 0.51275[4] & 0.51573[4] \\
188.6 & 0.53686[4] & 0.54095[5] \\
191.6 & 0.54883[4] & 0.55314[5] \\
195.5 & 0.56399[5] & 0.56891[4] \\
199.5 & 0.57935[5] & 0.58439[4] \\
201.6 & 0.58708[4] & 0.59243[4] \\
204.9 & 0.59905[4] & 0.60487[4] \\
206.6 & 0.61752[4] & 0.60819[4] \\
\hline
\end{tabular}
\renewcommand{\arraystretch}{1.}
\caption[]{%
Zee cross-section predictions (in pb) interpolated at the data 
\CoM\ energies,according to the \WPHACT~\protect\cite{4f_bib:wphact} and 
\Grace~\protect\cite{4f_bib:grace} predictions.
The numbers in brackets are the errors on the last digit and are coming
from the numerical integration of the cross-section only.}
\label{4f_tab:Zeetheo} 
\end{center}
\end{table}

\begin{table}[hbtp]
\begin{center}
\begin{small}
\begin{tabular}{|c|cccccc|c|c|}
\hline
\roots & & & {\scriptsize (LCEU)} & {\scriptsize (LCEC)} & 
{\scriptsize (LUEU)} & {\scriptsize (LUEC)} & & \\
(GeV) & $\rzee$ & 
$\Delta\rzee^\mathrm{stat}$ &
$\Delta\rzee^\mathrm{syst}$ &
$\Delta\rzee^\mathrm{syst}$ &
$\Delta\rzee^\mathrm{syst}$ &
$\Delta\rzee^\mathrm{syst}$ &
$\Delta\rzee$ &
$\chi^2/\textrm{d.o.f.}$ \\
\hline
\hline
\multicolumn{9}{|c|}{\Grace~\cite{4f_bib:grace}}\\
\hline
182.7 & 0.848 & $\pm$0.213 & $\pm$0.049 & $\pm$0.012 & $\pm$0.033 & $\pm$0.023 & $\pm$0.222 &
\multirow{8}{20.3mm}{$
  \hspace*{-0.3mm}
  \left\}
    \begin{array}[h]{rr}
      &\multirow{8}{6mm}{\hspace*{-4.2mm}12.2/16}\\
      &\\ &\\ &\\ &\\ &\\ &\\ &\\  
    \end{array}
  \right.
  $}\\
188.6 & 0.941 & $\pm$0.118 & $\pm$0.087 & $\pm$0.012 & $\pm$0.022 & $\pm$0.021 & $\pm$0.150 &\\
191.6 & 1.085 & $\pm$0.276 & $\pm$0.079 & $\pm$0.012 & $\pm$0.028 & $\pm$0.025 & $\pm$0.289 &\\
195.5 & 0.932 & $\pm$0.165 & $\pm$0.017 & $\pm$0.009 & $\pm$0.025 & $\pm$0.024 & $\pm$0.170 &\\
199.5 & 0.767 & $\pm$0.160 & $\pm$0.119 & $\pm$0.010 & $\pm$0.032 & $\pm$0.023 & $\pm$0.204 &\\
201.6 & 1.078 & $\pm$0.221 & $\pm$0.055 & $\pm$0.015 & $\pm$0.026 & $\pm$0.020 & $\pm$0.230 &\\
204.9 & 0.752 & $\pm$0.155 & $\pm$0.087 & $\pm$0.010 & $\pm$0.017 & $\pm$0.018 & $\pm$0.179 &\\
206.6 & 1.033 & $\pm$0.121 & $\pm$0.154 & $\pm$0.011 & $\pm$0.018 & $\pm$0.024 & $\pm$0.198 &\\
\hline
Average & 
0.914 & $\pm$0.060 & $\pm$0.033 & $\pm$0.011 & $\pm$0.009 & $\pm$0.022 & $\pm$0.073&
\hspace*{1.5mm}15.0/23\hspace*{-0.5mm}\\
\hline
\hline
\multicolumn{9}{|c|}{\WPHACT~\cite{4f_bib:wphact}}\\
\hline
182.7 & 0.846 & $\pm$0.214 & $\pm$0.083 & $\pm$0.011 & $\pm$0.033 & $\pm$0.022 & $\pm$0.233 &
\multirow{8}{20.3mm}{$
  \hspace*{-0.3mm}
  \left\}
    \begin{array}[h]{rr}
      &\multirow{8}{6mm}{\hspace*{-4.2mm}12.2/16}\\
      &\\ &\\ &\\ &\\ &\\ &\\ &\\  
    \end{array}
  \right.
  $}\\
188.6 & 0.962 & $\pm$0.118 & $\pm$0.036 & $\pm$0.013 & $\pm$0.022 & $\pm$0.022 & $\pm$0.128 &\\
191.6 & 1.094 & $\pm$0.278 & $\pm$0.120 & $\pm$0.012 & $\pm$0.028 & $\pm$0.025 & $\pm$0.305 &\\
195.5 & 0.903 & $\pm$0.168 & $\pm$0.129 & $\pm$0.009 & $\pm$0.026 & $\pm$0.027 & $\pm$0.216 &\\
199.5 & 0.784 & $\pm$0.161 & $\pm$0.095 & $\pm$0.010 & $\pm$0.032 & $\pm$0.023 & $\pm$0.192 &\\
201.6 & 1.051 & $\pm$0.224 & $\pm$0.107 & $\pm$0.015 & $\pm$0.025 & $\pm$0.019 & $\pm$0.251 &\\
204.9 & 0.763 & $\pm$0.156 & $\pm$0.063 & $\pm$0.010 & $\pm$0.017 & $\pm$0.019 & $\pm$0.170 &\\
206.6 & 1.044 & $\pm$0.121 & $\pm$0.035 & $\pm$0.011 & $\pm$0.018 & $\pm$0.023 & $\pm$0.130 &\\
\hline
Average & 
0.932 & $\pm$0.058 & $\pm$0.023 & $\pm$0.011 & $\pm$0.009 & $\pm$0.022 & $\pm$0.068&
\hspace*{1.5mm}15.0/23\hspace*{-0.5mm}\\
\hline
\end{tabular}
\end{small}
\caption[]{%
Ratios of LEP combined single-Z cross-section measurements
to the expectations, for different \CoM\ energies and for all energies combined.
The first column contains the \CoM\ energy,
the second the combined ratios,
the third the statistical errors.
The fourth, fifth, sixth and seventh columns contain
the sources of systematic errors that are considered as 
LEP-correlated   energy-uncorrelated (LCEU),
LEP-correlated   energy-correlated   (LCEC),
LEP-uncorrelated energy-uncorrelated (LUEU),
LEP-uncorrelated energy-correlated   (LUEC).
The total error is given in the eighth column.
The only LCEU systematic sources considered 
are the statistical errors on the cross-section theoretical predictions,
while the LCEC, LUEU and LUEC sources are those coming from
the corresponding errors on the cross-section measurements.
For the LEP averages, the $\chi^2$ of the fit is also given
in the ninth column.}
\label{4f_tab:rzeemeas} 
\end{center}
\end{table}

%% file: cr_app.tex
\chapter{Colour Reconnection Combination}

 \section{Inputs}
  \label{fsi:cr:app:inputs}
 \begin{table}[hbt]
  \center
  \begin{tabular}{|c||c|c|c|c|} \hline
               & \multicolumn{4}{c|}{Experiment} \\
   \Rn         & \multicolumn{1}{c}{\Aleph} & \multicolumn{1}{c}{\Delphi} &
                 \multicolumn{1}{c}{\Ltre}  & \multicolumn{1}{c|}{\Opal} \\ \hline\hline

   Data &   $1.0951\pm0.0135$   & $0.8996\pm0.0314$    & $0.8436\pm0.0217$ & $1.2570\pm0.0251$ \\
   \SKI\ (100\%)
        &   $1.0548\pm0.0012$   & $0.8463\pm0.0036$    & $0.7482\pm0.0033$    &     $1.1386\pm0.0027$   \\
  \Jetset
        &   $1.1365\pm0.0013$   & $0.9444\pm0.0039$    & $0.8622\pm0.0037$    &     $1.2958\pm0.0028$  \\
  \ARII
        &   $1.1341\pm0.0013$   & $0.9552\pm0.0041$    & $0.8696\pm0.0037$    &     $1.2887\pm0.0028$   \\
  \Ariadne
        &   $1.1461\pm0.0013$   & $0.9530\pm0.0039$    & $0.8754\pm0.0037$    &     $1.3057\pm0.0028$    \\
  \Herwig\ CR
        &   $1.1416\pm0.0013$   & $0.9649\pm0.0039$    & $0.8805\pm0.0037$    &     $1.3016\pm0.0029$    \\
  \Herwig
        &   $1.1548\pm0.0013$   & $0.9675\pm0.0040$    & $0.8822\pm0.0038$    &     $1.3204\pm0.0029$     \\
                                                                         \hline\hline
Systematics    &             &                 &                 &       \\ \hline\hline
 Intra-W BEC
              &  $\pm0.0020$ & $\pm0.0094$     & $\pm0.0017$     & $\pm0.0015$            \\
  \eeqq\ shape
              &  $\pm0.0012$ & $\pm0.0013$     & $\pm0.0086$     & $\pm0.0035$            \\
  $\pm10$\% $\sigma(\eeqq)$
              &  $\pm0.0036$ & $\pm0.0042$     & $\pm0.0071$     & $\pm0.0040$            \\
  $\pm15$\% $\sigma(\ZZtoqqqq)$
              &  $\pm0.0004$ & $\pm0.0001$     & $\pm0.0020$     & $\pm0.0013$            \\
 Detector effects
              &  $0.0040$    &         $-$     & $\pm0.0016$     & $\pm0.0072$            \\
 $E_{\mathrm{cm}}$ dependence
              &  $\pm0.0062$ & $\pm0.0012$     & $\pm0.0020$     & $\pm0.0030$            \\ \hline
  \end{tabular}
  \center
 \caption[Experimental inputs in particle flow.]{Inputs provided by the experiments for the combination.}
 \label{fsi:cr:tab:inputs}
 \end{table}

\begin{table}[bht]
\begin{center}
\begin{tabular}{||c|c|c|c|c|c||}
\hline
$k_{i}$  & $P_{reco}$ (\%) & ALEPH & DELPHI & L3 & OPAL \\
\hline
 \hline
 0.10 & \phantom{0}7.2& $1.1357\pm0.0057$  & $0.9410\pm0.0034$  & $0.8613\pm0.0037$ & $1.2887\pm0.0028$  \\
\hline
  0.15 & 10.2 & $1.1341\pm0.0057$ &  $0.9393\pm0.0032$  & 0.8598 $\pm$  0.0037 & $1.2859\pm0.0028$ \\
\hline
 0.20  & 13.4 & $1.1336\pm0.0057$ &  $0.9378\pm0.0031$  & 0.8585 $\pm$  0.0037 & $1.2823\pm0.0028$ \\
\hline
 0.25 & 16.1  & $1.1336\pm0.0057$ &  $0.9363\pm0.0030$  & 0.8561 $\pm$  0.0037 & $1.2800\pm0.0028$ \\
\hline
 0.35  & 21.4 & $1.1303\pm0.0057$ &  $0.9334\pm0.0028$  & 0.8551 $\pm$  0.0037 & $1.2741\pm0.0028$ \\
\hline
 0.45  & 25.9 & $1.1269\pm0.0057$ &  $0.9307\pm0.0027$  & 0.8509 $\pm$  0.0036 & $1.2693\pm0.0028$ \\
\hline
 0.60 & 32.1  & $1.1216\pm0.0057$ &  $0.9271\pm0.0025$  & 0.8482 $\pm$  0.0036 & $1.2639\pm0.0028$ \\
\hline
 0.80  & 39.1 & $1.1166\pm0.0056$ &  $0.9227\pm0.0024$  & 0.8414 $\pm$  0.0037 & $1.2576\pm0.0028$  \\
\hline
 1.00  & 44.9 & $1.1109\pm0.0056$ &  $0.9189\pm0.0024$  & 0.8381 $\pm$  0.0036 & $1.2499\pm0.0028$ \\
\hline
 1.50  & 55.9 & $1.1048\pm0.0056$ &  $0.9110\pm0.0025$  & 0.8318 $\pm$  0.0036 & $1.2368\pm0.0028$ \\
\hline
 3.00 &  72.8 & $1.0929\pm0.0056$ &  $0.8959\pm0.0028$  & 0.8135 $\pm$  0.0036 & $1.2093\pm0.0027$  \\
\hline
 5.00 &  82.5 & $1.0852\pm0.0056$ &  $0.8846\pm0.0030$  & 0.7989 $\pm$  0.0035 & $1.1920\pm0.0022$ \\
\hline
\end{tabular}
\caption[\SKI\ model predictions for \Rn.]
{\SKI\ Model predictions for $R_{N}$ obtained with the common LEP
  samples at 189 GeV. The second column gives the fraction of
  reconnected events in the common samples obtained for the different
  choice of $k_{I}$ values.}
\label{fsi:cr:tab:cetraro}
\end{center}
\end{table}

\clearpage

 \section{Example Average}
 \begin{table}[hbt]
  \begin{center}
  \begin{tabular}{|c||c|c|c|c|} \hline
     Model tested  & \multicolumn{4}{c|}{Experiment} \\
     \SKI\ (100\%)        & \multicolumn{1}{c}{\Aleph} & \multicolumn{1}{c}{\Delphi} &
                 \multicolumn{1}{c}{\Ltre}  & \multicolumn{1}{c|}{\Opal} \\ \hline\hline

\Rn\ (no-CR)      
    &   $1.1365\pm0.0013$   & $0.9444\pm0.0039$    & $0.8622\pm0.0037$    &     $1.2958\pm0.0028$  \\
\Rn\ (with CR)   
     &   $1.0548\pm0.0012$   & $0.8463\pm0.0036$    & $0.7482\pm0.0033$    &     $1.1386\pm0.0027$   \\
  weight &  19.688     &  7.054           &  18.250       &         28.202          \\ \hline\hline
  $r$ ($\equiv$\Rn(data)/\Rnnocr)
               &             &                 &                 &       \\
  Data                         
               &   0.9636    &   0.9526   &   0.9784   &  0.9701  \\
  Stat.\ error &  0.0119     &   0.0332   &   0.0252   &  0.0194  \\
  Syst.\ error &  0.0110     &   0.0206   &   0.0180   &  0.0121  \\ \hline\hline

Uncorrel.\ syst.
               &             &             &             &        \\
 Background
               &   0.0013    &   0.0035    &     0.0128  & 0.0029 \\
 Hadronisation
               &   0.0000    &   0.0094    &   0.0086    & 0.0051 \\
 Intra-W BEC
               &   0.0013    &   0.0099    &    0.0016   & 0.0000 \\
 Detector effects
               &   0.0035    &  $-$        &    0.0019   & 0.0056 \\
 $E_{\mathrm{cm}}$ dependence
               &    0.0055   &   0.0123    &    0.0023   & 0.0023 \\ \hline\hline
 Total uncorr. error
               &   0.0068    &  0.0187     &    0.0158   & 0.0084 \\ \hline\hline
 Correl.\ syst.
               &             &             &             &        \\
 Background
               &  0.0031     &  0.0031     &  0.0031     & 0.0031 \\
 Hadronisation
               & 0.0081      &  0.0081     &  0.0081     & 0.0081 \\
 Intra-W BEC
               &  0.0012     &  0.0012     &  0.0012     & 0.0012 \\ \hline\hline
 Total correl. error
               &  0.0087     &  0.0087     &  0.0087     & 0.0087 \\ \hline
 \hline
  \end{tabular}
  \end{center}
 \caption[Normalised results of particle flow to \SKI\ model.]
 {Normalised results of particle flow analysis, based on the
  predicted \SKI\ 100\% sensitivity.}
 \label{fsi:cr:tab:outputs}
 \end{table}

  \label{fsi:cr:app:results}

\begin{table}[hbt]
\begin{center}
\begin{tabular}{||c|c|c|c|c|c||}
\hline
$k_{i}$  & $P_{reco}$ (\%) &$\langle r\rangle^{MC}$  &
                           $\langle r\rangle^{ADLO}$ & data-MC ($\sigma$) \\\hline \hline
 0.10 & \phantom{0}7.2 & 0.9950& $ 0.9679 \pm 0.0167\pm 0.0087\pm 0.0076$   & -1.34  \\
\hline
  0.15 & 10.2 & 0.9935& $ 0.9677 \pm 0.0146\pm 0.0087\pm 0.0065$   & -1.42  \\
\hline
 0.20  & 13.4 & 0.9911& $ 0.9681 \pm 0.0148\pm 0.0087\pm 0.0066$   & -1.25  \\
\hline
 0.25 & 16.1  & 0.9895& $ 0.9687 \pm 0.0144\pm 0.0087\pm 0.0066$   & -1.15  \\
\hline
 0.35  & 21.4 & 0.9861& $ 0.9680 \pm 0.0136\pm 0.0087\pm 0.0062$   & -1.05  \\
\hline
 0.45  & 25.9 & 0.9834& $ 0.9681 \pm 0.0123\pm 0.0087\pm 0.0057$   & -0.98  \\
\hline
 0.60 & 32.1  & 0.9802& $ 0.9676 \pm 0.0112\pm 0.0087\pm 0.0053$   & -0.84  \\
\hline
 0.80  & 39.1 & 0.9757& $ 0.9678 \pm 0.0106\pm 0.0087\pm 0.0052$   & -0.54   \\
\hline
 1.00  & 44.9 &  0.9708& $ 0.9676 \pm 0.0103\pm 0.0087\pm 0.0051$   & -0.22 \\
\hline
 1.50  & 55.9 & 0.9626& $ 0.9676 \pm 0.0105\pm 0.0087\pm 0.0051$   & +0.34  \\
\hline
 3.00 &  72.8 &  0.9447& $ 0.9680 \pm 0.0107\pm 0.0087\pm 0.0053$   & +1.58   \\
\hline
 5.00 &  82.5 & 0.9324& $ 0.9683 \pm 0.0108\pm 0.0087\pm 0.0054$   & +2.42  \\
\hline
 10000 & 100 & 0.8909& $ 0.9687 \pm 0.0108\pm 0.0087\pm 0.0057$   & +5.20 \\
\hline
\end{tabular}
\caption[LEP Averages in $r$ for \SKI\ model.]{LEP Average values of $r$ in Monte
  Carlo, $\langle r\rangle^{MC}$
  ($\equiv\langle\Rn/\Rnnocr\rangle^{MC}$), and data, $\langle
  r\rangle^{ADLO}$ ($\equiv\langle\Rn/\Rnnocr\rangle^{ADLO}$), for
  various $k_I$ values in \SKI\ model.  The first uncertainty is
  statistical, the second corresponds to the correlated systematic
  error and the third corresponds to the uncorrelated systematic
  error.}
\label{fsi:cr:tab:average}
\end{center}
\end{table}

\begin{table}[hbt]
\begin{center}
\begin{tabular}{||c|c|c|c|c|c||}
\hline
Model & $ \langle r \rangle^{MC}$  & $\langle r\rangle^{ADLO}$ & data-MC ($\sigma$) \\
\hline
 \hline
 AR2 & 0.9888& $ 0.9589 \pm 0.0101\pm 0.0086\pm 0.0050$   & -2.10  \\
\hline
 \Herwig\ CR & 0.9874& $ 0.9498 \pm 0.0105\pm 0.0086\pm 0.0052$   & -2.59  \\
\hline
\hline
\end{tabular}
\caption[LEP Averages in $r$ for \ARII\ and \Herwig\ CR models.]
{LEP Average values of $r$ in Monte Carlo, $\langle r\rangle^{MC}$
  ($\equiv\langle\Rn/\Rnnocr\rangle^{MC}$), and data, $\langle r
  \rangle^{ADLO}$ ($\equiv\langle\Rn/\Rnnocr\rangle^{ADLO}$), for
  \Ariadne\ and \Herwig\ models with colour reconnection.  The first
  uncertainty is statistical, the second corresponds to the correlated
  systematic error and the third corresponds to the uncorrelated
  systematic error.}
\label{fsi:cr:tab:average2}
\end{center}
\end{table}

%% file: s03_ew.bbl
\begin{thebibliography}{100}

\bibitem{bib-EWEP-02}
The LEP Collaborations ALEPH, DELPHI, L3, OPAL and the LEP Electroweak Working
  Group, and the SLD Heavy Flavour Group, {\it A Combination of Preliminary
  Electroweak Measurements and Constraints on the Standard Model},
  CERN--EP-2002-91, hep-ex/0212036.

\bibitem{LEPLS}
The LEP~Collaborations ALEPH, DELPHI, L3 and OPAL and the LEP Electroweak
  working group, {\it Combination procedure for the precise determination of Z
  boson parameters from results of the LEP experiments}, CERN--EP-2000-153,
  hep-ex/0101027.

\bibitem{ref:lephf}
The LEP Experiments: ALEPH, DELPHI, L3 and OPAL, Nucl.~Inst.~Meth. {\bf A378}
  (1996) 101.

\bibitem{ALEPHLS}
ALEPH Collaboration, D.~Decamp \etal, Z.~Phys. {\bf C48} (1990) 365;\\ ALEPH
  Collaboration, D.~Decamp \etal, Z.~Phys. {\bf C53} (1992) 1;\\ ALEPH
  Collaboration, D.~Buskulic \etal, Z.~Phys. {\bf C60} (1993) 71;\\ ALEPH
  Collaboration, D.~Buskulic \etal, Z.~Phys. {\bf C62} (1994) 539;\\ ALEPH
  Collaboration, R.~Barate \etal, Eur. Phys. J. {\bf C 14} (2000) 1.

\bibitem{DELPHILS}
DELPHI Collaboration, P.~Aarnio \etal, Nucl.~Phys. {\bf B367} (1991) 511;\\
  DELPHI Collaboration, P.~Abreu \etal, Nucl.~Phys. {\bf B417} (1994) 3;\\
  DELPHI Collaboration, P.~Abreu \etal, Nucl.~Phys. {\bf B418} (1994) 403;\\
  DELPHI Collaboration, P.~Abreu \etal, Eur. Phys. J. {\bf C 16} (2000) 371.

\bibitem{L3LS}
L3 Collaboration, B.~Adeva \etal, Z.~Phys. {\bf C51} (1991) 179;\\ L3
  Collaboration, O.~Adriani \etal, Phys.~Rep. {\bf 236} (1993) 1; \\ L3
  Collaboration, M.~Acciarri \etal, Z.~Phys. {\bf C62} (1994) 551;\\ L3
  Collaboration, M.~Acciarri \etal, Eur. Phys. J. {\bf C16} (2000) 1-40.

\bibitem{OPALLS}
OPAL Collaboration, G.~Alexander \etal, Z.~Phys. {\bf C52} (1991) 175;\\ OPAL
  Collaboration, P.D.~Acton \etal, Z.~Phys. {\bf C58} (1993) 219;\\ OPAL
  Collaboration, R.~Akers \etal, Z.~Phys. {\bf C61} (1994) 19;\\ OPAL
  Collaboration, G.~Abbiendi \etal, Eur. Phys. J. {\bf C14} (2000) 373;\\ OPAL
  Collaboration, G.~Abbiendi \etal, Eur. Phys. J. {\bf C19} (2001) 587.

\bibitem{ref:QEDCONV}
F.A.~Berends \etal, in {\em Z Physics at LEP 1, Vol.~1}, ed. {G.~Altarelli,
  R.~Kleiss and C.~Verzegnassi}, (CERN Report: CERN 89-08, 1989), p.~89.\\
  M.~B{\"{o}}hm \etal, in {\em Z Physics at LEP 1, Vol.~1}, ed. {G.~Altarelli,
  R.~Kleiss and C.~Verzegnassi}, (CERN Report: CERN 89-08, 1989), p. 203.

\bibitem{ref:consoli}
See, for example, M.~Consoli \etal, in {\em Z Physics at LEP 1, Vol.~1}, ed
  G.~Altarelli, R.~Kleiss and C.~Verzegnassi, (CERN Report: CERN 89-08, 1989),
  p.~7.

\bibitem{ref:Jadach91}
S.~Jadach, \etal, Phys.~Lett.~{\bf B257} (1991) 173.

\bibitem{ref:Skrzypek92}
M.~Skrzypek, Acta Phys.~Pol.~{\bf B23} (1992) 135.

\bibitem{ref:Montagna96}
G.~Montagna, \etal, Phys.~Lett.~{\bf B406} (1997) 243.

\bibitem{bib-lumthopal}
G. Montagna \etal, Nucl. Phys. {\bf B547} (1999) 39; \\ G. Montagna \etal,
  Phys. Rev. Lett. {\bf 459} (1999) 649.

\bibitem{bib-lumth99}
B.F.L Ward \etal, Phys. Lett. {\bf B450} (1999) 262.

\bibitem{bib-PCP99}
D.{} Bardin, M.{} Gr{\"u}newald and G.{} Passarino, {\it Precision Calculation
  Project Report}, hep-ph/9902452.

\bibitem{bib-ALEPHTAU}
ALEPH Collaboration, D.~Buskulic \etal, Zeit.~Phys. {\bf C69} (1996) 183;\\
  ALEPH Collaboration, A.~Heister \etal, Eur. Phys. J. {\bf C20} (2001) 401.

\bibitem{bib-DELPHITAUnew}
DELPHI Collaboration, P.~Abreu \etal, Eur. Phys. J. {\bf C14} (2000) 585.

\bibitem{bib-L3TAUfin}
L3 Collaboration, M.~Acciarri \etal, Phys.~Lett.~{\bf B429} (1998) 387.

\bibitem{bib-OPALTAU}
OPAL Collaboration, G.~Abbiendi \etal, {\it Precision Neutral Current Asymmetry
  Parameter Measurements from the Tau Polarization at LEP}, CERN-EP-2001-023,
  Submitted to Eur. Phys. J. C.

\bibitem{bib-EWPPE187}
The LEP Collaborations ALEPH, DELPHI, L3, OPAL and the LEP Electroweak Working
  Group, {\it Combined Preliminary Data on $\Zzero$ Parameters from the LEP
  Experiments and Constraints on the Standard Model}, CERN--PPE/94-187.

\bibitem{ref:sld-s00}
SLD Collaboration, K.~Abe \etal, Phys.{} Rev.{} Lett.{} {\bf 84} (2000) 5945.

\bibitem{ref:sld-asym}
SLD Collaboration, K.~Abe \etal, SLAC-PUB-8618, (2000). Submitted to
  Phys.Rev.Lett.

\bibitem{ref:lephfnew}
The LEP Heavy Flavour Group, {\it Final input parameters for the LEP/SLD heavy
  flavour analyses,} LEPHF/01-01, \\
  http://www.cern.ch/LEPEWWG/heavy/lephf0101.ps.gz.

\bibitem{ref:alife}
ALEPH Collaboration, R.~Barate \etal, Physics Letters {\bf B 401} (1997) 150;\\
  ALEPH Collaboration, R.~Barate \etal, Physics Letters {\bf B 401} (1997) 163.

\bibitem{ref:drb}
DELPHI Collaboration, P.Abreu \etal, Eur. Phys. J. {\bf C10} (1999) 415.

\bibitem{ref:lrbmixed}
L3 Collaboration, M. Acciarri \etal, Eur. Phys. J. {\bf C13} (2000) 47.

\bibitem{ref:omixed}
OPAL Collaboration, G. Abbiendi \etal, Eur. Phys. J. {\bf C8} (1999) 217.

\bibitem{ref:SLD_R_B}
SLD Collaboration, K. Abe \etal, Phys. Rev. Lett. {\bf 80} (1998) 660; \\ see
  also~\cite{ref:groot2003}.

\bibitem{ref:arcd}
ALEPH Collaboration, R.~Barate {\em et~al.}, Eur. Phys. J. {\bf C4} (1998) 557.

\bibitem{ref:drcd}
DELPHI Collaboration, P.Abreu \etal, Eur. Phys. J. {\bf C12} (2000) 209.

\bibitem{ref:drcc}
DELPHI Collaboration, P.Abreu \etal, Eur. Phys. J. {\bf C12} (2000) 225.

\bibitem{ref:orcd}
OPAL Collaboration, K.~Ackerstaff \etal, Eur. Phys. J. {\bf C1} (1998) 439.

\bibitem{ref:arcc}
ALEPH Collaboration, R. Barate \etal, Eur. Phys. J. {\bf C16} (2000) 597.

\bibitem{ref:orcc}
OPAL Collaboration, G.~Alexander \etal, Z.~Phys.~{\bf C72} (1996) 1.

\bibitem{ref:SLD_R_C}
SLD Collaboration, {\it A Measurement of $R_c$ with the SLD Detector }
  SLAC--PUB--7880, contributed paper to ICHEP 98 Vancouver {\bf ICHEP'98 \#174
  };\\ see also~\cite{ref:groot2003}.

\bibitem{ref:afbqcd}
D.~Abbaneo {\em et~al.}, Eur. Phys. J. {\bf C4} (1998) 185.

\bibitem{ref:ZFITTER}
D.~Bardin \etal, Z.~Phys. {\bf C44} (1989) 493; Comp.~Phys.~Comm. {\bf 59}
  (1990) 303; Nucl.~Phys. {\bf B351}(1991) 1; Phys.~Lett. {\bf B255} (1991) 290
  and CERN-TH 6443/92 (May 1992); the most recent version of ZFITTER (6.21) is
  described in DESY 99-070, hep-ph/9908433 (Aug 1999) published in
  Comp.~Phys.~Comm. {\bf 133} (2001) 229.

\bibitem{ref:alasy}
ALEPH Collaboration, A. Heister, \etal, Eur. Phys. J. {\bf C24} (2002) 177.

\bibitem{ref:dlasy}
DELPHI Collaboration, J. Abdallah \etal, {\it Measurement of the
  Forward-Backward Asymmetries of $e^+e^-\ \rightarrow Z\ \rightarrow b
  \overline{b}$ and $e^+e^-\ \rightarrow Z\ \rightarrow c \overline{c}$ using
  prompt leptons}, CERN-EP-2003-083, Submitted to Eur. Phys. J.

\bibitem{ref:llasy}
L3 Collaboration, O.~Adriani \etal, Phys.~Lett. {\bf B292 } (1992) 454; \\ L3
  Collaboration, {\it L3 Results on \Abb, \Acc and $\chi$ for the Glasgow
  Conference,} L3 Note 1624;\\ L3 Collaboration, M. Acciarri \etal, Phys. Lett.
  {\bf B448} (1999) 152.

\bibitem{ref:olasy}
OPAL Collaboration, G. Abbiendi \etal {\it Measurement of Heavy Quark
  Forward-Backward Asymmetries and Average B Mixing Using Leptons in Hadronic Z
  Decays}, CERN-EP-2003-039, Submitted to Phys. Letts. B.

\bibitem{ref:ajet}
ALEPH Collaboration, A.~Heister \etal, Eur. Phys. J. {\bf C22} (2001) 201.

\bibitem{ref:dnnasy}
DELPHI Collaboration, {\it Determination of $A_{FB}^b$ using inclusive charge
  reconstruction and lifetime tagging at the Z pole}, DELPHI 2003-056-CONF-676.

\bibitem{ref:ljet}
L3 Collaboration, M. Acciarri \etal, Phys. Lett. {\bf B439} (1998) 225.

\bibitem{ref:ojet}
OPAL Collaboration, G. Abbiendi \etal, Phys. Lett. {\bf B546} (2002) 29.

\bibitem{ref:djasy}
DELPHI Collaboration, P.Abreu \etal, Eur. Phys. J. {\bf C9} (1999) 367.

\bibitem{ref:adsac}
ALEPH Collaboration, R.~Barate \etal, Phys. Lett. {\bf B434} (1998) 415.

\bibitem{ref:ddasy}
DELPHI Collaboration, P.Abreu \etal, Eur. Phys. J. {\bf C10} (1999) 219.

\bibitem{ref:odsac}
OPAL Collaboration, G.~Alexander \etal, Z.~Phys. {\bf C73} (1996) 379.

\bibitem{ref:SLD_AQL}
SLD Collaboration, K. Abe \etal, Phys. Rev. Lett. {\bf 88} (2002) 151801.

\bibitem{ref:SLD_ACD}
SLD Collaboration, K. Abe, \etal, Phys Rev. {\bf D63} (2001) 032005.

\bibitem{ref:SLD_ABJ}
SLD Collaboration, K. Abe, \etal, Phys Rev. Lett. {\bf 90} 141804 (2003).

\bibitem{ref:SLD_ABK}
SLD Collaboration, K. Abe et al, Phys. Rev. Lett. {\bf 83}, 1902 (1999).

\bibitem{ref:SLD_vtxasy}
SLD Collaboration, {\it Direct measurement of ${\cal A}_b$ using charged kaons
  at the SLD detector}, SLAC-PUB-8200, contributed paper to EPS 99 Tampere {\bf
  HEP'99 \#6\_473 };\\ SLD Collaboration, {\it Direct measurement of Ab using
  charged vertices}, SLAC-PUB-8542, Contributed Paper to ICHEP2000 , {\bf \#
  743};\\ see also~\cite{ref:groot2003}.

\bibitem{ref:SLD_ACV}
SLD Collaboration, {\it Direct measurement of ${\cal A}_c$ using inclusive
  charm tagging at the SLD detector}, SLAC-PUB-8199, contributed paper to EPS
  99 Tampere {\bf HEP'99 \#6\_474 };\\ see also~\cite{ref:groot2003}.

\bibitem{ref:abl}
ALEPH Collaboration, A. Heister, \etal, Eur. Phys. J., {\bf C22} (2002) 613.

\bibitem{ref:dbl}
DELPHI Collaboration, P.Abreu \etal, Eur. Phys. J. {\bf C20} (2001) 455.

\bibitem{ref:lbl}
L3 Collaboration, M. Acciarri \etal, Z Phys. {\bf C71} 379 (1996).

\bibitem{ref:obl}
OPAL Collaboration, G. Abbiendi \etal, Eur. Phys. J. {\bf C13} (2000) 225.

\bibitem{ref:ocl}
OPAL Collaboration, G.~Abbiendi \etal, Eur. Phys. J. {\bf C8} (1999) 573.

\bibitem{ALEPHcharge1996}
ALEPH Collaboration, D.~Buskulic \etal, Z.~Phys. {\bf C71} (1996) 357.

\bibitem{DELPHIcharge}
DELPHI Collaboration, P.~Abreu \etal, Phys.~Lett. {\bf B277} (1992) 371.

\bibitem{OPALcharge}
OPAL Collaboration, P.~D.~Acton \etal, Phys.~Lett. {\bf B294} (1992) 436.

\bibitem{JETSET}
T.~\hbox{Sj\"ostrand}, Comp.~Phys.~Comm. {\bf 82} (1994) 74.

\bibitem{HERWIG}
G.~Marchesini \etal, Comp.~Phys.~Comm. {\bf 67} (1992) 465.

\bibitem{ref:QED}
W.~Heitler.
\newblock {\em Quantum Theory of Radiation}.
\newblock Oxford University Press, 2 edition, 1944.
\newblock pages 204--207.

\bibitem{ref:radcor}
Frits~A. Berends and R.~Kleiss, Nucl. Phys. {\bf B186} (1981) 22.

\bibitem{gg:ref:LEPGG}
ALEPH Collab., Eur. Phys. J., C 28 (2003) 1 and ref. therein;\\ DELPHI Collab.,
  DELPHI 2001-093 CONF 521 and ref. therein; \\ L3 Collab., \PL {\bf B531}
  (2002) 28 and ref. therein;\\ OPAL Collab., Eur.Phys.J.C26 (2003) 331 and
  ref. therein.

\bibitem{gg:ref:drell}
S.~D. Drell, Ann. Phys. {\bf 4} (1958) 75.

\bibitem{gg:ref:low}
F.~E. Low, \PRL {\bf 14} (1965) 238.

\bibitem{gg:ref:eboli}
O.~J.~P. Eboli, A.~A. Natale, and S.~F. Novaes, \PL {\bf B271} (1991) 274.

\bibitem{gg:ref:estar}
P.~Mery, M.~Perrottet, and F.~M. Renard, Z. Phys. {\bf C38} (1988) 579.

\bibitem{gg:ref:g2_brodsky}
Stanley~J. Brodsky and S.~D. Drell, Phys. Rev. {\bf D22} (1980) 2236.

\bibitem{gg:ref:boudjema:1993}
F.~Boudjema, A.~Djouadi, and J.~L. Kneur, Z. Phys. {\bf C57} (1993) 425--450.

\bibitem{gg:ref:vachon}
B.~Vachon.
\newblock Excited electron contribution to the $\eegg$ cross-section.
\newblock hep-ph/0103132, 2001.

\bibitem{gg:ref:ad}
K.~Agashe and N.~G. Deshpande, \PL {\bf B456} (1999) 60.

\bibitem{ff:ref:ffbar_web}
LEPEWWG $\ff$ subgroup: http://www.cern.ch/LEPEWWG/lep2/~.

\bibitem{ff:ref:expts}
ALEPH Collaboration ``Study of Fermion Pair Production in $\ee$ Collisions at
  130-183 GeV'', Eur. Phys. J. {\bf{C12}} (2000) 183; \\ ALEPH Collaboration,
  ``Fermion Pair Production in $\ee$ Collisions at 189 GeV and Kimits on
  Physics Beyond the Standard Model'', ALEPH 99-018 CONF 99-013; \\ ALEPH
  Collaboration ``Fermion Pair Production in $\ee$ Collisions from 192 to 202
  GeV and Limits on Physics beyond the Standard Model'', ALEPH 2000-047 CONF
  2000-030; \\ ALEPH Collaboration ``Fermion Pair Production in $\ee$
  Collisions at high energy and Limits on Physics beyond the Standard Model'',
  ALEPH 2002-032 CONF 2002-021; \\ ALEPH Collaboration, ``Fermion Pair
  Production in $\ee$ Collisions at high energy and Limits on Physics beyond
  the Standard Model '', ALEPH 2001-019 CONF 2001-016; \\ DELPHI
  Collaboration,``Measurement and Interpretation of Fermion-Pair Production at
  LEP energies from 130 to 172 GeV'', Eur. Phys. J. {\bf{C11}} (1999), 383; \\
  DELPHI Collaboration, ``Measurement and Interpretation of Fermion-Pair
  Production at LEP Energies from 183 to 189 GeV'', Phys.Lett. {\bf{B485}}
  (2000), 45; \\ DELPHI Collaboration, ``Results on Fermion-Pair Production at
  LEP running from 192 to 202 GeV'', DELPHI 2000-128 OSAKA CONF 427; \\ DELPHI
  Collaboration, ``Results on Fermion-Pair Production at LEP running in 2000'',
  DELPHI 2001-094 CONF 522; \\ L3 Collaboration, ``Measurement of Hadron and
  Lepton-Pair Production at 161 GeV $< \sqrt{s} <$ 172 GeV at LEP'', Phys.
  Lett. {\bf{B407}} (1997) 361; \\ L3 Collaboration, ``Measurement of Hadron
  and Lepton-Pair Production at 130 GeV $< \sqrt{s} <$ 189 GeV at LEP'', Phys.
  Lett. {\bf{B479}} (2000), 101. \\ L3 Collaboration, ``Preliminary L3 Results
  on Fermion-Pair Production in 1999'', L3 note 2563; \\ L3
  Collaboration,``Preliminary L3 Results on Fermion-Pair Production in 2000'',
  L3 note 2648; \\ OPAL Collaboration, ``Tests of the Standard Model and
  Constraints on New Physics from Measurements of Fermion Pair Production at
  130 - 172 GeV at LEP'', Euro. Phys. J. {\bf C2} (1998) 441; \\ OPAL
  Collaboration, ``Tests of the Standard Model and Constraints on New Physics
  from Measurements of Fermion Pair Production at 183 GeV at LEP'', Euro. Phys.
  J. {\bf C6} (1999) 1; \\ OPAL Collaboration, ``Tests of the Standard Model
  and Constraints on New Physics from Measurements of Fermion Pair Production
  at 189 GeV at LEP'', Euro. Phys. J. {\bf C13} (2000) 553; \\ OPAL
  Collaboration, ``Tests of the Standard Model and Constraints on New Physics
  from Measurements of Fermion Pair Production at 192-202 GeV at LEP'', OPAL
  PN424 (2000); \\ OPAL Collaboration, ``Measurement of Standard Model
  Processes in e+e- Collisions at sqrt{s}~203-209 GeV'', OPAL PN469 (2001).

\bibitem{ff:ref:lepffwrkshp}
M. Kobel {\it et al.}, ``Two-Fermion Production in Electron Positron
  Collisions'' in S. Jadach {\it et al.} [eds] , ``Reports of the Working
  Groups on Precision Calculations for LEP2 Physics: proceedings'' CERN
  2000-009, hep-ph/0007180.

\bibitem{ff:ref:ZFITTER}
D.~Bardin {\it et al.}, CERN-TH 6443/92;
  http://www.ifh.de/$\sim$riemann/Zfitter/zf.html~.\\ Predictions are from
  ZFITTER versions 6.04 or later.\\ Definition 1 corresponds to the ZFITTER
  flags FINR=0 and INTF=0; definition 2 corresponds to FINR=0 and INTF=1 for
  hadrons, FINR=1 and INTF=1 for leptons.

\bibitem{common_bib:BLUE}
L. Lyons \etal, NIM A270 (1988) 110; A. Valassi, NIM A500 (2003) 391.

\bibitem{ff:ref:BHWIDE}
S.~Jadach, W.~Placzek and B.Ward, Phys. Lett. {\bf B390} (1997) 298.

\bibitem{ff:ref:lepff-osaka}
LEPEWWG $\ff$ Subgroup, C.~Geweniger {\it et. al.}, LEP2FF/00-03,ALEPH 2000-088
  PHYSIC 2000-034, DELPHI 2000-168 PHYS 881, L3 note 2624, OPAL TN673.

\bibitem{ff:ref:hfpublished}
ALEPH Collaboration, Euro. Phys J. {\bf{C12}} (2000) 183; \\ DELPHI
  Collaboration, P.Abreu {\it et al.}, Euro. Phys J. {\bf{C11}}(1999); \\ L3
  Collaboration, M.Acciarri {\it et al.}, Phys.\ Lett.\ {\bf B485} (2000) 71;
  \\ OPAL Collaboration, G.Abbiendi {\it et al.}, Euro.\ Phys.\ J.\ {\bf C16}
  (2000) 41.

\bibitem{ff:ref:hfpreliminary}
ALEPH Collaboration, ALEPH 99-018 CONF 99-013; \\ ALEPH Collaboration, ALEPH
  2000-046 CONF 2000-029; \\ ALEPH Collaboration, ICHEP2002, ABS388; \\ DELPHI
  Collaboration, DELPHI 2001-095 CONF 523; \\ L3 Collaboration, L3 Internal
  note 2640, 1 March 2001.

\bibitem{ff:ref:hfconfnote}
LEPEWWG Heavy Flavour at LEP2 Subgroup, ``Combination of Heavy Flavour
  Measurements at LEP2'', LEP2FF/00-02.

\bibitem{ff:ref:hfzfit}
ZFITTER V6.23 is used.\\ D. Bardin {\it et al.}, Preprint hep-ph/9908433. \\
  Relevant ZFITTER settings used are FINR=0 and INTF=1.

\bibitem{ff:ref:hflep1-99}
DELPHI Collaboration, P.Abreu {\it et al.}, Euro Phys J. {\bf{C10}}(1999) 415.
  \\ The LEP collaborations {\it et al.}, CERN-EP/2000-016.

\bibitem{ff:ref:ELPthr}
E. Eichten, K. Lane, and M. Peskin, Phys. Rev. Lett. {\bf 50} (1983) 811.

\bibitem{ff:ref:Kroha}
H. Kroha, Phys. Rev. {\bf D46} (1992) 58.

\bibitem{ff:ref:zprime-thry}
P. Langacker, R.W. Robinett and J.L. Rosner, Phys. Rev. {\bf D30} (1984) 1470;
  \\ D. London and J.L. Rosner, Phys. Rev. {\bf D34} (1986) 1530; \\ J.C. Pati
  and A. Salam, Phys. Rev. {\bf D10} (1974) 275; \\ R.N. Mohapatra and J.C.
  Pati, Phys. Rev. {\bf D11} (1975) 566.

\bibitem{ff:ref:sqsm}
G. Altarelli \etal, Z. Phys. {\bf{C45}} (1989) 109; \\ erratum Z. Phys.
  {\bf{C47}} (1990) 676.

\bibitem{ff:ref:lep1zprime}
DELPHI Collaboration, P. Abreu {\it et al.}, Zeit. Phys. {\bf{C65}} (1995) 603.

\bibitem{ff:ref:lq-thry}
W.~Buchm{\"{u}}ller, R.~R{\"{u}}ckl, D.~Wyler, Phys. Lett. {\bf B191} (1987)
  442; Erratum-ibid. {\bf B448} (1999) 320.

\bibitem{ff:ref:lq-squ}
J.~Kalinowski {\it et al.}, Z. Phys. {\bf C74} (1997) 595.

\bibitem{ff:ref:lq-h1}
H1 Collab., C.~Adloff {\it et al.}, Phys. Lett. {\bf B523} (2001) 234.

\bibitem{ff:ref:lq-zeus}
ZEUS Collaboration; J.~Breitweg {\it et al.}, Phys. Rev. {\bf D63} (2001)
  052002.

\bibitem{ff:ref:lq-tevatron}
CDF Collab., F. Abe {\it et al.}, Phys. Rev. Lett. {\bf 79} (1997) 4327;\\
  D{\O} Collab., B.~Abbott {\it et al.}, Phys. Rev. Lett. {\bf 80} (1998) 2051.

\bibitem{ff:ref:ADD}
N.~Arkani-Hamed, S.~Dimopoulos and G.~Dvali, Phys. Lett. {\bf B429} (1998) 263.

\bibitem{ff:ref:ADD2}
I.~Antoniadis, N.~Arkani-Hamed, S.~Dimopoulos and G.~Dvali, Phys. Lett. {\bf
  B436} (1998) 257.

\bibitem{ff:ref:ADD3}
N.~Arkani-Hamed, S.~Dimopoulos and G.~Dvali, Phys. Rev. {\bf D59} (1999)
  086004.

\bibitem{ff:ref:Hewett}
J.~Hewett, Phys. Rev. Lett. {\bf 82} (1999) 4765.

\bibitem{ff:ref:Rizzo}
T.~Rizzo, Phys. Rev. {\bf D59} (1999) 115010.

\bibitem{ff:ref:Giudice}
G.~Giudice, R.~Rattazi and J.~Wells, Nucl. Phys. {\bf B544} (1999) 3.

\bibitem{ff:ref:Lykken}
T.~Han, J.D.~Lykken and R-J.~ Zhang, Phys. Rev. {\bf D59} (1999) 105006.

\bibitem{ff:ref:Shrock}
S.~Nussinov and R.~Shrock, Phys. Rev. {\bf D59} (1999) 105002.

\bibitem{ff:ref:Bourilkov:1999}
D.~Bourilkov, J. High Energy Phys. {\bf 08} (1999) 006.

\bibitem{ff:ref:Bourilkov:2000}
D.~Bourilkov, Phys. Rev. {\bf D62} (2000) 076005.

\bibitem{ff:ref:Bourilkov:2001}
D.~Bourilkov, Phys. Rev. {\bf D64} (2001) 071701.

\bibitem{smat:ref:5and9}
Lineshape subgroup of the LEP EWWG,``Combination procedure for the precise
  determination of Z boson parameters from results of the LEP experiments'',
  CERN-EP/2000-153, hep-ex/0101027 (2000).

\bibitem{smat:ref:smat1997}
S-Matrix subgroup of the LEP EWWG,``An investigation of the interference
  between photon and Z boson exchange'', LEPEWWG/LS/97-02, ALEPH 97-92 PHYSICS
  97-82, DELPHI 97-153 PHYS 732, L3 Note 2164, OPAL TN510 (1997).

\bibitem{smat:ref:bcms}
A.~Borelli, M.~Consoli, L.~Maiani, and R.~Sisto, Nucl. Phys. {\bf B333} (1990)
  357.

\bibitem{smat:ref:rgs}
R.~G. Stuart, Phys.~Lett. {\bf B272} (1991) 353.

\bibitem{smat:ref:arr}
A.~Leike, T.~Riemann, and J.~Rose, Phys.~Lett. {\bf B273} (1991) 513.

\bibitem{smat:ref:tr}
T.~Riemann, Phys.~Lett. {\bf B293} (1992) 451.

\bibitem{smat:ref:smatasy}
S.~Kirsch and T.~Riemann, Comput.~Phys.~Commun. {\bf 88} (1995) 89.

\bibitem{smat:ref:lep2xsafbave}
LEP EWWG ``A combination of preliminary electroweak measurements and
  constriants on the Standard Model'', CERN-EP/2001-098, hep-ex/0112021 (2001).

\bibitem{smat:ref:aleph}
``S-Matrix fits to cross-section and forward-backward asymmetry measurements at
  LEP 1 and 2'', ALEPH 2000-042 PHYSICS 2000-015 (2002).

\bibitem{smat:ref:delphi}
DELPHI collaboration, ``S-Matrix fits to LEP 1 and LEP 2 data'', DELPHI 2002-04
  CONF 545 (2002).

\bibitem{smat:ref:l3lep1}
L3 Collaboration, M.~Acciarri et~al., Euro. Phys. J. {\bf C16} (2000) 1.

\bibitem{smat:ref:opallep1}
OPAL Collaboration, G.~Abbiendi et~al., Euro. Phys. J. {\bf C19} (2001) 587.

\bibitem{smat:ref:l3lep2}
L3 collaboration, ``Preliminary results on fermion-pair production in 2000'',
  L3 Note 2648 (2001).

\bibitem{smat:ref:opallep2}
OPAL Collaboration, ``Determination of S-Matrix parameters from fermion pair
  production at OPAL'', OPAL PN474 (2001).

\bibitem{smat:ref:venus}
VENUS Collaboration, K.~Yusa et~al., Phys. Lett. {\bf B447} (1999) 167.

\bibitem{smat:ref:topaz}
TOPAZ Collaboration, K.~Miyabuyaski et~al., Phys. Lett. {\bf B347} (1995) 171.

\bibitem{4f_bib:4f_s02}
The LEP WW Working Group, LEPEWWG/2002-02, note prepared for the Summer 2002
  conferences, {\tt http://lepewwg.web.cern.ch/LEPEWWG/stanmod/}.

\bibitem{4f_bib:fourfrep}
M.W.~Gr{\"{u}}newald, G.~Passarino \etal, {\it ``Four fermion production in
  electron positron collisions''}, Four fermion working group report of the
  LEP2 \MC\ Workshop 1999/2000, in {\it ``Reports of the working groups on
  precision calculations for LEP2 Physics''} CERN 2000--009, {\tt
  http://arXiv.org/abs/hep-ph/0005309}.

\bibitem{4f_bib:adloww161}
\Aleph\ Collaboration, R.~Barate \etal, Phys.~Lett.~{\bf B401} (1997) 347.\\
  \Delphi\ Collaboration, P.~Abreu \etal, Phys.~Lett.~{\bf B397} (1997) 158.\\
  \Ltre\ Collaboration, M.~Acciarri \etal, Phys.~Lett. {\bf B398} (1997) 223.\\
  \Opal\ Collaboration, K.~Ackerstaff \etal, Phys.~Lett. {\bf B389} (1996) 416.

\bibitem{4f_bib:adloww172}
\Aleph\ Collaboration, R.~Barate \etal, Phys.~Lett.~{\bf B415} (1997) 435.\\
  \Delphi\ Collaboration, P.~Abreu \etal, Eur.~Phys.~J.~{\bf C2} (1998) 581.\\
  \Ltre\ Collaboration, M.~Acciarri \etal, Phys.~Lett.~{\bf B407} (1997) 419;\\
  \Opal\ Collaboration, K.~Ackerstaff, \etal, Eur.~Phys.~J.~{\bf C1} (1998)
  395.

\bibitem{4f_bib:aleww183}
\Aleph\ Collaboration, R.~Barate \etal, \PLB{453}{1999}{107}.

\bibitem{4f_bib:delww183}
\Delphi\ Collaboration, P.~Abreu \etal, \PLB{456}{1999}{310}.

\bibitem{4f_bib:ltrww183}
\Ltre\ Collaboration, M.~Acciarri \etal, \PLB{436}{1998}{437}.

\bibitem{4f_bib:opaww183}
\Opal\ Collaboration, G.~Abbiendi \etal, Eur. Phys. Jour. {\bf C8} (1999) 191.

\bibitem{4f_bib:aleww189}
\Aleph\ Collaboration, R.~Barate \etal, \PLB{484}{2000}{205}.

\bibitem{4f_bib:delww189}
\Delphi\ Collaboration, P.~Abreu \etal, \PLB{479}{2000}{89}.

\bibitem{4f_bib:ltrww189}
\Ltre\ Collaboration, M.~Acciarri \etal, \PLB{496}{2000}{19}.

\bibitem{4f_bib:opaww189}
\Opal\ Collaboration, G.~Abbiendi \etal, \PLB{493}{2000}{249}.

\bibitem{4f_bib:delww}
\Delphi\ Collaboration, \Delphi\ 2003-053 \Conf\ 674, EPS 2003 abstract 240.

\bibitem{4f_bib:alewwsc01}
\Aleph\ Collaboration, \Aleph\ 2000--005 \Conf\ 2000--002, submitted to the
  Winter 2000 Conferences; \Aleph\ 2001--013 \Conf\ 2001--010, submitted to the
  Winter 2001 Conferences.

\bibitem{4f_bib:ltrwwsc02}
\Ltre\ Collaboration, \L3\ Note 2756, ICHEP 2002 abstract 450, and \L3\ Note
  2514.

\bibitem{4f_bib:opawwsc01}
\Opal\ Collaboration, \Opal\ Physics Note PN469, submitted to the Summer 2001
  Conferences, and PN437 and PN420.

\bibitem{4f_bib:valassi}
Andrea Valassi, Nucl. Instrum. Meth. {\bf A500} (2003) 391.

\bibitem{MINUIT}
F.~James and M.~Roos, Comput. Phys. Commun. {\bf 10} (1975) 343.

\bibitem{4f_bib:lepewwg97}
The LEP Collaborations \Aleph, \Delphi, \L3, \Opal, the LEP Electroweak Working
  Group and the SLD Heavy Flavour Working Group, {\it ``A Combination of
  Preliminary Electroweak Measurements and Constraints on the Standard
  Model''}, CERN--PPE/97--154.

\bibitem{4f_bib:lepewwg98}
The LEP Collaborations \Aleph, \Delphi, \L3, \Opal, the LEP Electroweak Working
  Group and the SLD Heavy Flavour Working Group, {\it ``A Combination of
  Preliminary Electroweak Measurements and Constraints on the Standard
  Model''}, CERN--EP/99--15.

\bibitem{4f_bib:yfsww}
S. Jadach, W. P{\l}aczek, M. Skrzypek, B.F.L. Ward, Phys. Rev. {\bf D54} (1996)
  5434. \\ S. Jadach, W. P{\l}aczek, M. Skrzypek, B.F.L. Ward, Z. W\c{a}s,
  Phys. Lett. {\bf B417} (1998) 326. \\ S. Jadach, W. P{\l}aczek, M. Skrzypek,
  B.F.L. Ward, Z. W\c{a}s, Phys. Rev. {\bf D61} (2000) 113010; preprint
  CERN-TH-99-222, hep-ph/9907346.\\ S. Jadach, W. P{\l}aczek, M. Skrzypek,
  B.F.L. Ward, Z. W\c{a}s, preprint CERN-TH/2000-337, hep-ph/0007012; submitted
  to Phys. Lett. B.\\ S. Jadach, W. P{\l}aczek, M. Skrzypek, B.F.L. Ward, Z.
  W\c{a}s, {\it The Monte Carlo Event Generator {\tt YFSWW3} version {\tt 1.16}
  for W-Pair Production and Decay at LEP2/LC Energies}, preprint
  CERN-TH/2001-017, UTHEP-01-0101, hep-ph/0103163, accepted for publication by
  Comput. Phys. Commun.\\ The \YFSWW\ cross-sections at 155--215 GeV have been
  kindly provided by the authors.

\bibitem{4f_bib:racoonww}
A.~Denner, S.~Dittmaier, M.~Roth and D.~Wackeroth, Nucl. Phys. {\bf B560}
  (1999) 33.\\ A.~Denner, S.~Dittmaier, M.~Roth and D.~Wackeroth, Nucl. Phys.
  {\bf B587} (2000) 67.\\ A.~Denner, S.~Dittmaier, M.~Roth and D.~Wackeroth,
  \PLB{475}{2000}{127}.\\ A.~Denner, S.~Dittmaier, M.~Roth and D.~Wackeroth,
  hep-ph/0101257.\\ The \RacoonWW\ cross-sections at 155--215 GeV have been
  kindly provided by the authors.

\bibitem{4f_bib:dpa}
See~\cite{4f_bib:fourfrep} and references therein for a discussion of complete
  $\oa$ radiative corrections to W-pair production in the LPA/DPA
  approximations.

\bibitem{4f_bib:dpaerr}
The theoretical uncertainty $\Delta\sigma/\sigma$ on the W-pair production
  cross section calculated in the LPA/DPA above 170 GeV can be parametrised as
  $\Delta\sigma/\sigma=0.4\oplus0.072\cdot t_1\cdot t_2$, where
  $t_1=(200-2\cdot\Mw)/(\roots-2\cdot\Mw)$ and $t_2=(1-(\frac{2\cdot
  M_{\mathrm{W}}}{200})^2)/ (1-(\frac{2\cdot M_{\mathrm{W}}}{\sqrt{s}})^2)$. In
  the threshold region, a 2\% uncertainty is assigned.

\bibitem{4f_bib:gentle}
D.~Bardin, J.~Biebel, D.~Lehner, A.~Leike, A.~Olchevski and T.~Riemann,
  \CPC{104}{1997}{161}. See also~\cite{4f_bib:fourfrep}.\\ The \Gentle\
  cross-sections at 155--215 GeV have been kindly provided by Eric Lan\c{c}on
  and Anne Ealet.

\bibitem{4f_bib:koralw}
M. Skrzypek, S. Jadach, M.~Martinez, W.~P{\l}aczek, Z.~W\c{a}s, Phys. Lett.
  {\bf B372} (1996) 289. \\ S. Jadach, W. P{\l}aczek, M. Skrzypek, Z. W\c{a}s,
  Comput. Phys. Commun. {\bf 94} (1996) 216. \\ S. Jadach, W. P{\l}aczek, M.
  Skrzypek, B.F.L. Ward, Z. W\c{a}s, Comput. Phys. Commun. {\bf 119} (1999)
  272. \\ S. Jadach, W. P{\l}aczek, M. Skrzypek, B.F.L. Ward, Z. W\c{a}s,
  preprint hep-ph/0104049, submitted to Comput. Phys. Commun.\\ The ``\KoralW''
  cross-sections at 155--215 GeV have been kindly provided by the authors. They
  have actually been computed using \YFSWW~\cite{4f_bib:yfsww}, switching off
  non-leading $\oa$ radiative corrections and the screening of the Coulomb
  correction, to reproduce the calculation from \KoralW.

\bibitem{bib:pdg02}
Particle Data Group Collaboration, K.~Hagiwara et~al., Phys. Rev. {\bf D66}
  (2002) 010001.

\bibitem{4f_bib:delzz}
\Delphi\ Collaboration, CERN--EP/2003-009, published in Eur. Phys. J C30 (2003)
  447.

\bibitem{4f_bib:ltrzz183a}
\Ltre\ Collaboration, M.~Acciarri \etal, \PLB{450}{1999}{281}. The Z-pair
  cross-section at 183 GeV therein follows the \Ltre\ definition: the
  corresponding {\sc NC02} cross-section is given in~\cite{4f_bib:ltrzz183b}.

\bibitem{4f_bib:ltrzz183b}
\L3\ Collaboration, \L3\ Note 2366, submitted to the Winter 1999 Conferences.

\bibitem{4f_bib:ltrzz189}
\Ltre\ Collaboration, M.~Acciarri \etal, \PLB{465}{1999}{363}.

\bibitem{4f_bib:ltrzz1999}
\Ltre\ Collaboration, M.~Acciarri \etal, \PLB{497}{2001}{23}.

\bibitem{4f_bib:ltrzzsc03}
\Ltre\ Collaboration, \L3\ Note 2805, EPS 2003 abstract 228.

\bibitem{4f_bib:opazz189}
\Opal\ Collaboration, G.~Abbiendi \etal, \PLB{476}{2000}{256}.

\bibitem{4f_bib:opazz}
\Opal\ Collaboration, CERN--EP/2003--049 submitted to Eur. Phys. J. C.

\bibitem{4f_bib:alezz189}
\Aleph\ Collaboration, R.~Barate \etal, \PLB{469}{1999}{287}.

\bibitem{4f_bib:alezzsc01}
\Aleph\ Collaboration, \Aleph\ 2001--006 \Conf\ 2001--003, submitted to the
  Winter 2001 Conferences.

\bibitem{4f_bib:yfszz}
S. Jadach, W. P{\l}aczek, B.F.L. Ward, \PRD{56}{1997}{6939}.

\bibitem{4f_bib:zzto}
G.~Passarino, in~\cite{4f_bib:fourfrep}.

\bibitem{4f_bib:alezeesc03}
\Aleph\ Collaboration, \Aleph\ 2003--012 \Conf\ 2003--008, EPS 2003 abstract
  176.

\bibitem{4f_bib:delswsc03}
\Delphi\ Collaboration, \Delphi\ 2003-055 \Conf\ 675, EPS 2003 abstract 247.

\bibitem{4f_bib:ltrzee}
\Ltre\ Collaboration, CERN--EP/2002--103 accepted for publication by Phys.
  Lett. B.

\bibitem{4f_bib:grace}
J.~Fujimoto \etal, \CPC{100}{1997}{74}.\\ Y.~Kurihara \etal,
  Prog.~Theor.~Phys.~{\bf 103}~(2000)~1199.\\ The \Grace\ cross-sections at
  160--210 GeV have been kindly computed by R.~Tanaka.

\bibitem{4f_bib:wphact}
E.~Accomando and A.~Ballestrero, \CPC{99}{1997}{270}.\\ E.~Accomando,
  A.~Ballestrero and E. Maina, \CPC{150}{2003}{166}.\\ The \WPHACT\
  cross-sections at 160--210 GeV have been kindly provided by A.~Ballestrero.

\bibitem{4f_bib:delwwg}
\Delphi\ Collaboration, CERN--EP/2003-045 submitted to Eur. Phys. J. C.

\bibitem{4f_bib:ltrwwg}
\Ltre\ Collaboration, M.~Acciarri \etal, \PLB{527}{2002}{39}.

\bibitem{4f_bib:opawwg}
\Opal\ Collaboration, CERN--EP/2003--043 submitted to Phys. Lett. B.

\bibitem{4f_bib:eewwg}
J.~W.~Stirling and A.~Werthenbach, \EPJC{14}{2000}{103}.

\bibitem{gc_bib:LEP2YR}
G. Gounaris \etal, in {\em Physics at LEP 2}, Report CERN 96-01 (1996), eds G.
  Altarelli, T. Sj{\"o}strand, F. Zwirner, Vol. 1, p. 525.

\bibitem{gc_bib:Montagna:2001ej}
G.~Montagna, M.~Moretti, O.~Nicrosini, M.~Osmo, and F.~Piccinini, Phys. Lett.
  {\bf B515} (2001) 197--205.

\bibitem{gc_bib:denner}
A. Denner \etal, Eur. Phys. J. {\bf C 20} (2001) 201.

\bibitem{gc_bib:budapest01}
The LEP-TGC combination group, LEPEWWG/TGC/2001-03, September 2001.

\bibitem{common_bib:racoonww}
A.~Denner, S.~Dittmaier, M.~Roth and D.~Wackeroth, Nucl. Phys. {\bf B560}
  (1999) 33.\\ A.~Denner, S.~Dittmaier, M.~Roth and D.~Wackeroth, Nucl. Phys.
  {\bf B587} (2000) 67.\\ A.~Denner, S.~Dittmaier, M.~Roth and D.~Wackeroth,
  \PLB{475}{2000}{127}.\\ A.~Denner, S.~Dittmaier, M.~Roth and D.~Wackeroth,
  hep-ph/0101257.\\ The \RacoonWW\ cross-sections at 155--215 GeV have been
  kindly provided by the authors.

\bibitem{common_bib:yfsww}
S. Jadach, W. P{\l}aczek, M. Skrzypek, B.F.L. Ward, Phys. Rev. {\bf D54} (1996)
  5434. \\ S. Jadach, W. P{\l}aczek, M. Skrzypek, B.F.L. Ward, Z. W\c{a}s,
  Phys. Lett. {\bf B417} (1998) 326. \\ S. Jadach, W. P{\l}aczek, M. Skrzypek,
  B.F.L. Ward, Z. W\c{a}s, Phys. Rev. {\bf D61} (2000) 113010. \\ S. Jadach, W.
  P{\l}aczek, M. Skrzypek, B.F.L. Ward, Z. W\c{a}s, preprint CERN-TH/2000-337,
  hep-ph/0007012; submitted to Phys. Lett. B.\\ S. Jadach, W. P{\l}aczek, M.
  Skrzypek, B.F.L. Ward, Z. W\c{a}s, Comput. Phys. Commun. {\bf 140} (2001)
  432.\\ The \YFSWW\ cross-sections at 155--215 GeV have been kindly provided
  by the authors.

\bibitem{gc_bib:ALEPH-cTGC}
ALEPH Collaboration, {\em Measurement of Triple Gauge-Boson Couplings in
  $e^+e^-$ collisions from 183 to 209GeV}, ALEPH 2003-015 CONF 2003-011.

\bibitem{gc_bib:DELPHI-cTGC}
DELPHI Collaboration, {\em Measurement of Charged Trilinear Gauge Boson
  Couplings }, DELPHI 2003-051 (July 2003) CONF-671.

\bibitem{gc_bib:L3-cTGC}
L3 Collaboration, {\em Preliminary Results on the Measurement of
  Triple-Gauge-Boson Couplings of the W Boson at LEP}, L3 Note 2820 (September
  2003).

\bibitem{gc_bib:OPAL-cTGC3}
\Opal\ Collaboration, {\em Measurement of charged current triple gauge boson
  couplings using W pairs at LEP}, submitted to Eur. Phys. J. C., CERN-EP
  2003-042, hep-ex/0308067.

\bibitem{gc_bib:QGC-Belanger}
G. B\'elanger \etal, Eur. Phys. J. {\bf C 13} (2000) 283.

\bibitem{gc_bib:GAEMERS}
K. Gaemers and G. Gounaris, Z. Phys. {\bf C 1} (1979) 259.

\bibitem{gc_bib:Hagiwara1987vm}
K.~Hagiwara, R.~D. Peccei, D.~Zeppenfeld, and K.~Hikasa, Nucl. Phys. {\bf B282}
  (1987) 253.

\bibitem{gc_bib:HAGIWARA}
K. Hagiwara, S. Ishihara, R. Szalapski, and D. Zeppenfeld, Phys. Lett. {\bf B
  283} (1992) 353; \\ K. Hagiwara, S. Ishihara, R. Szalapski, and D.
  Zeppenfeld, Phys. Rev. {\bf D 48} (1993) 2182; \\ K.~Hagiwara, T.~Hatsukano,
  S.~Ishihara and R.~Szalapski, Nucl. Phys. {\bf B 496} (1997) 66.

\bibitem{gc_bib:BILENKY}
M. Bilenky, J.L. Kneur, F.M. Renard and D. Schildknecht, Nucl. Phys. {\bf B
  409} (1993) 22;\\ M. Bilenky, J.L. Kneur, F.M. Renard and D. Schildknecht,
  Nucl. Phys. {\bf B 419} (1994) 240.

\bibitem{gc_bib:KUSS}
I. Kuss and D. Schildknecht, Phys. Lett. {\bf B 383} (1996) 470.

\bibitem{gc_bib:PAPADOPOULOSCP}
G. Gounaris and C.G. Papadopoulos, DEMO-HEP-96/04, THES-TP 96/11,
  hep-ph/9612378.

\bibitem{gc_bib:moriond01}
The LEP-TGC combination group, LEPEWWG/TGC/2001-01, March 2001.

\bibitem{gc_bib:Gounaris2000tb}
G.~J. Gounaris, J.~Layssac, and F.~M. Renard, Phys. Rev. {\bf D62} (2000)
  073013.

\bibitem{gc_bib:QGC-BelBou}
G. B\'elanger and F. Boudjema, Phys. Lett. {\bf B 288} (1992) 201.

\bibitem{gc_bib:ALEPH-nTGC}
ALEPH Collaboration, {\em Limits on anomalous neutral gauge couplings using
  data from ZZ and Z$\gamma$ production between 183-208 GeV}, ALEPH 2001-061
  (July 2001) CONF 2001-041.

\bibitem{gc_bib:DELPHI-nTGC}
DELPHI Collaboration, {\em Study of Trilinear Gauge Boson Couplings ZZZ,
  $ZZ\gamma$ and $Z\gamma\gamma$}, DELPHI 2001-097 (July 2001) CONF 525.

\bibitem{gc_bib:L3-hTGC}
L3 Collaboration, M. Acciari \etal, Phys. Lett. {\bf B 436} (1999) 187;\\ L3
  Collaboration, M. Acciari \etal, Phys. Lett. {\bf B 489} (2000) 55;\\ L3
  Collaboration, {\em Search for anomalous ZZg and Zgg couplings in the process
  ee$\rightarrow$Zg at LEP}, L3 Note 2672 (July 2001).

\bibitem{gc_bib:OPAL-hTGC}
OPAL Collaboration, G. Abbiendi \etal, Eur. Phys. J. {\bf C 17} (2000) 13.

\bibitem{gc_bib:L3-fTGC}
See references~\cite{4f_bib:ltrzz183a, 4f_bib:ltrzz189,4f_bib:ltrzzsc03}.

\bibitem{gc_bib:OPAL-fTGC}
See references~\cite{4f_bib:opazz189,4f_bib:opazz}.

\bibitem{gc_bib:ALEPH-QGC}
ALEPH Collaboration, {\em Constraints on Anomalous Quartic Gauge Boson
  Couplings}, ALEPH 2003-009 CONF 2003-006.

\bibitem{gc_bib:L3-QGC}
L3 Collaboration, {\em The e$^+$e$^- \rightarrow {\mathrm{Z} \gamma \gamma}
  \rightarrow {\mathrm{q} \bar{\mathrm{q}} \gamma \gamma} $ Reaction at LEP and
  Constraints on Anomalous Quartic Gauge Boson Couplings}, Phys. Lett. {\bf B
  540} (2002) 43.

\bibitem{gc_bib:OPAL-QGC}
\Opal\ Collaboration, {\em Constraints on Anomalous Quartic Gauge Boson
  Couplings using Acoplanar Photon Pairs et LEP-2}, \Opal\ Physics Note PN510.

\bibitem{gc_bib:LPA_A-B}
R. Bruneli{\`e}re \etal, Phys. Lett. {\bf B 533} (2002) 75 and references
  therein.

\bibitem{gc_bib:Alcaraz}
J. Alcaraz, {\em A proposal for the combination of TGC measurements}, L3 Note
  2718.

\bibitem{gc_bib:renaud}
R. Bruneli{\`e}re, {\em Tests on the LEP TGC combination procedures}, ALEPH
  2002-008 PHYS-2002-007 (2002).

\bibitem{gc_bib:klein}
O.Klein, {\em On the Theory of Charged Fields}, New Theories in Physics,
  Proceedings, Warsaw, 1938; reprinted in: Surveys of High Energ. Phys. {\bf 5}
  (1986) 269.

\bibitem{gc_bib:maiani}
L.Maiani and P.M.Zerwas, {\em W Static ELM Parameters}, Memorandum to the TGC
  Combination Group (1998).

\bibitem{bib:cr:GPZ}
Gosta Gustafson, Ulf Pettersson, and P.~M. Zerwas, Phys. Lett. {\bf B209}
  (1988) 90.

\bibitem{bib:cr:SK_MODELS}
Torbjorn Sjostrand and Valery~A. Khoze, Z. Phys. {\bf C62} (1994) 281--310.

\bibitem{bib:cr:ARIADNECR_MODEL}
Leif Lonnblad, Z. Phys. {\bf C70} (1996) 107--114.

\bibitem{HERWIG6}
G.~Corcella et~al., JHEP {\bf 01} (2001) 010.

\bibitem{bib:cr:GH_MODEL}
Gosta Gustafson and Jari Hakkinen, Z. Phys. {\bf C64} (1994) 659--664.

\bibitem{bib:cr:EG_MODEL}
John~R. Ellis and Klaus Geiger, Phys. Rev. {\bf D54} (1996) 1967--1990.

\bibitem{bib:cr:RATHSMAN}
Johan Rathsman, Phys. Lett. {\bf B452} (1999) 364--371.

\bibitem{bib:cr:OPAL_CR}
OPAL Collaboration, {\it Colour Reconnection Studies in \eeWW\ at
  $\roots=$189~GeV}, OPAL PN417.

\bibitem{bib:cr:DELPHI_CR}
DELPHI Collaboration, P.~Abreu et~al., Eur. Phys. J. {\bf C18} (2000) 203--228.

\bibitem{bib:cr:ALEPH_CR}
ALEPH Collaboration, {\it Preliminary Charged Particle Multiplicity in
  $\ee\rightarrow$ W-pairs}, ALEPH 2000-058 CONF 2000-038.

\bibitem{bib:cr:L3_CR}
L3 Collaboration, {\it Search for Colour Reconnection Effects in
  $\eeWW\rightarrow\textrm{hadrons}$}, L3 Note 2560.

\bibitem{bib:cr:SK_HEAVYHAD}
Valery~A. Khoze and Torbjorn Sjostrand, Eur. Phys. J. {\bf C6} (1999) 271--284.

\bibitem{bib:cr:OPAL_HEAVYHAD}
OPAL Collaboration, {\it Investigation of Colour Reconnection via Heavy
  Particle Production in \eeWW}, OPAL PN412.

\bibitem{bib:cr:JADE_STRING2}
JADE Collaboration, W.~Bartel et~al., Phys. Lett. {\bf B101} (1981) 129.

\bibitem{bib:cr:JADE_STRING3}
JADE Collaboration, W.~Bartel et~al., Z. Phys. {\bf C21} (1983) 37.

\bibitem{bib:cr:JADE_STRING4}
JADE Collaboration, W.~Bartel et~al., Phys. Lett. {\bf B134} (1984) 275.

\bibitem{bib:cr:JADE_STRING5}
JADE Collaboration, W.~Bartel et~al., Phys. Lett. {\bf B157} (1985) 340.

\bibitem{bib:cr:TPC2GAM_STRING1}
TPC/Two Gamma Collaboration, H.~Aihara et~al., Z. Phys. {\bf C28} (1985) 31.

\bibitem{bib:cr:TPC2GAM_STRING2}
TPC/Two Gamma Collaboration, H.~Aihara et~al., Phys. Rev. Lett. {\bf 57} (1986)
  945.

\bibitem{bib:cr:TASSO_STRING1}
TASSO Collaboration, M.~Althoff et~al., Z. Phys. {\bf C29} (1985) 29.

\bibitem{bib:cr:pflow1}
L3 Collaboration, {\it Colour Reconnection Studies in \eeWW\ Events at
  $\roots=189$~GeV}, L3 Note 2406.

\bibitem{bib:cr:pflow2}
D.\ Duchesneau, {\it New Method Based on Energy and Particle Flow in
  $\eeWW\rightarrow$ hadron Events for Colour Reconnection Studies},
  LAPP-EXP-2000-02.

\bibitem{bib:cr:OXFORD_WS}
A.~Ballestrero et~al., J. Phys. {\bf G24} (1998) 365--403.

\bibitem{bib:cr:ALEPH_PF}
ALEPH Collaboration, {\it Colour Reconnection Studies Using Particle Flow
  Between W Bosons at $\roots=$189--208~GeV}, ALEPH 2002-020 CONF 2002-009.

\bibitem{bib:cr:DELPHI_PF}
DELPHI Collaboration, {\it Update of the Investigation of Colour Reconnection
  in WW Pairs Using Particle Flow}, DELPHI 2002-047 CONF 581.

\bibitem{bib:cr:L3_PF}
L3 Collaboration, {\it Search for Colour Reconnection Effects in
  $\eeWW\rightarrow\textrm{hadrons}$ Through Particle-Flow Studies at
  $\roots=$189--208~GeV}, L3 Note 2748.

\bibitem{bib:cr:OPAL_PF}
OPAL Collaboration, {\it Colour Reconnection Studies in \eeWW\ at
  $\roots=$189--208~GeV Using Particle Flow}, OPAL PN506.

\bibitem{ALEPH-MW}
ALEPH Collaboration, R.~Barate et~al., Eur. Phys. J. {\bf C17} (2000) 241--261.

\bibitem{OPAL-MW}
OPAL Collaboration, G.~Abbiendi et~al., Phys. Lett. {\bf B507} (2001) 29--46.

\bibitem{bib:fsi:KORALW}
S.~Jadach, W.~Placzek, M.~Skrzypek, B.~F.~L. Ward, and Z.~Was, Comput. Phys.
  Commun. {\bf 140} (2001) 475--512.

\bibitem{ARIADNE}
L.\,Lonnblad, Comp. Phys. Comm. {\bf 71} (1992) 15.

\bibitem{be:DELPHI03}
DELPHI Coll., {\em Bose-Einstein Correlations in $W^+ W^-$ events at LEP2},
  DELPHI 2003-020-CONF-640.

\bibitem{bib:cr:ALEPH_BEC}
ALEPH Collaboration, {\it Further Studies on Bose-Einstein Correlations in
  W-pair Decays}, ALEPH 2001-064 CONF 2001-044.

\bibitem{be:L302}
L3 Coll., {\em Measurement of Bose-Einstein Correlations in $e^+ e^-
  \rightarrow W^+ W^-$ Events at LEP}, Phys. Lett. B547 (2002) 139.

\bibitem{bib:cr:OPAL_BEC}
OPAL Collaboration, {\it Bose-Einstein Correlations in \eeWW\ Events at 172,
  183 and 189 GeV}, OPAL PN393.

\bibitem{bib:LEP2_MCWS}
A.~Ballestrero et~al., hep-ph/0006259 (2000).

\bibitem{be:chekanov}
S.V. Chekanov, E.A. De Wolf, W. Kittel, Eur.Phys.J. C6 (1999) 403.

\bibitem{be:DELPHI97}
P. Abreu et al. (DELPHI Coll.), Phys.Lett. B401 (1997) 181.

\bibitem{be:ALEPH00}
ALEPH Coll., {\em Bose-Einstein Correlations in W-pair decays}, Phys.Lett.B478
  (2000) 50.

\bibitem{be:ALEPH03}
ALEPH Coll., {\em Sudies on Bose-Einstein correlations in W-pair decays}, ALEPH
  note ALEPH 2003-013 CONF 2003-009.

\bibitem{be:OPAL03}
OPAL Coll., {\em Study of Bose-Einstein Correlations in $e^+ e^- \rightarrow
  W^+ W^-$ Events at LEP}, OPAL Physics Note PN523.

\bibitem{be:PYTHIA57}
T. Sj{\"o}strand: PYTHIA 5.7 and JETSET 7.4, Computer Physics Commun. 82
  (1994)74.

\bibitem{be:LoSj}
L. L{\"o}nnblad and T. Sj{\"o}strand,Eur.Phys.J. C2 (1998) 165-180.

\bibitem{be:lep-be}
LEPWW FSI group http://lepewwg.web.cern.ch/LEPEWWG/lepww/fsi.html.

\bibitem{be:LEPW}
This report, section {\em W-Boson Mass and Width at LEP-II}.

\bibitem{common_bib:adloww161}
\Aleph\ Collaboration, R.~Barate \etal, Phys.~Lett.~{\bf B401} (1997) 347.\\
  \Delphi\ Collaboration, P.~Abreu \etal, Phys.~Lett.~{\bf B397} (1997) 158.\\
  \Ltre\ Collaboration, M.~Acciarri \etal, Phys.~Lett. {\bf B398} (1997) 223.\\
  \Opal\ Collaboration, K.~Ackerstaff \etal, Phys.~Lett. {\bf B389} (1996) 416.

\bibitem{mw:bib:ALEPHRevised}
ALEPH Collaboration, {\it Measurement of the W Mass in $\epem$ Collisions at
  $\roots$ between 183 and 209~$\GeV$}, ALEPH note 2003-005 CONF 2003-003.

\bibitem{common_bib:delww172}
DELPHI Collaboration, P.~Abreu \etal, Eur.~Phys.~J. {\bf C2} (1998) 581.

\bibitem{mw:bib:D-mw183}
DELPHI Collaboration, P.~Abreu et~al., Phys. Lett. {\bf B462} (1999) 410--424.

\bibitem{mw:bib:D-mw189}
DELPHI Collaboration, P.~Abreu et~al., Phys. Lett. {\bf B511} (2001) 159--177.

\bibitem{mw:bib:D-mw20X}
DELPHI Collaboration, {\it{Measurement of the mass and width of the W Boson in
  $\epem$ collisions at $\roots = 192-209~\GeV$}}, DELPHI 2001-103 CONF 531.

\bibitem{common_bib:ltrww172}
L3 Collaboration, M.~Acciarri \etal, Phys. Lett. {\bf B407} (1997) 419.

\bibitem{mw:bib:L-mw183}
L3 Collaboration, M.~Acciarri et~al., Phys. Lett. {\bf B454} (1999) 386--398.

\bibitem{mw:bib:L-mw189}
L3 Collaboration, {\it{Preliminary Results on the Measurement of Mass and Width
  of the W Boson at LEP}}, L3 Note 2377, March 1999.

\bibitem{mw:bib:L-mw19X}
L3 Collaboration, {\it{Preliminary Results on the Measurement of Mass and Width
  of the W Boson at LEP}}, L3 Note 2575, July 2000.

\bibitem{mw:bib:L-mw20X}
L3 Collaboration, {\it{Preliminary Results on the Measurement of Mass and Width
  of the W Boson at LEP}}, L3 Note 2637, February 2001.

\bibitem{common_bib:opaww172}
OPAL Collaboration, K.~Ackerstaff \etal, Eur.~Phys.~J. {\bf C1} (1998) 395.

\bibitem{mw:bib:O-mw183}
OPAL Collaboration, G.~Abbiendi et~al., Phys. Lett. {\bf B453} (1999) 138--152.

\bibitem{mw:bib:O-mw189}
OPAL Collaboration, G.~Abbiendi et~al., Phys. Lett. {\bf B507} (2001) 29--46.

\bibitem{mw:bib:O-mw19X}
OPAL Collaboration, {\it{Measurement of the Mass of the W Boson in \epem\
  annihilations at 192-202 GeV}}, OPAL Physics Note PN422 (updated July 2000).

\bibitem{mw:bib:O-mwlvlv}
OPAL Collaboration, {\it{Determination of the W mass in the fully leptonic
  channel using an unbinned maximum likelihood fit}}, OPAL Physics Note PN480,
  July 2001.

\bibitem{mw:bib:energy}
LEP Energy Working Group, LEPEWG 01/01, March 2001.

\bibitem{mw:bib:CRcomb}
LEP W Working Group, LEPEWWG/FSI/2002-01, July 2002.

\bibitem{mw:bib:ski}
Torbjorn Sj{\"o}strand and Valery~A. Khoze, Z. Phys. {\bf C62} (1994) 281--310.

\bibitem{mw:bib:LUBOEI}
L.~L{\"o}nnblad and T.~Sj{\"o}strand, Eur. Phys. J. {\bf C2} (1998) 165.

\bibitem{ref:CHARMIIgn}
CHARM II Collaboration, P.~Vilain \etal, Phys.~Lett. {\bf B335} (1994) 246.

\bibitem{bib-UA2MW}
UA2 Collaboration, J.~Alitti \etal, Phys.~Lett. {\bf B276} (1992) 354.

\bibitem{bib-CDFMW1}
CDF Collaboration, F.~Abe \etal, Phys.~Rev.~Lett. {\bf 65} (1990) 2243;\\ CDF
  Collaboration, F.~Abe \etal, Phys.~Rev. {\bf D43} (1991) 2070.

\bibitem{bib-CDFMW2}
CDF Collaboration, F.~Abe \etal, Phys.~Rev.~Lett. {\bf 75} (1995) 11;\\ CDF
  Collaboration, F.~Abe \etal, Phys.~Rev. {\bf D52} (1995) 4784.\\ A.~Gordon,
  talk presented at XXXIInd Rencontres de Moriond, Les Arcs, 16-22 March 1997,
  to appear in the proceedings.

\bibitem{bib-D0MW}
D\O\ Collaboration, S.~Abachi \etal, \PRL {\bf84} (2000) 222.

\bibitem{bib-CDFGW}
CDF Collaboration, T.~Affolder et~al., Phys. Rev. Lett. {\bf 85} (2000)
  3347--3352.

\bibitem{bib-D0GW}
D0 Collaboration, V.~M. Abazov et~al., Phys. Rev. {\bf D66} (2002) 032008.

\bibitem{bib-MWGWAVE-02}
Combination of CDF and D\O\ Results on W Boson Mass and Width, Tevatron
  Electroweak Working Group and the CDF and D\O\ Collaborations, CDF Note 5888,
  D\O\ Note 3963, FERMILAB-FN-0716, July 2002.

\bibitem{bib-topCDF}
CDF Collaboration, W.~Yao, {\it t Mass at CDF}, talk presented at ICHEP 98,
  Vancouver, B.C., Canada, 23-29 July, 1998.

\bibitem{bib-topD0}
D\O{} Collaboration, B.~Abbott \etal, \PRL{80} {(1998)} {2063}.

\bibitem{bib-Tevatop}
The Top Averaging Group, L.\,Demortier \etal, for the CDF and D\O{}
  Collaborations, FERMILAB-TM-2084 (1999).

\bibitem{bib-NuTeV-final}
NuTeV Collaboration, G.P. Zeller \etal, Phys. Rev. Lett. {\bf 88} (2002)
  091802.

\bibitem{QWCs:exp:1}
C.~S. Wood et~al., Science {\bf 275} (1997) 1759.

\bibitem{QWCs:exp:2}
S.~C. Bennett and C.~E. Wieman, Phys. Rev. Lett. {\bf 82} (1999) 2484--2487.

\bibitem{QWCs:theo:2003}
M. Yu. Kuchiev and V. V. Flambaum, {\em Radiative corrections to parity
  non-conservation in atoms}, hep-ph/0305053.

\bibitem{bib-Gmu}
T.{} van Ritbergen and R.G. Stuart, Phys.{} Rev.{} Lett.{} {\bf 82} (1999) 488.

\bibitem{bib-BP01}
H.~Burkhardt and B.~Pietrzyk, Phys.~Lett. {\bf B513} (2001) 46.

\bibitem{ref:sld-s99}
SLD Collaboration, J.~Brau, {\it Electroweak Precision Measurements with
  Leptons}, talk presented at EPS-HEP-99, Tampere, Finland, 15-21 July 1999.

\bibitem{bib-PCLI}
{\it Reports of the working group on precision calculations for the Z
  resonance}, eds.~D.~Bardin, W.~Hollik and G.~Passarino, CERN Yellow Report
  95-03, Geneva, 31 March 1995.

\bibitem{bib-twoloop}
G.~Degrassi, S.~Fanchiotti and A.~Sirlin, Nucl.~Phys.~{\bf B351} (1991) 49;\\
  G.~Degrassi and A.~Sirlin, Nucl.~Phys.~{\bf B352} (1991) 342;\\ G.~Degrassi,
  P.~Gambino and A.~Vicini, Phys.~Lett.~{\bf B383} (1996) 219;\\ G.~Degrassi,
  P.~Gambino and A.~Sirlin, Phys.~Lett.~{\bf B394} (1997) 188;\\ G.~Degrassi
  and P.~Gambino, Nucl.~Phys.~{\bf B567} (2000) 3.

\bibitem{bib-QCDEW}
A.~Czarnecki and J.~K{\"u}hn, Phys.~Rev.~Lett.~{\bf 77} (1996) 3955;\\
  R.~Harlander, T.~Seidensticker and M.~Steinhauser, Phys.~Lett.~{\bf B426}
  (1998) 125.

\bibitem{bib-SMNEW}
Electroweak libraries:\\ ZFITTER: see Reference~\citen{ref:ZFITTER};\\ BHM
  (G.~Burgers, W.~Hollik and M.~Martinez): W.~Hollik, Fortschr.~Phys. {\bf38}
  (1990) 3, 165; M.~Consoli, W.~Hollik and F.~Jegerlehner: Proceedings of the
  Workshop on Z physics at LEP I, CERN Report 89-08 Vol.I,7 and G.~Burgers,
  F.~Jegerlehner, B.~Kniehl and J.~K{\"u}hn: the same proceedings, CERN Report
  89-08 Vol.I, 55; \\ TOPAZ0 Version 4.0i: G.~Montagna, O.~Nicrosini,
  G.~Passarino, F.~Piccinni and R.~Pittau, Nucl.~Phys. {\bf B401} (1993) 3;
  Comp.~Phys.~Comm. {\bf 76} (1993) 328.\\ These computer codes have been
  upgraded by including the results of~\cite{bib-PCLI} and references therein.
  ZFITTER and TOPAZ0 have been further updated using the results of
  references~\citen{bib-twoloop} and~\citen{bib-QCDEW}. See, D.~Bardin and
  G.~Passarino, {\it Upgrading of Precision Calculations for Electroweak
  Observables}, CERN-TH/98-92, hep-ph/9803425.

\bibitem{FHWW-f2l-MW}
A. Freitas, W. Hollik, W. Walter and G. Weiglein, Phys. Lett. {\bf B495} (2000)
  338.

\bibitem{BGPWprivate}
D.~Bardin, P.~Gambino, G.~Passarino, G.~Weiglein, private communication, spring
  2001.

\bibitem{bib-SMALFAS}
T.~Hebbeker, M.~Martinez, G.~Passarino and G.~Quast, Phys.~Lett. {\bf B331}
  (1994) 165;\\ P.A.~Raczka and A.~Szymacha, Phys. Rev. {\bf D54} (1996)
  3073;\\ D.E.~Soper and L.R.~Surguladze, Phys. Rev. {\bf D54} (1996) 4566.

\bibitem{bib-alphalept}
M.~Steinhauser, Phys.~Lett.~{\bf B429} (1998) 158.

\bibitem{bib-JEG2}
S.~Eidelmann and F.~Jegerlehner, Z.~Phys. {\bf C67} (1995) 585.

\bibitem{BES_01}
BES Collaboration, J.~Z. Bai et~al., Phys. Rev. Lett. {\bf 88} (2002) 101802.

\bibitem{bib-Swartz}
M.~L.~Swartz, Phys.~Rev. {\bf D53} (1996) 5268.

\bibitem{bib-Zeppe}
A.D.~Martin and D.~Zeppenfeld, Phys.~Lett. {\bf B345} (1994) 558.

\bibitem{bib-Alemany}
R.~Alemany, \etal, Eur.~Phys.~J.~{\bf C2} (1998) 123.

\bibitem{bib-Davier}
M.~Davier and A.~H{\"o}cker, Phys.~Lett.~{\bf B419} (1998) 419.

\bibitem{bib-alphaKuhn}
J.H.~K{\"u}hn and M.~Steinhauser, Phys.{} Lett.{} {\bf B437} (1998) 425.

\bibitem{bib-Erler}
J.~Erler, Phys.~Rev.~{\bf D59}, (1999) 054008.

\bibitem{bib-ADMartin}
A.~D.~Martin, J.~Outhwaite and M.~G.~Ryskin, Phys.~Lett. {\bf B492} (2000) 69.

\bibitem{bib-jeger99}
F. Jegerlehner, {\it Hadronic Effects in (g-2) and $\alpha_{QED}$(M$_Z$) :
  Status and Perspectives}, Proc. of Int. Symp. on Radiative Corrections,
  Barcelona, Sept. 1998, page 75.

\bibitem{bib-TY0102}
J.~F. de~Troconiz and F.~J. Yndurain, Phys. Rev. {\bf D65} (2002) 093002.

\bibitem{common_bib:pdg2000}
Particle Data Group, D.E.~Groom \etal, \EPJ{15} {(2000)} {1}.

\bibitem{Siggi-Bethke-alpha-s}
S.~Bethke, hep-ex/0004021, J. Phys. G26 (2000) R27.

\bibitem{ref:TOPAZ0}
G.~Montagna \etal, Comput. Phys. Commun. {\bf 117} (1999) 278;\\ {\tt
  http://www.to.infn.it/$\sim$giampier/topaz0.html~}.

\bibitem{ref:LEP-HIGGS}
The LEP Collaborations and the LEP Working Group for Higgs Boson Searches, {\it
  Search for the Standard Model Higgs Boson at LEP}, CERN-EP/2003-011,
  submitted to PLB.

\bibitem{ref:groot2003}
N. de Groot (representing the SLD collaboration), Nucl. Phys. B (Proc.Suppl.)
  115 (2003) 267-271.

\end{thebibliography}
